# A holistic review on fatigue properties of additively manufactured metals

Min Yi∗, Wei Tang, Yiqi Zhu, Chenguang Liang, Ziming Tang, Yan Yin, Weiwei He, Shen Sun

*State Key Lab of Mechanics and Control for Aerospace Structures & Institute for Frontier Science & Key Lab for Intelligent Nano Materials and Devices of Ministry of Education & College of Aerospace Engineering, Nanjing University of Aeronautics and Astronautics (NUAA), Nanjing 210016, China*

**Abstract**

Additive manufacturing (AM) technology is undergoing rapid development and emerging as an advanced technique that can fabricate complex near-net shaped and light-weight metallic parts with acceptable strength and fatigue performance. A number of studies have indicated that the strength or other mechanical properties of AM metals are comparable or even superior to that of conventionally manufactured metals, but the fatigue performance is still a thorny problem that may hinder the replacement of currently used metallic components by AM counterparts when the cyclic loading and thus fatigue failure dominates. This article reviews the state-of-art published data on the fatigue properties of AM metals, principally including *S–N* data and fatigue crack growth data. The AM techniques utilized to generate samples in this review include powder bed fusion (e.g., EBM, SLM, DMLS) and directed energy deposition (e.g., LENS, WAAM). Further, the fatigue properties of AM metallic materials that involve titanium alloys, aluminum alloys, stainless steel, nickel-based alloys, magnesium alloys, and high entropy alloys, are systematically overviewed. In addition, summary figures or tables for the published data on fatigue properties are presented for the above metals, the AM techniques, and the influencing factors (manufacturing parameters, e.g., built orientation, processing parameter, and post-processing). The effects of build direction, particle, geometry, manufacturing parameters, post-processing, and heat-treatment on fatigue properties, when available, are provided and discussed. The fatigue performance and main factors affecting the fatigue behavior of AM metals are finally compared and critically analyzed, thus potentially providing valuable guidance for improving the fatigue performance of AM metals.

## 1. Introduction

Additive manufacturing (AM), an incremental layer-by-layer manufacturing, plays an essential ingredient in the Era of Industry 4.0 [1], which has brought great potential and development in the advanced manufacturing industry and metals [2]. Wherein metallic additive manufacturing (MAM) has recently attracted intensive attentions, because AM metallic components are generally used to substitute for the conventional metallic materials of intelligent manufacturing, rail transit, aerospace engineering, defense technology, etc. [2–6]. The MAM techniques include powder bed fusion (e.g., EBM, SLM, DMLS) and directed energy deposition (e.g., LENS, WAAM) depending on powder forming process [7, 8]. Today, it has become possible to reliably manufacture metallic parts with certain AM processes and for plenty of materials, including steel, aluminium, titanium, magnesium, etc. [9]. As well as, the static strength of AM metals is generally demonstrated to be comparable or even superior to that of conventional manufactured metals. But the fatigue properties under cyclic loading show wide dispersion resulting from various AM processes and are still a knotty issue that will hamper the engineering application of AM metallic parts when the fatigue loading prevails [7, 10–14]. Therefore, the fatigue performance is critical for using and designing, the research on fatigue properties has received sustained attentions along with rise of MAM.

∗Corresponding author
 *Email address:* yimin@nuaa.edu.cn (Min Yi)

A number of reviews have been reported over the last decade that chronicle the mechanical and fatigue properties of AM metals [7, 15–19], as shown in Fig. 1. Further, an available data of mechanical properties in AM metals is summarized in Ref [7]. Similarly, the use of published fatigue data could provide a convenient foundation for fatigue and damage tolerance (F&DT) design by alleviating expensive and durable fatigue tests or machine learning [20–22]. Then, the processing-material-performance relationship is established based on the big data to guide the application design for AM metal.

In conclusion, the processing, microstructure, fatigue performance of AM titanium alloys [23], aluminum alloys [24], stainless steel 316L [25], Mg-based alloy [26] are systematically reviewed. As well as the very high cycle fatigue (VHCF) response of Ti-6Al-4V, AlSi10Mg, AlSi7Mg, AlSi12, IN718, and 316L specimens [16], HCF of SLM [27] and EBM [28] Ti-6Al-4V, the LCF of SLM 316L [29] produced through different AM processes are reported. Under the cyclic loading, the dislocation density accumulates with the increase of cycle times and the stress concentration emerges at the location of surface or defects, then plastic deformation occurs at the microcrack tips, finally microcracks extend to the entire surface of fatigue samples. As well as, the influence of internal eigen factors, e.g., defects [30–32], microstructure [33–36], and residual stress[Effect of residual stress on fatigue strength of 316L stainless steel produced by laser powder bed fusion process] [23] in AM metals, on fatigue performance are also discussed in detail. Mishurova et al. [37] found that the defects are the primary effect controlling the fatigue life, which also leads to the fatigue performance of AM metals is lower than that of castings or forgings. Decreasing the porosity or defects, optimizing the microstructure or refining the grain, and introducing or releasing the residual stress are good ideas to improve fatigue performance of AM metals.

On the other hand, the fatigue properties of AM metals are subject to external manufacturing factors, e.g., particle, geometry of specimen, processing parameters, post-processing, heat treatment (HT), surface, and temperature, as shown in Fig. 2. The fatigue behavior of electron beam melting (EBM) [15] and PBF [38] Ti-6Al-4V has been investigated and the effects of build orientation, surface roughness, and hot-isostatic pressing (HIP) are linked to the fatigue properties highlighting microstructure, defects, and failure mechanisms. No statistical difference was observed in the fatigue properties based *S-N* data of HIPed and as-built EBM Ti-6Al-4V under the same as-built rough surface condition and anisotropic fatigue behavior was found in as-built parts [15]. Owing to the various combinations of manufacturing parameters, post-processing, and heat treatment, a large scatter exists inevitably in mechanical and fatigue properties. Thus, the affluent combinations can greatly meet the actual or engineering requirements and further improve fatigue performance. For instance, Kahlin et al. [39] demonstrated that centrifugal finishing and shot peening can improve the fatigue strength of L-PBF and E-PBF Ti-6Al-4V, but laser polishing decreases the fatigue strength compared to material with rough as-built surfaces. The surface roughness [39, 40] plays a role in fatigue properties owing to the fatigue crack initiation is mostly located at the surface of part. Kumar et al. [41] designed four heat treatments to optimize microstructures, thus improving the damage tolerance of SLM Ti-6Al-4V by enhancing both the fracture toughness and fatigue crack growth resistance. Furthermore, the microstructures, defects, and residual stresses may be changed or eliminated via HIP and/or HT [7, 42]. These changes affect both the anisotropy of fatigue properties and their magnitude, as this article documents.

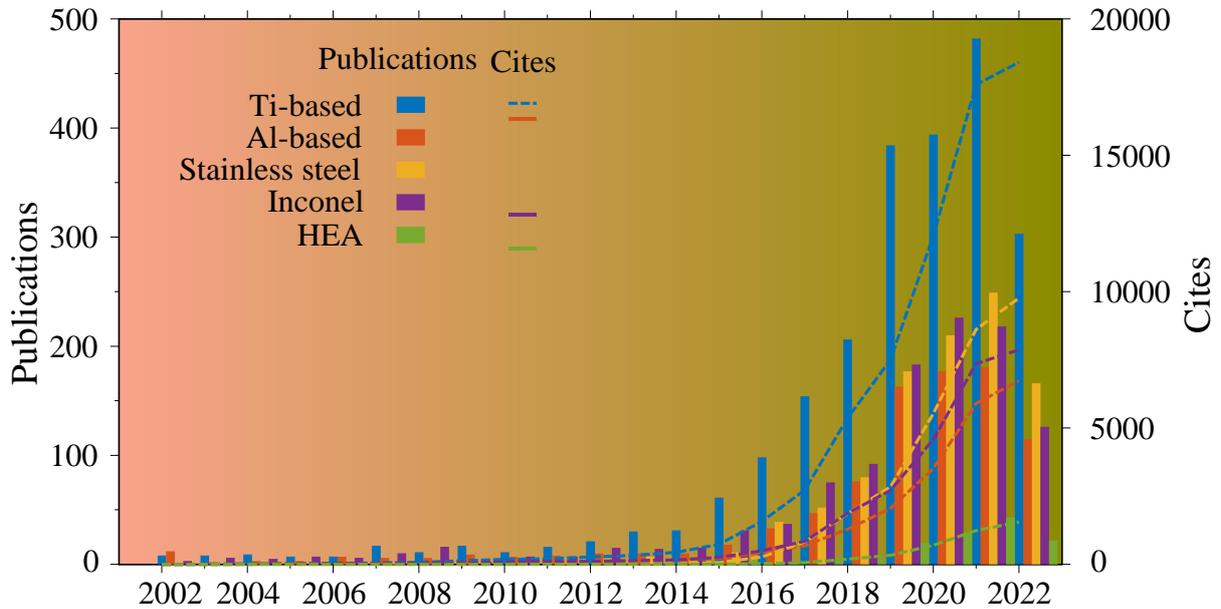

**Fig. 1.** The publications and cites for keywords of fatigue, additive manufacturing or 3D printing, metals (e.g., Titanium alloys, Aluminium alloys, and so on), data from web of science.

Because of such, it is still essential to introduce the impact of AM process parameters, particle, geometry, post-processing, HT and HIP on the microstructure, surface roughness, and defects to characterize the fatigue properties of AM metals. Thus, issues and relationships between the fatigue properties and influencing factors are worthwhile to further explore. There is still a lack of thorough overview on the fatigue properties of AM metals. A complete understanding of processing-material-performance relationship of AM components are critical from a design and application standpoint. As shown in Fig. 2, this review systematically summarizes the fatigue properties of the mainstream metals fabricated by kinds of AM processes. The extractive fatigue data including *S-N* curves data and fatigue crack data is classified along with the "external manufacturing factors", thus drawing the corresponding curves and analyzing the impact on fatigue properties, finally guiding the MAM processing and optimizing the fatigue performance. In the end, outlook and perspectives for the future research are presented. This timely review is anticipated to promote the development of novel fatigue characteristics and mature prediction models for AM metals in order to establish general guidelines for the design of fatigue resistant AM metallic parts. Further, the big data generation may become an effective way to success based on a large amount of fatigue data with AM metals.

## 2. Ti-6Al-4V

### 2.1. Introduction

As a lightweight structural metal, titanium is an excellent choice for many engineering applications due to its combination of excellent mechanical properties (maintained up to high operating temperatures), excellent corrosion resistance in highly corrosive environments, and biocompatibility. AM technologies have injected new ideas and potential into titanium alloys, such as high material yield, limited machining operations, and a high degree of freedom on alloy composition and structural complexity. Properties like fracture toughness and fatigue life, especially of AM metals, are critical parameters for materials selection and structural design optimisation. Large research on mechanical properties of AMed Ti-6Al-4V [7, 43, 44]. Fatigue performance [45] of EBMed [15, 46, 47], SLMed [48, 49], DEDed [50] Ti-6Al-4V are reported. Especially, literatures [7, 51] in detail introduced the additive manufacture of titanium alloys, involving mechanical properties, comparison of various technologies. It is difficult to directly discuss fatigue properties from different sources, as the choice of method, particle, processing parameters, post-processing, temperature, and specimen preparation will affect the results greatly. A summary of relevant papers on Ti-6Al-4V are made below.

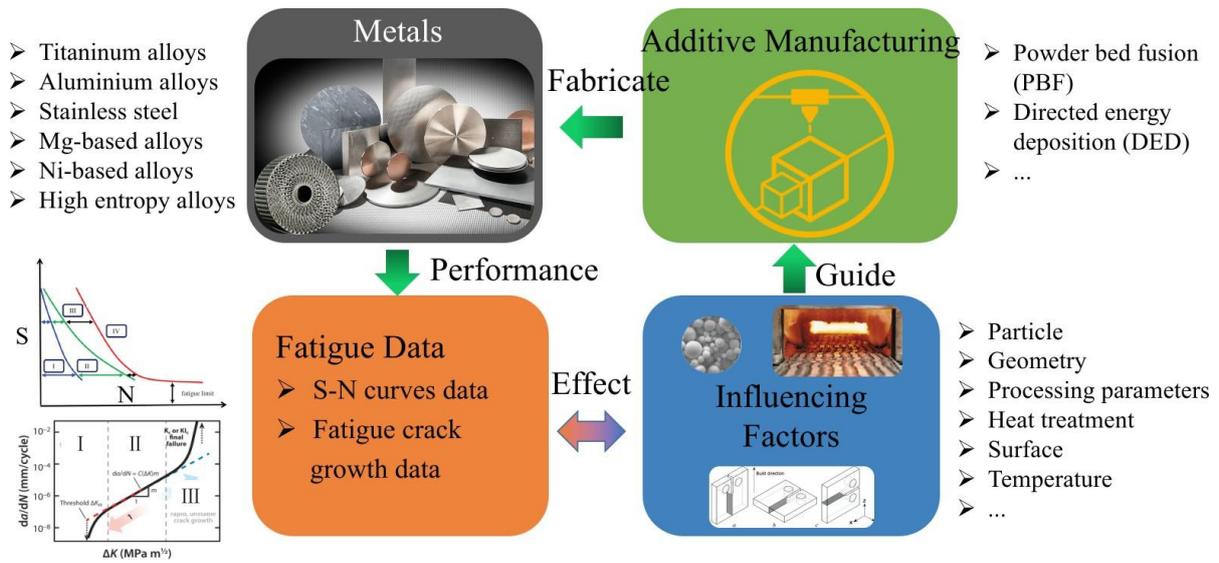

**Fig. 2.** The schematic of processing-material-performance relationships in additive manufacturing.

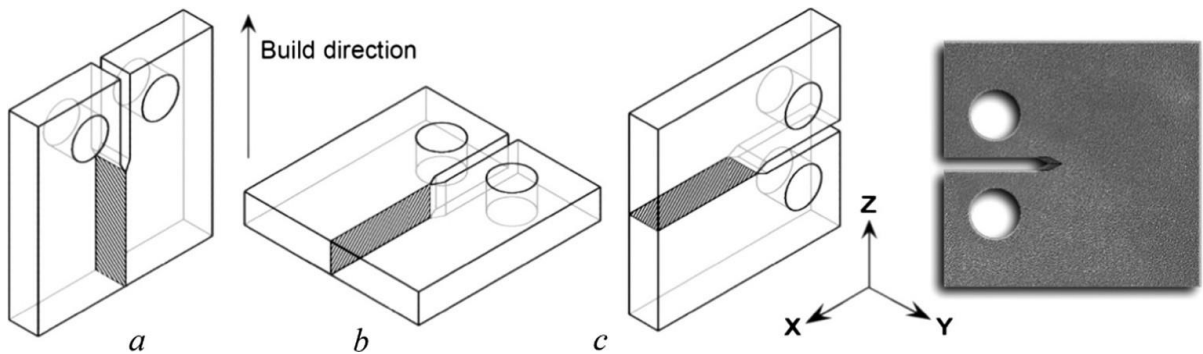

**Fig. 3.** Three orientations of CT specimens for crack growth tests with respect to the build direction Z, the macroscopic crack plane for the orientation a coincides with the plane *z-y*, b with plane *z-x* and c with plane *x-y*. Reproduced with permission from [52].

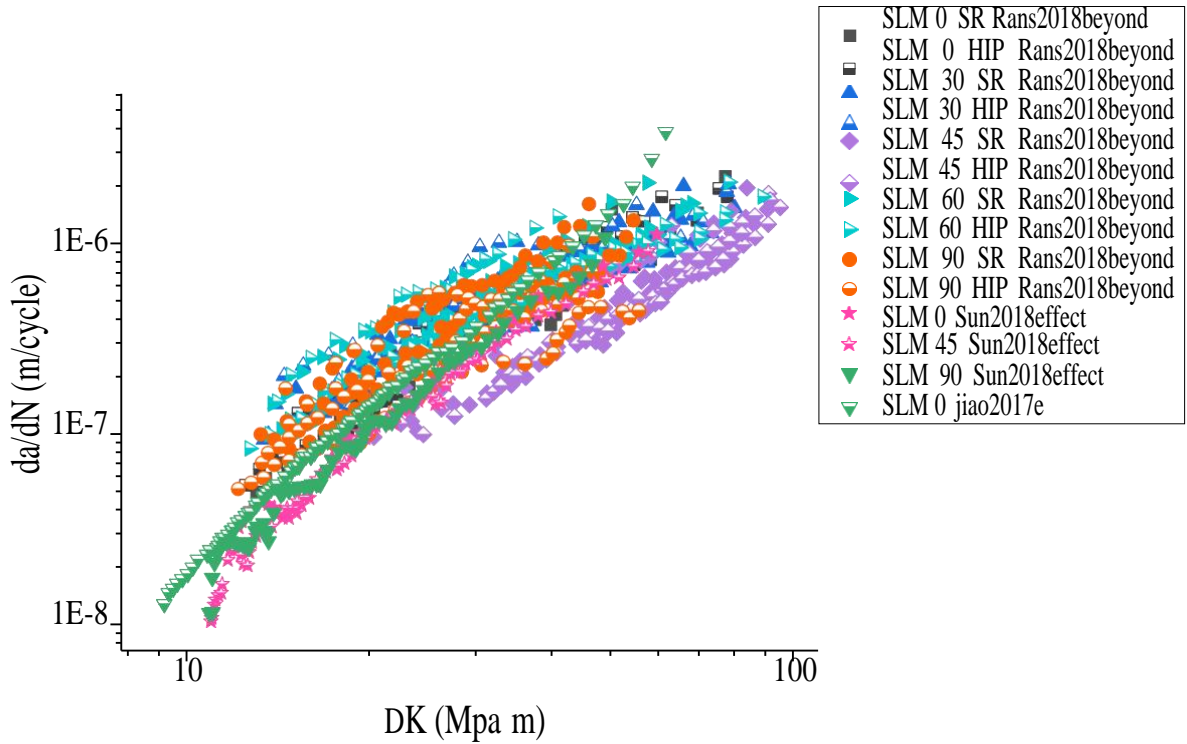

**Fig. 4.** Comparison of calculated fatigue crack growth rates for different build orientations with *R* = 0.1, room temperature, and frequency=10 Hz, data from [54, 57, 65].

## 2.2. Fatigue crack growth

### 2.2.1. Built orientation

A great deal of research investigated the built orientation effect [49, 52–58] on FCG of AMed Ti-6Al-4V. Fig. 3 show three basic orientations of CT specimens for crack growth tests. 45 CT specimen is not shown. CT specimen width and CT specimen thickness are undefined. The stress ratio is usually 0.1 and frequency is usually 20 Hz.

Literatures [52, 54, 57, 59–63] reported the anisotropy of build direction (H-0 , V-90 , or 45 ) effect on FCG, they found that build orientation had a small, but repeatable influence on FCG. Sun et al. [57, 59] found that build direction can affect the fatigue crack growth life and the 45 sample has the highest fatigue crack growth life, as shown in Fig. 4. Beyond the orthogonal, Rans et al. [54] investigated the influence of off-axis build direction (0, 30, 45, 60, 90 ) in thin SLMed Ti-6Al-4V plates, see Fig. 4, with the objective of the influence of columnar grain orientation on the FCG behaviour. The results show lager scatter of build orientation effects and insignificant influence on FCGR. In addition, there is a strong influence of specimen orientation on the fatigue crack growth behaviour of SLM Ti6Al4V [62] and L-PBF Ti-6Al-4V [64]. The greatest improvement in properties after heat treatment was demonstrated when the fracture plane is perpendicular to the SLM build direction. This behaviour is attributed to the higher anticipated influence of tensile residual stress for this orientation.

However, the FCGR of WAAMed Ti-6Al-4V did not show anisotropy on built orientation when $\Delta K < \Delta K_T$ [33, 66], as shown in Fig. 5. $\Delta K_T$ is stress intensity factor transition point. The FCGRs of the 900-

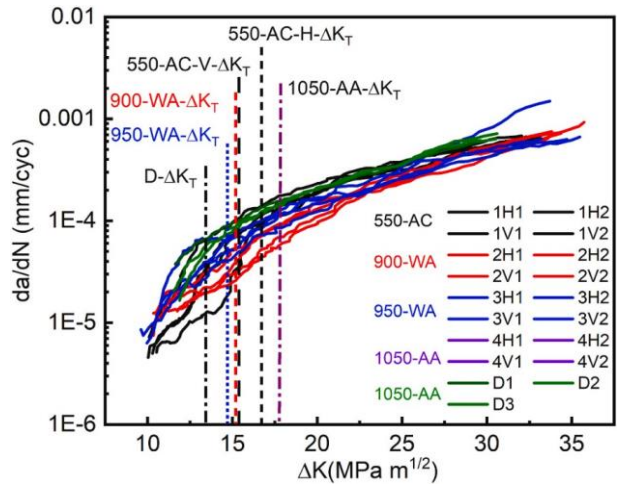

**Fig. 5.** The detail comparison of FCGR (da/dN) versus Δ$K$ for WAAMed Ti-6Al-4V and the forged Ti-6Al-4V with different samples. Note that 500, 900, 950 and 1050 mean heat treatment temperature, and WC-water cooling, AC-air cooling, WA-water cooling and aging, AA-air cooling and aging. Reprinted from [33, 66], Copyright(2021), with permission from Elsevier.

WA, the 950-WA, and the 1050-AA WAAM Ti-6Al-4V smaples were 87.4%, 85.6%, and 94.5% lower than that of the standard sample, respectively. Under the same Δ$K$, the FCGRs of V-sample and H-sample were almost the same. Fig. 6 shows the fatigue crack growth process of various samples. The continuous grain boundary $\alpha_{GB}$ plays a dominant role in the anisotropy of crack growth direction in WAAM Ti-6Al-4V. For the sample with continuous $\alpha_{GB}$ (550-AC), the H-sample and the V-sample showed the anisotropic FCG behavior (see Fig. 6 (a) and (b)), the crack in the H-sample passed through the continuous $\alpha_{GB}$ and grew along with mixed region while the crack in the V-sample grew along with basket-weave microstructure. For the sample with discontinuous $\alpha_{GB}$ (see Fig. 6 (c) and (d)), the H-sample and the V-sample showed the same FCG behavior.

In addition, the heat treatment, such as $\beta$ annealed, would have an important influence on the anisotropy of built orientation. As shown in Fig. 7, Galarraga et al. [63] found that $\beta$ annealed could eliminate the effect of built orientation on FCG under different stress ratio (R=0.1 and 0.8). The vertical EBMed Ti-6Al-4V has lower FCG-rate (FCGR) than the horizontal one during as-fabricated state. For the $\beta$ annealed conditions, FCG behavior in both orientations is identical at both stress ratios as expected due to the equiaxed microstructure. Comparing the results to the as-fabricated and $\beta$ annealed conditions, a considerable improvement of the FCG resistance of the $\beta$ annealed material is observed for both stress ratios and at all growth stages. The improvement of the FCG properties of the $\beta$ annealed microstructure, compared to the as-fabricated microstructure, can be attributed to the higher FCG resistance of a coarse lamellar structure versus a fine lamellar structure.

On the other hand, there is no influence of the building orientation (and thus microstructure direction-ality) of the Ti6Al4V alloy produced by DMLS with subsequent stress relieving heat treatment at 380 °C for 8 hours on the growth of long fatigue cracks [67]. And non-orthogonal build orientation had negligible effect on fatigue crack growth [64]. Analogously, The EBMed Ti-6Al-4V specimens in both orientations displayed essentially equivalent behavior with respect to each other. This indicates that there is no noticeable difference in crack growth rates as a function of specimen, crack, or loading orientation [46]. This result is in agreement with the finding published in [64, 67].

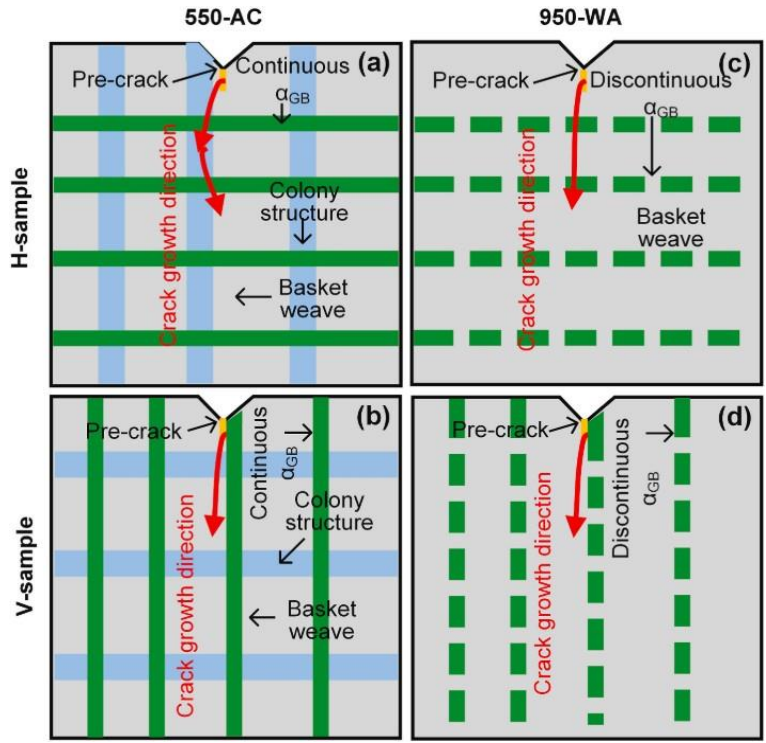

**Fig. 6.** Fatigue crack growth of different samples in low ΔK region, (a) crack growth mechanism with continuous $α_{GB}$ of the H-sample, (b) crack growth mechanism with continuous $α_{GB}$ of the V-sample, (c) crack growth mechanism with discontinuous $α_{GB}$ of the H-sample, (d) crack growth mechanism with discontinuous $α_{GB}$ of the V-sample. Reprinted from [33], Copyright(2021), with permission from Elsevier.

The building orientation effect on Short/Small FCG [68]. Small fatigue crack growth (SFCG) behavior is examined for additively manufactured Ti-6Al-4V specimens with optimal and trial build conditions (representing an increased degree of defects) through a combination of in situ tomography and in situ energy dispersive X-ray diffraction. The results showed slower crack growth rates for the SFCG samples compared to the long FCG [67, 69–71], as shown in Fig. 8. Ren et al. [69] studied the FCG behavior of long cracks in Ti-6Al-4V alloy fabricated by high-power laser directed energy deposition. The results show the deposited parts under heat treated conditions have superior FCG resistance than that of the other reported additive manufactured Ti-6Al-4V, owing to the specimens with coarse prior-$β$ grains and interior thick $α$-laths. In addition, small fatigue crack growth resulted from foreign object damage in Ti-6Al-4V are investigated in detail [72].

### 2.2.2. Processing parameters

Various manufacture technologies could result in different fatigue crack growth behaviors, Greitemeier et al. [73] compared the d$a$/d$N$ -Δ$K$ curves between EBM and DMLS Ti-6Al-4V under various heat treatments and $R$ = 0.1, 0.7. During the same condition, $R$ = 0.1, build direction: XZ, heat treatment: 710 °C/2 h, surface: milled, the Δ$K_{th}$ of DMLS is lower than the threshold for EBM, and there is a same sloop both the two. Dhansay et al. [74] found that with stress relieving heat treatments, the fracture mechanics parameters of fatigue behaviour (Paris curve) and fracture toughness show strong correlation between SLM

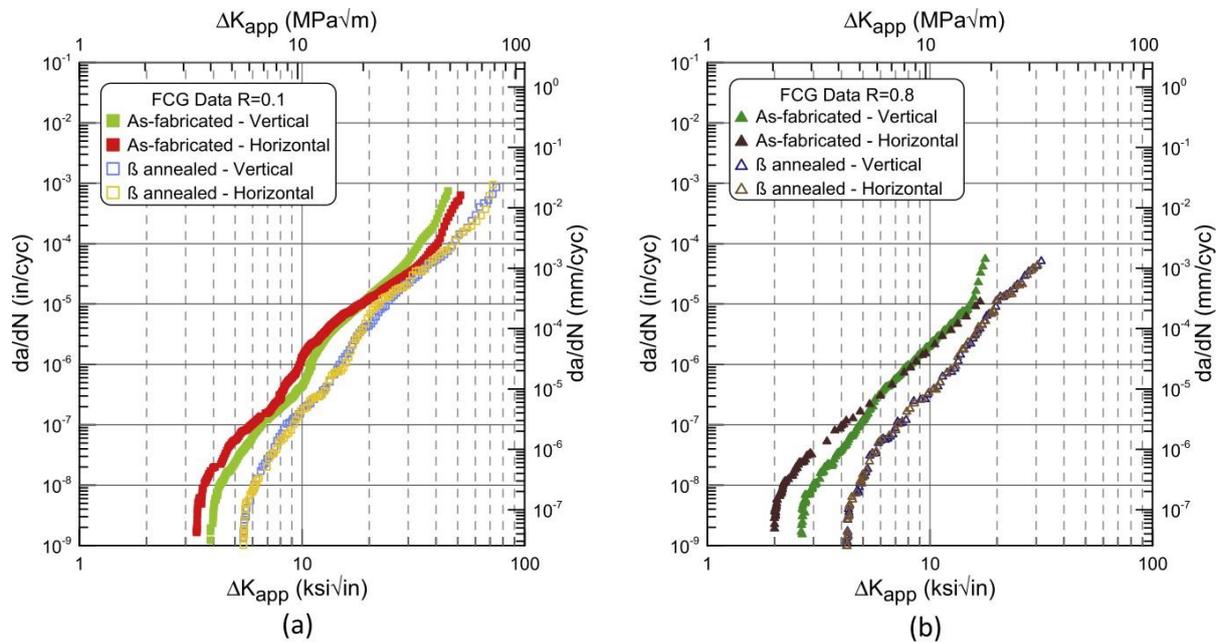

**Fig. 7.** FCG of as-fabricated and *β* annealed Ti-6Al-4V ELI in horizontal and vertical directions tested at: (a) R = 0.1 and (b) R = 0.8. Reprinted from [63], Copyright(2017), with permission from Pergamon.

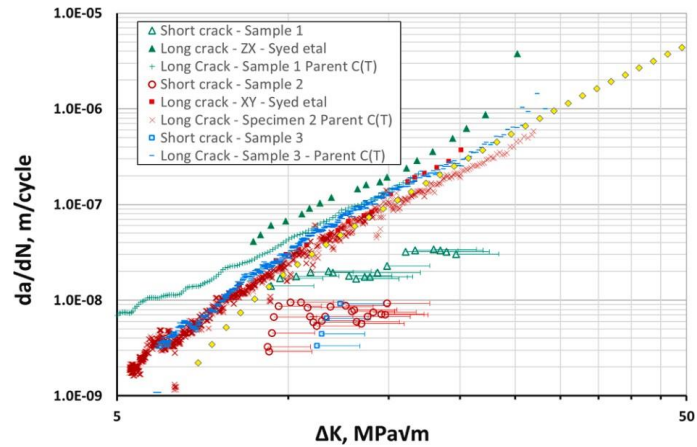

**Fig. 8.** Comparison of small fatigue crack growth rates for the miniature samples with associated long crack data for similar SLM conditions from the original C(T) samples and from Ref [71]. Error style bars show maximum stress intensity value if the crack adopted a semicircular fan like shape. Reprinted from [68], Copyright(2020), with permission from Elsevier.

manufactured and conventional Ti-6Al-4V.

Interestingly, Zhang et al. [75, 76] investigated the fatigue crack growth behavior in WAAM Ti-6Al-4V. They found that the difference in crack growth rate and pattern in the substrate and WAAM is attributed to the different microstructure characteristics. Crack propagates in straight and smooth line in the wrought condition that has equiaxed structure, but in tortuous path in the WAAM alloy owing to the lamellar structure. Consequently, WAAM alloy has lower crack propagation rate than the substrate. As shown in

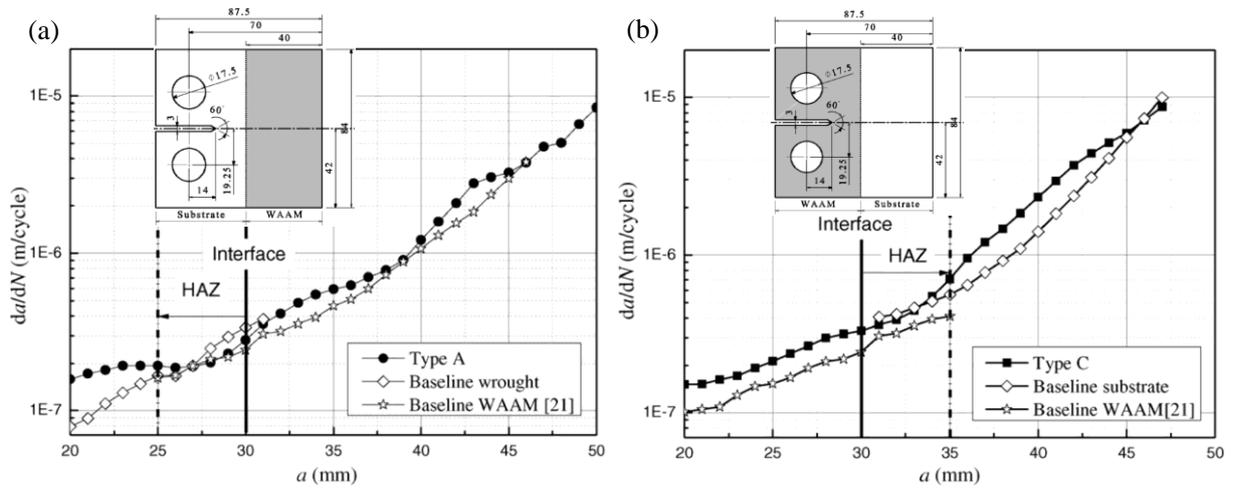

**Fig. 9.** Comparison of fatigue crack propagation rate versus crack length (a): (a) Type A: baseline wrought and WAAM alone, (b) Type C: baseline wrought and WAAM alone. Reprinted from [75], Copyright(2016), with permission from Elsevier.

Fig. 9, the crack growth rate as a function of the crack length and comparison with that of the two baseline materials (wrought and WAAM only). For the Type A specimen, crack propagates in the substrate first before reaching the WAAM-substrate interface; after the crack has passed the interface, it enters the WAAM alloy. For the Type C specimen, crack growth rate is compared with the baseline WAAM prior to it reaching the interface since the crack propagates in the WAAM alloy in this stage. Crack growth rate is faster than that in the baseline WAAM. After it crossing the interface, crack enters the substrate and still propagates faster than that of the baseline wrought alloy. And besides, Gao et al. [77] obtained superior fatigue crack growth resistance of Damage-tolerant Ti-6Al-4V alloy by in-situ rolled wire-arc additive manufacturing compared WAAMed Ti-6Al-4V, which is mainly attributed to its finer defects and finer $\alpha$ phase.

Recently, a fractographic analysis of AM Ti-6Al-4V has shown that microstructural faceting due to slip can cause fatigue failure, especially in the presence of porosity [78]. The likelihood of pore related failure was found to be dependent on microstructure size and morphology. Additionally, a minimum pore size threshold was found to exist for each microstructure, under which pores will not cause fatigue failure. The effect of manufacturing defects on the fatigue crack propagation and fast fracture should be insignificant [58, 79, 80]. Akgun et al. [81] investigated the behaviour of fatigue cracks initiated from process-induced porosity under axial loading in L-PBFed Ti-6Al-4V. A positive skewed distribution was observed for pore size distribution. Large pores located at the extreme tail of the distribution acted as crack initiation sites. Note that the pores were not randomly located, they had a tendency towards the free surface. According to the replica technique, crack initiation life occupies at least 50% of overall life, measured as the first crack detectable on the surface. After crack initiation, with pore size being added to the crack length, a similitude with the long crack growth rate was demonstrated. In the near-threshold region, small cracks initiated from pores showed a faster crack growth rate than long cracks. Nevertheless, small crack growth rates were not as significant as in conventional manufactured products.

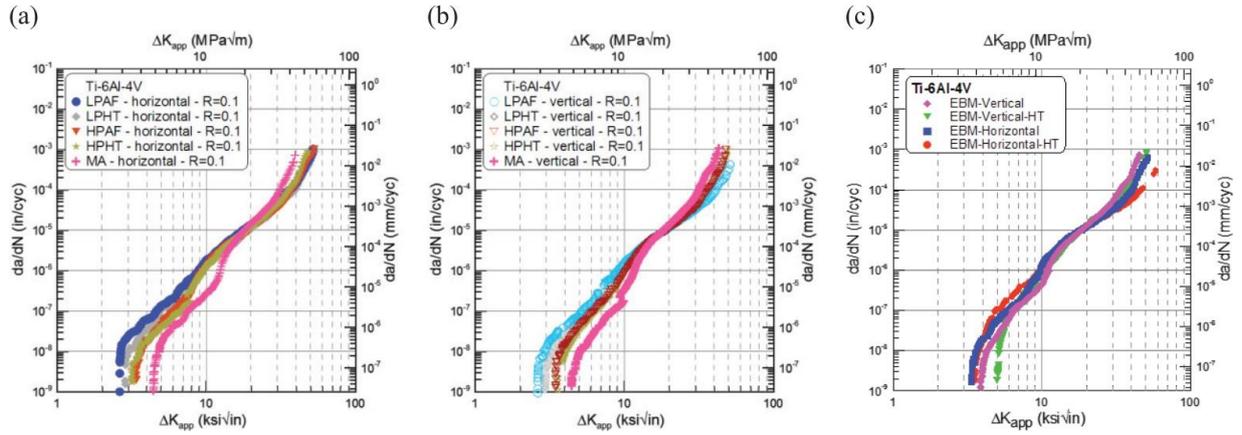

**Fig. 10.** FCG data of: (a) LENS fabricated Ti-6Al-4V at horizontal fatigue crack propagation direction; (b) LENS fabricated Ti-6Al-4V at vertical fatigue crack propagation direction; (c) EBM fabricated Ti-6Al-4V. Low power (LP) for 330 W and high power (HP) for 780 W. Reprinted from [70], Copyright(2016), with permission from Elsevier.

### 2.2.3. Post-Processing

As shown in Fig. 10, for laser engineered net shaping (LENS) horizontal propagation cases, Fig. 10(a), LENS fabricated Ti-6Al-4V alloys in general showed lower threshold value $\Delta K_{th}$ but higher fracture toughness $\Delta K_{FT}$ than mill-annealed (MA) Ti-6Al-4V (substrate), indicating a better high cycle fatigue performance in mill-annealed Ti-6Al-4V, and a better low cycle fatigue performance in LENS Ti-6Al-4V. Comparisons between LENS fabricated alloys indicated that LP fabrication yielded slightly lower threshold value than HP fabrication due to the presence of martensitic $\alpha'$ phases. After annealing, slight increase in $\Delta K_{th}$ was found in LP fabricated Ti-6Al-4V as a result of $\alpha'$ decomposition. Similar observations were found in vertical propagation cases, Fig. 10(b). In EBM fabricated Ti-6Al-4V alloys, B1 and B2 batches showed similar threshold and toughness values for both horizontal and vertical propagation directions, thus only data from B2 batch are plotted here in Fig. 10(c) against the heat treated conditions. Compared to LENS, EBM as-deposited alloys showed similar threshold to LENS HP fabricated alloys, i.e. higher threshold than LENS LP fabricated alloys, due to their similar microstructures. Fracture toughness values in as-deposited EBM Ti-6Al-4V alloys are comparable to those in LENS. After solutionizing and aging, significant increase in threshold was found in vertical propagation direction, achieving a threshold value comparable to the mill-annealed Ti-6Al-4V; while significant increase in fracture toughness was found in horizontal propagation direction.

To evaluate the influence of post-processing and surface machining, fracture toughness and FCG tests were performed for SLM Ti-6Al-4V in three types of post processing status: as-built, heat treated and HIPed, respectively [82]. They found that as-built SLM Ti-6Al-4V presents poor ductility and FCG behavior due to martensitic microstructure and residual stresses. Both heat treatment and hot isostatic pressing improve the plane-stress fracture toughness and FCG performance considerably, while surface machining shows slight effect (see Fig. 11). As can be seen from Fig. 11(a) and (b), SLM Ti-6Al-4V specimens show quite inferior FCG performance, and relatively wider scatter band is observed. After heat treatment or HIP processing, the FCG behavior is improved significantly, better than that of mill-annealed material, and comparable to that of cast material when the stress intensity factor range $\Delta K \geq 21$ MPa$\sqrt{m}$. For specimens in SLM condition, no significant improvement in FCG behavior is found after surface machining, as shown in Fig. 11(c), but the scatter seems to be reduced. For specimens in heat-treated condition, FCG rate for $\Delta K < 15$ MPa$\sqrt{\text{m}}$ decreases via surface machining, as shown in Fig. 11(d). For specimens in HIPed condition, surface machining hardly improves the FCG behavior, as shown in Fig. 11(e).

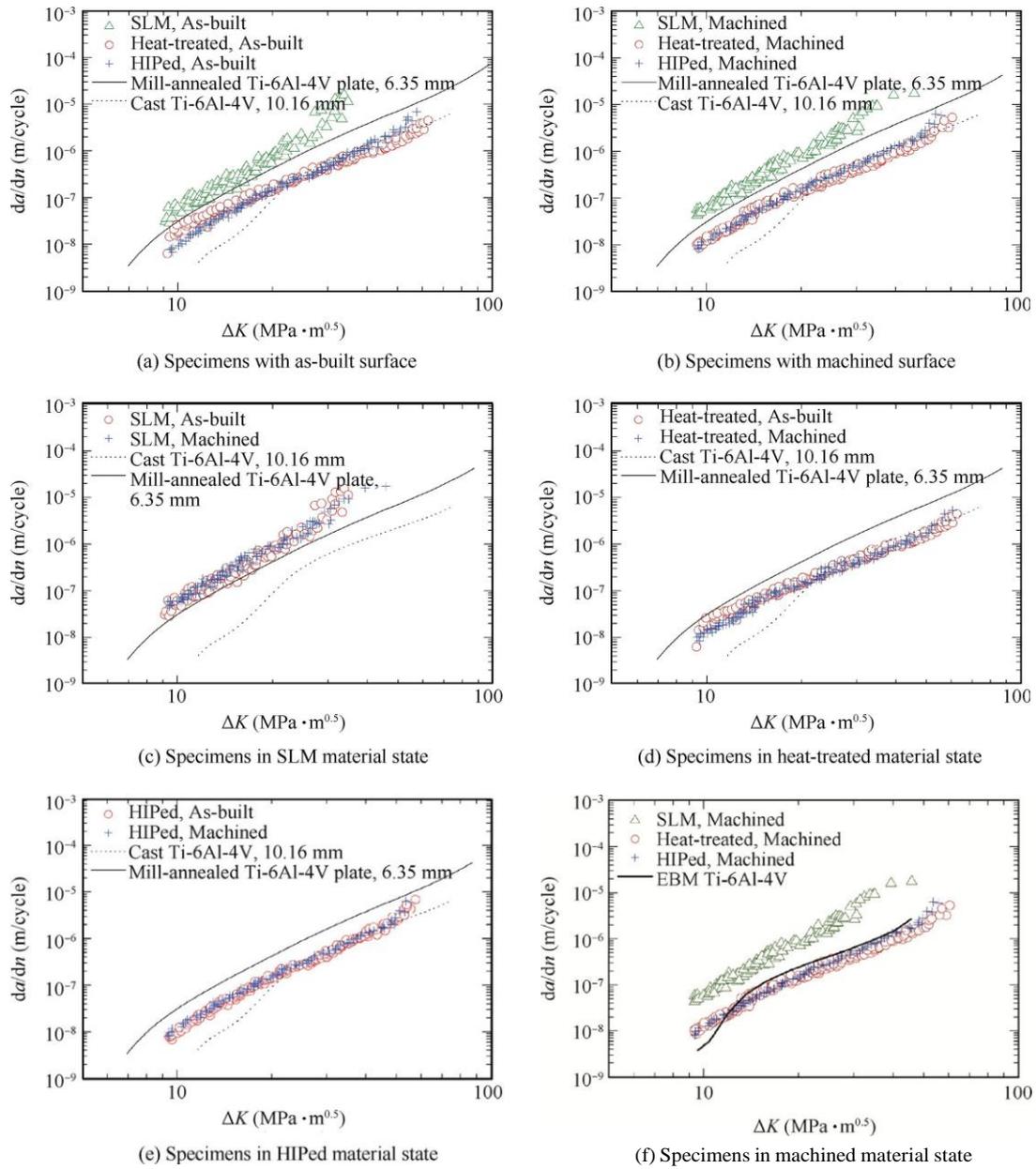

**Fig. 11.** FCG rate versus stress intensity factor range for Ti-6Al-4V in different conditions. Reprinted from [82], Copyright(2019), with permission from Elsevier.

As well as, Qiu et al. [83] discussed effect of rolling on fatigue crack growth rate of WAAM processed Ti-6Al-4V. The main characteristics of rolling is the reduced β grain size to refine the alloy's microstructure. Zhang et al. [84] investigated fatigue crack growth behavior in residual stress fields induced by laser shock peening (LSP). Therefore, even if the size of a surface crack is smaller than a subsurface crack and the crack growth rate is slower at the beginning, the growth rate of the surface crack quickly overtakes the growth of a larger subsurface crack. The crack size effect on the stress intensity factor for surface cracks and subsurface cracks must be noted as the crucial factor, even if the environment effect, air or vacuum, exists [85, 86].

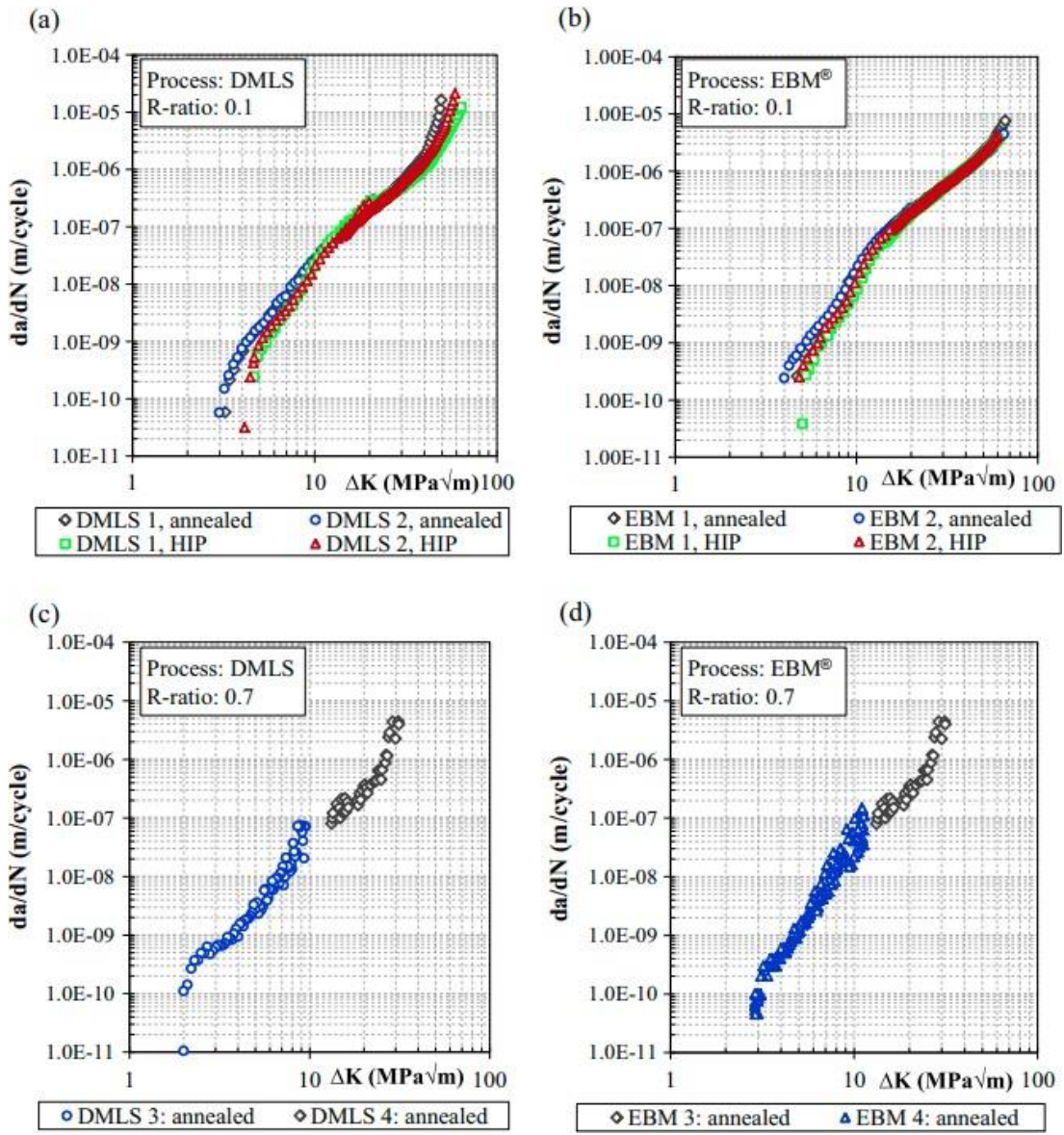

**Fig. 12.** Fatigue crack growth properties of additive manufactured TiAl6V4: (a) DMLS, R = 0.1, annealed & HIPed; (b) EBM, R = 0.1 annealed & HIPed; (c) DMLS, R = 0.7 and (d) EBM, R = 0.7. Reprinted from [73, 87], Copyright(2017), with permission from Elsevier.

### 2.2.4. HT and HIP

Greitemeier et al. [73, 87] investigated the FCG of DMLSed and EBMed Ti-6Al-4V with annealed or with HIP at R=0.1 and 0.7, as shown in Fig. 12. The comparison of annealed and HIPed material is displayed in Fig. 12a for DMLS and Fig. 12b for EBM material with an R-ratio of 0.1. The crack propagation rates differed within the threshold regime ($\Delta K_{th}$) for both processes subjected to the applied heat treatment. The crack growth resistance was higher for the HIPed material (DMLS: 4.4 MPa$\sqrt{m}$, EBM: 4.8 MPa$\sqrt{m}$) compared to the annealed condition (DMLS: 3.2 MPa$\sqrt{m}$, EBM: 4.2 MPa$\sqrt{m}$). In contrast similar crack propagation values can be seen within the Paris regime, independently from the process or applied heat treatment. Crack growth measurements were obtained at an R-ratio of 0.7 to reduce the influence of roughness-induced crack closure. The results for the annealed specimens (DMLS: Fig. 12c, EBM: Fig. 12d) showed variations within the threshold regime (DMLS: 2.0 MPa$\sqrt{m}$, EBM: 2.9 MPa$\sqrt{m}$) similar to the results obtained for an R-ratio of 0.1.

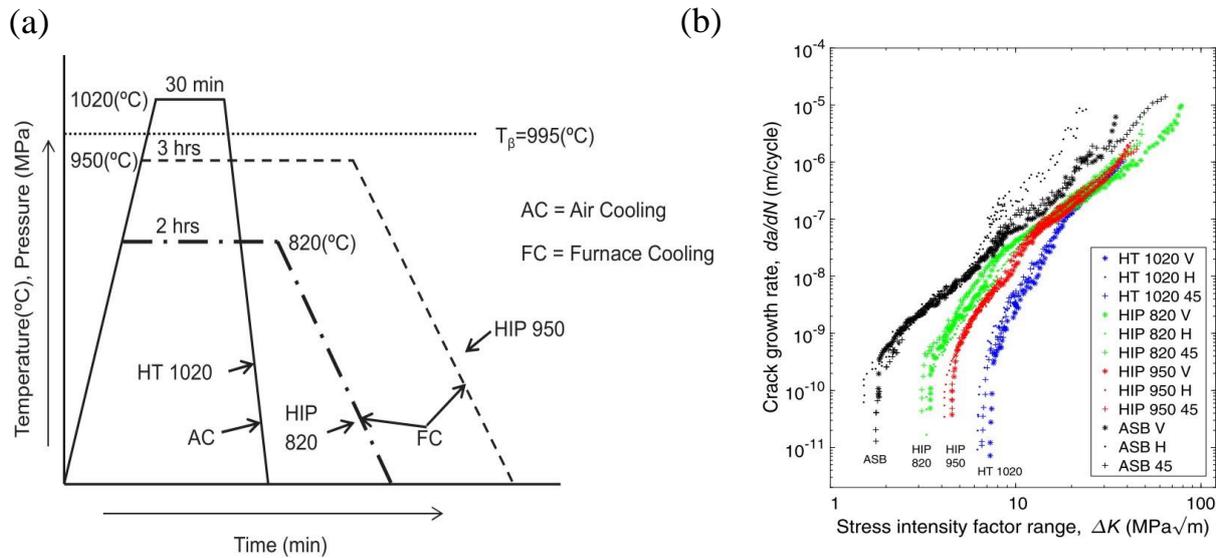

**Fig. 13.** (a) Schematic diagram of the post heat treatments used. (b) Fatigue crack growth rates, da/dN, plotted as a function of the applied stress intensity range, ΔK, for samples tested at a load ratio of R = 0.1 and a sine wave frequency of 60 Hz [87].

Clearly, a significant improvement in crack growth behaviour can be achieved by an appropriate treatment [60]. Effect of α and α/β annealing and crack orientation on fatigue crack propagation (FCP) of Ti-6Al-4V alloy fabricated by DED [88] was investigated. The FCP resistance of as-built DEDed Ti-6Al-4V specimen was inferior to that of conventional manufacturing counterpart, due to the fine microstructural feature associated with DED process. Acicular α platelets became coarsened with annealing, either in α/β or β region, and the resistance to FCP of Ti-6Al-4V specimen increased substantially. The increase was particularly significant with β anneal, along with approximately 25% increase in fracture toughness. The crack orientation effect on FCP is neglected. Jesus et al. [89] yielded the FCGR in CT samples of Ti-6Al-4V alloy produced by L-PBF and submitted to HIP, and found that the HIP treatment showed a relevant influence in the FCGR strength increasing the fatigue threshold, the critical stress intensity factor and the FCGR resistance. In addition, effects of build orientation (0 , 45 , 90 ) and post heat treatments (hot isostatic pressing 820 °C, 950 °C or annealing 1020 °C) on the FCGRs of Ti-6Al-4V fabricated by laser powder bed fusion were examined [64], as shown is Fig. 13. On the contrary, the stress relieving heat treatment and HIP treatment did not have a significant influence on FCGR [54].

### 2.2.5. Other factors

Fatigue crack growth takes place in a transgranular mode by cyclic damaging of acicular α′-martensite within the plastic zone. The original β grain boundaries play a supporting role in the crack growth process [67]. Besides, the applied overloads caused a FCGR retardation [89–91]. The crack growth retardation was due to the increase of crack closure levels induced by plasticity resulting from the overloads, When underloads were applied, no change in the FCGR behaviour and the crack closure level were observed. Finally, the effect of hydrogen embrittlement on FCG was investigated by carrying out crack propagation tests in air and high-pressure H2 environment [92]. By exposing the EBM Ti-6Al-4V material to a hydrogen-rich environment the FCG rate increased significantly above Δ$K$= 23 MPa$\sqrt{m}$ compared with the air environment. Below Δ$K$= 23 MPa$\sqrt{m}$ the hydrogen-tested material fluctuated, whereas the air-tested material followed Paris law throughout all the Δ$K$. Relative to already published FCG results of wrought and cast Ti-6Al-4V, the EBM Ti-6Al-4V sample was found to have superior FCG properties in high-pressure hydrogen compared to cast material while being slightly lower than wrought.

## 2.3. Fatigue life or fatigue limit

### 2.3.1. Particle

EBM specimens than in DMLS [93], powder [94, 95]

The characteristics of powder used in the additive manufacturing process have a great impact on the quality and the mechanical properties of the target product. The oxygen content of the Ti-6Al-4V powder increased with powder reuse, so the yield strength and the ultimate tensile strength of EBM samples increased as well[96].

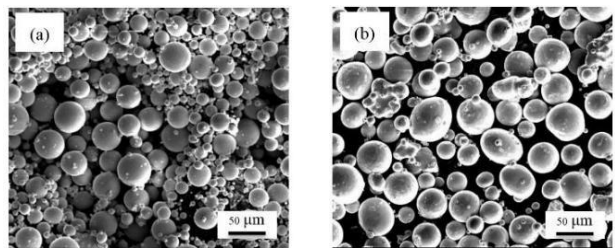

**Fig. 14.** SEM of (a) AP&C Ti-6Al-4V Powder and (b) EOS Ti-6Al-4V Powder. [97]

Gong et al. [97] studied two Ti-6Al-4V powders with different levels of fines in Fig. 14. Based on the powder provided by SLM machine vendor, some small particle inclusion named AP&C were mixed to explore the influence of particle size on fatigue performance of AM specimens. The results in Fig. 15 showed that the reduction of powder particle size had a noticeably negative effect on fatigue performance. Because the finer particles tend to increase the laser absorption of the powder bed, which could in turn lead to the instability of the meltpool and to the appearance of lack-of-fusion (LOF) defects in the solidified regions.

Brika et al. [98] compared the different powder production techniques which named plasma atomization and gas atomization to investigated the impact of powder particle morphology on the fatigue properties of PBF Ti-6Al-4V components. Fig. 16 shows the particle size distributions of the two powders were significantly similar, but the plasma-atomized powder particles were more spherical and regular shaped than their gas-atomized equivalents. The presence of aligned pores in specimens was horizontal to the loading direction for gas-atomized and perpendicular to loading direction for plasma-atomized, respectively(Fig. 17).

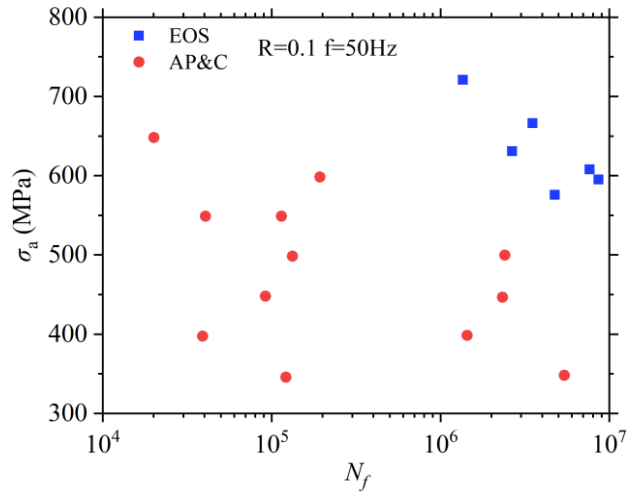

**Fig. 15.** Maximum stress amplitude *vs.* cycles to failure of specimens produced by AP&C Ti-6Al-4V powder and EOS Ti-6Al-4V powder. (The data comes from Ref. [97].)

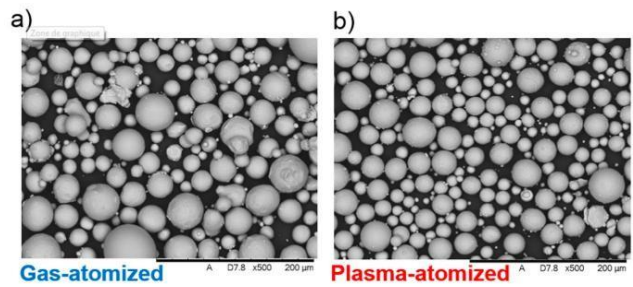

**Fig. 16.** SEM micrographs of two powder lots. [98]

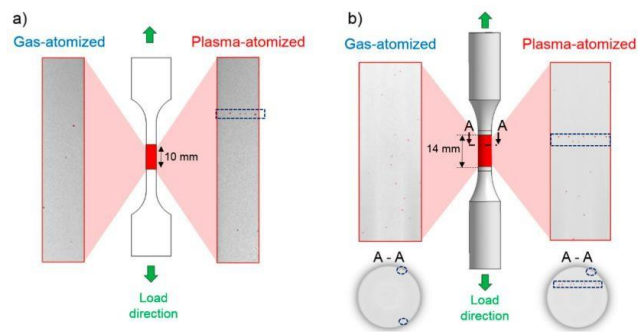

**Fig. 17.** Processing-induced pores characterization ( CT): (a) 2D vertical sections of the gauge regions of tensile specimens printed with two powders and (b) 2D horizontal and vertical sections of the gauge regions of fatigue specimens printed with two powders. Aligned pores could be seen in the tensile and fatigue specimens produced with the plasma-atomized powder. [98]

7% greater ultimate and 4% greater yield strength were observed in plasma-atomized powder specimens, while showed 3% lower elongation to failure. The same printing parameters were used for both powders but these parameters were specifically optimized for the gas-atomized powder. The gas-atomized powder specimens showed uniform pore distributions, which reduced fatigue crack propagation and initiation caused by processing-induced porosity.(Fig. 18).

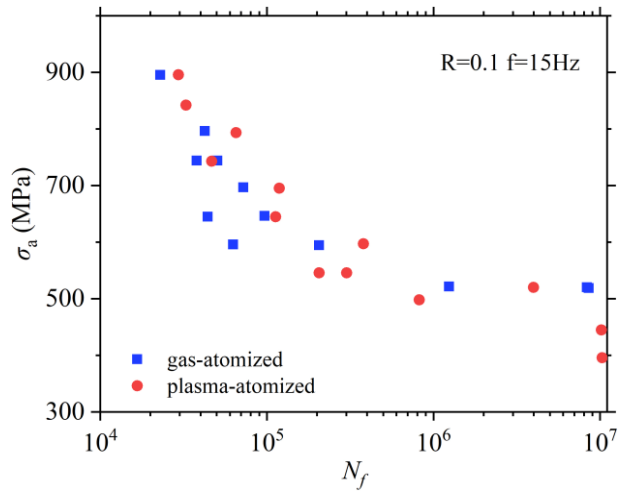

**Fig. 18.** Maximum stress amplitude *vs.* cycles to failure of specimens produced by gas-atomized powder and plasma-atomized powder. (The data comes from Ref. [98].)

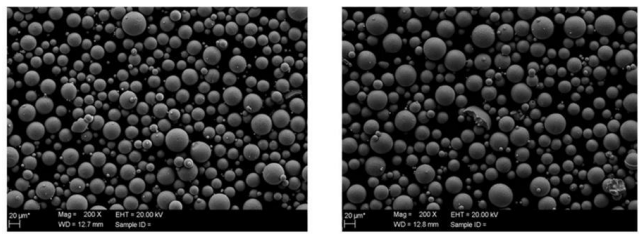

**Fig. 19.** Comparison of particle morphologies for new(left) and used(right) Ti-6Al-4V powder obtained via SEM [99].

However, there are fewer comprehensive studies on how powder recycling affects the fatigue behavior of the manufactured part. Carrion et al. [99] gave an exploratory research on the effect of powder recycling on fatigue behavior of L-PBF manufactured Ti-6Al-4V parts. The improved flowability and the less compressibility of the used powder may result in smaller internal pores, and recycling the powder may improve the fatigue resistance of machined L-PBF Ti-6Al-4V specimens in the high cycle regime(Fig. 20). Under the stress amplitude of 500 MPa, the life of fatigue can improve from $10^6$ to $10^7$, while the stress amplitude values of new and used powder specimens in the as-built condition showed no differences.

Popov et al. [100] and Soundarapandiyan et al. [101] investigated the effect of powder reuse on fatigue performance in electron beam powder bed fusion additive manufacturing process. The Ti-6Al-4V powders were checked after at least 10 build cycles and even up to 69 cycles. As a result of the increased LOF defects, the high cycle fatigue performance deteriorated with reuse.(Fig. 21). This might be attributed to the voids formed in the powder bed due to decrease in the number of fine particles coupled with an increase in the number of high-aspect ratio particles and lack of fusion in the EBM printing process by powder surface oxidation.

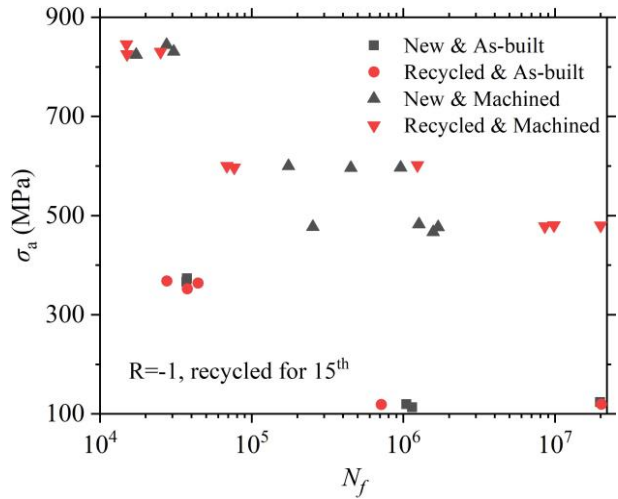

**Fig. 20.** Maximum stress amplitude *vs.* cycles to failure of as-built and machined specimens produced by new and recycled powders. (The data comes from Ref. [99].)

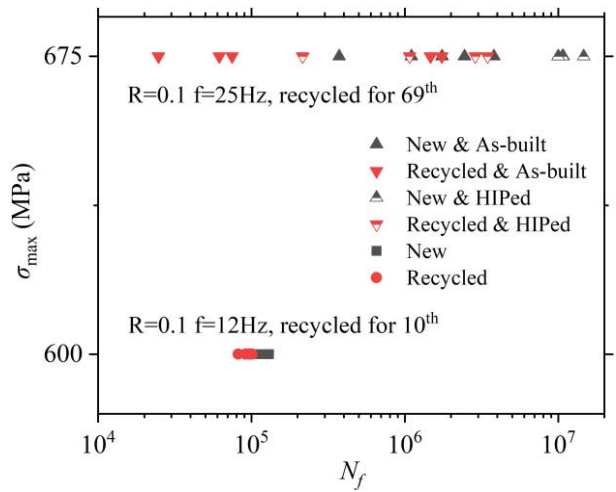

**Fig. 21.** Maximum stress *vs.* cycles to failure of as-built and HIPed specimens produced by new and recycled powders. (The data comes from Ref. [100, 101].)

### 2.3.2. Sample geometry

Besides the influence of particle on the fatigue performance, geometries of the notched sample for AM Ti-6Al-4V fatigue tests are also significant to contribute to the fatigue properties. Despite the fact that the fatigue characteristics of flat additive produced specimens with rough as-built surfaces have been extensively investigated, few aerospace components have a basic flat shape with no corners or radii that would act as stress concentrations in practice. Therefore, the combined effect on fatigue life of the specimen size and a geometrical notch needs to be established. Fig. 22 shows the various geometries of different size and notched AM Ti-6Al-4V specimens. Zhao et al. [109] in detail compared the mechanical properties of EBM and SLM Ti-6Al-4V parts with different part size. The thickness of the $\alpha$ lamellas decreases with decrease in part size during EBM fabrication, which results in an improvement in strength and decrease in ductility. For SLM samples, the part size has almost no effect on the mechanical properties. Wang et al. [110] investigated the fatigue properties of Ti-6Al-4V alloy fabricated by EBM. The fatigue strength decreases as the increasing length of the stem, because of the higher concentrated stress on the surface of a longer stem.

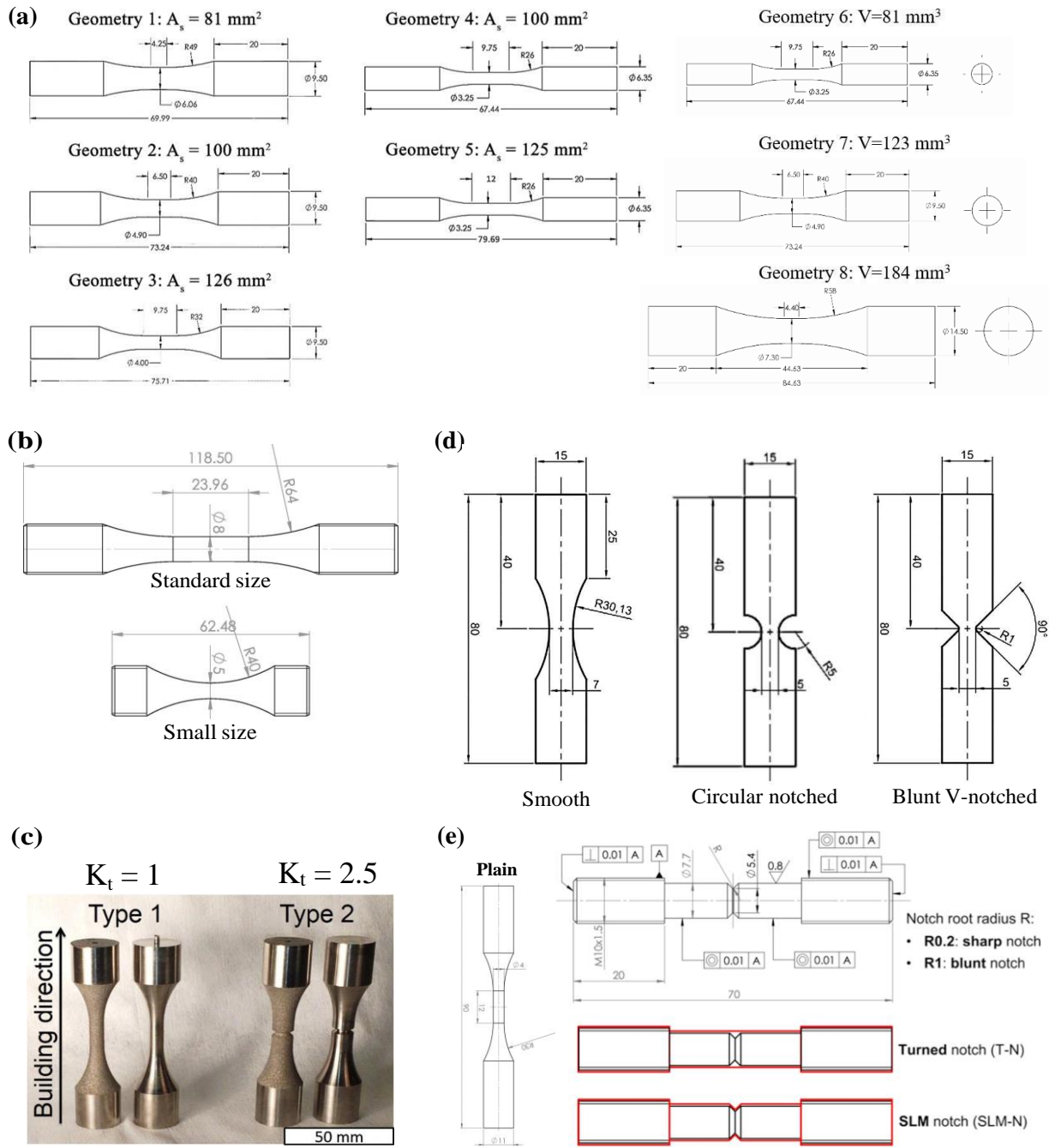

**Fig. 22.** Geometries of different size and notched specimens. (a) Geometries of different sizes [31, 102– 104]. (b) Geometries in standard size and small size [31]. (c) Specimens of type 1(smooth) and 2(notched) with both rough as-build surfaces and machined-and-polished surfaces [105]. (d) Geometries of smooth, circular notched and blunt V-notched specimens [106, 107]. (e) Geometry of the plain specimens(left), V- notched axisymmetric specimens (right top) and schematic illustrations of the two manufacturing routes(right bottom) [108].

Due to their high strength-to-weight ratio and biocompatibility, alloys like Ti-6Al-4V are used in the aerospace and biomedical sectors. Part sizes in these two sectors can range from tiny surgical implants to massive structural components. Comparing with wrought counterparts, the fatigue performance of additively manufactured parts may be more sensitive to part size because of the flaws inherent in the fabrication process. Therefore, some research have investigated the sensitivity of additively manufactured Ti-6Al-4V parts to surface area, volume size and gage diameter[31, 102–104].

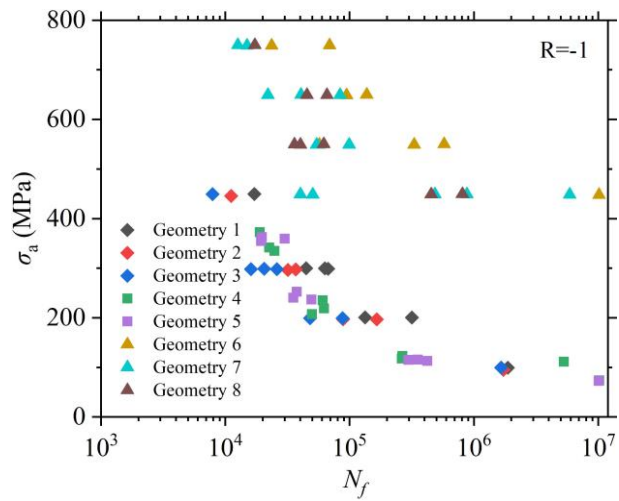

**Fig. 23.** Stress-life curves for each sample geometry (Fig. 22(a)). The data comes from Ref. [102–104].

As shown in Fig. 23, the systematical work by Pegues et al. [102–104] presented the stress-life curves for each sample geometry (see Fig. 22(a)). For the geometry 1-3 specimens with the intention of boosting gage surface area from 81 mm$^2$ to 126 mm$^2$ while keeping gage volume constant at around 123 mm$^3$, the result of S-N behavior shows that the fatigue life is harmed by increasing surface area for a fixed volume size. For the geometry 4-5 with 6.35 mm while incrementally increasing the gage surface area from 100 mm$^2$ to 125 mm$^2$, results indicate that the fatigue behavior is more sensitive to part diameter than surface area, that is, smaller cross-sectional diameter specimens were more influenced by the surface roughness in the HCF regime resulting in higher applied stresses and lower fatigue lives. Due to an increase in partially fused powder particles on the down-skin surfaces, the surface roughness on the down-skin surfaces was increased, resulting in a lower fatigue resistance. For the geometry 6-8, the three geometries were implemented to maintain a constant surface area in the uniform gage section while incrementally increasing the gage volume by 50%, a significant improvement in fatigue resistance is shown in the data for the smallest geometry compared to the larger specimens.

Le et al. [31] focused on the influence of statistical size effect of Ti-6Al-4V alloy fabricated by the SLM process. A vast fatigue test campaign has been undertaken, for two surface conditions (as-built and machined surfaces) and two specimen geometries (Fig. 22b) with different highly loaded volume sizes. The standard size geometry is proposed in the standard ISO-NF EN 6072 while the small size geometry was chosen so that the loaded volume is much smaller than the standard size geometry without introducing a strong stress concentration. For the S-N curve of the machined specimens, as shown in Fig. 24, the fatigue strength at $2\times10^6$ cycles of the small size specimens is higher than the standard size specimens (700 MPa *vs.* 500 MPa). This is because the probability of occurrence of LOF pores for the standard specimens is much higher than for the small specimens.

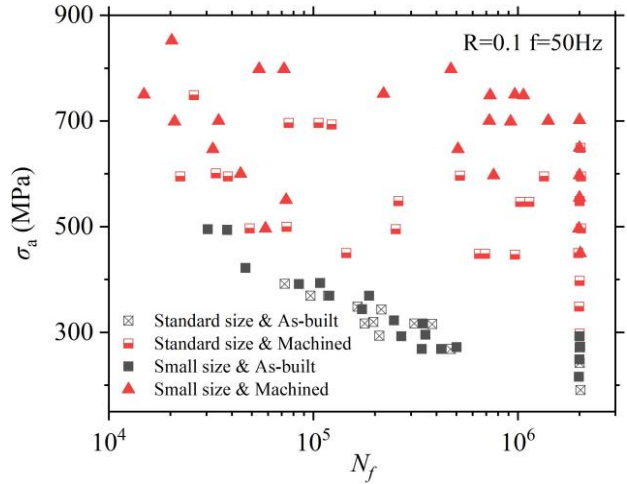

**Fig. 24.** Comparison between the machined specimens and the As-built specimens for standard size specimens and small size specimens (Fig. 22(b)). The data comes from Ref. [31].

    The fatigue properties of both SLM and EBM Ti6Al4V have been investigated and a combined effect of a rough as-built surface and a geometrical notch has been determined in Kahlin's work [105, 111]. For the fatigue specimens, as shown in Fig. 22c, the type 1 with a smooth surface $K_t = 1$ ($K_t$ is the theoretical stress concentration factor) and the type 2 with a 0.85 mm radius geometrical notch $K_t = 2.5$ are manufactured by AM. Both specimens were produced with either a rough as-build surface or a machined surface. The corresponding un-notched samples ($K_t = 1$) showed drastically different fatigue behaviour, but the specimens with geometrical notches ($K_t = 2.5$) demonstrated identical fatigue behaviour for both machined surfaces and rough as-built surfaces. There is most likely a volume effect at work in geometrical notches specimens, where crack initiation occurs near the constructed notch region, rather than near the most severe surface stress concentration. Furthermore, even though the fatigue characteristics and local stress concentrations, $K_t$, are considerably different, the notch sensitivity, *q*, for SLM and EMB Ti6Al4V with rough as-built surface is comparable.

    Razavi et al. [106, 107] assessed the fatigue strength of Ti-6Al-4V smooth, circular notched and blunt V-notched samples produced by SLM. The circular notched samples have a radius of 5 mm at the notch and a thickness of 3 mm, while the radius of the blunt V-notch sample at the notch tip is 1 mm as shown in Fig. 22d. The result of fatigue tests, as shown in Fig. 25, shows that the fatigue strength of double circular notched and blunt V-notch specimens at 1 million cycles was 213 MPa and 144 MPa compared to a value of 243 MPa related to the fatigue strength of smooth samples. A very low notch sensitivity was measured even though fatigue specimens were weakened by the notch.

    Benedetti et al. [108] explored the effect of manufacturing-induced defectiveness on the notch fatigue and crack growth resistance of SLM Ti-6Al-4V ELI. To study the influence of notch, four batches of axisymmetric V-notched coupons are fabricated via SLM adopting same machine, raw material and process parameters used to build the plain specimens in Fig. 22e. Note that the notch is either directly manufactured via SLM or machined by turning AMed bars. It's worth mentioning that the T-N variant's fatigue strength is lower than the SLM-N variant's, particularly in the medium cycle fatigue regime, thanks to the higher hardness

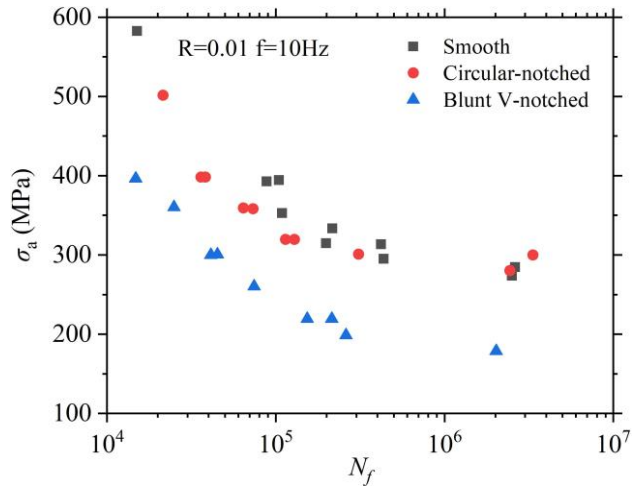

**Fig. 25.** Fatigue life of smooth, circular notched and blunt V-notched samples (Fig. 22(d)). The data comes from Ref. [106, 107].

and lower defectiveness of the SLM-N with respect to the T-N sample.

In addition, Razavi et al. [107, 112] investigated that effect of build thickness and geometry on quasi-static and fatigue behavior of Ti-6Al-4V produced by EBM. As shown in Fig. 26, a scatter ($T_\sigma$) of unnotched specimens can be observed regardless of the build thickness $t$. Lower surface roughness values were obtained for EBM specimens with bigger (5 mm) thickness. As a result of the consistency of the surface condition, the scatter bands of fatigue data in these specimens were smaller than those with the lowest build thickness. On the other hand, the specimens with 1 mm thickness have the highest scatter index, $T_\sigma$ (1.38 for unnotched specimens, and 1.28 for notched specimens), while the scatter values of the specimens of 3 and 5 mm thickness with the same geometry were comparable. In all cases, the inverse slope of notched specimens were lower than that of the unnotched specimens due to the geometrical discontinuities, indicating a steeper slope of S-N data.

### 2.3.3. Print cell or mesh

By optimizing the buckling and bending deformation through cell shape design, Ti-6Al-4V cellular solids with high fatigue strength and high strength can be fabricated by the AM technique. The print cell or mesh effect on fatigue life has attracted a great deal of attention [113–117] since the combination of AM technologies and topology. As shown in Figs. 27 and 28, the porous structures (cell) and S-N behavior of AM Ti-6Al-4V are summarized.

Yavari et al. [113] studied the fatigue behavior of porous structures made of SLM Ti6Al4V ELI. The specifications of these porous structures including the pore size and strut thickness were different between the four different porous structures, resulting in four different values of porosity (Fig. 27a, truncated cuboctahedron type). This study found that low-thickness struts may accumulate strain within porous structures, resulting in a much lower fatigue life than solid Ti6Al4V where there are no weak links. An intuitive result that low-porosities samples have a superior fatigue property than high porosities can be observed. Note that, the residual thermal stresses caused by selective laser melting process could lower the fatigue resistance of porous structures. One could use heat treatment for removal of residual thermal stresses. Further, the S-N behavior of SLM Ti6Al4V ELI with three different types of repeating unit cells, namely cube, diamond, and truncated cuboctahedron (Fig. 27c) is investigated [114]. As shown in Fig. 28, there is a strong correlation between the type of unit cell and the porosity of porous biomaterials when it comes to fatigue properties. For the same level of normalized applied stress, the truncated cuboctahedron unit cell resulted in a longer fatigue life as compared to the diamond unit cell. Similarly, porous structures based on truncated cuboctahedra and diamond unit cells have a longer fatigue life than the porous structures based on rhombic dodecahedra (determined in a previous study [113]).

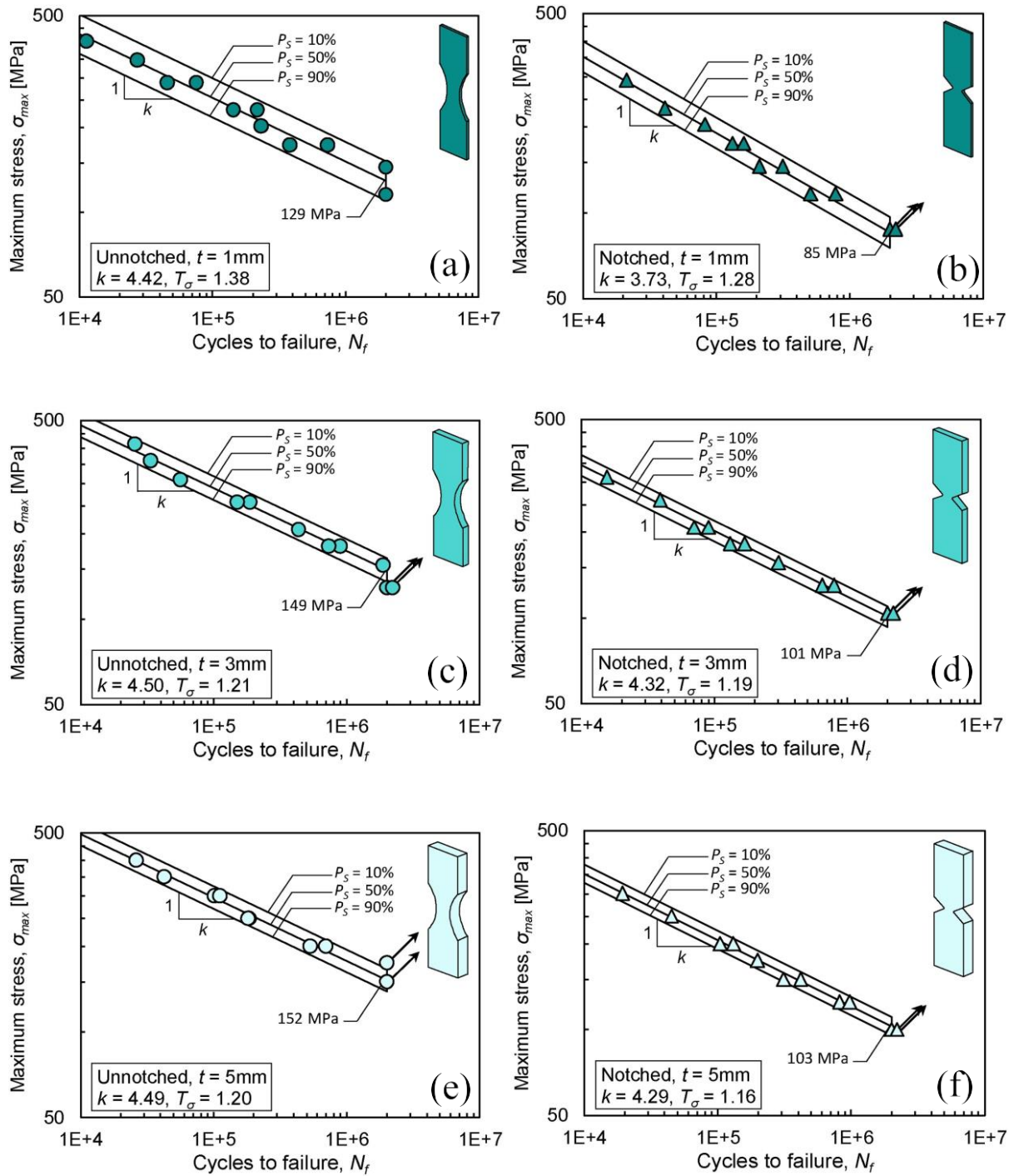

**Fig. 26.** S-N curves from different specimen geometries made by EBM process tested under loading ratio of R = 0 and 30 Hz loading frequency (*k* is the inverse slope, and $T_\sigma$ is scatter index.) Reprinted from [112], Copyright(2020), with permission from Elsevier.

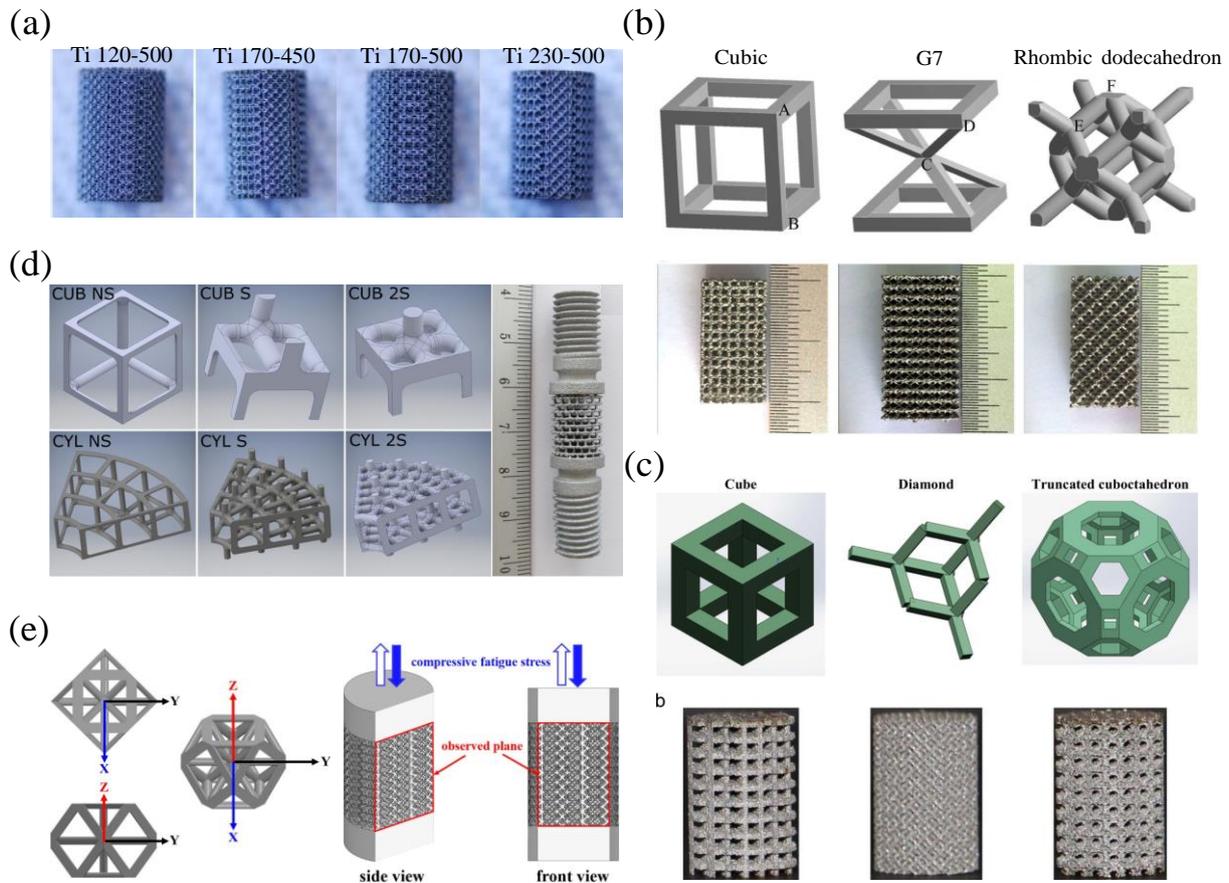

**Fig. 27.** Geometries of different built structures and notched specimens. (a) Geometries of different porosities[113]. (b) Geometries in structures [115]. (c) Cubic, G7 and rhombic dodecahedron element in the materialize software (top) and the corresponding Ti-6Al-4V prototype blocks fabricated by EBM method (bottom) [114]. (d) CAD models of the unit cells of the six cellular structures studied[116]. (e) The unit cell of the new cuboctahedron COH-Z structure (left). The schematics of the sandwiched cylindrical specimens used for the compressive fatigue test (right) [117].

Zhao et al. [115] studied the influence of cell shape on the compressive fatigue behavior of EBM Ti-6Al-4V mesh arrays the three kinds of meshes (cubic, G7 and rhombic dodecahedron), see Fig. 27b. Cyclic ratcheting was the dominating mechanism that determined the compressive fatigue of the studied meshes. The effect of rough surfaces and pores in the mesh struts can be reduced by increasing the buckling contribution or decreasing the bending component. In addition, for the compression-compression fatigue strength, Hrabe et al. [118] manufactured a regular EBM periodic porous Ti-6Al-4V to reduce the effects of stress shielding in load-bearing bone replacement implants (e.g., hip stems). The normalized fatigue strengths from 0.15 to 0.25 are achieved during the compression-compression fatigue testing (15 Hz, R = 0.1), which is lower than the expected value of 0.4 for solid material of the same acicular $\alpha$ microstructure. As a result, these microstructural changes significantly alleviated the stress/strain concentration effects and improved the compressive fatigue performances.

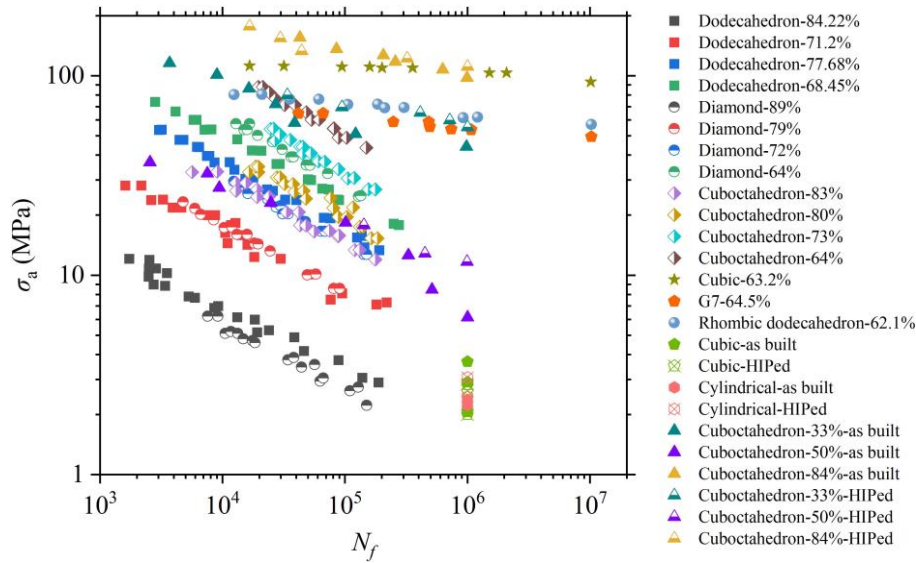

**Fig. 28.** Fatigue life of various AM porous-structure Ti-6Al-4V samples for Fig. 27. (The data comes from Ref. [113–117].)

Dallago et al. [116, 119] investigated the effect of HIP and unit cells (see Fig. 27d) on the fatigue properties of cellular lattice structures SLM Ti-6Al-4V. Surface irregularities, such as sharp notches at the strut junctions, seem to have a greater impact on fully reversed fatigue resistance than internal porosity. Thus, the approaches eliminating the porosity, such as HIP, are seen to not work in the fatigue resistance. FE analyses of the stress distribution at the strut junctions indicate that the as-built junctions induce a higher stress concentration, although not as high as expected from the experimental analyses. It is likely that fatigue strength is negatively affected by other factors, such as tensile residual stress. Furthermore, shifting the unit cells (staggering) to introduce bending actions in the struts improves fatigue resistance because thicker struts and fillet radii are necessary in order to maintain a constant elastic modulus, and thus lower stress concentrations are achieved.

In 2020, Wu et al. [117] investigated the effects of porosity (33 vol%, 50 vol%, and 84 vol%) on the compressive fatigue performance and fracture mechanism of an SLM Ti-6Al-4V cellular material with a new cuboctahedron COH-Z unit cell (see Fig. 27e). This unit cell is based on the traditional cuboctahedron structure. Three additional straight struts along the X, Y, and Z axes and four diagonal struts were added into the cuboctahedron structure. To investigate the effect of porosity on the compressive fatigue properties of the new cuboctahedron structure, cylindrical specimens with a sandwiched architecture were prepared, as illustrated in Fig. 27e (right). As a function of fatigue cycle, the ratcheting strain rate was slow when the stress cycle was less than 70% fatigue life. As a function of fatigue cycle, the ratcheting strain rate was slow when the stress cycle was less than 70% fatigue life. Strain localization along a direction inclined at 45 to the stress axis was observed at about 95% fatigue life. The shear band was extended with fatigue cycles increases and finally led to the high ratcheting strain rate and rapid fatigue failure. Consequently, the more stress/strain concentrators is introduced with increasing the porosity of the SLM Ti-6Al-4V cellular material. The fatigue stress/strain could not easily be relieved by plastic deformation due to the low ductility of $α_0$-martensite, which explains the deterioration of the fatigue endurance performance with increases in porosity.

### 2.3.4. Processing parameters

Processing parameters include manufacturing technologies, building orientation, print parameters, and print strategy. The effect of processing parameters on fatigue life or limit is investigated in detail and as follows.

**Printing parameters.** Kumar et al. [41, 120, 121] have studied the microstructure, mechanical properties, and HCF of different combinations of SLM process parameters utilized for SLMed Ti-6Al-4V. Four various combinations are 3090, 3067, 6090, 6067 that meaning Layer thickness 30 mm and scan rotation 90 (3090) and so on. Results shows that the overall porosity gets reduced when the layer thickness is increased from 30 to 60 mm (with a consequent reduction in energy density), the spatial distribution of pores changes from aligned to random when the scan rotation is changed from 90 to 67. Further, the fatigue strengths of 3090 and 3067 samples are considerably lower than those of 6090 and 6067, possibly due to the fact that the probability of finding a pore, whose size exceeds a critical value, near the surface is significantly lower in the later two.

Further, Du and Gong et al. [80, 122] investigated the effect of processing parameters of SLM Ti-6Al-4V on fatigue life, as shown in Fig. 29. Fig. 29 presents the obtained S-N data for the 10 groups of specimens in terms of stress amplitude versus fatigue life $N_f$ . It is seen that the S-N data for the 10 groups vary greatly. The cycles to fatigue failure are between 105 and 109 for all specimens, which cover the HCF and VHCF regimes. A sharp decrease in fatigue strength occurred when the porosity of the specimens becomes larger. This implies that the porosity has a significant effect on the HCF and VHCF behaviour of the SLMed Ti-6Al-4V alloy (see Fig. 29b). According to the S-N data for various processing parameters, the optimization result of fatigue performance in Group 10 is obvious, the lower porosity and surface roughness are dominant reasons for higher fatigue limit and goals to design in SLM Ti-6Al-4V.

**Different AM techniques**. Additive manufacturing technology is mainly divided into two categories: electron beam melting and powder bed melting technology, such as PBF *vs.* DED [40, 131], SLM *vs.* EBM [80, 109, 132, 133]. Literatures [40, 87, 93, 111, 131] reported that the fatigue strength/properties of as- built DMLS Ti6Al4V is significantly higher than that of as-built EBM Ti6Al4V whether or not HIP [93, 131], because of the lower surface roughness compared with EBMed Ti-6Al-4V, as shown in Fig. 30. Chastand et al. [133] compared the fatigue properties of Ti-6Al-4V specimens built by EBM and SLM, and found that EBM and SLM parts have approximately the same fatigue properties, which are equivalent to conventional casting processes, as shown in Fig. 30. Due to the pores contained in the as-fabricated samples, the fatigue strength of both EBM and SLM Ti-6Al-4V samples is lower than those of cast and annealed alloys. After HIP treatment, most of pores in EBM and SLM samples are closed and the fatigue strength is significantly improved to above 550 MPa [109]. HCF properties of as-built EBM [28] and stress relieved SLM specimens are very similar [132], SLM and EBM samples show comparable fatigue strength. But the polished specimens with HIP in both process have almost the same fatigue strength, which suggests the post-processing could eliminate fatigue gap between EBM and DMLS additive manufactured technologies, as shown in Fig. 31.

In addition, EBM sample was inferior to SLM one [123] because of the lamellar phase microstructure in EBMed Ti-6Al-4V. Compared to the AM processes utilizing laser as an energy source, EBMed parts are expected to have lower residual stresses due to the much higher temperature environment during fabrication.

On the other hand, the comparison in fatigue properties between AMed and conventional Ti-6Al-4V

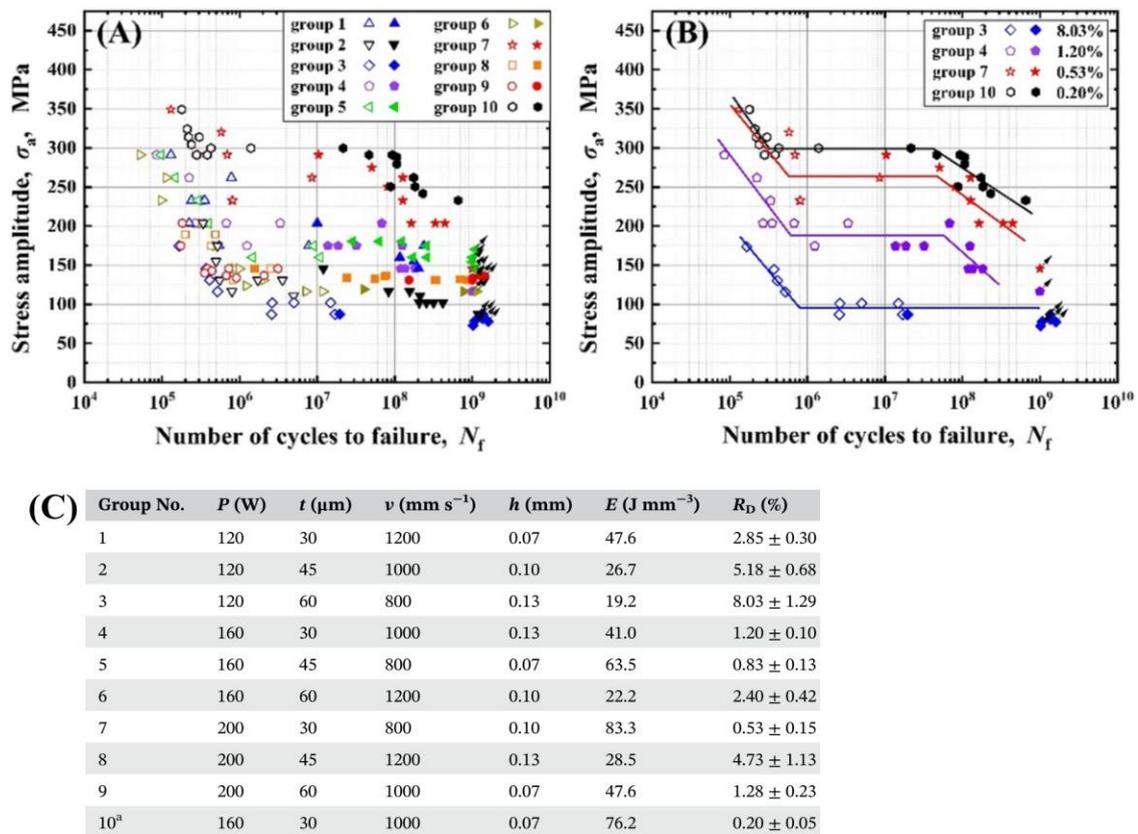

**Fig. 29.** (A) $S - N$ data of 10 group specimens in terms of stress amplitude versus number of cycles to failure and (B) $S - N$ curves of Groups 3, 4, 7 and 10; hollow symbols indicating surface crack initiation, solid symbols indicating internal (subsurface) crack initiation and symbols with arrow indicating run-out specimens; Group 10 being the validation test [122].

is necessary. In general, the fatigue properties of traditional manufactured Ti-6Al-4V are superior to the lots of AMed Ti-6Al-4V, Hot-rolled *vs.* SLM [125], wrought annealed *vs.* SLM [79], wrought annealed *vs.* EBM [126], wrought *vs.* SLM [59], wrought *vs.* machined SLM [48], wrought *vs.* EBM [50] because of lager porosity. However, Kasperovich et al. [134] conducted a careful optimization procedure of the SLMed process parameters to obtain a high-quality Ti-6Al-4V, which reduces the porosity by 6-10 times. Further, the hardness, tensile properties and high cycle fatigue resistance of all samples were tested and compared with wrought TiAl6V4 alloy. As a result of this optimization, a significant improvement of fatigue resistance compatible to the wrought TiAl6V4 for the SLM produced material (including machined and HIP) was achieved.

Where, laser-sintered (as-built) [135], machined L-PBF [136], HIPed and polished Ti-6Al-4V with EBM [105] are comparable in fatigue to conventional wrought Ti-6Al-4V. This is could because the reduction of internal defects through HIP and improved surface finish through machining and polishing. In addition, Molaei et al [137] considered that the fatigue behavior of the AM machined surface HIPed material under different loading conditions, and found that is similar to that of the wrought material and no additional cyclic hardening due to out-of-phase loading was observed for the AM HIPed material. Further, they [138–140] studied that fully-reversed torsional fatigue tests on machined AM Ti-6Al-4V with different surface condi-tions, build orientations, and heat treatments. As well as, favourable post-processing could help to make fatigue performances superior or possible for AMed Ti-6Al-4V to overtake the traditional Ti-6Al-4V. The fatigue properties of SLM Ti-6Al-4V (machined+HIP) [141] was not lose to wrought sample. These data can be summarized in Figs. 30 and 31, which presents a relatively large scatter. We found that the SLM and DMLS have more potential to obtain high fatigue properties.

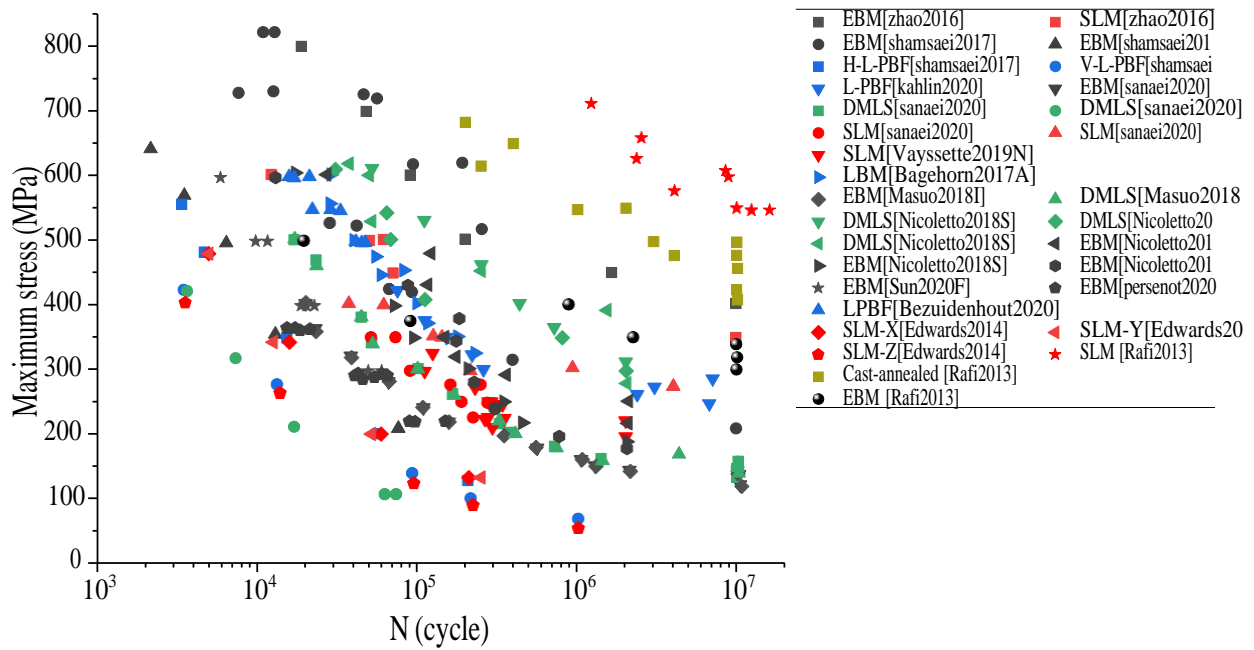

**Fig. 30.** S-N curves of ABed Ti-6Al-4V sample manufactured by various technologies, data from [40, 45, 48, 109, 123–128].

**Built orientation.** A great deal of research [40, 56, 59, 79, 137, 145–148] indicate the effect of built orientation on fatigue performance is a matter of opinion. The effect of three built orientation, including horizontal, vertical and 45 [40, 56, 59, 79, 137] on fatigue properties of AMed Ti-6Al-4V are investigated.

Zhao et al. [109] investigated the mechanical and fatigue properties of EBM and SLM Ti-6Al-4V parts. EBM and SLM Ti-6Al-4V samples in the vertical orientation possess higher ultimate tensile strength, yield strength, and better ductility than those in the horizontal orientation. The difference is mainly attributed to a change in tensile axial direction relative to the orientation of the prior $\beta$ grains. Qian et al. [130] investigated that the effect of building orientation (0, 45, and 90) on the very-high-cycle fatigue (VHCF) response of SLMed Ti-6Al-4V. As shown in Fig. 32 (partly), the fatigue strength of specimens manufactured at 0 is the highest, followed by those manufactured at 45 and 90. In HCF regime the difference among the three orientation specimens is evident, but there is almost a negligible difference in the VHCF regime owing to the similar microstructure. Further, VHCF response [16] of EBMed Ti-6Al-4V also was investigated and concluded that, HIP enhances the VHCF strength, becoming comparable with that of traditionally manufactured Ti-6Al-4V alloy. EBM process induces higher anisotropy along the building direction. And the VHCF strength of vertically built specimens is in average 40% lower than that of horizontally built specimens. Sun et al. [59] indicate build direction has a significant influence on fatigue performance of SLM Ti6Al4V titanium alloy. Fatigue strength of the 0 and the 90 sample at $10^7$ cycles are 300 MPa and 350 MPa respectively.

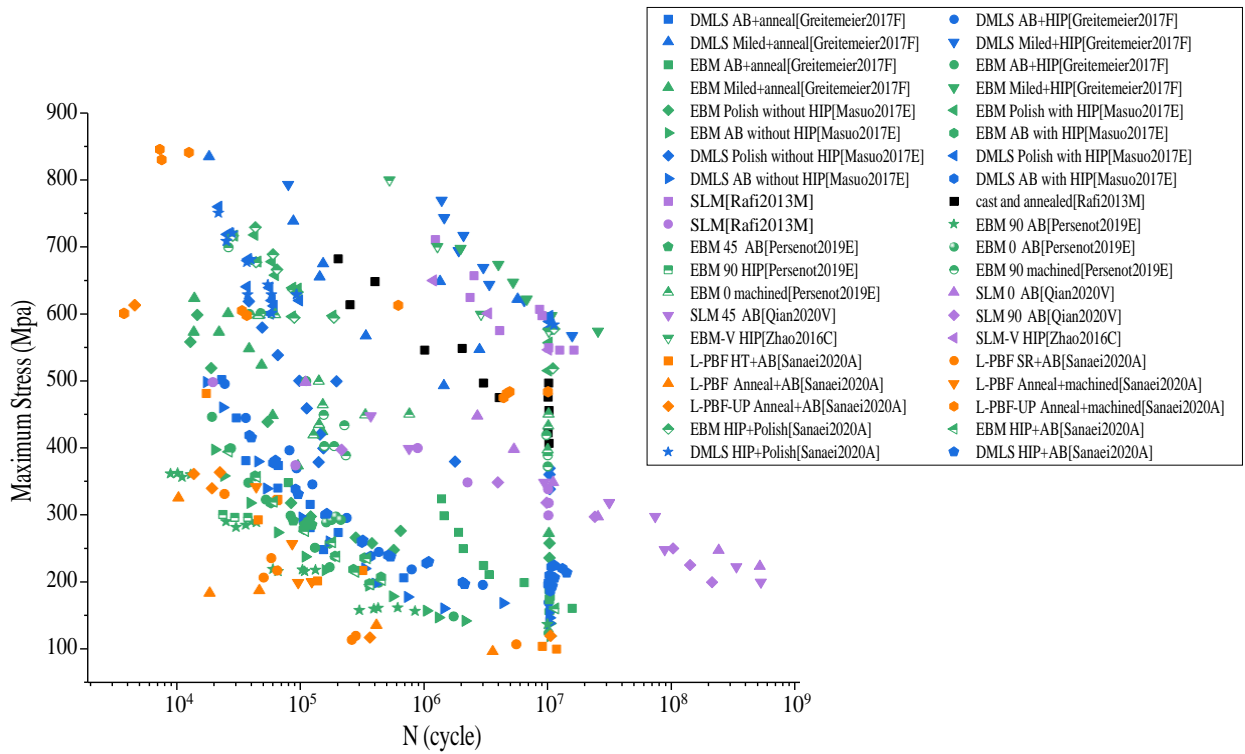

**Fig. 31.** S-N curves of treated Ti-6Al-4V sample manufactured by various technologies, data from [42, 56, 87, 109, 123, 124, 129, 130].

The DMLS and EBM[40, 149], SLM [79, 130],laser beam additive layer manufacturing (ALM) [143], L-PBF [48] Ti-6Al-4V has been observed superior fatigue properties (fatigue life and fatigue limit) on the horizontal orientation. These can be summarized in Figs. 32 and 33. Noticeably, the subsequent post-processing plays a significant roles in fatigue properties of AMed Ti-6Al-4V. Fatigue performance of as-produced or AB L-PBF Ti-6Al-4V test coupons are mostly influenced by the surface condition which in turn is influenced by the sample's building orientation with vertically built samples over performing the 50 and 0 ones [150]. However, Machining post-process yield the highest fatigue performance by completely removing the typical L-PBF induced surface waviness. In this case, 0 specimens outperformed 90 and 50 ones. For machining coupons the failure mechanism is driven by surface and sub-surface pores, as shown in 34.

As well as, anisotropic fatigue behavior was observed in the four-point bend fatigue tests [151]. In the vertical orientation (where the long axis of the sample is perpendicular to the build plate), the flexural fatigue strength was limited by the inter-layer melt defects that served as early crack-initiation sites. Samples built in the horizontal orientation (long axis parallel to the build plate) displayed comparable fatigue performance to the wrought Ti-6Al-4V, indicating that fatigue is affected by the orientation of LOF melt-related defects, also reported consistent with [149] The fatigue limit of horizontal samples is quiet higher than the vertical one under same stress level [152] Further, fatigue test of [144, 153] results showed that the specimens subject to stress directed parallel to the layers exhibited basically identical fatigue behavior. The fatigue lifetime of specimens oriented in the vertical direction is significantly lower. This result is in agreement with the finding published in [48, 148, 154] where the strong effect of interlayer defects on the initiation of fatigue cracks was pointed out is assumed.

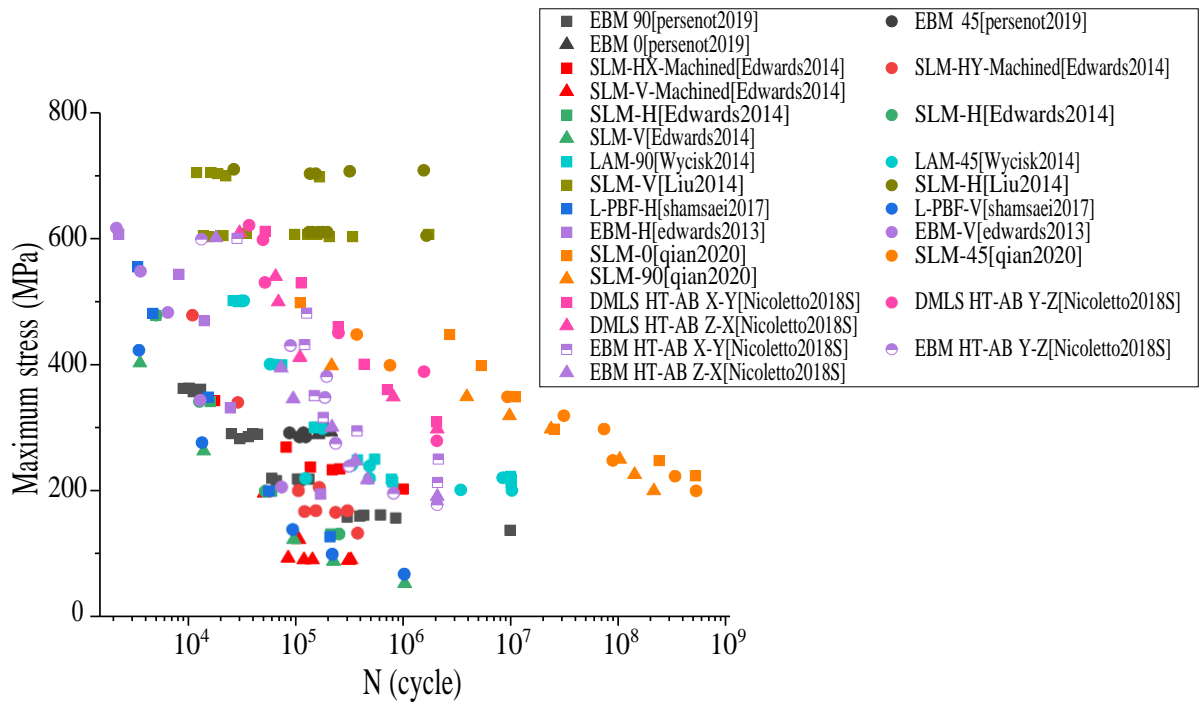

**Fig. 32.** S-N curves of built orientation in ABed Ti-6Al-4V sample, data from [40, 45, 46, 48, 56, 79, 130, 142].

The effect of different loadings on orientation anisotropy is discussed, Molaei et al. [34] studied the axial, torsion, and multiaxial fatigue behavior of the AMed Ti-6Al-4V specimens. Under all loading conditions, the AM Ti-6Al-4V without any further surface and/or thermal post processing treatment exhibited significantly short fatigue lives, whether or not the build orientation. Sun et al [145] investigated effects of the build direction (0 sample, 45 sample, and 90 sample) on mechanical performance of L-PBFed Ti6Al4V under different loadings (torque, bending, and shear loading). In addition, Edwards et al [46] found that peening of the vertical build orientation specimens provided a noticeable improvement in fatigue strength at higher cycles, but this improvement was not seen on the horizontal peened specimens. Walker et al. [58] found that SLMed Ti-6Al-4V specimens with different build orientation under constant amplitude cyclic loading displayed very significant scatter/variability in total fatigue life. Renzo et al. [155] studied multiaxial fatigue behavior of SLMed Ti-6Al-4V alloy under different loading conditions. Small differences in lifetime are observed between horizontal and vertical building directions because of different defect shapes regarding the loading axis [133]. As well as, the notch fatigue strength of heat treated Ti6Al4V was found to depend on build direction [153]. Chern et al. [15] found that build orientation effects are not evident after machining, but may exist due to inclusions of partially-sintered powders. The fatigue life gap between various built orientation may be weakened by post-processing.

**Scanning strategies.** Syed et al. [156] investigated the influence of deposition strategies (oscillation and parallel pass) on tensile and fatigue properties in WAAMed Ti-6Al-4V. As shown in Fig. 35, the fatigue strength was comparable with its wrought counterpart and greater than typical material by casting. At $10^7$ cycles, fatigue strength of 600 MPa was achieved for the oscillation build vertical samples and the parallel pass build in both orientations. Only the oscillation build horizontal samples had lower fatigue strength of 500 MPa.

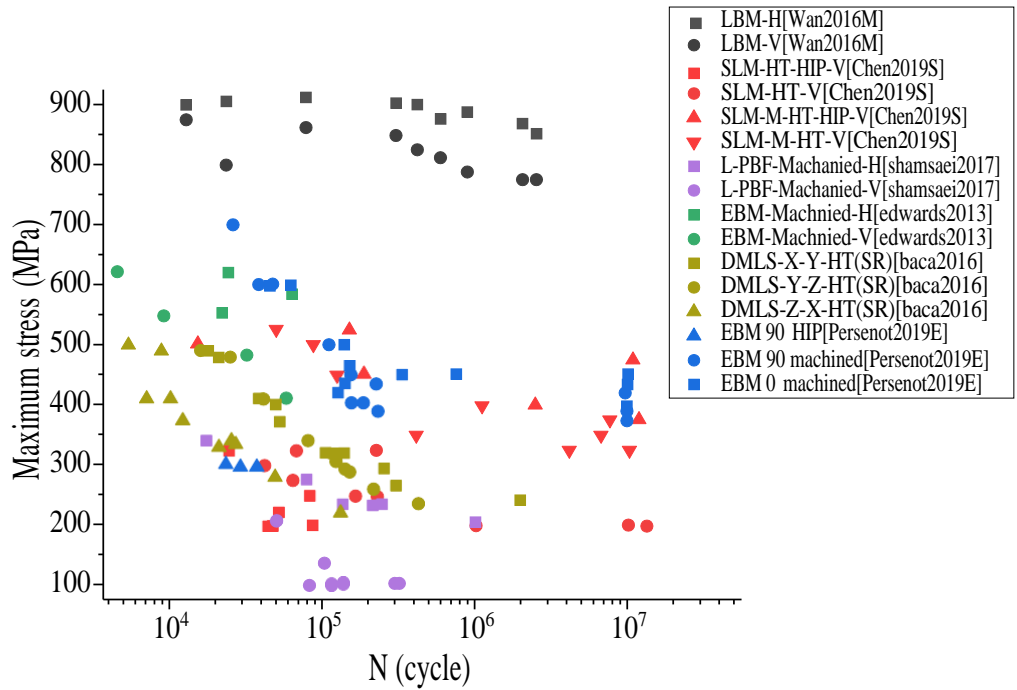

**Fig. 33.** S-N curves of built orientation in post-processing Ti-6Al-4V sample, data from [45, 46, 56, 141, 143, 144]. SR represents stress release

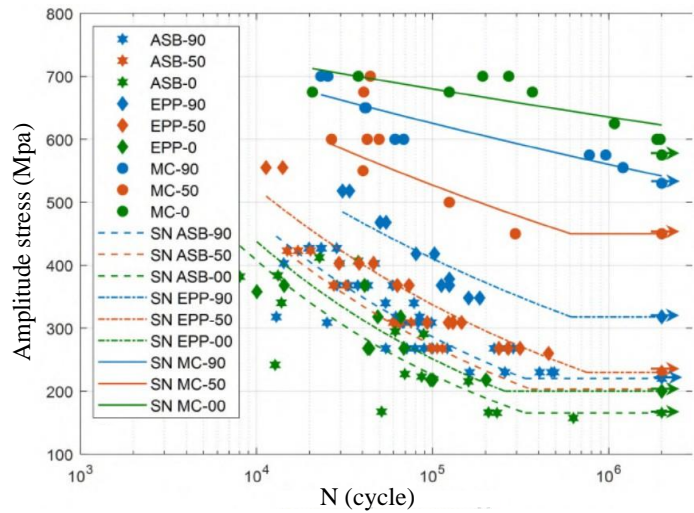

**Fig. 34.** S-N curves of built orientation effect in post-processing Ti-6Al-4V sample. ASB:As-produced, EPP:Electro-plasma polish, MC:Machined. Reprinted from [150], Copyright(2020), with permission from Elsevier.

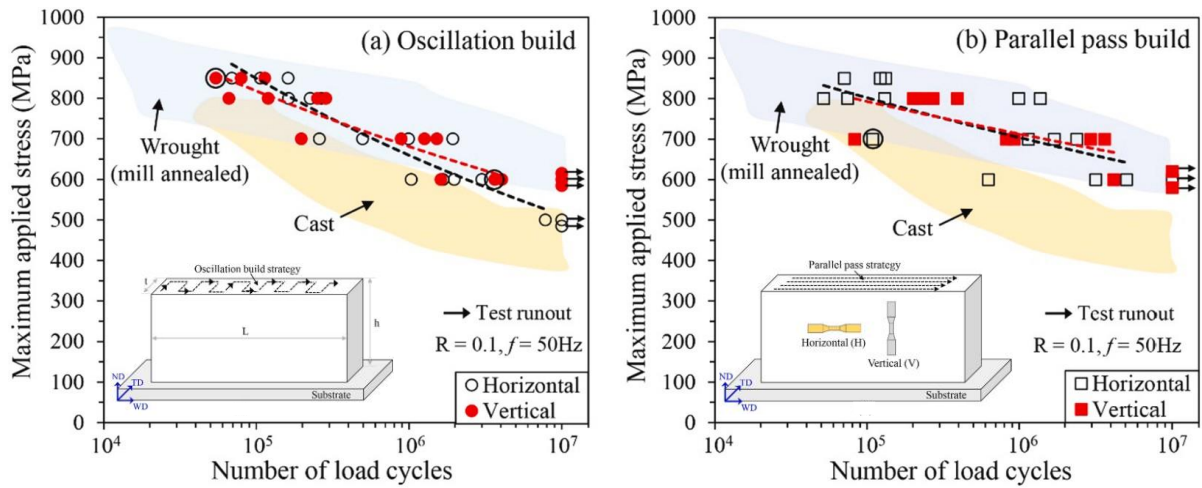

**Fig. 35.** Maximum applied stress *vs.* load cycles (S-N curves) for oscillation and parallel pass build strategies Encircled data points indicate crack initiation from a surface pore. Reprinted from [156], Copyright(2021), with permission from Elsevier.

The effect of processing parameters, involving print parameters, built orientation, and scanning strategies [156, 157], are analysed from the viewpoint of surface roughness and S-N curves. There is no apparent difference and large scatter in fatigue strength between the various AMed technologies. This is because the influence of processing parameters and post-processing are more crucial in fatigue properties. Selecting appropriate processing parameters and build orientations and/or post fabrication heat treatment processes could eliminate or significantly decrease residual stresses produced during the fabrication process, and therefore, their effects on fatigue performance.

### 2.3.5. Post-processing

In this part, the post-processing include the thermal treatment, (e.g. hot isostatic pressing (HIP), heat treatment (HT)), and surface treatment (polish, machined, as-built, peening). The fatigue properties of AMed Ti-6Al-4V are greatly depend on the microstructure, surface roughness, residual stress, and porosity or defects/pores after post-processing.

Molaei and Fatemi et al [136–140, 158, 159] conducted lots of researches on fatigue behavior of AMed Ti-6Al-4V under different condition, including variable amplitude service loading conditions (torsional and uniaxial fatigue behavior), notched, build orientation, surface roughness and HIP. Li et al. [158] have attempted to catalog and analyze the published fatigue performance data of an additively manufactured alloy of significant technological interest, Ti-6Al-4V. Focusing on uniaxial fatigue performance, they compare to traditionally manufactured Ti-6Al-4V, discussing failure mechanisms, defects, microstructure, and processing parameters. On the other hand, for torsional fatigue behavior, they [136, 137, 159] found that all AM samples with different conditions (as-built, as-built annealed, and machined annealed) showed higher monotonic shear strength (yield and ultimate), but lower shear ductility than the wrought material. This was consistent with the findings in the literature for axial loading. At the same time, these specimens showed shorter torsional fatigue lives compared to the wrought material, when compared based on shear strain amplitude. Due to the significant softening of the wrought material in LCF, the AM annealed material had longer lives than the wrought material in this life regime, when compared based on shear stress amplitude. Annealing or other heat treatment were found to improve torsional fatigue life of AM specimens by more than an order of magnitude. Effects of defects, surface roughness, HT, and HIP [160, 161] on fatigue properties of AMed Ti-6Al-4V are also investigated.

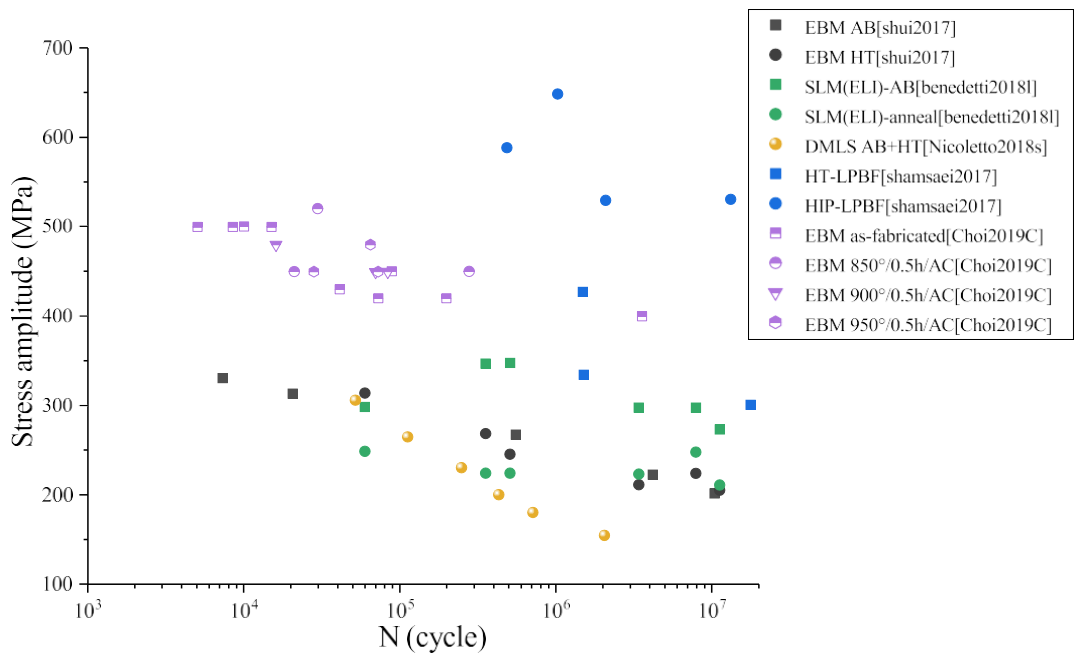

**Fig. 36.** S-N curves with anneal and HT AMed Ti-6Al-4V samples, AC: air cooling, data from [40, 45, 162–164].

#### 2.3.6. Heat treatment

Additively manufactured fatigue samples failed predominantly due to process-induced defects, and porosity and internal defects play important roles on the fatigue behavior of AM parts. They might be preventable, if best-practice measures are taken during manufacturing. HT has positive influence on eliminating internal defects and further improving fatigue properties [41, 121]. The influence mechanism is reducing the detrimental effect of internal defect, as well as relaxing or reducing tensile residual stresses generated during the fabrication process, e.g. SLM/SLS [44]. However, report in literature [162] indicate that the HT produced no apparent improvement in the fatigue properties, as shown in Fig. 36.

Previous studies have suggested heat treatment plays a significant role in fatigue properties of AMed Ti-6Al-4V, Choi et al. [164] advised that appropriate post heat treatment could have positive effects on enhancing the high cycle fatigue of additive manufacturing parts by control the surface roughness and residual stress around the pore, as shown in Fig. 36. However, Benedetti et al. [163] observed the annealed SLM Ti-6Al-4V has significantly lower fatigue strengths compared to ABed SLM Ti-6Al-4V.

**Pores and Lack-of-Fusion (LOF) defects.** Pores, defect, LOF, and notched are the most straightforward factors of fatigue. Defects or pores in AMed components usually are inevitable during the manufactured processing, as well as the effect of defects or pores on the fatigue properties is not neglected. The defects were mostly defects made by LOF [42]. Therefore, investigating the defects influence on fatigue behavior is required for robust designs and engineering applications of high performance AMed components [30]. The experimental results [30, 42, 165] revealed that the fatigue cracks initiated from LOF defects in the AMed Ti-6Al-4V specimens. Due to its morphology the LOF defect had the most detrimental influence on fatigue life. It is expected that the elimination of LOF defects would substantially increase the fatigue life of the SLMed Ti-6Al-4V specimens. In addition, various energy input have a strong bearing on the defect in SLM [80] and EBM [166]. Wu et al. [167] found that two types of defects including gas pores and LOF can be clearly distinguished inside SLMed Ti-6Al-4V. Fatigue crack with a typical semi-ellipse usually initiates from the defects at the surface and near the surface. Besides, the defects less than 50 $\mu$m and sphericity of 0.4-0.65 dominate for the SLMed Ti-6Al-4V alloys. It is also found that the larger the characteristic size of the defect, the lower the fatigue life. Compared with the internal pores, the LOP has a larger/greater influence on fatigue properties of AMed Ti-6Al-4V owing to the fatigue crack initiation mechanism that usually origins from surface defects. HT was found to reduce the sharpness of the LOF defects and to improve fatigue performance in L-PBFed Ti-6Al-4V [168]. In particular, fatigue is a localized phenomenon in which its failure mechanisms are largely driven from the stress concentration caused by manufacturing-induced defects. Therefore, understanding the correlation between the manufacturing parameters/post-processing and the manufacturing-induced defect distribution is considered to be an important step to minimize and control these material anomalies within AM parts, which eventually alleviate the scatter in their fatigue resistance.

Akgun et al. [169] investigated the fatigue of WAAMed Ti-6Al-4V in presence of process-induced porosity defects and found that pores larger than the 85 $\mu$m diameter reduced fatigue life. But, they did not act as cracks obeying the Kitagawa–Takahashi diagram, which is a common assumption in the field. Commonly seen dispersion in S-N data (see Fig. 37) of AMed Ti-6Al-4V were associated with the defect location, such as surface pore or embedded pore, rather than the pore size for this particular alloy. Accordingly, individual S-N curves are recommended for engineering design based on defect type, which is used in the welding industry based on the joint type Wycisk et al. [142] found that, with consistent defect type the fatigue properties of additive manufactured titanium experiences low scatter (Fig. 37). High scatter in the region near the endurance limit, however, is caused by variants of defect type, sizes and location. Le et al. [31] investigated the fatigue behaviour of SLMed Ti-6Al-4V and role of defects on scatter and statistical size effect. In addition, Cheng et al. [170] investigated that effect of thermal induced porosity on high-cycle fatigue and very high-cycle fatigue behaviors of hot-isostatic-pressed Ti-6Al-4V powder components. The results show that the residual pores in the as-HIPed powder compacts present no obvious effect on the HCF life.

Importantly, the artificial/surface notched impact on the fatigue properties of AMed Ti-6Al-4V [105] are also discussed, including circular notched [106], V-notched [107], porous specimens [171], and overall notched fatigue analysis [138, 140], as shown in Fig. 38. Kahlin et al. [105] found that the fatigue reduction of AM materials could be quantified by fatigue notch factor ($K_f$). The combined effect of a rough as-built surface and a geometrical notch ($K_t = 2.5$) gives a fatigue notch factor, $K_f$, of 6.15 for LS material and 6.64 for EBM material. The notch which is built in the specimens during the fabrication process can significantly lower the fatigue performance of the material. The fatigue crack growth behavior at notches in AM parts can be very different from that in wrought metals due to the presence of surface and/or near surface defects and their interactions with the mechanically induced notch, anisotropy, ductility, and build orientation [138, 140]. AM Parts with optimized process parameters and machined or polished surfaces have a higher sensitivity to the presence of a notch when it comes to fatigue performance, whereas parts with similar surface roughness but defects are less sensitive. Notched fatigue behavior may be similar to that of notched wrought metal, and the effect of internal defects may become small when stress concentrations are caused by notches or geometry complexities.

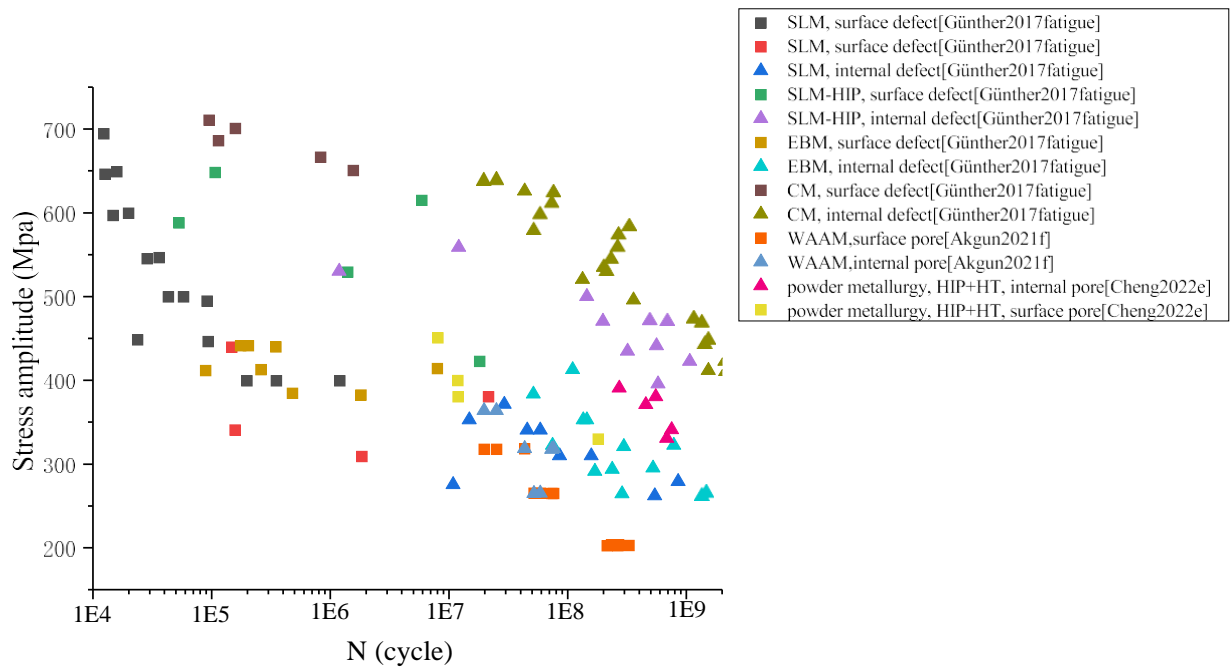

**Fig. 37.** S-N curves of AMed Ti-6Al-4V for surface defect and internal defect, data form [132, 169, 170]

**Hot isostatic pressing.** HIP is an effective thermo-mechanical treatment to minimize the volume of internal voids and porosity, which are known to be the major life limiting factors for AM parts under cyclic loading. This treatment involves applying high temperature and high pressure simultaneously to remove or reduce the size of internal defects. Yan et al [172] investigated the influence of HIP on microstructure of SLMed Ti-6Al-4V. The results show that the microstructure of SLMed Ti-6Al-4V alloy is composed of acicular martensite $α′$ phase, and the sample exhibits high microhardness and strength, but low plasticity. After HIP, acicular martensite $α′$ phase transforms into $α+β$ phase. So that the mechanical and fatigue properties shall make a difference. Fig. 39 shows the microstructures made by EBM and DMLS without HIP and with HIP. It can be observed and confirmed that most of the internal defects which were visible before HIP disappeared due to the effect of HIP.

Molaei et al. [137] and others [173, 174] investigated the HIP effect on the fatigue life of AMed Ti-6Al-4V. Comparison of the monotonic tensile and torsion behaviors of the wrought and AM HIPed materials indicates that the HIPed material follows the same trend as the wrought material. This is mainly due to the microstructural transformation and more uniformity of the AM microstructural grains after HIP process, but also attributed to the shrinkage and fusing of internal defects. The improvements after HIP process were conclusively related to closure of internal defects. Analogously, the HIP process did not alter the chemical composition of the material but did dramatically reduce the pore and void density of the material and induce minimal microstructural coarsening. Compared to as-built and stress relieved conditions, HIPing led to a more than 100% increase in the fatigue strength at $10^7$ cycles (approximately 550-600 MPa) [47]. Fatemi et al. [136] summarized the fatigue behaviour of AMed Ti-6Al-4V considering various HTed processing and loading direction effects. Experimental fatigue data for L-PBF Ti-6Al-4V after the HIP treatment indicates a similar performance to that of the wrought material. Nonetheless, the significant effect of HIP treatment

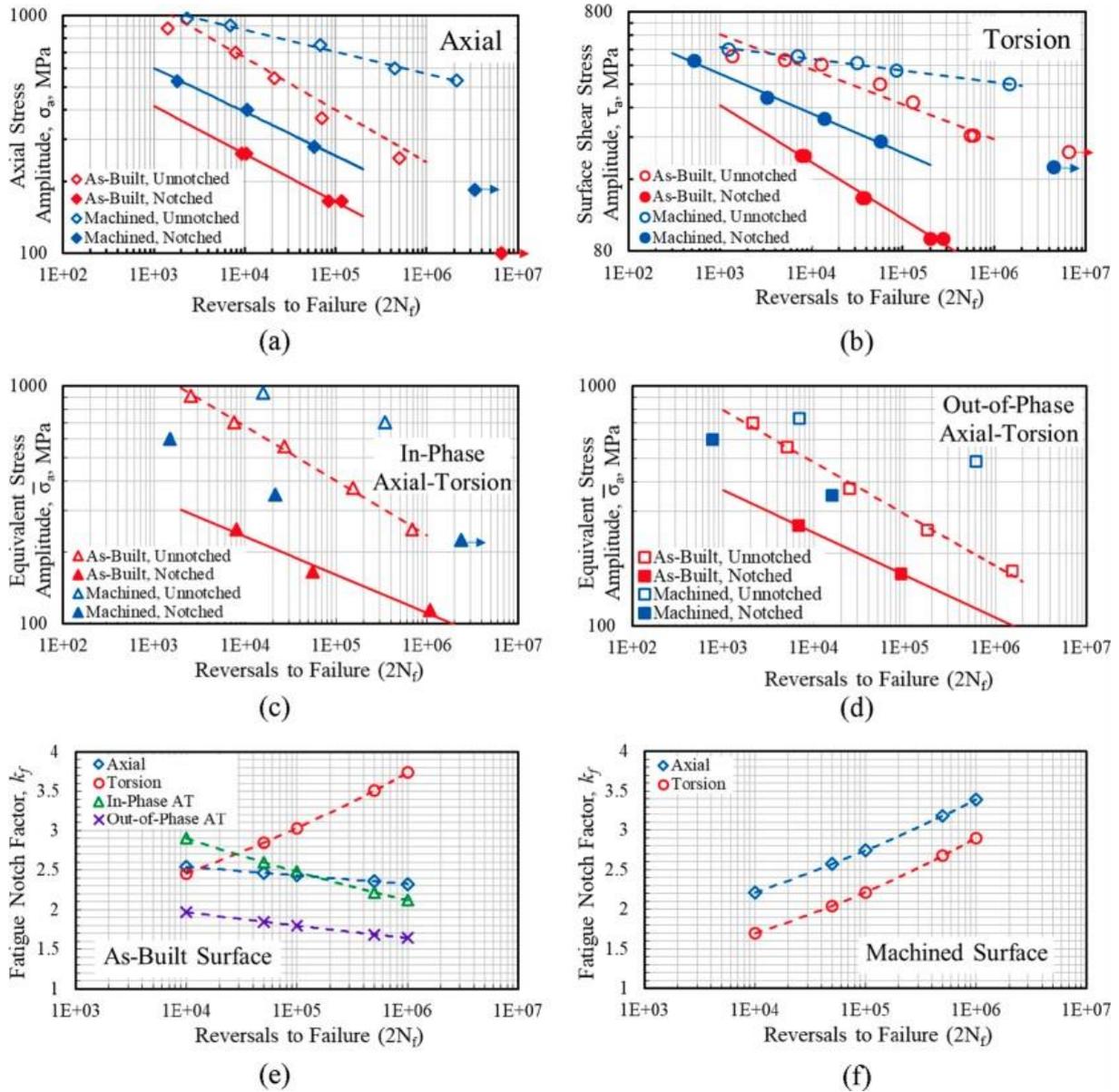

**Fig. 38.** Comparison of fatigue test results of as-built surface and machined/polished surface LB-PBF annealed Ti-6Al-4V specimens under (a) axial, (b) torsion, (c) in-phase axial-torsion, and (d) 90 out-of-phase axial-torsion loading conditions. Comparison of fatigue notch factors *vs.* life for the same materials under different loading conditions for (e) as-built surface and (f) machined/polished surface conditions. The notch was built in for as-built surface specimens and drilled for machined/polished surface specimens. Reprinted from [138]. Copyright (2021), with permission from Elsevier

can only be observed after subsequent machining since the rough surface of as-built HIPed specimens still plays a key role in shortening their fatigue lives, especially in HCF regime. Further, Li et al. [175] and Childerhouse et al. [176] investigated the mechanisms by which HIP improves the fatigue performance of PBF and EBM Ti-6Al-4V. The results suggest that HIP may act most significantly by decreasing the fraction of the defect population that can initiate fatigue cracks, both by decreasing defect sizes below a threshold and by changing the microstructure that surrounds defects. Longer HIP soak times and HIP chambers with greater pressure capacities might be more useful for improving fatigue properties. In addition, HIP treatment results in more improvements at longer lives due to the higher significance of defects.

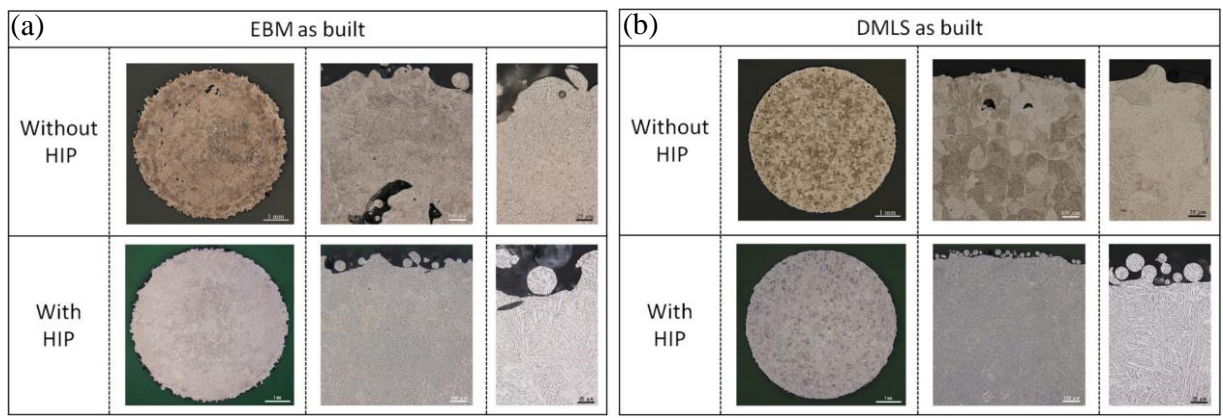

**Fig. 39.** (a) Microstructure of EBMed Ti-6Al-4V without HIP and with HIP. (b) Microstructure of DMLSed Ti-6Al-4V without HIP and with HIP [93, 131]. White area for $\alpha$ phase (hcp) and dark area for $\beta$ phase (bcc). Reprinted from [150], Copyright(2020), with permission from Elsevier.

However, the microstructure coarsening during the HIP process could have some deleterious effects compared to the beneficial effect of internal defects closure/shrinkage. The specimens in the HIPed condition had significantly longer fatigue lives as compared to the annealed samples (by about an order of magnitude in HCF). This was due to the shrinkage of the internal defects and also reducing their volume fraction after HIP treatment. Meanwhile, there were still some residual defects (unmelted particles, gas pores, and surface defect) left after this treatment which can act as crack initiation sites.

As shown in Fig. 40, [93, 111, 126, 131, 132, 137, 162, 177] investigated the HIP effect on fatigue properties of AMed Ti-6Al-4V, and found that the HIP can reduce porosity mainly owing to eliminating the internal defect (see Fig. 39). It must be noted that HIP does not improve the surface roughness. Although the reduction of surface roughness plays a critical role in improving fatigue limit. In addition, Kahlin et al. [105] also observed that no effect of HIP treatment on improving the fatigue behavior as compared to the annealed condition in the negative presence of rough surface. Similar to previous reports in the literature [87, 137], HIP treatment was found not to be an effective process in improving the fatigue behavior of notched specimens when the rough surface is not removed.

### 2.3.7. Surface post processing

The surface of additive manufactured parts can be treated with mechanical polishing, electrochemical polishing, chemical etching, and laser polishing. Mechanical polishing methods, such as milling, polishing, and blasting, have limited accessibility to geometrically complex structures made by AM processes. Laser polishing/remelting technology is another option to enhance surface quality of AM parts. A pulsed or continuous wave laser beam is used to re-melt the surface of additive manufactured parts to improve surface finish [141]. An optimum linear energy density was a key factor, SLMed machine parameters, other factors, such as particle size distribution (see secs for more details) and part spacing, can also affect surface roughness. Surface defects serve as the stress concentrators in fatigue testing of Ti-6Al-4V, which limits the fatigue life of the specimens, especially for high-cycle fatigue. Surface finishing treatments can significantly improve the fatigue life owing to the reduction in surface roughness. Compared with closing internal defects, improving the surface condition is more effective in increasing fatigue performance of selective electron beam melting (SEBM) Ti-6Al-4V [126] and SLM Ti-6Al-4V [141]. In addition, Chen et al. [141] studied the correlation between surface quality and production, as well as found that higher production of SLM process inevitably leads to a higher surface roughness. Surface roughness is the most dominant fatigue life-limiting factor for AMed Ti-6Al-4V.

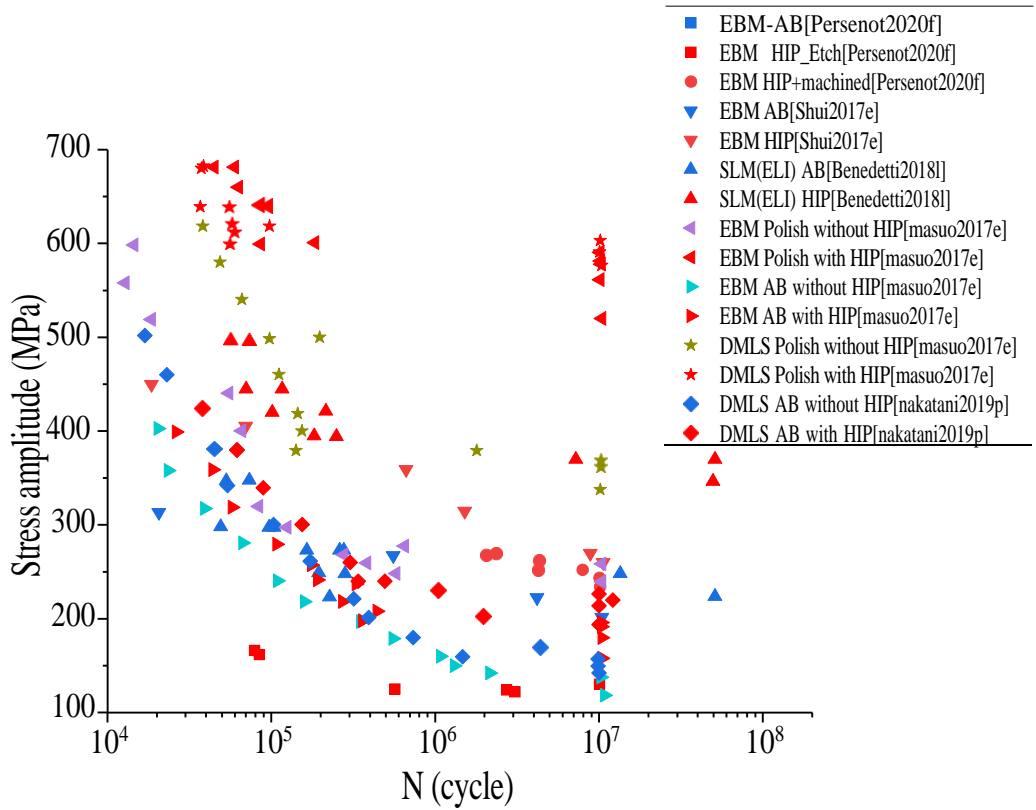

**Fig. 40.** S-N curves of AMed Ti-6Al-4V with HIP and without HIP, ELI means extra low interstitials, data from [42, 128, 131, 162, 163].

Sanaei et al. [124] and Pegues [103] in detail analysed the effect of surface roughness on fatigue performance of powder bed fusion additive manufactured Ti-6Al-4V. Arithmetic mean surface roughness parameter Ra is used here as a representation of surface roughness in work [124], since this parameter has been presented in most of the studies dealing with metal AM surface roughness. Surface roughness effect on fatigue performance is shown to be dominant even in the presence of relatively large internal defects and various microstructures. Ra decreases with an increase in power. This consequently increases fatigue limit, although scatter is observed due to the secondary effect of internal defects and microstructure. For similar microstructures, considering synergistic effect of Ra along with average critical internal defect size in combination with the stress amplitude, resulted in improved correlation of the S-N fatigue data.

**Polish, LP, LPSR (Laser polished and secondary stress relieved), M/P (Machined and polished).** Laser polishing has been widely used for polishing metals (steels, nickel alloys, titanium al-loys [178, 179], and aluminum alloys, for example). due to its high efficiency, low environmental impact, high controllability of surface roughness, high flexibility, and capability for treating a localized area. The result of literatures [93, 131, 180] show that polished Ti-6Al-4V superior to as-built one.

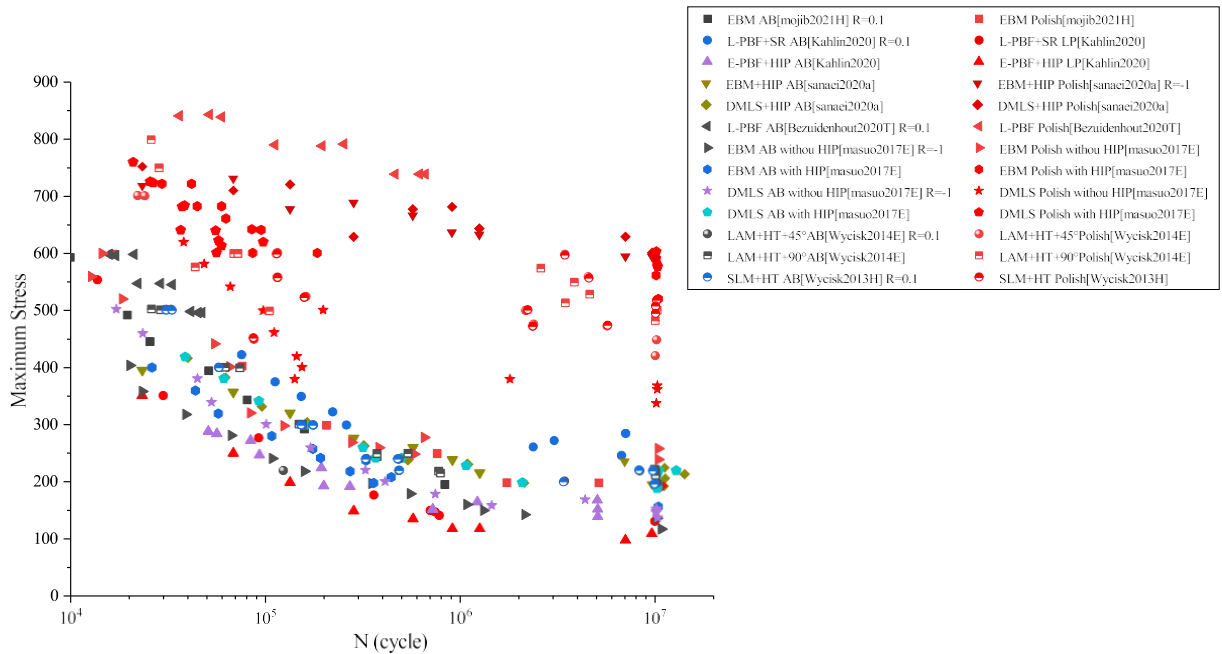

**Fig. 41.** S-N curves with ABed or polished Ti-6Al-4V samples, data from [27, 28, 39, 42, 124, 127, 142].

Wycisk et al. [27] observed that mechanical polishing alone increased the fatigue strength of SLMed Ti-6Al-4V from 250 to 500 MPa ($10^7$ cycles), even though the fatigue life data showed a high degree of scatter (see Fig. 41). Masuo et al. [93, 131] found that surface polished specimens have superior fatigue strength for DMLS and EBMed Ti-6Al-4V, whether with HIP or without HIP. The common trend of fatigue strength for all series of specimens is further expressed as follows: As-built without HIP < As-built with HIP < Surface polish without HIP < Surface polish with HIP. (see Fig41). Surface roughness has strong a detrimental influence on fatigue strength. This is right in the HCF owing to the lower surface roughness after laser polishing [180]. However, Lee et al. [180] found that laser polished AM Ti-6Al-4V had lower fatigue strengths in the low cycle fatigue regime compared to specimens with as-built surfaces because of the dominant residual stresses. Therefore, stress relieving during the laser polished specimens is an effectual approach to improved fatigue resistance for both low and high cycle fatigue regimes. Moreover, Moreover, both L-PBF and E-PBF Ti-6Al-4V [39] subjected to laser polishing showed lower fatigue strength with 30% (E-PBF) to 50% (L-PBF) compared to the AB specimen. A large number of spherical pores and subsurface defects were observed at the interface between the re-melted material and the base material, which is the main reason for decreasing fatigue strength.

Besides, Bezuidenhout et al. [127] investigated the beneficial effect of HF-HNO3 chemical polishing on the surface roughness and fatigue life of laser powder bed fusion produced Ti6Al4V. They found the process efficiencies should also be considered, especially when intending to scale production. When time is the only resource of importance, the 4 M (HF concentration) solution was the most efficient, but when other resources such as titanium alloy material and HF are considered, the efficiencies vary greatly with polishing time. Persenot et al. [181] demonstrated that chemical polishing can significantly reduce the surface roughness of EBMed Ti-6Al-4V, as a result, increase the fatigue strength by 60% (to 200 MPa). These suggest that the surface roughness is a determinant factor in a degradation of fatigue strength for as-built specimen, compared with polish one. It is noted that the defect also should contribute to fatigue fracture if there is defect near subsurface.

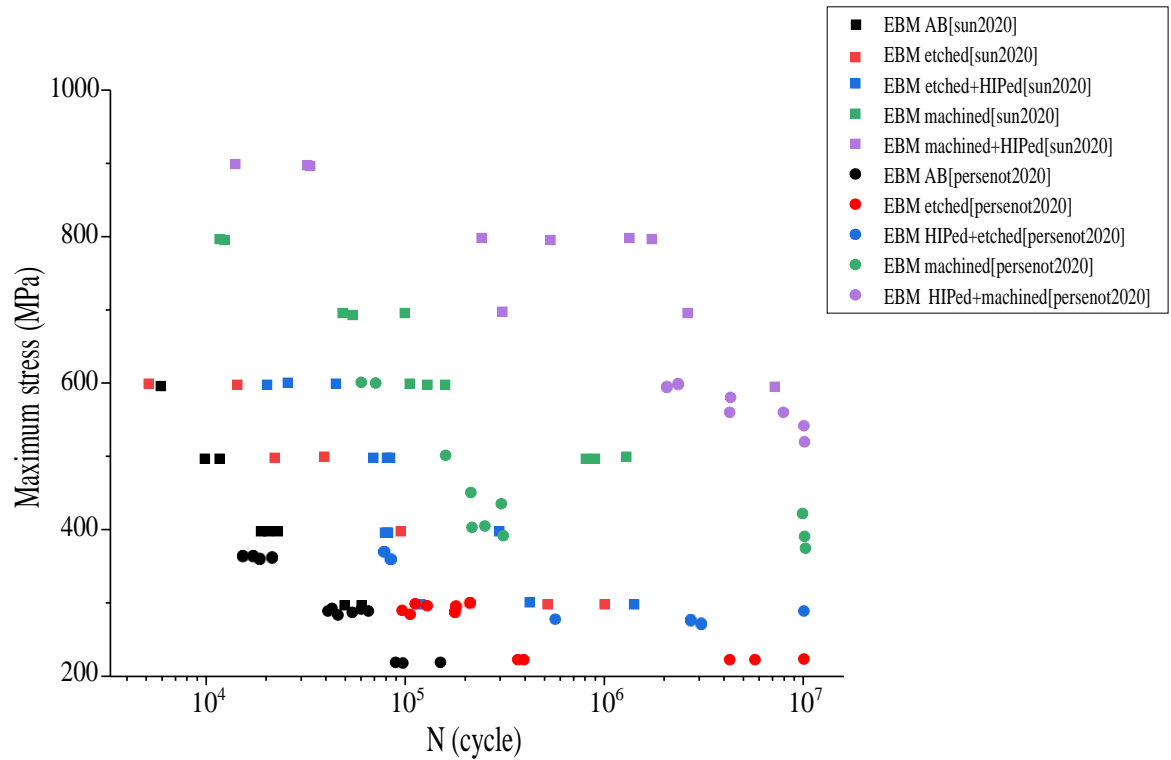

**Fig. 42.** S-N curves for EBM Ti-6Al-4V samples, data form [126, 128].

Chemical polishing/etching can penetrate into intricate internal structure, such as SLMed lattice; however it requires careful control of chemical solutions and etching time to achieve effective reduction in roughness while retaining the desired geometry of the parts [182]. In particular to chemical etching of titanium alloys, some content of hydrofluoric acid is needed, which raises safety concerns and is not environmentally friendly. Persenot et al. [181] demonstrated that chemical polishing can significantly reduce the surface roughness of EBM Ti-6Al-4V and, as a result, increase the fatigue strength by 60% (to 200 MPa). They [128] further concluded that HIPed+Machined≥Machined≥HIPed+Etched ≥Etched≥As-built, see Fig. 42. Also, Sun et al. [126] observed that chemical etching is able to enhance the fatigue performance to about 2×10$^4$ cycle with 500 MPa.

**Shot Peening (SP).** Shot peening is a cold work process used to finish metal parts to prevent fatigue and stress corrosion failures and prolong product life for the components which are subject to high alternating stress [183]. In shot peening, small spherical shot bombards the surface of the part to be finished. The shot acts like a peen hammer, dimpling the surface and causing compression stresses under the dimple. As the media continues to strike the part, it forms multiple overlapping dimples throughout the metal surface being treated. The surface compression stress strengthens the metal, ensuring that the finished part will resist fatigue failures, corrosion fatigue and cracking, and galling and erosion from cavitation.

The effect of SP on fatigue properties is mainly reflected in the following aspects: surface morphology, micro-cracking will reduce its fatigue performance, roughness, and residual compressive stress [184]. The decreases of roughness and residual compressive stress play a key role in the improvement of fatigue properties. Soyama et al. [185, 186] investigated effect of various peening methods on the fatigue properties of DMLS and EBM Ti-6Al-4V. Hackel et al. [184] demonstrated that AM material can be formed and the shape can be modified by the laser peening. The laser peening process is highly controlled and the application discrete allowing precise shape corrections by means of adding beneficial compressive stress, which becomes a major approach to improve fatigue lifetime. The fatigue performance optimization through LS/SP is also demonstrated by Ref [121, 129, 187] of the SLM Ti-6Al-4V and L-PBF Ti-6Al-4V [39], as shown in Fig. 43. The effect of shot peening on fatigue performance, including shot peening power, stress and size was investigated [188]. The fatigue life and fatigue strength of the DMLS Ti-6Al-4V specimen are higher than that of EBM specimen whether AB or SP condition [185, 186]. In the case of the DMLS specimen, the fatigue life at $\sigma_a \approx$ 450 MPa was improved by 10 times through cavitation peening after grinding, 5.9 times through cavitation peening, 4.7 times through shot peening, 3.5 times through laser peening, and 1.5 times through grinding, compared with the as-built specimen. Similarly, the fatigue life of EBM specimen was improved by 1.75 times through cavitation peening, 1.87 times through laser peening, and 1.95 times through shot peening.

Morita et al. [188] found that the fatigue strength of Ti-6Al-4V alloy was enhanced by the SP treatments. Further, as the size of collision particles and injection pressure were increased, the position where the microstructures were plastically deformed reached deeper. The improvement rates in fine particle bombarding materials were much higher than those in conventional shot peening materials because the maximum height waviness was smaller and the absolute values of compressive residual stress were higher [188]. In addition, a combination of SP and chemical assisted surface enhancement produces an almost equal compressive stress field to SP but a smoother surface, improving the fatigue life relative to SP. The improvement of the surface roughness caused all the initiation points to be located in the interior of the specimen [129, 190]. Denti et al. [191] combined tumbling and SP with the objective of improving the fatigue properties of PBFed Ti-6Al-4V, then the fatigue life improved of 150% respect to AB.

However, the persistent high roughness and possible damage due to the SP cause many tests to produce failure from a crack initiated on the surface. Wycisk et al. [27] found that SP is detrimental to fatigue performanc when SP was carried out following mechanical polishing, possibly due to the poor surface finish caused by shot peening. Noticeably, cold spray coatings (surface treatment) also have shown the ability to effectively contribute to the fatigue performance and repair of additive manufacturing cases [192].

**Laser Shock Peening (LSP).** Laser peening (LP) (differentiate to the laser polish), or laser shock peening (LSP), is a surface engineering process used to impart beneficial residual stresses in materials. The deep, high-magnitude compressive residual stresses induced by laser peening increase the resistance of materials to surface-related failures, such as fatigue, fretting fatigue and stress corrosion cracking [184, 194]. As shown in Fig. 44, the compressive stress ($\sigma_c$) is observed in the surface or subsurface, the grain size, twin, and dislocation may begin to present gradient distribution along with depth direction. The effects of LSP on fatigue properties in AMed Ti-6Al-4V, especially for fatigue limit and fatigue life, have been widely investigated in the past decade. Guo et al. [194] found that LSP modified the tensile residual stress into a compressive residual stress with the maximum value of around 200 MPa and an affected depth of 700 $\mu$m in LAMed Ti-6Al-4V. Refinement of microstructure in the surface layer after peening was observed. High density of dislocations in $β$ grain and multi-directional mechanical twins in $α$ grain account for the grain refinement and microhardness improvement. Elongation was enhanced after peening while yield strength and ultimate tensile strength showed insignificant changes. Further, Jin et al. [195] researched the effects of LSP on microstructure and fatigue behavior of EBM Ti-6Al-4V. They found the $\alpha$ and $\beta$ lamellar structure in the as-prepared sample changed to a microstructure containing equiaxed nanograins, deformation twins, and equiaxed submicrongrains of $\alpha$ phase after LSP. Then, the run-out fatigue strength (2,000,000 fatigue cycles) of the as-prepared EBM sample was 600 MPa, whereas that of the LSP-treated sample was significantly increased, at 700 MPa. Literatures [23, 84, 196] also agreed that the LSP/SP can improve fatigue properties of EBMed Ti-6Al-4V. Residual compressive stress and grain refinement during LSP play a critical role in the improvement of fatigue performance.

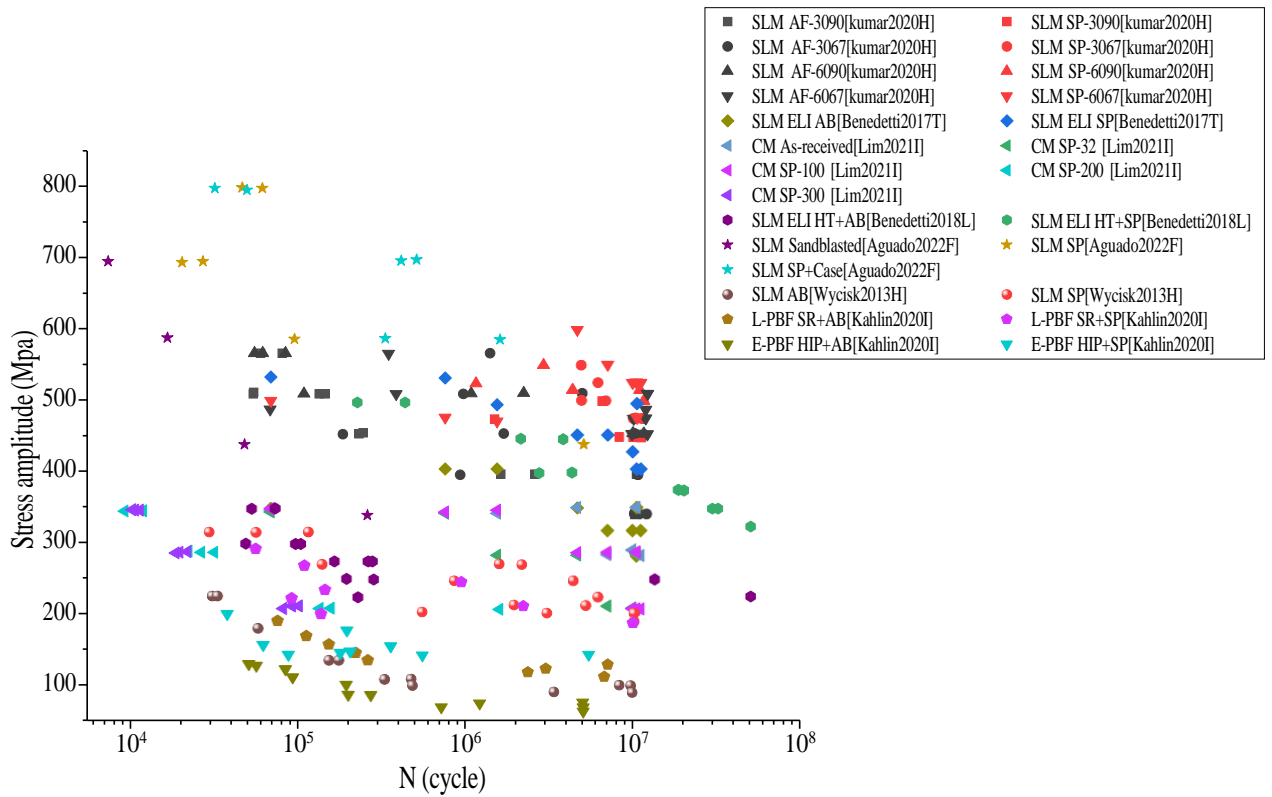

**Fig. 43.** S-N curves of SP effect on AMed Ti-6Al-4V fatigue properties, data from [27, 39, 121, 129, 163, 187, 189]. CM is conventional metals.

Similarly, Kahlin et al. [39] investigated systematically the fatigue properties of L-PBF and E-PBF Ti6Al4V have been subjected to five surface processing methods, shot peening, laser shock peening, centrifugal finishing, laser polishing and linishing. (see Fig. 45). The increases of fatigue strength (compared with AB sample) at 5×10$^6$ cycles and LSP are +5% (10 MPa) and +20% (50 MPa) for E-PBF and L-PBF, respectively, where is a little improvement for fatigue properties. In addition, [197] mentioned the effect of LSP on properties of heat-treated L-PBF Ti-6Al-4V. Aguado-Montero et al. [129] compared that the effect of surface treatments such as shot peening (SP), SP+chemical-assisted vibration finishing and LP on the fatigue behavior of SLMed Ti-6Al-4V specimens. (see Fig. 45) The three treatments improved the fatigue strength. LP does not significantly modify the surface roughness and produces a residual stress distribution with a slightly lower peak but located deeper than that of SP, resulting in the highest fatigue strength of the three treatments analyzed. The high roughness of the specimens made two of them fail due to a crack initiated on the surface. In a nutshell, the larger compressive residual stress caused by LSP is the main reason causing the superior fatigue properties. Besides, [198, 199] demonstrated the fatigue strength improvement in Ti-6Al-4V subjected to foreign object damage by combined treatment of laser shock peening and shot peening.

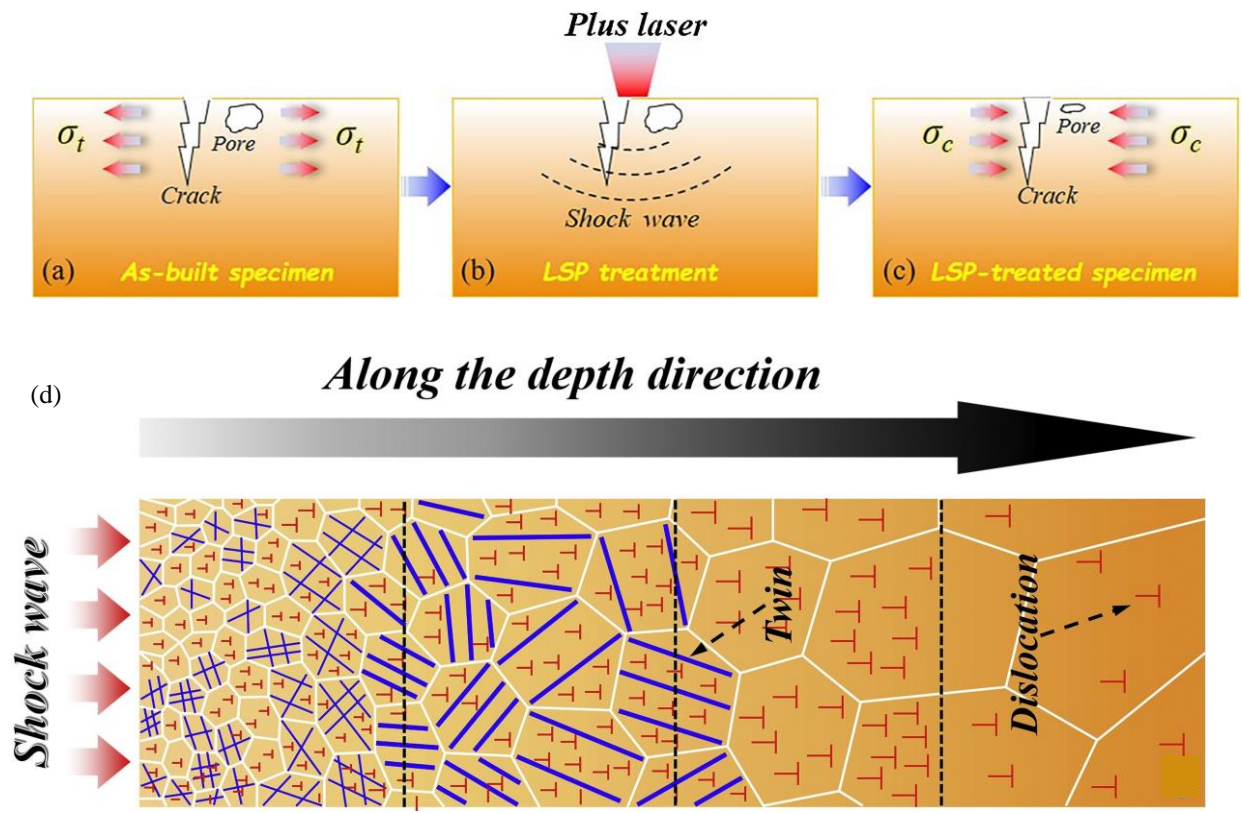

**Fig. 44.** Schematic illustration of the evolution of the metallurgical defects and residual stress in specimens prepared by LDED during LSP. (d) schematic illustration of the entire microstructure along the depth direction. Reprinted from [193], Copyright(2020), with permission from Elsevier.

On the other hand, it is important to point out that AMed Ti-6Al-4V subjected to shot peening or laser shock peening failed generally by cracks starting from subsurface defects, originally from the as-built surface, that had been compressed and re-located to underneath the surface layer. So, the surface roughness alone is not a sufficient indicator of fatigue properties for surface post-processed material since prior surface defects could be hidden below a smooth surface. The final fatigue strength after post-processing depends on a combination of surface roughness, surface residual stress, microstructure and remaining defects. Beretta et al. [85] found that surface defects are more detrimental than subsurface defects on AMed Ti-6Al-4V. So the competition mechanism between surface and subface defects for fatigue properties is worth to discuss, especially for various surface post processing.

However, Jiang et al. [200] investigated the effects of LSP on the ultra-high cycle ($10^9$cycles) fatigue performance of SLMed Ti-6Al-4V and found that the LSP reduce the fatigue limit during the ultra-high cycle compared with HT, which was caused by inherent defects, as well as increased surface roughness and non-uniform residual stresses.

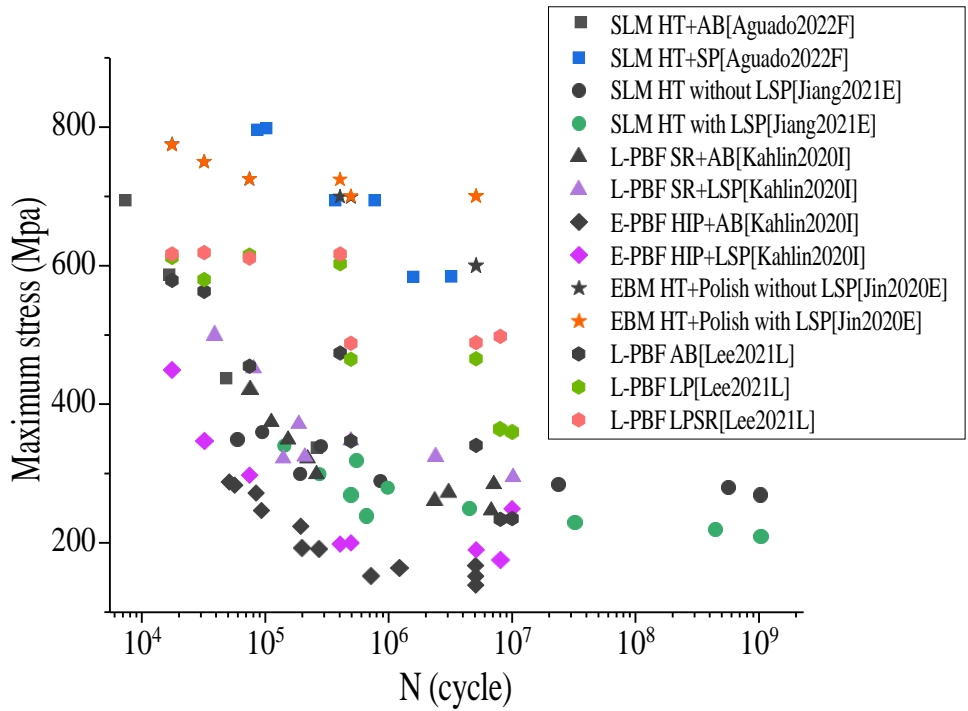

**Fig. 45.** S-N curves of LSP and non-LSP AMed Ti-6Al-4V specimens, data from [39, 129, 180, 195, 200].

**Machining.** Much is known about machined *vs.* as-built surface[40, 125, 131, 137, 138, 141, 177, 201, 202] effect on fatigue properties, as shown in Fig. 46. Molaei and Fatemi et al. [137] found that, in spite of shrinkage of the internal defects with the HIP treatment, the rough surface and micro-cracks on the surface unaffected by this treatment are still the governing factors in controlling the fatigue life. Machined AMed Ti-6Al-4V has significantly lower surface roughness compared to AB Ti-6Al-4V. Denti et al [191] found that the machined AMed Ti-6Al-4V has lowest surface roughness compared to other finishing processes. Bagehorn et al [201] investigated as-built *vs.* milled, blasted, vibratory ground, micro machined effect on $S-N$ curves, as shown in Fig. 48. Out of the applied mechanical surface finishing processes, milling achieved excellent results with respect to both, surface roughness and fatigue life. In this case a decrease in surface roughness (Ra=0.3 lm) improved the fatigue performance to 775 MPa after $3\times10^7$ cycles. Okura et al. [196] found that as the fatigue life of as machined specimen was larger than that of as built specimen, the surface roughness was an important factor of the fatigue life of EBM Ti6Al4V. Besides, the influence of finish machining depth [176] on defect distribution and fatigue behaviour of SLM Ti-6Al-4V is discussed. Employing a sufficient material removal depth during machining was shown to significantly improve fatigue performance. This was attributed to be a direct result of removing material rich in LOF defects, concentrated in the sub-surface. Scatter of stress-life data of machined surface specimens is considerable as opposed to as-built surface specimens stress-life data due to the dominance of surface roughness.

**USMAT/SMAT, SMRT, UNSM.**

Surface mechanical attrition treatment (SMAT) is an emerging post-treatment technology used to introduce a nanostructured layer to metal surface and improve the fatigue resistance of the metal [203]. In this process, small particles repetitively impact on a metal surface using an ultrasonic transducer to accelerate the projectiles. The resultant high-strain-rate deformation causes nanocrystallization in the surface. The nanostructured layer can thus have superior fatigue resistance as compared to the coarse-grained counterparts. Similar ones are surface mechanical rolling treatment (SMRT) [204], surface mechanical grinding treatment (SMGT) [205]. The fatigue strengthening mechanism of them on the Ti-6Al-4V alloy could be ascribed to the comprehensive effects of the compressive residual stress, gradient nanostructured surface layer and surface work hardening, similar to LSP/SP. The compressive residual stress is a key factor in determining the fatigue performance. Simultaneously, the gradient nanostructured (GNS) surface layer and surface work hardening have a synergistic effect that accompanies the effect of compressive residual stress.

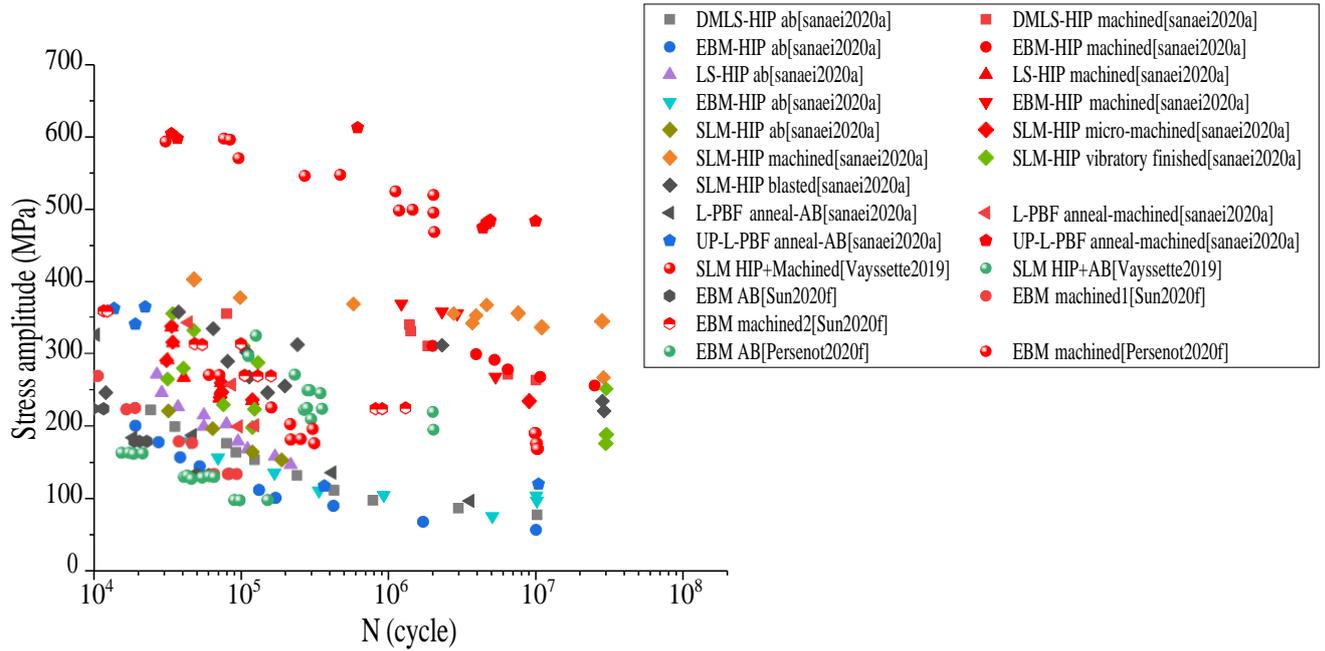

**Fig. 46.** S-N curves for machined and non-machined Ti-6Al-4V specimens, data from [124–126, 128].

Yan et al. [206] found that, a nanostructured layer is generated on the surface of the SLM Ti6Al4V fatigue specimen after SMAT. This nanostructured layer can improve the sample fatigue (see Fig 47) resistance in comparison with the HIP and as-fabricated samples in both LCF (675 MPa *vs.* 430 MPa *vs.* 350 MPa) and HCF regime (580 MPa *vs.* 377 MPa *vs.* 290 MPa) owing to the relief of stress concentrations, improved mechanical strength and compressive residual stress in the surface layer. As compared with other surface treatment technologies, SMAT demonstrates superior fatigue improvement, provides better surface quality and is suitable for processing parts with complex structure. Taken overall, SMAT is a promising surface treatment process for fatigue strength improvement of SLM materials. In addition, Cheng et al. [207] investigated the effect of numbers of ultrasonic surface process on the fatigue properties of Ti3Zr2Sn3Mo25Nb and found that the fatigue limit increases with the increasing numbers of vibration strike. However, Liu et al. [204] detected that the fatigue enhancement degree decreases gradually with increasing ultrasonic surface rolling processing (USRP) number. Interestingly, under HCF condition (≥5×10$^4$ cycles), USRP yields greater improvement in the fatigue life than under the LCF condition. This trend results mainly from the fact that the level of compressive residual stress relaxation is lower under the HCF condition than under the LCF condition

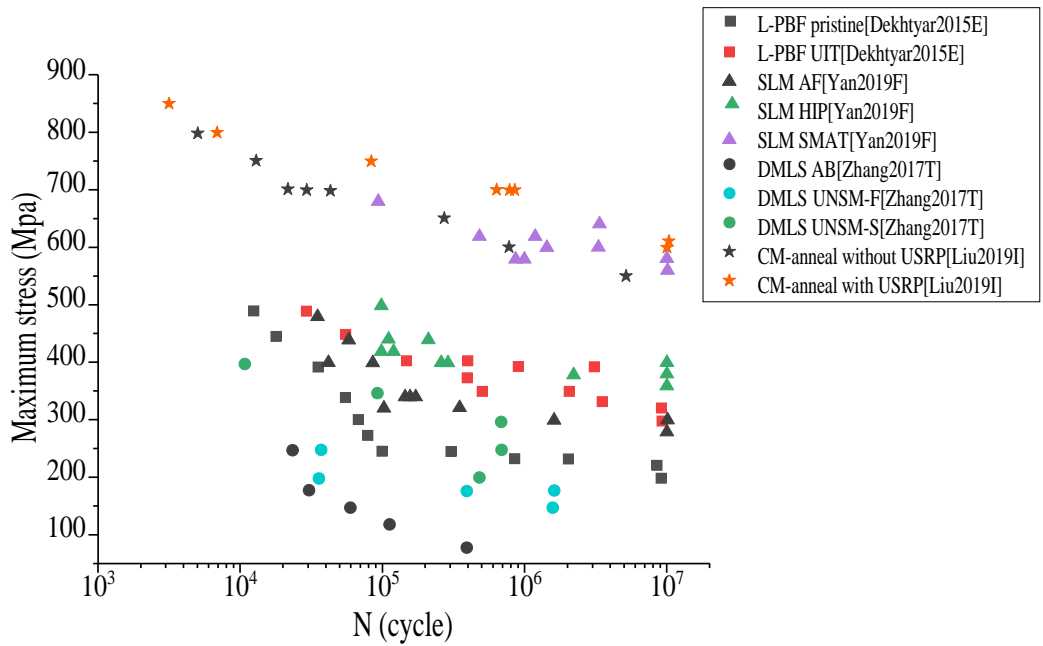

**Fig. 47.** S-N data for SMAT, data from [206, 208, 211, 212]. UNSM-F (fast), UNSM-low (low).

Zhang et al. [208] found that the ultrasonic nanocrystal surface modification (UNSM) treatment resulted in much better surface finish, lower subsurface porosity, and a high magnitude of compressive residual stresses, leading to significant improvement in rotation bending fatigue performance, which also agrees with the DEDed AISI M4 [209]. Meanwhile, owing to the high hardness of AMed Ti-6Al-4V, the UNSM effect is sometimes not effective to reduce surface roughness and further improve fatigue performance. So, Zhang et al. [210] advanced electrically assisted UNSM (EA-UNSM) to process AMed Ti-6Al-4V sample. Compared with conventional UNSM treatment, EA-UNSM was found to be more efficient in improving the surface finish and eliminating pores in the 3D-printed metals. Further, the combination of SMAT/SMRT and polish or machined (other traditional surface treatments) is more efficient to improve the fatigue properties [211]. Surface self nanocrystallization (SSN) by mechanical process transforms the surface coarse grains of a bulk material into nano-sized grains by severe plastic deformation (SPD). These mechanical processes include SMAT, ultrasonic shot peening, LSP, ultrasonic surface rolling processing, ultrasonic nanocrystal surface modification (UNSM) and ultrasonic cold forging technology. In summary, the high energy of ultrasonic, laser and squeezing causes severe plastic deformation, which is reported that the fatigue properties can be improved by these methods.

Here, the post-processing methods, including HT, HIP, LP, LSP, SP, machined, and USMAT/SMAT, play a dominant role in fatigue properties during the numerous factors (e.g., manufacturing parameters, particle, geometry), which can improve the fatigue performance resulting from the elimination of porosity by HIP, the decreasing of surface roughness by various surface post-processing, or the refinement of microstructure by HT. As shown in Figs. 48 and 49, a significant improvement in fatigue performance has been observed. The AM Ti-6Al-4V parts subjected to the post-processing have a parallel fatigue performance compared to the conventional one [15]. Further, Chern et al. [15] and literatures [23, 162, 177] made a comparison between the conventional and AM Ti-6Al-4V for machined and non-machined, HIP and non-HIP, HT and non-HT. As shown in Fig. 49, Machined/HIPed and Machined/Non-HIPed materials yielded a statistically higher average fatigue life than the Non-Machined/Non-HIPed and Non-Machined/HIPed conditions. Machined/HIPed material displayed statistically higher fatigue life then the Machined/Non-HIPed material. EBM Ti-6Al-4V parts with HIPed and machined post-processing exhibit comparable-to-superior fatigue life compared to traditional lamellar and bimodal microstructures. The failure mechanisms and endurance limits are similar to those of traditionally manufactured lamellar Ti-6Al-4V.

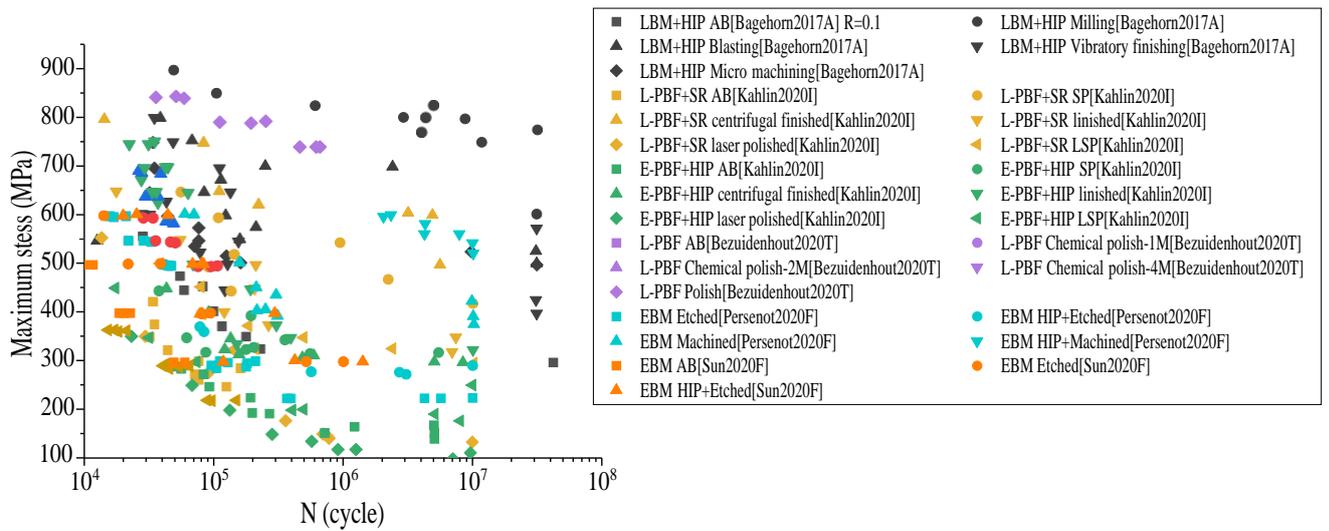

**Fig. 48.** S-N data for other surface treatment specimens, data from [39, 126–128, 201]. The initial nitric acid concentration, $[HNO_3]_0$ was kept constant at 3.17 M and the initial HF concentration, $[HF]_0$ varied at 1, 2, and 4 M ($[HF]:[HNO_3]$ wt% ratios, 1:10, 1:5, and 1:2.5).

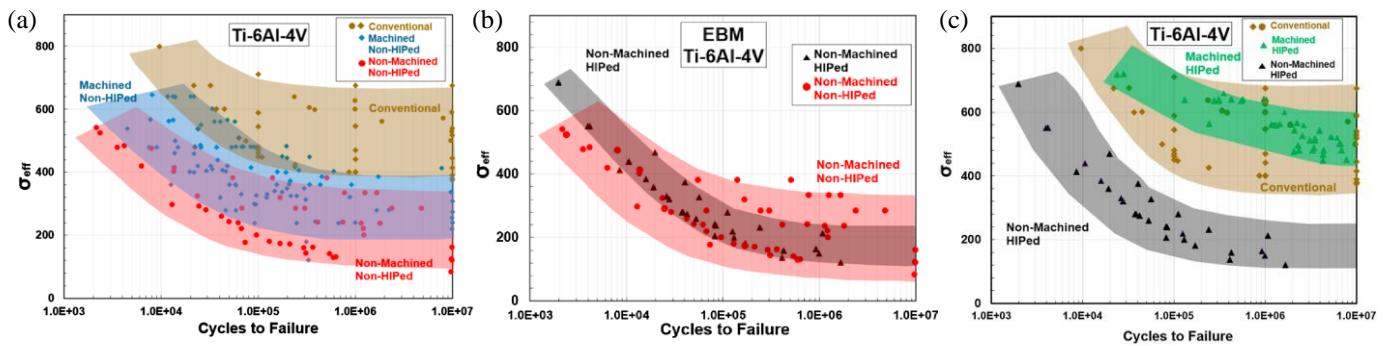

**Fig. 49.** (a) Fatigue behavior of Machined specimens compared to Non-Machined (Non-HIPed material). (b) Fatigue behavior of HIPed parts compared with Non-HIPed with Non- Machined surfaces. (c) Fatigue behavior of the HIPed material with the machined surface compared with the as-built surface and reference material. Reprinted from [15], Copyright(2019), with permission from Elsevier.

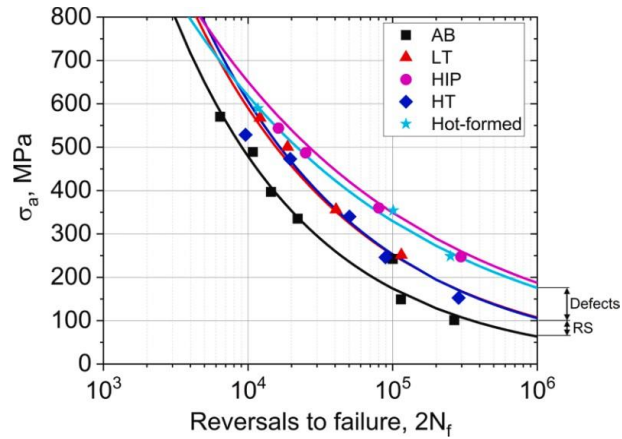

**Fig. 50.** S-N curves of L-PBF Ti-6Al-4V. The fit by Basquin's equation is shown with solid line. LT means low temperature, HT means high temperature. Reprinted from [37], Copyright(2021), with permission from Elsevier.

### 2.3.8. High temperature fatigue

Mishurova et al. [37, 213] investigated the fatigue performance of L-PBF Ti-6Al-4V at elevated temperature (300 °C). The HIP and heat treatment are considered and the effects of defects and RS on fatigue properties are discussed in Ref. [37]. As shown in Fig. 50, the high stress HCF performance of the low-temperature heat treatment condition (LT) specimen almost coincides with that of the high-temperature one (HT). Due to the presence of compressive RS, the fatigue life increases at 300 °C. In the same way, an increase of the tensile RS, for instance by decreasing the volumetric energy density during production [213], will lead to a lower fatigue life. Thus, potentially, a HCF fatigue life improvement could be achieved by tuning the RS via the proper. Besides, the fatigue life at high temperature is primary controlled by the defects with the absence of tensile RS.

## 3. Other Ti alloys

### 3.1. TC17 (Ti-5Al-2Sn-2Zr-4Mo-4Cr)

TC17 (Ti-5Al-2Sn-2Zr-4Mo-4Cr) alloy was researched and developed by GE in 1970, which is classified as "β-rich" α+β phase titanium alloy. It is widely used for aeroengine component, such as low-pressure compressor blades and blisk due to its excellent comprehensive service performances. Microstructure has critical influence on the fatigue properties and mechanical behavior. And the microstructure evolution greatly depends on the thermal history of the laser additive manufactured process [214]. Chi et al. [215] employed a post-treatment method combining HT and LSP to alter the microstructure and mechanical properties of WAAM TC17 titanium alloy, as shown in Fig. 51. By combining HT and LSP, severe plastic deformation was induced in the surface layer, resulting in a high-level surface compressive residual stress (-763 MPa). As well as, the phase structure (Fig. 51 b) The primary α (blue) phase decreases after LSP. In WAAM TC17 alloys, LSP can transform residual stresses from tensile to compressive, in addition to significantly increasing the hardness since the refine grain and work-hardening.

Wang et al. [216] investigated the fatigue failure mechanism for CM TC17 with surface roughness. Zhuo et al. [217] discussed effect of diameter and content of zirconium dioxide on the microstructure and mechanical properties of the TC17 titanium alloy repaired by wire arc additive manufacture and found the ultimate strength of the multi-layer deposition with 0.5 wt% + 50 nm $ZrO_2$ was highest, 1116 MPa, which is 100 MPa higher than that of the deposition without $ZrO_2$ addition. Little is known about the fatigue properties of AMed TC17. Liu et al [218, 219] investigated the mechanical behavior and fatigue properties of the EBMed TC17 alloy joint and found that the fatigue strength can also be enhanced to 240 MPa at $10^7$ cycles by direct aging treatment.

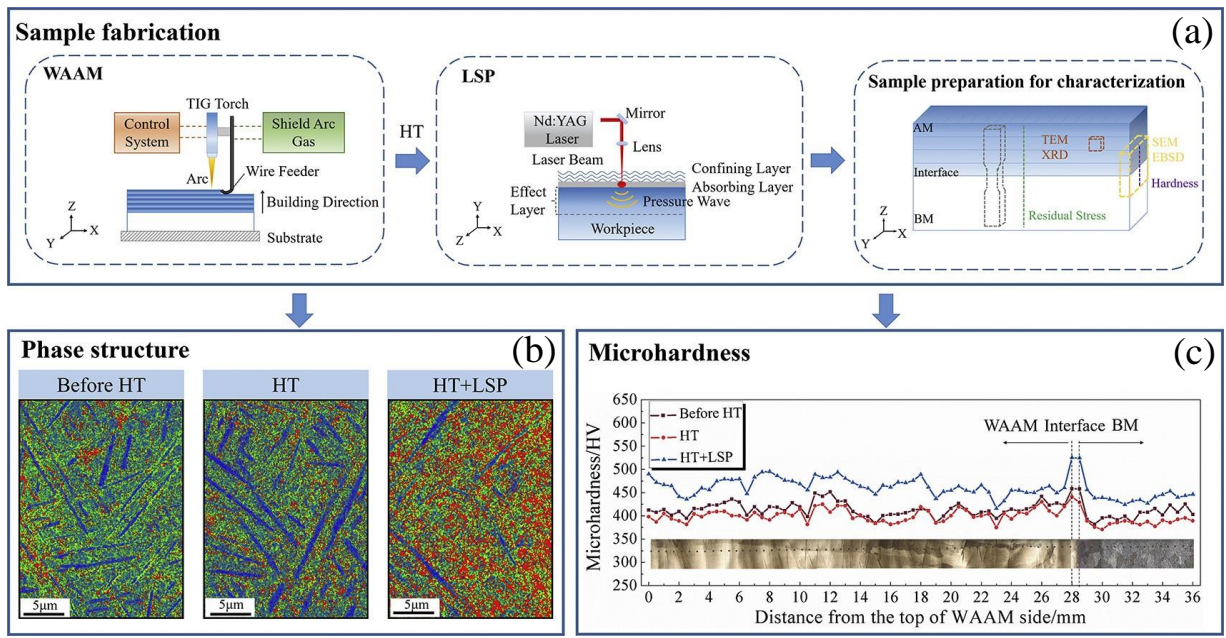

**Fig. 51.** (a) Schematic of WAAM and LSP processes. (b) EBSD analysis at interface region after different treatments. KAM maps for before HT; HT; HT + LSP. (c) Microhardness profile. Reprinted from [215], Copyright(2020), with permission from Elsevier.

In addition, regaining or repairing the fatigue properties of laser additive manufactured Ti alloy via additive manufacture technologies also is an interesting aspect. Liu et al. [220] researched TC17 titanium alloy laser melting deposition repair process and properties. The results show that the laser melting deposition process could realize the form restoration of groove defect. The repaired zone is typically columnar and dendrite crystal, and the 0.5-1.5 mm deep heat-affected zone in the groove interface is coarse equiaxial crystal. Xue et al. [221] found that the fatigue strength of the specimens treated by the micro-scale laser shock processing is improved by 32% at the high cycle fatigue tests, which mainly attributes to the fatigue strips and quadric cracks in the propagation zone of the fatigue fractures caused by the compressive stress during LSP zone. Similarly, Zhao et al. [222] studied the microstructure, micro-hardness and room temperature tensile properties of laser additive repaired (LAR) specimen. Luo et al. [223, 224] investigated the fatigue strength of laser additive manufactured TC17 and Ti-3.5Mo-6.5Al-1.5Zr-0.25Si titanium alloy via laser shock peening (LSP). The results shows that both the fatigue strength of TC17 and Ti-3.5Mo-6.5Al-1.5Zr-0.25Si titanium alloy improved from 365 MPa to 451 MPa and from 438 MPa to 544 MPa, respectively, after the compound process treatment. The regain mechanism of fatigue strength was attributed to the elimination of tensile residual stress and microstructure change in the laser additive layer after LSP treatment. Grain refine after LSP treatment and the grain sizes presented gradient distribution, which contributes to that the plastic deformation occurred in the surface layer. And then dislocation was accumulated and then the sub-boundary formed, the fatigue strength increases with the increasing dislocation density.

Usually, the grain boundary strength is superior to the grain strength, so the fracture mechanism is transgranular at room temperature. However, the fine rib-like $\alpha$ phase improves the grain strength significantly and results in that the grain strength is superior to the grain boundary strength, so the cracks tend to initiate and expand along the grain boundary $\alpha_{GB}$, resulting in intergranular fracture.

Liu et al. [225] found that the continuous the grain boundary $\alpha(\alpha_{GB})$ with the accompanying precipitate free zone (PFZ) is the main reason for the low ductility of LAM TC18 since the crack prefers to nucleating and propagating along $\alpha_{GB}$, resulting in intergranular fracture. It is important to point out that the ultra-fine basket-weave $\alpha$ phase was obtained, which plays a critical role in high strength. The vary fine basket-weave $\alpha$ phase also can be observed in Zhang et al. [214]. Then, Zhu et al. [226] found the equiaxed grains, continuous $\alpha_{GB}$ and PFZ might cause intergranular fracture, and column grain along with discontinuous $\alpha_{GB}$ might cause transgranular fracture from fractography of LAM TC17. Besides, the dimension of $\alpha$ lath might affect the mechanical properties, as wider $\alpha$ laths probably brought about lower strength in tensile test. These revealed the additive manufactured microstructure variability and its sensitivity to the fracture models. In addition, Sun et al. [227] investigated the LSP induced fatigue crack retardation in Ti-17 titanium alloy. The fatigue life was increased up to 2.4 times that of the unpeened counterpart, and the fatigue crack lives of basic metal, LSP-20 J×3 and LSP-30 J×3 specimens are 168000, 405000, 290000, respectively. LSP-induced residual stresses play a significant role in the fatigue crack propagation. Further, they [228, 229] investigated the effect of drilling of TC17 alloy prior and post to LSP on its fatigue property. There is a high density of dislocations and significant increase in fatigue life with LSP, which is due to the lack of detrimental tensile residual stress at the hole edge when drilling post to LSP is used. This indicates that drilling after LSP is an optimal processing procedure.

### 3.2. TC18 (Ti-5Al-5Mo-5V-1Cr-1Fe)

Alloy Ti-5Al-5Mo-5V-1Fe-1Cr (TC18 or Ti-55511 or VT22), is a typical near $β$ titanium alloy with high strength and excellent fatigue property [230] and is successfully utilized in landing gears and other structural parts for aircraft [231]. In addition, beta titanium alloys were also widely utilized as implant materials in the fields of orthodontics because of its low elastic modulus, excellent biocompatibility and minimum cytotoxicity [232].

Li et al. [233] investigated the low cycle fatigue behavior of LMD TC18 titanium alloy at room temperature. They obtained the microstructure of double annealed LMD TC18 titanium alloy consists of fine lamella-like primary $\alpha$ phase and transformed $β$ matrix and further predicted the low cycle fatigue curve. When the crack grows along the grain boundary (intercrystalline fracture), a continuous grain boundary $\alpha$ phase ($\alpha_{GB}$) leads to a straight propagating crack, whereas a discontinuous grain boundary phase results in a flexural propagating mode. The discontinuous $\alpha_{GB}$ can improve the propagation resistance and reduces propagation speed, which provides a good idea to increase the fatigue life via reducing the continuity of the grain boundary. Meanwhile liu et al. [225] studied systematically the evolution of macrostructure and microstructure of LMDed TC18 during LMD process. The relationship between the path of crack propagation and grain boundary is revealed. The continuous $\alpha_{GB}$ causes the low ductility and intergranular fracture. Meanwhile, the cracks are prone to propagate along the interface between the columnar and equiaxed regions. Fig. 52 shows the microstructure of LMDed TC18, and the fracture propagates along the interface(see Fig. 52b) between the columnar and equiaxed regions can be observed obviously. During the tensile test, the possible crack propagation path in LMD TC18 is schematically illustrated in Fig. 52c, which includes the intercrystalline fracture in equiaxed region (see the blue line 3) and interface (see the red line 2), and transcrystalline fracture in columnar region (see the green line 1). The crack will always choose to propagate along the path with minimum energy. Compared to the intercrystalline fracture, transcrystalline fracture will take more energy to over the grain boundary if the crack propagates in the columnar region. For the path 2 and 3, we can find that path 2 is more direct, which means smaller crack propagation energy. As a result, the intercrystalline fracture along with interface (path 2) naturally becomes the path of crack propagation. This suggest that the continuous $α_{GB}$ is detrimental to the ductility. The treatment processes which can eliminate the continuous $α_{GB}$ by HIP and HT will be necessary to improve the ductility [225].

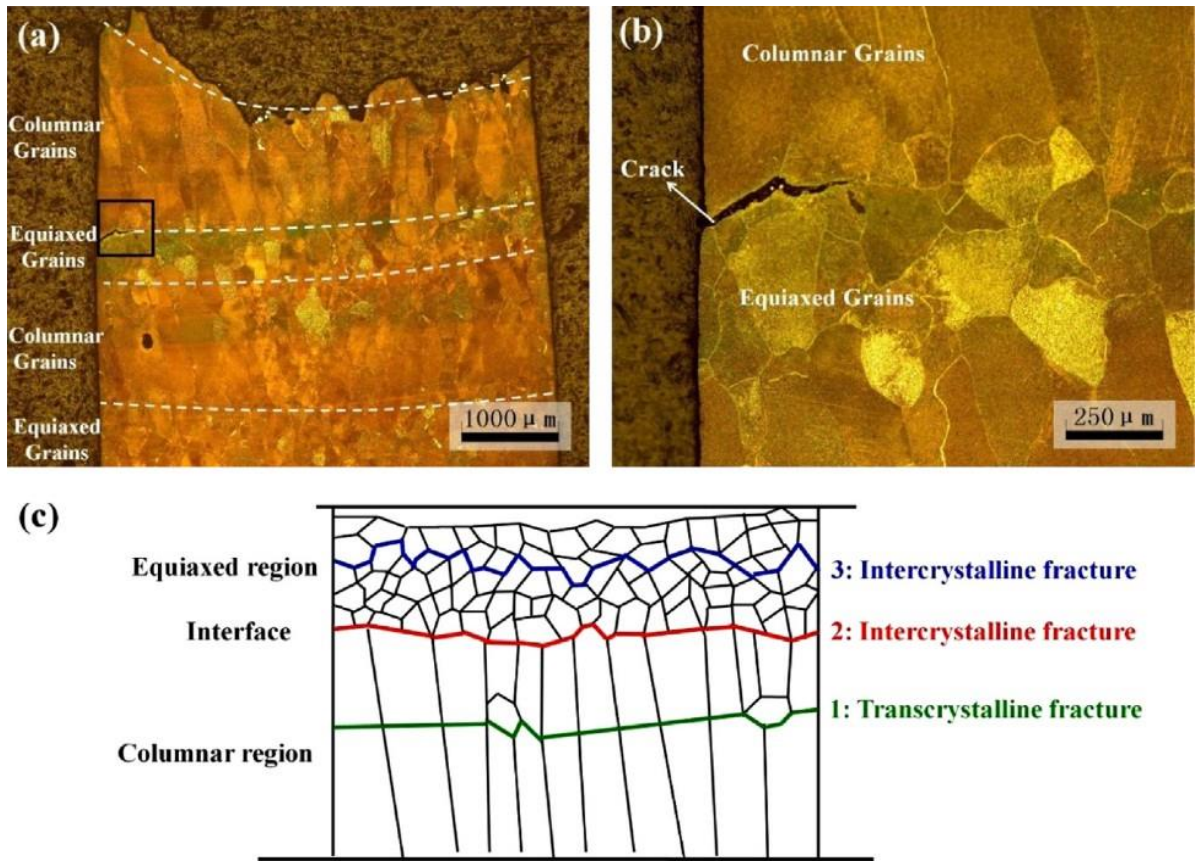

**Fig. 52.** (a) Optical microscopy (OM) image showing the cross section images of room temperature tensile specimens. White dashed lines represent the interface between the columnar and equiaxed regions. (b) OM image showing the microcrack initiation and propagation along the grain boundary between the columnar and equiaxed grains. (c) Schematic illustration of the possible crack propagation path in laser melting deposited Ti-5Al-5Mo-5V-1Cr-1Fe alloy. Reprinted from [225], Copyright(2013), with permission from Elsevier.

Further, Feng et al. [234] studied the effect of ceramic shot peening on fatigue properties of TC18 titanium alloy formed by laser direct deposition. The middle fatigue life of the unpeening specimens was about 83,600 cycles. With the increase of shot peening strength, the middle fatigue life of the specimen increases first and then decreases. The reason is that the combined effect of residual stress on the specimen surface and surface roughness. As the increase of shot peening intensity, the surface roughness and residual compressive stress increased. The fatigue life of the sample with 0.20-0.25 mm shot peening intensity was the highest and reached 226,600 cycles, which is about 2.7 times of that of the unpeening specimen.

Wang et al. [235–237] studied the effect of primary $\alpha$ phase ($\alpha_p$) on the fatigue crack path, fatigue crack branching and the effects of $\alpha/\beta$ phase interfaces on fatigue crack deflections of LMD TC18. For LMD TC18 titanium alloy, compared with the C(T)-0-2.5 and C(T)-45-2.5 with a relatively straight path, the deflective crack path of C(T)-90-2.5 greatly extended the fatigue crack growth life due to the inflection points and overall deflection of the crack path, as shown in Fig. 53. Due to the significant reduction of FCG rates caused by the inflection points and overall deflection of the crack path, the crack propagation life of C(T)-90-2.5 is much longer than those of C(T)-0-2.5 and C(T)-45-2.5, as shown in the a-N curve in Fig. 53b. The fracture morphology [235] indicate that the deflection of the fatigue crack path is caused by the orderly distributed primary $\alpha$ phase laths in grains, especially in the columnar grains, and the deflective extent varies accompanied by the angle between the initial crack path and primary $\alpha$ phase laths. After, fatigue crack branching was observed in constant amplitude fatigue crack growth test [236]. The fatigue crack branching results from complicated interactions of many factors, such as $\alpha_p$ laths, continuous $\alpha_{GB}$, voids and lack of fusion, especially the effect of orderly distributed $\alpha_p$ laths.

## 3.3. TA15 (Ti-6.5Al-2Zr-1Mo-1V)

TA15 (Ti-6.5Al-2Zr-1Mo-1V) has the high strength, toughness and hardenability of metastable $\beta$ type titanium alloy, as well as the tensile ductility and elastic modulus of $\alpha+\beta$ type titanium alloy. He et al. [238] investigated the fatigue behavior of DLD TA15 alloy under a stress level of 800 MPa with $R$ = 0.06. And a novel bimodal lognormal model was established, which has a good fit with the experimental data, to educe the scatter of the fatigue life significantly. They found that no significant difference can be identified in the fatigue lives of two categories of specimens that divided into two categories for crack Initiated from the Surface (Case I) or Subsurface and crack Initiated from the Inner (Case I).

The fatigue life are from 256709 to 30715 for Case I, and from 544102 to 30293 for Case II, and the scatter is great. Zhan et al [239] built a novel TA2-TA15 titanium alloy fabricated by the LMD. According to the results of the uniaxial tensile and fatigue experiments for the LMD TA2-TA15 titanium alloy, the values of Young's module, yield stress, ultimate stress and fatigue limit of the LMD TA2-TA15 are between that of TA2 and TA15, which can be referred in the application of the novel material. As shown is Fig. 54, it is obvious that the dispersion of data exists in the fatigue lives of the LMD TA2-TA15 specimens, the smooth specimens have significantly higher fatigue life compared to the notched specimens. SEM tests of fracture surfaces reveal two reasons for dispersion, one is the multiple sources of cracks at the fracture surface of specimens, and the other is manufacturing defects inside the specimens. As well as, at $R$ = 0.1, the smooth LMD TA2-TA15 specimens yields greater improvement in the fatigue life than LMD TA2 specimens.

## 3.4. CP Ti (commercially pure titanium)

For AM CP Ti alloy, Okazaki et al. [135] examined the examined microstructures and fatigue properties of Laser-Sintered CP Ti for dental applications. The tensile and fatigue properties of CP Ti grade (G) 2 annealed at 700 °C for 2 h after laser sintering were close to those of wrought CP Ti G 2 annealed at the same temperature after hot forging. The fatigue strengths ($\sigma_{FS}$) at $10^7$ cycles of the 90 - and 0 -direction-built CP Ti G 2 rods after laser sintering 10 times were 320 and 365 MPa, respectively. Fig. 55 shows S-N curves of annealed CP Ti G 2 rods ((a) EOS and (b) TIROP powders) after laser sintering and dental-cast CP T G 2 (shown in Fig. 55a). Fig. 55b also shows S-N curves of CP Ti G 2 annealed after hot forging taken from the literature for comparison. The fatigue strengths of the 90 -, 45 -, and 0 -direction-built CP Ti G 2 rods were 330, 290, and 380 MPa, respectively. This result shows the anisotropy of built orientation.

Note that CP Ti G 2 (EOS (EOS GmbH Electro Optical System, Krailling, Germany) and TILOP (Osaka Titanium Technologies Co., Ltd., Osaka, Japan) powders were prepared by plasma and Ar gas atomization processes, respectively.

Hasib et al. [240] investigated the effects of build orientation and HT on the tensile and FCG behavior of CP Ti manufactured by L-PBF. They found the HIP treatments roughened the $\alpha$ grain structure and increased the fatigue crack growth threshold ($\Delta K_{th}$). Build orientation had a negligible dependence on fatigue crack growth resistance of L-PBF fabricated CP-Ti compared with wrought sample which greatly depends on the sample orientation. In addition, at higher growth rates, the fatigue crack growth resistance became microstructure insensitive, and the transition to microstructure insensitive fatigue crack growth occurred when the cyclic plastic zone size was on a similar order of magnitude to the grain size. Importantly, Liu et al. [241] advanced a topology-optimized structure fabricated by selective laser melting using CP-Ti exhibits excellent fatigue properties. As shown in Fig. 56, The fatigue life increases with decreasing the applied cycle stress. The fatigue strength and normalized fatigue property (i.e., fatigue stress/yield stress) of the porous rhombic dodecahedron samples are significantly lower than those of the topology-optimized samples. For example, the topology-optimized CP-Ti samples exhibit excellent normalized fatigue stress/yield stress of 0.65 at 106 cycles, which is much higher than that of the rhombic dodecahedron structure (0.13 at $10^6$ cycles).

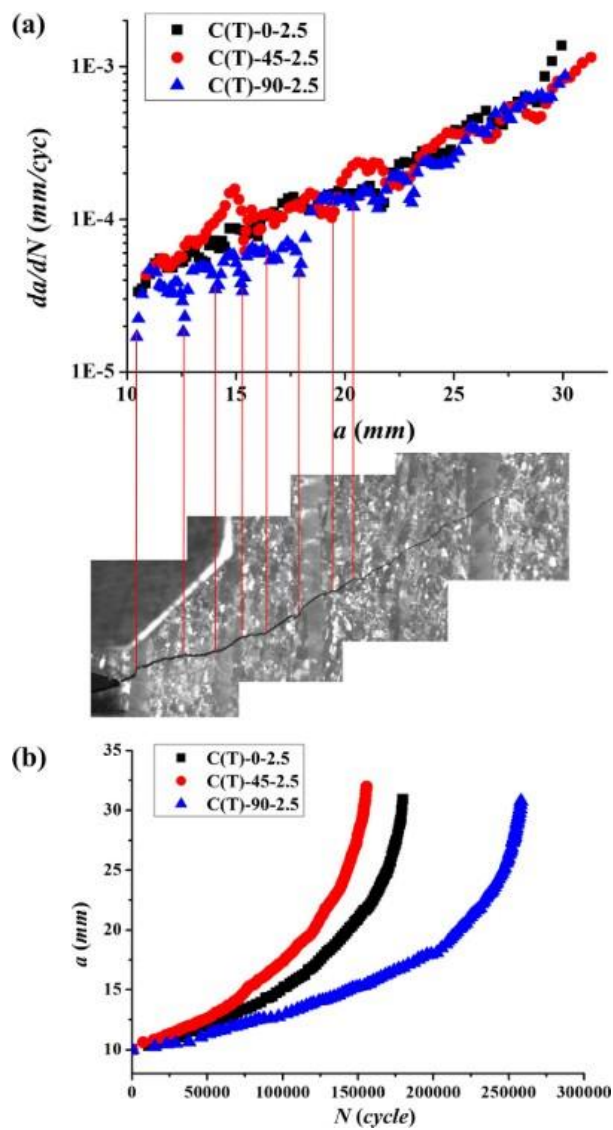

**Fig. 53.** (a) Da/dN-a data of C(T) specimens and corresponding relationship with crack path of C(T)-2.5-90, (b) a-N curve of C(T) specimens. C(T) means compact tension specimen. 0, 45, 90 stand for the angle between the sample notch and the deposition direction. 2.5 (KN) is maximum load. Reprinted from [235], Copyright(2018), with permission from Elsevier.

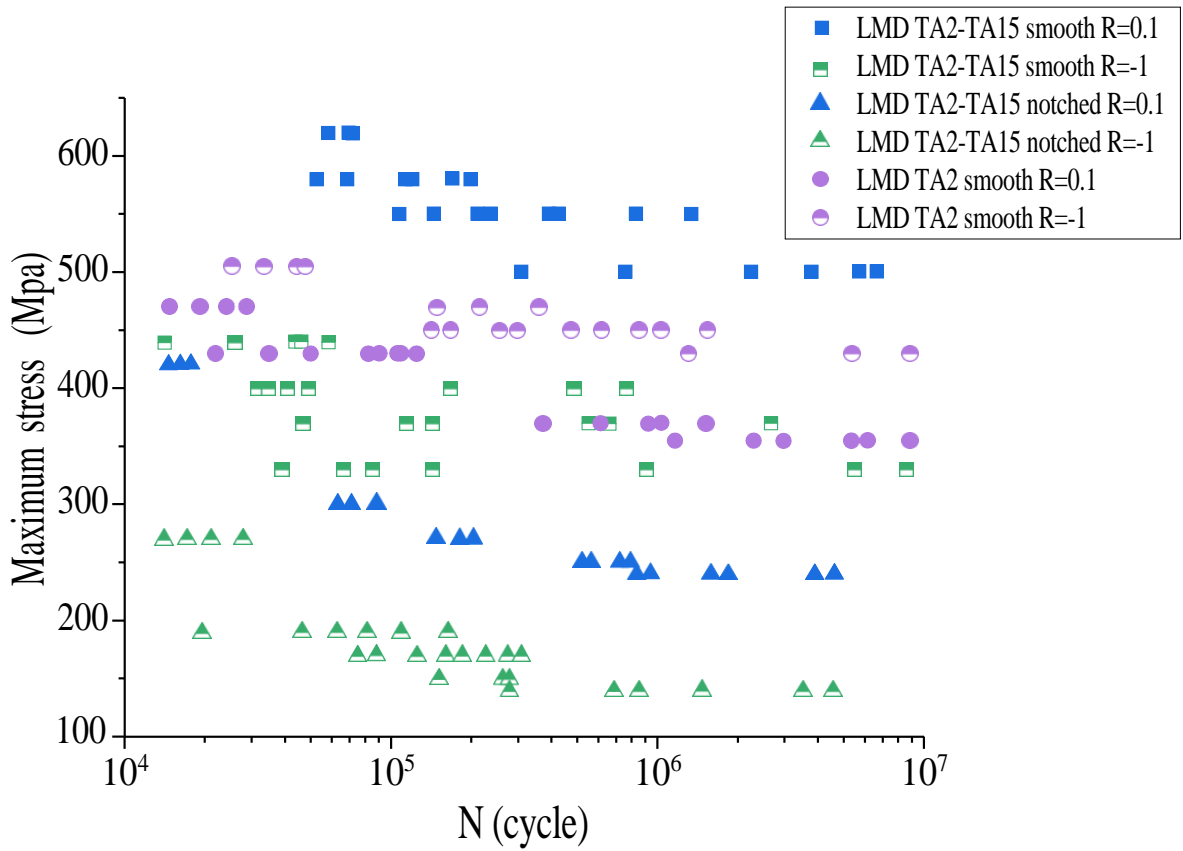

**Fig. 54.** S-N curves of TA2-TA15

## 3.5. Others

The fatigue properties of other AM Ti-based alloys are also investigated, which included TC21 (Ti-6Al-2Zr-2Sn-3Mo-1.5Cr-2Nb) [242], Ti-6Al-2Mo-2Sn-2Zr-2Cr-2V [243], TC11 (Ti-6.5Al-3.5Mo-1.5Zr-0.3Si) [244, 245], Ti5553(Ti-5Al-5Mo-5V-3Cr) [246, 247], Ti2448 (Ti-24Nb-4Zr-8Sn) [248, 249], Ti6242/TA19 (Ti-6Al-2Sn-4Zr-2Mo) [250]. Zhang et al. [242] studied the build direction effect on microstructure and mechanical and fatigue properties of laser additive manufactured TC21 titanium alloy. Perevoshchikova et al. [246] researched the HIP schemes for Ti-5553 manufactured by powder metallurgy and the HCF behavior. The HCF properties of the HIPped materials are shown to be only slightly inferior that those reported in the literature for ingot Ti-5553 subjected to an optimized heat treatment process, and significantly greater than ingot Ti-6Al-4V material. In addition, the mechanical properties of the near-β titanium alloy (SLMed Ti-5553) was investigated [247]. Bulk samples with a density of 99.95% were built and exhibit a tensile strength of about 800 MPa and a strain up to 14%. Further its roughness ($R_z$) values depending on grain size distributions [251] and detailed evolution of microstructure [252] are studied. HCF damage behavior of Ti-5Al-5Mo-5V-3Cr-1Zr (Ti-55531) titanium alloy with lamellar microstructure [253] and bimodal microstructure [254] was systematically investigated at room temperature.

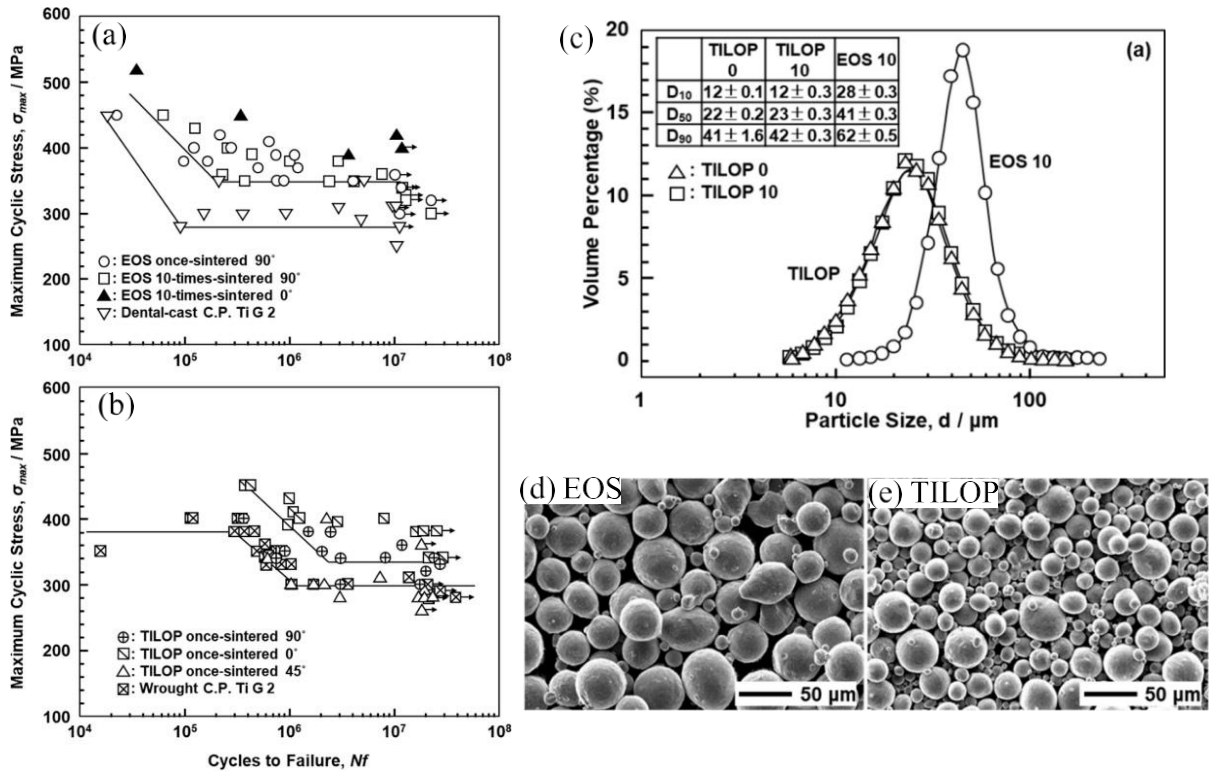

**Fig. 55.** S-N curves of laser-sintered (a) EOS (once- and 10-times-sintered) and (b) TILOP (once-sintered) CP Ti G 2 rods; dental-cast CP Ti G 2 in (a) and wrought CP Ti G 2 in (b). (EOS (EOS GmbH Electro Optical System, Krailling, Germany) and TILOP (Osaka Titanium Technologies Co., Ltd., Osaka, Japan). Reprinted from [135], Copyright(2020), with permission from MDPI.

Liu et al. [248] investigated the differences in the microstructure, defects and mechanical behavior of porous structures for β-type EBM and SLM Ti-24Nb-4Zr-8Sn. The fatigue strength of the annealed EBM or SLM samples is sensitive to the stress amplitudes, as shown in Fig. 57. At the lower stress levels, the fatigue behavior of the meshes is mainly determined by the cyclic ratcheting and surface properties of the struts, resulting in similar properties for both manufacturing processes. However, at higher stress levels, the crack initiation and propagation from the pores tends to occur and therefore the SLM samples (which contain a higher number of defects) have a lower and more variable fatigue life. Note that β-type Ti-24Nb-4Zr-8Sn is a great AM metal for producing high quality, artificial joints. Therefore, the applied max stress is a compressive stress of 12 MPa, simply introduce it.

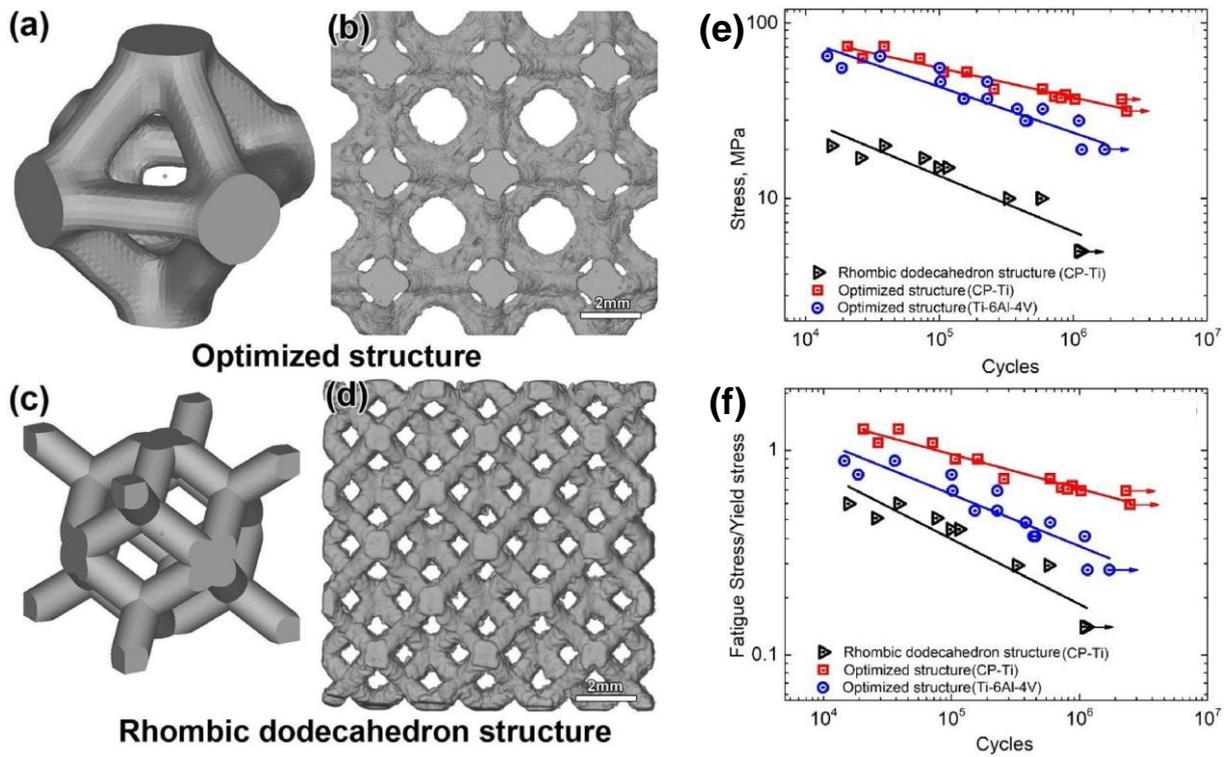

**Fig. 56.** cell structure and S-N curves SLMed CP-Ti. (e) Fatigue S-N curves of topology-optimized and rhombic dodecahedron CP-Ti specimens, and (f) normalized S-N (fatigue stress/yield stress) curves of porous topology-optimized and rhombic dodecahedron specimens. Reprinted from [241], Copyright(2020), with permission from Elsevier.

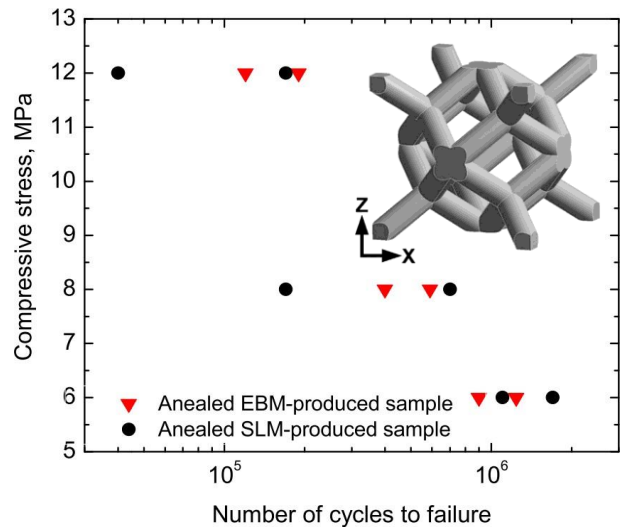

**Fig. 57.** The S-N of annealed Ti-24Nb-4Zr-8Sn samples manufactured by SLM and EBM. Reprinted from [248], Copyright(2016), with permission from Elsevier.

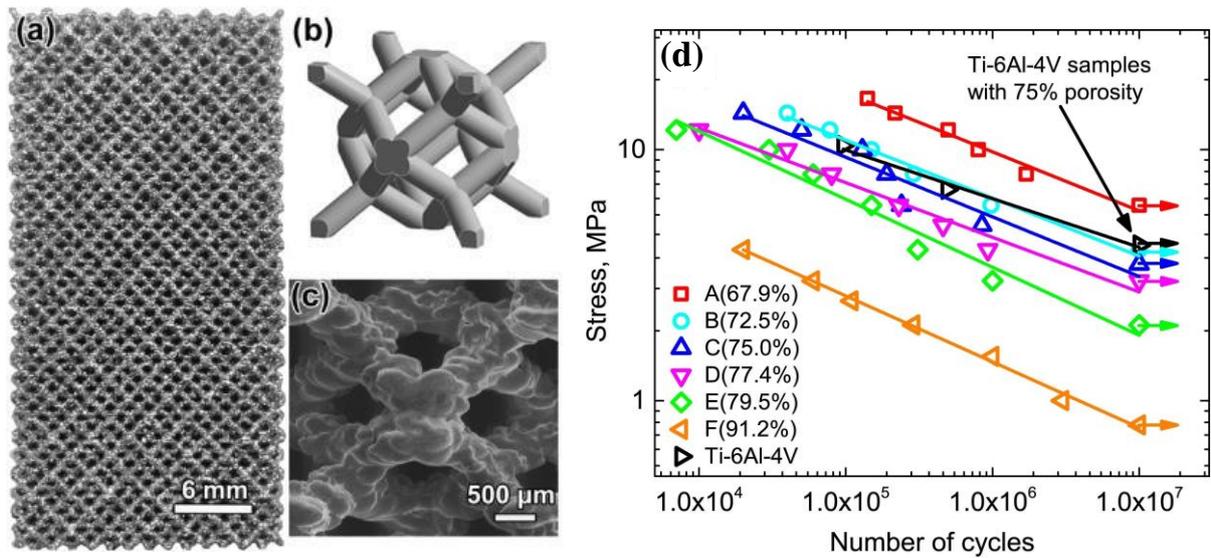

**Fig. 58.** S-N curve of Ti2448 alloy with EBM. Reprinted from [249], Copyright(2017), with permission from Elsevier.

Lu et al. [244] investigated the fatigue crack growth behaviour along the deposition direction in LMD TC11 alloy under constant amplitude loading with R=0.06. They also investigated Their fatigue properties [255]. The fatigue limit of the alloy perpendicular to and parallel to the deposition direction at specified life of $10^7$ and stress ration of 0.1 is 337 MPa and 365 MPa, respectively. The porosities of specimens are among 0.014-0.028%, averaged porosity is 0.02%. As the maximum stress increases, the fatigue life decreases rapidly. The fatigue life is down to only 56,185 cycles at the maximum stress of 600 MPa. And the complete data is as follows: middle fatigue life ($N_{50}$) and safe life ($N_{95}$) are 157,647 and 56185 for 600 MPa, $N_{50} = 162,223$ and $N_{95} = 72,528$ for 550 MPa, $N_{50} = 569,571$ and $N_{95} = 238,390$ for 500 MPa. Wu et al. [245] focused on fatigue crack tip strain evolution and crack growth prediction under single overload in laser melting deposited TC11 titanium alloy.

Liu et al. [249] investigated the influence of porosity variation in electron beam melting (EBM)-produced β-type Ti2448 alloy samples on the mechanical properties including super-elastic property, Young's modulus, compressive strength and fatigue properties (see Fig. 58). Preparation of detail: The powder had a nominal composition of Ti-23.9Nb-3.9Zr-8.2Sn-0.19O (in wt%) and had a spherical shape with a particle size range of 45-106 μm. Starting from the Ti2448 powder, porous rhombic dodecahedron structures containing 7×7×14 unit cells were fabricated by an Arcam A1 EBM system (Fig. 58) with powder layer thickness of 70 μm. The powder in the selected area was melted by electron beam with a spot size of 200 μm generated from a tungsten filament in an EBM system. The annealing treatment for all porous samples was conducted at 750 °C for 1 h followed by air cooling. Compression fatigue tests were performed using an Instron E10000 machine with a stress ratio R of -0.1 and a frequency of 10 Hz. The fatigue tests were controlled by varied applied stresses, i.e. 1, 1.5, 2, 2.5, 3, 4, 6, 8, 10, 12, 14 and 16 MPa. Note that this applied stress is much lower than the fatigue test stress of typical ti-based alloys (Ti-6Al-4V), mainly owing to that, as an alternative biomaterial for next generation implants, Ti2448 works in a low stress case. The S-N curves of all samples with absolute and normalized stress values are plotted in Fig. 58. For Ti2448 samples, the fatigue life is dominated by the porosity and applied stress level; the fatigue life decreases with increasing porosity for both low and high stress levels. The high-porosity samples (such as F group with 91.2% porosity) present much lower fatigue strength than low-porosity groups. Other information see [249].

Sui et al. [250] investigated the cyclic heat treatment (CHT) procedures suitable for the LAAM-built near-$\alpha$ titanium alloy Ti-6Al-2Sn-4Zr-2Mo (Ti6242/TA19) to attain the globular $\alpha$ phases. The results show that 980 °C is the most suitable upper temperature limit for CHT. However, it is difficult to achieve a high volume fraction of the globular $\alpha$ phases in the LAAM-built Ti6242 alloys through CHT, which is ascribed to the low composition gradient caused by more $\alpha$-stabilizing elements and fewer $\beta$-stabilizing elements. The as-built sample demonstrate elongation of 6.3%, which is lower than the AMS 4919J standard (elongation ≥ 10%). After 980 °C CHT and 980 °C CHT with solution heat-treatment, the formation of the globular $\alpha$ phases significantly increased the elongation to 13.5% and 12.9%, respectively. Although the mechanical strength is reduced after heat-treatment, the room-temperature tensile properties still exceed the AMS 4919J standard. Fractography examination showed that the as-built sample exhibit a mixed brittle and ductile fracture behavior, while the 980 °C CHT and 980 °C CHT with solution heat-treated samples displayed ductile fracture. Its high temperature (500 °C) performances including tensile, yield strength and elongation are far superior to optical Ti-6Al-4V.

## 4. AlSi10Mg

### 4.1. Introduction

AlSi10Mg has pretty welding and cutting performance, because it is close to the eutectic alloy [256], the phase diagram of Al-Si is illustrated in Fig. 59. Although aluminum-based materials are difficult to fabricate in additive manufacturing, the relatively high density and low shrinkage rate of AlSi10Mg in aluminum alloys weaken this trend [257, 258], therefore AlSi10Mg is the most typical in the research of AM aluminum alloys. General researches mainly focus on the technological parameters, microstructure, and mechanical properties of AlSi10Mg. The AM manufacturing process for AlSi10Mg is focused on SLM and L-PBF, LMD and DMLS have also received attention. Compared with the AlSi10Mg produced by traditional manufacturing methods such as die-casting and forge, this aluminum alloy fabricated using AM processes has a finer microstructure and better static and dynamic mechanical properties. There is a notable difference between SLM and die-cast samples in microstructure, it can be seen in Fig. 60 (a) that the needle-shaped Al-Si eutectic alloys (A), coarse hypo-eutectic solidification structure (B), and Fe-rich phase (C) constitute a typical die-cast AlSi10Mg microstructure [259, 260]. In contrast, the microstructure of SLM AlSi10Mg (Fig. 60 (b)) exhibits an overall fine honeycomb morphology, albeit sightly coarse at the melt pool boundaries (red dotted area).

For a part of static tensile mechanical properties such as tensile strength and yield strength [17, 261], the improvement of AM AlSi10Mg relative to traditional processes sample is significant. However, due to the differences in process parameters and dispersion of experimental data, it is hard to get a brief answer as to which process would have better fatigue property between AM and traditional process. Fig. 61 illustrates the fatigue life data of AM AlSi10Mg, and the fitted curve of the traditional process Al-alloy is attached for comparison. We could be informed that there are many AM AlSi10Mg whose fatigue strength is still lower than that of die-cast parts. This is because the interaction between powder and laser leads to more serious defects that reduce the fatigue performance of AM AlSi10Mg [261]. Meanwhile, the AM samples with fatigue performance stronger than traditionally manufactured samples could be achieved by modifying specific process parameters and performing suitable post-processing. However, it is is remarkable that compared with the most widely used aerospace Al-alloy Al-6061 are still gap, there are not many AM AlSi10Mg that could reach the fatigue limit of traditional Al-6061. The conclusion can be drawn: the overall fatigue performance of AM AlSi10Mg is still not a par with the outstanding AL-alloy, but has great potential to improve. The influencing factor of fatigue life can be discussed from many aspects as discussed below.

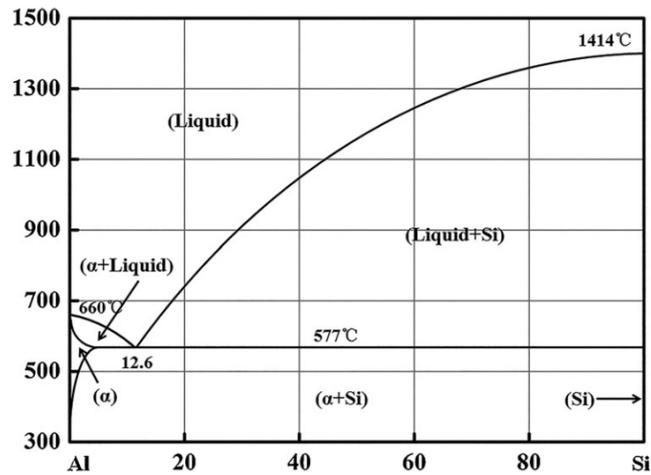

**Fig. 59.** Phase diagram of Aluminium-Silicon alloy. Reprinted from [24], Copyright (2019), with permission from Elsevier.

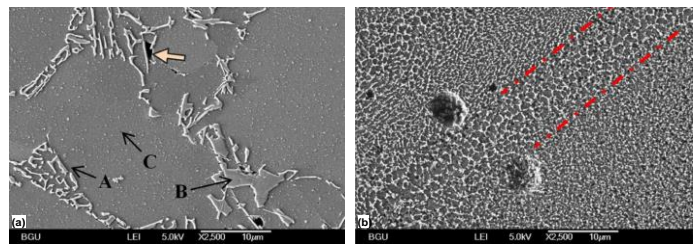

**Fig. 60.** High-resolution scanning electron microscopy image of (a) SLM sample; (b) die-cast sample. Reprinted from [259], Copyright (2018), with permission from Elsevier.

## 4.2. Particle effect

As the raw material of MAM, the nature of metal spherical powder primarily affects the quality of the finished product. In general, the indicators of powder include surface roughness, size, chemical composition and sphericity, etc. However, since the powder characteristics is not simple to control as other processes, there is not much attention paid to the influence of powder properties on fatigue life. With regard to the size of the powder, Jian et al. [269] reported that the smaller defects in sample are observed as the powder size increases. As shown in Fig. 62(a), the average fatigue defects size of samples with 50 $\mu$m are much lower than that of 20 $\mu$m, which makes the resulted density higher. The S-N results are summarized in Fig. 62(b), small-sized powder products represent worse fatigue performance and greater dispersion. Besides the geometry size factor, in the research of Muhammad et al. [291], the effects of chemical composition are explained. Two products made of powder with different chemical element content (the specific composition difference are shown in Table 1), LPW AlSi10Mg and EOS AlSi10Mg, having a very different microstructures, LPW sample contains obvious refined Si particles, and there is a complete Si grid structure in EOS AlSi10Mg as shown in Fig. 62(c) and (d). However, the formation of Si network only improves the tensile strength of EOS AlSi10Mg, the discrepancy in fatigue strength of two samples is not much obvious due to their similar surface roughness. By introducing nano-$TiB_2$ particles into Al alloy powder, Li et al. [310] successfully increased the laser absorption rate of AlSi10Mg powder by 50%, which would effectively improve the manufacturability of AM aluminum-based materials. Meanwhile, the combination of nano-$TiB_2$ and rapid solidification promotes grain boundary modification and grain dislocation in the microstructure, allowing to make a prediction of improving the fatigue performance. In order to increase the utilization rate and reduce the cost of metal powder, the study on the reuse of residual powder have been derived. Del Re et al. [309] assessed the influence of powder reuse on the fatigue strength of AM parts. As the number of reuses increased, the overall fatigue behavior showed a slight decline and finally AlSi10Mg parts fatigue life has been reduced by 12% after 8 reuses times.

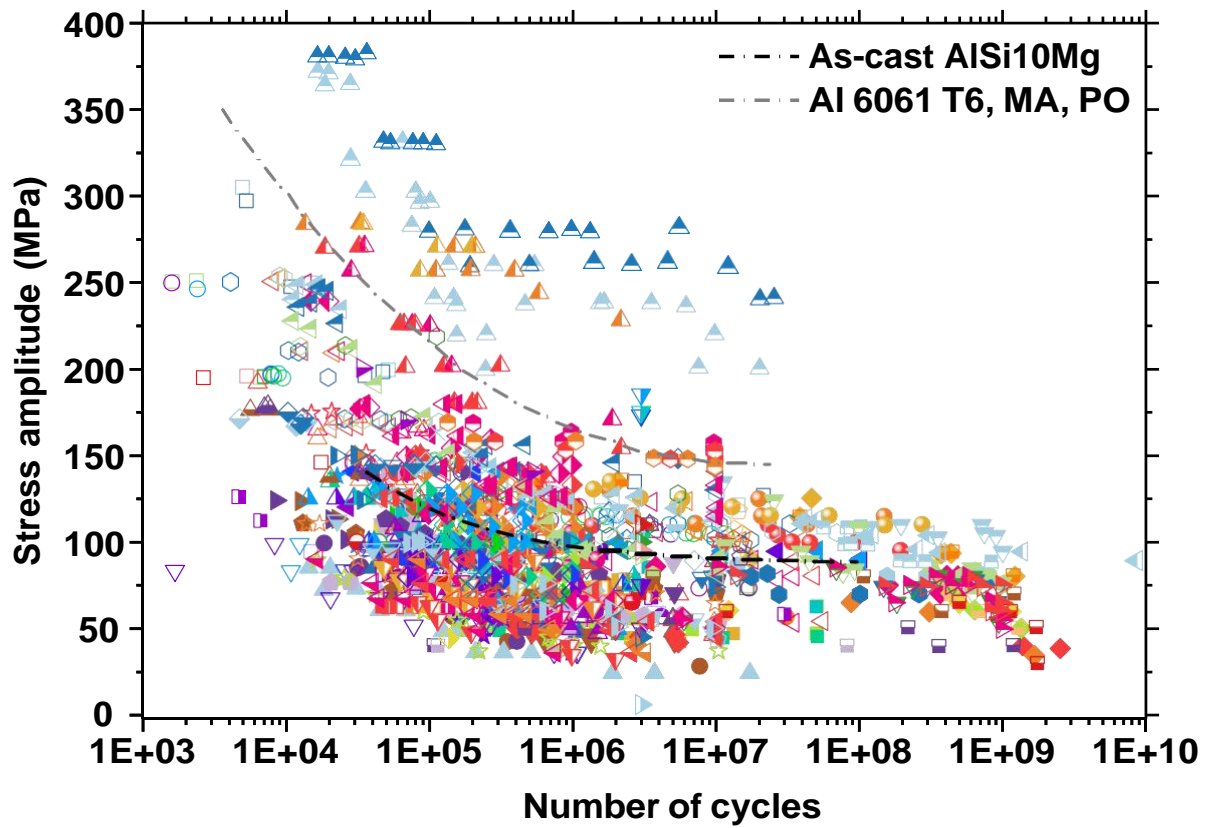

**Fig. 61.** The fatigue *S-N* data. The point represents AM AlSi10Mg, the black dotted line represents the as-cast AlSi10Mg, the gray dotted line represents the Al-6061 alloy after T6, machined and polished [17, 259, 261–309].

Overall, the fatigue life of AlSi10Mg could be adjusted by modifying the powder size and composition. But this change is not significant and may be covered by other more influential factors, such as the process parameters and post treatment that would be discussed below.

### 4.3. Manufacturing parameters

During the manufacturing process, when the heat source is applied, the metal powder melts into the liquid state, normally after the scanning laser is transfer, the material is in a non-equilibrium coagulation condition. Essentially, with the continuous repetition of the melting and solidification, the parts undergoes the complex thermal cycle during manufacture. The technical parameters that control the heat input include but are not limited to energy density and platform temperature, which usually impact the performance by changing the surface roughness and porosity. However, there are few studies on the relationship between the scanning parameters and the surface state in AM AlSi10Mg alloy, then we will mainly discuss the aspect of porosity.

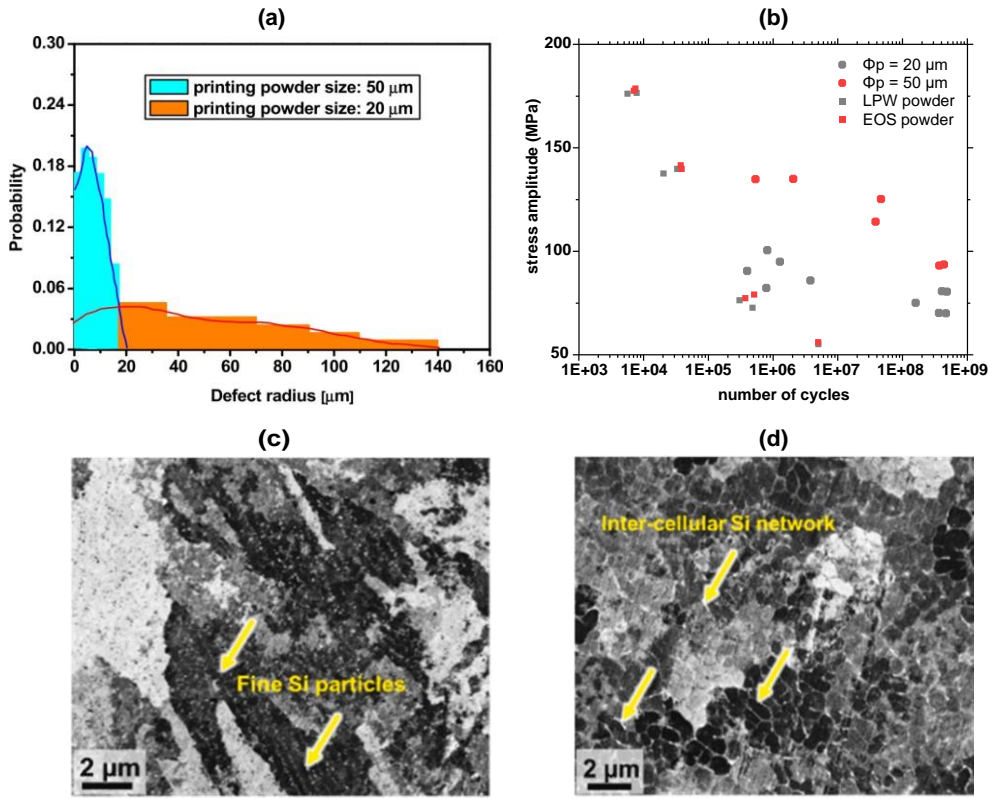

**Fig. 62.** (a) Defect density distributions for the specimens with powder sizes of 50 μm and 20 μm. Reprinted from [269], Copyright(2021), with permission from Elsevier; (b) comparison of S-N data for AlSi10Mg samples with different powder conditions [269, 291]; (c) the ECCI micrographs of LPW AlSi10Mg; (d) the ECCI micrographs of EOS AlSi10Mg. Reprinted from [291], Copyright(2021), with permission from Elsevier.

**Table 1.** Main chemical composition for LPW and EOS AlSi10Mg powder batches [291].

| (Wt.%) | Si | Fe | Mn | Zn |
|---|---|---|---|---|
| LPW AlSi10Mg | 9.60 | 0.18 | <0.01 | <0.01 |
| EOS AlSi10Mg | 9-11 | <0.55 | <0.45 | <0.10 |

### 4.3.1. Scanning energy density

The energy input during the manufacturing process affects various properties of the finished product. Combing the contribution of laser power $P$ and scanning speed $V$, the "line energy density" ($\lambda = P/V$) could be obtained, upon further considering the influence of hatch space $h$ and layer thickness $t$, we get the "bulk energy density" ($\varepsilon = P/V\,ht$) that are more commonly accepted [311]. We could discuss the role of energy density in fatigue strength according to these four parameters.

For scanning power, higher laser power could effectively promote powder fusion and eliminate defects, Rhein et al. [277] used the cumulative distribution function (CDF) to represent the distribution of defect sizes, as can be seen in Fig. 63(a), the defect size parameters of high-power specimens is concentrated in the range of 50-150 $\mu$m, while the defect sizes of lower-power specimens are scattered in the area of 100-300 nm. This shows that the increase in scanning power reduces the pore size scatter while obtaining smaller defects, thereby strength fatigue behavior. With regard to scanning speed, Wang et al. [312] collected the corresponding crack length data of AlSi10Mg under different cycle times, and regarded the average crack growth rate as the indicator of fatigue resistance. Fig. 63(b) shows that the relationship between crack length and fatigue cycle times in different scanning speed. Due to the initial slow growth rate, the origin time of the microcrack is unimportant, what is really worth pay attention to is the rapid development of cracks. The rapid propagation of cracks in the sample with appropriate scanning speed (V = 1000 m/s) appears very late, in contrast, the samples with too high or low speed will have fast crack extension earlier, which increases the average rate of crack growth. At the same time, the phenomenon that too low scanning speed brings lower fatigue performance implies that excessive energy density will play a negative role as discussed later.

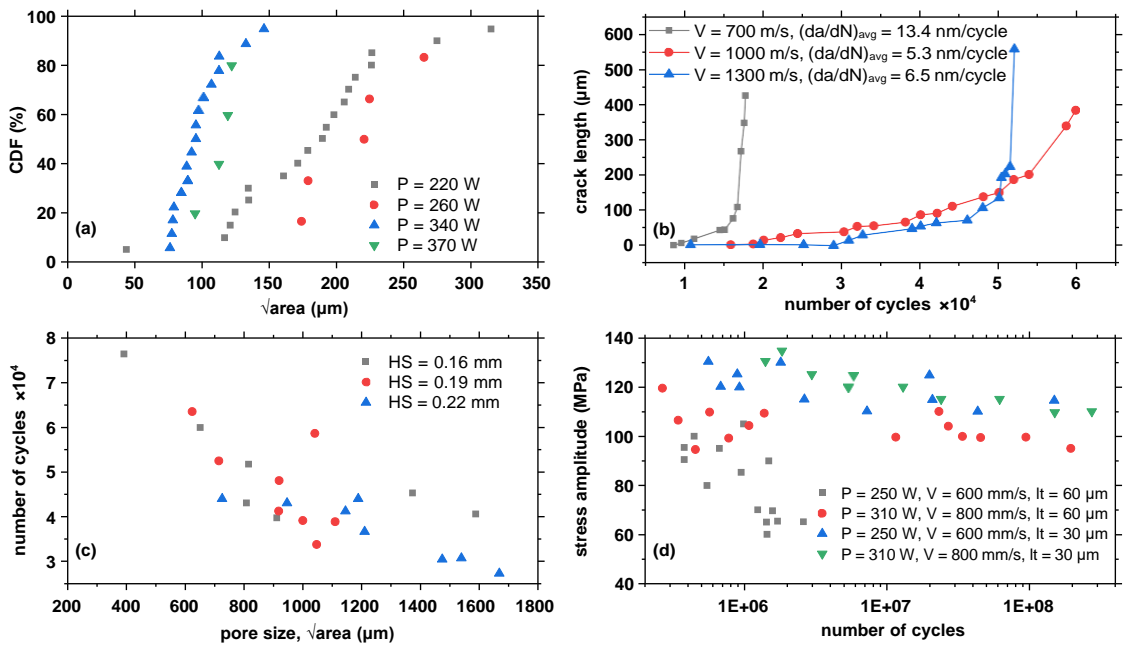

**Fig. 63.** (a) The CDF of defects sizes for different power condition [277]; (b) the relationship between crack length and fatigue cycle times in different scanning speed [312]; (c) the relationship between fatigue cycle times and pore sizes in different hatch spacing [285]; (d) *S-N* data points for different scanning parameters [277].

Hatch spacing usually refers to the distance between two adjacent lasers, narrowing the spacing could reduce porosity and shrink columnar grains by promoting the remelting cycles process. Tang et al. [285] calculated the melting times of per unit corresponding to three different scan spacing, it is found that each unit was melted 4 times on average under the condition of smallest hatch spacing (H = 0.16 mm), whereas the melting times only reached about 3 times when H = 0.22 mm. As shown in Fig .63(c), although there are many overlapping areas due to the dispersiveness of the sample datas, the overall trend is still that the AlSi10Mg parts built with smaller hatch spacing would own smaller defect sizes related to the better fatigue resistance. The thickness of the powder layer with stronger adjustability has received relatively more research. It is found that the reduction of layer thickness could efficiently lower the size of killer defects [277, 313, 314]. As shown in Fig. 63(d), in comparison with the 60 $\mu$m layer thickness samples, the fatigue life of specimens built with 30 $\mu$m layer thickness is improved. It is worth noting that as energy density increases, the role of reduction of layer thickness on improvement of fatigue life gradually weakens. The disparity in fatigue life of samples with different layer thickness at high energy density ($P$ = 310 W, $V$ = 800 m/s) is not as significant as low energy density ($P$ = 250 W, $V$ = 600 m/s). Beever et al. [313] investigated the impact of the layer thickness on the fatigue properties of net shaped and milled samples. They found that for milled condition, the fatigue life of the specimens with thinner powder layer is about three times that of the specimens with thicker powder layer. However, the fatigue life of net shaped samples with different lay thickness is basically the same. This is due to the fatigue cycle strength greatly depends on the local characteristics [300], so in the net shaped samples, the fatigue life is mainly determined by the poor surface condition, ignoring the effect of fewer defects caused by reducing layer thickness. Based on the above research, it can be found that it is not feasible to pursue enhancement of fatigue strength only through increasing energy density.

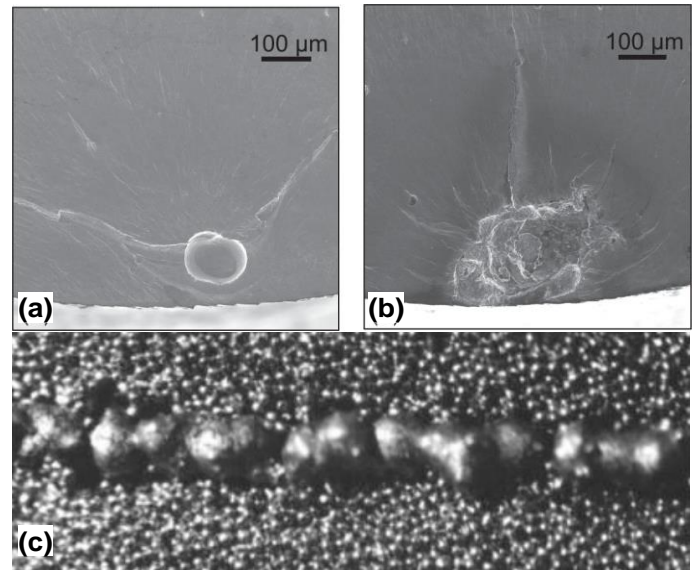

**Fig. 64.** (a) The lack of fusion defects; (b) the keyhole defects. Reprinted from [277], Copyright(2021), with permission from Elsevier; (c) spheroidized particles. Reprinted from [315], Copyright(2004), with permission from Elsevier.

Insufficient energy input could result in incomplete powder consolidation as depicted in Fig. 64(a) and weak connection between powder layers, it does not mean that the energy density should be as high as possible. Undue energy density correspondingly brings about large-volume melt pool and balling effect [316] (Due to the rapid melting and shrinkage, the powder becomes loosely connected spherical particles under surface tension [317], as depicted in Fig 64(c).) to distort the structure. In the case of overmelting, the molten pool is extremely unstable and easily collapse due to small changes in heat put leaving entrapped gas [300], forming the keyhole(Fig 64 (b)). This is verified in Fig. 65 (a) showing that when the energy density exceeds the threshold of 60 J/mm$^3$, the porosity raises again and the dispersity increase. Zhan et al. [318] took this energy density threshold with the lowest porosity as an indicator and make a systematic study of the relationship between fatigue life and energy density. Fig. 65 (b) illustrates that the increased energy density drives a rapid improvement in fatigue life, if the energy density is less than $0.8E_{d0}$. Once the relative density is over $E_{d0}$, the fatigue life will essentially remain unchanged or even exist a downtrend. Therefore, it is necessary to find an scanning energy density that can provide the maximum material density. This range is about 60-75 J/mm$^3$ for Al-Si alloy [319].

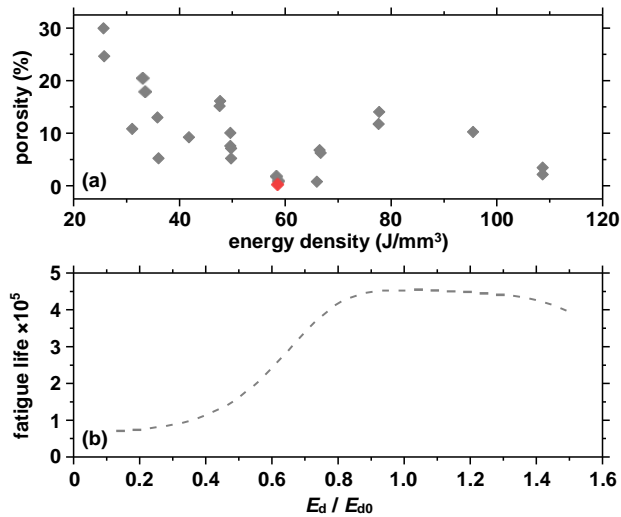

**Fig. 65.** (a) Porosity variation versus the energy density, the red dot represents the lowest porosity [320]; (b) variation of the fatigue life with the relative energy density [318].

### 4.3.2. Platform heating

Platform heating plays a role as the additional heat source, upward conduction of heat energy delays the cooling rate. It could be expected that the slower solidification process and higher energy density promote the formation of a homogeneous and dense microstructure, which could remarkably hinder early crack propagation and eliminate residual stress [262, 321]. Awd et al. [271] compared the microstructure of the two samples batch A (no platform heating) and batch B (platform heating), showing that larger and more uniform melt tracks appears after platform heating (Fig. 66(a), (b)). Batch B correspondingly has the coarsened and homogenized microstructure [322]. Meanwhile, this research explained well some phenomenon that appeared to damage the fatigue properties after the platform temperature increases. First, for batch B, the relative content of Mg that can used as the MgSi strengthening phase seems to be low, but this is actually caused by the higher relative percentage of Si, the real content of Mg should not be greatly interrupted. A relative Si content of at least 1% higher in the batch with platform heating suggests that lower cooling rate eliminates the original semi-coherent Si agglomerates, allowing the Si to segregate to the grain boundaries to attain the compositional homogeneity. Secondly, the average relative density of batch B is light lower, a naive conclusion would be that the high platform temperature has a negative impact on pore defects. However, a comparison of pore size distribution of two batches in Fig. 66(c) reveals when the pore diameter is about 40 $\mu$m, batch A has more defects. The extra heat energy from the high platform temperature drives the initial small semi-spherical or elliptical pores to expand into a larger spherical pore. This is why the batch B has more pores in the diameter range of 50-120 $\mu$m. Although the size of pores increases, the spherical shaped pores are difficult to serve as the initiation of cracks and the path of crack propagation, the decrease of subsurface defects that are more prone to fracture in batch B is also considered. Platform heating still effectively reduces the effect of pores in fatigue. In general, owing to the change improvement of the defect morphology and the densification and homogenization of the microstructure, the fatigue resistance of the platform heated specimens is appreciably improved as shown in Fig. 66(d).

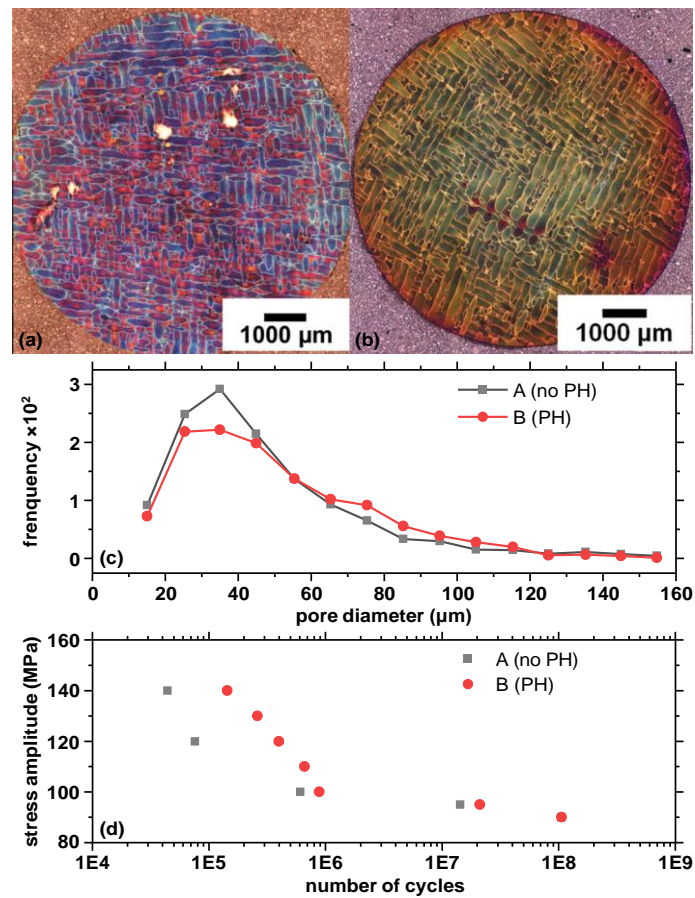

**Fig. 66.** Melt tracks in the light microscope for (a) sample without platform heating; (b) sample without platform heating. (c) frequency counts of pores with respect to the discrete diameter class; (d) mean *S-N* data points for sample without or with platform heating. Reprinted from [271], Copyright(2019), with permission from Elsevier.

### 4.3.3. Scanning strategy

In addition to the amount of overall heat input, changes in the scanning strategies would also influence material fatigue life. Paul et al. [323] investigated the effect of relative scanning rotation angle between adjacent layers (Fig. 67(a)) on fracture resistance. Compared to 90° scan rotation, there is a more random distribution of melt pools in AlSi10Mg built by 67° scan angle, which requires more deflection and twisting of crack as it propagates as demonstrated in Fig. 67(b), (c). This is why 67° scan rotation is used in the most researches. Beever et al. [313] try to add the contour scan after the internal scan to improve the surface quality (Fig. 67(d), (e)). Unfortunately, the supplementary contour scan in fact harm the fatigue life. This is because non-contoured sample have greater surface residual stress, and extra contour scan brings a rougher surface and larger subsurface pores. Under the influence of these elements, the fatigue life of specimens with contour dropped by an order of magnitude.

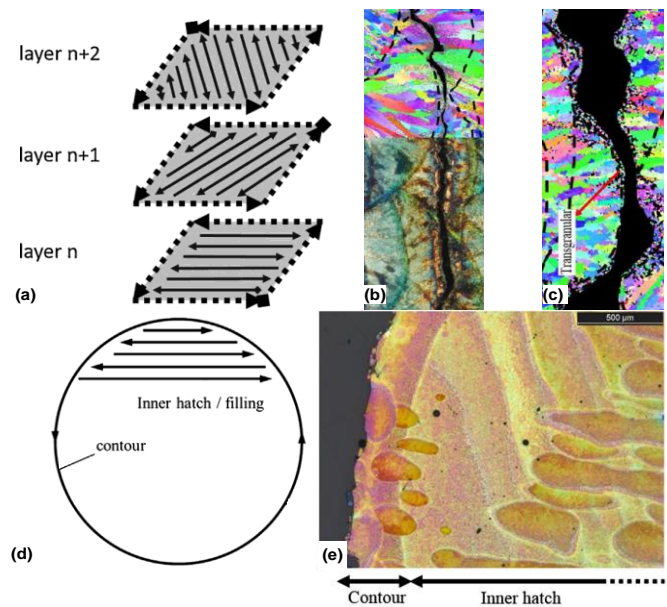

**Fig. 67.** (a) Scanning rotation angle between adjacent layers. Reprinted from [324], Copyright(2020), with permission from Elsevier; (b) crack propagation path in the material with 90° scan rotation; (c) crack propagation path in the material with 67° scan rotation. Reprinted from [323], Copyright(2021), with permission from Elsevier; (d) scanning strategy of regarding inner hatch/filling and contour; (e) etched tranversal cross-subsection of sample with contour scan added. Reprinted from [313], Copyright(2018), with permission from Elsevier.

### 4.3.4. Building direction

Simply put, the building direction refers to the angle between the long axis of the specimen and building platform as plotted in Fig. 68(g). Since the laser scanning is concentrated on a specific plane, the finished AM AlSi10Mg is a typical layered structure with clear anisotropy. This feature is well reflected in Fig. 68(a), the molten pool in the Z plane parallel to build orientation (perpendicular to build platform) appears as the stacked fish-scale. For the XY plane that normal to build orientation (parallel to build platform), the long melt pool is rotated by an angle and overlaps the bottom layer (Fig. 68(b)). Increase the magnitude to the crystal level, Fig. 68(c) shows that along the building orientation, a large number of columnar dendrites growth is observed. From the view perpendicular to the building direction (Fig. 68(d)), the eutectic Si phase gathers at the grain boundary, dividing the Al matrix into independent island-like regions, thus forming the honeycomb network melt pools [325, 326]. The discrepancy of the microstructure in different orientations implies that the building direction will have an bearing on the fatigue life of AM parts.

Affected by other factors, it is unreasonable to simply summarize which building direction the samples produced have the best fatigue performance. The most research on the comparison of vertical (90°) and horizontal (0°) samples, a main trend is that horizontal samples have better fatigue resistance, at the same stress level, the fatigue life could be extended several times or even dozens of times. From the perspective of microstructure, vertical samples whose long axis of columnar crystals are parallel to the direction of applied stress are more likely to be damaged [279]. There are the largest number of defects at the boundary of the molten pool, and the flat defects perpendicular to the loading direction are prone to high stress concentration as the crack propagation path. For vertical samples, when cracks are initiated on the surface, the weak connection between layers is difficult to resist the crack growth, crack could quickly and smoothly extend along the boundary of melt pool without deflection and twisting (Fig. 68(e)). The cracks of the horizontal sample produce a large amount of deflection and secondary crack which consume plenty of energy during the growth. Apart from this, the need to propagate across the pool center with fine structure and superior fatigue resistance also extend the crack growth life as shown in Fig. 68(f) [273, 281]. This trend is even more pronounced under HCF and VHCF, because the lower stress amplitude makes the plastic damage appear late [288]. In the actual use process, it should be avoided that the building orientation of the product is parallel to the force direction. On the contrary, the horizontal sample have the maximum time to failure due to its microstructure and defects morpholgy relative to the loading orientation.

Surely that there are some cases that throw a wrench on this trend, many research [265, 275, 290] found that the fatigue life of vertical and horizontal samples is basically the same. One possible explanation is that though the LOF defects distributed along the fracture surface would weaken the performance of the vertical sample, these large LOF defects with complex morphology may cause local stress concentration and affect the fatigue life of horizontal samples [275]. Another widely accepted argument is that the impact of building direction is covered by other parameters such as surface roughness. A pretty instance is the research of Brandl et al. [262], they found when the platform temperature is maintain at 30 °C, the fatigue resistance of specimens in horizontal direction is much higher than that in vertical direction. In contrast, when platform temperature reaches 300 °C, there is no difference between the fatigue life in two direction samples. The reason for the change is that the horizontal sample is not sensitive to defects at high temperature, the high temperature to eliminate the defects is of little significance to it. Julius et al. [297, 328, 329] found a similar phenomenon, the fatigue life of the samples with different building direction was no longer the same after T6 treatment. The overall trend of the impact of the building direction of the fatigue life is described in Fig. 68(h), the fatigue resistance of the horizontal sample is lightly higher, but there is a large overlap area.

### 4.3.5. Other factors

Ch et al. [280] investigated the role of protective atmosphere and found that the melt pool morphology and defect distribution of the AlSi10Mg manufactured under argon and nitrogen are very similar, which means the influence of gas could be ignored.

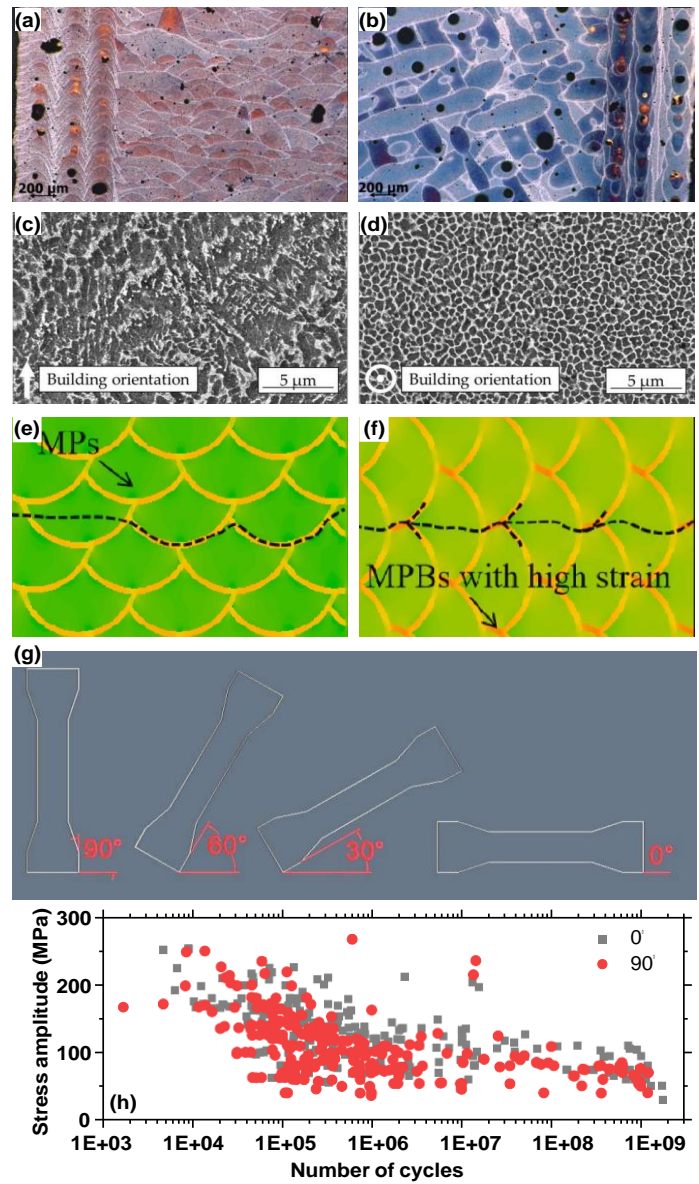

**Fig. 68.** Microstructure of SLM AlSi10Mg: in the Z plane (a and c); in the XY plane (b and d). Reprinted from [265], Copyright(2016), with permission from Elsevier. Reprinted from [283], copyright(2018), with permission from MDPI. Crack propagation path in (e) vertical sample; (f) horizontal sample. Reprinted from [281], Copyright(2021), with permission from Elsevier. (g) different building direction relative to the SLM plane. Reprinted from [287], Copyright(2019), with permission from Politehnium Publishing House. (h) *S-N* data points for sample with different building direction [265, 267, 273, 275, 280, 281, 283, 285, 287, 288, 293, 295, 298, 301, 327].

## 4.4. Post-treatment

Post processing has a received a lot of attention in AM process due to its the mature technology and huge influence. In many cases, post-processing with abundant traditional technique references is more popular than cubersome adjustment of manufacturing parameters. But this does not mean that post-treatment could be used arbitrarily to improve performance. For actual sample with various fabrication process parameters, types and the specific parameters of the post-processing need to be carefully considered. Improper technique would even cause the damage to fatigue strength. Baek et al. [276] investigated the influence of two different heat treatments, T6 and DA on SLM AlSi10Mg, showing that the fatigue performance of T6 alloy are inferior to those of the as-built sample, and in contrast DA treatment effectively improves fatigue performance due to the different morphology of the precipitated Si particles. Maleki et al. [306] found that the smaller and harder ceramic shot peening is better at enhance the fatigue life by improving the surface quality compared to the steel shot. Therefore, the following would discuss in detail how the post-processing adjusts the fatigue properties from the aspects of surface quality, microstructure, and residual stress.

### 4.4.1. Heat treatment

Heat treatment mainly affects the microstructure, ductility and residual stress of the material [264]. The heat treatment for Al-Si alloys is generally based on several widely used methods such as T6 and annealing, and the parameters are modified according to the actual conditions of the metal.

**T6.** The most studied T6 heat treatment could eliminate laser scanning traces and homogenize the Si phase [306], its step is to perform artificial aging after solid solution. As shown in Fig. 69(a), origin as-built microstructure presents a honeycomb shape, and the Si is distributed in the grid. T6 treatment eliminated the molten pools trace and makes the overall structure more homogeneous. The Si phase wrapped around the $\alpha$-Al matrix gradually coarsens into uniformly dispersed spherical particles (Fig. 69(b)). Solution treatment also brings $Mg_2Si$ precipitations that could improve AlSi10Mg performance. However, a too long solution time may cause the formation of easily platelet-like phase $\beta$-$Al_5FeSi$ thus damage the life [314]. Most researches believe that due to the elimination of $\alpha$-Al subcrystals and the coalescence of Si particles enhance the AlSi10Mg matrix, coupled with the reduction of tensile residual stress, ductility of peak-hardened sample is enhanced. This phenomenon is also evidenced by the comparison of fracture surface of samples as depicted in Fig 69(c), (d), the larger dimples in T6 suggest that the fracture mechanism after T6 has evolved from quasi-brittle to ductile fracture behavior [262, 264, 294]. Some researchers have come to the opposite conclusion that the original continuous network dendrite Si has the greatest contribution to fatigue life, therefore separation and coarsening of Si phase caused by T6 plays an adverse effect on fatigue performance [276, 330]. This very different conclusion may be induced by the discrepancies in specific parameters (e.g. ageing time) that make the size of Si particle after T6 unequal, the influence of T6 needs more in-depth discussion, but it can indeed be used as an approach to enhance the fatigue properties of AlSi10Mg.

**Solid solution and aging.** Two steps of T6, solid solution and aging, have been individually studied for their role. Long-term high temperature solution treatment could greatly coarsen Si particles [331]. But since a single solution treatment cannot form too much $Mg_2Si$, the improvement in fatigue strength is not as good as T6 [330]. Baek et al. [276] investigated the effect of direct aging on the fatigue of AlSi10Mg, indicating that such low temperature heat treatment leads to a number of compact fine Si precipitates and the more uniform kernel average misorientation value. Eventually, the fatigue life of the aging specimens is superior to that of the as-built specimens.

**Hot isostatic pressing.** HIP refers to the sintering or densification of products through high temper-ature and pressure, the process parameters are usually above 500 °C and about 100 MPa for AlSi10Mg. The consensus on HIP is that it can relieve the residual stress and spheroidize irregularly shaped voids, decreasing the porosity [325]. It can be seen from Fig. 69(e) the porosity is reduced to one-third of the initial, and feature size is also decreased after HIP. Schneller et al. [332, 333] also added that applied high pressure and temperature could positively affects surface characteristics during HIP. And the Si agglomerations formed after the HIP would slow down crack growth, thus about a 14% increase in fatigue life could be observed. However, Uzan et al. [263] showing the opposite conclusion that HIP leads to the coarsening of the microstructure and the reduction of mechanical performances, resulting in a decline in fatigue properties. Apart from this, an investigation [267] has compared the response of T6 samples with different building directions to HIP, and found the fatigue strength of vertical parts could be improving by HIP. For horizontal samples, HIP negatively influenced fatigue life as given in Fig. 69(f). They speculated this is because the pores that were originally insensitive to cracks were spheroidized by HIP and the stress level increased. Kan et al. [275] observed a similar phenomenon but a completely reverse trend, the fatigue life of vertical samples will drop after HIP, while the life of horizontal specimens would ascend. The above discussions indicate that the impact of HIP as an additional heat-treatment most of the time on AlSi10Mg may be interfered by other factors, more study is required.

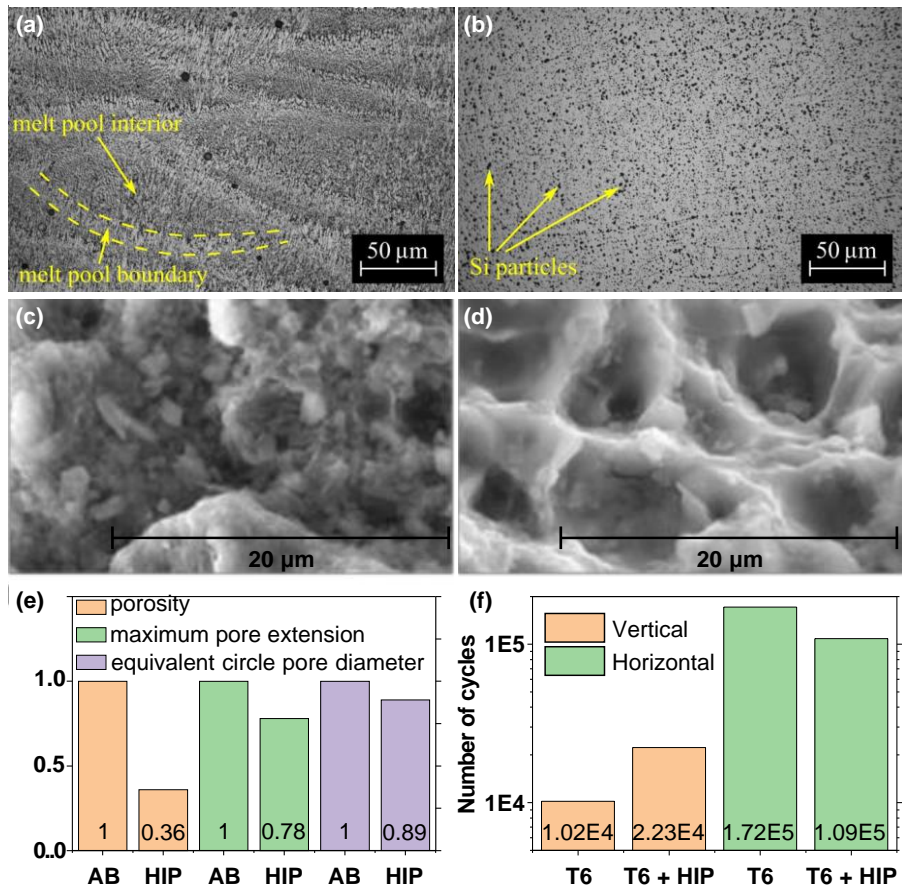

**Fig. 69.** Microstructure of (a) as-built AlSi10Mg; (b) T6 AlSi10Mg. Reprinted from [266], Copyright(2016), with permission from Elsevier. fracture surface of (c) as-built AlSi10Mg; (d) T6 AlSi10Mg. Reprinted from [294], Copyright(2021), with permission from MDPI. (e) normalized comparison of porosity and pore size characteristics between the AB and HIP condition [332]. (f) Fatigue lives for one each build direction and heat treatment combination in specific applied stress amplitude [267].

**Annealing.** Anneal steps in the research for AlSi10Mg are all kept at a certain temperature for 2 h then cooled, it works best for samples without platform heating [274]. After anneal, Si particle begin to precipitate and coarsen, but the original grid structure is still retained as shown in fig. 70. The temperature used for annealing has a great influence on the performance of the sample. Tridello et al. [274] compares the fatigue strength of samples with annealing temperatures of 244 °C and 320 °C. The fatigue limit of AB sample is 65 MPa, runout stress amplitude of the 244 °C sample is increased to 80 MPa, whereas it was equal to 60 MPa for 320 °C sample. The reason is that break and spheroidization of Si network happened at anneal temperature of 320 °C, while the 244 °C anneal could retain the original structure as much as possible and minimize residual stress. Thus only the anneal in 244 °C allow to enhance fatigue response. Annealing could eliminate stress, but it is not limited to this. Some views believe that anneal could promote the formation of high strength microstructure and improve the ductility [334–336]. Whereas, the performance of anneal sample is not as good as that of as-built sample in many fatigue tests [263, 313, 330]. The reason may be the decrease of solid solubility and the destruction of the original structure [330]. Another explanation is that the yield strength lower after anneal [292].

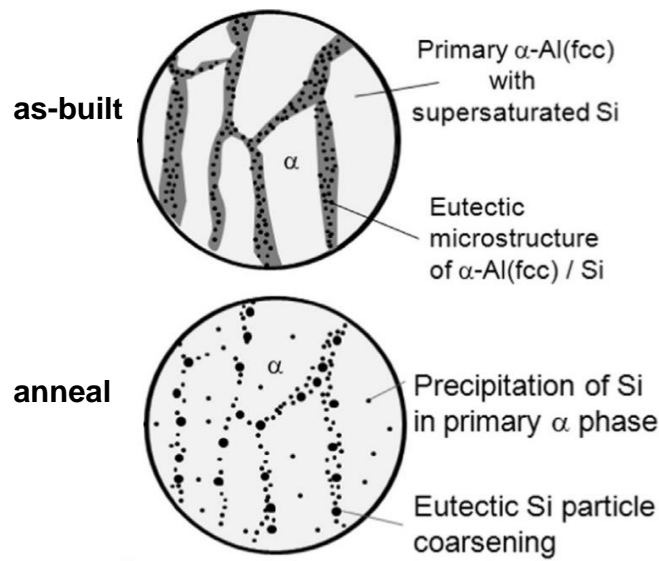

**Fig. 70.** Comparison diagram of AlSi10Mg microstructure, Reprinted from [331], Copyright(2017), with permission from Elsevier.

Fig. 71 lists the number of literatures about heat treatment that affect fatigue life in *S-N* chart. It can be seen there are disagreements in the views on the influence of heat treatment of three heat treatments, especially for HIP and annealing that have a larger number statistics of negative effects. The reason for this divergence may be the difference in specific parameters used makes the morphology and size of Si that affect the fatigue properties diverse. Furthermore, multiple reinforcement mechanisms including fine-grain and dispersion strengthening make it tough to discuss the role of Si [337, 338]. On the whole, the exact process parameters need to be carefully selected to optimized procedure if the improvement of fatigue life

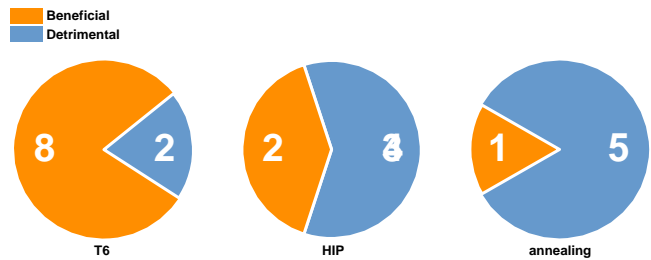

**Fig. 71.** According to the *S-N* diagrams, the number of documents discussing the effect of heat treatment on the fatigue life of AlSi10Mg, orange represents the improvement of fatigue performance, and blue represents the damage of fatigue performance [17, 262–264, 266, 267, 274–276, 292, 294, 297, 306, 313, 329, 330].

through heat treatment is pursued.

### 4.4.2. Surface treatment

Two-thirds of fatigue fractures are induced by surface cluster of pores [339], so surface treatment that brings improvement of surface and sub-surface characteristics is essential. In addition to conventional processing methods such as SP and polishing, novel methods including friction stir processing (FSP) and vibra polishing have also received attention.

### 4.4.3. Shot peening

Low-cost and easy-to-operate shot peening could also effectively raise the fatigue life of AM AlSi10Mg. According to the experimental results of Bagherifard et al. [17], the strength of the sample after SP reaches 370% of fatigue strength of the as-built product in 3E06 times cycles. There are various contributory factors in the fatigue performance advancement brought by SP, Fig. 72(a) demonstrates the high kinetic energy of SP realized the significant reduction of multiple surface roughness indicators [340]. Although there are opinions that roughness will increase, the overall surface morphology still develops more regular and homogeneous [306]. The violent surface deformation caused by SP directly increased the surface hardness, while the rise of plastic deformation and dislocations could also drive the microstructure refinement and further strengthen the microhardness [341]. In terms of influence of pores, SP not only tremendously reduces the porosity of near-surface area (up to a depth of 0.5 mm), but also make pores turn to a more favorable direction for stress intensity [270]. In addition, SP mainly targets the pores in the surface and sub-surface areas, and it also has a well elimination effect for small-sized pores. Compared with HIP that mainly eliminates large voids in the entire material, SP could improve the fatigue performance in a targeted manner [342]. Finally, the stretched surface layer after SP has a compressive residual stress due to the interaction with the inner layer (Fig. 69(b)), which would alleviate the effect of tensile fatigue [306]. Under the synergistic effect of improvement of toughness, surface hardness, reduction of porosity, and compressive residual stress, the exaltation in fatigue life by SP is quite considerable. However, Fig. 72(b) does not simply reflect the promotion of SP to the compressive residual stress, it also illustrates the difference in the response of samples to SP. It is not difficult to find that the outermost and maximum compressive residual stress of HT specimens are relatively small, whereas the overall stress depth is larger than that of AB sample. The reason is that the heat-treated samples with lower hardness downscale the surface treatment effect, which will weak the role of SP in fatigue strength. The fatigue strength of AlSi10Mg are summarized in Table 2. SP is applied to the T6 sample, and the strength is only increased by 37 MPa compared with only-T6 sample, which is far inferior to the 135 MPa performance improvement of SP on the no-T6 sample. This proves that SP still has a strengthening impact on HT specimens, but the effect is not satisfactory. SP process parameters are also worth considering, at the same coverage rate, ceramic shots with larger hardness and smaller size are more effective than steel shots in the surface morphology modification [306]. The trade-off of SP intensity would affect the strengthening focus, the high SP intensity would result in a surface damage and structure refinement, low-strength conditions protect the surface but coarsen the crystal grain [341].

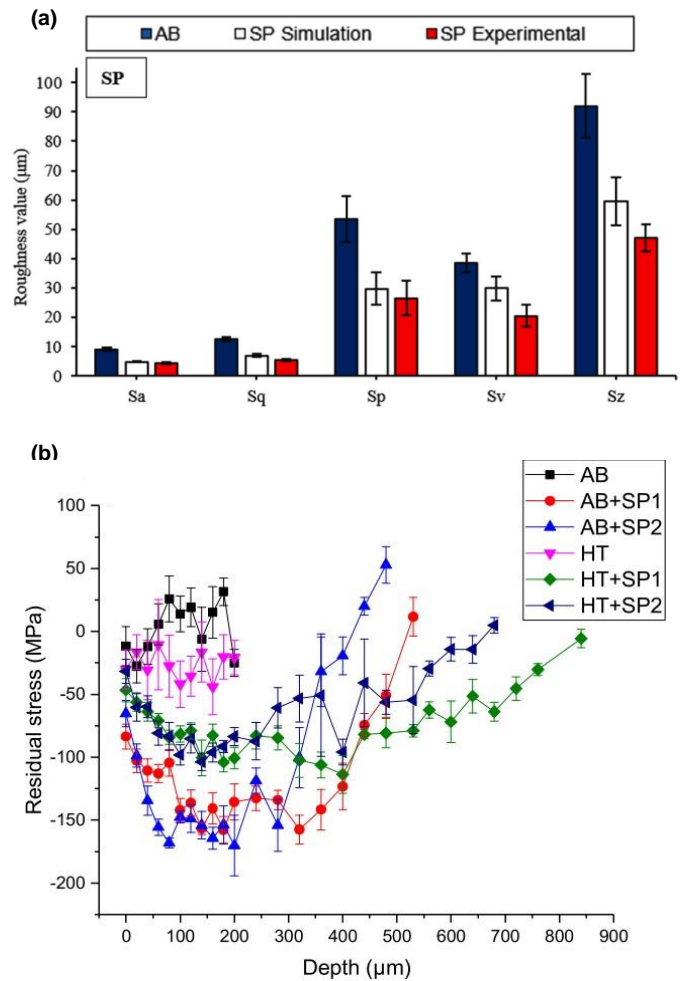

**Fig. 72.** Comparison of the surface roughness data. Reprinted from [340], Copyright(2021), with permission from Elsevier. $S_a$: areal arithmetic average height; $S_q$: areal root mean square deviation; $S_p$: areal maximum height of peaks; $S_v$: areal maximum depth of valleys; $S_z$: areal maximum valley to peak height. Residual stresses distributions of specimens. Reprinted from [306], Copyright(2021), with permission from Elsevier. SP1 represents cast steel shot material; SP2 represents ceramic shot material.

### 4.4.4. Sand blasting

Sand blasting (SB) is very similar to shot peening processes, the only difference is that the medium particles used in sand blasting (e.g. glass and corundum) are angular, while the round particles are used in shot peening. SB could stably promote the fatigue strength of AM AlSi10Mg, Nasab et al. [268] believed this promotion is particularly prominent in HCF situation. This view is based on the impact kinetic energy of sand blasting is relative low, and the improvement is mainly concentrated on the surface roughness that is more sensitive to HCF [313]. Although Pola et al. [286] thought that SB also facilitates the transition from tensile to compressive residual stress to a certain extent. Because of the lower kinetic energy, SB could be regarded as a gentler SP treatment, and has a better life improvement on HT samples with low hardness than SP [17]. This is proved in Table 2 showing that the fatigue strength of T6 + SB sample reaches 175 MPa and is close to that of SB sample, whereas the strength of T6 + SP is only 102 MPa.

**Table 2.** Fatigue strength of AM AlSi10Mg corresponding to 3E06 cycles [17].

| Sample series | Fatigue strength (MPa) |
|---|---|
| AB | 50 |
| SP | 185 |
| SB | 173 |
| T6 | 75 |
| T6 + SP | 102 |
| T6 + SB | 175 |

### 4.4.5. Machining

The treatment methods for removing the surface material including machining and polishing. The machining effect of AM products discussed is generally micro-machined type, and the main function is to reduce the roughness. Aboulkhair et al. [264] concluded that the enhancement of fatigue performance induced by machining is concentrated on HCF, and it is hard to improve the fatigue strengthen at high stress levels. This is exactly the same as the above conclusion about SB, and the improvement in fatigue strength by decreasing the roughness is more prominent under HCF condition. Corresponding, whether the AlSi10Mg alloy has been heat-treated or not, MA both could bring a pretty life extension [297].

### 4.4.6. Polishing

The most intuitive thing about polishing is that it can reduce the roughness of material, the decline of $R_a$ of AlSi10Mg could attain two-third after polishing [263]. As the surface profile become flat and large defects that prone to multiple cracks are eliminated, the cracks need to initiate at deeper interior position, which means more energy is consumed [259, 314]. Even for the finished product that has been machining, polishing could still slightly increase the fatigue respond [263], and show improvement in the whole range of fatigue lives [268]. However, according to the researches of Mower et al. [265] and Nicoletto et al. [302], the increase in fatigue life caused by polishing seems to be unstable. The fatigue life of sample is illustrated in Fig .73(a). For the polished conditions, the fatigue strength is indeed enhanced, but it can be seen that data scatter has also become larger. This could be explained by the surface morphology before and after polishing (Fig. 73(b), (c)), the high points of the original uneven scaly surface are indeed removed, but the reduction of materials also makes the internal pores emerge on the surface, which has an adverse effect on the fatigue life. This means that the depth of removal surface needs to be carefully considered. The study of Uzan et al. [259] illustrates that as the depth increases, the fatigue life gradually rises. The life reaches a maximum value when the depth is 25 $\mu m$ and then begins to decrease. When the depth reaches 60 $\mu m$, the fatigue life is basically not increased compared with that of unpolished samples. The type of polishing are generally mechanical and electrolytic polishing. Mower et al. [265] believe that electrolytic polishing is hard to reduce the surface roughness. There are also experiment [259] found that even though the removal depth is same, the performance of mechanical polishing samples is slightly better than that of electrolytic polishing. Besides that, the vibratory polishing has gradually attracted attention, the surface roughness could be reduced to a relatively low value, and the fatigue strengthen induced by vibratory polishing is still not as good as that of sand blasting and the combination of machining and polishing [268].

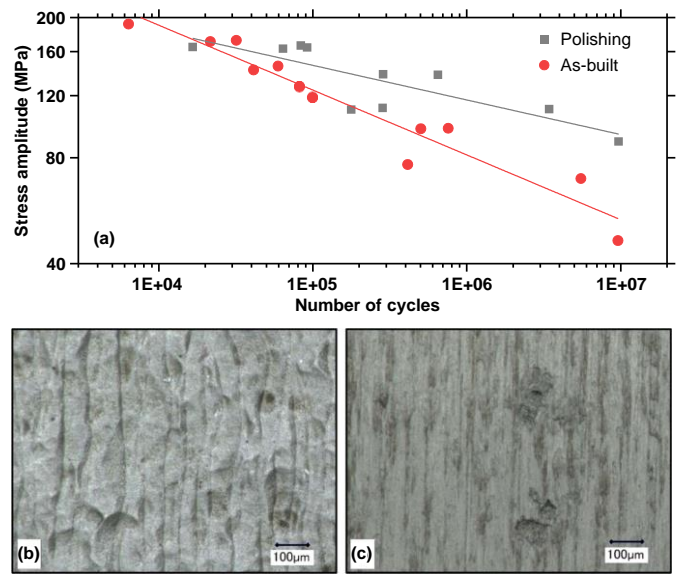

**Fig. 73.** (a) *S-N* data points and fitted line for sample with or without polishing. Optical microscopy of SLM AlSi10Mg specimen surfaces for (b) as-built and (c) mechanically polished. Reprinted from [265], Copyright(2016), with permission from Elsevier.

#### 4.4.7. Other methods

SantosMacias et al. [292, 335] investigated the effect of friction stir processing (FSP) on the fatigue bahavior of AlSi10Mg, showing that FSP not only lessen the porosity from 0.09% to 0.02%, but also change the metal microstructure. The interconnected network is broken in the area where FSP impacts, which greatly improves the ductility, and the heat affected zones are eliminated thus reducing the influence of local strain. Hence a improvement in fatigue resistance after FSP is observed.

By using LSP, it could play a role similar to SP that enhances residual compressive stress and lower porosity. At the same time the potential disadvantages of SP, that is the increased roughness and micro crack, could be avoided [342].

## 4.5. Workpiece types

Since the size of defects in AM products are generally larger than of those in traditional process products, AM products may be more sensitive to the workpiece size effect. However, there are few studies on the influence of geometry of AM AlSi10Mg workpiece on fatigue at present. Tridello et al. [293] compared that property of hourglass specimens and Gaussian specimens shown in the Fig. 74(a), (b), respectively, suggesting that fatigue resistance of hourglass specimens is always stronger than that of Gaussian sample (Fig. 74(c)). They analyzed that the maximum defect size of the hourglass sample was only half of that of Gaussian sample regardless of the building orientation.

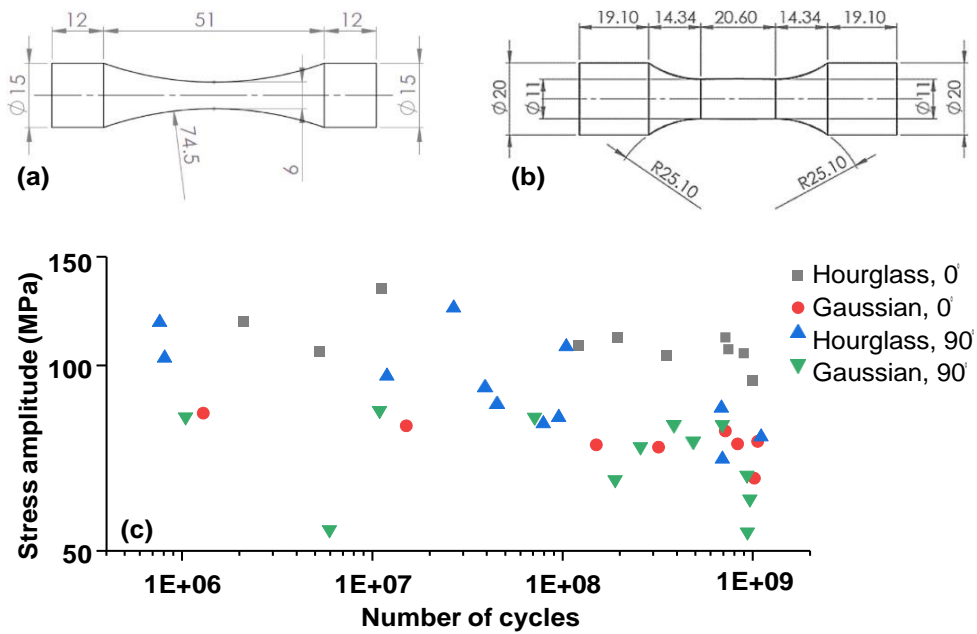

**Fig. 74.** (a) Hourglass specimens geometry; (b) Gaussian specimens geometry; (c) *S-N* data points for sample with different geometry characteristic. Reprinted from [293], Copyright(2021), with permission from Elsevier.

## 4.6. Work temperature

AlSi10Mg alloy is not suitable for working in high temperature condition. There are many aspects to the weakening of the resistance of AlSi10Mg due to high temperature. When the AlSi10Mg specimen is cyclically loaded at a lower temperature (about 100 °C), the internal high-density $\beta''$-$Mg_2Si$ precipitates could well pin the dislocations and generate cyclic hardening. Whereas the high temperature makes $\beta''$-$Mg_2Si$ transformed to the $\beta'$-$Mg_2Si$, activating dislocations and reducing the threshold of dislocation climbing, and eventually give rise to cyclic softening [343]. In terms of microstructure, cyclic loading at low temperature will promote the formation of subgrains, as the temperature rise, the grain is considerably coarsened, voids started to enlarge, and the coarsen Si particles precipitated around the grain boundary. Different from the strengthen Si particles formed during HT, the coarse Si particles precipitated at high temperature are much larger as shown in Fig. 75(a), (b) and a amount of pores occurring around the Si. The high-density voids grow and coalesce rapidly, seriously weakens the fatigue life [334]. In conclusion, the high temperature causes the crack growth rate to accelerate and the corresponding fatigue life is also shortened. It can be seen from Fig. 75(c) that the fatigue life is only 25% of that at room temperature when the test temperature reaches 400 °C.

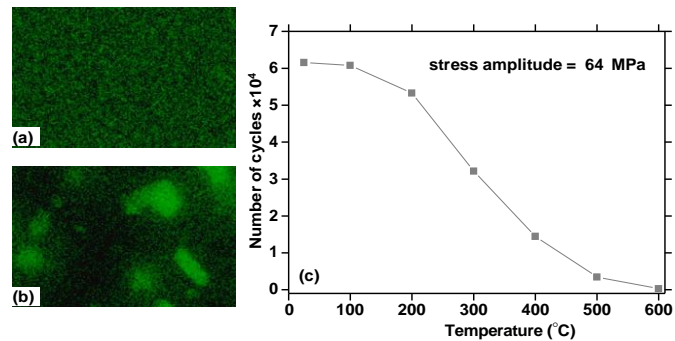

**Fig. 75.** Energy dispersive spectrometry maps for (a) SR material; (b) the material test at 400 °C. Reprinted from [334], Copyright(2021), with permission from Elsevier. (c) the relationship between fatigue life and the temperature [312].

### 4.7. Stress ratio

The influence of stress ratio should be considered, as the stress ratio increases, the fatigue life would decrease [279]. Under HCF condition, the fatigue strength of the sample with stress ratio of -1 is 3 times that of the sample with stress ratio of 0.5.

### 4.8. Fatigue crack growth

The fatigue crack growth (FCG) behavior could illustrate the fatigue resistance of AlSi10Mg from another aspect. At the same stress intensity factor ΔK, the larger the rate of change of the crack length $a$ with the cycle times $N$, the worse the fatigue resistance. As shown in Fig. 77, the overall dispersion of the FCG rate is greater than of the fatigue characteristics, and the data on the left and right shows different characteristics. The threshold stress intensity factor (SIF) of the data on the right is generally above 5 MPa$\sqrt{m}$. Most of the curves on the right are consistent with the NASGRO model, that is, as the intensity factor increases the expansion rate decreases, and them have a log-linear relationship, and the crack size expands rapidly when fracture occurs. On the contrast, the threshold stress intensity factor of data on the left is only 1.5 MPa$\sqrt{m}$. The explanation given by Rhein et al. [278] is that it may be due to ultrasonic fatigue (f = 20 kHz) allows small crack increments to be observed. However, the frequencies used in the other two references [289, 296] did not reach the ultrasonic level, only 15 and 10 Hz. These data on the bottom left even has a downward trend at the beginning, the reasons [278] may be due to the temporary hindering effect of Si particles on crack propagation, thus the curve as a whole dropped and showed an irregular jagged shape.

SantosMacias et al. [292] analyzed the FCG results under different post-treatments as depicted in Fig. 76. Compared with the AB specimens, both SR and FSP treatments reduce the propagation rate, but there are also difference between them. It can be seen that the propagation rate of the FSP sample is lower at the beginning, and the crack occur later, but the subsequent propagation is faster. In the early crack propagation, the pores with larger size is the main influence, the smaller porosity of the sample after FSP delays the extension of the crack. As the stress intensity increases, the existence of Si-rich eutectic phase and the elimination of the molten pool boundary accelerate the expansion, so the FSP specimens have a steeper fitting result.

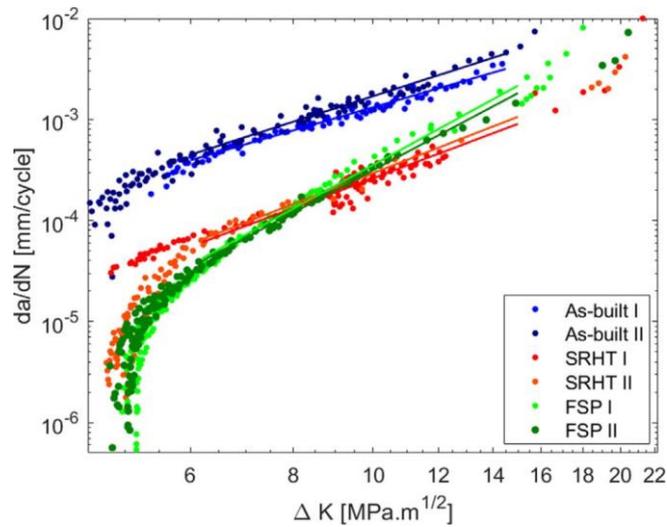

**Fig. 76.** FCG rate curves and fitting lines for AlSi10Mg in AB, SRHT, and FSP [292].

DiGiovanni et al. [344] investigated the sensitivity of AlSi10Mg to the direction of building before and after T6. The FCG response of the original sample is extremely dependent on the building orientation, the cracks in vertical (Z-Y) sample could move quickly along the heat-affected zone, while the horizontal (Y-X, Y-Z) sample forces the crack to propagate in the tough $a$-Al matrix and bear the residual compressive residual stress. The eutectic segregate Si, which is easy to be the crack propagation path, precipitates and coarsens into spherical particles after T6 treatment, and the residual stress is also homogenized. The crack propagation rate of specimens in different directions remains consistent. Rhein et al. [278] modified the aging heat treatment of the standard T6 to obtain a lower crack growth rate than of the original T6 condition.

Xu et al. [281] observed the obvious FCG behaviour difference of samples with different building orientations. 0° and 15° samples not only have smaller propagation rates, their threshold SIF and fracture toughness are relatively larger. This means that the sample with a smaller building angle would occur crack and eventually fracture later, while samples with larger angle (45° and 90°), that is, closer to the vertical direction, will represent rapid fatigue fracture characteristics.

### 4.9. Hybrid parts

Tommasi et al. [307] used casting A356-T6 as a substrate, and performed different surface pretreatments on the machined surface of the substrate. AlSi10Mg was then deposited onto the machined surface of cast aluminum base by DMLS technique, the obtained uncut raw sample is shown in Fig. 78(a), additional T6 heat treatment is applied to homogenize the hardness. Undoubtedly, the contact interface of aluminum alloys is the key part (Fig. 78(b), (c)), and the surface preparation that affects the contact properties will determine the fatigue performance of the hybrid component. They performed fatigue tests on three surface-pretreated samples, founding that the sand blasted and laser textured specimens were fractured at the DMLS part, and the fracture of the grinded samples occur at the interface. Correspondingly, the fatigue strength of laser textured and sand blasted samples is already on a par with of homogeneous A356-T6 and even slightly better than of pure DMLS AlSi10Mg, expecting for the very low lifetime of the grounded sample. The DMLS process on the ground surface results in a large number of lack of fusion defects thus impairing the joint performance. Sand blasting and laser texturing provide random and regular textures, respectively, increasing the contact surface, thus helping to promote compact and in-place powder layers during DMLS processing of the first few layers, strengthening the tightness of the connection.

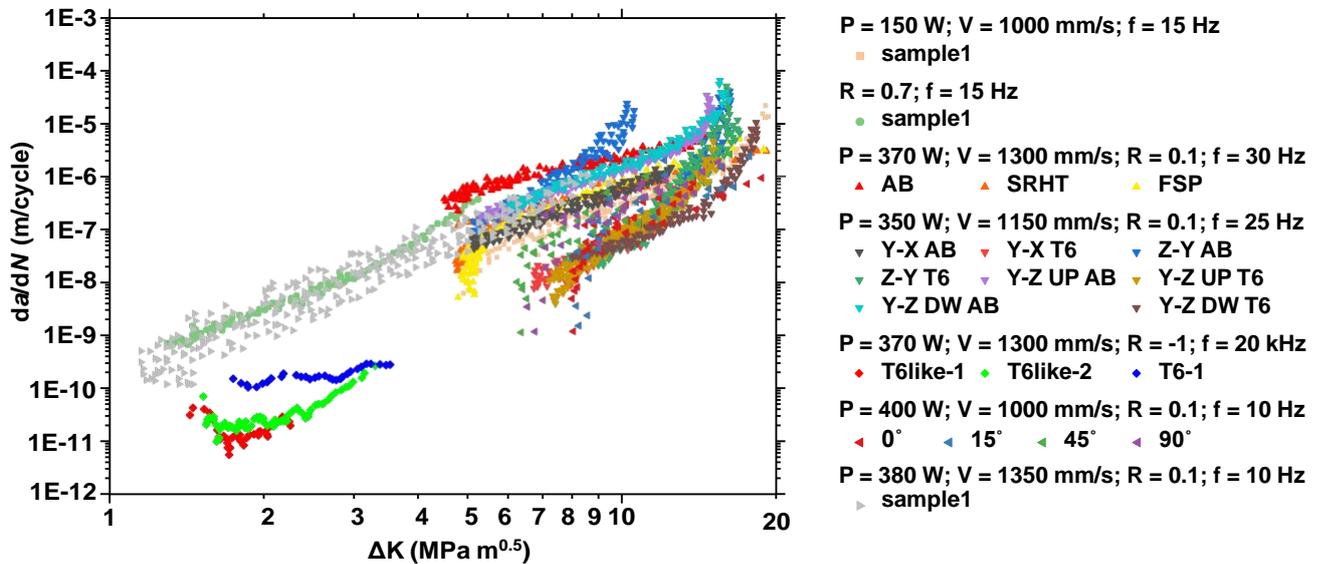

**Fig. 77.** Fatigue crack growth rate data for AM AlSi10Mg [278, 281, 289, 292, 296, 336, 344].

## 5. Other Al alloys

### 5.1. AlSi12 alloys (AA 4xxx)

Aluminum and silicon (Al-Si) alloys is widely synthesized by AM due to its lower melting point and narrower solidification temperature range. Selective laser melting (SLM) of AlSi12 leads to the formation of a 3D interconnected network of Si dendrites infiltrating an aluminum matrix resulting in a eutectic morphology [345]. During the process, Si precipitates is easily rejected due to the fast cooling rates. And the ultrafine eutectic microstructure is beneficial for the mechanical properties. It has been found that the SLM processed (SLMed) AlSi12 is very brittle but rather tough. The ultra tensile strength (220–380 MPa) and yield strength (180–220 MPa) of SLMed AlSi12 alloys is higher than cast specimens (≈200 MPa) but the ductility (≈3%) is lower than the cast ones (≈10%) [346, 347]. Rashid et al. [348] found different building orientations (parallel, inclined and perpendicular of the substrate) had great influence of quasi-static mechanical properties on SLMed AlSi12. Ma and coworkers has paid attention to the Si content (12, 20 and 35 wt%) influence of Al-Si alloys [349, 350]. The precipitate Si particles in lower Si-content Al alloys (AlSi12) were smaller and more evenly distributed in the Al matrix after hot pressing and hot extrusion, compared to higher Si content Al alloys. This led to the better ductility in AlSi12. Tensile and yield strength, conversely, had positive correlation to the Si content. UTS (YS) increased from 219.8 ± 1.8 to 287.5 ± 2.7 MPa (from 127.5 ± 1.1 to 199.4 ± 1.6 MPa) with the Si content of 12–35 wt%, while the opposite trend of elongation indicated high brittle in high Si content Al-Si alloy.

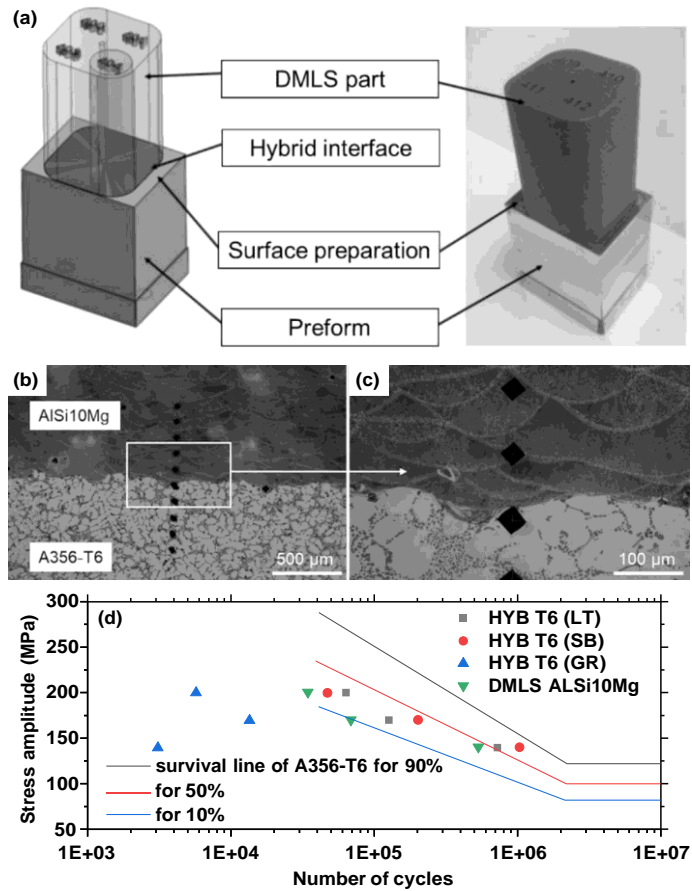

**Fig. 78.** (a) The structure of raw hybrid block that need to be cut. (b) microscope image and (c) relative high magnification of hybrid interface. (d) *S-N* data points of three hybrid parts and pure DMLS AlSi10Mg, and the survival line of homogeneous A356-T6 for 90%, 50%, 10%. LT represents laser texture, SB represents sand blasting, GR represents grinding. Reprinted from [307], Copyright(2021), with permission from MDPI.

Fatigue performance of AM AlSi12 has been investigated by previous researches, and the fatigue cycles were shown in Fig. 79 (a). The very high cycle fatigue response (VHCF) could even up to $10^9$ with the stress amplitude ($\sigma_a$) of 100 MPa [351, 352], while the cycle lives were around $10^3$ with $\sigma_a \geq 200$ MPa. This was attributed to the different manufacturing and testing measures. The effective factors of cycle fatigue and fracture performance could be separated in at least four parts, e.g. particle information itself, manufacturing parameters, testing specimen size and post-processing ways. It had been reported that the particle size was used as 20-63 $\mu$m with an average diameter of 33 $\mu$m [352, 353]. Siddique et al. [347] found that high laser power and energy density, low scan speed and building rate are beneficial to the growth of fine microstructure of AlSi12 and great quasi-static mechanical properties which is even higher than cast specimens. Therefore, the results was taken into account in subsequent AM process and fatigue performance tests, and more attention had been paid into other influence factors, such as platform heating

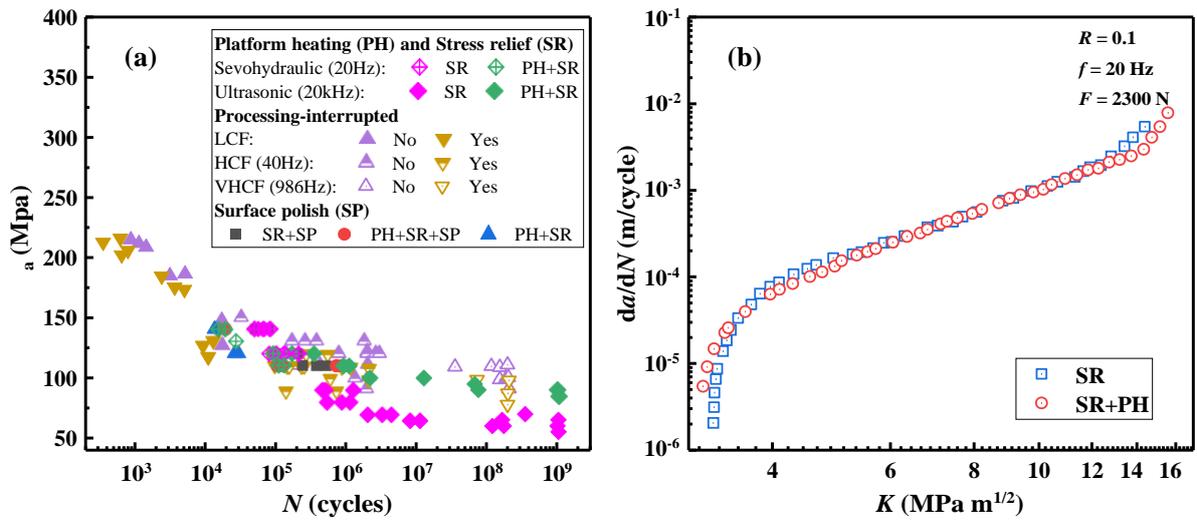

**Fig. 79.** (a) Woehler (*S-N*) curves and (b) fatigue crack growth behavior of AlSi12. Data from [345, 352, 354–356]

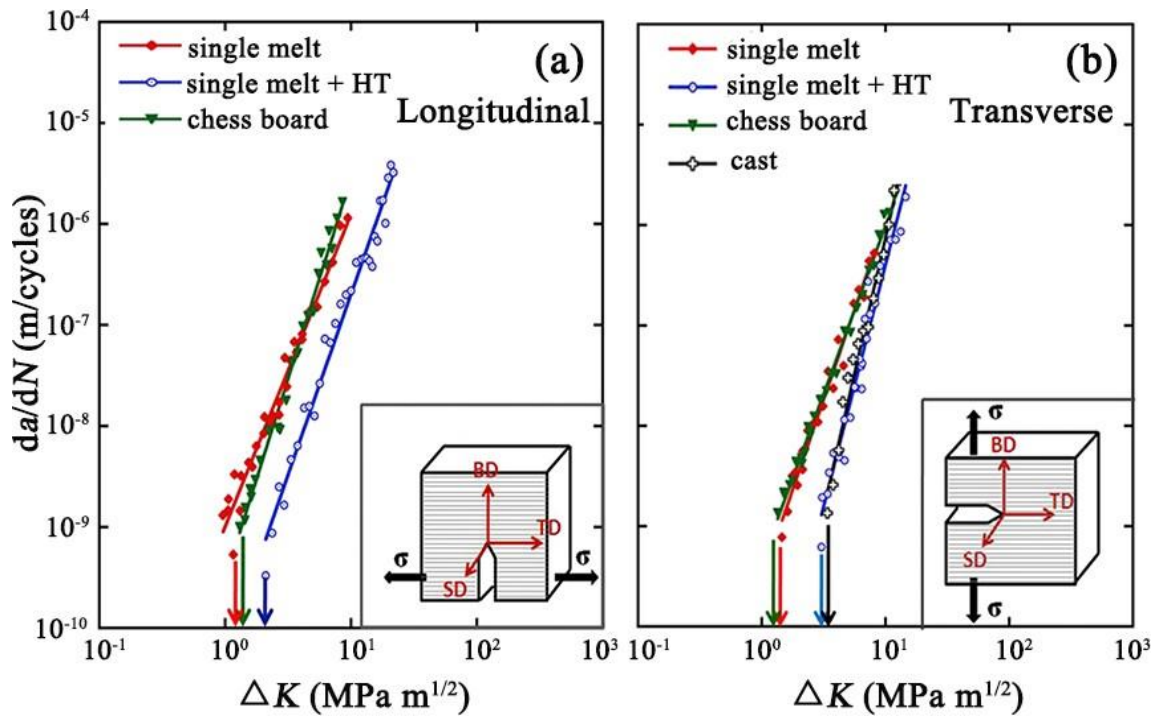

**Fig. 80.** Fatigue crack growth behavior of AlSi12 with (a) longitudinal and (b) transverse building direction. Single melt and chess board represented the different scanning strategies. Reprinted from [357], Copyright (2016), with permission from Elsevier.

and post-processing solution.

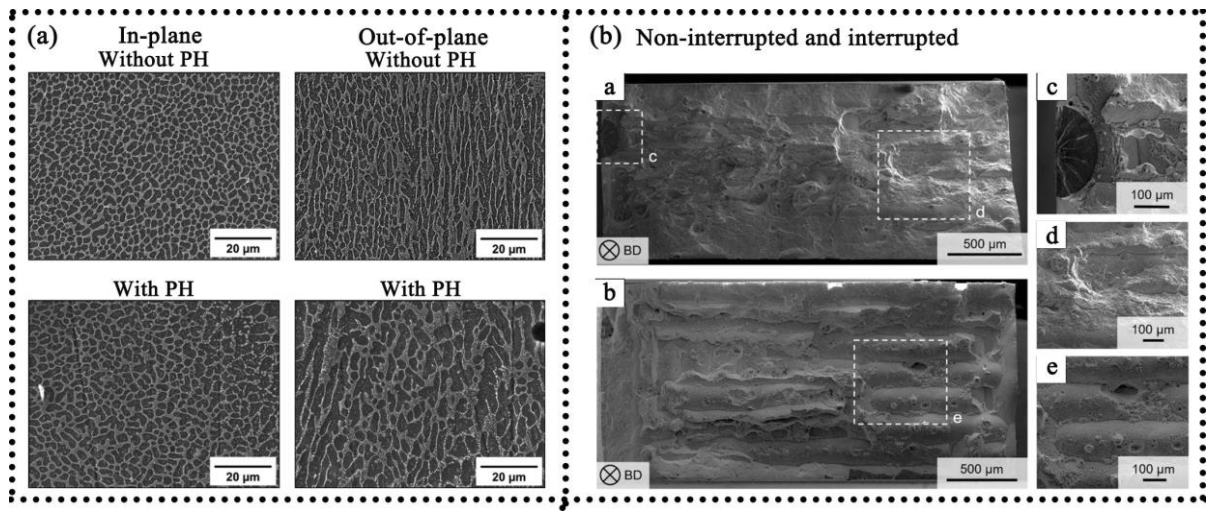

**Fig. 81.** (a) Microstructures of AlSi12 without or with platform heating (PH). Reprinted from [322], Copyright (2020) with permission from Elsevier. (b) Crack fracture of non-interrupted and interrupted AlSi12. Reprinted from [355], Copyright (2021), with permission from Elsevier.

### 5.1.1. Manufacturing parameters

Firstly, scanning direction would be an important manufacturing parameter that influence the anisotropy of crack growth in AlSi12 alloy. Suryawanshi et al. [357] considered the effect of scanning strategies and building directions on fatigue performance in AlSi12 alloys. The scanning strategies included single melt (SM) and checker board (CB) hatch styles, while building directions contained longitudinal and transverse to the laser tracks. First, fatigue toughness of as-built SM and CB samples in the same build direction was improved (Fig. 80), especially the samples with longitudinal loading direction whose were three to four times than that of the cast alloy ($\approx 11$ MPa$\sqrt{m}$). The marked improvement in the toughness values of the SLMed alloys was somewhat surprising due to their ductility was substantially low as compared to the cast alloy. However, the fatigue toughness with different scanning strategies has little discrepancy. On the other hand, the anisotropy of fatigue properties was revealed by different building directions that longitudinal building direction led to about 20% higher fatigue toughness ($K_q$ = 46.7 MPa$\sqrt{m}$) than transverse direction ($K_q$ = 37.9 MPa$\sqrt{m}$).

When considering gas porosity, it was proposed that applying heat (200 °C) to the deposition platform could effectively reduce heat gradient and thus inflict more stable melt pools to enhance degassing [322, 352, 354, 356, 358]. The microstructures of AlSi12 without or with platform heating (PH) were shown in Fig 81(a). Dendritic walls became thickened and saturated Si formed as agglomerates with bigger sizes. From the in-plane view, the thickening of dendritic walls in samples with PH were more observable. From the out-of-plane view, more textures in specimens with PH than that in specimens with PH indicated that the anisotropy of the microstructure in samples with PH was reduced. Grain refinements, as well as supersaturation of Si particles in the Al matrix, were the main contributors to the achieved strength.

Platform heating improved the very high cycle fatigue strength of AlSi12 alloy. The results of the experiments [322, 354] indicated that fatigue strength in very high cycle regime ($\geq 10^6$ cycles) of specimen manufactured with PH was 45 % higher than without PH, while only slight difference in low cycle regime. This indicated mutual interactions of porosity and microstructure in SLMed AlSi12 alloy with PH. The high melting state of powder particles due to platform heating resulted in the reduction of remnant porosity. Fracture surface investigation showed that multiple as well as internal crack initiation occurred in samples without PH; whereas only surface crack initiated in the samples with PH which increase the fatigue strength. Especially, in very high-cycle fatigue where plastic saturation of grains did not occur, fine grains combined with reduced porosity were a more effective combination to resist fatigue crack initiation and result in higher fatigue strength.

However, the influence of base platform heating on fatigue crack growth only exhibited at initial crack propagation phase. Results of a crack propagation test could be classified in different regions as shown in Fig. 79(b) where a relation between crack propagation rate with respect to number of cycles (d$a$/d$N$) and stress intensity factor $\Delta K_{th}$ can be realized [352]. Region I ($\Delta K < 4$ MPa$\sqrt{m}$) represented an initial crack propagation phase, Region II (4 MPa$\sqrt{m} \leq \Delta K \leq 12$ MPa$\sqrt{m}$) was the Paris region with stable crack propagation and Region III ($\Delta K < 12$ MPa$\sqrt{m}$) was the unstable crack propagation. Threshold stress intensity factor $\Delta K$ for samples without and with PH were 3.16 MPa$\sqrt{m}$ and 3.36 MPa$\sqrt{m}$, respectively, indicating the resistance of samples with PH to crack propagation was higher. This could be ascribed to that the microstructure of samples with PH was coarser than those of samples without PH by microstructural analysis. Therefore, the resistance to crack growth could be improved by building the samples with PH. In the Paris region, both the samples have nearly similar crack propagation behavior due to their coincident slopes, which implied weak dependent relationship between crack growth and PH treatment in the stable crack propagation region. Besides, the theoretical fitting results exhibited that the variation of the final fatigue life of both batches was likely to result also from differences in fatigue crack propagation, not only crack initiation properties.

Given manufacture of large-size and time-consuming components, some emergencies (e.g. power outage or issues in process gas supply) occurred, which result in the process instabilities. Richter et al. [355] investigated the effect of processing interrupt on AM AlSi12. As for quasi-static properties, the specimens with or without interrupted showed slight differences. In the LCF regime the effect of processing interrupt on cyclic stress responses could be ignored as well. However, The HCF fatigue limit for the non-interrupted and interrupted built specimens are about 121 ± 11 MPa and 118 ± 7 MPa, respectively. The runouts in the VHCF regime occur below a level of 104 ± 8 MPa and 94 ± 10 MPa for the non-interrupted and interrupted built specimens, respectively. Thus, only a minor impact of process interruptions on fatigue properties is found regardless of the fatigue regime considered.

They concluded that the fracture surfaces were characterized by the presence of well-known defects like pores and other process-induced features. The mismatch in terms of local microstructure was rationalized based on residual stresses and non-uniform thermal expansion, which directly influenced the crack initiation. Fracture surfaces for the non-interrupted and interrupted building conditions after LCF testing were revealed in Fig. 81(b). Crack initiations could be both found near the surface of non-interrupted (Fig 81(b)-a) and interrupted (Fig 81(b)-b) building specimen, and easily distinguished. In non-interrupted built specimen (Fig. 81(c)-d) some dimple-like facets indicated the ductile fracture and areas exhibited an even surface even without any signs of proper bonding, while in interrupted built specimen (Fig. 81(c)-e) the surface fraction of the even area without proper connection was obviously higher than in case of the non-interrupted built specimen. However, a severe detrimental effect on the mechanical properties could not be found.

### 5.1.2. Post-processing

Contrary to the platform heating which reduced the density of gas porosity, the post-processing of stress relief heat treatment (SR) increases the pores. The stress-relieved samples had a higher percentage of porosity as compared to those non-stress-relieved samples [352]. Porosity percentage for the samples turned out to be 0.25% for AM samples with SR. The application of stress relief treatment allowed further dendritic width growth in AM samples due to more thermal energy to drive the segregation of Si particles from the Al matrix to the grain boundaries.

Tenkamp et al. [345] found SR had positive feedback to fatigue cycles of SLMed AlSi12. With load ratio ($R$) of -1 and frequency ($f$) of 20 Hz, in the HCF and VHCF regime, the fatigue strength difference was distinct that samples with SR had a significantly increment of fatigue strength and lifetime compared to those without SR. However, in the LCF regime, the difference was small since the interaction between porosity and microstructure. They further proposed that porosity was not the main influence factor behind fatigue failure, but the defect size. Compared to samples with SR, the defect size in samples without SR was larger with non-regular shape which acted as internal micro notches so that it led to significantly less fatigue lifetime, especially at a stress amplitude of 70 MPa.

On the other hand, the less gas porosity in samples without SR rendered to form the finer microstructure which resisted plastification at crack tips, hindered propagation and delayed fracture. With $R = 0.1$, the $\Delta K$ value of samples without or with SR were 3.2 and 3.5 MPa$\sqrt{m}$, respectively. This revealed that SR had slightly negative influence on the fatigue crack growth of SLMed AlSi12. Suryawanshi and coworkers [357] tested the Paris region of SLMed AlSi12 without or with heat treatment (HT). The d$a$/d$N$ curves were shown in Fig. 80. Heat treatment of the SLM alloys reduced the fracture resistance compared to as-built samples, especially in the longitudinal direction. Even then, the $K_q$ value measured at 19.7 MPa$\sqrt{m}$ was nearly twice of that of the cast alloy. This manifested that SLMed samples with or without HT possessed higher fatigue toughness but substantially lower ductility compared to the cast samples.

In addition, bending fatigue strength of SLMed as-built and heat-treated samples were 60 and 110 MPa, respectively, while that of cast sample was 94.5 MPa [357]. This indicated the heat treatment could improve the bending fatigue strength of SLMed AlSi12 alloys as well. However, in Siddique's research, the effect of stress relief was diminished after platform heating as compared to the effect without platform heating [353], which provided the guidance for follow-up researches. Until now, most published reports about SLMed AlSi12 had taken PH and SR effect into account.

The role of process-induced surface defects in fatigue performance was investigated [356, 359]. Under the same manufacturing processing, it was found the surface roughness of samples with and without surface polish (SP) craft were 16.27 and 45.25 $\mu$m [356]. The samples with SP exhibited longer fatigue life but stronger fatigue scatter than those without SP. A drop of fatigue life caused by high stress concentrations at surface roughness which significantly outweighed that at internal defects. When surface roughness was substantially high, crack initiation from surface became the dominant failure mechanism. And the fatigue scatter occurred due to interacting phenomena like defect size, location as well as the magnitude and type of load.

## 5.2. AlSi7Mg alloys (A356/A357)

AlSi7Mg aluminium alloy, a typical cast alloy material, had many advantages in structural durability and corrosion resistance for potential applications in aerospace and automotive industries. The industry classification of AlSi7Mg includes A356 and A357, where the differences are that the contents of Mg in A356 and A357 are 0.3–0.45 % and 0.45–0.6 %, respectively. Recently, this kind of aluminium alloy has been widely synthesized by AM technology. The main combination of AM AlSi7Mg is Al, Si and Mg, in which the content of Si is lower than AlSi10Mg and AlSi12. Similar with AlSi10Mg and AlSi12, AlSi7Mg can be relatively easy to process by laser fabrication due to its relatively small difference between liquidus and solidus temperatures compared to other high strength aluminium alloys.

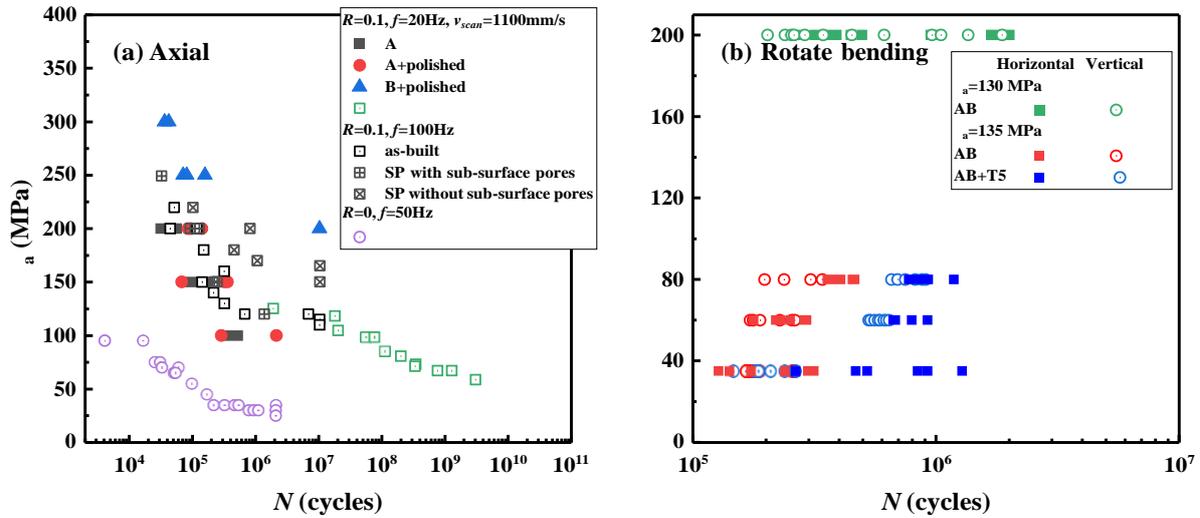

**Fig. 82.** Woehler (*S-N*) curves of SLMed AlSi7Mg for (a) axial fatigue sponse. Data from [360–363]. (b) Effect of platform temperature on rotate bending fatigue cycles. Data from [364].

Several researches had reported that the quasi-static mechanical properties of SLMed AlSi7Mg (e.g. ultimate tensile strength of 400 MPa, yield strength of 200 MPa and breaking elongation of 12–17 %, respectively) were even better than cast alloys [345, 365, 366]. Besides, under different process and post-process parameters, the tensile mechanical properties exhibited obvious discrepancies as well [366, 367]. For example, AlSi7Mg performed mechanical anisotropy under different building direction (vertical or horizontal direction). The mechanical properties along the horizontal direction were better than that along the vertical direction. Besides, manufacturing substrate temperature rendered to the relatively low quasi-static tensile properties.

For dynamic mechanical properties of AlSi7Mg, *S-N* curves in Fig. 82 showed the total trend of fatigue performance of AM AlSi7Mg with different pro- and post-processing treatment. Compared to AlSi10Mg, AlSi7Mg exhibited worse fatigue response to be subjected to higher stresses under similar range of life cycles [361]. The highest fatigue limit strength was up to 300 MPa with $10^5$ cycles. Strong fatigue scatter could be observed due to different load ratios and frequency, and other factors as well, which should be further described. Rotate bending fatigue properties had been investigated by previous research [364]. The fatigue cycles were in the region of $10^5$–$10^6$ at the stress amplitude of 135 MPa. The influence factors, e.g. platform temperature and heat treatment, were discussed in detail in the following subsections. In addition, a little researches reported the three-point bending fatigue of AlSi7Mg [368, 369]. Boniotti et al. [369] designed three micro-lattice structure from SLMed AlSi7Mg and found that the initial ratcheting strain rate was significantly correlated to the life of each sample.

### 5.2.1. Manufacturing parameters

The process parameters (e.g. energy power and density, laser scanning speed, building direction, hatch distance and layer thickness) are considered to be the most important factors in regulating fatigue performance of AM fabricated parts.

Rao et al. [360] tested the effect of several process parameters (laser power, hatch distance, layer thickness and energy density) on the axial fatigue performance of SLMed AlSi7Mg. The machined samples under two different fabricating parameters were investigated under $f$ = 20 Hz and $R$ = 0.1. Samples with higher energy density ($E_l$ = 60.6 J/mm$^3$), smaller hatch distance ($d_H$ = 0.15 mm) and layer thickness ($h$ = 0.04 mm) showed higher fatigue life than those with opposite conditions ($E_l$ = 37.9 J/mm$^3$, $d_H$ = 0.3 mm and $h$ = 0.06 mm), which was corresponding to the result of AlSi12 mentioned above and indicated the relatively high energy density was more preferred. The cycle fatigue of the former samples at 200 MPa was up to 10$^7$, which is twice of that of cast A357 alloy. When the stress level was about 250 MPa, the cycles of failure were more than 5×10$^4$.

The effect of building direction on rotating-bending fatigue of AlSi7Mg was investigated in Ref. [364], including horizontal and vertical directions. Horizontal specimens had longer fatigue life than those of vertical specimens with the same manufacturing parameters and testing conditions, while the fatigue scatter of the former was stronger as well. At 200°C, the average fatigue life obtained from the horizontal condition (9.63×10$^5$ cycles) was 1.67 times longer than the vertical condition (5.77×10$^5$ cycles), indicating the construction of AlSi7Mg along horizontal direction was prone to better fatigue performance in terms of average workpiece quality. The authors found that the difference in scatter of diverse built direction led to the lower probability of survival of horizontal specimens, but this might be the marginal trend that could be changed by more tests.

In addition, platform heating is one of the most important processing steps to effectively reduce the porosity and defects, which could improve the axial cycle fatigue strength. Though this was ignored in Rao's work, some other researchers had considered. Bonneric et al. [370] used the load ratio (R=0.1) and platform heating of 150°C, and found the cycle failure was at least 2×10$^6$. With platform heating of 180°C, the HCF and VHCF limit were 2×10$^6$ with 125 MPa and 2×10$^9$ with 60 MPa, respectively [361]. However, the samples without platform heating also had high cycle failure (1×10$^7$) with the stress amplitude of 110–115 MPa (the test frequency is 100 Hz) [362]. This might be attributed to the different testing measure but also indicated the high fatigue scatter of AlSi7Mg.

Martins et al. [364] discussed the effect of platform heating on the rotating-bending fatigue of AlSi7Mg. The samples were fabricated at four different platform temperature (35°C, 60°C, 80°C and 200°C). The first three temperature conditions were considered as the cold temperature and the last one was high temperature. With R=-1, the fatigue scatter at high substrate temperature was obviously higher than that at low temperature, which was attributed to the coarser grain forming at high temperature. The rotating-bending fatigue life at 200°C of 130 MPa was beyond 3×10$^6$, which was comparable with the axial fatigue life (2×10$^6$ with stress amplitude of 125 MPa at platform temperature of 180°C), indicating the comparable torsion and tensile strength. Besides, the average rotating-bending fatigue life of 200°C was the lowest (9.63×10$^6$ cycles along horizontal direction and 5.77×10$^6$ cycles along vertical direction), while that of 80°C was the highest (3.39 ×10$^6$ cycles and 2.11×10$^6$ cycles). This was quite different from the common axial fatigue testing parameter that the platform temperature was often used as 200°C. At the cold platform temperature range (35°C–80°C), however, the results of the comparison of the average fatigue lives revealed a increasing trend of warmer conditions, which indicated the positive effect of PH on rotating-bending fatigue performance, to some degree.

### 5.2.2. Post-processing

The irregular shape formed by surface defects and internal LOF pores might be detrimental to fatigue performance, which indicates the significance of post-processing of AM samples.

Stress relief and heat treatment often play positive roles in fatigue performance of AM aluminium alloys,

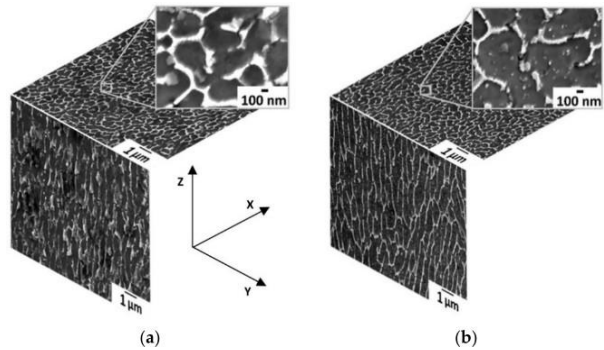

**Fig. 83.** High-magnification SEM images (10,000× and 50,000×) of the typical microstructure at 35 °C (a) without and (b) with T5 treatment. The Al matrix is the dark phase and the lighter phase surrounding the Al matrix is Si particles. Reprinted from [364], Copyright (2021), with permission from MDPI.

which had been taken into account by most previous reports. For the rotating-bending fatigue of AlSi7Mg, the samples with T5 heat treatment elongated the fatigue life compared to those without T5, regardless of different platform heating temperature or different building directions [364]. The T5 heat treatment resulted in the precipitation of Si particles (Fig. 83), which might act as microstructural barriers to the mobility of dislocations, and benefit to fatigue performance by delaying the accumulation of plastic damage.

Herein, the effect of stress relief and heat treatment on axial fatigue cycles of AlSi7Mg was neglected. The reasons were as follows: 1. lack of sufficient experimental data so that the comparison between diverse tests might be unreasonable due to the discrepant parameters (e.g. testing frequency and load ratio); 2. advantages of SR and HT on fatigue properties on other Al-Si alloys proposed by most previous reports.

Another controllable measure is surface polish which could effectively eliminate big pores and reduce the porosity. Yang et al. [362] proposed a further investigation about process parameters and porosity development in AlSi7Mg parts. This was, specifically, the effect of polishing degree on the axial fatigue performance. They compared the axial fatigue behavior at R = 0.1 with a frequency of 100 Hz of three different kinds of samples, namely (i) as-built samples (no sub-surface porosity removal and no surface finishing); (ii) surface machined samples (no sub-surface porosity removal); and (iii) machined samples without sub-surface porosity. After printing, each sample was thermally treated through stress relief, solution heat treatment, and artificial aging. The results showed that the fatigue performance of surface polished samples were better than that as-built samples. The fatigue limit of machined samples without sub-surface porosity were up to $10^7$ cycles at a stress level of 175 MPa, but the fatigue life decreased with the presence of sub-surface pores. Thus, they proposed that sub-surface porosity can be more influence on the fatigue performance than surface roughness. The sub-surface porosity were attributed to contour and core edge keyhole type defects which could be reduced by lowering laser energy and changing scan tracks of laser. At the same stress amplitude (200 MPa), the fatigue life of AlSi7Mg with surface polish ($1.4 \times 10^5$) was about 2.5 times of that without surface polish ($5.6 \times 10^4$) [360]. However, the fatigue scatter of polished samples was stronger than those non-polished samples.

### 5.2.3. High temperature fatigue

Thermo-mechanical fatigue (TMF) reflects fatigue performance under the condition combined with mechanical and thermal loads. TMF effects on the low cycle fatigue properties of cast A356-T6 were reported that TMF life could increase with the increasing aging time and decreasing strain range [372]. There were large differences in the fatigue lives due to the strain range and aging time. Recently, the TMF effects on SLMed AlSi7Mg alloys had been investigated as well. When the temperature range of thermal cycle was fixed at 100°C-280°C, the low cycle fatigue response of AlSi7Mg deteriorated significantly with the increasing mechanical load from 90 to 120 MPa [373, 374]. This result of SLMed AlSi7Mg alloys was comparable of the cast A356-T6.

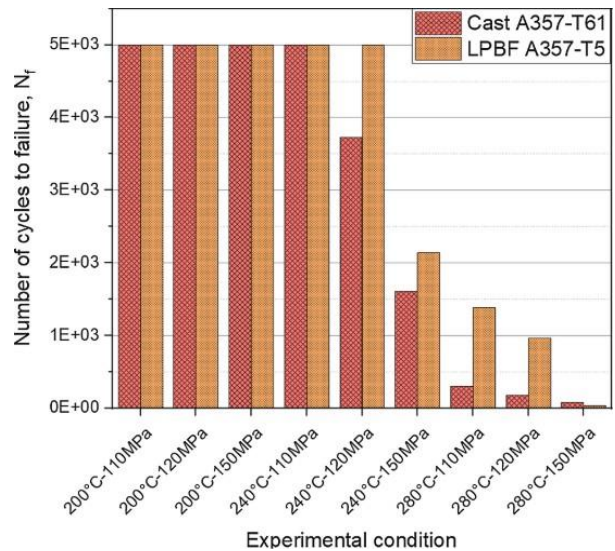

**Fig. 84.** The comparison of thermal cycle fatigue between SLMed AlSi7Mg and sand cast Al357-T61. Reprinted from [371], Copyright (2021), with permission from Elsevier.

Then, Sajedi et al. [371] took a comparison of the thermal fatigue behavior between SLMed AlSi7Mg and sand cast A357-T61. The results of the TMF experiments under the tensile load at the different temperature intervals were shown in Fig. 84 for both cast and L-PBF alloys. The thermal cycling tests of both L-PBF and cast samples could reach the run-out limit (setted as 5000 cycles) under three stress load of 110, 120 and 150 MPa for the temperature range of 100–200°C. However, when the maximum temperature was 240°C, the cast alloy failed after 3730 cycles under the stress level of 120 MPa and the L-PBF sample still reached the run-out limit. When the same thermal analysis was performed on the samples subjected to temperature cycles between 100–200°C and 100–240°C in three consecutive cycles, there was no evidence of phase transformation for the over-aged samples, suggesting that the precipitation was already completed and confirming no any further phase transformation presented. The A357 alloys processed by both technologies failed systematically at the stress of 150 MPa, in the range of 100–240°C and under all three stress levels in the range of 100–280°C ($T_{mean}$ = 190°C). It was obvious that by increasing the upper temperature limit of the thermal cycling and stress level, the fatigue life decreases significantly. Both cast and L-PBF samples tested under the stress of 150 MPa within the range of 100–280°C failed only after few tens of cycles.

## 5.3. Al-Mg-Sc-Zr alloys (Al5024)

Weldable Al-Mg 5xxx alloys cannot be strengthened through heat treatment, which brought the barriers for its mechanical performance improvement. An high-strength Al-Mg alloy modified by a small amount of scandium and zirconium, namely Scalmalloy, has been produced for fusion based additive manufacturing because of the alloy being aged hardenable [382–385]. In melt pool areas close to the pool base, numerous Al3(Sc,Zr), Al-Mg-oxides and mixed particles act as nuclei for Al matrix solidification, leading to the formation of a very fine grained microstructure. The participation of Sc and Zr would finally form as Al3(ScZr) particles, which could act in different ways to refine microstructures, and improve mechanical properties. The ultra tensile stress (≥ 400 MPa) and yield strength (≥280 MPa) of as-built samples were even higher than those of AlSi10Mg. Besides, the static mechanical properties of Al-Mg-Sc-Zr alloys could be improved after an optimize heat treatment (between 325°C/4h and 350°C/4h): the ultra tensile stress ≥ 400 MPa, yield strength ≥ 280 MPa) and elongation to fracture ≥ 8.6±1.9% [386, 387].

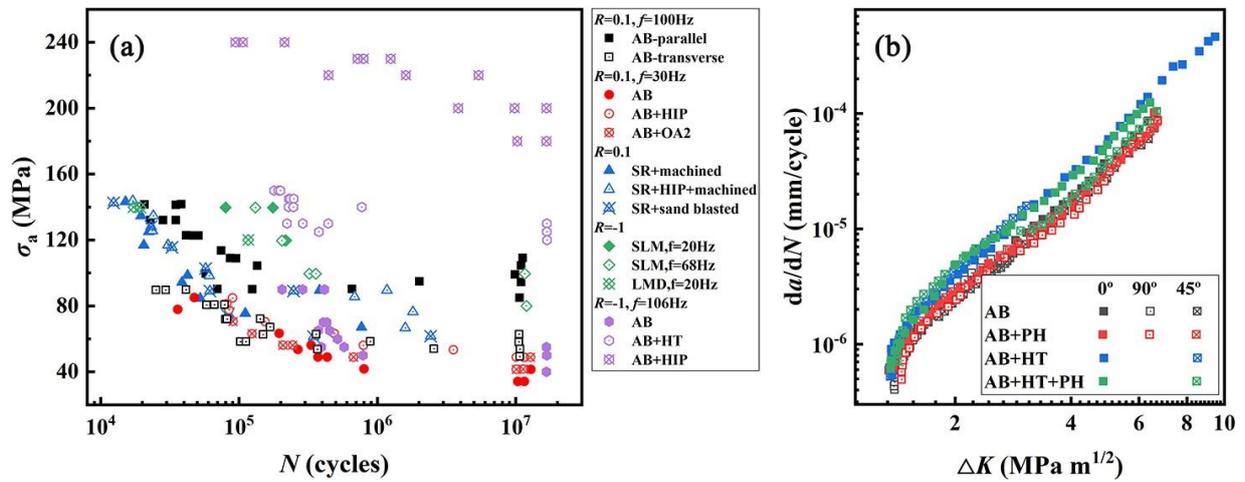

**Fig. 85.** (a) Woehler (*S-N*) curves and (b) fatigue crack growth of AM Al-Mg-Sc-Zr alloy. Data from [375–381]

Compared to traditional manufacturing methods, additive manufacturing technology enables a high reduction of waste by avoiding chips typically generated in 'subtractive'. The utilization of residual unfused powder can further promote the economic feasibility and sustainability of L-PBF process. Griffiths et al. [388] investigated the microstructure of Al-Mg-Zr alloys after laser rescanning. Li et al. [389] proposed that a few Si addition into the Al-6Mg-0.2Sc-0.1Zr alloys could effectively inhibited hot cracking and multiply strengthened the mechanical properties. Additionally, Cordova et al. [390] revealed that the powder reuse has a slight influence on the quasi-static mechanical behaviour of Al-Mg-Sc-Zr alloy, where the differences were less than 10%. It was concluded that Al-Mg-Sc-Zr powders were suitable for reuse in the studied number of building cycles, if proper powder sieving and powder rejuvenation steps were included to preserve the required powder properties without detrimental effects on the repeatability and microstructural/mechanical properties of the manufactured L-PBF parts. The relative studies about powder reusing or laser rescanning effect on fatigue of high strength Al-Mg-Sc-Zr alloys were rare till now, which remained to be further investigated.

Fig. 85(a) showed the fatigue cycles of Al-Mg-Sc-Zr alloys from different studies [375–380]. In general, the wide range of cycle life from $10^4$ to $10^7$ indicated the strong fatigue scatter. Meanwhile, most tests of fatigue performance of Al-Mg-Sc-Zr alloys were performed with the stress amplitude below 160 MPa. These must be traced back to different pro- and post-process and testing measure of fatigue performance examinations, which would be listed and discussed as follows.

### 5.3.1. Manufacturing parameters

The effect of different manufacturing methods on fatigue performance of Scalmalloy was revealed by Awd and coworkers [380], which contained SLM and LMD. LMD can be considered an alternative to SLM due to larger deposition rates and larger average size of particles (92.5 $\mu$m of LMDed samples and 45 $\mu$m of SLMed samples). In their tests, the stress load ratio and frequency were applied as -1 and 20 Hz for both SLMed and LMDed samples. The LCF and HCF performances of SLMed samples were obviously better than LMDed samples, especially at 140 MPa (the former was one order of magnitude higher than the latter). At least two SLMed samples with load frequency of 68 Hz would exceed the upper bound of cycle limit at stress amplitude of 100 and 80 MPa, while the others coincided with the calculated boundaries. The high fatigue scatter of both SLMed and LMDed samples, additionally, indicated that no repeatability of properties was possible with the process setup.

Qin et al. [376] investigated the fatigue properties of Al-Mg-Sc-Zr alloys with different building directions (vertical and horizontal). The results showed that, in all tests, horizontally deposited (HD) samples exhibited better fatigue strength than vertically deposited (VD) samples, while their tensile and hardness properties were similar. The superior fatigue strength (100.5 MPa) of HD and VD samples were 100.5 and 57 MPa, respectively. The anisotropic fatigue properties were attributed to the different defects and microstructure induced by discrepancy of building directions. For the defects, the lack of fusion resulted in a higher stress concentration in the VD samples than that of the HD samples in the crack initiation stage. For the microstructure, owing to the indistinctive crack deflection and possible cyclic softening of the columnar grains, a decrement in fatigue resistance appears. Thus, the different crack propagation paths led to different columnar grain or equiaxed grain area ratios for the anisotropic fatigue properties.

Chernyshova et al. [381] investigated the crack growth of Scalmalloy and revealed the influence of the notch orientation and platform heating. The fatigue crack growth curves were shown in Fig. 85(b), where the data was averaged and smoothed to form a curve. First, as-built samples exhibited the slight difference of three orientations (the angle between crack direction and building direction of 0°, 90° and 45°). The $\Delta K_{Th}$ values all close to 1.36 MPa$\sqrt{m}$ and in Paris law region $C \approx 3.3 \times 10^{-10}$ MPa$\sqrt{m}$ and $m \approx 2.9$. This indicated the independent relationship between crack growth and building direction. Additionally, the influence of platform heating was investigated as well (red in Fig. 85(b)). The fatigue behavior had no change with $\Delta K_{Th}$ = (1.34–1.42) MPa$\sqrt{m}$ and the curves in Paris law region were nearly coincident. Thus, fatigue crack growth had no obvious dependence with building orientation and platform heating treatment.

However, there exist no report about how the platform heating affects the fatigue cycles of Al-Mg-Sc-Zr alloys.

### 5.3.2. Post processing

Considering the influence of secondary Al$_3$Sc precipitates in an in-situ manner during L-PBF fabrication, Li et al. [377] applied post heat treatment to produce secondary strengthening precipitates and obtain the required tensile strength and other important mechanical properties. Three conditions were performed on the Scalmalloy samples: as built (AB, treated as a pre-heating temperature of 35°C), hot isostatic pressing (HIP, performed for 4 h at 325°C with 100 MPa pressure in high purity nitrogen atmosphere) and two-step over aging (OA2, an aging temperature at 300°C for a duration of 5 min, followed by water quenching and then another aging step at 350°C for 18 h in ambient atmosphere). The *S-N* curves showed that, in LCF range, OA2 samples exhibited comparable fatigue response with AB samples, but HIP samples possessed the highest fatigue cycles among all conditions. Besides, the AB, OA2 and HIP samples exhibited the 10$^7$ cycle fatigue strength of 34, 41, 47 MPa, respectively, which was consistent with the increasing yield strength of 236, 380, 414 MPa. HIP samples possessed the best tensile strength and fatigue life at $10^7$ fatigue cycles while sacrificing structural ductility.

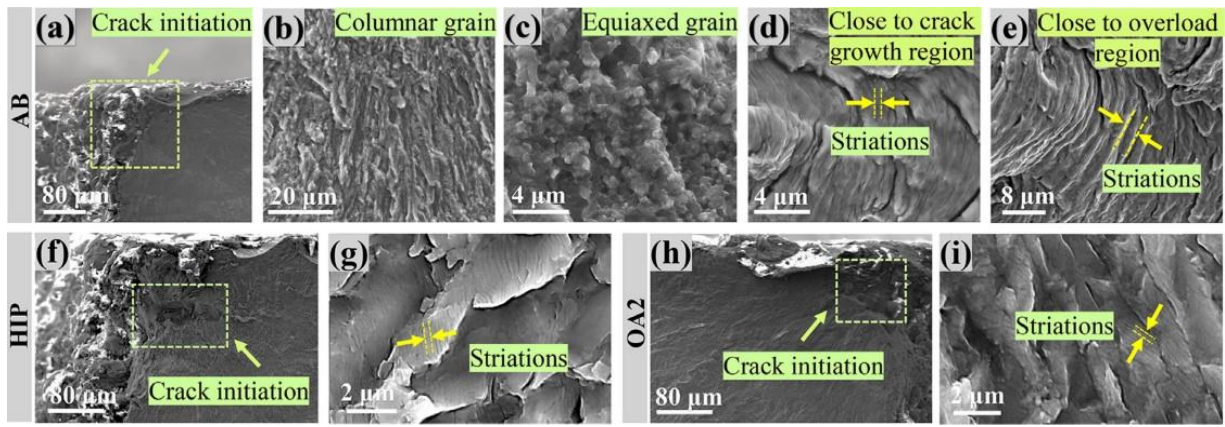

**Fig. 86.** Fatigue fracture surface in L-PBF AB (a-e), HIP (f, g), OA2 (h, i) Al-Mg-Sc-Zr sample tested at 135 MPa. Reprinted from [377], Copyright (2021), with permission from Elsevier.

For fatigue fracture in Fig. 86, the observed initiation sites in AB, HIP, OA2 samples were quite similar, and cracking initiates corners with large surface defects and high roughness that provide stress concentrators [377]. In AB samples (Fig. 86(d) and (e)), the fracture striations close to overload region was obviously wider than that closer to crack growth region, which implied the reduction of effective cross-section area promoted the crack growth. On the other hand, the samples after heat treatment (HIP and OA2) exhibited more shallow and less apparent striations than the AB samples in crack growth area, indicating the weak crack growth phenomenon after heat treatment and explaining the good fatigue performance. Fatigue crack growth curves for heat-treated samples were shown in Fig. 85(b) with blue and green blocks. The threshold stress intensity factor $\Delta K_{Th}$ of heat-treated samples was close to those without heat treatment. In Paris law region $R \approx 5.0 \times 10^{-10}$ MPa$\sqrt{m}$ and $m \approx 3.0$. This indicated that the fatigue crack growth rate of heat-treated samples increased lower, and heat treatment was beneficial for fatigue performance of Al-Mg-Sc-Zr alloys though it was widely considered that microstructure of alloys had slight influence on $R$ and $m$ values.

Lasagni et al. [378] considered not only the effect of heat treatment on fatigue performance of Scalmalloy, but the influence of surface polishing. There existed two group for convenient comparison: one was the samples with or without HIP, and the other was the samples with different surface treatment (machined or sand blasted). The stress relieve were applied in all samples. The results showed that Scalmalloy samples easily cracked in high stress level (≥ 200 MPa). Among all tested samples, those with HIP and aging (machined) performed the best fatigue cycles, especially for stresses below 200 MPa, which were several times larger than those without HIP treatment. This was attributed to the few porosity (a final porosity of 0.10%) in HIP samples which approximately 60% less than in samples without HIP treatment. In Ref. [379], the HIP treatment and surface modified measure were considered as well, where the results showed small difference with Lasagni's conclusion. With R=-1, the fatigue performance in HIP samples was 220 MPa at about $1.0 \times 10^6$ load cycles, while the maximum stress amplitude of machined samples was 130 MPa. It must be pointed that different sample geometry was used with continuously decreasing cross section at a minimum, which exhibit less stress concentration within the testing section in comparison with uniform cross section samples used in different works.

Regarding the mechanical performance, both surface machine and sand blasted were responsible of inducing surface compressive stresses that improved the performance and prolong the life of the material. This took relative importance from the industrial point of view, where for the fabrication of parts with complex and/or topology optimized geometries, surfaces might be nor feasible to machine, but also increased the final cost of a part. As for the effect of different surface treatment, there was no relatively large discrepancy between the machined and sand blasted samples [378].

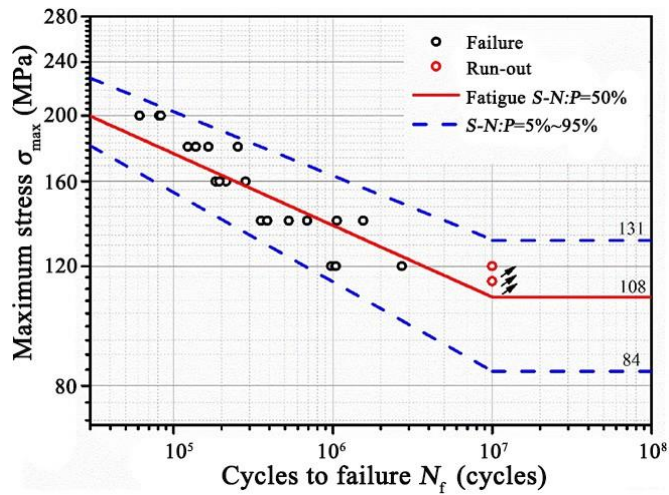

**Fig. 87.** Wochler (*S-N*) curves of WAAMed Al-Mg4.5Mn alloys. Reprinted from [391], Copyright (2021), with permission from Elsevier.

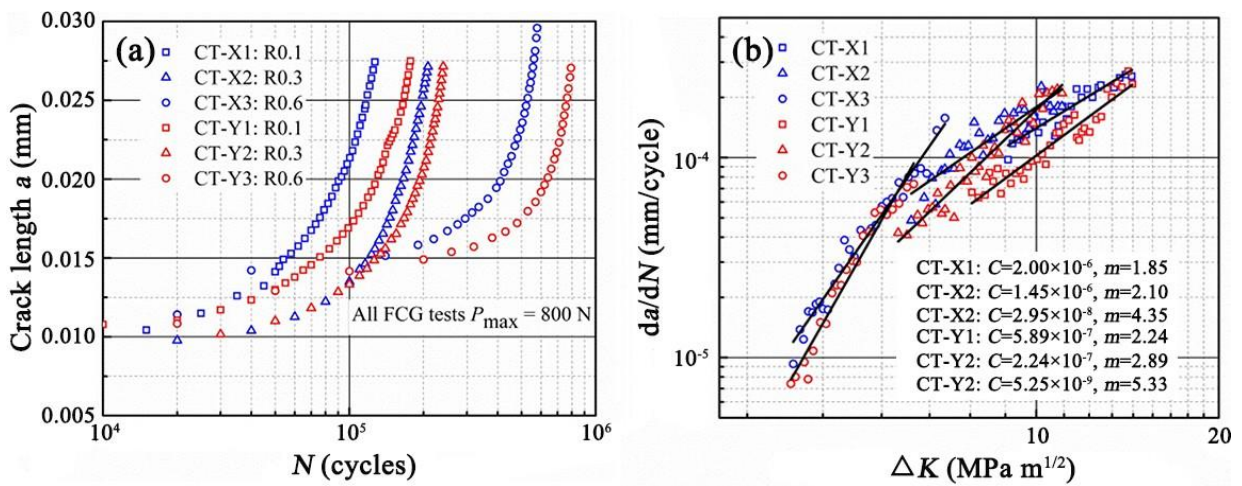

**Fig. 88.** (a) crack length as the function of fatigue cycles and (b) fatigue crack growth of WAAMed Al-Mg4.5Mn alloys, X (transverse) and Y (longitudinal) notch orientations related to building direction, respectively. Reprinted from [392], Copyright (2021), with permission from Elsevier.

## 5.4. Al-Mg-Mn alloys (AA5087)

Al-Mg4.5Mn alloys (AA5087), one kind of Al-Mg alloys, have been widely applied in the shipping, military, automotive, and aviation industries due to their favorable weldability and excellent fatigue resistance. Since the non-heat-treatable wrought and large-scale complexity, as reported, wire arc additive manufacturing (WAAM) is another choice for rapid manufacturing Al-Mg4.5Mn alloys [391–393]. The mean yield strength, ultra tensile strength and elongation for WAAMed AA5087 alloy were 142 MPa, 291 MPa and 22.4%. With increasing loads (15–45 kN) and deformation rates, yield strength increased with 19.7%–69% and ultra tensile strength with 3.4%–18.2% [393]. YS and UTS of the alloys with loading 45 kN reached to 240 MPa and 344 MPa, respectively. These values were close to or higher than standard commercial Al-Mg alloys with similar compositions, implying that the investigated alloy could meet the industrial requirements for practical implementation.

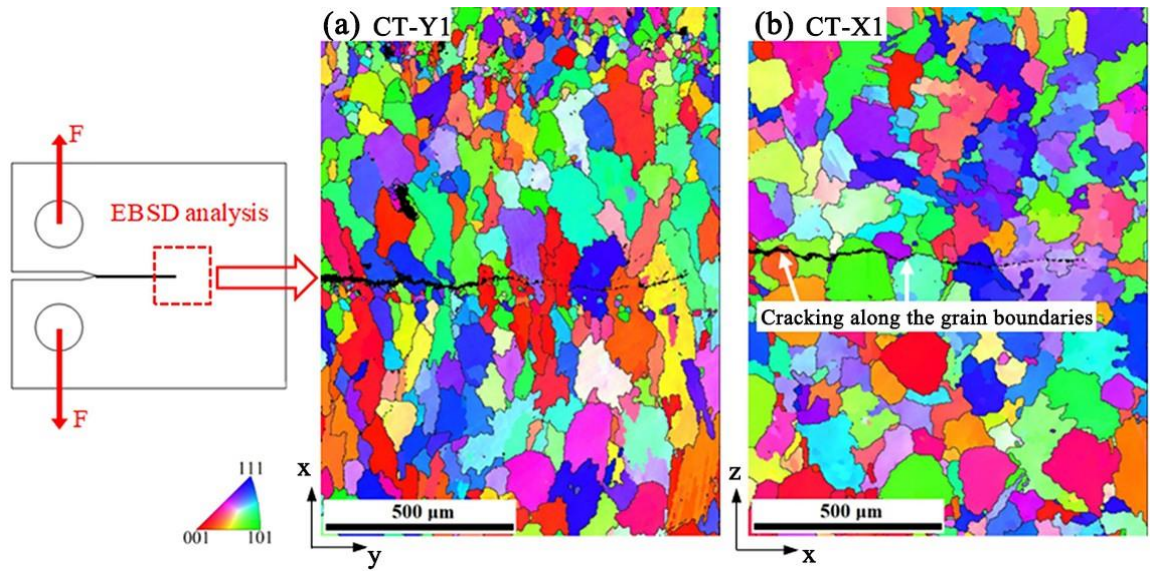

**Fig. 89.** Inverse pole figures (IPFs) of WAAMed Al-Mg4.5Mn alloys with (a) Y and (b) X notch under $R$=0.1. Reprinted from [392]. Copyright (2021), with permission from Elsevier.

Xie et al. [391] used constant-amplitude with a stress ratio ($R$) of 0.1 and frequency ($f$) of 100 Hz to investigate the fatigue performance of Al-Mg4.5Mn alloys. The $S$-$N$ curve (Fig. 87) exhibited the strong fatigue scatter. The maximum stress of the fatigue limit was approximately 108 MPa at a 50% survival rate, corresponding to the stress amplitude of 48.6 MPa. For a conservative and reliable evaluation of the fatigue performance, the maximum stress of the fatigue limit was approximately 84 and 131 MPa at 95% and 5% survival rates, respectively. The results showed that the fatigue failure of all Al-Mg4.5Mn specimens originated from LOF defects or from surface or subsurface pores. First, LOF defects, especially larger and more irregular defects, would arouse greater stress concentrations and accelerate the fatigue failure. This highlighted both the size and morphology of defects were the significant factors of fatigue life. Besides, there was no obvious evidence that the surface or subsurface pores accelerate the crack growth rate, though the crack was prone to deflect when they encountered these defects.

The properties of fatigue crack growth were dependent on different notch and stress load ratio, which were systematically revealed by Liao and coworkers [392]. Under the same fixed notch direction (Fig. 88(a)), a larger stress ratio led to a longer crack growth life. The Paris coefficient ($C$) of the alloy decreased with increasing stress ratio, but the Paris exponent ($m$) increased. The separate FCG rates caused by stress ratio were attributed to the constant maximum stress intensity factor ($K_{max}$) and the increasing minimum stress intensity factor ($K_{max}$). Thus, the stress intensity factor range ($\Delta K = K_{max} - K_{min}$) decreased. When the crack length was the same, the value of initial $\Delta K$ under $R = 0.6$ was much less than that under $R = 0.1$ and 0.3, causing the damage to the reducing crack tip and the increasing FCG resistance. Therefore, a larger stress ratio resulted in a lower FCG rate and an increased crack growth life. Meanwhile, with the increase of $R$, the FCG rate was less affected by the crack closure effect, which leads to the separation of FCG rates with the increase of $\Delta K$.

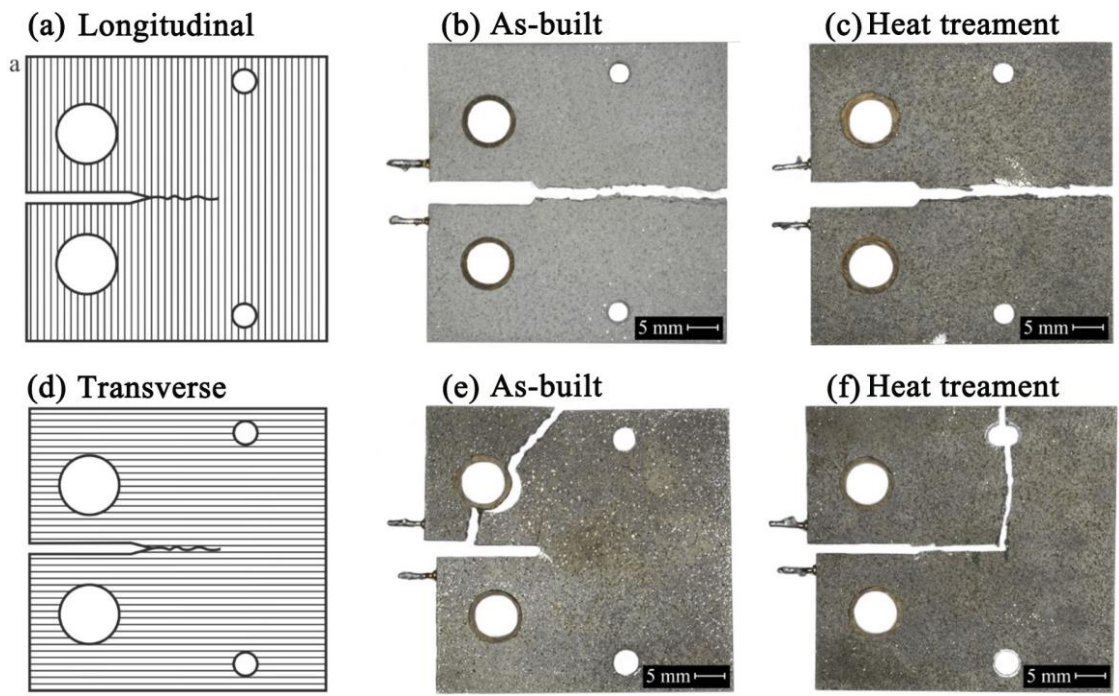

**Fig. 90.** Crack path of SLMed Al-Zn-Mg-Cu alloy with different loading direction, (a)–(c) parallel and (d)–(f) perpendicular direction. Reprinted from [394], Copyright (2016), with permission from Elsevier.

On the other hand, under the same stress load ratio, the anisotropy of fatigue performance was obvious. The crack growth life of the CT-Y series specimens was longer than that of the CT-X series specimens in the Paris region. The Paris coefficient ($C$) and Paris exponent ($m$) in CT-Y specimens were larger than in CT-X specimens as well. In addition, the fatigue crack growth routes were different in two notch direction (Fig. 89) that in CT-Y1 specimen the crack propagation route only perpendicularly crossed the columnar grains while in the CT-X1 specimen the main route crossed the equiaxed grains. This was ascribed to grain boundary misorientation that more low-angle grain boundaries in CT-Y1 than in CT-X1, which was beneficial for improving ductility [392]. Therefore, the CT-Y1 sample underwent a large amount of plastic deformation, and the crack tip was more likely to close, which led to a decrease in the driving force for crack propagation and thus retarded crack propagation.

## 5.5. High strength aluminum alloys: 2xxx, 6xxx and 7xxx alloys

High strength Al alloys (e.g. 2xxx, 6xxx and 7xxx alloys) are well known since the potential utilization in aircraft structural components and automobile industries.

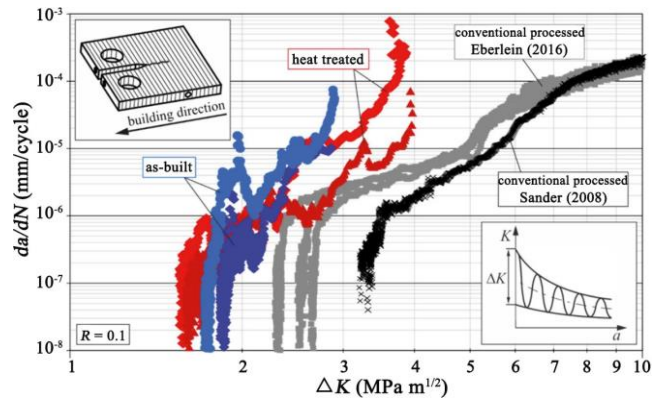

**Fig. 91.** Fatigue crack growth of SLMed Al-Zn-Mg-Cu alloy with (red) or without (blue) heat treatment, and the conventional processed materials as comparison. Reprinted from [394], Copyright (2016), with permission from Elsevier.

### 5.5.1. Al-Zn-Mg-Cu (AA7xxx) alloys

Wrought Al-Zn-Mg-Cu (AA7xxx) alloys were reported with great quasi-static mechanical properties that the tensile strength and elongation could be 462–538 MPa and 3–9% [395]. However, the non-fusion weldability and poor corrosion resistance result in the difficulty to additively manufacture using fusion-based techniques. In recent years, researches focused on AM (e.g. SLM and WAAM) Al-Zn-Mg-Cu alloy have been continuously pushed forward [396–399]. It was found Si elements promoted the intergranular microsegregation so that the reduction of the Si content in the AA7075 powder would benefit to decreasing the cracking susceptibility [396]. However, Li et al. [397] revealed Si and Zr could efficiently modify the microstructure of SLMed Al-Zn-Mg-Cu alloys and prevent the hot cracking, where Si-rich eutectics backfill the cracks and the $Al_3Zr$ significantly decreases the grain sizes. The Si- and Zr-modify Al-Zn-Mg-Cu possessed the tensile strength of 446 MPa and elongation of 6.5%, which were better than those without Si and Zr and comparable with cast alloy. Meanwhile, functionalization with the nucleates, i.e. the reduction of lattice mismatch, was in advantages to achieve crack-free, equiaxed and fine-grained microstructures [11]. Recently, the superior mechanical properties of SLMed Al-Zn-Mg-Cu alloys was proposed by an architectured microstructure consisting of a multimodal grain structure and a hierarchical phase morphology, where the yield strength and elongation were even up to 647 MPa and 11.6% [400]. In addition, fatigue failure and fracture strength were other performance targets of alloy materials for practical applications.

In 2016, Reschetnik et al. [394] reported the tensile strength and fatigue crack growth behaviour of SLMed ion of fatigue properties of SLMed Al-Zn-Mg-Cu alloy. First, they found heat treatment had no significant effect on the quasi-static mechanical properties but building direction indicated noticeable anisotropic fatigue behaviour, i.e. the tensile strength, yield strength and elongation in specimens with vertical direction were higher. Then, the crack path of specimens with different building direction was shown in Fig. 90. The area of failure appeared in the clamping or force transmission point (Fig. 90(e) and (f)), indicating no possibility of crack initiation by cyclic loading in transverse direction. Furthermore, the fatigue crack growth properties had no obvious dependence of heat treatment as well. The threshold value $\Delta K_{Th}$ for the as-built condition (blue curves in Fig. 91) was 1.77 MPa$\sqrt{m}$ ± 0.08 MPa$\sqrt{m}$. For the heat-treated condition (red curves in Fig. 91) the threshold value $\Delta K_{Th}$ was 1.58 MPa$\sqrt{m}$ ± 0.03 MPa$\sqrt{m}$. Compared to those conventional processed AA7075 alloys, SLMed Al-Zn-Mg-Cu alloy showed fracture mechanical properties and higher

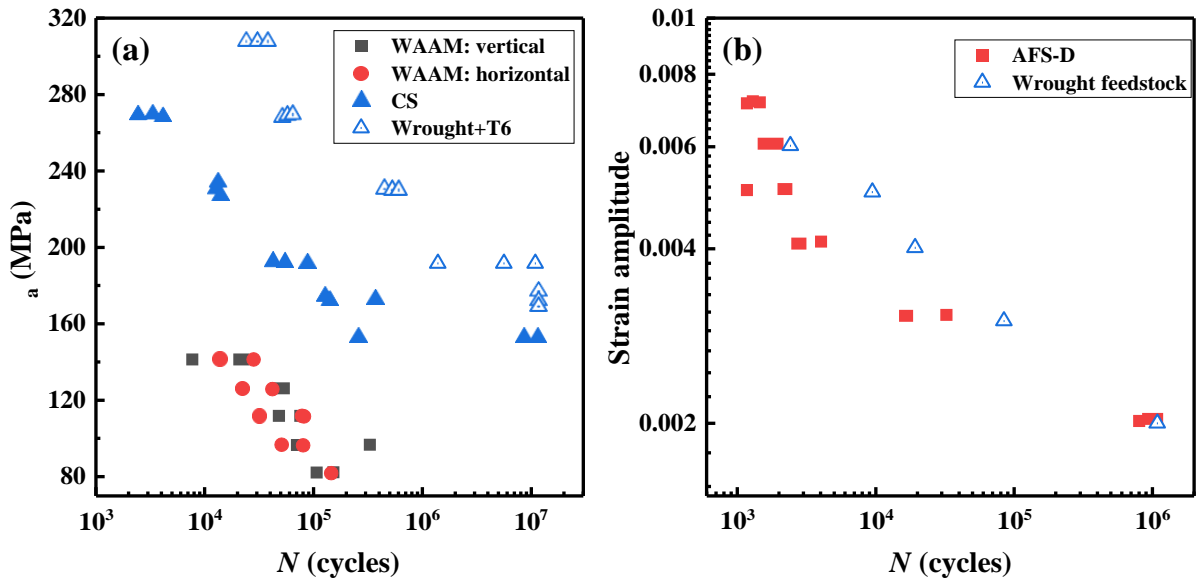

**Fig. 92.** Fatigue cycles of Al-Zn-Mg-Cu alloys with different AM technologies, e.g. (a) WAAM [401], CS [402], and (b) AFS-D [403].

threshold value.

The further investigation of fatigue properties (e.g. fatigue cycles) of SLMed Al-Zn-Mg-Cu alloys has been obstructed because of the technology gap. On the other hand, several researchers found the substituted manufacturing technologies, e.g. additive cold-spray, friction stir welding (FSW), WAAM and additive friction stir-deposition (AFS-D) [401–404].

WAAM is one AM method that benefit for materials which easily engender solidification cracks [399]. The heat sources is electric arcs and the feedstock is wires or powders. As-built Al-Zn-Mg-Cu alloy fabricated by WAAM showed high quasi-static mechanical strength and anisotropy due to different building directions. For example, tensile strength, yield strength and elongation in vertical direction were 197.4 ± 8.5 MPa, 292.2 ± 4.4 MPa and 3.2 ± 0.7 MPa, while those in horizontal direction were 207.8 ± 0.2 MPa, 278.0 ± 2.8 MPa and 2.2 ± 0.1 MPa. The properties were further intensified by heat treatment. After two stage of heat treatment (stage 1: annealing at 120°C for 24 h, stage 2: annealing at 160°C for 24 h after stage 1), the quasi-static mechanical properties were obviously strengthened (e.g. YS ≥ 339 MPa, UTS ≥ 430 MPa) [405]. Morais et al. [401] investigated the fatigue performance of WAAMed Al-Zn-Mg-Cu alloy under heat treatment measures (T73). T73 is a heat treatment based on T6 treatment that adds another stage aging process, which further improved the quasi-static mechanical properties. The fatigue test results exhibited a comparable scatter for both deposition directions (vertical and horizontal). No obvious tendency could distinguished the optimize direction with better fatigue cycles. The scatter was concluded to the material structural heterogeneities and residual porosity.

Cold spray (CS) is one solid-state material deposition technology that at relatively low temperature high pressure gas is utilized to deposit powders to the substrate accompanied by plastic deformation. It is an ideal and promising process for additive repair, especially for significant damage to the substrate, and retard subsequent crack growth. However, AM Al alloys fabricated by CS were rare. White and coworkers performed the first study to investigate the fatigue behavior of CSed Al-Zn-Mg-Cu alloy [402]. The ultra tensile strength in CSed samples (385 MPa) was about 10% lower than in wrought samples. The elongation of CSed samples was 0.64–1.2%, indicating the low ductility as well. The fatigue cycles of CSed samples (Fig. 92(b)) were quite lower than those of wrought samples at the same stress amplitudes, which manifested CS technology led to the bad fatigue toughness of Al-Zn-Mg-Cu alloy. On the other hand, CSed samples exhibit the comparable scatter of wrought AA7075. In the high stress amplitudes (225–275 MPa), the fatigue scatter was weak, while the strong scatter appeared with the stress amplitudes below 200 MPa. It was proposed that CS deposition of Al-Zn-Mg-Cu alloy formed a relatively uniform distribution of layer defects and further resulted in a consistent strength. When normalized by the UTS of each sample, the stress-life of CSed samples overlaid nearly perfectly on the wrought *S-N* results, although the endurance limit exhibited a higher runout stress at $10^7$ cycles. The promising results of the normalized AA7075 CS data suggested that the fatigue performance of CSed materials could exhibit a linear correlation to the ultra tensile stress if the layer defects could be further minimized.

Additive friction stir-deposition (AFS-D) is another solid-state additive manufacturing technology that produces fully dense depositions which can be used for repair, coatings, and the fabrication of structural components [406]. Avery et al. [403] fabricated the AFS-D Al-Zn-Mg-Cu alloy by choosing solid AA7075-T651 rods as the deposited feedback. The ultra tensile strength, yield strength of AFS-D Al-Zn-Mg-Cu alloy were about 1–2 times lower than heat-treated wrought feedback materials, while these of naturally aged materials were even lower. The comparison of fatigue cycles between heat-treated feedback and as-built AFS-D samples was shown in Fig. 92 (b). In LCF regime (around $10^3$ cycles), the fatigue responses of the two were coincident. However, the overall fatigue resistance of AFS-D sample was lower compared to feedback since the discontinuous precipitate growth in microstructure of as-built AFS-D samples weakened the mechanical strength.

Heat treatment, normally, improves the fatigue cycles in SLMed Al alloys though the quasi-static mechanical properties decrease. The researches correlated to post-processing treatment effect on fatigue performance based on AFS-D were lacking. Furthermore, the viewpoint that, other factors (e.g. grain refinement, constituent particle refinement, and uniform dispersion of constituent particles) should increase the fatigue crack resistance of the material, remains to be confirmed. Overall, it is impossible to identify the fatigue performance from the limited testing data, which remains to be penetrated whether fatigue properties of AM AA7075 is better than traditional cast and wrought alloy or not.

### 5.5.2. Al-Cu-Mg alloys (AA2xxx series)

High strength Al-Cu-Mg alloys, classified as 2xxx series, have been widely applied in automotive, aerospace and defense industries. It has been reported that the ultra tensile strength, yield strength and elongation of as-built SLMed Al-Cu-Mg alloys were 402 MPa, 276 MPa and 6%, respectively. The quasi-static mechanical properties could be further increased by solution treatment. When the solution temperature increased from 480°C to 540°C, the UTS, YS and EL increased by approximately 15%, 22% and 47% [407].

The positive effect of the addition of Zr or Sc element in SLMed Al-Mg alloys had been discussed further. Zeng et al. [408, 409] found that Zr element modified SLMed Al-Cu-Mg alloys possess fine microstructure, great mechanical properties (UTS ≈ 451 ± 3.6 MPa, YS ≈ 446 ± 4.3 MPa) and low cracking sensitivity as well. These were attributed to the the formation of heterogenous nucleation (Al3Zr and ZrO particles) during solidification processing promotes the the expansion of processing window as well as the transformation of grain type from columnar to equiaxed. With the Zr contents increasing (0–2.5%), the ultra tensile strength and yield strength had obvious improvement, especially at 2.0%, while the elongation had the opposite trend.

White et al. [402] had investigated the fatigue performance of CSed Al-Cu-Mg alloy (Fig. 93). Compared to the traditional wrought 2024-T3, the fatigue life of CSed samples was about an order of magnitude lower as well as strong fatigue scatter, which was ascribed to the larger layer interface defects. However, The fatigue cycle and fracture properties of SLMed Al-Cu-Mg alloys, there was a serious lack of experimental data, as well as other AM technology.

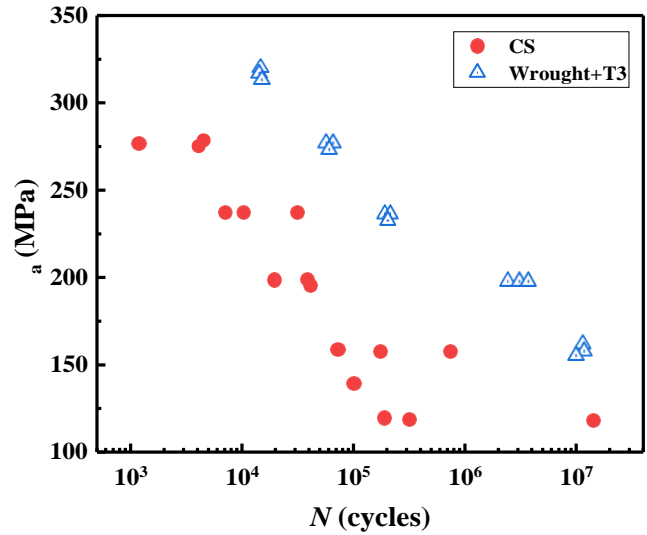

**Fig. 93.** Fatigue cycles of CSed and wrought Al-Cu-Mg alloys. Data from [402].

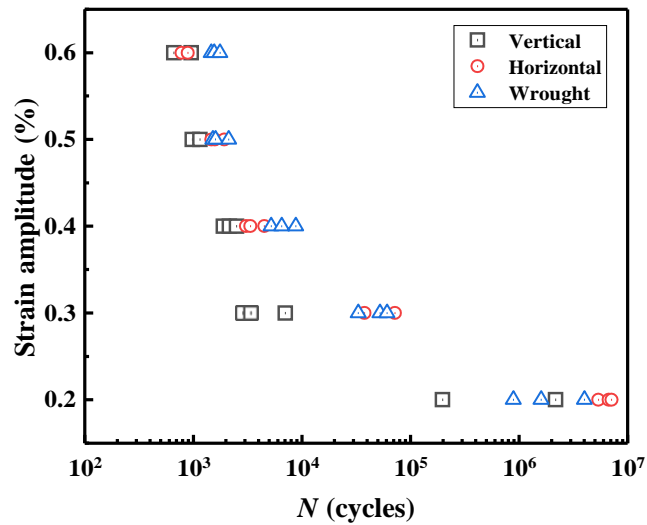

**Fig. 94.** Fatigue cycles of AFS-D Al-Mg alloys with different building directions. Data from [410].

### 5.5.3. Al-Mg (AA6xxx) alloys

Al-Mg alloys (AA6xxx) are heat-treatable alloys with high strength corrosion resistance and ease of conventional fabrication. AM technology would solve the problems of manufacturing complexity and optimize the internal structures. Al-Mg alloy, however, has been a difficult alloy to process using L-PBF because of high crack sensitivity of the material during solidification [411]. Uddin et al. [412] applied powder-bed-preheating measure for crack-free fabrication of Al-Mg alloy. They found platform heating improved the quasi-static mechanical properties of L-PBF Al-Mg alloy (66–75 MPa of YS, 133–144 MPa of UTS) compared to those of wrought AA6061 (55 MPa of YS, 124 MPa of UTS), but weaken the elongation. Furthermore, platform heating combined with heat treatment could further increase about 3–4 times of YS/UTS.

Dynamically mechanical performance (fatigue failure and fracture) of AFS-D Al-Mg alloy had been revealed under the fatigue testing of R = -1 and $f$ = 1 Hz [410], shown in Fig. 94. The effect of building orientation (vertical or horizontal) on fatigue performance of Al-Mg alloy were investigated as well. At LCF regime (< $10^3$ cycles), there was no anisotropy of fatigue in Al-Mg alloy, where the strain amplitudes were 0.6%. The LCF performance of AFS-D Al-Mg alloy was slightly lower than wrought samples. Then, the fatigue anisotropy showed that horizontal samples showed higher fatigue cycles than vertical samples, and even comparable to wrought samples. In addition, horizontal samples experienced run-outs (defined as $5 \times 10^6$ cycles in this study) in all samples tested at 0.2% strain amplitude.

## 6. Stainless steel

### 6.1. Introduction

The typical AM fabricated steels include austenitic stainless steel (SS), precipitation hardening (PH) SS, martensitic SS, maraging steels and tool steels, et al. Over the past years, several excellent reviews on AM metals have been published, in which they were mentioned with a focus on stainless steels [413–417]. The austenitic stainless steels can be simplified as Fe-Cr-Ni based alloys. Martensite is transformed from austenite during rapid cooling, e.g. quenching. For welded components, martensite usually forms directly from the solidified austenite due to the high cooling rate during solidification [418]. In addition, for martensite SS with high carbon content (>0.1 wt-%), coarse primary eutectic carbides and fine secondary carbides usually form during solidification.

Amongst the steels, common austenitic stainless steels (SS, all the designations are as per AISI) 316L, 304L, precipitation hardenable 17-4PH steel, and maraging steel 18Ni300 received considerable attention. The direct AM techniques (i.e., LB-PBF, EB-PBF, and LB-DED) is used to fabricated alloys with solidification cellular structure [419–421]. Most steels processed by AM have successfully been fabricated via LB-PBF, EB-PBF, LB-DED, and BJP with superior mechanical properties. In addition, tool steel has been produced by LB-PBF [422–424] and LB-DED [425, 426]. The steels produced by AM technique with unique characteristics such as a high-density of dislocation, cellular structure, fine grains, adjustable texture and phase constituents [427–431]. Also, AM enables the near-net-shape forming of steels components with complex geometries. Thus, AM is a pathway for many industrial applications such as ultra-high-strength components, complex structured moulds [424, 432, 433], turbine parts, machining tools [434–436], porous parts with complex geometric features [437, 438], and functionally graded components. Recently, a large amount of research on steel AM particularly focuses on those with high and/or special performance, such as high strength and toughness, high hardness/wear resistance, high corrosion resistance, and high weldability.

Here, several steels are summarised with tensile properties under different manufactured processes and heat treatment conditions in table. 3. Obviously, AM technologies with the PBF method is most commonly used [420, 439, 440]. The LB-PBF process may largely promote the Yield strength of steels with 415-462 MPa. The ultimate tensile strength and elongation at failure can up to 450-705 MPa and 35%-64%. L-DED 316L exhibits a mechanic properties with the yield strength of 352-580 MPa, the ultimate strength of 414-941 MPa. Meanwhile, some technologies with method were used for AM steels with 193 MPa yield strength, 548 MPa ultimate strength and 76% elongation at failure. From table. 3, the mechanics of AM austenitic steels is better than the conventional manufactured one. Thus, 316L is the most commonly studied austenitic stainless steel due to its comprehensive mechanical properties, corrosion resistance and high AM processability.

### 6.2. austenitic stainless steels

#### 6.2.1. manufacturing parameters

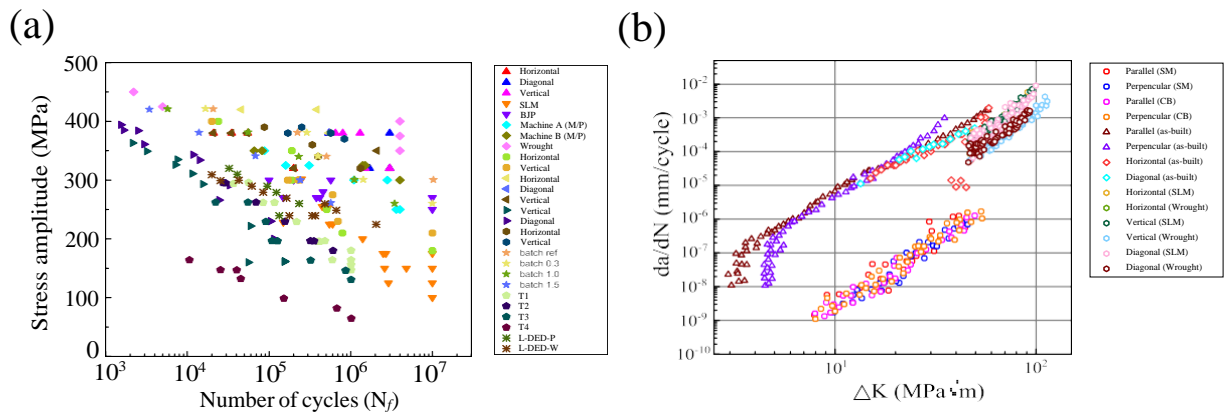

**Fig. 95.** (a) $\sigma_a - N_f$ diagram and (b) fatigue crack growth data of AM-316L generated by different manufactured parameters. Fatigue data was conducted by Stern [441], [442], Kumar [443], Shrestha [444], [445], Yu [446], Blinn [447], [448], Zhang [449], kotzem [450]; FCG data was conducted by Suryawanshi [420], Riemer [451], Fergani [452], Kluczynski [453]

For all the AM techonologies, LB-PBF (for example, selective laser melting) revealing a wide palette of material applicable of metals [454, 455]. The SLM-316L exhibit a great fatigue behavior with different manufacturing parameters. Thus, Fig. 95(a) shows the fatigue results via $\sigma_a - N_f$ diagram with different manufacturing parameters. As depicted, the diagram exhibit a very anisotropic behavior for SLM-316L. The fatigue data in Fig. 95(a) illustrated that horizontal specimens generally exhibited the highest fatigue strength followed by vertical specimens, while diagonal specimens had the lowest fatigue resistance, regardless of the surface condition. Furthermore, the fatigue data exhibits a large scatter in specimens for the tests performed at lower strain amplitudes (i.e. long-life regime) as compared to the tests at higher strain amplitudes (i.e. short-life regime). This can be explained by the fact that fatigue behavior is highly dependent on the size, shape, orientation, and location of the defects, which mostly influences the crack initiation stage [456, 457]. For stern et al. [441, 442], vertical specimens show a drastically shorter fatigue life, and the average fatigue life of the vertical specimens at 380 MPa (ca.36000) is less than 7% compared with the horizontal specimens (ca.547000) and only 4% of the fatigue life of diagonal specimens, respectively. Meanwhile, the stress amplitude of horizontal and vertical samples were conducted by Yu et al [446], the S-N data exhibit that the fatigue strength coefficient of horizontal sample was higher than that of vertical sample. Shrestha et al. [445] investigated the effects of layer orientation on fatigue behavior. To investigate the dependentment of orientation, a cylindrical specimen fabricated by Shrestha [445] with the vertical, diagonal and horizontal one is shown in Fig. 96(a). Figure 96(b) shows the optical micrographs for microstructural features of three orientations. The increasing cooling rate increase the directional grain growth, and results in the more epitaxial grains and lager aspect ratio, which may lead to anisotropic mechanical and fatigue properties. Yu et al. [446] find that a relatively flat LOF perpendicular to the loading direction, and a narrow LOF parallel to the loading direction, were found to be the defects responsible for crack initiation in horizontal and vertical samples, respectively. The specimens fabricated in the vertical direction may have more detrimental effect as compared to specimens fabricated in the horizontal direction. Plastic strain range for vertical specimens as compared to the ones for horizontal specimens in most cases is shorter. The fatigue data tested by Wood et al. [458] reported that both horizontal direction with as-built and machined condition display a 37% and 136% higher fatigue limit compared to vertical specimens, respectively. It is also worth noting that, going from the large-stress and low-cycles region towards the low-stress and high-cycles region, both horizontal and vertical test pieces exhibit a comparatively sharp decline of fatigue resistance. This is largely attributed to the large surface roughness of these samples, which results in the opening of surface defects and immature fatigue fracture. Thus, as seen in Fig. 95, horizontal specimens exhibited the highest fatigue strength followed by vertical specimens, while diagonal specimens had the lowest fatigue. For these fatigue data in Fig. 95, the horizontal AM sample always have a great fatigue behavior with highest fatigue strength and fatigue resistance, regardless of the surface condition.

Another greatly influences of manufacturing parameters is the laser power. Zhang et al.[459] fabricated five sets of specimens, denoted as $0.5P_0$, $0.7P_0$, $1.0P_0$, $1.3P_0$ and $1.5P_0$ with no post-processing heat treatment, which is shown in Fig. 97(a). The highest volume energy density was found at $1.3P_0$. Reduction in the laser powder led to a gradual drop in part density from $1.3P_0$ also resulted in a small decrease in the density for $1.5P_0$. In Fig. 97(b), the fatigue data shows a lower fatigue life for $0.5P_0$ specimens. The pore sizes of $0.5P_0$ 316L are far beyond the range of the conventional alloy steels, which results in a poor fatigue strength. The SEM fracture images of $0.7P_0$, $P_0$ and $1.3P_0$ are shown in Fig. 97(c)-(e). The fatigue life of $0.7P_0$ is longer despite the low-energy intergranular fracture mode. This is because LB-PBF parts contain a high density of differently oriented grains [460]. They are formed due to local grain nucleation and competitive grain growth at the solid-liquid interface of the melt pool, where grains with less favourable orientation relative to the temperature gradient stop growing upon encountering the melt pool boundary [461]. The grain boundaries act as obstacles to dislocation movement and the crack length is longer in the case of intergranular fracture. This explains the richer morphology of the fracture surface of $0.7P_0$ shown in Fig. 97(d). Moreover, as similar factors, i.e. small dendrite size and intergranular fracture, could have been responsible for the higher ductility and longer fatigue life of $0.7P_0$, the direct relationship between ductility and fatigue life could be explained.

Additionally, the research of Nezhadfar et al.[462] in Fig. 95(a) shows some improvement in fatigue performance of LB-PBF 316L SS across life regimes by preheating the build platform up to 150°C(P150), which is most likely due to the lower volumetric defects, specifically LOF ones, in the P150 specimens as compared with the non-preheated(NP) counterparts. Moreover, it has been shown that the larger the defects are, the shorter the fatigue lives become[463]. Accordingly, the longer fatigue lives for P150 specimens in the HCF regime are attributed to the fewer and smaller volumetric defects as compared with NP specimens. The fatigue tests performed by Kotzem et al.[450] showed that a reduction in fatigue strength can be detected with increasing defect size. Here, different batches infer to different side length of the Boolean operation $a$ with 0.3mm, 0.5mm, 1.0mm. For batch 0.3, where an intended defect volume of 0.3 0.3 0.3 $mm^3$ was introduced into the specimens' center, it could be shown that the investigated alloy has a high defect tolerance against internal defects as no internal cracking occurred in the fatigue tests and much smaller surface defects were crack initiating or microstructural cracking with no visible defect occurred. When comparing all batches, the results show that not only the defect size, but also the defect position obviously must have a significant influence on the resulting fatigue strength. Specifically, Kumar [443] fabricated 316L austenitic stainless steel using binder jet printing (BJP) and selective laser melting (SLM) with investigation and compared with those of the conventionally manufactured alloy, with particular emphasis on the unnotched fatigue re-sistance. Fig. 98 shows the optical micrographs and fracture surface of 316L microstructures using SLM and BJP . Results show that work hardening behavior, ductility, and fatigue strength ($\sigma_f$) of the BJP specimens, which contain significant amounts of pores, are surprisingly comparable to those of the wrought alloy. In contrast, the SLM specimens are considerably stronger, especially in terms of the yield strength, less ductile, and far inferior in terms of $\sigma_f$ although the porosity in them is relatively smaller as com- pared to the BJP specimens. The fatigue lifetimes conducted by Blinn et al. [448] in Fig. 95(a) showed that wire-based L-DED (L-DED-W) exhibits longer fatigue lifetimes at higher $\sigma_a$, which is less pronounced with decreasing $\sigma_a$. At stress amplitudes lower than 270 MPa, L-DED-W exhibits generally shorter fatigue lifetimes compared to Powder-based (L-DED-P). meanwhile, Fig. 99 shows the optical micrographs of microstructure of 316L based on L-DED-P and L-DED-W. Zhang et al. [449] tested four sets of samples were prepared by varying the powder layer thickness at 20 $\mu m$ (t1), 40 $\mu m$ (t2), 60 $\mu m$ (t3), and 80 $\mu m$ (t4). They found that for as-built L-PBF stainless steel 316L, monotonic and fatigue properties are most strongly affected by porosity. A linear relationship was found between the fatigue endurance limit and ductility as both properties are sensitive to porosity.

In addition, crack growth is another index of fatigue properties. The results for FCG of AM-316L is shown in Fig. 95(b). Obviously, the threshold value increases from 3.0±0.2 MPa m$^{1/2}$ for horizontal to about 4.3±0.2 MPa m$^{1/2}$ which is an improvement of about 40% without post-processing. The different FCG rates, especially in the near threshold regime and at higher $\Delta K$ values, are observed. Kluczynski [453] investigate that after selective laser melting, the smallest cracking velocity value was registered in vertically oriented reference specimens. Meanwhile Suryawanshi [420] found that the fatigue crack growth characteristics of the SLM alloys are similar, overall, to those of wrought alloys. A slightly lower $\Delta K$ and higher values of m in the SLM alloys are due to reduced plasticity-induced crack closure and lowered ductility respectively. Riemer et al. [451] fabricate the AM-316L which is depending on the relation of crack growth direction and grain long axis shown in Fig. 101(a). The magnification factography is shown in Fig. 101(b) and (c). The vertically oriented sample were characterized by a low fatigue life of the tested parts, and diagonal samples were characterized by the highest value of cracking velocity and a highest fatigue life. However, in the experiments of Fergani et al. [452], there is no distinct observable difference between horizontal and diagonal samples in their as-built condition. For high values of $\Delta K$, the slope of the diagonal samples is slightly higher,giving slightly higher crack growth rates for high values of $\Delta K$.

## 6.3. post-processing

Firstly, a $\sigma_a$ – $N_f$ diagram for post-processing of 316L was shown in Fig. 100(a). Elangeswaran et al. [480] investigate the fatigue results in semi-log plot for as-built and post-processing specimens. Meanwhile, Fig. 102 exhibits the optical microscopic and LOF defects with different post-processing method. The as-built(AB) and stress-relieved(SR) in Fig. 100(a) exhibit similar fatigue behavior, which show superior results compared to the reference. However, the significant performance decrease is observed for FA and HIP conditions. This may because the coarsened microstructure and absence of sub-grained cellular boundaries after post-treatment. However, the fatigue data tested by Leuders et al.[486] and Spierings et al.[487] illustrated that heat treatments appear to have little influence on fatigue behavior. Polishetty et al. [482] showed that the heat-treated specimens have higher number of cycles to failure than the as-built samples at same stress levels. The fatigue data with Horizontal-M-SR(machined and stress relieved) and Horizontal-M-NSR(machined and non-stress relieved) samples by wood [458] show the highest fatigue strength among all the studied samples. Stress-relief does not have a considerable effect on the fatigue behavior of Horizontal-

**Table 3.** Tensile properties of several steels generated by different manufactured processes and heat treatment conditions.

| Alloy | Process | Source | Condition | YS (MPa) | UTS (MPa) | EL (%) |
|---|---|---|---|---|---|---|
| 316L | Wrought | ASTM A276 | AN | 170 | 485 | 40 |
| | LB-PBF | [464] | AB | 430 (90°) | 550 | 64 |
| | | [451] | AB | 462 | 565 | 53.7 |
| | | [465] | AB | 590 17 | 705 15 | 44 7 |
| | | | AN(1095 for 1 h in vacuum, FC in argon) | 375 11 | 635 17 | 51 3 |
| | | [443] | AG (550 for 1 h) | 511 14 | 621 11 | 20 2 |
| | | | | 430 11 | 510 20 | 12 1 |
| | | [466] | AB | 325 | 450 | 8.6 7.3 |
| | | | | 415 | 565 | 35 3.1 |
| | | | HIP (1125 for 4 h at 137 MPa, FC) | 225 | 515 | 28.6 17 |
| | | | | 225 | 570 | 46.7 4.5 |
| | EB-PBF | [427] | AB | 334 16 | 572 19 | 29.3 5.2 |
| | | | | 396 9 | 652 8.5 | 30.6 3.0 |
| | | [421] | AB | 253 3 | 509 5 | 59 3 |
| | LB-DED | [467] | AB | 530 5 | 670 6 | 34 1 |
| | | | | 469 3 | 628 7 | 31 2 |
| | | [468] | AB | 533 23 | 235 6 | 48 2 |
| | | | AN | 549 18 | 255 25 | 41 0 |
| | | | AN | 474 4 | 215 12 | 57 8 |
| | | | AN | 494 6 | 204 10 | 70 5 |
| | | [469] | AB | 410 5 | 640 20 | 34 4 |
| | | | AN | 340 15 | 610 5 | 42.5 0.5 |
| | BJP | [443] | AG | 193 2.8 | 548 3.8 | 67.6 2.8 |
| | | | | 197 3.2 | 553 5.7 | 76 2.8 |
| 304L | Wrought | ASTM A276 | | 170 | 485 | 40 |
| | LB-PBF | [470] | AB | 540 15 | 660 20 | 36 12 |
| | | [471] | AB | 568 2 | 716 1 | 41.7 1.1 |
| 17-4 PH | Wrought | ASTM A276 | ST+AG | 1170 | 1310 | 10 |
| | LB-PBF | [472] | AB | 1190 | 1370 | 8.3 |
| | | [473] | AB | 661 24 | 1260 3 | 16.2 2.5 |
| | | | AG (480 for 1 h, AC) | 945 12 | 1420 6 | 15.5 1.3 |
| | | | AG (620 for 1 h, AC) | 1010 15 | 1320 2 | 11.1 0.4 |
| | | | ST (1040 for 0.5 h, AC) | 939 9 | 1190 6 | 9 1.5 |
| | | | ST (1040 for 0.5 h, AC)+AG(480 for 1 h, AC) | 1350 18 | 1440 2 | 4.6 0.4 |
| | | [474] | AG (600 for 2 h) | 600 | 1300 | 28 |
| 15-5 PH | LB-PBF | | AN | 1297 1.0 | 1450 2.1 | 12.5 1.1 |
| 18Ni300 | Wrought | [414] | SA | 828 68 | 1000 130 | 28 |
| | | | SA+AG | 1930 140 | 1960 150 | 8 3 |
| | LB-PBF | [475] | AB | 713 | 1060 | 11.3 |
| | | | | 851 | 1020 | 9.63 |
| | | | ST(815 for 1 h, AC)+AG(520 for 5 h, AC) | 920 | 1530 | 10.6 |
| | | | | 886 | 1550 | 10.7 |
| | | [476] | AB | 915 7 | 1170 7 | 12.4 0.14 |
| | | | AG (490 for 6 h, AC) | 1970 11 | 2020 9 | 3.28 0.05 |
| | | | ST (840 for 1 h, AC) | 962 6 | 1030 5 | 14.4 0.35 |
| | | [477] | ST (840 for 1 h, AC)+AG(490 for 6 h, AC) | 1880 14 | 1940 8 | 5.60 0.08 |
| | | | AB | 915 13 | 1190 10 | 6.14 1.3 |
| | | | ST (815 for 0.5 h, AC)+AG(460 for 8 h, AC) | 1960 43 | 2020 58 | 1.51 0.2 |
| | | | ST (815 for 0.5 h, AC)+AG(540 for 8 h, AC) | 1550 100 | 1660 88 | 1.77 0.1 |
| | | [478] | AB | 1060 20 | 1160 42 | 12 0.1 |
| | | | | 975 75 | 1060 50 | 10.2 1.9 |
| | | | ST (840 for 1 h, AC) | 808 7.5 | 970 20 | 13 0.5 |
| | | | | 805 5 | 975 25 | 11.9 0.1 |
| | | | ST (840 for 1 h, AC)+AG(490 for 6 h, AC) | 1750 35 | 1820 20 | 4.5 0.1 |
| | | | | 1750 | 1850 | 5.1 |
| | | [479] | AB | 1210 99 | 1290 110 | 13.3 1.9 |
| | | | AG (480 for 5 h, AC) | 1200 32 | 2220 73 | 1.6 0.26 |
| | DED | | AB | - | 959 20 | - |
| | | | ST (830 for 1 h, AC)+AG(490 for 10 h, AC) | - | 1560 17 | 0.12 0.05 |
| H13 tool steel | Wrought | [426] | AB | 1569 | 1999 | 7.5 |
| | LB-PBF | [423] | AB | 1236 178 | 1712 103 | 4.1 1.2 |
| | LB-PBF | [424] | AB | 1003 8.5 | 1370 175.1 | 1.7 0.6 |

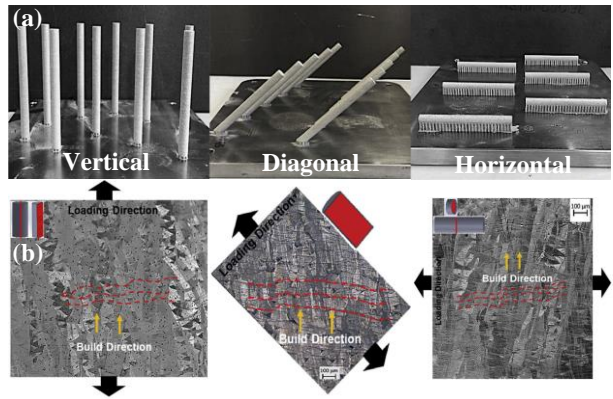

**Fig. 96.** (a)The cylindrical AM-316L parts fabricated in vertical, diagonal and horizontal directions. (b)Optical micrographs showing microstructural features on the plane cut perpendicular to the build plate for vertical, diagonal and horizontal AM-316L specimens. Reprinted from [445], Copyright (2019), with permission from Elsevier.

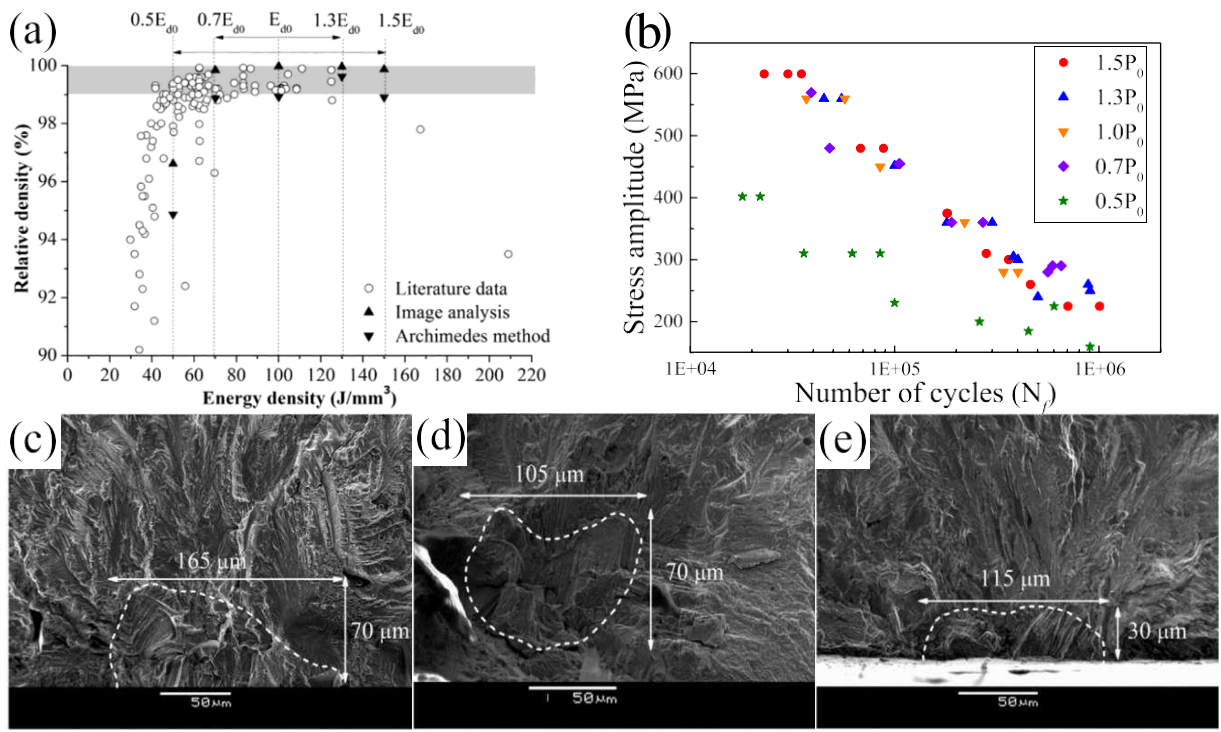

**Fig. 97.** (A) Relative density against volumetric energy density results from the literature for the processing of AM-316L. (B) $\sigma_a$ – $N_f$ diagram of 316L samples made with different laser powers. SEM images of fatigue initiation regions for $0.7P_0$(c), $P_0$(d) and $1.3P_0$(e). Reprinted from [459], Copyright (2017), with permission from Elsevier.

M samples. Compared to Horizontal-AB-NSR(as-built and stress relieved) samples, Horizontal-M-SR and Horizontal-M-NSR samples show a considerably higher fatigue strength trend without a steep decrease.

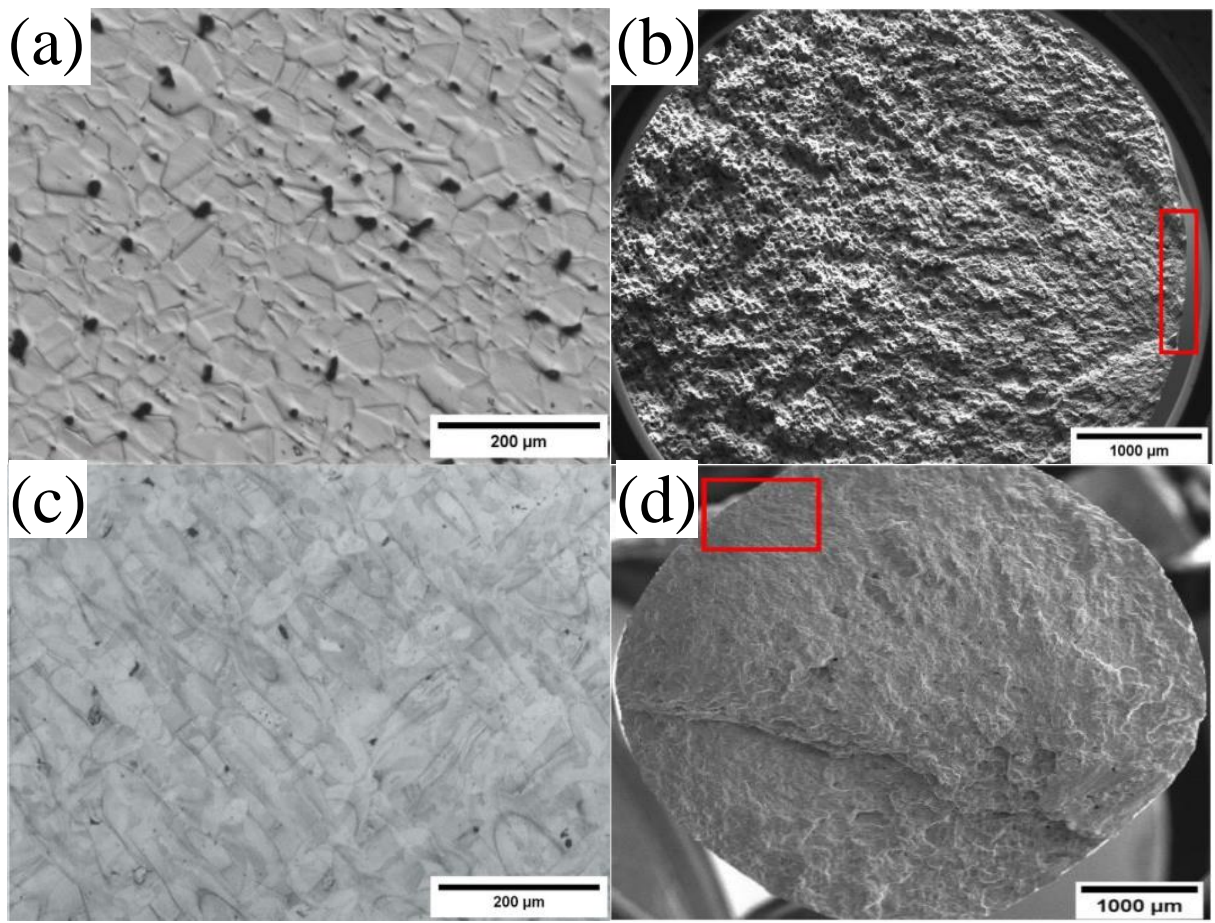

**Fig. 98.** Optical micrographs showing the representative microstructures of manufactured 316L steel using (a) binder jet printing (BJP), and (c) selective laser melting. The fracture surfaces of (b) BJP Specimen fatigued at stress amplitude of 300 MPa and number of cycles to failure, 98368 and (d) SLM specimen fatigued at stress amplitude of 152 MPa and number of cycles to failure, 250000. Reprinted from [443], Copyright (2020), with permission from Elsevier.

This is due to the detrimental prevalence of LOF and surface defects in as-built samples resulting in crack initiation and a lower fatigue strength. Meanwhile, Vertical-M-SR samples show much larger scatter and a considerably lower fatigue strength compared to Vertical-M-NSR samples. Specially, the fatigue data conducted by Shrestha [444] showed that in comparison of the fatigue behavior of R-B and axial specimens at lower applied stress amplitudes (i.e. 250 and 300 MPa), specimens with higher surface roughness (for AB surface condition), or larger volumetric defects (for M/P surface condition) exhibited lower fatigue performance. Kluczyński et al.[488] exhibited that additively manufactured material, without any heat treatment, weakened throughout the entire load cycle range. At the same time—conventionally made material (and also additively manufactured after heat treatment) were characterized by a visible stabilization at some cycles range. Otherwise, laser shock peening(LSP) is another post-processing method which widely used. Here, Fig. 100(a) the fatigue data results for parts in the AB-SLM state, post-processed by HIP, 2D LSP, and 3D LSP is given, respectively [483]. The measured fatigue limit was 250 MPa in the AB state, and 300 MPa when post processed with 2D LSP. It was lower than the fatigue limit in conventional manufactured 316L, which can be attributed to the significantly smoother machined surface state compared to SLM samples. The achieved fatigue limit of 3D LSP was 360 MPa, which represents a modest increase compared to conventional manufacturing. Fig. 100(a) compares fatigue lives of SLM AB, HIP post processed, 2D LSP and 3D LSP processed with conventionally made 316L after machined on the top surface. In this machined state, SLM AB outperforms the conventionally made 316L. The 3D LSP condition is better than all other treatments and achieved a fatigue limit of 480 MPa. In HCF and VHCF regime, specimens tested by Voloskov [489] showed higher endurance at similar stress levels obtained for the AM specimens in the HCF and VHCF regimes. Interestingly, the effects of surface roughness on initiating fatigue cracks were minor for LB-PBF 316L specimens in Shrestha's research [445]. In the study of Blinn [447], no significant influence of heat treatment on the fatigue behavior of vertical building direction can be observed. However, the heat treated horizontally built specimens show significantly higher fatigue life in the whole HCF-regime compared to Horizontal specimen, which is significantly more pronounced at lower stress amplitudes. When comparing different surface integrities (polished, pre-corroded and with a EDM defect), see fatigue data tested by Merot [484], an important decrease of the fatigue resistance of the material, and a fortiori, of its fatigue strength, when surface defects are introduced can be noticed. The scatter significantly increased since various surface states were tested. Meanwhile, Leuders et al. [486] used machined samples, Blinn et al. [490] used machined and polished specimens and Mower and Long [43] studied net-shape samples. Loadings were also alternative and uniaxial, except for Mower and Long [43] who used rotating bending loadings. In comparison of the specimens by Leuders [486], the polished batch by Merot [484] is in good agreement with the literature data. Mower and Long [43] data shows (as-expected) that as-built specimen surface roughness highly decrease the fatigue behaviour of the material, particularly in the high cycle fatigue (HCF) regime. Blinn et al. [490] data also shows a decreased resistance. This may be due to a higher porosity rate in their tested material (almost 1%), and probably bigger defects presence, compared to tested material by Merot [484]. machining has proven to be an easy and reliable post-treatment processes of stainless steels. In Braun's study [485], increases in fatigue strength of PBF-LB/M 316L stainless steel of 87% and 17.5% compared to the as-built condition were determined for machining and heat treatment, respectively. Similarly, in the study of Zhang et al. [449], in comparison with T3, the higher ductility of HT1 and HT2 compensates the lower strength and their fatigue properties at short lives are almost equal. Comparing with T4, whose monotonic strength is about the same as HT1 and HT2, the heated-treated parts sustained much longer fatigue life.

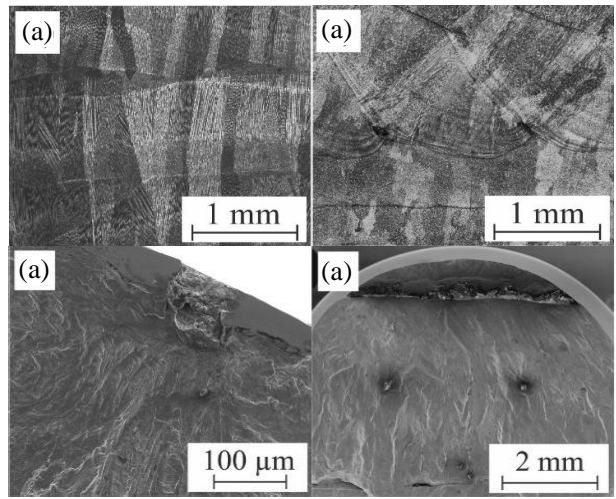

**Fig. 99.** Light optical micrographs of the microstructure of 316L resulting from (a)powder-based and (b)wire-based L-DED. (c) and (d) is the examples of fatigue fracture surfaces obtained by L-DED-P and L-DED-W, respectively. Reprinted from [448], Copyright (2021), with permission from Elsevier.

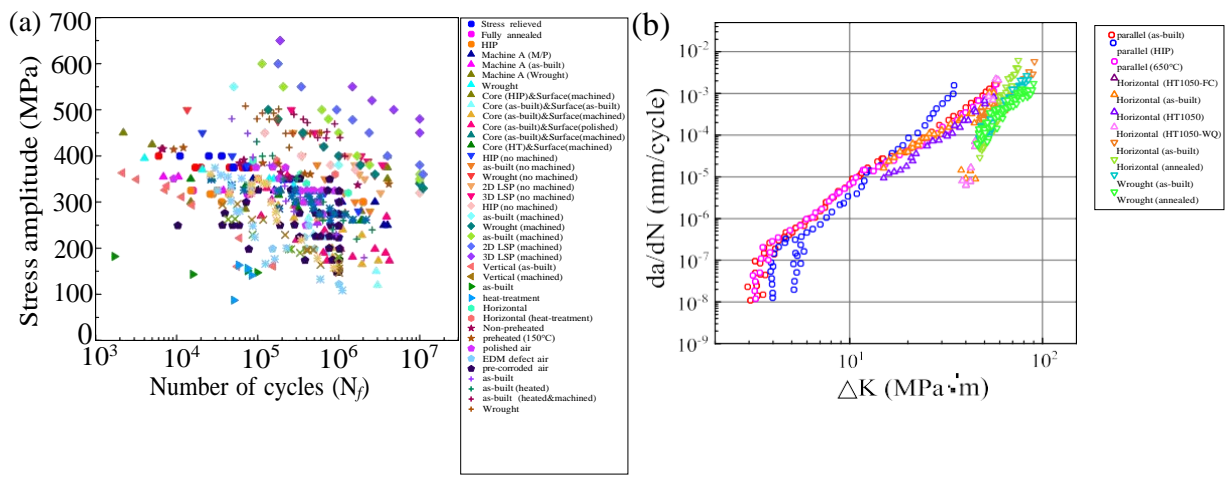

**Fig. 100.** $\sigma_a - N_f$ diagram and (b) fatigue crack growth data of AM-316L generated by post-treatments. Fatigue data was conducted by Elangeswaran[480],[481], Shrestha[445], [444], Polishetty[482], Blinn[447], Nezhadfar[462] Kalentics[483], Merot[484], Braun[485] and Zhang[449]; FCG data was conducted by [451], Fergani[452], Kluczynski[453].

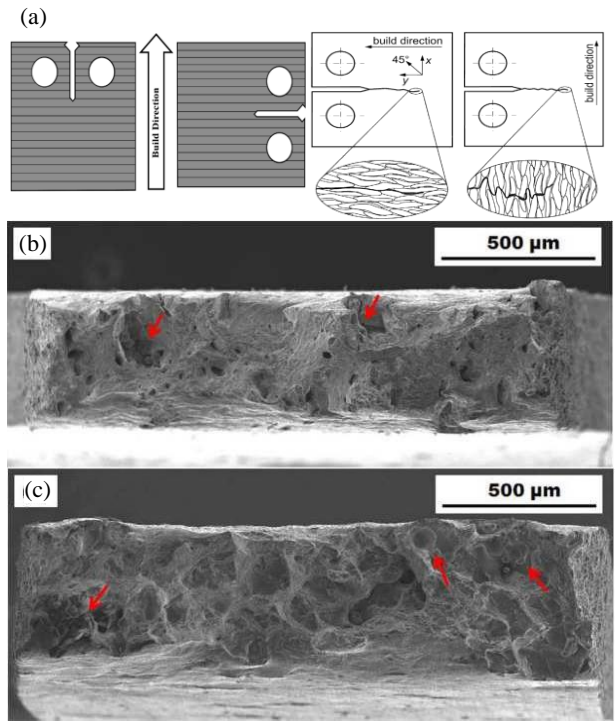

**Fig. 101.** (a) Schematic illustration depicting (preferential) grain orientation and morphology dependent on building direction [451]. And low magnification factography images of the tensile tested samples of AM-316L tested in (b)transverse and (c) longitude loading conditions [420].

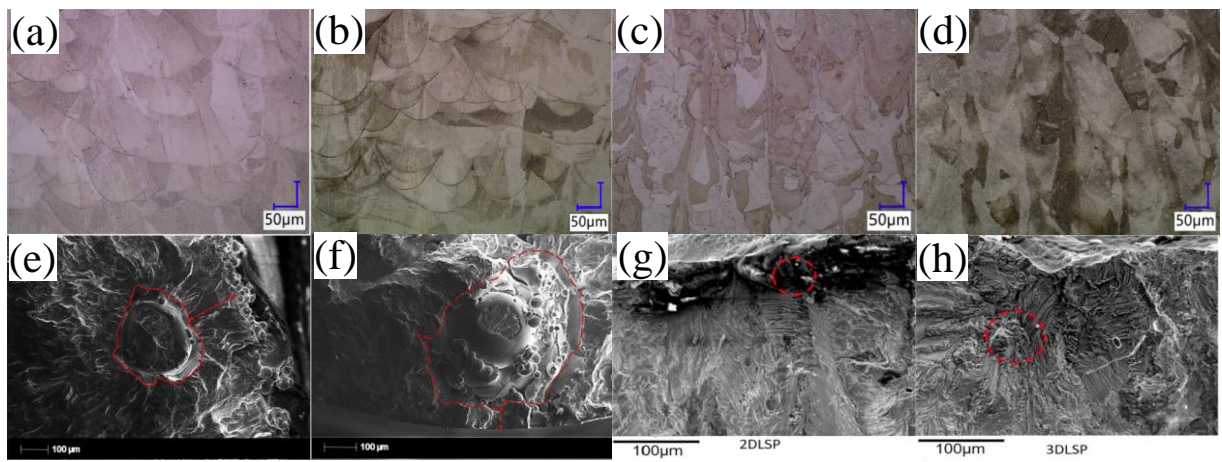

**Fig. 102.** Optical microscopic images showing microstructures in as-built(a), stress relieved(b), fully annealed(c) and HIP(d) conditions.[480] Fracture surfaces showing LOF defects responsible for crack initiation in vertical specimens fabricated in (e)as-built and (f)machined surface conditions subjected to a strain amplitude of 0.003 mm/mm. [445] Cross-sections of fractured surfaces fot the AM-316L with 2D LSP(g) and 3D LSP(h) samples.[483]

The results of 316L for FCG was shown in Fig. 100(b). The data tested by Riemer et al. [451] exhibit that the FCG curve for crack growth parallel to building direction, which is shown in Fig. 101(a). It increases from 3.0 MPa m$^{1/2}$ for as-built condition to 4.7 MPa m$^{1/2}$. Meanwhile, HIP processing of SLM-316L attributes to an isotropic material behavior with respect to crack growth. Kluczynski [453] investigate that solution heat treatment in the case of selective laser melted samples caused the reduction of microcracks and lowered the number of cracks in the different layers. Thus, after solution annealing, the cracking velocity increased slightly, which could be connected with porosity growth. It increases the fatigue life of produced parts, but completely changes the cracking growth behavior. The SLM-316L with heat-treated and as-built are compared by Fergani et al. [452] in Fig. 100(b) shows the 650°C stress relieving has no obviously effect on the crack growth behavior. In contrast, The 1050°C furnace cooling improves the crack growth behavior slightly. The samples with this heat treatment have lower crack growth rates than the as-built samples for most parts of the Paris law.

Another widely used austentinic stainless steel is 304L, Here, Zhang et al. [440] showed in Fig. 103 investigate that the fatigue life of horizontal AM-304L is generally higher than that of vertical parts. The fatigue performance of 304L was dominated by building direction, which cause the laser power and microstructure variable, induce the volumetric energy density, residual stress. Meanwhile, Parvez et al. [491] illustrate that the specimens prepared at horizontal direction has a higher fatigue strength than the specimens prepared at diagonal, and vertical direction. This may because the specimens at horizontal direction have a lower probability of defects within its volume since smaller number of layers results in a smaller number of interlayer zones. Interlayer bonding is weak and it has a higher possibility of creating defects during the fabrication. However, Gordon et al. [492] convinced that the as-built vertical specimens displayed a longer fatigue life as compared to the horizontal specimens for the 85% UTS transitional fatigue experiments under directed energy deposition method. Meanwhile, Zhang et al. [440] convinced that SLMed 304 L exhibited superior fatigue performance at a high-stress level. Increasing scanning speed increased the surface roughness and porosity of SLMed 304L, which is shown in Fig. 103. However, applying too low laser power will violate this principle.

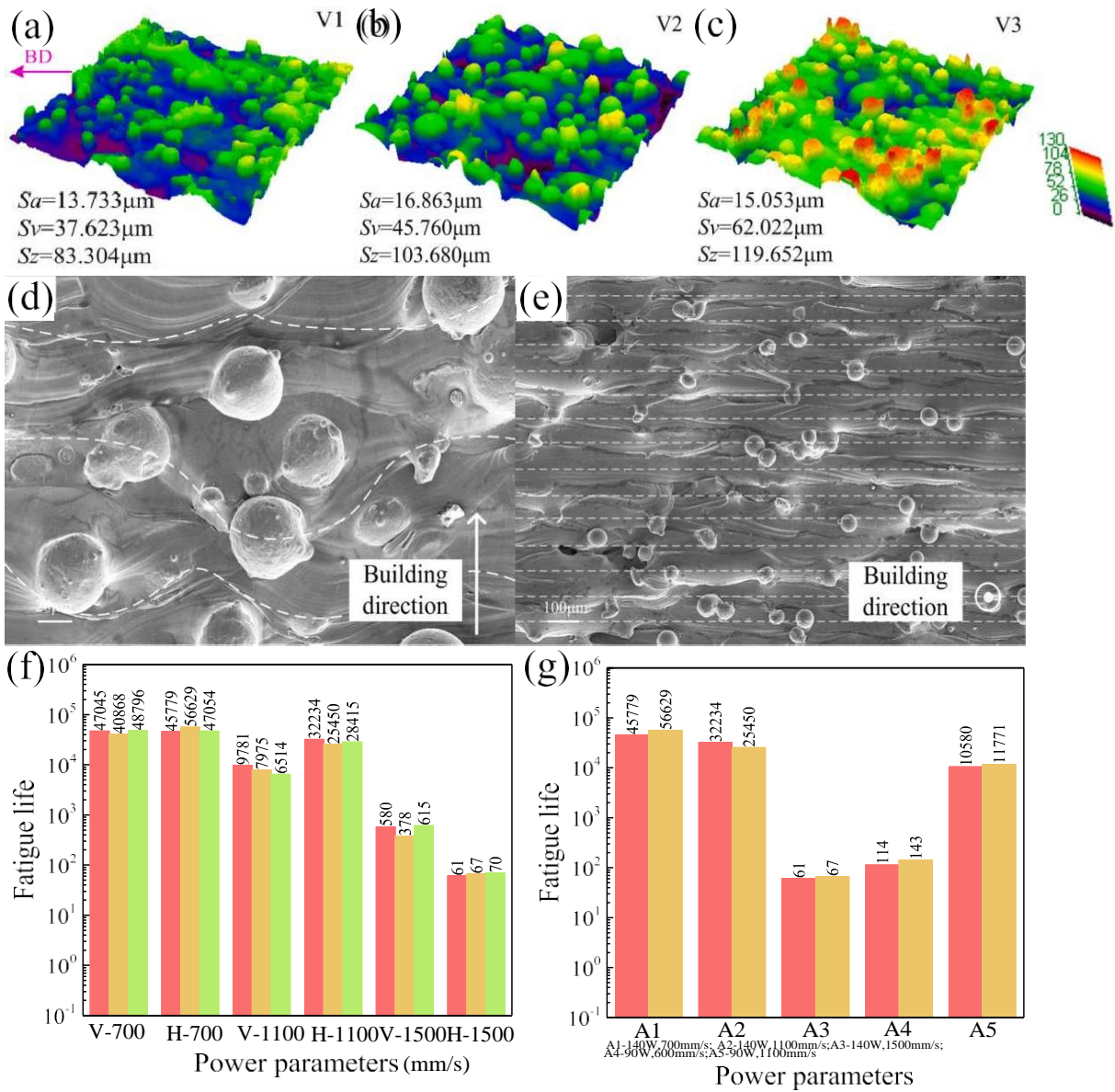

**Fig. 103.** 3D surface topography and surface roughness of as-built surfaces of vertical parts with scanning speeds of 700 mm/s(a), 1000 mm/s(b), and 1500 mm/s(c), respectively. Surface morphology of longitude (d) and transverse(e) direction with layer tracks overlay. Fatigue life with different manufactured scanning speed and power energy [440, 440]

Interestingly, Gordon et al. [493] investigate the fatigue crack growth of 304L by wire and arc additive manufacturing was shown in Fig. 104(b). Horizontally orientated as-printed WAAM specimens of both treatments showed lower intercept and comparable slope values compared to horizontally oriented SLM processed 316L builds in. Also, as in, vertical specimens showed improved FCG compared to vertical specimens in. The differences between horizontal and vertical specimens FCG can be attributed to long columnar grains and strong texture in the building direction.

For the AM austenitic 304L, the S-N curves of as-built specimens and some post-processing specimens were shown in Fig. 105(a). As it can be seen, in the low cycle fatigue(LCF), there is no significant influence by post-processing but great effect under HCF via Lee et al. [494], which means LCF shows little to no effect from surface roughness. For the crack growth of 304L shown in Fig. 105(b), Gordon et al. [493] investigate that the WAAM produced 304L shows that this novel material compares favourably to conventional wrought 304L. The retained compressive residual stresses positively affect for as-built 304L. However, the stress relieving heat treatment change the texture and residual stress, which results in the dissimilarities between WAAM as-built and heat treated FCG.

### 6.4. Precipitation hardening steel

#### 6.4.1. Introduction

Precipitation hardening (PH) stainless steel is a promising aerospace material with high strength steel and good corrosion resistance. Thus, it attracted a lot interest for researchers in fabrication, especially additive manufacturing technology. The most commonly studied PH steels in recent years is 17-4 PH and 15-5 PH steels. Here, Fig. 106 shows the EBSD and phase maps of 17-4PH[456]. Meanwhile, AM-PH steels reveals a great performance in fatigue behavior because their great mechanic properties and microstructure. Thus, this section focuses on the fatigue performance and fatigue crack growth behavior on the most widely studied 17-4 PH and 15-5 PH steel.

#### 6.4.2. manufacturing parameters

Fig. 107(a) shows the fatigue behavior of the 17-4 PH produced by additive manufactured with different processing parameters. Similar to the data of austenitic steels, the fatigue results of samples were fabricated by L-PBF, L-DED et al. The fatigue properties of AM-steel are comparable to their wrought counterparts. According to Fig. 107(a), building orientation obviously influences fatigue behavior. Horizontal specimens show higher fatigue strengths relative to vertical ones both AB and HT conditions and noticeable in both short life and long life regimes-though, it is more pronounced in HCF [456]. Similarly, another fatigue results reported by Yadollahi et al. [495] show that horizontal specimens, regardless of heat treatment, show higher fatigue strengths relative to vertical ones in both short life and long life regimes. Thus, the AM-17-4 PH steel always exhibit a anisotropic fatigue behavior with higher fatigue strength in horizontal axis in spite of heat treatment. Different factors can be responsible for the observations like level of porosity and residual stress, as well as the orientation of deposited layers with respect to applied load direction. The great fatigue resistance during HCF and LCF in horizontal specimens may owing to the higher strength and higher elongation to failure for both AB and HT samples. Meanwhile, the horizontal specimens generally have smaller voids and a higher level of porosity. In HCF range, a smaller number of large voids can be more detrimental on fatigue performance than a larger number of small voids. Thus, the differences of fatigue behavior can be seen obviously in short-life and long-life in HCF regime [495]. Additionally,Soltani-Tehrani et al. [496] investigated the powder characteristics and location dependences of L-PBF 17-4 PH shown in Fig. 107(a) illustrate that no effect of powder re-use was observed on the fatigue life of specimens with as-built surface condition.

The FCG behavior of PH steels were reported these years. In order to consider the effect of material anisotropy on the FCG behavior, crack propagation direction perpendicular to the build direction (trans-verse crack) and crack propagation direction parallel to the build direction (longitudinal crack), as shown in Fig. 107(b). In Yadollahi's research [456], AM 17-4 PH steel with different crack orientations displayed distinct FCG behavior with respect to each other. Comparing the FCG results of transverse as-built conditions and longitudinal as-built conditions specimens indicates that different crack orientations displayed distinct FCG behavior with respect to each other. The FCG rate is slightly lower for longitude specimens as compared to that of transverse counterparts, indicating the crack growth resistance is weaker when crack is perpendicular to the build direction. This can be attributed to greater tendency of crack to propagate along lower strength melt pool boundary layers in transverse specimens[456]. However, in the study of Nezhadfar [497] shown in Fig. 107(b), there is an insignificant variation in FCG rate of LPBF 17-4 PH with transverse and longitudinal orientation with one of the wrought counterpart, all heat treated utilizing CA-H900 procedure. As seen from this figure, the wrought steel exhibited superior crack growth resistance in the near threshold regime. Interestingly, FCG rates of L-PBF 17-4 PH steel and wrought materials in the Paris regime are similar. There is also an insiginificant variation in FCG rate of L-PBF set 1 and set 2 specimens in this regime, which is attributed to the homogenous fine microstructure obtained through CA-H900 heat treatment procedure, as shown in Fig. 107(b).

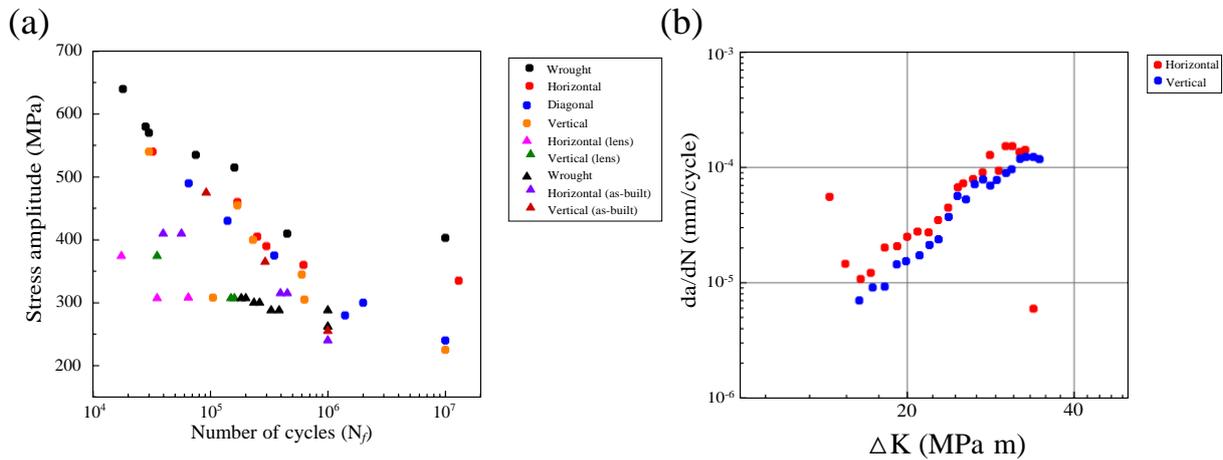

**Fig. 104.** $\sigma_a - N_f$ diagram and (b) fatigue crack growth data of AM-304L generated by different manufactured parameters. Fatigue data was conducted by Parvez [491], Gordon [492]; FCG data was conducted by Gordon [493].

### 6.4.3. post-processing

For the post-processing of AM-17-4 PH steel, a lower fatigue life was displayed as compared to the wrought counterpart reported by Carneiro et al. [498] shown in Fig. 108(a). Nezhadfar et al. [499] reported in Fig. 108(a) show that removing the surface roughness through machining process significantly improves the fatigue resistance of L-PBF 17-4 PH. Meanwhile, heat treatment had some influence on the fatigue strength of L-PBF 17-4 PH. A comparison between H1100 and H900 also presented in that H900-17-4 PH exhibit a lower fatigue performance. Samples subjected to H900 procedure had the lowest fatigue strength. The higher heat treating temperature and longer soaking time may have resulted in partial elimination of micro-segregation of elements and coarsening of precipitates, which improve the fatigue resistance for H1025 specimens. Meanwhile, primary solution treating cycle (CA) enhanced the fatigue resistance of L-PBF 17-4PH specimens. The data reported by Yadollahi [456] showed that the HT samples-regardless of building orientation - demonstrate higher fatigue strengths relative to their AB counterparts in LCF, but the opposite trend occurs for HCF. Expect to the heat treatment, Akita et al.[500] pronounced a method to give a post-treatment at 1050°C followed by water quenching. The SLM and SLM-quenched 17-4 PH exhibited much lower fatigue strengths compares to the CM one, which could be attributed to their lower hardness. Meanwhile, some fatigue life prediction of 17-4 PH was investigated by Yadollahi et al [495]. and Romano et al. [501] .

For the FCG rate of AM 17-4 PH steel, as seen the data in Fig. 108(b) tested by Yadollahi et al [502]. FCG test results are very similar for as-built and heat-treated specimens from near-threshold to fracture. Furthermore, it can be seen from Fig. 108 that the FCG behavior in the Paris regime for AM 17-4 PH specimens with a transverse crack in both as-built and heat-treated conditions are comparable to that of wrought 17-4 PH steel in the H900 condition. This implies that as opposed to fatigue life, which is considerably lower for AM 17-4 PH steel as compared to its wrought counterparts due to presence of process-induced defects, FCG behavior of AM 17-4 PH steel is not significantly influenced by such defects. Meanwhile, the fatigue life is predicted by Yadollahi et al with a crack-growth approach [502]. As mentioned by Nezhadfar et al. [499] after a comparison between H1025 sample and CA-H900 sample in Fig. 108(b), the fracture surface of the CA-H900 specimen seems to be slightly rougher. This may because FCG through the uniformly distributed $\delta$-ferrite strings on the grain boundaries shown in Fig. 108(b). Figure 109 exhibits the optical micrographs of 17-4 PH specimens in transverse direction with different post-treatment parameters. In addition, although the final fracture regions of the CA-H900 specimens appearing slightly smoother than those of H1025 specimens, they do share the similar morphology as H1025 specimens, i.e. a combination of transgranular and intergranular features.

## 6.5. 15-5 PH steel

### 6.5.1. manufacturing parameters

Similar results for another part of precipitation hardening steel, 15-5 PH, is reported by Sarkar et al [503]. Fig. 110(a) shows that fatigue life for horizontally build SLM specimens, as-built SLM as well as heat treated are more than vertical counterpart. However, the S-N data presented by Croccolo et al. [504] exhibited that for horizontally and vertically built specimens are quite close, while the slated ones is much higher, indicating a greater estimated strength for the same fatigue life. This may because scan errors or to powder residuals and of increasing the resistance against crack propagation due to layered structure, which results in the higher fatigue behavior.

### 6.5.2. post-processing

In addition, results in Fig. 110(b) shows the fatigue properties of 15-5 PH steel with post-processing. Sarkar et al. [505] reported the fatigue performance of SLMed 15-5 PH steel in as-built condition as well as undergone different surface treatments. In comparison to surface treated specimens, As-built SLM specimen has relatively a poor fatigue life. In both low cycle fatigue and high cycle regime the data are scattered because of the high surface roughness and sub-surface flaws for as-built specimen. Meanwhile, the electro-polished specimens have almost similar fatigue performance in LCF regime with machined ones although electro-polished samples have lower surface roughness. Meanwhile, machined specimens have high fatigue life in HCF regime. In the mean time, another study of Sarkar et al. [503] for aging and overaging specimens of 15-5 PH steel, which is shown in Fig. 110(b). Low cycle fatigue life is more for H900 as compared to as-built SLM specimens. However, reverse trend is observed in HCF regime where crack initiation stage dominates the total fatigue life. Meanwhile, the overaged(H1150) specimens exhibit more LCF life for H1150 as compared to as-built specimens. The effect of defects in HCF regime is less prominent because of the ductile of H1150 specimens. Relatively more stress concentration can be accommodated around a defect than for H900 ones through localized plastic deformation, which results in a higher fatigue strength.

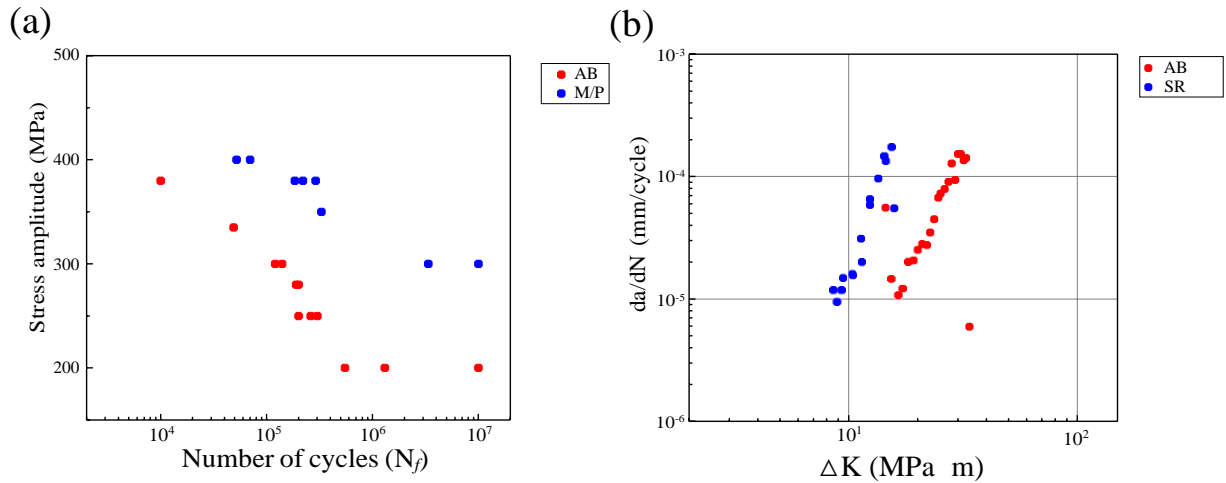

**Fig. 105.** $\sigma_a - N_f$ diagram and (b) fatigue crack growth data of AM-304L generated by post-treatments. Fatigue data was conducted by Lee[494]; FCG data was conducted by Gordon[493].

### 6.6. maraging steel

#### 6.6.1. Introduction

Similar to the austenitic and PH stainless steels, cellular/dendritic structure is common in AM processed maraging steels (e.g. 18Ni300 and 18Ni250). Table. 3 lists the tensile properties of the majority of maraging steels produced by AM. Similar to the data for austenitic and PH steels, the tensile results of L-PBF and L-DED fabricated samples were reported. The tensile properties of as-built maraging steels are comparable to their wrought counterparts. The ultimate tensile strength of most as-built maraging steels are over 1 GPa. Meanwhile, LB-PBF 18Ni300 maraging steel also exhibits a hierarchical microstructure, with solidification cells inside a martensitic microstructure. In the AB condition, the microstructure consists of both martensite and austenite, and no precipitates or small clusters of atoms are observed, indicating a cooling rate high enough to suppress precipitation. This leads to a comparably soft and ductile alloy in the AB state. However, there are indications of early stages of precipitation in the L-DED material, accompanied by an increased hardness. Solutionising and age hardening heat treatments (815–840 °C for 1 h, AC, 490–530 °C for 8 h, AC), which are similar to those used for conventionally produced maraging steels, are necessary for imparting high strength. These heat treatments result in the formation of martensite with intermetallic precipitates. With increasing ageing temperature and time, the volume fraction of retained austenite increases by the process of austenite reversion.

#### 6.6.2. manufactured parameters

The results of AM-18Ni300 steels fatigue tests for all the materials states are shown in Fig. 111. Fatigue strength at HCF regime is inferior to the bulk material for both horizontal and vertical samples reported by

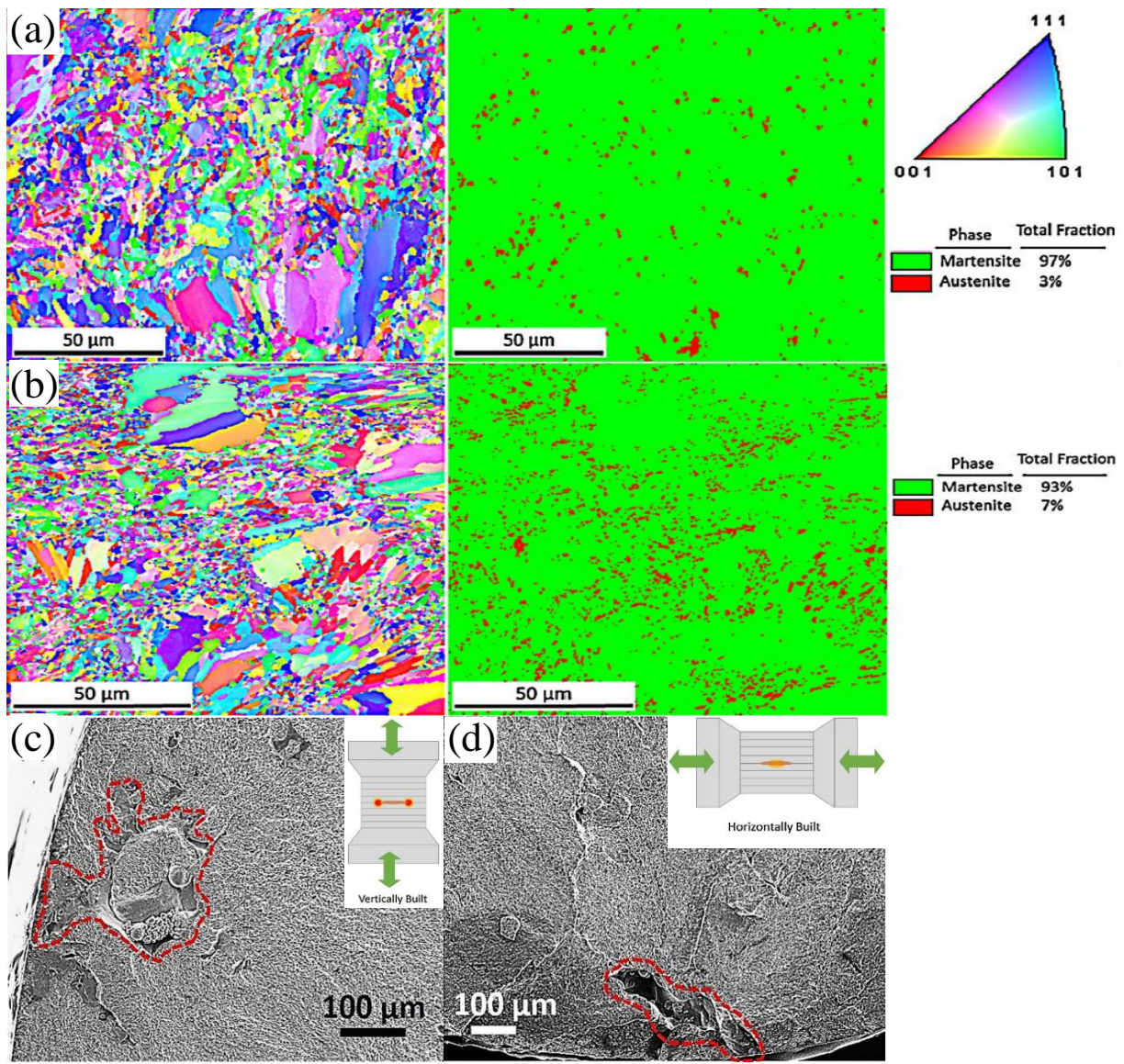

**Fig. 106.** EBSD and phase maps of selected areas of (a)vertically and (b)horizontally AM built 17-4 PH steel samples; Comparison of un-melted regions serving as crack initiation sites in (c)vertical and (d)horizontal specimens. Reprinted from [456], Copyright (2017), with permission from Elsevier.

Damon et al. [506] All SN curves show a transition between $10^5$ and $10^6$ cycles evolving in a plateau without any decrease of strength until the fatigue life of $10^7$ cycles is reached. Additionally, a distinctive anisotropy in fatigue performance of additively manufactured specimens can be seen. While vertical specimens show comparable values with bulk material in the range up to $10^5$ cycles, horizontally build specimens demonstrate significantly lower fatigue strength. Interestingly, a test series referred to horizontal building direction exhibit a lower fatigue strength than those referred to vertical orientation in the fatigue data tested by Meneghetti et al. [507], both for as-built and aged conditions in spite of the large scatter of the AM results. Then, they used a new axial fatigue test with using short crack-corrected stress intensity factors to improve the correlation of experimental data [508]. Moreover, it is worth noting that the age hardening treatment improved the fatigue strength for the series having horizontal building direction. Conversely, for the specimens built at vertical orientation, the aging heat treatment did not lead to significant fatigue strength improvement. The fatigue strength of all additively manufactured test series is lower than the vacuum melted maraging steel 300. Meanwhile, Croccolo et al. [509] found similar fatigue performance for both horizontal and vertical orientations for bending fatigue. Then, Solberg et al. [510] tested the fatigue behavior with build angle from 0° to 135° with a increasing of 15° angle shown in Fig. 113(a). Fig. 113(b) exhibits a clear trend in the fatigue data; the lowest fatigue life is found close to the horizontal orientation, then the fatigue life is increased when approaching the vertical orientation, and decreasing when entering the down-skin region. The highest fatigue life was found for the 60°-specimens. In general, the down-skin specimens displays lower fatigue life than their up-skin counterparts. Although considering the scatter, the difference is striking, e.g. by comparing 45° and 135°, both orientations display a fatigue life in a similar range. The clearest difference is observed between 60° and 120°.

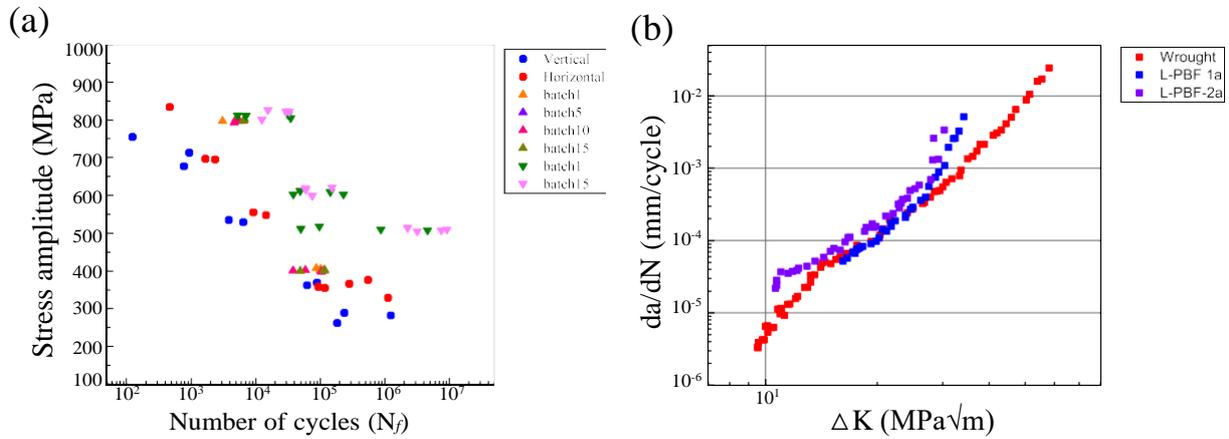

**Fig. 107.** $\sigma_a - N_f$ diagram and (b) fatigue crack growth data of AM-17-4PH generated by different manufactured parameters. Fatigue data was conducted by Soltani [496]; FCG data was conducted by Nezhadfar [497].

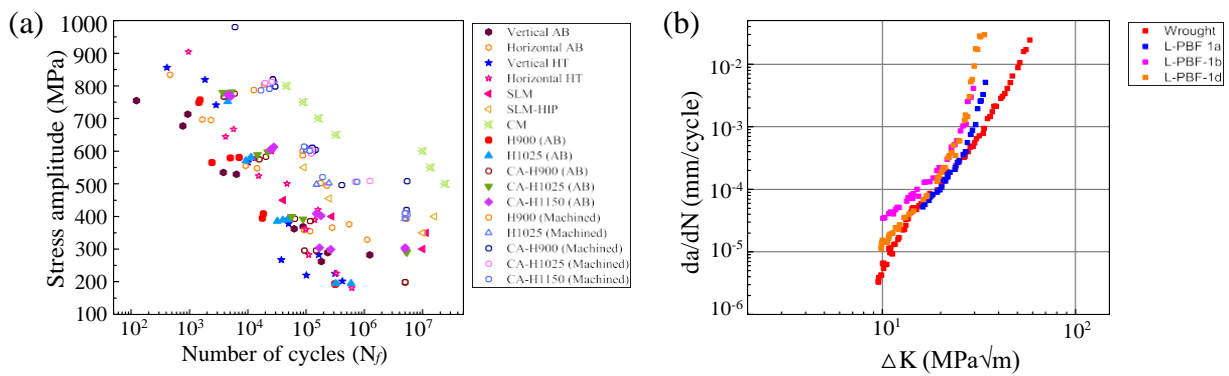

**Fig. 108.** $\sigma_a - N_f$ diagram and (b) fatigue crack growth data of AM-17-4PH generated by post-treatment. Fatigue data was conducted by Nezhadfar [497], Yadollahi [456], akita[500]; FCG data was conducted by Nezhadfar [497].

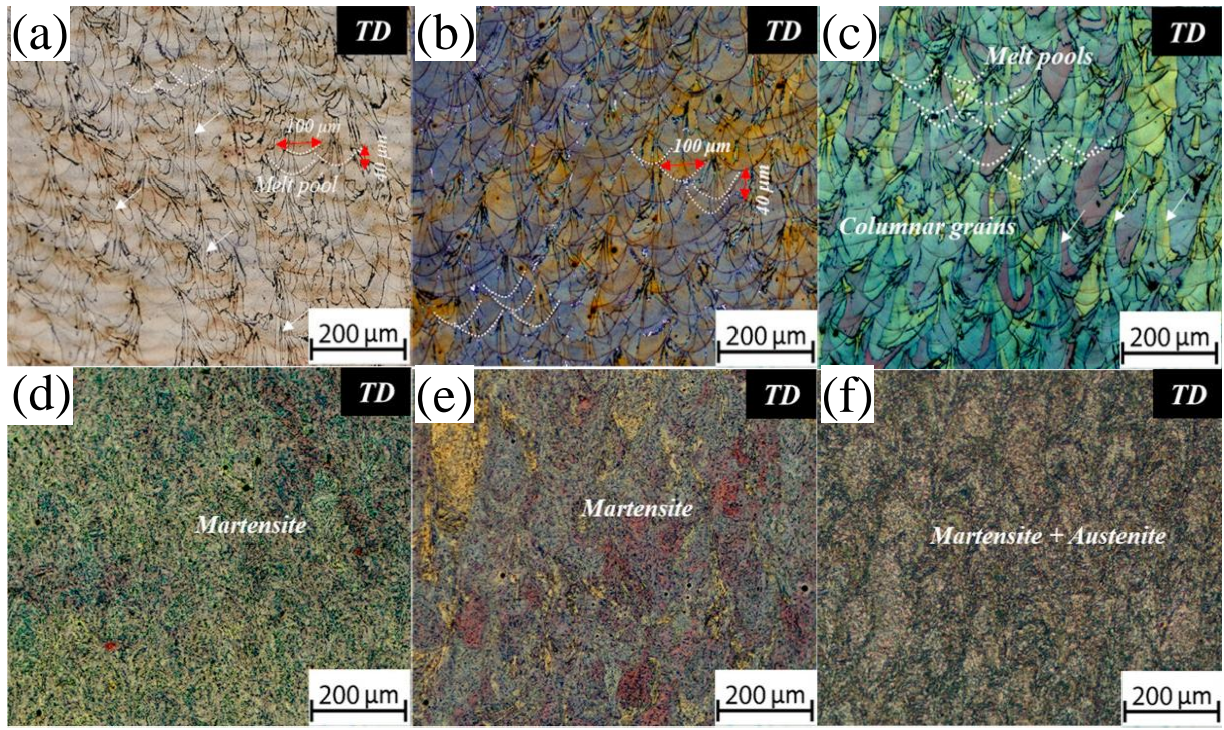

**Fig. 109.** Optical micrographs of AM-17-4PH specimens in transverse direction for (a)non heat-treated, (b)H900, (c)H1025, (d)CA-H900, (e)CA-H1025 and (f)CA-H1150 procedures. Reprinted from [499], Copyright (2019), with permission from Elsevier.

### 6.6.3. post-processing

For the post-processing of maraging steel 18Ni300, the fatigue data of Meneghetti et al. [507], as mentioned above, reported that age hardening treatment improved the fatigue strength for the series having horizontal building direction, which is shown in Fig. 112(a). The fatigue test results of the samples with a variety of heat treatments applied by Tezel et al. [511] showed the decreasing fatigue life of the alloy as stress was increased. The fatigue life obtained fort the H900 samples was longer than the fatigue life of the as-built fabricated samples. Compared to the samples that were solution-heat-treated at 840°C for 1h, the as-built samples had longer fatigue life at high stress. In contrast, solution-heat-treated samples had longer fatigue strength due to decreased stress. The previously mentioned solution-treated samples and subsequently, the solution 1 age treated samples had longer fatigue strength than the as-fabricated samples. These precipitate particles were also semi-coherent with martensite, which resulted in minimal deformation while increasing the amount of dislocation. Thus, strain strengthening occurred, which was the main reasons for the increased fatigue strength of the alloy due to the heat treatments. In the investigation of Elangeswaran et al. [480], nonsurface-treated with as-built condition (AB NT) samples exhibit the lowest fatigue behaviour, followed by sand-blasted (AB SB) and vibratory-finished (AB VF) ones in increasing order. After heat treatment, nonsurface-treated (HT NT) samples show a similar fatigue behaviour like AB NT, but with a steeper slope and reduced performance at low stresses. The two surface-treated samples show marked increase in fatigue performance. A different trend is observed after heat treatment, that is, sand-blasted samples perform better than vibratory-finished ones. Reference test results for conventionally manufactured (vacuum melted and aged) and polished maraging steel reported by Van Swam et al. [512] are also plotted in Fig. 112(a). HT VF sample state is considered as comparable with the reference condition because of similar age-hardened microstructure and fine surface finish. It is seen that the fatigue performance of HT VF samples is lower than the reference. However, after sand blasting, the fatigue response of HT SB specimens approaches that of conventionally manufactured samples. Thus, with no surface treatment, both as-built and heat-treated samples exhibit similar fatigue behaviour. Influence of micronotches from rough surface dominates the fatigue life leading to early failures, and fatigue performance is significantly enhanced by surface treatments, namely, vibratory finishing and sand blasting. Meanwhile, After heat treatment, sand blasting introduced higher magnitudes of beneficial compressive stresses in the samples, enhancing the fatigue life compared with vibratory-finished specimens. Specially, a maraging steel with few paper reported named maraging steel CX was reported by Ćirić-Kostić et al. [513] The outcomes indicate that both heat treatment and machining have a significantly beneficial effect, even if performed singularly. The effect of the machining is particularly relevant even without heat treatment. It indicates this material is highly sensitive to surface irregularities triggering cracks: running machining leads to a four-time incremented fatigue strength. When heat treatment and machining are applied together (in this order and with subsequent shot-peening), they have a synergic beneficial effect, which leads to a further enhancement of the fatigue strength (up to five times).

Figure .112(b) summarized by Santos et al. [514] showed the constant amplitude tests, comparing da/dN-$\Delta K$ curves in the Paris regime and also in the near threshold regime, for the as built SLM samples and the post-built heat-treated SLM specimens. For each batch two tests were performed, one with crescent $\Delta K$ (constant load) and another with $\Delta K$ successively decreasing, both starting from $\Delta K=8$ MPa$\sqrt{m}$. This methodology justifies some irregularity found in the crack propagation rate around $\Delta K=8$ MPa$\sqrt{m}$, which is not a feature of the material, but is due to the stabilization of the crack propagation rate. The results presented in the figure show that the post-manufacturing heat treatment has a significant influence on the fatigue crack propagation rate, decreasing da/dN not only in the stable Paris Law regime, but also near threshold. The main causes for this behaviour are the microstructure changes and the hardness increase, as stated before. In addition, the residual stresses for SLM heat-treated specimens are much lower than for as-built specimens, reducing the effective loading stress ratio. The microstructure modification also causes both different crack path and failure mechanisms for the two types of samples.

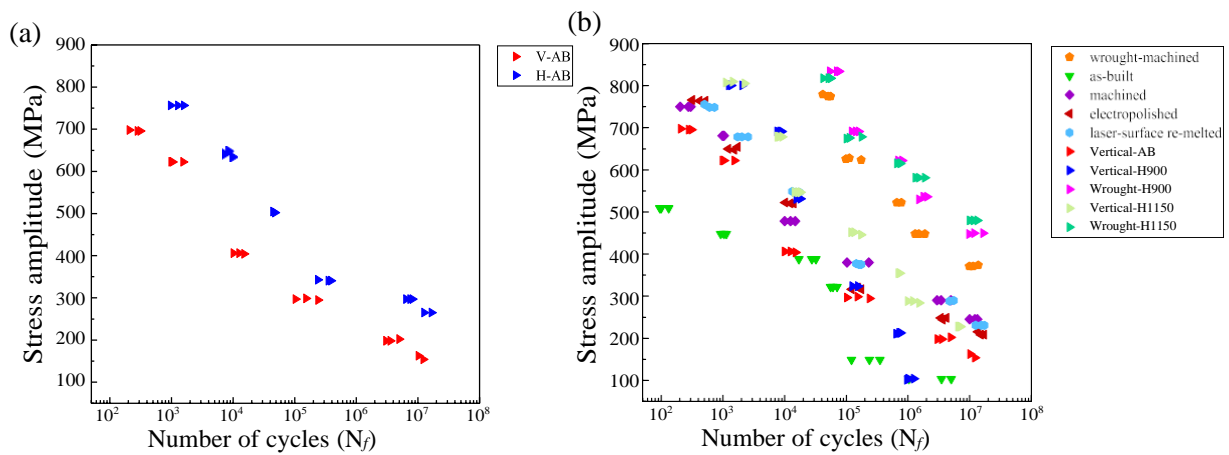

**Fig. 110.** $\sigma_a - N_f$ diagram of AM-15-5 PH generated by different manufactured parameters and post-treatment conducted by Sarkar [503], [505].

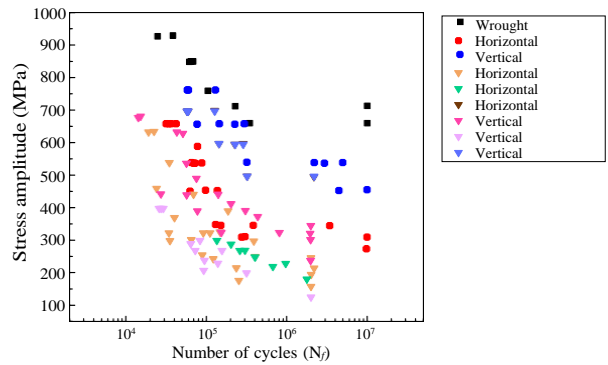

**Fig. 111.** $\sigma_a - N_f$ diagram of AM-18Ni300 generated by different manufactured parameters. Fatigue data was conducted by Damon [506], Meneghetti [507, 508].

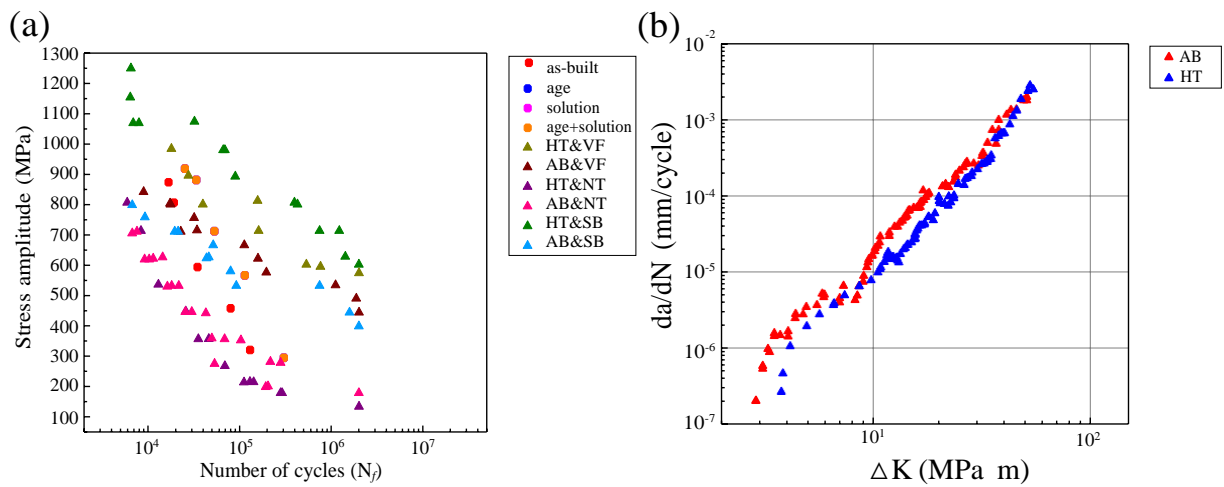

**Fig. 112.** $\sigma_a - N_f$ diagram and (b) fatigue crack growth data of AM-18Ni300 generated by post-treatment. Fatigue data was conducted by Tezel [511], Elangeswaran [480]; FCG data was conducted by Santos [514].

### 6.7. H13 steel

#### 6.7.1. manufacturing parameters

The tested fatigue data by Pellizzari et al. [516] reported that The fatigue strength of the horizontal is lower than that of the diagonal and especially vertical inclined variants. This behavior can be attributed to different factors, such as the difference in residual stresses and the orientation of defects and deposited layers with respect to the applied load. As depicted in Fig. 114, the horizontal oriented samples are characterized by LOF defects with a split shape perpendicular to the loading axis, which produces a stress concentration factor that is several times higher compared to the defects in the vertical oriented samples. The variation in defect orientation according to the building direction produces anisotropy in the mechanical properties of SLM parts. The building direction leads to anisotropy in fatigue performance, which is mainly attributed to the orientation of layers with respect to the loading axis. The defects in the horizontal oriented samples are perpendicular to the loading direction and characterized by large critical size and sharp tips. Hence, the results of the fully reversed axial fatigue tests (R = -1) show a slightly lower fatigue resistance for these samples. Fatigue resistance is influenced by the defect size and location. In all samples, LOF is the dominant defect that triggers the main crack leading to eventual failure. The critical defect in the horizontal samples is a defect located immediately below the outer surface, viz. its distance from the surface is comparable with its characteristic size. The dominant defect in the diagonal and vertical is a superficial flaw in most cases. Meanwhile, The AM 1.2344 tool steel reported by Dörfert et al. [515] with PBF-surface suffer from the notch effects of the surface itself as well as tensile residual stresses, which result in the very low fatigue strength. In the investigation of Tshabalala et al. [517] The different build direction did not show substantial difference on fatigue properties of the material. The fatigue samples experience strain hardening and broke without significant plastic deformation as experienced during static tensile tests as shown in Fig. 114. All XY specimens broke closest to the top of the gauge while the Z specimens fractured within the gauge length. Tool steels tend to develop compressive mean stress under axial cyclic conditions. The stress limit in compression is found to change slightly while the tensile stress limit increases resulting in a net hardening effect Furthermore, residual stresses occurred during the manufacturing and post-processing have an influence on the resulting fatigue strength. Due to the limited amount of the produced samples, the fatigue limits are in certain scattering ranges. However, the fatigue influencing factors and failure mechanisms can be acquired consistently.

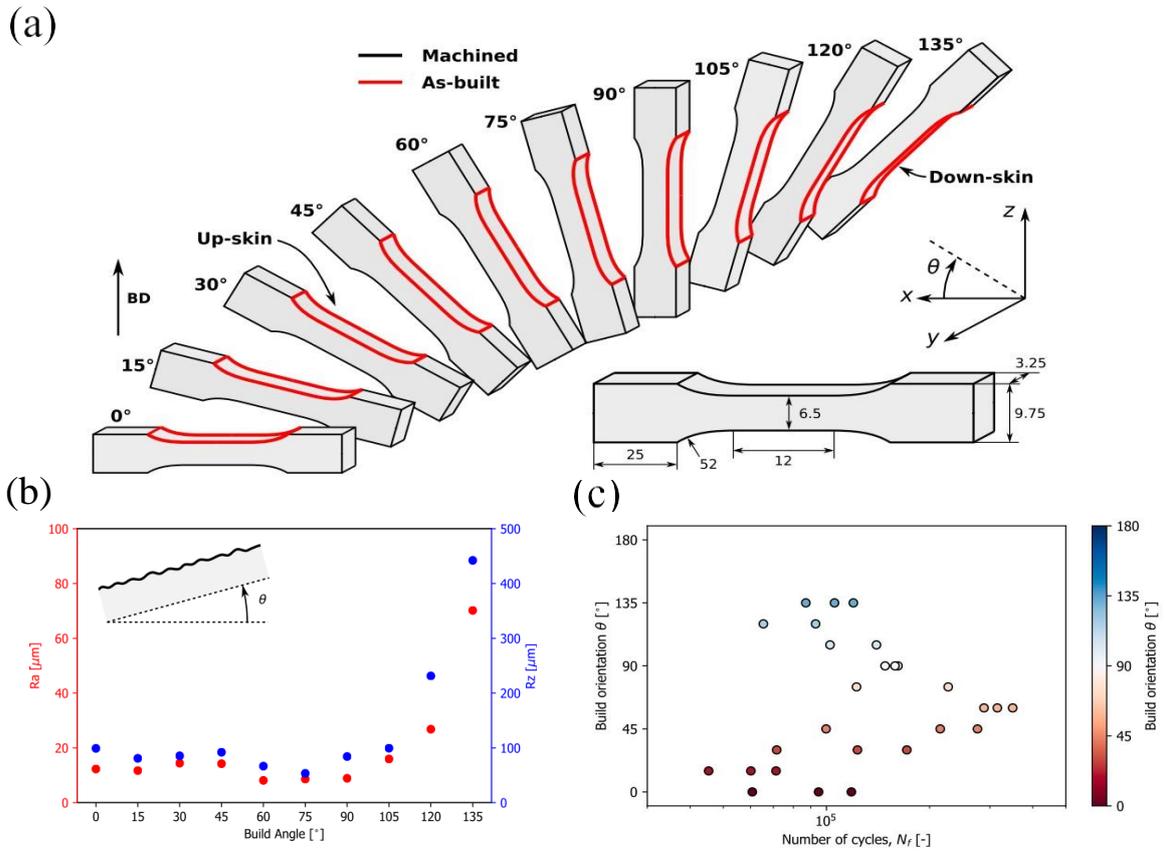

**Fig. 113.** (a)Specimen orientations and dimensions of AM18Ni300;(b)Correlation between surface roughness (from SEM) and build angle of 18Ni300; (c) Fatigue data for different orientations of 18Ni300. Reprinted from [510], Copyright (2021), with permission from Elsevier.

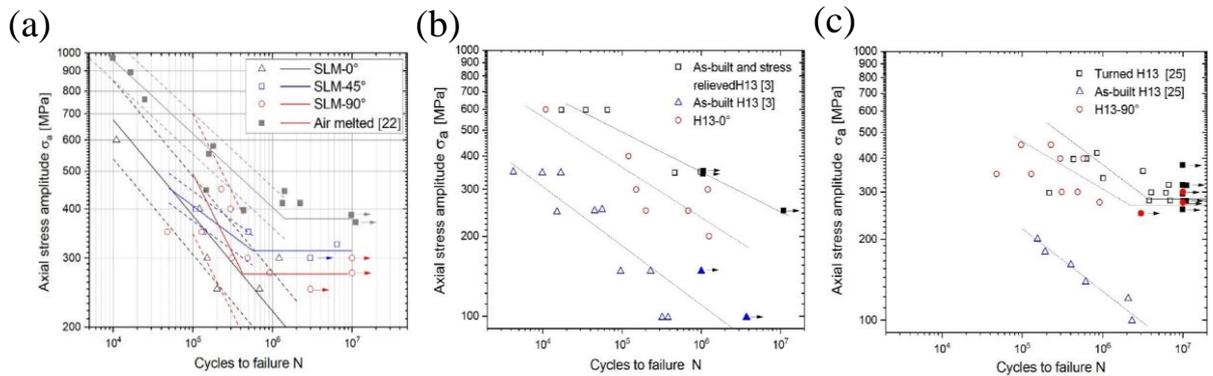

**Fig. 114.** (a)Fatigue test results (R = -1) for tool steels oriented at 0°, 45°, and 90°. Double-logarithmic scale using the Basquin law to fit the data. (b)Comparison of fatigue test results for samples oriented at 0° and 90° with those found in literature by Mazur [424] and R. Dörfert [515].

### 6.7.2. post-processing

For the post-processing of H13 steels, R. Dörfert et al. [515] studied the effect of surface roughness on the fatigue resistance of H13 steel (HV1 = 571 15) in the as-built condition and proved that the high surface roughness in the as-built samples acted as a small notch. In addition, a detectable improvement in fatigue strength was observed under post-processing operations such as turning. M. Mazur et al. [424] investigated the effect of stress-relieving treatment (600 °C, 2 h) on the fatigue resistance of AISI H13 steel in the as-built condition (Rz = 31.3  3.9 $\mu$m). The hardness values before and after the treatment were 674 HV and 528 HV, respectively. After the treatment, fatigue strength improved noticeably, and it was slightly higher than the fatigue resistance achieved in this study, considering the same building direction. Similar results were obtained for the turned samples in the as-built condition [515] without further heat treatment. As expected, surface and heat treatments improve fatigue resistance. Meanwhile, the experimental results tested by Tshabalala [517] compared well with results of maraging steel obtained by Croccolo [509] who investigated the impact of aging heat treatment and sub-sequential machining of additively manufactured samples using EOSInt M280 machine. The results from the set of specimen which did was not heat treated but machined before fatigue testing recorded the fatigue limit of 363 MPa with 90% confidence level. This fatigue limit was extrapolated from the fatigue data in Fig. 114 for the stress corresponding to an expected life of $10^7$ cycles.

## 7. Inconel 718

### 7.1. Introduction

Inconel 718 (IN718) is a kind of nickel-based superalloy known for its high creep and corrosion resistance that can withstand loading at an operating temperature up to 700 °C [518]. IN718 is typically used in industries with harsh and high-temperature environment, such as jet engines and gas turbines in the field of aviation. The conventional fabricating technology of nickel-based superalloy components focuses on casting, forging and powder metallurgy. However, it is difficult to machine the IN718 materials due to their low thermal conductivity and material removal rates, resulting in excessive tool wear, premature cracking, and built-up edge formation [519, 520]. On the other hand, IN718 parts are very costly in the application of complex shapes or cavities. With the development of additive manufacturing, such as PBF and DED, it has been generally used to product high quality metal parts of IN718 in recent few years. For example, General Electric (GE) Aviation fabricates a new fuel nozzle based IN718 by SLM technology, which is five times more durable than the previous model [521].

Nickel is the primary constituent element of IN718. To achieve the superior characteristics, elements like Cr, Fe, Nb, Mo, and so on are also used. Commonly, the microstructure of IN718 AM parts commonly consists of $\gamma'$ (intermetallic compound with stoichiometric composition $Ni_3AL$ characterized by face-centered-cubic, fcc) and $\gamma''$ (stoichiometric composition $Ni_3Nb$ characterized by body-centered-tetragonal, bct) in the solid solution $\gamma$ matrix, $\delta$ ($Ni_3Nb$) phase and the brittle Laves (Ni, Fe, Cr)$_2$(Nb, Mo, Ti) phase. The fine uniform metastable $\gamma''$ precipitates formed by age hardening is the main strengthening mechanism of IN718. At high temperatures above 750 °C, the $\gamma''$ precipitations reach their limit of stability, causing a significantly reduced yield stress and tensile strength [522]. In addition, the $\gamma''$ phase can transform to $\delta$ phase, which could reduce the fracture toughness and increase the creep resistance of the alloy specimen at elevated temperature [523, 524].

In general, $\delta$ phase in the form of intragranular acicular precipitates limits ductility. At elevated temperature, $\delta$ phase of spherical morphology in grain boundaries prevents from grain growth [525]. The laves phase is formed on the grain boundaries due to the low diffusion rate of element Nb to the matrix. Whereas, the hard or brittle Laves phases generally have bad effects on tensile strength, stress rupture and fatigue life of AM Inconel 718 [526]. Sui et al. [527] reported that an unbroken Laves phase plays a primary role in the resistance to high-cycle fatigue in IN718. Therefore, $\delta$ phase, Laves phase as well as carbides/oxides should be avoided for the overall improvement in mechanical properties [36].

As shown in Table 4, fatigue is the major factor contributing to failure of engineering and aircraft components. In the context of application of metal AM parts in aerospace, understanding the fatigue properties of Inconel 718 material is crucial.

**Table 4.** Frequency of failure mechanisms [528].

| Failure mechanism | Percentage of Failures | |
|---|---|---|
| | Engineering Components | Aircraft Components |
| Fatigue | 25 | 55 |
| Corrosion | 29 | 16 |
| Brittle fracture | 16 | - |
| Overload | 11 | 14 |
| High temperature corrosion | 7 | 2 |
| SCC/Corrosion fatigue/HE | 6 | 7 |
| Creep | 3 | - |
| Wear/abrasion/erosion | 3 | 6 |

## 7.2. Fatigue performance

### 7.2.1. Fatigue life

According to a report, the failture of 55% of components in aerospace is attributed to fatigue. Actual fatigue life evaluations on AM materials still mostly rely on conventional fatigue tests. According to the loading cycles (i.e., cyclic strain rate), the fatigue tests of IN718 can be divided into low cycle fatigue (LCF), high cycle fatigue (HCF) and very high cycle fatigue (VHCF).

Conventional fatigues tests are typically conducted in LCF, where loading frequency is at 20 Hz or lower. Gribbin et al. [529] investigated the low cycle fatigue behavior of IN718 counterpart fabricated by DMLS. They found an interesting phenomenon that most DMLS specimens underwent a sudden brittle failure with rapid crack propagation for the strain amplitude less than 1.4%. However, failure of the wrought specimens was not so abrupt. Moreover, LCF behavior of DMLS material was found to be better than that of wrought material at lower strain amplitudes, while the wrought material had longer life at higher strain amplitudes. Johnson et al. [530] also found that the fatigue lives of AM Inconel 718 specimens are shorter than those of wrought counterparts (Fig. 115) and the AM specimens possess lower fatigue resistance than wrought Inconel 718 when they underwent the same heat treatment process. Meanwhile, they pointed out that AM Inconel 718 specimens were not comparable to those of wrought in the long life regime because they can be an order of magnitude shorter than the wrought counterpart, which may be a very high cost in practical application. AM defects are most likely the major cause for the reduced fatigue life.

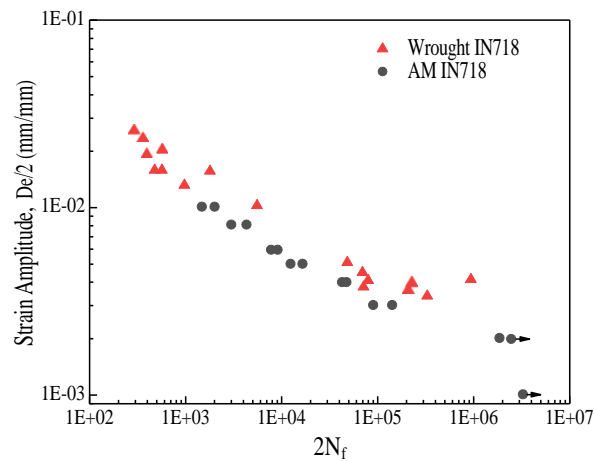

**Fig. 115.** Strain-life fatigue behavior of AM Inconel 718 and its comparison to wrought Inconel 718 [530]. Arrows indicate runout tests.

However, due to low load frequency, the number of cycles of LCF tests is typically less than $1 \times 10^7$ to save time and money. In recent few years, HCF and VHCF tests have received much more attention [531]. Especially, Ni-base superalloy is commonly subjected to cyclic loading at very high frequency and service lives beyond ten-million cycles in the application of aerospace engineering (gas turbine, jet engine et al.). Smaller defects are able to become potential crack initiation sites because of the high sensitivity of VHCF resistance to defects. Therefore, a thorough understanding of high/very high cycle fatigue behavior of IN718 parts is critical.

Muhammad et al. [532] focuses on the HCF and VHCF behavior of both wrought and L-PBF fabricated Inconel 718 in machined or polished surface condition. It were obtained from Fig. 116 that the L-PBF specimens happened to have a longer fatigue lives at higher stress amplitudes under both loading frequencies of 20 kHz, which attenuated at lower stress amplitudes. It is also quite notable that, under conventional test frequency (i.e., 5 Hz), the L-PBF specimens exhibited significantly higher fatigue resistance in all stress levels compared to wrought specimens, which is contrary to the conclusion from ref[529]. Moreover, there was an improvement in fatigue life of roughly an order of magnitude for almost all specimens tested under ultrasonic frequency. so that the stresses should be corrected in the ultrasonic fatigue tests considering test frequency effects (i.e., cyclic strain rate) on fatigue behavior. The fatigue performance of SLM IN718 and forged IN718 was also compared in the literature[533]. It is obviously that SLM IN718 shows a lower fatigue strength compared with forged IN718 in both LCF and HCF regions, which is reduced to the low ductility and the micro-defects in the material. Especially in high cycle regime, the fatigue life of SLM IN718 only reaches the half of the forged material value.

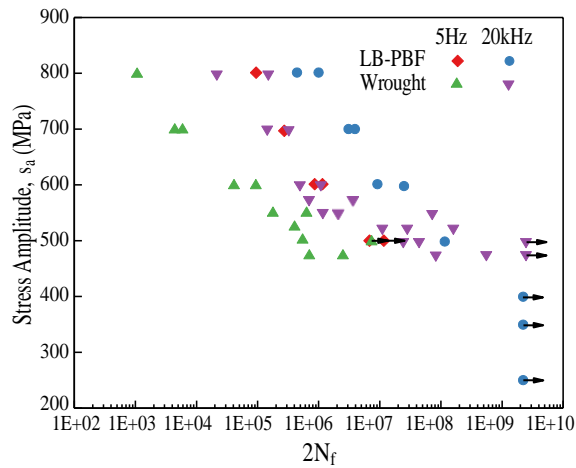

**Fig. 116.** Stress-life plot for wrought IN718 and vertical LB-PBF IN718 tested using conventional and ultrasonic fatigue test methods [532]. Wrought IN718 data from MMPDS [534] is included for comparison.

The correlations between applied strain amplitude and fatigue life can be described approximatively by the Coffin-Manson model, which considers the sum of elastic and plastic portions of strain to calculate the total strain amplitude:

$$\varepsilon_a = \frac{\Delta\varepsilon}{2} = \varepsilon_e + \varepsilon_p \tag{1}$$

where $\varepsilon_a$ is the total strain amplitude, $\varepsilon_e$ is the elastic amplitude and $\varepsilon_p$ is the plastic strain amplitude. The elastic and plastic components of strain amplitude are plotted against $N_f$ in a log-log scale as follow:

$$\varepsilon_e = \frac{\sigma'_f}{E}(2N_f)^b \tag{2}$$

$$\varepsilon_p = \varepsilon'_f(2N_f)^c \tag{3}$$

where $\sigma'_f$ is the fatigue strength coefficient, $E$ is the elastic modulus, $b$ is the fatigue strength exponent, $\varepsilon'_f$ is the fatigue ductility coefficient and $c$ is the fatigue ductility exponent, $2N_f$ is the reversals to failure. Since the data is plotted in a log-log scale, the Coffin-Manson model becomes linear, with $b$ and $c$ as the slopes and $\sigma'_f/E$ and $\varepsilon'_f$ as the intercepts for the elastic and plastic curves, respectively. In the higher cycle regime the life values of present SLM IN718 deviate from the Manson-Coffin model's prediction obviously [532].

In addition, the fatigue properties obtained at room temperature and elevated temperatures are different as listed in Table 5. Obviously, temperature has a great effect on the fatigue properties.

Table 5. Fatigue properties of LDED IN718 at room [535] and elevated temperatures [536].

| Test condition | $\sigma'_f$ (MPa) | b | $\varepsilon'_f$ (mm/mm) | c |
|---|---|---|---|---|
| RT | 2090 | -0.111 | 1.8 | -0.78 |
| 650 °C | 1708 | -0.092 | 0.1 | -0.48 |

### 7.2.2. Fatigue crack growth

According to current research, the local crack growth rate is mostly determined by the applied stress intensity factor, local microstructure and residual stresses. A regression analysis is performed on the FCGR of different specimens according to a Paris law:

$$da/dN = C(\Delta K)^m \tag{4}$$

where a is crack length (mm), N is the number of fatigue cycles, da/dN represents fatigue crack growth rate (mm/cycle), $\Delta K$ is stress intensity factor (SIF) range (MPa$\sqrt{m}$), and C, m are material parameters fitted by the experimental data. The bigger the m is, the finer the fatigue striations (FS) become. Kin et al. [537] investigated fatigue crack propagation (FCP) behavior of L-PBF Inconel 718 alloy. In the FCP tests, the crack propagation direction (CD) of L-PBF specimen was either parallel or perpendicular to the building direction (BD), i.e. CD//BD *vs.* CD⊥BD. The fatigue crack propagation rate of conventionally manufactured (CM) IN718 specimen with a CL (circumferential and longitudinal) orientation at room temperature is also considered to compare the difference. As shown in Fig. 117), the FCP rates of L-PBF Inconel 718 specimen were significantly higher than those of CM counterpart in low and intermediate $\Delta K$ regimes, while lower in high $\Delta K$ regime. Moreover, it's seen that the orientation effect on the FCP behavior of L-PBF Inconel 718 specimen was slight over the entire $\Delta K$ regime.

Konecna et al. [538] found that SLM IN718 in the as-built condition has a similar long crack growth resistance to the conventionally manufactured Inconel 718 for loading in the high $\Delta K$ region, but exhibits very severe difference in the threshold region. Pei et al. [533] also got the same conclusion by comparing IN718 produced by SLM or forge. As shown in Fig. 118, in high load region the crack growth rates of SLM IN718 approach the forged materials' speed. But in lower load regime, the fatigue crack growth rates of SLM IN718 specimen are much higher than those of the forged materials.

Ye et al. [542] proposed a crystal plasticity framework based on the finite element method to quantitatively assess the mechanistic drivers existing in columnar IN718 polycrystals for fatigue crack nucleation. By revealing the relationship between the stored plastic strain energy density and the number of cycles for fatigue life through simulation, the microstructure-level critical stored energy was used to evaluate and predict the fatigue life of the investigated material and the results are consistent with experimental findings. Under the low cycle loading conditions, the simulated results shows that grain boundaries are the perffered crack initiation sites.

Laser Directed Energy Deposition (LDED) feeds metal powder onto a melt pool, which is prone to produce bulk defects. It was reported that FCP resistance of LDED built sample is in between to that of forged and L-PBF samples and the forged samples have the highest fatigue resistance [543]. Yu et al. [544] found that there is a decrease in FCG resistance of LDED IN718 due to the coarse Laves phases with uneven $\gamma''/\gamma'$ precipitates and the soft precipitate-free zone around short-acicular $\delta$ phase. For as-built specimen, it can be seen from Fig. 119a that some long-striped Laves phases break up while some granular Laves phases maintain completely their shapes, and the cracking planes are mainly parallel to the direction of main crack advance. As shown in Fig. 119b, a lot of dislocations stack around Laves phases for uneven slip deformation, resulting in stress concentration at the interface of Laves and $\gamma$ phase. Some cracks belonging to slip-band cracking also appear in $\gamma$ matrix.

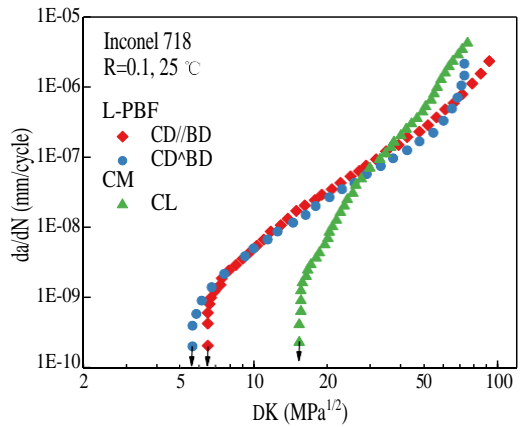

**Fig. 117.** The da/dN *vs.* ΔK curves of L-PBF and CM Inconel 718 specimens at an R ratio of 0.1 and a testing temperature of 25 °C.

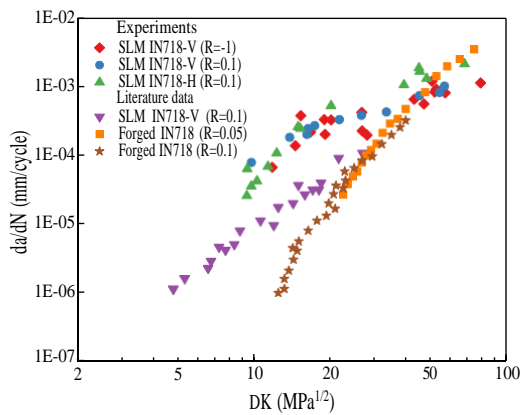

**Fig. 118.** Comparison of fatigue crack growth rates of the present SLM Inconel 718 with literature data[539–541]

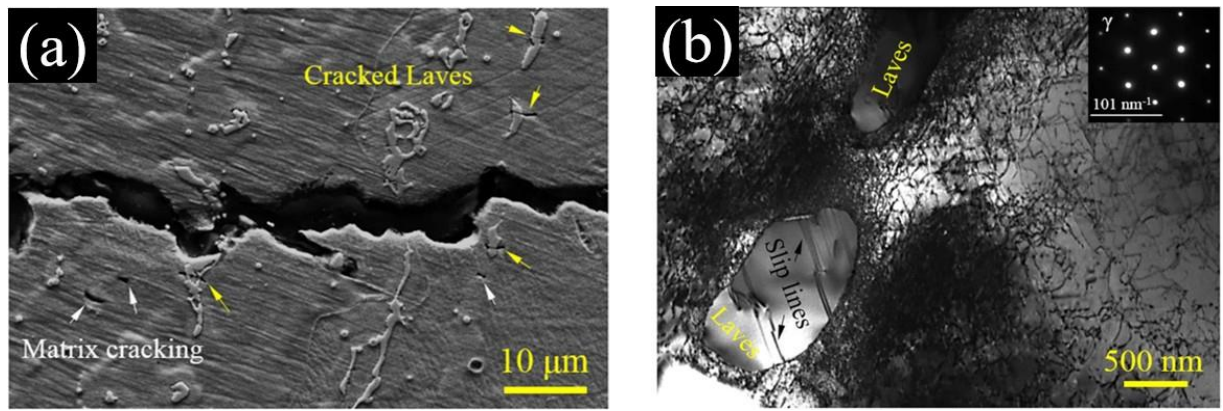

**Fig. 119.** (a) FCG paths in near-threshold regime. (b) TEM micrographs showing crack-tip vicinity in near-threshold regime. Reprinted from [544]. Copyright (2021), with permission from Elsevier.

### 7.3. Effects on fatigue behaviour

In general, the fatigue strength for additive manufactured samples can be reduced due to several reasons, for example due to a rough as-built surface, an internal defect or a geometrical notch. In addition to factors intrinsic to the material itself, specimen geometry and loading conditions are likely to influence fatigue results. Here, several factors including post processing, temperature, orientation and notch are investigated, which significantly affecting the fatigue behaviour of AM Inconel 718. Other processing parameters are introduced briefly.

#### 7.3.1. Post processing

The Ni-based Superalloy 718 manufactured by AM contained various irregularly shaped defects which were observed by microstructural investigation on the as-built materials. Fatigue performance of AM Inconel 718 parts is typically poorer than that of the cast and wrought alloy parts. This is because of detrimental effects arising from the microstructures such as brittle phases and anisotropy as well as porosities. The bad surface roughness and porosity would increase number of crack initiation sites. It's reported that the fatigue life are most closely related to the location of the pores by machine learning method [545]. In other words, pores with smaller size and farther from the surface may lead to a higher fatigue life. Post-processing can enhance the mechanical properties of AM IN718 part in both directions and at room/evaluate temperature [546]. In industry, the common post-processing consists of hot isostatic pressing (HIP), heat treatment (HT), surface treatment and et al..

**Hot Isostatic Pressing**

HIP has been widely used to remove the internal porosity, relieve residual stresses and alter microstructure through the application of both high pressures and high temperatures in order to consolidate materials to full density [547, 548]. It has been confirmed that HIP can considerably reduce porosity in AM IN718 material and led to high-quality parts with fine and homogeneous microstructure [549]. And HIP can substantially improve fatigue crack growth thresholds by reducing residual stresses. Balachandramurthi et al. [32] reported that HIP improves fatigue life for both EBM and SLM IN718. However, in comparison with the HT condition, the HIP+HT treated condition only increases the fatigue performance slightly. Even in the case of a high initial porosity and residual argon content, high fatigue strength of of IN718 samples can also be achieved after HIP post-processing treatment [549].

It is notable that pores can't be completely eliminated even when the pressure is increased beyond 100 MPa [550] and defects open to the surface will not be closed as well [551]. Moreover, HIP may reduce the fatigue resistance for the increased surface roughness, especially for VHCF resistance which is extremely sensitive to the internal defects and the surface roughness [552]. Thus, it will be best to combine HIP with other post-processing.

**Heat Treatment**

After HIP, heat treatment is generally carried out to create a smaller and more uniform grain structure of Inconel 718. As shown in Table 6, the microstructure of IN718 AM parts commonly consists of different phases by different heat treatment. The variation of strengthening phase in the preparation of Inconel 718 alloy can improve the comprehensive mechanical properties [553, 554].

However, dendrites and the interdendritic Laves phase cannot be completely dissolved after heat treatment. The $\delta$ phase may also form during heat treatment, which could result in the breakdown of the dendritic structures. There is noticeable strengthen of UTS, YS and microhardness observed for samples after heat treatment. However, the decrease of plastic properties accompanied with heat treatment [525]. It was reported that heat treatment could cause increase of yield strength by 72-95%, of tensile strength by 30-46% and of hardness by 48% compared with the as-built condition [555]. Periane et al. [556] investigated the influence of heat treatment on the fatigue resistance of Inconel 718 fabricated by SLM. The as-built Inconel 718 samples were heat treated under Hot Isostatic Pressing (HIP) and Aeronautic Heat Treatment (AHT) conditions. The result from the fatigue testing of SLM heat-treated samples both HIP and HIP+AHT with reference to the cast and wrought (C&W) sample is presented in Fig. 120. There is a notable difference in the fatigue strength between the SLM and C&W specimens. The C&W samples have better fatigue performance compared to SLM samples. Within the SLM samples, the HIP+AHT SLM has survived more fatigue cycles at high stress levels compared to the SLM HIP samples, which is own to the presence of more porosity at the surface acting as the crack initiation site and the presence of $\delta$ phase in the SLM HIP samples.

In Yu's study [544], Inconel 718 (IN718) specimens manufactured by laser directed energy deposition (LDED) are subjected to three different heat-treatments as shown in table 7 and Fig. 121 shows the corresponding curves of fatigue crack growth rate. For a constant $\Delta K$ in Pairs regime, the FCGR of SA and HSA specimens is lower than that of as-deposited and DA specimens. In the whole da/dN regime, the FCGR of HSA specimen could be in good consistency with results of wrought IN718. They found that the coarse Laves phases with uneven $\gamma''/\gamma'$ precipitates severely decrease the FCGR resistance of DA specimen.

**Surface Treatment**

It is well known that surface morphology has a significant impact on the high cycle fatigue performance of a near-net shape metal part. Surface treatment can remove the outer layer of the material with high surface roughness and defects. Thus, mechanical properties, such as hardness, tensile and fatigue, have been significantly improved as compared to as-built specimens once finishing post-processing are deployed [558, 559]. Especially, it was reported that the fatigue lives of the machined specimens were at least seven times greater than those of as-built specimens [560].

**Table 6.** The qualitative phase composition of SLM Inconel 718 by X-ray diffractograms (XRD) [557].

| Sample | Phase | Quantitative composition/vol% |
|---|---|---|
| Powder material | γ-Ni | 90.0 |
|  | γ″-Ni₃Al | 3.5-3.9 |
|  | γ″-Ni₃Nb | 4.3-4.5 |
|  | δ-Ni₃Nb | 1.8-2.0 |
| As-built SLM part | γ-Ni | 86.8 |
|  | γ′-Ni₃Al | 1.9 |
|  | γ″-Ni₃Nb | 8.0 |
|  | δ-Ni₃Nb | 3.3 |
| Homogenisation | γ-Ni | 90.1 |
|  | γ′-Ni₃(Al,Ti) | 1.9 |
|  | γ″-Ni₃Nb | 8.0 |
| Homogenization and aging | γ-Ni | 67.3 |
|  | γ′-Ni₃(Al,Ti) | 8 |
|  | γ″-Ni₃Nb | 4 |
|  | δ-Ni₃Nb | 3.5 |
|  | γ′-Ni₃Al | 17.2 |

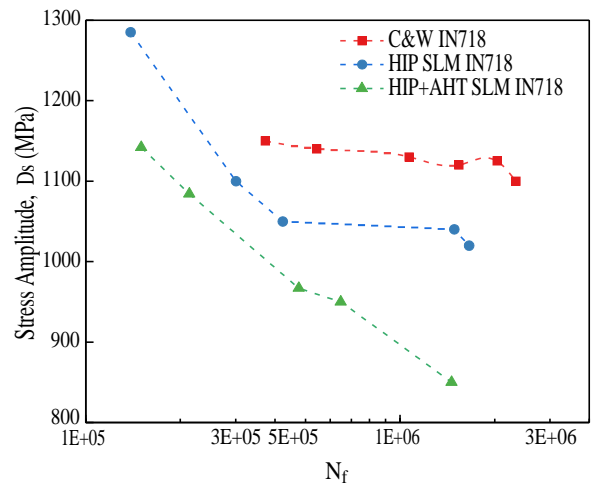

**Fig. 120.** Stress-life curves for HIP and HIP+AHT SLM samples in comparison to C&W IN718.

**Table 7.** Heat-treatment details in Yu's study[544].

|  | details |
|---|---|
| Direct aging(DA) | 720 °C×8 h/FC at 50 °C/h to 620 °C×8 h/AC |
| Solution+aging(SA) | 980 °C×1 h/WC+720 °C ×8 h/FC at 50 °C/h to 620 °C×8 h/AC |
| Homogenization+SA(HSA) | 1100 °C ×1.5 h/WC+980 °C×1 h/WC+720 °C×8 h/FC at 50 °C/h to 620 °C×8 h/AC |

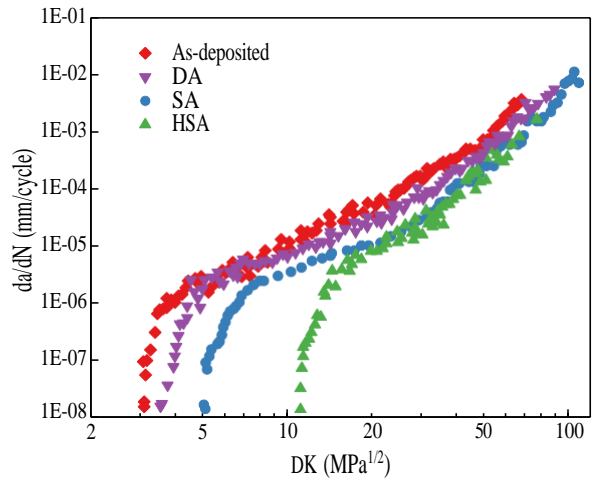

**Fig. 121.** Comparison FCGR of AMed IN718 alloys at different conditions.

Kantzos et al. [561] thought that the value of conventionally measured surface roughness is not directly correlated with fatigue behavior. Witkin et al. [562] researched the influence of surface conditions on high cycle fatigue properties of nickel-based superalloy Inconel 718 made by L-PBF. They found that stress ratio doesn't have a significant effect on fatigue properties. However, surface condition, specifically near-surface process defects, influenced fatigue properties. Comparing specimens with as-built or machined surfaces (Fig. 122), it is evident that as-built surfaces in L-PBF materials give rise to lower fatigue lives and fatigue limits than machined counterparts. Balachandramurthi et al. [32] reported that the machined specimens show a larger scatter in the fatigue life curve than the unmachined specimens due to different number of crack initiation sites.

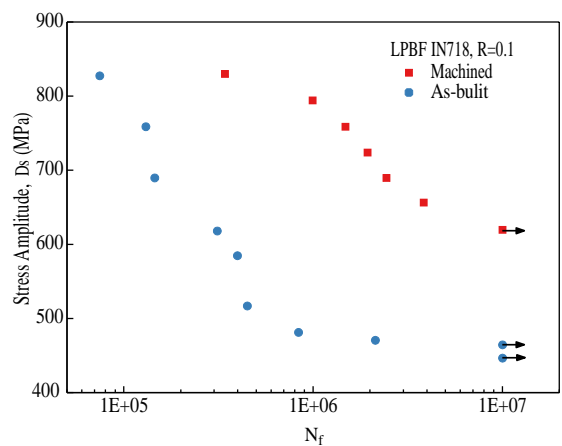

**Fig. 122.** Fatigue results for samples with machined surfaces.

The surface roughness may affect fatigue properties with reducing the build thickness. Wan et al. [563] investigated the combined effects of build thickness and surface roughness on the fatigue behavior of SLM Inconel 718 parts at 650 °C. They found that fatigue strength was enhanced by about 50% after surface machining and polishing when their fatigue lifetimes were at the same order of magnitude, and the surface roughness Ra was reduced from ∼14 μm to ∼110 nm. Both the surface conditions and the microstructures have effects on the thickness-dependent fatigue lifetime. As for as-built thin-wall specimens, the thinner specimens exhibited a longer fatigue lifetime than the thicker ones, while the opposite trend appeared in the machined specimens. In addition, there is no obvious influence of stress distribution on the fatigue lifetime of specimens with the thickness at the millimeter scales.

Lee et al. [564] investigeted the effects of various post surface treatments on the surface roughness and fatigue properties of LB-PBF IN718, including drag-finishing (DF), grinding (G), grinding + drag-finishing (GDF), sand-blasting (SB), and turning (T), compared with the as-built(AB) specimens. The standard and hybrid surface roughness of all surface conditions are listed in Table 8. It is confirmed that all surface treatments can improve the surface roughness greatly expect for the sand-blasting condition. All surface roughness values after GDF among all conditions are lowest. As shown in Fig. 123, the fatigue lives of specimens with different surface treatments were significantly improved when $\varepsilon_a$<0.006 mm/mm. It's seen that surface roughness is not the main factor of the fatigue behavior of LB-PBF IN718 in the LCF regime. The subtractive treatments G, T and GDF significantly improved fatigue performance both in the low and high cycle regime.

**Table 8.** Standard and hybrid surface roughness parameters generated by height data measured by optical microscope [564].

| Surface Condition | $R_a$ (μm) | $R_p$ (μm) | $R_v$ (μm) | $R_{sk}$ | $R_{ku}$ | $R_{mode}$ (μm) |
|---|---|---|---|---|---|---|
| AB | 4.35 | 13.73 | 12.42 | 0.39 | 2.82 | -2.08 |
| SB | 3.14 | 9.68 | 9.91 | 0.05 | 2.84 | -0.65 |
| DF | 2.71 | 7.68 | 7.85 | 0.00 | 2.59 | -0.94 |
| T | 1.08 | 3.55 | 3.45 | 0.02 | 2.64 | -0.25 |
| G | 0.37 | 1.31 | 1.29 | 0.07 | 3.13 | 0.02 |
| GDF | 0.31 | 1.11 | 1.10 | -0.14 | 2.81 | 0.02 |

However, conventional surface machining techniques (milling, turning, grinding, etc) can hardly be implemented for geometrically complex AM components. Moreover, such treatments on complex shapes cost quite a lot due to the poor machinability of IN718. Thus, non-conventional surface treatment alternatives are much needed and the effect of various surface treatments on fatigue properties of AM IN718 should be investigated thoroughly [565].

#### 7.3.2. Temperature

Most of fatigue tests are conducted at room temperature limited to technical difficulties and experimental efforts. Considering the fact that Inconel 718 is often used in applications prone to cyclic loading at elevated temperatures (e.g. turbine blades, jet engines, etc.), it is crucial to understand the fatigue behavior of AM IN718 at elevated temperature.

Concerning temperature effect on fatigue properties, previous investigations have poing out some principal tendencies in initial crack growth and fatigue life on the conventional IN718 parts. At high temperature, the fatigue strength of the wrought IN718 coarse-is lower than fine-grain alloy in a high-cycle region [566].

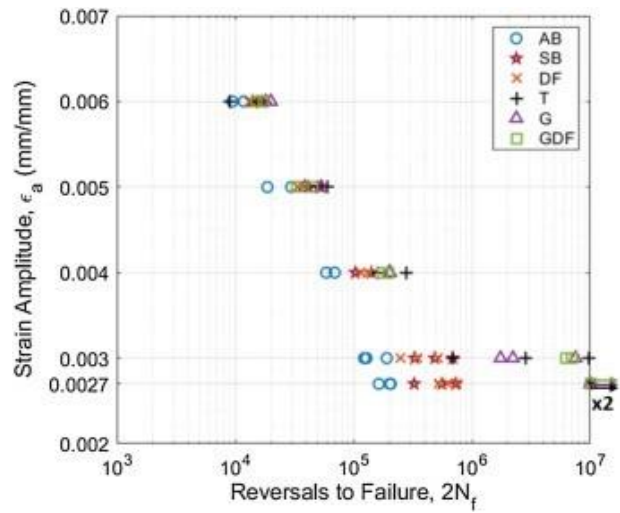

**Fig. 123.** Strain-life fatigue tests results of various post surface treatments. Reprinted from [564]. Copyright (2022), with permission from Elsevier.

Under rotating bending fatigue tests and at different temperature ranging from 25 °C to 600 °C, Chen et al. [567] found that the crack growth resistance rapidly decreases when crack grows beyond 50 μm at higher temperature, resulting faster growth speed. However, it is reported that crack growth rate of conventional IN718 parts is $7 \times 10^{-8}$ m/cycle at 550 °C, while decreasing to $3 \times 10^{-8}$ m/cycle at 450 °C and increasing to $3 \times 10^{-7}$ m/cycle at 650 °C [568]. Crack growth rate at room and elevated was estimated on Inconel 718 fabricated by additive manufacturing. For the loading with the stress intensity factor range exceeding 20 MPa$\sqrt{m}$, the crack growth resistance of SLM and wrought alloys is nearly identical at room temperature [538].

Fig. 124 compares the fatigue performance of LB-DED IN718 machined specimens with that of the wrought specimens both in room temperature and elevated temperature at 650 °C. The room temperature fatigue data for LDED IN718 material are adopted from ref[530], and the fatigue data at elevated temperature are adopted from ref[536]. For the wrought specimens, the fatigue data in both room temperature and high temperature conditions are adopted from ref[569]. It's obvious that the LDED IN718 possesses slightly lower fatigue resistance at room temperature as compared to the wrought counterpart, either in the low cycle or high cycle fatigue regimes, which is consistent with the previous conclusion. However, in high temperature condition, the overall fatigue behavior of LDED IN718 is somewhat similar to that of the wrought counterpart. In low cycle fatigue regime (i.e. high strain amplitude level), it's obvious that the LDED IN718 material tested at room temperature possesses higher fatigue resistance than the high temperature fatigue tests conducted at 650 °C. However, the effect of test temperature was observed to be minimum with the decrease of the strain amplitude, especially in the high cycle fatigue region [570]. Interestingly, it has been reported that in the high cycle fatigue regime the wrought IN718 possesses a higher fatigue strength at elevated temperature than the room temperature due to the γ-matrix softening and surface oxidation occurring at elevated temperature [571].

Johnson et al. [535] attributed the lower fatigue resistance of LDED IN718 at room temperature compared to the wrought IN718 to the heterogeneous microstructure as well as the presence of defects (i.e. gas entrapped pores, or inclusions) in the LDED specimens. On the contrary, the cracks initiated from the surface for the LDED IN718 specimens tested at elevated temperature. This means that the defects (gas

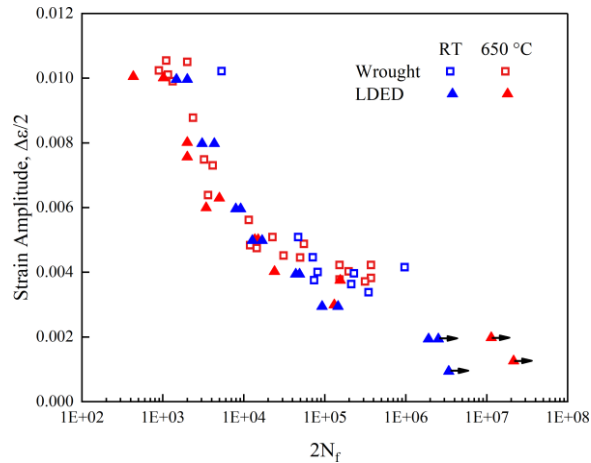

**Fig. 124.** Comparison of strain-life fatigue data of LDED and wrought IN718 at room temperature and elevated temperature (650 °C). All the wrought data are adopted from [569]

entrapped pores, or inclusions) did not affect the fatigue behavior of LDED IN718 specimens at elevated temperature; therefore, the fatigue resistance of LDED IN718 is similar to that of the wrought material.

The FCGR of nickel-based superalloy is also remarkably affected by temperature. In the research of Ma et al. [572], the fatigue crack propagation (FCP) behavior of Inconel 718 alloy manufactured by L-PBF or conventionally processe was studied at 25 °C and 650 °C. The FCP rates of L-PBF Inconel 718 alloy were found to be significantly higher than those of conventionally manufactured (CM) counterpart at 25 °C in low and intermediate $\Delta K$ regimes. The FCP rates of L-PBF Inconel 718 at 650 °C was lower than those at 25 °C in near-threshold $\Delta K$ regime because that the microstructure evolution at 650 °C can increase the FCP rates. Furthermore, it is obvious that the FCGR is much higher at 650 °C than that of 25 °C unrelated to orientation. The reduced fatigue crack growth resistance at elevated temperature can be generally attributed to many factors, such as strength material decrease, cyclic softening, and oxidation-induced degradation etc [573–575].

Fig. 125 also shows that the FCP resistance of L-PBF Inconel 718 specimen was higher at 650 °C than those at 25 °C in near-threshold $\Delta K$ regime, which is due to crack blunting with fine striations observed on the fracture surface [537]. With the increase of $\Delta K$ regime in elevated temperature at 650 °C, blunting no longer prevailed.

### 7.3.3. Orientation

Rapid solidification of thin layers results in the distinct directional grain growth in the AM process, which makes alloys have the obvious anisotropic mechanical properties, especially along the building direction and the scanning orientation. For example, Young's modulus is associated with the crystallographic orientation that grains with ⟨001⟩ orientation possess the lower Young's modulus than that of the grains with ⟨110⟩ and ⟨111⟩ orientations. As a result of the anisotropy, it's difficult to evaluate the fatigue strength and failure mechanism of AM IN718 alloy.

As shown in Fig. 126, fatigue specimens with the loading axis along 0°, 45°, and 90° off the build orientation were respectively marked as Z, XZ, and X. In other words, X represents the scanning/horizontal orientation and Z is the building/vertical direction in many literatures. For tensile test samples fabricated by SLM, Trosch et al. [576] reported that the horizontal built SLM specimens are characterized by a higher ultimate tensile strength (UTS) and lower elongation to failure ($\varepsilon_f$) at room temperature than vertical built samples. Interestingly, XZ oriented SLM specimens achieve the highest tensile strengths and show lowest ductility due to the combined effect of columnar grain growth and the layered structure. Kelley et al. [560] verified the anisotropy of DMLS IN718 alloy that the yield strength and tensile strengths were lower in the Z orientation than in the X orientation and the elastic modulus in the Z orientation was less than half of that in the X orientation. Kuo et al. [577] found that the X-oriented specimens exhibited inferior creep life and worse ductility than the Z-oriented specimens, which is attributed to the interdendritic $\delta$-phase precipitates arrayed perpendicular to the stress axis in the X-oriented specimens.

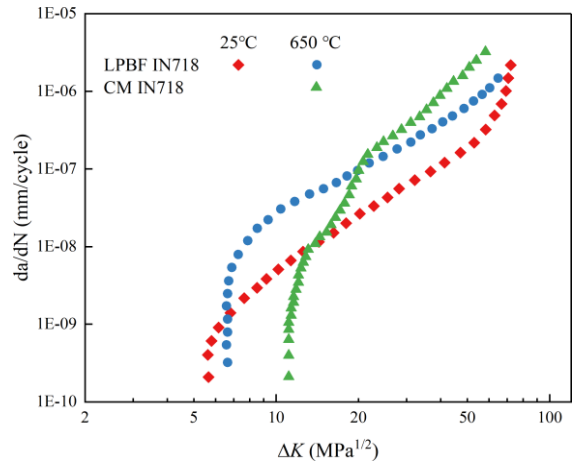

**Fig. 125.** The da/dN *vs.* ΔK curves of L-PBF and CM Inconel 718 specimens at an R ratio of 0.1 and a testing temperature of 25 °C and 650 °C.

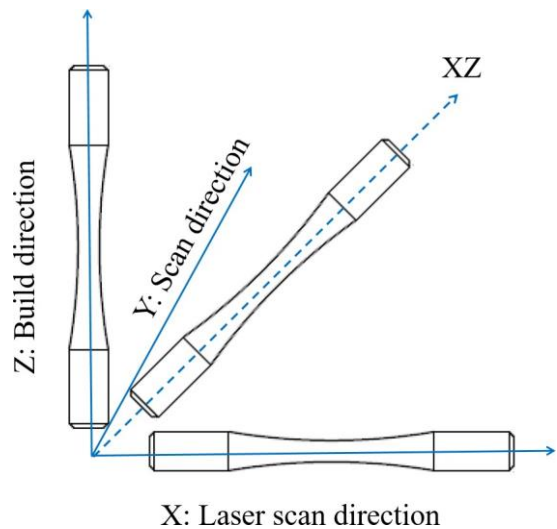

**Fig. 126.** Build Orientations

The effects of orientation on the fatigue behavior of IN718 alloy by SLM were investigated by many re-searchers. The measured fatigue strength of horizontal build orientations was greater than that of specimens built in a vertical orientation. Konecna et al. [539] found that the direction parallel to the build direction is associated with the lowest fatigue strength of IN718 parts produced by SLM. However, it was also reported that the horizontal and vertical specimens with machined surfaces were not significantly different in fatigue limit at stress ratio R = 0.1 [562]. However, Yamashita et al. [578] observed that there was no apparent difference in fatigue properties between the built directions because the influence of defects prevails over the microstructural difference made by built direction. Zhou et al. [579] verified the existence of anisotropic LCF properties for SLMed Inconel 718 samples with columnar grains. The columnar grains oriented parallel to the build direction showed the highest fatigue life. While the anisotropic LCF properties of samples with equiaxed grains was insensitive to orientations. Tomasz et al. [525] reported that XZ-oriented samples had the best overall mechanical properties determined in both tensile and FCG tests.

As shown in Fig. 127, the anisotropic high-cycle fatigue behavior of SLM IN 718 is obvious. The stress direction in C specimens is orthogonal to the material layers and its fatigue lives are shorter than the other two specimen orientations. The other two specimen orientations (A and B specimens) show similar fatigue behaviour with a slight higher performance of orientation A [539]. However, layer thicknesses may change the fatigue performance of specimens oriented along different directions. It is reported that C specimens exhibit the highest fatigue strength equal to 450 Mpa fabricated with 50 $\mu$m layer thicknesses, while A specimens exhibit the best fatigue performance equal to 350 Mpa fabricated with 30 $\mu$m layer thickness [580]. This problem should be further studied because of the above experimental results of specimens fabricated with different equipments, which makes great difference to fatigue strength.

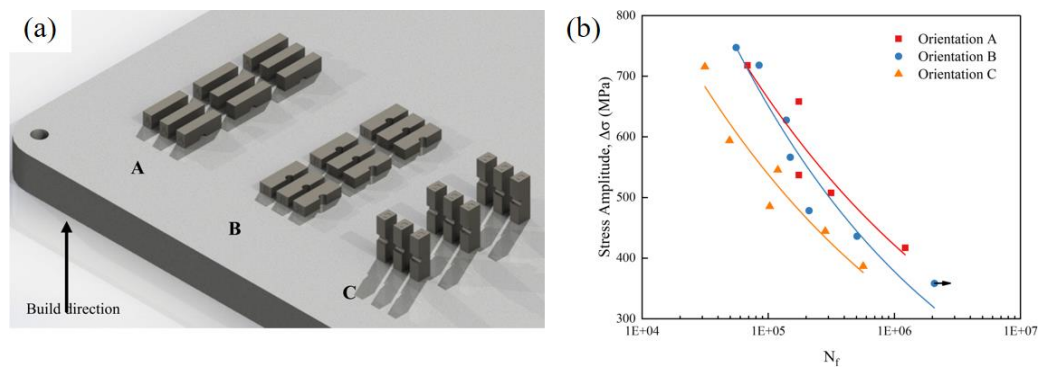

**Fig. 127.** Fatigue life of SLM IN 718 specimens in dependence on their orientation with respect to build orientation. (a)Rendering of specimens layup in the build. (b) S-N curves. Reprinted from [539]. Copyright (2016), with permission from Elsevier.

In addition, Zhou et al. [579] studied the impact of the induced texture (columnar or equiax grain structure) on the LCF behavior of Inconel 718 fabricated by SLM. As shown in Fig. 128, obvious anisotropy exhibited for samples with columnar grains and the columnar grains oriented parallel to the build direction exhibited the highest life on average. However, there is no obvious sensitivity to orientations for samples with equiaxed grains, namely the mechanical anisotropy nearly disappeared.

Fig. 129 represents the fatigue crack growth rate curves of IN718 alloy in three different orientations at different temperature [572]. In Fig. 129a, the Z-oriented specimens obviously provides less resistance to fatigue crack propagation than those of X and XZ oriented specimens at room temperature. The maximum stress intensity factor before final facture is smallest for the specimen in Z orientation, indicating the lowest fracture toughness accordingly. At 650 °C, it's seen that the difference between the data of X and XZ orientation was narrowed compared with that of 25 °C, while the effect of orientation on FCGR follows the same trend (XZ >X >Z). It should be noted that all the FCGR curves commonly show obvious fluctuations, which is quite different than the fatigue long crack growth behavior [581].

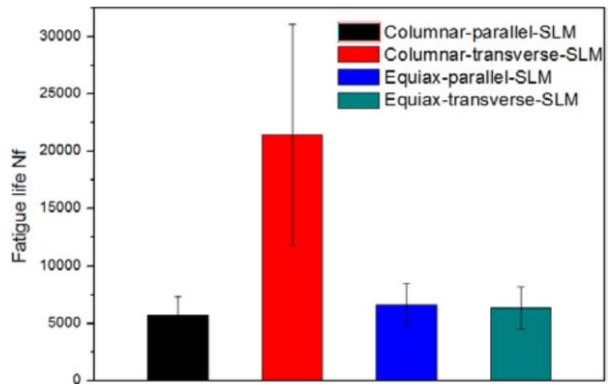

**Fig. 128.** The LCF life of samples with different textures. Reprinted from [579]. Copyright (2018), with permission from Elsevier.

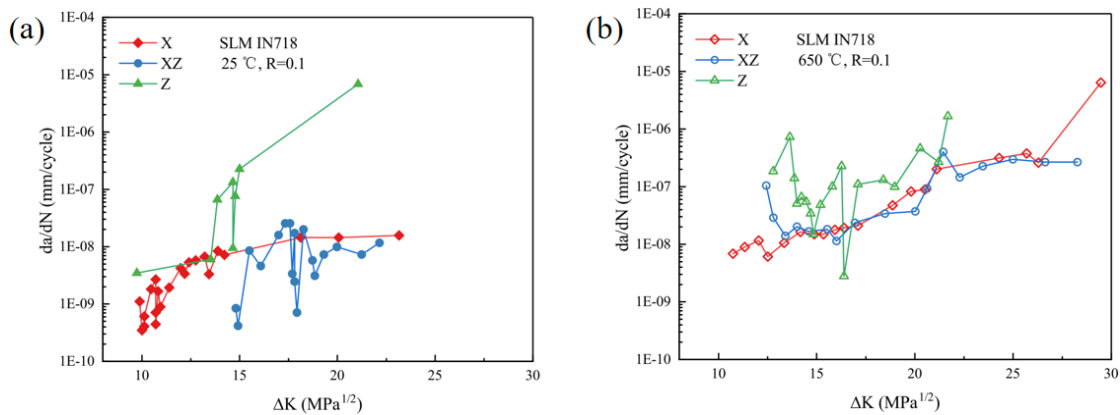

**Fig. 129.** Fatigue crack growth rate da/dN versus stress intensity range $\Delta K$ curves of IN718 alloy (X, XZ, and Z orientations)): (a) at 25 °C, (b) at 650 °C [572].

### 7.3.4. Notch

Components with highly notched regions can strongly affect the fatigue life, while the notch geometrical induced failure of AM Inconel 718 is not well understood. Chen et al. [582] studied notched fatigue behaviour and the influence of temperature on the notch sensitivity of conventional Inconel 718 parts under rotary bending, obtaining results showing low fatigue notch sensitivity at room temperature. Recently, samples with different notch geometries have been investigated, which is often used to evaluate the influence of such features or stress concentrations. As shown in Fig. 130, there are mainly three different notch geometries considered: unnotched specimens, semi-circular notch and v-shaped notch respectively, which is widely used in the fatigue test to compare the fatigue properties. Furthermore, v-shaped notch can be divided according to the angle. For the research about notch sensitivity, it has been reported that the semi-circular specimens displayed a notch sensitivity higher than full notch sensitivity, while the v-shaped notch is lower [583].

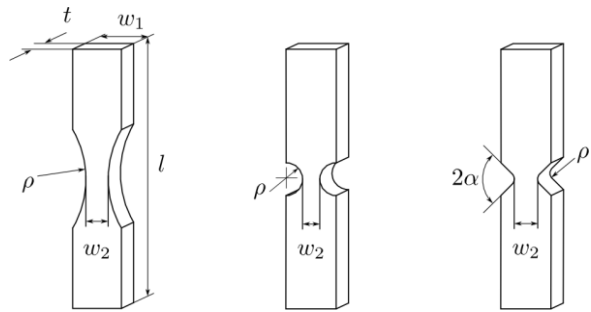

**Fig. 130.** Geometry of different test specimens. From left unnotched, semi-circular notch and v-shaped notch. Reprinted from [583]. Copyright (2018), with permission from Elsevier.

As we known, notched fatigue specimens have been used to evaluate the influence of such features or stress concentrations. There are mainly two kinds of specimens for testing the notched fatigue behaviour: machined specimens and as-built specimens. For the machined specimens, the notch geometries might interact with internal defects. For the as-built specimens, the surfaces of the notches are populated with smaller notches/defects during the AM process [584]. The work of Witkin et al. [585] focused on notch fatigue behaviour of as-built SLM Inconel 718. The results not only show that the surfaces of SLM metal parts influence the fatigue behaviour in the presence of a macroscopic designed notch, but also that the final as-built notch dimensions are dependent on both the notch geometry and specimen orientation.

Solberg and Berto [583, 586] had determined a criterion based on the relationship between notch geometries and local defects through a thorough analysis the fracture properties of the AM parts, indicating that the stress concentration is much higher on the as-built surface than the machined notched one. The fatigue testing results of the three different notch geometries were normalized with the ultimate tensile strength and compared in Fig. 131. From the fatigue data, the fatigue strength at $2 \times 10^6$ cycles for the unnotched specimens was 26.0% of UTS, while decreasing due to the existence of semi-circular or v-shaped notch. Fatigue life curve of the two different notch geometries had the similar slope and scatter, while the unnotched one is less steep and has a larger scatter.

It should be noted that the defects in the surface region act as notches within the actual notch geometry. The surface quality of the notches is affected by the specimen orientation due to the layer-by-layer fabrication, which might influence the test results. Konečná et al. [587] studied the directional notch fatigue behavior with four directional specimens (shown in Fig. 132a). From the S/N curve in Fig. 132c, it is obvious that Type A- specimens showed the worst notch fatigue performance while Type B had the best fatigue performance. Moreover, the directional quality of notch surfaces of the four specimen types was studied by using magnified optical microscopy, as shown in Fig. 132b. For Type B specimens with the semicircular notch, the build direction does not affect the entire notch geometry. However, the actual notch geometry of the Type C and Type A- specimens showed evident large global deviations from notch circularity. In addition, Konečná thought that the local quality at the notch root is also important for the fatigue behavior. Although notch geometry of the Type C is globally poor, its notch root fits well with the theoretical geometry. Therefore, Type C specimens determine an intermediate fatigue behavior. Conclusions mentioned above were also verified in their next paper [588].

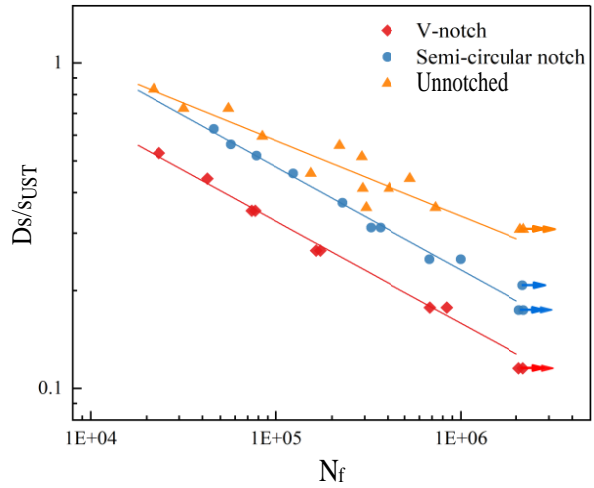

**Fig. 131.** The effect on fatigue behavior of different notch geometries.

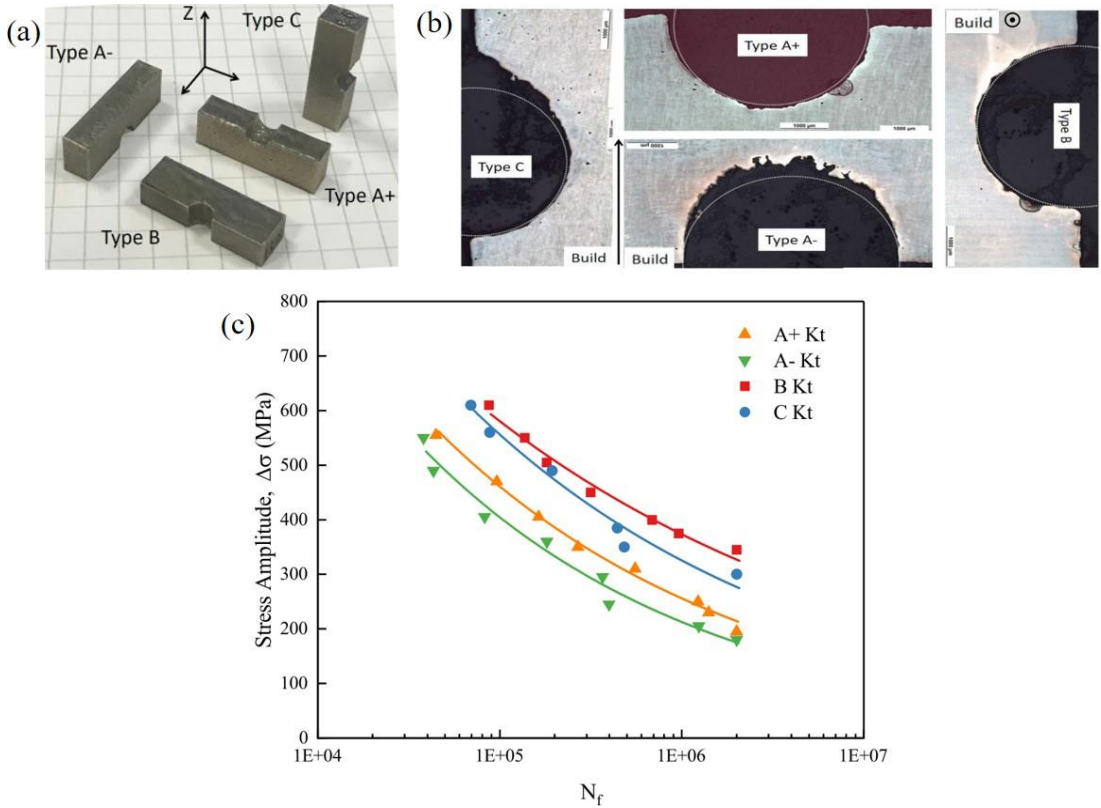

**Fig. 132.** Notch fatigue behavior of Inconel 718 produced by SLM. (a) Orientation of the notched miniature specimens. (b)Effective semicircular notch in four types of specimens. (c) S-N curves of directional notch fatigue data. Reprinted from [587]. Copyright (2019), with permission from Elsevier.

Similarly, the surface roughness of different regions in different orientations was studied by Solberg and Berto [586]. Fig. 133 shows SEMs of upward (45°) and downward (45°) facing surfaces of v-notch specimens built in vertical orientation. The upward and downward facing surface of the notch was measured to be 3.20 $\mu$m and 20.95 $\mu$m respectively. Fig. 134 shows the notched region of a v-notch specimen loaded until 2 × $10^6$ cycles. The geometry in the notch root is shown in Fig. 134b, where the surface is rougher and defects are apparent in the overhanging. There is a fatigue crack as defect in the overhanging region, which is amplified in Fig. 134c. The fatigue crack is growing with about 200 $\mu$m of depth, which may contribute to the break of specimens. Fig. 134d shows the defects of porosity and lack of fusion in the downward facing surface, which have the potential to develop into fatigue cracks.

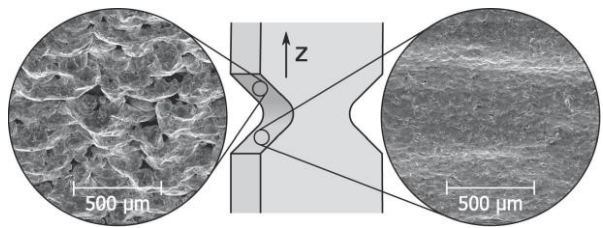

**Fig. 133.** SEM of surface build facing downwards and upwards. Reprinted from [586]. Copyright (2019), with permission from Elsevier.

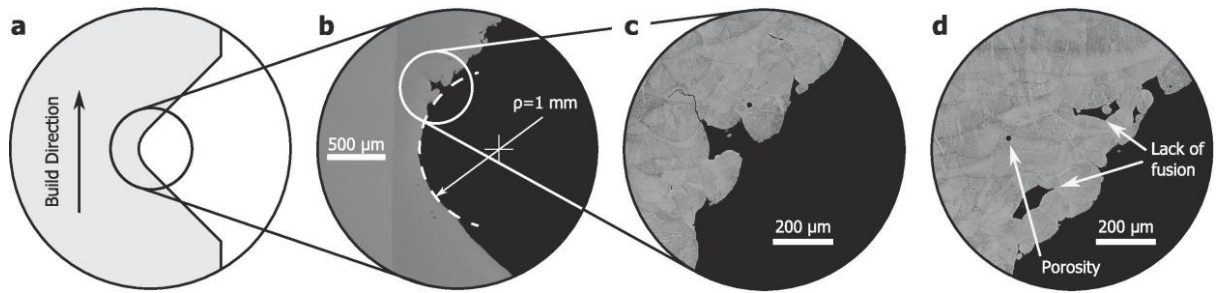

**Fig. 134.** V-notch specimen with 1 mm radius tested until 2 ×$10^6$ cycles; (a) schematic illustration of notch and build direction; (b) optical micrograph of centre plane of specimen, polished; (c) microstructure and defect initiating fatigue crack growth; (d) microstructure and defects in downward facing region. Reprinted from [586]. Copyright (2019), with permission from Elsevier.

#### 7.3.5. Other factors

In addition to the factors mentioned above, much effort need to be devoted to optimize the processing parameters for the final components with high density and optimal mechanical properties. Process parameters have a strong influence on microstructure and surface morphology of components for surface texture and grain generated by powder melting and layer stratification [589]. There are over one hundred kinds of process parameters, most commonly-considered among which include scanning strategy, laser power, hatch distance (distance between two scanning lines), exposure time, layer thickness, and so on.

Scanning strategy is the spatial moving pattern of the energy beam over the surface of each layer, which plays an important role in controlling the grain structure and crystal structure of nickel-based superalloy during the AM processing [590–592]. Mechanical properties of the fabricated part can also be affected [593]. Zhang et al. [594] compared the fatigue properties of SLM IN718 alloy with two scanning strategies, i.e. bidirectional scanning without and with a 90° rotation between the successive layers, suggesting that superior fatigue strength of the former one is attributed to the variation of grain size and texture feature.

Powder size distribution (PSD) can also influence the fatigue properties, but there isn't relevant research about this. The study of Farzadafar [595] showed that the tensile properties of samples printed using the blend or single powder are different. The blend powder made the printed samples show higher strength (YS and UTS), lower ductility (El% and RA%), and less anisotropic behavior. Guzmán-Tapia et al. [596] focused on the pre-corrosion effects on the ultrasonic fatigue endurance failure of N718, as well as the analysis of crack initiation and propagation. The results of ultrasonic fatigue endurance show that pre-corroded specimens present a decrease on fatigue endurance with a factor of 10 approximately, compared with the non-pre-corroded specimens. It is also verified by numerical simulation that pre-corrosion pits make an increase of stress concentration factors, which is at the origin of the drastic reduction in fatigue endurance.

## 8. Inconel 625

Nickel-based superalloy Inconel 625 (IN625) is widely applied in aeronautical, aerospace, chemical, petro-chemical and marine industries. The alloy receives its name from the trademark "Inconel", which belongs to Special Metals Corporation, a company based in New Hartford, USA. IN625 belongs to a family of austenitic nickel-chrome alloys with different compositions. This alloy is considered as a modification of IN718. IN625 has a good combination of yield strength, tensile strength, creep strength, excellent process ability, weldability and good resistance to high temperature corrosion on prolonged exposure to aggressive environments[597–599]. The excellent mechanical and fatigue properties mainly depend upon its specific composition.

The chemical composition of IN625 is listed in Table 9. Molybdenum (Mo) and chromium (Cr) are responsible for a high corrosion resistance and strength properties. Elements such as niobium (Nb) and ferrum (Fe) are used to obtain a higher degree of strength in the material. Other elements including titanium (Ti) and aluminum (Al) are used for refining and for welding purposes [600].

**Table 9.** Chemical composition of IN625 alloy.

| Element | | Ni | Cr | Mo | Nb | Fe | Al | Ti | C | Mn | Si | Co | P | S |
|---|---|---|---|---|---|---|---|---|---|---|---|---|---|---|
| Wt. % | Min | Rem. | 20 | 8 | 3.15 | | | | | | | | | |
| | Max | Rem. | 23 | 10 | 4.15 | 5 | 0.4 | 0.4 | 0.1 | 0.5 | 0.5 | 1 | 0.015 | 0.015 |

IN625 is a solid solution strengthened, face centered cubic (FCC) alloy. The material contains Mo and Nb on its Ni-Cr matrix [601] which is the so-called γ phase showing FCC crystal structure. Fig. 135 depicts that IN625 displays the elongated Ni rich austenitic grains with streams of Nb, Mo rich phases consecutively formed along the grain boundaries. The primary phases that can form include γ″, Laves and δ. Precipitation hardening in this alloy is mainly derived from the precipitation of fine metastable phase γ″-Ni$_3$Nb after annealing over a long period at 550 – 850 °C. Laves phase is common in IN625, which is the intermetallic compound taking the form of (Ni,Cr,Fe)$_2$(Nb,Ti,Mo). The approximate composition of Laves phase can be found in Fig. 136. The δ phase which is also shown in Fig. 136 is known to form in conventional solid solution strengthening when subjected to high temperatures for an extended time period. This alloy may also contain carbides in the form of MC and M$_6$C (rich in Ni, Nb, Mo and silicon (Si)). An example for carbides is presented in Fig. 137.

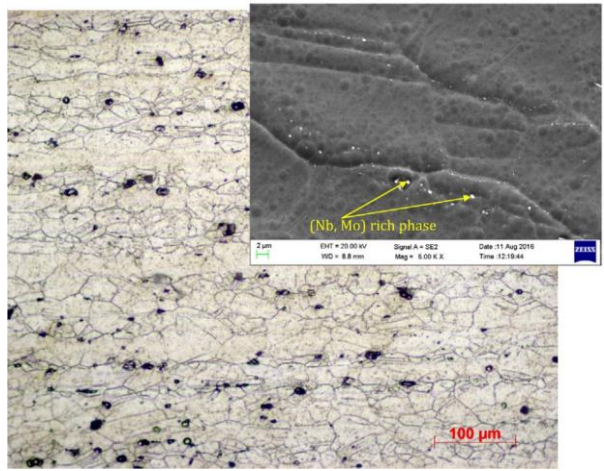

**Fig. 135.** Microstructure of the base IN625. Reprinted from [601]. Copyright(2017), with permission from Elsevier.

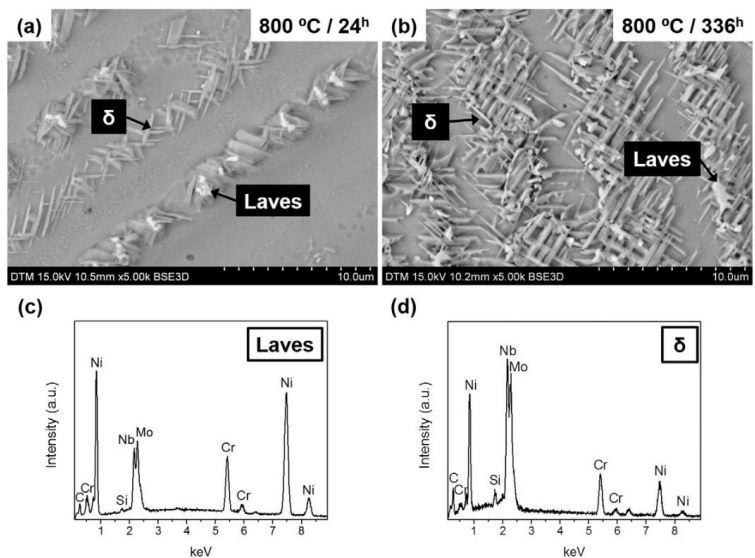

**Fig. 136.** Back scattered electron (BSE) images of the $\delta$-$Ni_3$(Nb, Mo) and the Laves phases formed in the IN625 clads during exposure at 800 C, (a) after 24 h; and (b) after 336 h. energy dispersive X-ray (EDX) spectra (c) of the Laves phases; and (d) of the $\delta$-$Ni_3$(Nb, Mo) phase. Reprinted from [602]. Copyright(2017), with permission from Elsevier.

Most of the laser additively manufacturing (AM) processed parts exhibit columnar dendritic microstructure [603]. The finer microstructure was obtained in the powder bed fusion (PBF) process over the direct energy deposition (DED) [604]. As is shown in Table 10, a mount of mechanical tests on IN625 which were made through Selective Laser Melting (SLM) and Laser Melting Deposition (LMD) have been carried out. The influence of heat treatment was also evaluated. The DED processed samples showed superior mechanical properties to PBF [604]. The preferable texture is <100> along the build direction due to its high creep and low Young modulus in PBF [605].Electron beam melting (EBM) is another form of additive manufactury used for producing complex parts in aerospace and medical implants. The mechanical properties of EBM IN625 are highly dependent on the electron beam parameters [606]. Wire and Arc Additive Manufacturing (WAAM) is an effective technique to produce medium to large-sized components as well. The microstructure of WAAM parts varied with layers from bottom to top owing to the cooling rate. The bottom layer was consisted of cellular grains while the top layer had columnar grains. Mechanical properties changed from bottom to top due to variations in the microstructure. It is depicted in Table 11 that the ultimate tensile strength (UTS) and yield stress (YS) increased with a rise in the deposition travel speed. Compared with parts manufactured by casting, the elongation and YS of the specimens deposited by WAAM are better, with the UTS being slightly lower than the cast parts [607].

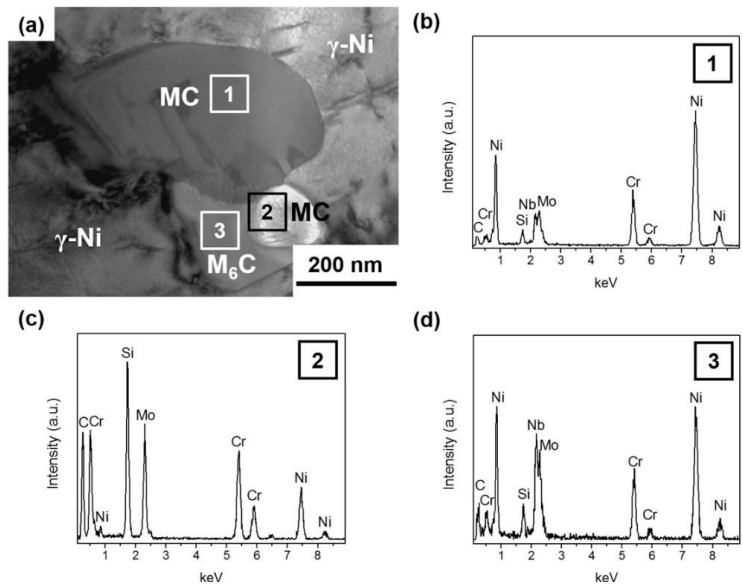

**Fig. 137.** Example of carbides detected at the dendritic boundaries of the γ-Ni crystals of the Inconel 625 clads after heat treatment at 520 °C for 48 h. (a) bright-field (BF) transmission electron microscopy (TEM) image. EDX spectra of the three carbides: (b) MC, mark 1; (c) MC, mark 2; and (d) $M_6C$, mark 3. Reprinted from [602]. Copyright(2017), with permission from Elsevier.

Due to the microstructure and mechanical properties that are effected by a large number of factors such as process parameters, post-pocessing, the fatigue property of AM IN625 is worth discussing. The S-N data is shown in Fig. 138 while the crack propagation plot is shown in Fig. 139.

A novel AM process, commercially known as MELD, can produce a refined microstructure resulting in higher mechanical properties compared with other fusion-based processes. The progress using high-shear deformation to achieve solid-state depositions can produce fully-dense, near-net shape IN625 AM components from solid or powder feedstock. The MELD progress is shown in Fig. 140. Avery [608] compared the MELD samples to the feedstock samples. The MELD samples exhibited generally greater fatigue resistance compared to feedstock ones. The fatigue peoperties were even better than cast, laser consolidation (LC) IN625.

Ganesh [609] investigated fatigue crack propagation behaviour of laser rapid manufactured (LRMed) IN625 compact tension specimens whose thicknesses were 12 and 25 mm. The fabrication of the CT test specimens involved machining of a V-groove in a block of type 304L stainless steel (SS), followed by laser deposition of IN625 in the V-grooved region. The fabrication process is shown in Fig. 141. The specimens

Table 10. Mechanical properties of IN625 fabricated through SLM and LMD processes [604].

|  | Young Modulus (GPa) | Yield Stress (MPa) | Ultimate Tensile Strength (MPa) | Elongation (%) |
|---|---|---|---|---|
| Wrought Samples | 184 ± 16 | 482 ± 42 | 955 ± 6 | 41 ± 1 |
| Samples Processed by SLM |  |  |  |  |
| As-built (AS) | 145 ± 4 | 652 ± 10 | 925 ± 13 | 32 ± 3 |
| AS + 900°C/1h | 142 ± 11 | 567 ± 15 | 869 ± 7 | 38 ± 1 |
| AS + 1100°C/1h | 114 ± 8 | 409 ± 14 | 886 ± 11 | 56 ± 5 |
| Samples Processed by LMD |  |  |  |  |
| AS | 223 ± 24 | 723 ± 23 | 1073 ± 5 | 26 ± 2 |
| AS + 900°C/1h | 224 ± 19 | 654 ± 15 | 1084 ± 2 | 27 ± 2 |
| AS + 1100°C/1h | 213 ± 22 | 532 ± 22 | 991 ± 13 | 43 ± 1 |

Table 11. Tensile properties of WAAM IN625 under different deposition travel speeds [607].

| Sample | Travel Speed (mm/s) | UTS(MPa) | YS(MPa) | Elongation(%) |
|---|---|---|---|---|
| 1 | 8 | 647.9 | 376.9 | 46.5 |
| 2 | 9 | 675.6 | 391.4 | 44.45 |
| 3 | 10 | 687.7 | 400.8 | 43 |
| Casting |  | 710 | 350 | 40 |

with different thicknesses exhibited similar FCG rate in the Paris' regime.

## 8.1. Manufacturing parameters

The energy source used during the AM process is divided through the laser beam, electron beam, and electric arc. When it comes to IN625, the Laser-based AM technique is promising and widely employed. Selective laser melting (SLM) is commonly accessed in the field of L-PBF. Each step in the AM experiment counts. The fatigue resistance of AM IN625 has something to do with all kinds of parameters including the building orientation, the scanning velocity and so on.

### 8.1.1. Building directions

Because of the various sizes of grains and the existence of columnar grains, the fatigue property of AM IN625 shows great anisotropy. For a vertically build sample, the crack growth is perpendicular to the grain growth. In this case, the grain boundary spacing is very small and the crack path is tortuous. On the other hand, for horizontally build sample, the crack growth is parallel to the grain orientation. The crack growth resistance seems to be lower and a smoother crack surface should be observed. What's more, the porosity and the voids in the samples also play an important role in the crack propagation. As a result, the higher fatigue strength of vertical samples may not be confirmed.

Anam [610] fabricated forty cylindrical bars in horizontal (XY) and vertical (Z) direction. Half of the samples were heat treated at 1038 C for 1 hour in an argon-filled furnace for stress relief and water quenched to room temperature. The cylindrical bars were then machined in the gauge section to form the dog-bone shape of the fatigue specimen. Fatigue tests were conducted at room temperature. A sinusoidal load with a frequency of 40 Hz was applied while the stress ratio was R = -1 and the maximum stress levels ranged between 350 to 750 MPa. Compared with the Z samples, the XY samples have longer lifetime under the same stress no matter whether the samples experienced the heat treatment and machining procedure. In the Z samples, the direction of the un-melted interlayer porosity is perpendicular to the loading axis. However, the interlayer porosity of the XY samples is parallel to the loading axis. The schematics representing is shown in Fig. 142. As a result, the stress concentration in Z samples is significantly higher compared to XY samples. The cracks in Z samples tend to propagate earlier.

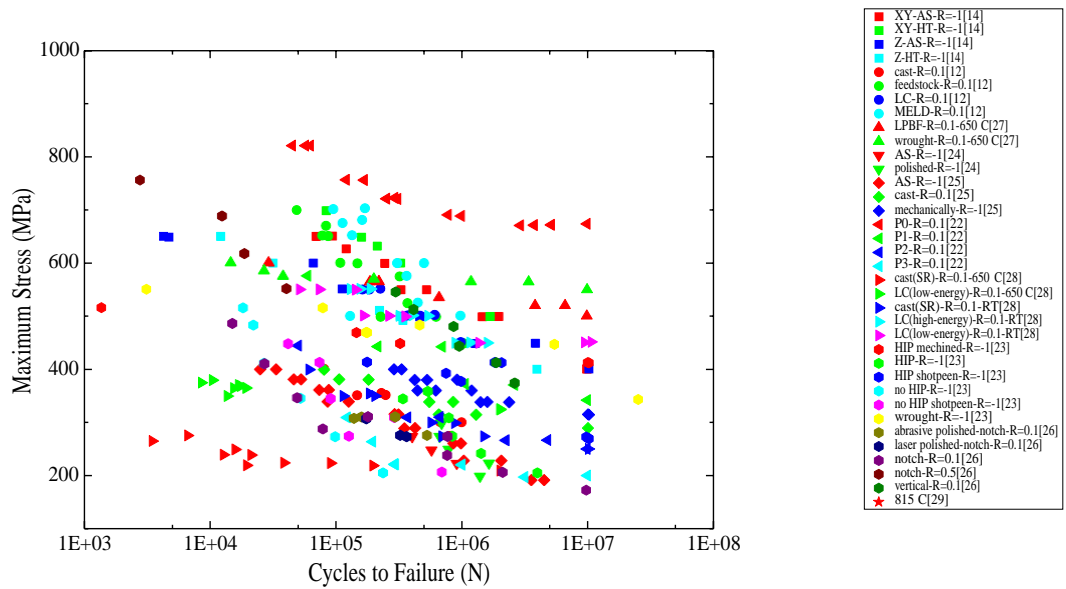

**Fig. 138.** S-N plot of IN625.

Abela [612] applied two L-PBF systems in the experiment that fabricated the fatigue samples. They are the Concept M2 and the Renishaw AM250. Six builds of samples were made by the Concept M2 while other five builds were made by the Renishaw AM250. The Z and XY samples were both tested in the as-is condition and the fatigue strength at $2 \times 10^6$ cycles was recorded. For the builds fabricated by both L-PBF systems, a significant difference of fatigue strengths was found between the two orientations and there are better fatigue strength properties in the Z-direction. It seems that there is a larger anisotropy in fatigue strengths between Z and XY specimens for the Renishaw AM250 than the Concept M2 machine.

Poulin [613–615] has conducted abundant experiments on the fatigue crack propagation behavior of AM IN625. The compact tension (CT) specimens were manufactured with four different configurations that are observed in Fig. 143. From the four crack propagation specimen configurations, three configurations corresponded to build orientations of 0°, 45° and 90° and contained starter notches oriented parallel to the build plane. The fourth configuration was obtained in 90° specimens by orienting their starter notches perpendicular to the build orientation (90°⊥ specimen). The compact crack propagation specimens were manufactured according to ASTM E647-13a standard, with a characteristic length (W) of 38mm and a thickness (b) of 9.5 mm. In all cases, to help with the visual observations of cracks, the side surfaces were polished with silicon carbide paper up to grit 1200 before each test. Stress relief (SR) annealing was performed on the build plate with specimens (870 °C for 1 h, under argon atmosphere). Half of the specimens were tested in this state (SR), while the other half were additionally subjected to hot isostatic pressing (HIP): 1120 °C and 100 MPa for 4 h, under argon atmosphere. All the fatigue crack propagation tests are operated under a stress ratio of $R$ = 0.1. The resistance to crack propagation of the L-PBF IN625 specimens was found to be similar to that of a wrought alloy [613]. For the SR samples, the threshold values are varied because of the building orientations. However, in the Paris region, the most of the SR L-PBF specimens showed no crack orientation dependence. When it comes to HIP L-PBF specimens, the crack propagation behavior was almost insensitive to the build and the crack orientations.

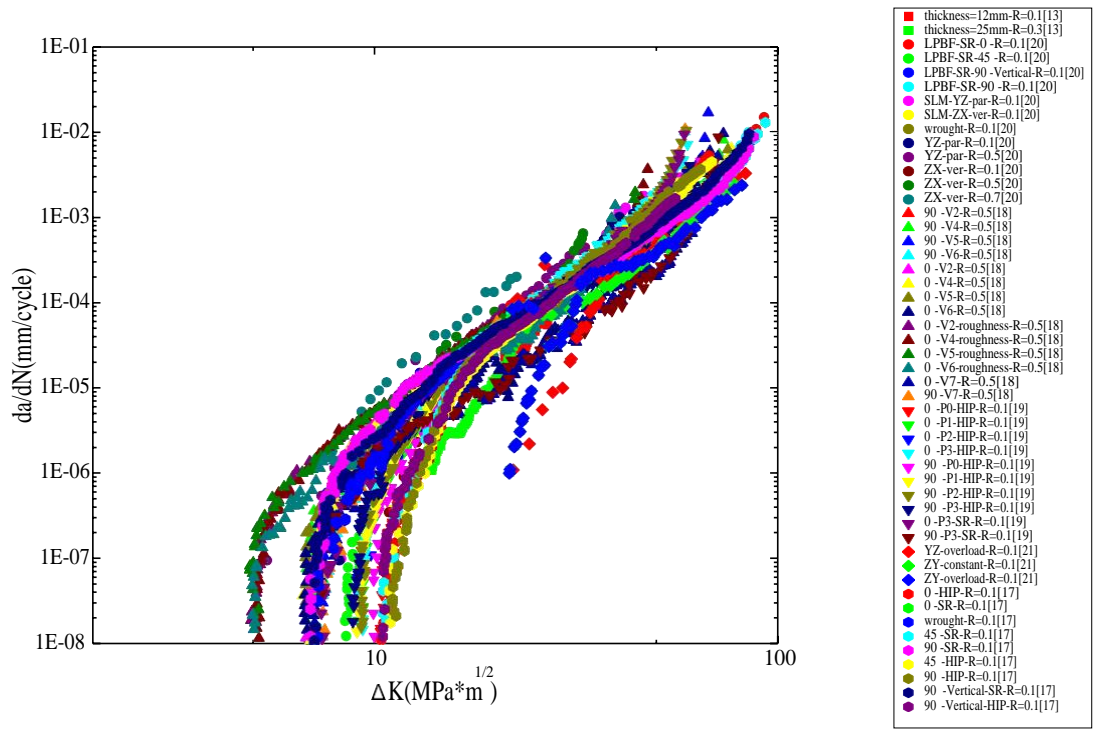

**Fig. 139.** Crack propagation plot of IN625.

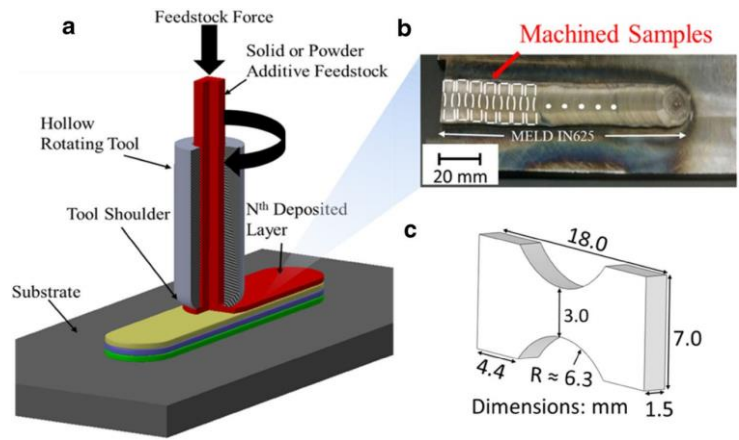

**Fig. 140.** (a) Schematic of the MELD process with a solid rod extruded through the hollow stirring tool. (b) As-deposited IN625 sample on a HY80 substrate. (c) Fatigue specimen geometry used for both as-deposited IN625 and feedstock IN625 specimens. Reprinted from [608]. Copyright(2018), with permission from Springer.

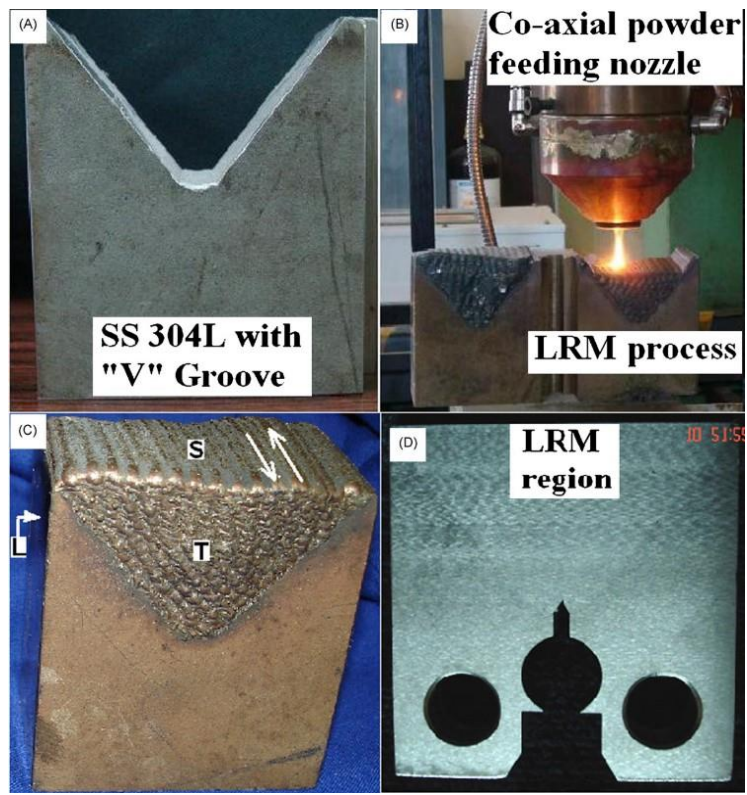

**Fig. 141.** Sequence of steps followed for laser rapid manufacturing (LRM) of compact tension test samples of Inconel 625. (A): Stainless steel 304L block with "V" groove; (B) LRM process with co-axial powder-feeding nozzle; (C) Stainless steel block with laser deposit on central region. Arrows indicating the direction of laser scanning in each layer. S represents the plane parallel to laser scanning direction and top surface; L represents the plane parallel the laser scanning direction and normal to top surface (S-plane); T represents the transverse plane normal to top surface (S) and the laser scanning direction. (D): Compact tension specimen extracted from laser-deposited block shown in (C). Reprinted from [609]. Copyright(2010), with permission from Elsevier.

Hu [616] has carried out many investigations on the building directions of AM IN625. The fatigue crack growth resistance in the near threshold region of SLM Inconel 625 after solution annealing heat treatment was better than that of the wrought material [616]. The fatigue crack growth (FCG) rate was highly affected by the material orientation in the near threshold region while the orientation factor could be neglected in the Paris region. In the near threshold region, the threshold values of ZX-ver samples are higher compared to those of YZ-par samples. In the Paris region, there is a slight difference between the two samples and it's difficult to confirm which sample has the lower FCG rate.

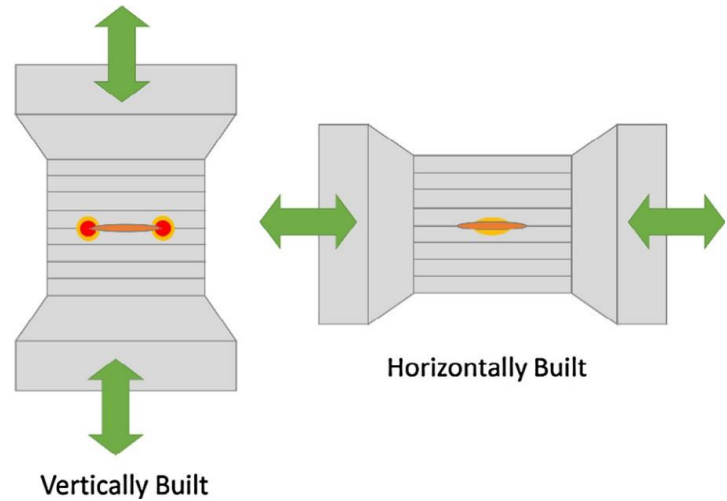

**Fig. 142.** Schematics representing the orientation of an un-melted region formed during fabrication of vertical and horizontal specimens with respect to the loading direction and the resultant stress concentrations. Reprinted from [611]. Copyright(2017), with permission from Elsevier.

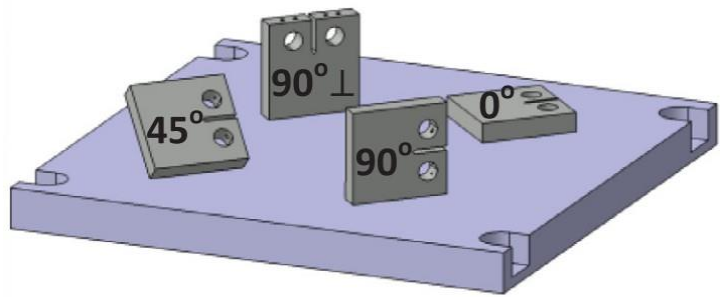

**Fig. 143.** Schematic representation of CT specimens with different build orientations. Reprinted from [613]. Copyright(2018), with permission from Elsevier.

Ren [617] studied the the single tensile overload influence on crack growth of SLM IN625. At the same $\Delta K$ value, the FCG rate of ZY specimen which is the same as the ZX sample in Hu's work was quicker than that of YZ specimen. Under the same overload ratio $R_{OL}$ = 1.55, the crack growth retardation span of YZ specimen was large than that of ZY specimen. The scatter influence of orientation can be ignored in Paris region. In crack growth retardation region, the different microstructure in two directions would cause the slight difference of $\Delta a_{OL}$.

### 8.1.2. Scanning speeds

During the L-PBF fabrication, the scanning speed will affect the porosity of the samples which has a great influence on the fatigue property.

Table 12. Laser exposure parameters [615].

| Parameter set | Scanning Speed (mm/s) | Power (W) | Hatching Space (mm) | Energy Density (J/mm$^3$) | Expected Porosity (%) |
|---|---|---|---|---|---|
| $P_0$ | 960 | 285 | 0.11 | 67.5 | ≤ 0.1 |
| $P_1$ | 1440 | 285 | 0.11 | 45 | 0.3 |
| $P_2$ | 1680 | 285 | 0.11 | 38.6 | 0.9 |
| $P_3$ | 1920 | 285 | 0.11 | 33.7 | 2.7 |

Poulin [614, 615, 618] applied different scanning speeds to achieve samples with various porosities. As the increase of sacnning speed, the lower porosity is obtained with the higher energy density. The L-PBF parameters are shown in Table 12. Under the stress ratio of $R$ = 0.5, there appears to be a limited effect of porosity on the long crack propagation behavior in the near threshold region. It is also the case in the Paris regime, with the exception of the 0°-oriented $P_2$ and $P_4$ specimens [614]. As for the $R$ = 0.1, the $P_0$ specimens have a bit higher threshold values [615]. The rest behaviour is the same as the case of $R$ = 0.5. There appears to be an interaction of the crack front with defects which translates to a lower, while more erratic, growth rate in the Paris region. What's more, the higher the porosity, the lower the SIF of the specimens at final fracture. The fatigue lifetime of L-PBF IN625 made with different scanning speeds was tested as well. The tests were conducted at a frequency of 30 Hz using a stress ratio of $R$ = 0.1 applied in a sinusoidal waveform. The number of cycles to failure was also recorded with a runout set at 10$^7$ cycles. The samples with slower sacnning speed have higher fatigue strength due to the lower porosity.

## 8.2. Post-processing

Post-processing methods are very effective for improving the fatigue performance of AM samples by improving their surface finish, decreasing porosity and eliminating tensile residual stresses. Common post-processing methods include heat treatment (HT), HIP, SR and polishing.

Witkin [619] first studied the fatigue lifetime of SLM IN625 under different post-processing. All samples are SLM 625 in which the fatigue stress axis is the vertical direction of SLM growth and his tests presented are for a stress ratio of $R$ = −1. It can be inferred from the data that HIP machined samples have the best fatigue properties which are similar to wrought IN625. Shot-peening and HIP can both improve the fatigue behavior of IN625.

Koutiri [620] also investigated the influence of the surface roughness on the high cycle fatigue (HCF) of IN625 with results suggesting an endurance limit around 200 MPa and that polishing the surfaces to remove the influence of surface roughness generally increases the fatigue life. Despite the poor surface roughness of as-built specimen, the polished and the as-built fatigue results exhibit similar fatigue strength.

The heat treatment did not have a large impact on fatigue performance in Anam's experiment [610]. The HT XY samples tested showed improved fatigue lives compared to as-built samples. Fatigue life of Z samples after heat treatment were found lower compared to as-built samples tested below 500 MPa. The defects contained in the samples tends to be the main factor that influences the fatigue life.

Mostafaei [621] evaluated the fatigue properties of binder-jet 3D-printed IN625. Standard fatigue specimens were printed and sintered, then half of the samples were mechanically ground, while the other half were left in their as-sintered state. The samples were fatigue tested in air under a stress-controlled, sinusoidal waveform with a tension-compression condition at stress ratio $R$ = −1 with 5 Hz and 7 Hz frequency (for the as-sintered and mechanically ground samples, respectively. The unground as-sintered sample would show poorer performance than the ground cast samples. However, the mechanically ground samples had superior fatigue life compared to both the cast alloy and as-sintered samples.

In another study carried out by Witkin [622] in 2019, SLM IN625 was treated with both an abrasive polishing method and laser surface remelting. Both methods led to improvements in surface roughness, but these did not lead to improvements in fatigue properties of SLM 625.

As for the fatigue crack growth behaviour, Poulin [615] compared the HIP spicemens to the SR ones (before HIP). The HIP samples had lower porosity. HIP reduces the defect and size of defects and increases the threshold for long fatigue crack growth. In the Paris region, there is no difference between the FCG rate of HIP samples and that of SR samples.

## 8.3. Working temperature

High temperature tensile and high cycle fatigue properties of HIP L-PBF IN625 were investigated and then compared to conventional wrought IN625 [623]. HIP L-PBF and wrought alloys show similar microstructural characteristics in average grain size, defects, fraction of special grain boundaries and grain morphologies. However, $Al_2O_3$ and TiN phase was found (without carbides) in HIP L-PBF IN625 and various stoichiometry of carbides (without $Al_2O_3$) were identified in wrought IN625. The YS, UTS and elongation at fracture of both alloys have similar values at room temperature. At 650 °C, elongation at fracture of HIP L-PBF 625 significantly decreases compared to that of wrought IN625. HIP L-PBF IN625 also has a lower high cycle fatigue limit of 500 MPa (at $10^7$ cycles without fracture) and shorter life in low stress conditions compared to wrought IN625. The relatively higher sulfur content in HIP L-PBF IN625 (33 ppm) is suggested as the reason for lower fatigue endurance at 650 °C. The L-PBF, which uses powder feedstock, is vulnerable to impurity concentration increases, elemental contamination control is critical in the initial powder and L-PBF process.

Theriault [624] evaluated the fatigue behavior of LC IN625 at both room temperature and 650 °C and compared with wrought and investment cast IN625 materials. At room temperature, LC IN625 in the SR condition tested in the build direction has a fatigue resistance higher than the investment cast material but lower than the wrought material. There was no significant difference in fatigue resistance between the LC samples produced from low energy parameters and high energy parameters. At 650 °C, the fatigue resistance of all three forms of IN625 material decreased but the trend was the same as the room temperature results. The LC IN625 produced with low-energy parameters in the SR condition tested in the build direction has a fatigue resistance higher than the investment cast material but lower than the wrought material.

Zhang [625] tested low-cycle fatigue and creep-fatigue properties of laser-welded SLM IN625 at 815 °C. It can be found that fatigue and creep-fatigue life decrease as the increase in stress level and have a linear relationship with logarithmic coordinate. The overall fatigue life of welded specimens is higher than that of non-welded specimens, especially at 300 MPa.

## 8.4. Others Inconel alloys

### 8.4.1. Inconel 939

Inconel 939 (IN939) is wildly used in the gas turbine industry for the construction of baldes and vanes. It has also been considered for building large aircraft engine structures as well as turbine airfoils. The alloy shows excellent oxidation resistance and high creep strength at elevated temperatures. IN939 exhibits higher corrosion resistance than Inconel 738LC alloy. IN939 has good mechanical properties which is capable of operating at temperatures up to 850°C for long periods of time. The chemical composition of IN939 is listed in Table 13. The excellent performance is caused by coherent $\gamma'$-Ni$_3$(Al,Ti) particles present after solution annealing and single-step ageing of the cast material. The $\gamma'$-Ni$_3$(Al,Ti) particles have a high amount of Ti and Al. The high level of $\gamma'$-forming elements (Al and Ti) as well as different carbides-forming elements(e.g. W, Ta, Ti, Nb) can affect the weldability of this alloy.

Table 13. Chemical composition of IN939 alloy.

| Element | | Ni | Cr | Co | W | Ti | Nb | Ta | Zr |
|---|---|---|---|---|---|---|---|---|---|
| Wt. % | Min | 46.4 | 21.4 | 16 | 1.5 | 3.2 | 0.7 | 0.9 | |
| | Max | 56.3 | 23.4 | 20 | 2.5 | 4.3 | 1.3 | 1.9 | 0.2 |

Philpott [626] studied the differences between cast and SLM IN939 as a function of heat treatment and subsequent ageing. In the SLM IN939, as there are no carbides in the as-produced state, carbides form very quickly during the first step of the heat treatment, and then continue to precipitate up until the end of the heat treatment. Moreover, the microstructure undergoes almost full recrystallization, with the initial high-aspect ratio 'stressed' grains becoming larger, more equiaxed, 'relaxed' grains.

Marchese [627] evaluated the effect of the process parameter on the densification and porosity of L-PBF IN939 alloy. The most appropriate hatching distance and scanning speed have been determined. The microstructure revealed elongated grains along the building direction (z-axis) characterized by a strong texture along the ⟨001⟩ orientation. Inside the grains, a network of dendritic/cellular architectures with sub-micrometric phases mainly located along the grain boundaries and interdendritic areas were observed.

Kanagarajah [628] tested the fatigue life of SLM IN939 under the stress amplitude of about 700 MPa. The experiment has also been carried out with the aged samples and under the elevated temperature of 750 °C. The results which is shown in Table 14 reveal that the aging process does nothing good to the fatigue life of SLM IN939. In its initial condition SLM IN939 shows significantly better performance under cyclic loading at room temperature than the cast material while the aged SLM IN939 has far shorter lifetime compared to the aged cast IN939. At the elevated temperature, the fatigue property of SLM IN939 is much worse than the cast ones.

Table 14. Fatigue lives of different conditions tested at RT and at 750 °C [628].

| Condition | SLM,as-built | SLM,aged | Cast,as-cast | Cast, aged |
|---|---|---|---|---|
| RT | 4702 | 1598 | 313 | 2677 |
| 750 °C | 209 | 73 | 230 | 272 |

### 8.4.2. Hastelloy X

Hastelloy X is a solid solution strengthened, nickel-based superalloy that is widely applied in gas turbine engines because of its exceptional combination of oxidation resistance, formability and high-temperature strength. The alloy is also called K536 in China. Hastelloy X is resistant to stress-strain cracking in petrochemical applications. Because of Hastelloy X's increased hardness and tensile strength, however, fabricating the customised engineering parts used in the aerospace and marine industries is a challenge when using traditional manufacturing techniques. Hastelloy X alloy has excellent forming and welding characteristics. It can be forged and cold worked because of its good ductility, even at temperatures as high as 1200 °C. It can also be welded by both manual and automatic methods including shielded metal arc, gas tungsten arc, and gas metal arc. The chemical composition of Hastelloy X is listed in Table 15. Fig. 144 shows a scanning electron microscopy (SEM) image of the Hastelloy X particles having an average size of less than 50 μm. The S-N data is shown in Fig. 145.

**Table 15.** Chemical composition of Hastelloy X.

| Element | | Ni | Cr | Fe | Mo | Co | W | Mn | Si | C | P | S |
|---|---|---|---|---|---|---|---|---|---|---|---|---|
| Wt. % | Min | Rem. | 20.5 | 17 | 8 | 0.5 | 0.2 | | | 0.05 | | |
| | Max | Rem. | 23 | 20 | 10 | 2.5 | 1 | 1 | 1 | 0.15 | 0.04 | 0.03 |

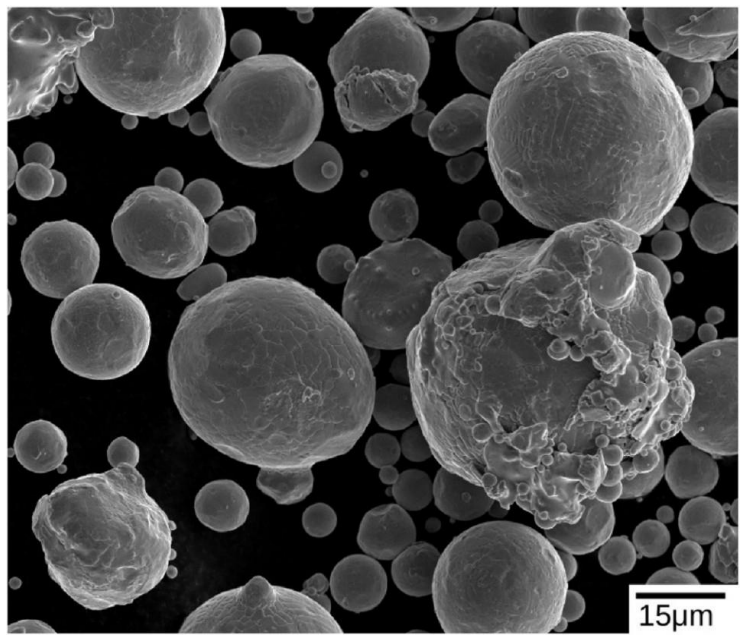

**Fig. 144.** SEM image of pre-alloyed gas atomized Hastelloy-X powder. Reprinted from [629]. Copyright(2017), with permission from Elsevier.

Wang [630] considered the fatigue life under two loading conditions that the material may suffer. The specimen geometry and the loading modes for four-point bend and tension-tension fatigue tests are shown in Fig. 146. Both tests were conducted under a stress ratio of $R = 0.1$ with a sinusoidal waveform at frequencies of 110 Hz for four-point bend and 117 Hz for tension-tension. Samples were built in the vertical and horizontal directions. Half of ones for four-point bend were given a HIP treatment. It's obvious that the data shows great anisotropy of SLM Hastelloy X at high stress regimes above 600 MPa while there is no difference between samples built under two directions at low stress regimes below 600 MPa. What's more, the HIP process dramatically improved the four-point bend fatigue property. When it comes to the tension-tension test, the HIP treatment slightly improves the fatigue limit.

Han [631] also made investigations on the effect of HIP process. However, a much lower loading frequency than that used by Wang [630] is adopted. The frequency was set to 10 Hz and the stress ratio was maintained

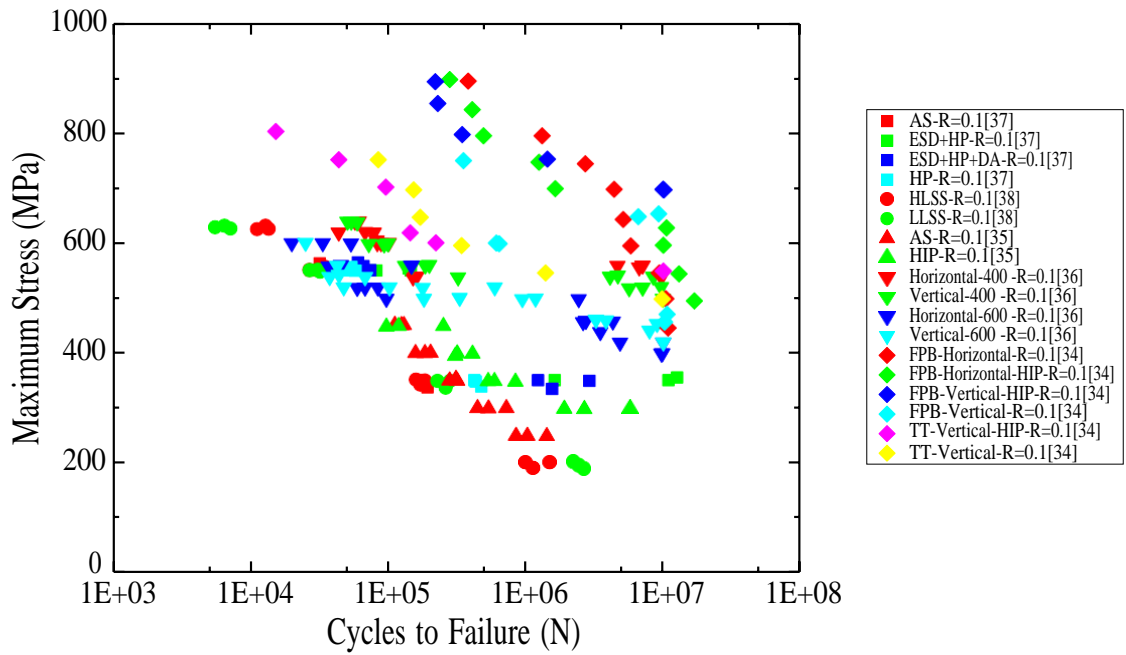

**Fig. 145.** S-N plot of Hastelloy-X.

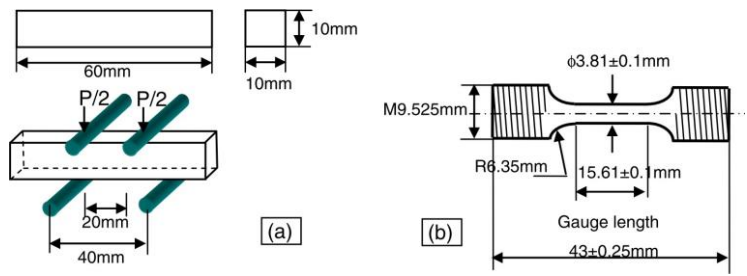

**Fig. 146.** Specimen geometry for (a) four-point bend and (b) tension-tension fatigue tests. Reprinted from [630]. Copyright(2012), with permission from Springer.

at 0.1 as well. The HIP specimens exhibited significant improvement in fatigue life.

Jiao [632] compared the fatigue property of horizontal and vertical samples. A sine wave loading was conducted, with the frequency of about 100 Hz. The stress ratio is $R = 0.1$. There is no difference between the two samples. The influence of elevated temperatures was studied as well. The higher elevated temperature made the poor fatigue property.

More post-pocessing approaches were studied by Enrique [633]. All samples were divided into four groups, as-built L-PBF Hastelloy X, the hammer peened (HP), ESD and hammer peened (ESD+HP), and ESD and hammer peened with a direct aging heat treatment (ESD+HP+DA). Two stress levels, 550 MPa for low cycle fatigue (LCF) and 350 MPa for high cycle fatigue (HCF), were chosen for comparison, and three samples per each group were tested at each stress level. A frequency of 5 Hz was used for all samples except the post-processed samples tested at HCF conditions. These samples were tested at a frequency of 5 Hz until 10$^6$ cycles, and then switched to 30 Hz due to the long test durations. Under both stress levels, the post-processing had an improvement on the fatigue life compared to the as-built samples. ESD+HP samples had the best performance while the enhancement of HP samples is the least obvious. The rise of fatigue life under low stress level is greater than that under high stress level.

Esmaeilizadeh [634] devoted efforts on the effect of laser scanning speed during the L-PBF process. 850 mm/s was considered as low laser scanning speed (L-LSS) while 1150 mm/s was high laser scanning speed (H-LSS). The fatigue tests were carried out under four stress levels which were 200, 350, 550, 625 MPa. Under each level three samples were tested. Basquin equation was fitted to the S-N data. The H-LSS samples shown longer fatigue life under the stress level of 625 MPa while the L-LSS ones had better fatigue property. There were no clear difference between the fatigue life of the two kinds of samples when the stress level was 550 MPa.

Saarimäki [635] explored the fatigue crack propagation behaviour of Hastelloy X. The infuence of building direction, heat treatment and the dwell time was taken into consideration. The direction setting is depicted in Fig. 147. All loading conditions are listed in Table 16. The FCG results are depicted in Fig. 148. The introduction of a longer dwell-time accelerated crack propagation. The S direction tended to be a strong direcion built under which the FCG rate was lower. The higher HT temperature would increase the fatigue crack resistance.

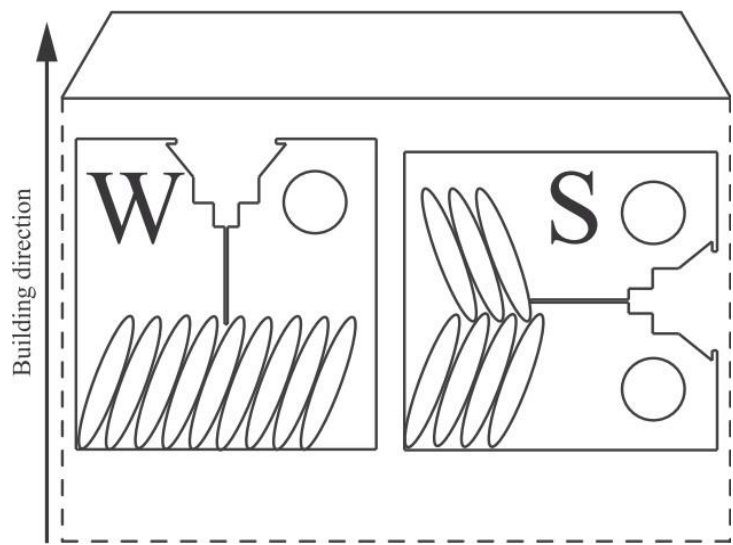

**Fig. 147.** An illustration of the sample cutting orientation. Reprinted from [635]. Copyright(2018), with permission from Elsevier.

### 8.4.3. CMSX-4

The second-generation single-crystal Ni-base superalloy CMSX-4 has excellent thermal and mechanical stability at high temperature, which allows it to operate at temperatures close to its melting point. The extraordinary properties are associated with its complex chemical composition, which contains high amount of refractory elements such as W, Ta and Re, as well as the microstructure, consisting of coherent secondary $\gamma'$ precipitates embedded in the solid-solution $\gamma$ matrix. The chemical composition of CMSX-4 is listed in Table 17. CMSX-4 exhibits the typical microstructural heterogeneities associated with dendritic growth during casting, such as strong segregation between dendritic and interdendritic regions, large eutectic areas, topologically close-packed (TCP) phases as well as porosity. To reduce these heterogeneities, complex and long heat treatments are required to achieve a homogeneous chemical composition, to dissolve large γ/γ′ eutectics and to develop a coherent γ/γ′ microstructure with about 70 vol.% of γ′ [636].

**Table 16.** Summary of loading conditions with the constants:temperature 700 °C and load ratio $R$ = 0.05 [635].

| HT | Direction | Dwell-time (s) | ΔP (N) | $a_n$ |
|---|---|---|---|---|
| 1177 °C | W | 2160 | 4000 | 8.5 |
| 1177 °C | S | 2160 | 4500 | 8.5 |
| 1177 °C | W | 90 | 2500 | 8.5 |
| 1177 °C | S | 90 | 4500 | 8.5 |
| 900 °C | W | 2160 | 2200 | 8.5 |
| 900 °C | S | 2160 | 3500 | 8.5 |
| 900 °C | W | 90 | 1500 | 8.5 |
| 900 °C | S | 90 | 2700 | 8.5 |
| AS | W | 2160 | 1000 | 10.5 |
| AS | S | 2160 | 2100 | 10.5 |
| AS | W | 90 | 1000 | 10.5 |
| AS | S | 90 | 1500 | 10.5 |
| cast | | 2160 | 4500 | 10.5 |

**Table 17.** Chemical composition of CMSX-4 alloy.

| Element | | Ni | Co | Cr | W | Mo | Ta | Re | Ti | Hf |
|---|---|---|---|---|---|---|---|---|---|---|
| Wt. % | Min | 61 | 9 | 5.5 | 5 | 0.3 | 5.5 | 2 | 0.5 | |
| | Max | 72 | 11 | 7.5 | 7 | 0.9 | 6.5 | 4 | 1.5 | 0.2 |

Körner [637] took investigation into the low cycle fatigue property of SEBM CMSX-4. Load-controlled LCF tests have been carried out until rupture at a temperature of 950 °C, with a frequency of 0.25 Hz (triangular signal shape) in tension-tension loading ($R$ = 0.62 − −0.65). The samples were tested in the as-built state and after heat treatment. The results are compared with results deduced from conventional cast and heat-treated material. Table 18 shows the results. HT SEBM CMSX-4 specimens achieved a lifetime as high or even higher than the cast specimens. The as-built SEBM specimen had the worst LCF properties.

### 8.4.4. Inconel 100

Among nickel-base superalloys, IN100 is used mainly as jet engine parts such as turbine blades and wheels operating in the intermediate temperature regime. Due to the high Ti/Al content (>11%), the two major phases present in IN100 are ordered γ′ ($Ni_3Al$-type) phase embedded in a fcc solid-solution γ-Ni matrix [638]. The chemical composition of IN100 is listed in Table 19. The total Al + Ti content for IN100 is nearly 11.0 wt.%. Hence, it is considered to be one of the most difficult-to-weld alloys whose Al + Ti contents are over 4.0 wt.%. Acharya [639] foucued on the AM IN100 component repair and found appropriate process parameters for better microhardness and less crack deposition compared to cast substrate. Basak [640] did the similar work and studied the microstructure of scanning laser epitaxy (SLE) IN100.

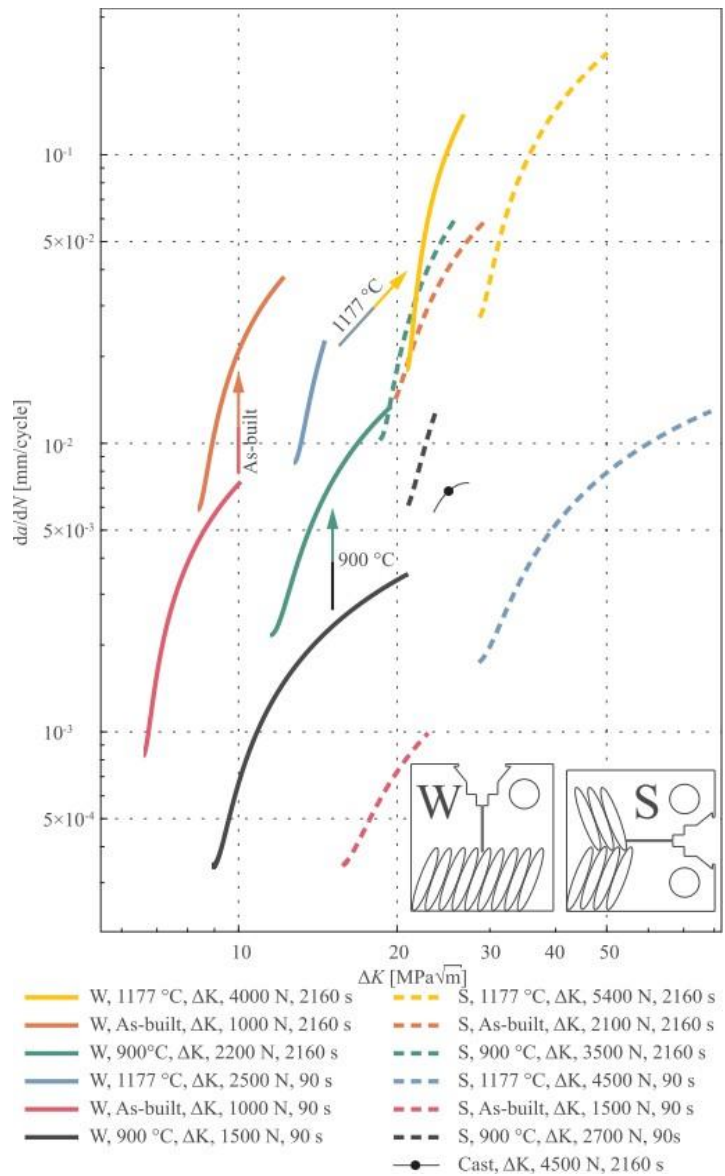

**Fig. 148.** Crack propagation curves of all samples. Reprinted from [635]. Copyright(2018), with permission from Elsevier.

Weng [641] took efforts in the high cycle vibration fatigue (HCVF) behavior of micro-laser aided additive manufacturing (micro-LAAM) IN100. The post heat treatment (solution treatment and aging, STA) samples were tested as well. The results were also compared to the cast IN100. It is obvious that the cast IN100 shows better HCVF performance than the as-built and STAed specimens. And the as-built IN100 exhibits slightly longer fatigue life than the STAed IN100, especially at high acceleration (30 g).

**Table 18.** Fatigue lives of CMSX-4 at 950 °C [637].

| Condition | Mean Stress (MPa) | $N_f$ |
|---|---|---|
| cast | 550 | 2765 |
| cast | 590 | 2018 |
| SEBM,as-built | 550 | 908 |
| SEBM,HT | 550 | 3341 |
| SEBM,HT | 590 | 2047 |

**Table 19.** Chemical composition of IN100 alloy.

| Element | | Ni | Co | Cr | Mo | Fe | Al | Ti | V | C | Si | S | B | Mn | Zr |
|---|---|---|---|---|---|---|---|---|---|---|---|---|---|---|---|
| Wt. % | Min | Rem. | 13 | 8 | 2 | | 5 | 4.5 | 0.7 | 0.15 | | | 0.01 | | 0.03 |
| | Max | Rem. | 17 | 11 | 4 | 1 | 6 | 5 | 1.2 | 0.2 | 0.2 | 0.015 | 0.02 | 0.2 | 0.09 |

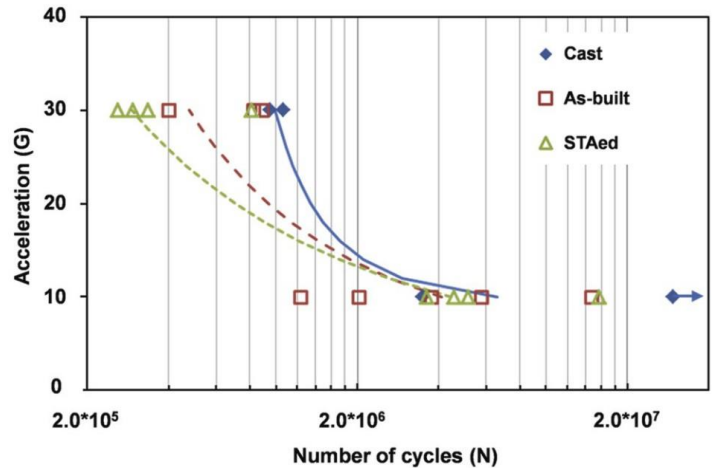

**Fig. 149.** HCVF test results of IN100. Reprinted from [641]. Copyright(2020), with permission from Elsevier.

## 9. Mg-based alloys

Magnesium is one of the most highly important lightweight metals used in automotive or aerospace application to reduce vehicle or aircraft weight [26]. Also, it is gaining acceptance in the orthopedic implant sector. Currently, by means of the AM technologies, the Mg alloys becomes a better choice for orthopedic implants with superior qualities like ductility, strength and biodegradability. There is very little stress shielding produced by biodegradable magnesium, so secondary surgery is not required. As well as, AM Magnesium alloy is getting popularity in structural, aviation and automobile sectors and they can be a good replacement for expensive Aluminum [642], Cobalt and Titanium based alloys. Allavikutty et al. [643] discussed the importance of Mg in biomedical applications, feasibility of manufacturing Mg and its alloys through AM technology, challenges in microstructural engineering to achieve improved mechanical properties and corrosion behaviour. As a new sector, very little work has been done on evaluating mechanical properties and fatigue properties of the materials.

## 9.1. Fatigue properties

The investigation about fatigue properties of AM Mg-based alloys focuses on the porous scaffolds. The small stress range and HCF are striking characteristics for AM Mg-based alloys during working. Li et al. [644] found that fatigue crack propagation took place transgranularly in SLMed porous magnesium alloy scaffolds based on diamond unit cells. At the macro-scale, the cracks tend to initiate at strut junctions where tensile stress is concentrated, especially for the struts positioned on the periphery of the specimens. Deng et al. [645] fabricated a high-strength L-PBFed Mg-10Gd-3Y-1Zn-0.4Zr (GWZ1031K) alloy and systematically studied the microstructure and mechanical properties of the as built, LPBF-T5, LPBF-T4, and LPBF-T6 states (T4, T5, T6 is a post-processing approach). Interestingly, a high-strength GWZ1031K alloy with YS of 316 MPa and UTS of 400 MPa was first fabricated by L-PBF technology with favourable post-processing. Fine and homogeneous equiaxed grains (4.1 $\mu$m) and eutectic phase were found in the as built GWZ1031K alloy, which results in the excellent mechanical properties. In a word, compared with cast and cast-T6 magnesium alloys (AM50hp and AZ91hp [646]), the LPBF-T6 GWZ1031K parts have obviously higher tensile strengths and slightly higher ductility.

In 2021, Wang et al. [647] designed a SLMed biodegradable metallic scaffolds (see Fig. 150b). The applied stress has an evident influence on the fatigue life of all scaffolds, where the fatigue life increased with the decrease in normalized applied stress or absolute applied stress. The substantial influence of the porous structure on the fatigue properties of the AM Mg-based scaffolds is observed, as shown in Fig. 150a. Compared to the B scaffolds with stochastic structure and D scaffolds with beam-based lattice structure, the sheet-based G scaffold had the highest fatigue strength. As a result, the fatigue strength at $10^6$ cycles was 2.0 MPa for B scaffold, 4.3 MPa for D scaffold, and 15.1 MPa for G scaffold, respectively. In addition, fatigue crack initiation and distribution after $10^6$ cycles was observed. However, there was no fatigue failure for all scaffolds after fatigue run-out, and local cracks were observed. For fatigue failure mode, G scaffolds presented lay-wise deformation characteristics without apparent integral fracture, which is an explanation that G scaffolds have superior fatigue performance. In contrast, the B scaffold exhibited fragile fatigue and the D scaffold exhibited shear fracture mode with nearly 45° fracture bands. Further, for dynamic biodegradation behavior of the three AM Mg scaffolds, there was a strong correlation between the geometrical design of the three scaffolds and the degradation behavior, including the structural loss of the Mg matrix and the volume increments of biodegradable products, as shown in Fig. 150c.

Among traditional manufacturing techniques, the die casting process developed voids and pores causing adverse effect on fatigue properties. The combination of AM technologies and Mg alloys brings the AM Mg alloys with understanding mechanical properties and superior fatigue performance [643, 645, 647]. The microtexture of the implants or scaffolds can be controlled and engineered via the AM fabrication process. Besides, LSP decreased stress shielding of samples based on Mg alloy and led to the formation of micro-dents on the surface and decelerated corrosion process, and improved fatigue life.

## 10. High entropy alloy

## 10.1. Introduction

High-entropy alloys (HEAs) are "novel materials" created using the multi-principal element blending approach, and they have reviewed a lot of attention in the materials science world in the last decade [648, 649]. In accordance with AM Ti-6Al-4V and AlSi10Mg, AM HEAs are regarded a future material and are hopeful to exhibits superior mechanical and fatigue properties owing to the presence of multi-principal elements [19], as shown in Fig. 151. Therefore, HEAs have been introduced as promising advanced materials with vast potential applications in aerospace [650] and anticipated to replace traditional metallic materials. Furthermore, various approaches for HEAs, such as the development of interstitial HEAs, oxide dispersion strengthened HEAs [651], non-equiatomic HEAs, multi-phase HEAs [652], metastable HEAs [653, 654], and medium-entropy alloys (MEAs) [655], have recently been attempted with the aim of improving fatigue properties of AMed HEAs. More studies need to be conducted to further investigate the fatigue behavior as well as prediction models for HEAs [656].

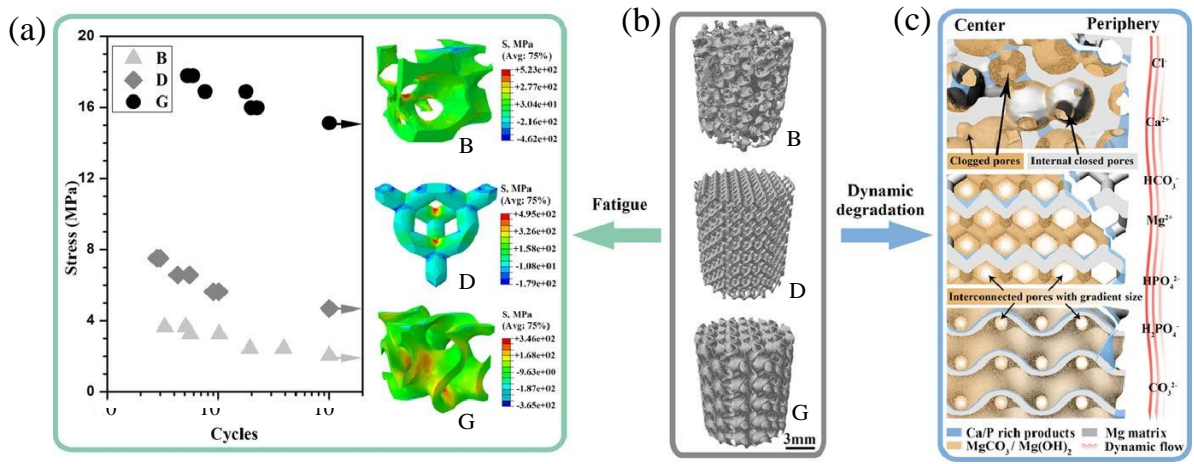

**Fig. 150.** (a) Fatigue S-N curves of three L-PBFed Mg scaffolds. (b) Three different AM Mg scaffolds (from top to bottom: biomimetic (B), diamond (D), and sheet-based gyroid (G)). (c) Schematic illustration of the dynamic degradation behavior of the 3 AM Mg scaffolds. Reprinted from [647], Copyright(2021), with permission from Elsevier.

The additive manufacturing of HEAs has also gradually emerged in recent years with the rise of additive manufacturing technology. The most prominent AM techniques for HEAs are DLD and SLM. Some reports have been published on SEBM of HEAs [19, 657, 658]. The process of DLD may also be known as LMD, DMD, LENS and laser cladding (see Fig. 151). Similarly, LBM, DMLS, laser metal fusion and industrial 3D printing are some frequently used synonyms for SLM process [659].

## 10.2. Fatigue life or fatigue limit

Fatigue occurs in almost all applications, and in order to maintain it, materials must necessarily be manufactured. HEAs go through low and high fatigue cycles in many studies aimed at quantifying their properties and performance. An increase in the strain amplitude leads to the formation of higher misorientation, which leads to the accumulation of damage near the grain boundaries and a decrease in the fatigue life at low cycles. Fatigue life decreases as the applied stress amplitude increases prior to cracking. Nano-twinning is a prime deformation mechanism and occurs during fatigue crack propagation, which results in strengthening [660, 661]. Another study pointed out that defect in HEAs affects fatigue resistance, but the nano-twinning effect strengthens it and increases fatigue endurance. The defects of the HEA-fatigue samples influence the fatigue resistance. Nano-twinning bolsters HEAs' fatigue resistance with higher fatigue-endurance limits. HEAs show greater fatigue-endurance limits, fatigue-crack-growth properties and fatigue

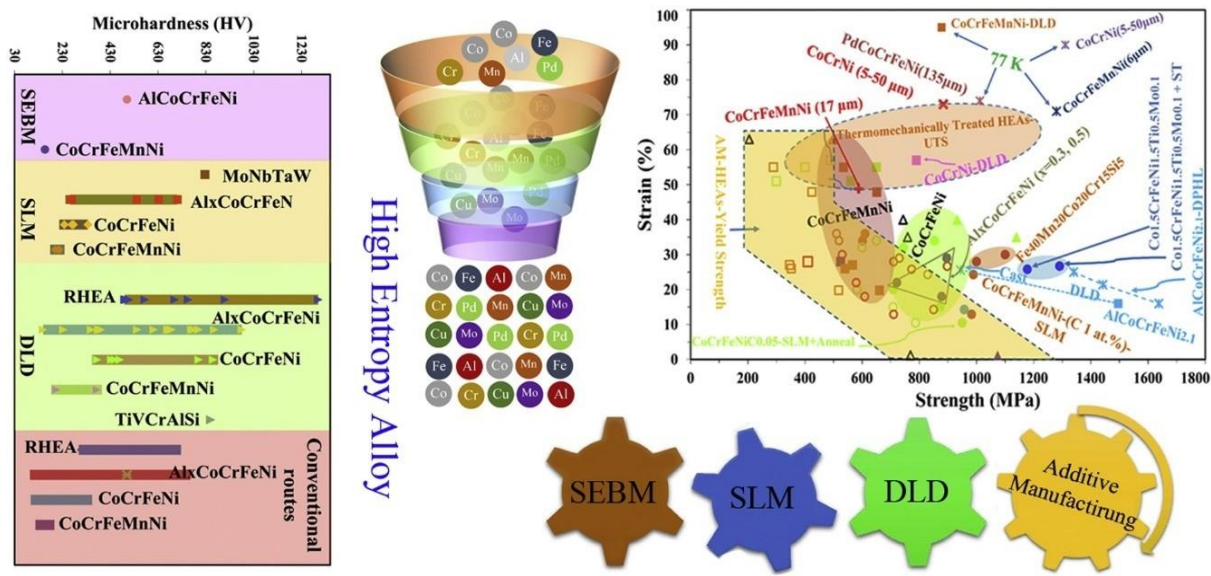

**Fig. 151.** Illustrations of additive manufacture and mechanical properties of HEA. Reprinted from [19], Copyright(2021), with permission from Elsevier.

ratio [662].

Metastability is regarded as an effective approach to enhance fatigue properties of the material which involves localized crack nucleation and growth. Liu et al. [652] found that multi-pass friction stir processing (FSP) of FeMnCrCoSi HEA resulted in substantial grain refinement and phase redistribution throughout the nugget, thereby significantly increasing its YS and UTS. This grain-refined HEA further demonstrated unconventional fatigue response owing to extreme metastability of the $\gamma$ matrix. $\gamma$ to $\varepsilon$ transformation provides localized work hardening in the vicinity of nucleated fatigue cracks and thus delays crack propagation by transformation induced crack retardation, thereby providing good fatigue resistance to these new alloys [653, 663]. Further, They designed a Cu containing FeMnCoCrSi high entropy alloy for improved fatigue strength [663]. Ultra fine grained microstructure in combination with localized martensitic transformation is a critical factor to improved fatigue properties. It is important to point out that Cu or other proper element could improve the mechanical performance and fatigue resistance of HEAs with an additive. Thapliyal et al. [664] printed the transformation induced plasticity assisted $Fe_{38.5}Mn_{20}Co_{20}Cr_{15}Si_5Cu_{1.5}$ L-PBFed HEA (Cu-HEA). The Pre-alloyed Cu-HEA exhibited highest tensile strength (–1235 MPa) among additively manufactured HEAs as well as a good ductility of 17.2%.

The HCF properties and deformation behavior of SLM equiatomic CoCrFeMnNi HEAs, strengthened by in-situ formed oxides, were investigated in Ref [651]. As shown in Fig. 152, the fatigue limit of SLM CoCrFeMnNi (570 MPa) and SLM defect-free CoCrFeMnNi HEA is higher than the one of homogenized (conventional casting + hot rolling + heat treatment) HEA (280 MPa). A major reason for the outstanding fatigue resistance of the SLM-built HEA is its unique microstructures (heterogeneous grain structures, dislocation networks, and in-situ oxides) and the twin deformations caused by cyclic loading. In addition, the defect-free samples have superior fatigue properties, the results further show that eliminating defect or LOF though HT/HIP is a effective way to improve fatigue performance for AM metals. Agrawal et al. [654] investigated fatigue response of metastable $Fe_40Mn_{20}Co_{20}Cr_{15}Si_5$ high entropy alloy (CS-HEA) to obtain high fatigue-resistance with L-PBF, as shown in Fig. 152. As a result, the ab L-PBF exhibited excellent fatigue properties despite its pores since a decrease in mean free path resulted from crack branching along slip bands and strain accommodation via twinning. Noticeably, The fatigue properties of most AM (e.g. HEA SLMed CoCrFeMnNif [651], and machined annealed CrFeCoNi MEA [655]) are superior to the cast samples.

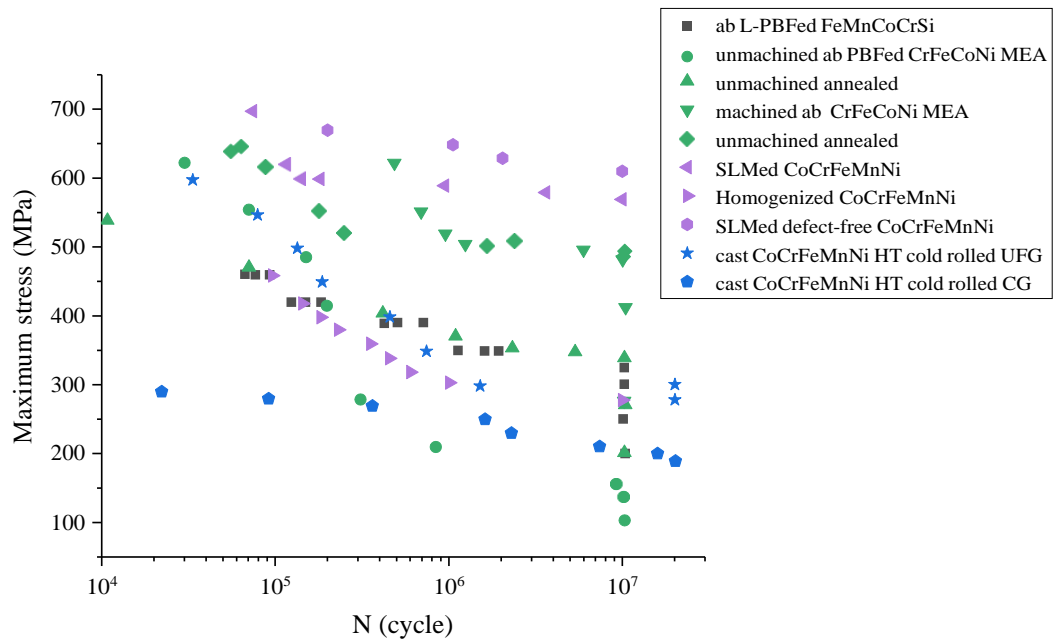

**Fig. 152.** The S-N curve of AMed HEA. Data from ab L-PBFed FeMnCoCrSi [654], machined annealed CrFeCoNi MEA [655], SLM CoCrFeMnNi [651], cast CoCrFeMnNi HEA [665].

### 10.2.1. Post-processing

Several post-treatment processes have been developed to alleviate the defects and to improve the fatigue properties of the AMed components. Heat treatment [666] is generally used to relieve residual tensile stresses to improve fatigue properties and ductility, it cannot remove pores and cracks from specimens. Furthermore, various surface-treatment technologies, such as ultrasonic nanocrystal surface modification [209] and LSP [193] have been employed to remove the tensile residual stress and refine grains on the surface, both of which are beneficial for improving the mechanical and fatigue properties.

Inevitably, heat treatment, as a stronger post-processing, is applied to enhance the mechanical and fatigue performance on AM HEAs. The effect of heat treatment and HIP on the phase evolution and mechanical properties of L-PBF CrFeCoNi MEA [655, 667], $Al_x$CoCrFeNi HEAs [668], and atomized AlCoCrFeNi HEAs [669] are investigated in detail. As shown in Fig. 152, the annealed L-PBF CrFeCoNi MEA sample with unmachined as-built exhibit higher fatigue limit compared with the unannealed one [655, 667]. The as-built samples without annealing are able to withstand $10^7$ cycles at maximum fatigue stress of 414 MPa after machining, in contrast to the unmachined samples showing maximum fatigue stress of 138 MPa. With or without machining, the heat treatment improves the fatigue life of the alloy under HCF conditions. Besides, the heat treatment provides more contribution to the fatigue strengths from 138 MPa up to 336 MPa after annealing in the unmachined samples compared with the machining process. Also, the machining process plays a critical role in the fatigue properties of the L-PBFed CrFeCoNi MEA. [655] by removing of near-surface defects in the samples, which increases the total fatigue life by the complication of the crack initiation. Note that annealing does not improve the fatigue characteristics of the material due to the formation of the $\sigma$ phase that facilitates crack propagation.

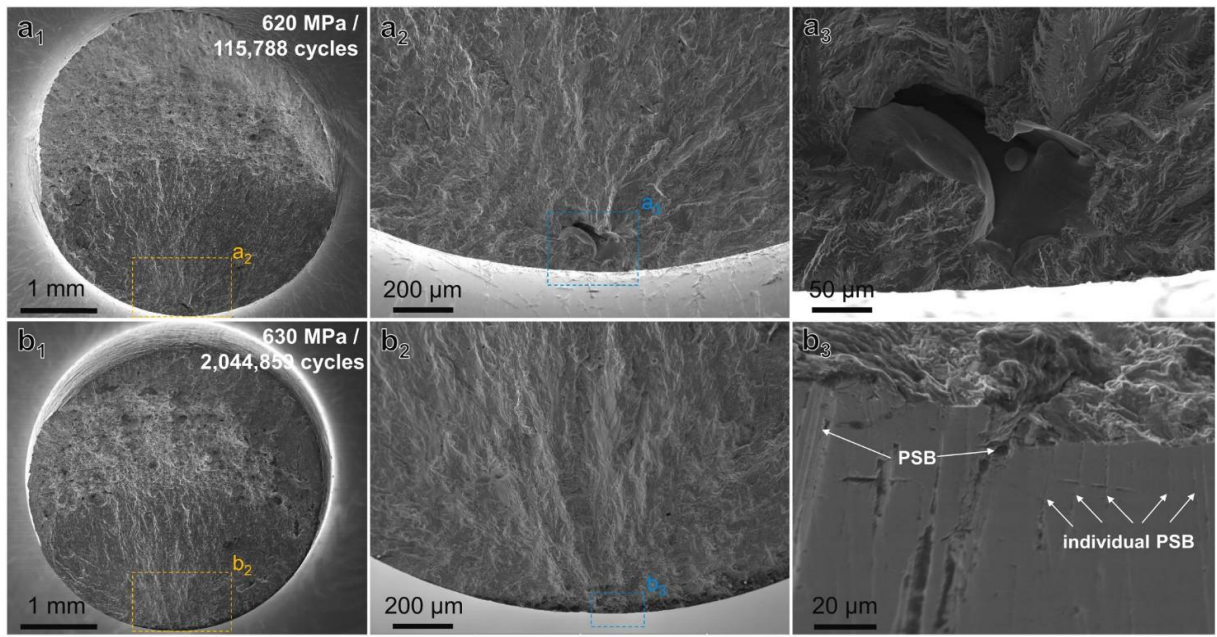

**Fig. 153.** Fatigue fractographies showing the different failure behavior between samples with ($a_1$-$a_3$) and without ($b_1$-$b_3$) defects. Reprinted from [651], Copyright(2021), with permission from Elsevier.

Further, the microstructure, residual stress, and mechanical properties of LAMed FeCrCoMnNi HEA were investigated [666]. As a result, there is a single FCC solid solution in the AB specimen, which exhibits both epitaxially grown dendrite columnar grain and equiaxed grain microstructures. After heat treatment, the dendritic microstructure of AB sample evolves into recrystallized grain structure of sample with heat treated at 1100 °C. The HTed specimen exhibits much higher ductility than that of the AB specimen without distinct sacrifice of strength. As well as, eliminating or relaxing residual tensile stress in AM materials can enhance mechanical and fatigue properties such as tensile and fatigue resistance.

LSP is an effective method of inducing severe plastic deformation (SPD) on the surface layer of materials and can generate high-amplitude compressive residual stress [193] to a depth of several millimeters while counteracting some of the tensile residual stress in the surface layer of specimens generated by LAM [670]. The SPD can close the existing pores in LAM-fabricated specimens, thereby decreasing the stress intensity factors and improving the tensile and fatigue performances. Some studies [223, 228, 670, 671] also agreed that LSP is an effective surface modification technique for further improving the integrated fatigue properties of components fabricated via AM. There is a consensus that the LSP can refine microstructure, suppress residual stresses, and delay crack propagation. The LSP strengthening trend and mechanism for AM HEA is similar to that for AM Ti-6Al-4V.

## 10.3. Fatigue crack growth

It is noteworthy that the research on the fatigue crack growth (long crack) of HEAs fabricated by AM is scarce, and most investigate crack initiation and propagation during fatigue tests. Similarly to the most AM metals, defect during manufacturing process has a critical influence on fatigue performance and failure behavior. Fig. 153 shows the fatigue failure behavior of SLM CoCrFeMnNi HEA samples with ($a_1$-$a_3$) and without ($b_1$-$b_3$) defects. The $\sigma_{max}$ is 620 MPa and $N_f$ = 115,788 cycles for $a_1$-$a_3$, and $\sigma_{max}$ = 630 MPa and $N_f$ = 2,044,859 cycles for $b_1$-$b_3$. Both cases show fatigue propagation-final failure behavior resulting in HCF after crack initiation on the surface. The un-melted particles or defect near the fractured surface are observed in the samples with defect, which leads to the lower $N_f$ = 115,788 cycles under $\sigma_{max}$ = 620 MPa. In contrast, no un-melted particles or defects were found in Fig. 153($b_1$-$b_3$). Besides, the image (Fig. 153$b_3$) revealed that the fracture was initiated by a persistent slip band (PSB), which is generally considered to be the mechanism whereby the HCF of AM metals is initiated. As a result, the initiation of fatigue cracks occurs at the particle boundaries and the resistance to HCF significantly decreases as the presence of un-melted particles generated during the SLM process [651], rather than the PSB. Previous research on the fatigue behavior of AM alloys has confirmed that processing defects such as pores are the potential crack initiation sites [672] (In718). Local work hardening enabled by $\gamma$ to $\varepsilon$ transformation near the pores, a decrease in mean free path of the crack due to crack branching along the slip bands, and strain accommodation via twinning enabled excellent fatigue properties to the as-printed CS-HEA [654].

### 10.3.1. Processing parameters

Scanning strategies can effectively regulate grain morphology, texture and the distribution of residual stress to control density and morphology of cracks [673], further regulating the fatigue properties of AM HEA. Jin et al. [674] investigated the significant contribution of the scan strategies to grain arrangements and grain dimensions, thus leading to noticeable effects on fatigue crack growth for L-PBF CrMnFeCoNi HEA. Fig. 154 shows the maximum tensile stress responses of the HEA fabricated by the considered scan strategies during cyclic loading. During the first five cycles, all samples displayed cyclic hardening, followed by cyclic softening. There is some influence of the print strategies on the cyclic plastic deformation, but a negligible effect on the fatigue life of the samples. In particular, the highest resistance crack propagation and the largest fatigue life was seen in the meander scan strategy with 0° rotation because of the most columnar grains and the smallest spacing of grain boundaries along the crack propagation path. In addition, no sub-surface pores were observed in the machined specimen thanks to the surface machining. This resulted in a significant improvement in the fatigue performance of a machined sample fabricated with the chessboard 67° scan, as shown by an increase in maximum stress response and by a longer fatigue life. SLMed equiatomic CoCrFeMnNi HEA sample [673] with 0-scan exhibits the largest crack density while 67-scan shows the least. Furthermore, fracture surface investigations revealed transgranular quasi-cleavage fractures to be more dominant for porous CoCrFeMnNi HEA using binder jetting additive manufacturing.

It is important to point out that intergranular hot cracks occur during SLM CoCrFeNi HEA regardless of processing parameters and are present throughout the builds. Sun et al. [675] found out that hot cracking tends to occur beyond a characteristic depression pressure limit due to severe residual stress caused by the formation of large grains. Similarly, the SLM (0°) AlCrCuFeNi specimen exhibits quasi-cleavage fracture while the SLM (90°) specimen shows intergranular fracture [676]. Wang et al. [677] found that SLM printed AlCoCrCuFeNi HEA shows high crack sensitivity with faced centered cubic (FCC) and body-centered cubic (BCC) phases. High-angle grain boundary, segregation of Cu element and the misfit between BCC and FCC phase account for the crack formation during SLM process. Bahadur et al. [660] investigated micro-mechanisms of microstructural damage due to low cycle fatigue in CoCuFeMnNi HEAs.

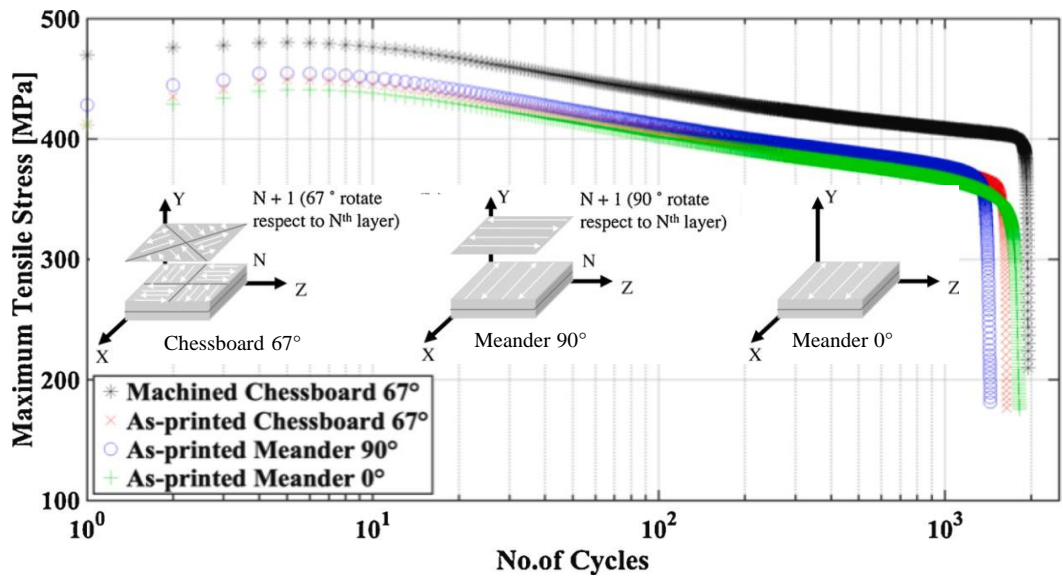

**Fig. 154.** Cyclic plastic deformation response of L-PBF CrMnFeCoNi HEA samples fabricated using three printing strategies and with machined and AB. Reprinted from [674], Copyright(2020), with permission from Elsevier.

### 10.3.2. Post-Processing

Sun et al. [678] found that within a certain range of compositions, such as $Al_{0.5}CoCrFeNi$, the hot crack density was drastically decreased. Contrary to the previous works which either try to reduce solidification range or introduce grain refinement, the current work presents a new approach of employing segregation engineering to alter the residual stress states at the interdendritic and grain boundary regions and consequently prevent hot crack. This result mainly contributes to that the inherent residual strain is accommodated and transformed from a maximum 0.006 tensile strain in CoCrFeNi to a compressive strain of 0.001 in $Al_{0.5}CoCrFeNi$. In addition, Peng et al. [679] observed that cracks easily initiate and propagate along with the matrix FCC and precipitated B2 phase boundaries in LMDed $Al_{0.3}CoCrFeNi$ HEA. These cracks damage the tensile properties of the LMDed samples leading to their poor ductility during tensile tests. The ductility and strength of Fe-Co-Cr-Ni-Al HEA present contradictory trends with the increasing Al element content. Further, the LMD FeCoCrNi+FeCoCrNiAl-laminated HEA [680] exhibits excellent ductility and strength and superior crack resistance. For $Al_{0.3}CoCrFeNi$, HIP considerably increased the tensile ductility and changed the fracture mode from grain boundary cracking along the loading direction to cracking perpendicular to the loading direction (necking), which was attributed to the dissolution of grain boundary particles. The HIP-processed SLMed $Al_{0.6}CoCrFeNi$ HEA exhibited brittle fracture in tension due to the precipitation of coarse B2 phases in the interphase and grain boundaries. For $Al_{0.85}CoCrFeNi$, HIP coarsened the microstructure and led to the precipitation of brittle σ-phase and consequently remarkable loss of ductility [193].

Niu et al. [681] found that large thermal stress is generated due to the high cooling rate during the SLM process, resulting in the formation of cracks in SLMed equimolar CoCrFeMnNi HEA. Most probably the hot cracking cannot be inhibited by simply adjusting SLM parameters. Moreover, compared with the conventionally cast HEAs, the CoCrFeNiMn fabricated using LAAM possesses significantly higher yield strength (518 MPa) and ultimate tensile strength (660 MPa) [682]. The strengthening effect is attributed to finer grains and could be explained quantitatively through grain boundary strengthening. On the other hand, Amar et al. [683] found that tensile properties of them could be adjusted by controlling the quantity of the TiC addition. The CrMnFeCoNi HEA with 5 wt% TiC addition has exhibited 723 MPa tensile strength and 32% tensile strain. Nitrogen induced heterogeneousy [684] structures also shows the work in increasing both strength and ductilit of SLMed FeCoNiCrN HEA.

## 11. Summary

In summary, we have reviewed the state-of-art progress on the fatigue properties of AM metals in terms of fatigue life, fatigue limit and fatigue crack growth. The mentioned AM metals in this work include the usual titanium alloys (e.g., TC4, TC17, TC18, TA15, and CP Ti), aluminum alloys (e.g., AlSi10Mg, AlSi12, AlSi7Mg, AlMgScZr, and AlMgMn), stainless steel (e.g., 316L, and 17-4H), nickel-based alloys (e.g., Inconel 718, 625 and others), mg-based alloys, and high entropy alloys. Although AM continues to show the potential for full-scale production of customized and/or complex components, the mechanical and fatigue behavior, and thus, trustworthiness of these components are not yet clear. To understand fully the processing-material-performance relationships for AM metals, the effects of particle, geometry, manufacturing processing parameters (e.g., AM technologies, laser power and speed, built orientation, and scanning strategies), post-processing (e.g., LP, SP, LSP, machined, and USMAT), HT, HIP, and high temperature on the fatigue properties for AM metals are investigated systematically.

Metal additive manufacturing technologies, built direction, scanning strategies, powder quality will have a direct impact on the fatigue properties of AM metals. On the contrary, by investigating the influence law and optimizing the printing strategy, the fatigue performance of AM metal can be improved, the service life can be prolonged and the damage tolerance can be reached. Furthermore, the current variability of properties (controlled by microstructure, residual stress, defects, etc.) within and between builds in one machine and across different machines and techniques, as well as the presence of process-induced defects and location/orientation-dependent properties, limits the more widespread use of these processing techniques for fracture-critical applications. In general, damage evolution of AM metals under cyclic loading conditions is directly affected by impurities sourced from the AM process itself. Among the many different sources of damage evolution under cyclic loadings, voids are the major life limiting factor and the most dominant mechanism for fatigue crack initiation in AM metals. Variations in location, shape, and size of voids resulted from various external manufacturing factors are found to be the main reason for the large scatter in the AM fatigue data. The HIP process can be used to improve the durability and HCF performance of AM parts by fusing un-melted particles, decreasing the voids size and smoothening their sharp angles, and even closing some voids. However, it should be noted that employing the HIP process in order to improve the fatigue resistance may not necessarily lead to the same outcome for different AM materials.

Compared with the traditional casting or forging metals, the fatigue performance of the most AM metals is reduced because of the porosity, bad surface roughness and inferior microstructure. Certainly, the fatigue performance of AM Ti-6Al-4V parts subjected to the post-processing can reach levels comparable to, or even better than, those of conventional Ti-6Al-4V [15, 23, 162, 177, 685]. Herein, the post-processing methods, including HT, HIP, LP, LSP, SP, machined, and USMAT/SMAT, play a dominant role in fatigue properties during the numerous manufacturing factors (e.g., manufacturing parameters, particle, geometry, etc.), which can improve the fatigue performance resulting from the elimination of porosity by HIP, the decreasing of surface roughness by various surface treatments, or the improvement of microstructure by HT.

A great scatter in fatigue properties is observed during the post-processing, the combination or optimization between numerous post-processing methods is a great idea to design and further improve fatigue performance for AM metals. In addition, owing to the various forming features in AM and the effect mechanism of the post-processing, the application of the post-processing should be careful. For instance, the HIP has a positive effect on fatigue life in EBM Ti-6Al-4V, but it does not work on the SLM Ti-6Al-4V with same pretreatment. Also, the cracks initiate from the surface and fatigue life is sensitive to surface roughness, thus the surface treatment is essential for AM metals to advance fatigue life. The machined AM components have lower surface roughness and the machined method is always a effective way to improve fatigue performance. The LSP, LP, and SP can repair the damaged parts or even refine the microstructure in influence zone, but the resulting large residual stress has a significant effect on fatigue properties. Then the novel USMAT/SMAT shows a good potential to decrease surface roughness without residual stress and increase fatigue life of AM metals.

The fatigue properties of AM metals are temperature-dependent. Aim at the high-temperature work environment, especially for the application in the aeroengine, the fatigue performance under high temperature is investigated. The HCF performance of the low-temperature heat treatment condition specimen almost coincides with that of the high-temperature one for L-PBF Ti-6Al-4V [37]. As well as, the fatigue life is only 25% of that at room temperature when the test temperature reaches 400 °C for AlSi10Mg, similar to AlSi7Mg [371], the high temperature largely limits the fatigue life of AM metals. Whereas the high temperature makes $\beta''$-$Mg_2Si$ transformed to the $\beta'$-$Mg_2Si$, activating dislocations and reducing the threshold of dislocation climbing, and eventually give rise to cyclic softening [334, 343]. Otherwise, The LDED IN718 displays a similar fatigue performance in both room temperature and high temperature conditions, which also similar to that of the wrought counterpart [569]. The fatigue resistance of SLM IN625 [625] and LC IN625 [624] decreases with the increasing of temperature.

Notoriously, the most significant factors driving the F&DT of AM metals are the surface roughness, porosity-related defects, and microstructure. Elimination of both the surface roughness by machining and porosity by HIP yielded statistically improved fatigue resistance. Also, it is demonstrated that as compared with the present F&DT evaluations for traditionally manufactured materials, anisotropy of material resistance to fatigue is of great significance for AM materials [237]. Both FCG driving-force and resistance are of great significance in F&DT evaluations for AM metals due to their unique microstructure features, while the resistance to FCG is quasi-isotropic in traditionally processed materials and thus be omitted in the present F&DT evaluation methods. This may be one of important considerations if alternatives to conventional qualification methods for AM metals must be found.

The fatigue behavior of AM metals has been the subject of extensive experimental research over the last decade. Further research is needed to enable more accurate and reliable methods for estimating fatigue life for AM parts [686]. Literatures [687, 688] developed a computational framework, integrating discrete element method, phase-field simulation, crystal plasticity, and finite element method, to investigate the process-structure-property relationship for AM metals and their fatigue dispersity. Furthermore, the fatigue life of AM HEAs was also estimated using a probabilistic model [289, 689] based on the AM characteristics, the statistical size distribution of pores, solid-state inclusions, grains, and their mutual interactions.

Currently there have been a great deal of experimental fatigue data of AM metals scattered in the literatures. These data provide the possibility to use the advances in machine learning to enhance the prediction capabilities and efficiency of fatigue properties of AM metals. For example, Zhan et al. [22, 690] present a damage mechanics based machine learning framework for the data-driven fatigue life prediction of AM titanium alloy. A machine learning method is adopted to explore the influence of defect location, size, and morphology on the fatigue life of SLM-AM Ti-6Al-4V [691]. Machine learning and data-driven approaches should be promising for the efficient prediction of fatigue performance of AM metals. By the way, the ultrasonic fatigue testing was discussed as one of the key enabling technologies to optimize fatigue performance [531].

## CRediT authorship contribution statement

**Min Yi:** Conceptualization, Resources, Supervision, Project administration, Funding acquisition, Writing - original draft & review & editing. **Wei Tang:** Conceptualization, Investigation, Data curation, Writing - original draft & review & editing. **Yiqi Zhu:** Conceptualization, Investigation, Data curation, Writing - original draft. **Chenguang Liang:** Conceptualization, Investigation, Data curation, Writing - original draft. **Ziming Tang:** Conceptualization, Investigation, Data curation, Writing - original draft. **Yan Yin:** Conceptualization, Investigation, Data curation, Writing - original draft. **Weiwei He:** Conceptualization, Investigation, Data curation, Writing - original draft. **Shen Sun:** Conceptualization, Investigation, Data curation, Writing - original draft.

## Declaration of Competing Interest

The authors declare that they have no known competing financial interests or personal relationships that could have appeared to influence the work reported in this paper.

## Acknowledgements

The authors acknowledge the support from National Key Research and Development Program of China (2022YFB4600700), National Science and Technology Major Project (J2019-IV-0014-0082), National Youth Talents Program of China, and a project Funded by the Priority Academic Program Development of Jiangsu Higher Education Institutions.

## Data Availability

Data will be made available on request.

## References

[1] U. M. Dilberoglu, B. Gharehpapagh, U. Yaman, M. Dolen, The role of additive manufacturing in the era of industry 4.0, Procedia Manufacturing 11 (2017) 545–554. doi:10.1016/j.promfg.2017.07.148.
[2] K. V. Wong, A. Hernandez, A review of additive manufacturing, International Scholarly Research Notices 2012 (2012) 208760. doi:10.5402/2012/208760.
[3] J. Beaman, D. L. Bourell, C. Seepersad, D. Kovar, Additive manufacturing review: Early past to current practice, Journal of Manufacturing Science and Engineering 142 (11) (2020) 110812. doi:10.1115/1.4048193.
[4] D. Li, R. Qin, J. Xu, J. Zhou, C. B, Improving Mechanical Properties and Energy Absorption of Additive Manufacturing Lattice Structure by Struts'Node Strengthening, Acta Mechanica Solida Sinicadoi:10.1007/s10338-022-00341-4.
[5] S. H. Huang, P. Liu, A. Mokasdar, L. Hou, Additive manufacturing and its societal impact: a literature review, The International Journal of Advanced Manufacturing Technology 67 (5) (2013) 1191–1203. doi:10.1007/s00170-012-4558-5.
[6] N. Li, S. Huang, G. Zhang, R. Qin, W. Liu, H. Xiong, G. Shi, J. Blackburn, Progress in additive manufacturing on new materials: A review, Journal of Materials Science & Technology 35 (2) (2019) 242–269. doi:10.1016/j.jmst.2018.09.002.
[7] J. J. Lewandowski, M. Seifi, Metal additive manufacturing: a review of mechanical properties, Annual review of materials research 46 (2016) 151–186. doi:10.1146/annurev-matsci-070115-032024.
[8] W. E. Frazier, Metal additive manufacturing: a review, Journal of Materials Engineering and performance 23 (6) (2014) 1917–1928.

[9] L. E. Murr, S. M. Gaytan, D. A. Ramirez, E. Martinez, J. Hernandez, K. N. Amato, P. W. Shindo, F. R. Medina, R. B. Wicker, Metal fabrication by additive manufacturing using laser and electron beam melting technologies, Journal of Materials Science & Technology 28 (1) (2012) 1–14.

[10] A. Yadollahi, N. Shamsaei, Additive manufacturing of fatigue resistant materials: Challenges and opportunities, International Journal of Fatigue 98 (2017) 14–31.

[11] J. H. Martin, B. D. Yahata, J. M. Hundley, J. A. Mayer, T. A. Schaedler, T. M. Pollock, 3D printing of high-strength aluminium alloys, Nature 549 (7672) (2017) 365–369. doi:10.1038/nature23894.

[12] F. Liu, C. He, Y. Chen, H. Zhang, Q. Wang, Y. Liu, Effects of defects on tensile and fatigue behaviors of selective laser melted titanium alloy in very high cycle regime, International Journal of Fatigue 140 (2020) 105795. doi:10.1016/j.ijfatigue.2020.105795.

[13] R. Biswal, X. Zhang, A. K. Syed, M. Awd, J. Ding, F. Walther, S. Williams, Criticality of porosity defects on the fatigue performance of wire + arc additive manufactured titanium alloy, International Journal of Fatigue 122 (2019) 208–217. doi:10.1016/j.ijfatigue.2019.01.017.

[14] S. Wang, L. Zhan, H. Xi, H. Xiao, A unified approach toward simulating constant and varying amplitude fatigue failure effects of metals with fast and efficient algorithms, Acta Mechanica Solida Sinica 34 (2021) 53–64. doi:10.1007/s10338-020-00187-8.

[15] A. H. Chern, P. Nandwana, T. Yuan, M. M. Kirka, R. R. Dehoff, P. K. Liaw, C. E. Duty, A review on the fatigue behavior of Ti-6Al-4V fabricated by electron beam melting additive manufacturing, International Journal of Fatigue 119 (September 2018) (2019) 173–184. doi:10.1016/j.ijfatigue.2018.09.022.
URL https://doi.org/10.1016/j.ijfatigue.2018.09.022

[16] R. Caivano, A. Tridello, G. Chiandussi, G. Qian, D. Paolino, F. Berto, Very high cycle fatigue (VHCF) response of additively manufactured materials: A review, Fatigue and Fracture of Engineering Materials and Structures 44 (11) (2021) 2919–2943. doi:10.1111/ffe.13567.

[17] S. Bagherifard, N. Beretta, S. Monti, M. Riccio, M. Bandini, M. Guagliano, On the fatigue strength enhancement of additive manufactured AlSi10Mg parts by mechanical and thermal post-processing, Materials and Design 145 (2018) 28–41. doi:10.1016/j.matdes.2018.02.055.

[18] S. Liu, Y. C. Shin, Additive manufacturing of ti6al4v alloy: A review, Materials & Design 164 (2019) 107552.

[19] A. Ostovari Moghaddam, N. A. Shaburova, M. N. Samodurova, A. Abdollahzadeh, E. A. Trofimov, Additive manufacturing of high entropy alloys: A practical review, Journal of Materials Science & Technology 77 (2021) 131–162. doi:10.1016/J.JMST.2020.11.029.

[20] C. Wang, X. Tan, S. Tor, C. Lim, Machine learning in additive manufacturing: State-of-the-art and perspectives, Additive Manufacturing 36 (2020) 101538.

[21] Z. Zhan, H. Li, Machine learning based fatigue life prediction with effects of additive manufacturing process parameters for printed ss 316l, International Journal of Fatigue 142 (2021) 105941.

[22] Z. Zhan, H. Li, A novel approach based on the elastoplastic fatigue damage and machine learning models for life prediction of aerospace alloy parts fabricated by additive manufacturing, International Journal of Fatigue 145 (2021) 106089. doi:10.1016/J.IJFATIGUE.2020.106089.

[23] C. Ye, C. Zhang, J. Zhao, Y. Dong, Effects of Post-processing on the Surface Finish, Porosity, Residual Stresses, and Fatigue Performance of Additive Manufactured Metals: A Review (jul 2021). doi:10.1007/s11665-021-06021-7.
URL https://link.springer.com/article/10.1007/s11665-021-06021-7

[24] J. Zhang, B. Song, Q. Wei, D. Bourell, Y. Shi, A review of selective laser melting of aluminum alloys: Processing, microstructure, property and developing trends, Journal of Materials Science and Technology 35 (2) (2019) 270–284. doi:10.1016/j.jmst.2018.09.004.

[25] H. Zhang, C. Li, M. Xu, W. Dai, P. Kumar, Z. Liu, Z. Li, Y. Zhang, The fatigue performance evaluation of additively manufactured 304L austenitic stainless steels, Materials Science and Engineering A 802 (December 2020) (2021) 140640. doi:10.1016/j.msea.2020.140640.
URL https://doi.org/10.1016/j.msea.2020.140640

[26] M. N. Jahangir, M. A. H. Mamun, M. P. Sealy, A review of additive manufacturing of magnesium alloys, in: AIP Conference Proceedings, Vol. 1980, AIP Publishing LLC, 2018, p. 030026.

[27] E. Wycisk, C. Emmelmann, S. Siddique, F. Walther, High cycle fatigue (hcf) performance of ti-6al-4v alloy processed by selective laser melting, in: Advanced materials research, Vol. 816, Trans Tech Publ, 2013, pp. 134–139. doi:10.4028/WWW.SCIENTIFIC.NET/AMR.816-817.134.

[28] M. Mojib, R. Pahuja, M. Ramulu, D. Arola, High Cycle Fatigue Behavior of Recycled Additive Manufactured Electron Beam Melted Titanium Ti6Al4V, ASME International Mechanical Engineering Congress and Exposition, Proceedings (IMECE) 2A-2020. doi:10.1115/IMECE2020-24194.

[29] J. Kluczyński, L. Śnieżek, K. Grzelak, J. Torzewski, I. Szachogłuchowicz, A. Oziebło, K. Perkowski, M. Wachowski, M. Ma-lek, The influence of heat treatment on low cycle fatigue properties of selectively laser melted 316l steel, Materials

13 (24) (2020) 5737. doi:10.3390/ma13245737.

[30] N. Sanaei, A. Fatemi, Defects in additive manufactured metals and their effect on fatigue performance: A state-of-the-art review, Progress in Materials Science 117 (2021) 100724. doi:10.1016/j.pmatsci.2020.100724.
URL https://doi.org/10.1016/j.pmatsci.2020.100724

[31] V. D. Le, E. Pessard, F. Morel, S. Prigent, Fatigue behaviour of additively manufactured Ti-6Al-4V alloy: The role of defects on scatter and statistical size effect, International Journal of Fatigue 140 (2020) 105811. doi:10.1016/J.IJFATIGUE.2020.105811.

[32] A. R. Balachandramurthi, J. Moverare, N. Dixit, R. Pederson, Influence of defects and as-built surface roughness on fatigue properties of additively manufactured Alloy 718, Materials Science and Engineering A 735 (June) (2018) 463–474. doi:10.1016/j.msea.2018.08.072.
URL https://doi.org/10.1016/j.msea.2018.08.072

[33] Y. Xie, M. Gong, Q. Zhou, Q. Li, F. Wang, X. Zeng, M. Gao, Effect of microstructure on fatigue crack growth of wire arc additive manufactured Ti-6Al-4V, Materials Science and Engineering: A 826 (2021) 141942. doi:10.1016/J.MSEA.2021.141942.

[34] R. Molaei, A. Fatemi, N. Sanaei, J. Pegues, N. Shamsaei, S. Shao, P. Li, D. Warner, N. Phan, Fatigue of additive manufactured ti-6al-4v, part ii: The relationship between microstructure, material cyclic properties, and component performance, International Journal of Fatigue 132 (2020) 105363.

[35] Y. Zhai, H. Galarraga, D. A. Lados, Microstructure, static properties, and fatigue crack growth mechanisms in ti-6al-4v fabricated by additive manufacturing: Lens and ebm, Engineering Failure Analysis 69 (2016) 3–14, special issue on the International Conference on Structural Integrity. doi:https://doi.org/10.1016/j.engfailanal.2016.05.036.
URL https://www.sciencedirect.com/science/article/pii/S1350630716303752

[36] M. Komarasamy, S. Shukla, S. Williams, K. Kandasamy, S. Kelly, R. S. Mishra, Microstructure, fatigue, and impact toughness properties of additively manufactured nickel alloy 718, Additive Manufacturing 28 (July 2018) (2019) 661–675. doi:10.1016/j.addma.2019.06.009.
URL https://doi.org/10.1016/j.addma.2019.06.009

[37] T. Mishurova, K. Artzt, B. Rehmer, J. Haubrich, L. Ávila, F. Schoenstein, I. Serrano-Munoz, G. Requena, G. Bruno, Separation of the impact of residual stress and microstructure on the fatigue performance of LPBF Ti-6Al-4V at elevated temperature, International Journal of Fatigue 148 (2021) 106239. doi:10.1016/J.IJFATIGUE.2021.106239.

[38] C. Romero, F. Yang, L. Bolzoni, Fatigue and fracture properties of Ti alloys from powder-based processes - A review, International Journal of Fatigue 117 (2018) 407–419. doi:10.1016/j.ijfatigue.2018.08.029.
URL https://doi.org/10.1016/j.ijfatigue.2018.08.029

[39] M. Kahlin, H. Ansell, D. Basu, A. Kerwin, L. Newton, B. Smith, J. J. Moverare, Improved fatigue strength of additively manufactured Ti6Al4V by surface post processing, International Journal of Fatigue 134 (2020) 105497. doi:10.1016/J.IJFATIGUE.2020.105497.

[40] G. Nicoletto, R. Konečná, M. Frkáň, E. Riva, Surface roughness and directional fatigue behavior of as-built EBM and DMLS Ti6Al4V, International Journal of Fatigue 116 (2018) 140–148. doi:10.1016/J.IJFATIGUE.2018.06.011.

[41] P. Kumar, U. Ramamurty, Microstructural optimization through heat treatment for enhancing the fracture toughness and fatigue crack growth resistance of selective laser melted Ti–6Al–4V alloy, Acta Materialia 169 (2019) 45–59. doi:10.1016/j.actamat.2019.03.003.
URL https://doi.org/10.1016/j.actamat.2019.03.003

[42] H. Masuo, Y. Tanaka, S. Morokoshi, H. Yagura, T. Uchida, Y. Yamamoto, Y. Murakami, Effects of Defects, Surface Roughness and HIP on Fatigue Strength of Ti-6Al-4V manufactured by Additive Manufacturing, Procedia Structural Integrity 7 (2017) 19–26. doi:10.1016/J.PROSTR.2017.11.055.

[43] T. M. Mower, M. J. Long, Mechanical behavior of additive manufactured, powder-bed laser-fused materials, Materials Science and Engineering: A 651 (2016) 198–213. doi:10.1016/j.msea.2015.10.068.

[44] Y. Zhai, H. Galarraga, D. A. Lados, Microstructure Evolution, Tensile Properties, and Fatigue Damage Mechanisms in Ti-6Al-4V Alloys Fabricated by Two Additive Manufacturing Techniques, Procedia Engineering 114 (2015) 658–666. doi:10.1016/J.PROENG.2015.08.007.

[45] N. Shamsaei, J. Simsiriwong, Fatigue behaviour of additively-manufactured metallic parts, Procedia Structural Integrity 7 (2017) 3–10, 3rd International Symposium on Fatigue Design and Material Defects, FDMD 2017. doi:https://doi.org/10.1016/j.prostr.2017.11.053.
URL https://www.sciencedirect.com/science/article/pii/S2452321617304031

[46] P. Edwards, A. O'conner, M. Ramulu, Electron beam additive manufacturing of titanium components: properties and performance, Journal of Manufacturing Science and Engineering 135 (6).

[47] N. Hrabe, T. Gnäupel-Herold, T. Quinn, Fatigue properties of a titanium alloy (ti–6al–4v) fabricated via electron beam melting (ebm): Effects of internal defects and residual stress, International Journal of Fatigue 94 (2017) 202–210. doi:10.1016/j.ijfatigue.2016.04.022.

[48] P. Edwards, M. Ramulu, Fatigue performance evaluation of selective laser melted ti-6al-4v, Materials Science and Engineering: A 598 (2014) 327–337. doi:https://doi.org/10.1016/j.msea.2014.01.041.
URL https://www.sciencedirect.com/science/article/pii/S0921509314000720

[49] P. Edwards, M. Ramulu, Effect of build direction on the fracture toughness and fatigue crack growth in selective laser melted Ti-6Al-4-V, Fatigue and Fracture of Engineering Materials and Structures 38 (10) (2015) 1228–1236. doi:10.1111/ffe.12303.

[50] A. Sterling, N. Shamsaei, B. Torries, S. M. Thompson, Fatigue behaviour of additively manufactured ti-6al-4 v, Procedia Engineering 133 (2015) 576–589. doi:10.1016/J.PROENG.2015.12.632.

[51] B. Dutta, F. S. Froes, The additive manufacturing (AM) of titanium alloys, Metal powder report 72 (2) (2017) 96–106. doi:10.1016/j.mprp.2016.12.062.

[52] R. Konečná, L. Kunz, A. Bača, G. Nicoletto, Resistance of direct metal laser sintered Ti6Al4V alloy against growth of fatigue cracks, Engineering Fracture Mechanics 185 (2017) 82–91. doi:10.1016/J.ENGFRACMECH.2017.03.033.

[53] P. Edwards, M. Ramulu, Effect of build direction on the fracture toughness and fatigue crack growth in selective laser melted ti-6al-4 v, Fatigue & Fracture of Engineering Materials & Structures 38 (10) (2015) 1228–1236. doi:https://doi.org/10.1111/ffe.12303.

[54] C. Rans, J. Michielssen, M. Walker, W. Wang, L. 't Hoen-Velterop, Beyond the orthogonal: on the influence of build orientation on fatigue crack growth in slm ti-6al-4v, International Journal of Fatigue 116 (2018) 344–354. doi:https://doi.org/10.1016/j.ijfatigue.2018.06.038.
URL https://www.sciencedirect.com/science/article/pii/S0142112318302731

[55] M. Seifi, M. Dahar, R. Aman, O. Harrysson, J. Beuth, J. J. Lewandowski, Evaluation of Orientation Dependence of Fracture Toughness and Fatigue Crack Propagation Behavior of As-Deposited ARCAM EBM Ti-6Al-4V, Jom 67 (3) (2015) 597–607. doi:10.1007/s11837-015-1298-7.

[56] T. Persenot, A. Burr, G. Martin, J. Y. Buffiere, R. Dendievel, E. Maire, Effect of build orientation on the fatigue properties of as-built Electron Beam Melted Ti-6Al-4V alloy, International Journal of Fatigue 118 (2019) 65–76. doi:10.1016/j.ijfatigue.2018.08.006.
URL https://doi.org/10.1016/j.ijfatigue.2018.08.006

[57] W. Sun, Y. Ma, X. Ai, J. Li, Effects of the building direction on fatigue crack growth behavior of Ti-6Al-4V manufactured by selective laser melting, Procedia Structural Integrity 13 (2018) 1020–1025. doi:10.1016/j.prostr.2018.12.190.

[58] K. Walker, Q. Liu, M. Brandt, Evaluation of fatigue crack propagation behaviour in ti-6al-4v manufactured by selective laser melting, International Journal of Fatigue 104 (2017) 302–308. doi:https://doi.org/10.1016/j.ijfatigue.2017.07.014.
URL https://www.sciencedirect.com/science/article/pii/S0142112317303043

[59] W. Sun, Y. Ma, W. Huang, W. Zhang, X. Qian, Effects of build direction on tensile and fatigue performance of selective laser melting Ti6Al4V titanium alloy, International Journal of Fatigue 130 (August 2019). doi:10.1016/j.ijfatigue.2019.105260.

[60] S. Leuders, M. Thöne, A. Riemer, T. Niendorf, T. Tröster, H. Richard, H. Maier, On the mechanical behaviour of titanium alloy tial6v4 manufactured by selective laser melting: Fatigue resistance and crack growth performance, International Journal of Fatigue 48 (2013) 300–307. doi:https://doi.org/10.1016/j.ijfatigue.2012.11.011.
URL https://www.sciencedirect.com/science/article/pii/S014211231200343X

[61] T. H. Becker, N. M. Dhansay, G. M. T. Haar, K. Vanmeensel, Near-threshold fatigue crack growth rates of laser powder bed fusion produced Ti-6Al-4V, Acta Materialia 197 (2020) 269–282. doi:10.1016/J.ACTAMAT.2020.07.049.

[62] V. Cain, L. Thijs, J. Van Humbeeck, B. Van Hooreweder, R. Knutsen, Crack propagation and fracture toughness of Ti6Al4V alloy produced by selective laser melting, Additive Manufacturing 5 (2015) 68–76. doi:10.1016/j.addma.2014.12.006.

[63] H. Galarraga, R. J. Warren, D. A. Lados, R. R. Dehoff, M. M. Kirka, Fatigue crack growth mechanisms at the microstructure scale in as-fabricated and heat treated Ti-6Al-4V ELI manufactured by electron beam melting (EBM), Engineering Fracture Mechanics 176 (2017) 263–280. doi:10.1016/J.ENGFRACMECH.2017.03.024.

[64] M. Tarik Hasib, H. E. Ostergaard, X. Li, J. J. Kruzic, Fatigue crack growth behavior of laser powder bed fusion additive manufactured ti-6al-4v: Roles of post heat treatment and build orientation, International Journal of Fatigue 142 (2021) 105955. doi:https://doi.org/10.1016/j.ijfatigue.2020.105955.
URL https://www.sciencedirect.com/science/article/pii/S0142112320304874

[65] Z. Jiao, R. Xu, H. Yu, X. Wu, Evaluation on tensile and fatigue crack growth performances of ti6al4v alloy produced by selective laser melting, Procedia Structural Integrity 7 (2017) 124–132.

[66] Y. Xie, M. Gao, F. Wang, C. Zhang, K. Hao, H. Wang, X. Zeng, Anisotropy of fatigue crack growth in wire arc additive manufactured ti-6al-4v, Materials Science and Engineering: A 709 (2018) 265–269. doi:https://doi.org/10.1016/j.msea.2017.10.064.
URL https://www.sciencedirect.com/science/article/pii/S0921509317313904

[67] R. Konečná, L. Kunz, A. Bača, G. Nicoletto, Long fatigue crack growth in ti6al4v produced by direct metal laser sintering, Procedia Engineering 160 (2016) 69–76.

[68] M. Waddell, K. Walker, R. Bandyopadhyay, K. Kapoor, A. Mallory, X. Xiao, A. C. Chuang, Q. Liu, N. Phan, M. D. Sangid, Small fatigue crack growth behavior of Ti-6Al-4V produced via selective laser melting: In situ characterization of a 3D crack tip interactions with defects, International Journal of Fatigue 137 (2020) 105638. doi:10.1016/j.ijfatigue.2020.105638.
URL https://doi.org/10.1016/j.ijfatigue.2020.105638

[69] Y. Ren, X. Lin, Z. Jian, H. Peng, W. Huang, Long fatigue crack growth behavior of Ti-6Al-4V produced via high-power laser directed energy deposition, Materials Science and Engineering: A 819 (2021) 141392. doi:10.1016/J.MSEA.2021.141392.

[70] Y. Zhai, D. A. Lados, E. J. Brown, G. N. Vigilante, Fatigue crack growth behavior and microstructural mechanisms in Ti-6Al-4V manufactured by laser engineered net shaping, International Journal of Fatigue 93 (2016) 51–63. doi:https://doi.org/10.1016/j.ijfatigue.2016.08.009.
URL https://www.sciencedirect.com/science/article/pii/S0142112316302420

[71] A. K. Syed, B. Ahmad, H. Guo, T. Machry, D. Eatock, J. Meyer, M. E. Fitzpatrick, X. Zhang, An experimental study of residual stress and direction-dependence of fatigue crack growth behaviour in as-built and stress-relieved selective-laser-melted ti6al4v, Materials Science and Engineering: A 755 (2019) 246–257. doi:https://doi.org/10.1016/j.msea.2019.04.023.
URL https://www.sciencedirect.com/science/article/pii/S0921509319304691

[72] L. Zhu, X. Hu, R. Jiang, Y. Song, S. Qu, Experimental investigation of small fatigue crack growth due to foreign object damage in titanium alloy TC4, Materials Science and Engineering: A 739 (2019) 214–224. doi:10.1016/J.MSEA.2018.10.031.

[73] D. Greitemeier, C. Dalle Donne, F. Syassen, J. Eufinger, T. Melz, Effect of surface roughness on fatigue performance of additive manufactured Ti-6Al-4V, Materials Science and Technology (United Kingdom) 32 (7) (2016) 629–634. doi:10.1179/1743284715Y.0000000053.

[74] N. M. Dhansay, R. Tait, T. Becker, Fatigue and fracture toughness of ti-6al-4v titanium alloy manufactured by selective laser melting, in: AMI Light Metals Conference 2014, Vol. 1019 of Advanced Materials Research, Trans Tech Publications Ltd, 2014, pp. 248–253. doi:https://doi.org/10.4028/www.scientific.net/AMR.1019.248.

[75] J. Zhang, X. Wang, S. Paddea, X. Zhang, Fatigue crack propagation behaviour in wire+arc additive manufactured ti-6al-4v: Effects of microstructure and residual stress, Materials & Design 90 (2016) 551–561. doi:https://doi.org/10.1016/j.matdes.2015.10.141.
URL https://www.sciencedirect.com/science/article/pii/S026412751530719X

[76] X. Zhang, F. Martina, J. Ding, X. Wang, S. W. Williams, Fracture toughness and fatigue crack growth rate properties in wire+ arc additive manufactured ti-6al-4v, Fatigue & Fracture of Engineering Materials & Structures 40 (5) (2017) 790–803.

[77] Y. Gao, C. Wu, K. Peng, X. Song, Y. Fu, Q. Chen, M. Zhang, G. Wang, J. Liu, Towards superior fatigue crack growth resistance of tc4-dt alloy by in-situ rolled wire-arc additive manufacturing, Journal of Materials Research and Technology 15 (2021) 1395–1407. doi:https://doi.org/10.1016/j.jmrt.2021.08.152.
URL https://www.sciencedirect.com/science/article/pii/S2238785421009777

[78] M. K. Dunstan, J. D. Paramore, Z. Z. Fang, The effects of microstructure and porosity on the competing fatigue failure mechanisms in powder metallurgy Ti-6Al-4V, International Journal of Fatigue 116 (2018) 584–591. doi:10.1016/J.IJFATIGUE.2018.07.006.

[79] Q. C. Liu, J. Elambasseril, S. J. Sun, M. Leary, M. Brandt, P. K. Sharp, The effect of manufacturing defects on the fatigue behaviour of Ti-6Al-4V specimens fabricated using selective laser melting, in: Advanced Materials Research, Vol. 891-892, Trans Tech Publications, 2014, pp. 1519–1524. doi:10.4028/www.scientific.net/AMR.891-892.1519.

[80] H. Gong, K. Rafi, H. Gu, G. D. Janaki Ram, T. Starr, B. Stucker, Influence of defects on mechanical properties of Ti–6Al–4 V components produced by selective laser melting and electron beam melting, Materials & Design 86 (2015) 545–554. doi:10.1016/J.MATDES.2015.07.147.

[81] E. Akgun, X. Zhang, T. Lowe, Y. Zhang, M. Doré, Fatigue of laser powder-bed fusion additive manufactured Ti-6Al-4V in presence of process-induced porosity defects, Engineering Fracture Mechanics 259 (September 2021). doi:10.1016/j.engfracmech.2021.108140.

[82] H. ZHANG, D. DONG, S. SU, A. CHEN, Experimental study of effect of post processing on fracture toughness and fatigue crack growth performance of selective laser melting Ti-6Al-4V, Chinese Journal of Aeronautics 32 (10) (2019) 2383–2393. doi:10.1016/j.cja.2018.12.007.

[83] X. Qiu, Effect of rolling on fatigue crack growth rate of wire and arc additive manufacture (waam) processed titanium, Ph.D. thesis, Cranfield University (2013).

[84] H. Zhang, Z. Cai, J. Chi, R. Sun, Z. Che, H. Zhang, W. Guo, Fatigue crack growth in residual stress fields of laser shock

peened ti6al4v titanium alloy, Journal of Alloys and Compounds 887 (2021) 161427. doi:https://doi.org/10.1016/j.jallcom.2021.161427.
URL https://www.sciencedirect.com/science/article/pii/S092583882102836X

[85] S. Beretta, S. Romano, A comparison of fatigue strength sensitivity to defects for materials manufactured by am or traditional processes, International Journal of Fatigue 94 (2017) 178–191. doi:10.1016/j.ijfatigue.2016.06.020.

[86] I. Serrano-Munoz, J.-Y. Buffiere, R. Mokso, C. Verdu, Y. Nadot, Location, location & size: defects close to surfaces dominate fatigue crack initiation, Scientific reports 7 (1) (2017) 1–9. doi:10.1038/srep45239.

[87] D. Greitemeier, F. Palm, F. Syassen, T. Melz, Fatigue performance of additive manufactured tial6v4 using electron and laser beam melting, International Journal of Fatigue 94 (2017) 211–217. doi:10.1016/j.ijfatigue.2016.05.001.

[88] S. Kim, H. Oh, J. G. Kim, S. Kim, Effect of Annealing and Crack Orientation on Fatigue Crack Propagation of Ti64 Alloy Fabricated by Direct Energy Deposition Process, Metals and Materials International 2021 (2021) 1–11 doi:10.1007/S12540-021-01087-3.
URL https://link.springer.com/article/10.1007/s12540-021-01087-3

[89] J. Jesus, L. Borrego, J. Ferreira, J. Costa, A. Batista, C. Capela, Fatigue crack growth in ti-6al-4v specimens produced by laser powder bed fusion and submitted to hot isostatic pressing, Theoretical and Applied Fracture Mechanics 118 (2022) 103231. doi:https://doi.org/10.1016/j.tafmec.2021.103231.
URL https://www.sciencedirect.com/science/article/pii/S0167844221003268

[90] L. Borrego, J. Jesus, J. Ferreira, J. Costa, C. Capela, Overloading effect on transient fatigue crack growth of ti-6al-4v parts produced by laser powder bed fusion, Procedia Structural Integrity 37 (2022) 330–335, iCSI 2021 The 4th International Conference on Structural Integrity. doi:https://doi.org/10.1016/j.prostr.2022.01.092.
URL https://www.sciencedirect.com/science/article/pii/S2452321622001007

[91] F. Wang, X. Zhang, P. Kong, J. Zhang, R. Luo, J. Zhang, Y. Wang, An improved small-time-scale crack growth rate model considering overloading and load-sustaining effects for deep-sea pressure hulls, Ocean Engineering 247 (2022) 110361. doi:https://doi.org/10.1016/j.oceaneng.2021.110361.
URL https://www.sciencedirect.com/science/article/pii/S0029801821016577

[92] M. Neikter, M. Colliander, C. de Andrade Schwerz, T. Hansson, P. Ãkerfeldt, R. Pederson, M.-L. Antti, Fatigue crack growth of electron beam melted ti-6al-4v in high-pressure hydrogen, Materials 13 (6) (2020) 1287. doi:10.3390/ma13061287.

[93] H. Masuo, Y. Tanaka, S. Morokoshi, H. Yagura, T. Uchida, Y. Yamamoto, Y. Murakami, Influence of defects, surface roughness and HIP on the fatigue strength of Ti-6Al-4V manufactured by additive manufacturing, International Journal of Fatigue 117 (2018) 163–179. doi:10.1016/J.IJFATIGUE.2018.07.020.

[94] A. T. Sutton, C. S. Kriewall, M. C. Leu, J. W. Newkirk, Powder characterisation techniques and effects of powder characteristics on part properties in powder-bed fusion processes, http://dx.doi.org/10.1080/17452759.2016.1250605 12 (1) (2016) 3–29. doi:10.1080/17452759.2016.1250605.
URL https://www.tandfonline.com/action/journalInformation?journalCode=nvpp20

[95] P. Nandwana, M. M. Kirka, V. C. Paquit, S. Yoder, R. R. Dehoff, Correlations Between Powder Feedstock Quality, In Situ Porosity Detection, and Fatigue Behavior of Ti-6Al-4V Fabricated by Powder Bed Electron Beam Melting: A Step Towards Qualification, JOM 2018 70:9 70 (9) (2018) 1686–1691. doi:10.1007/S11837-018-3034-6.
URL https://link.springer.com/article/10.1007/s11837-018-3034-6

[96] H. P. Tang, M. Qian, N. Liu, X. Z. Zhang, G. Y. Yang, J. Wang, Effect of powder reuse times on additive manufacturing of ti-6al-4v by selective electron beam melting, JOM 67 (3) (2015) 555–563. doi:10.1007/s11837-015-1300-4.
URL https://doi.org/10.1007/s11837-015-1300-4

[97] H. Gong, J. J. Dilip, L. Yang, C. Teng, B. Stucker, Influence of small particles inclusion on selective laser melting of ti-6al-4v powder, IOP Conference Series: Materials Science and Engineering 272 (1) (2017) 2–8. doi:10.1088/1757-899X/272/1/012024.

[98] S. E. Brika, V. Brailovski, Influence of powder particle morphology on the static and fatigue properties of laser powder bed-fused ti-6al-4v components, Journal of Manufacturing and Materials Processing 4 (4) (2020) 107. doi:10.3390/jmmp4040107.
URL https://doi.org/10.3390/jmmp4040107

[99] P. E. Carrion, A. Soltani-Tehrani, N. Phan, N. Shamsaei, Powder recycling effects on the tensile and fatigue behavior of additively manufactured ti-6al-4v parts, Jom 71 (3) (2019) 963–973. doi:10.1007/s11837-018-3248-7.
URL https://doi.org/10.1007/s11837-018-3248-7

[100] V. V. Popov, A. Katz-Demyanetz, A. Garkun, M. Bamberger, The effect of powder recycling on the mechanical properties and microstructure of electron beam melted ti-6al-4v specimens, Additive Manufacturing 22 (2018) 834–843. doi:10.1016/J.ADDMA.2018.06.003.

[101] G. Soundarapandiyan, C. Johnston, R. H. Khan, C. L. A. Leung, P. D. Lee, E. Hernández-Nava, B. Chen, M. E. Fitzpatrick, The effects of powder reuse on the mechanical response of electron beam additively manufactured ti6al4v

parts, Additive Manufacturing 46 (2021) 102101. doi:10.1016/J.ADDMA.2021.102101.

[102] J. Pegues, M. Roach, R. S. Williamson, N. Shamsaei, Effect of specimen surface area size on fatigue strength of additively manufactured ti-6al-4v parts, 28th International Solid Freeform Fabrication Symposium-An Additive Manufacturing Conference (2017) 122–133.

[103] J. Pegues, M. Roach, R. Scott Williamson, N. Shamsaei, Surface roughness effects on the fatigue strength of additively manufactured ti-6al-4v, International Journal of Fatigue 116 (2018) 543–552. doi:10.1016/J.IJFATIGUE.2018.07.013.

[104] J. Pegues, M. Roach, R. Scott Williamson, N. Shamsaei, Volume effects on the fatigue behavior of additively manufactured ti-6al4v parts, Solid Freeform Fabrication 2018: Proceedings of the 29th Annual International Solid Freeform Fabrication Symposium - An Additive Manufacturing Conference, SFF 2018 (2020) 1373–1381.

[105] M. Kahlin, H. Ansell, J. J. Moverare, Fatigue behaviour of notched additive manufactured ti6al4v with as-built surfaces, International Journal of Fatigue 101 (2017) 51–60. doi:10.1016/J.IJFATIGUE.2017.04.009.

[106] S.-M.-J. Razavi, P. Ferro, F. Berto, et al., Fatigue assessment of ti–6al–4v circular notched specimens produced by selective laser melting, Metals 7 (8) (2017) 291. doi:10.3390/met7080291.

[107] S. Razavi, P. Ferro, F. Berto, J. Torgersen, Fatigue strength of blunt v-notched specimens produced by selective laser melting of ti-6al-4v, Theoretical and Applied Fracture Mechanics 97 (2018) 376–384. doi:https://doi.org/10.1016/j.tafmec.2017.06.021.
URL https://www.sciencedirect.com/science/article/pii/S0167844217302884

[108] M. Benedetti, C. Santus, Notch fatigue and crack growth resistance of ti-6al-4v eli additively manufactured via selective laser melting: A critical distance approach to defect sensitivity, International Journal of Fatigue 121 (2019) 281–292. doi:10.1016/J.IJFATIGUE.2018.12.020.

[109] X. Zhao, S. Li, M. Zhang, Y. Liu, T. B. Sercombe, S. Wang, Y. Hao, R. Yang, L. E. Murr, Comparison of the microstructures and mechanical properties of Ti–6Al–4V fabricated by selective laser melting and electron beam melting, Materials & Design 95 (2016) 21–31. doi:10.1016/J.MATDES.2015.12.135.

[110] L. Wang, S. Li, M. Yan, Y. Cheng, W. Hou, Y. Wang, S. Ai, R. Yang, K. Dai, Fatigue properties of titanium alloy custom short stems fabricated by electron beam melting, Journal of Materials Science and Technology 52 (2020) 180–188. doi:10.1016/j.jmst.2020.02.047.
URL https://doi.org/10.1016/j.jmst.2020.02.047

[111] M. Kahlin, H. Ansell, J. J. Moverare, Fatigue behaviour of additive manufactured Ti6Al4V, with as-built surfaces, exposed to variable amplitude loading, International Journal of Fatigue 103 (2017) 353–362. doi:10.1016/j.ijfatigue.2017.06.023.
URL http://dx.doi.org/10.1016/j.ijfatigue.2017.06.023

[112] S. M. Razavi, B. Van Hooreweder, F. Berto, Effect of build thickness and geometry on quasi-static and fatigue behavior of Ti-6Al-4V produced by Electron Beam Melting, Additive Manufacturing 36 (June) (2020) 101426. doi:10.1016/j.addma.2020.101426.
URL https://doi.org/10.1016/j.addma.2020.101426

[113] S. Amin Yavari, R. Wauthle, J. van der Stok, A. Riemslag, M. Janssen, M. Mulier, J. Kruth, J. Schrooten, H. Weinans, A. Zadpoor, Fatigue behavior of porous biomaterials manufactured using selective laser melting, Materials Science and Engineering: C 33 (8) (2013) 4849–4858. doi:https://doi.org/10.1016/j.msec.2013.08.006.
URL https://www.sciencedirect.com/science/article/pii/S0928493113004694

[114] S. Amin Yavari, S. Ahmadi, R. Wauthle, B. Pouran, J. Schrooten, H. Weinans, A. Zadpoor, Relationship between unit cell type and porosity and the fatigue behavior of selective laser melted meta-biomaterials, Journal of the Mechanical Behavior of Biomedical Materials 43 (2015) 91–100. doi:https://doi.org/10.1016/j.jmbbm.2014.12.015.
URL https://www.sciencedirect.com/science/article/pii/S1751616114003944

[115] S. Zhao, S. J. Li, W. T. Hou, Y. L. Hao, R. Yang, R. D. Misra, The influence of cell morphology on the compressive fatigue behavior of ti-6al-4v meshes fabricated by electron beam melting, Journal of the Mechanical Behavior of Biomedical Materials 59 (2016) 251–264. doi:10.1016/J.JMBBM.2016.01.034.

[116] M. Dallago, V. Fontanari, E. Torresani, M. Leoni, C. Pederzolli, C. Potrich, M. Benedetti, Fatigue and biological properties of ti-6al-4v eli cellular structures with variously arranged cubic cells made by selective laser melting, Journal of the Mechanical Behavior of Biomedical Materials 78 (2018) 381–394. doi:10.1016/J.JMBBM.2017.11.044.

[117] M.-W. Wu, J.-K. Chen, B.-H. Lin, P.-H. Chiang, M.-K. Tsai, Compressive fatigue properties of additive-manufactured ti-6al-4v cellular material with different porosities, Materials Science and Engineering: A 790 (2020) 139695. doi:https://doi.org/10.1016/j.msea.2020.139695.
URL https://www.sciencedirect.com/science/article/pii/S0921509320307735

[118] N. W. Hrabe, P. Heinl, B. Flinn, C. Körner, R. K. Bordia, Compression-compression fatigue of selective electron beam melted cellular titanium (ti-6al-4v), Journal of Biomedical Materials Research Part B: Applied Biomaterials 99 (2) (2011) 313–320. doi:10.1002/jbm.b.31901.

[119] M. Dallago, B. Winiarski, F. Zanini, S. Carmignato, M. Benedetti, On the effect of geometrical imperfections and defects

on the fatigue strength of cellular lattice structures additively manufactured via selective laser melting, International Journal of Fatigue 124 (2019) 348–360. doi:10.1016/j.ijfatigue.2019.03.019.

[120] P. Kumar, O. Prakash, U. Ramamurty, Micro-and meso-structures and their influence on mechanical properties of selectively laser melted Ti-6Al-4V, Acta Materialia 154 (2018) 246–260. doi:10.1016/J.ACTAMAT.2018.05.044.

[121] P. Kumar, U. Ramamurty, High cycle fatigue in selective laser melted ti-6al-4v, Acta Materialia 194 (2020) 305–320. doi:https://doi.org/10.1016/j.actamat.2020.05.041.
URL https://www.sciencedirect.com/science/article/pii/S1359645420303888

[122] L. Du, G. Qian, L. Zheng, Y. Hong, Influence of processing parameters of selective laser melting on high-cycle and very-high-cycle fatigue behaviour of Ti-6Al-4V, Fatigue & Fracture of Engineering Materials & Structures 44 (1) (2021) 240–256. doi:10.1111/FFE.13361.
URL https://onlinelibrary.wiley.com/doi/full/10.1111/ffe.13361https://onlinelibrary.wiley.com/doi/abs/10.1111/ffe.13361https://onlinelibrary.wiley.com/doi/10.1111/ffe.13361

[123] H. K. Rafi, N. V. Karthik, H. Gong, T. L. Starr, B. E. Stucker, Microstructures and Mechanical Properties of Ti6Al4V Parts Fabricated by Selective Laser Melting and Electron Beam Melting, Journal of Materials Engineering and Performance 2013 22:12 22 (12) (2013) 3872–3883. doi:10.1007/S11665-013-0658-0.
URL https://link.springer.com/article/10.1007/s11665-013-0658-0

[124] N. Sanaei, A. Fatemi, Analysis of the effect of surface roughness on fatigue performance of powder bed fusion additive manufactured metals, Theoretical and Applied Fracture Mechanics 108 (November 2019) (2020) 102638. doi:10.1016/j.tafmec.2020.102638.
URL https://doi.org/10.1016/j.tafmec.2020.102638

[125] B. Vayssette, N. Saintier, C. Brugger, M. El May, E. Pessard, Numerical modelling of surface roughness effect on the fatigue behavior of Ti-6Al-4V obtained by additive manufacturing, International Journal of Fatigue 123 (October 2018) (2019) 180–195. doi:10.1016/j.ijfatigue.2019.02.014.
URL https://doi.org/10.1016/j.ijfatigue.2019.02.014

[126] Y. Y. Sun, S. L. Lu, S. Gulizia, C. H. Oh, D. Fraser, M. Leary, M. Qian, Fatigue Performance of Additively Manufactured Ti-6Al-4V: Surface Condition vs. Internal Defects, Jom 72 (3) (2020) 1022–1030. doi:10.1007/s11837-020-04025-7.

[127] M. Bezuidenhout, G. Ter Haar, T. Becker, S. Rudolph, O. Damm, N. Sacks, The effect of HF-HNO3 chemical polishing on the surface roughness and fatigue life of laser powder bed fusion produced Ti6Al4V, Materials Today Communications 25 (2020) 101396. doi:10.1016/J.MTCOMM.2020.101396.

[128] T. Persenot, A. Burr, R. Dendievel, J. Y. Buffière, E. Maire, J. Lachambre, G. Martin, Fatigue performances of chemically etched thin struts built by selective electron beam melting: Experiments and predictions, Materialia 9 (January) (2020) 100589. doi:10.1016/j.mtla.2020.100589.
URL https://doi.org/10.1016/j.mtla.2020.100589

[129] S. Aguado-Montero, C. Navarro, J. Vázquez, F. Lasagni, S. Slawik, J. Domínguez, Fatigue behaviour of PBF additive manufactured TI6AL4V alloy after shot and laser peening, International Journal of Fatigue 154 (2022) 106536. doi:10.1016/J.IJFATIGUE.2021.106536.

[130] G. Qian, Y. Li, D. S. Paolino, A. Tridello, F. Berto, Y. Hong, Very-high-cycle fatigue behavior of Ti-6Al-4V manufactured by selective laser melting: Effect of build orientation, International Journal of Fatigue 136 (2020) 105628. doi:10.1016/J.IJFATIGUE.2020.105628.

[131] M. Nakatani, H. Masuo, Y. Tanaka, Y. Murakami, Effect of Surface Roughness on Fatigue Strength of Ti-6Al-4V Alloy Manufactured by Additive Manufacturing, Procedia Structural Integrity 19 (2019) 294–301. doi:10.1016/j.prostr.2019.12.032.
URL https://doi.org/10.1016/j.prostr.2019.12.032

[132] J. Günther, D. Krewerth, T. Lippmann, S. Leuders, T. Tröster, A. Weidner, H. Biermann, T. Niendorf, Fatigue life of additively manufactured ti-6al-4v in the very high cycle fatigue regime, International Journal of Fatigue 94 (2017) 236–245, fatigue and Fracture Behavior of Additive Manufactured Parts. doi:https://doi.org/10.1016/j.ijfatigue.2016.05.018.
URL https://www.sciencedirect.com/science/article/pii/S0142112316301207

[133] V. Chastand, P. Quaegebeur, W. Maia, E. Charkaluk, Comparative study of fatigue properties of ti-6al-4v specimens built by electron beam melting (ebm) and selective laser melting (slm), Materials Characterization 143 (2018) 76–81, metal Additive Manufacturing: Microstructures and Properties. doi:https://doi.org/10.1016/j.matchar.2018.03.028.
URL https://www.sciencedirect.com/science/article/pii/S1044580317331741

[134] G. Kasperovich, J. Hausmann, Improvement of fatigue resistance and ductility of tial6v4 processed by selective laser melting, Journal of Materials Processing Technology 220 (2015) 202–214. doi:https://doi.org/10.1016/j.jmatprotec.2015.01.025.
URL https://www.sciencedirect.com/science/article/pii/S0924013615000278

[135] Y. Okazaki, A. Ishino, Microstructures and Mechanical Properties of Laser-Sintered Commercially Pure Ti and Ti-6Al-4V

Alloy for Dental Applications, Materials 13 (609).

[136] A. Fatemi, R. Molaei, J. Simsiriwong, N. Sanaei, J. Pegues, B. Torries, N. Phan, N. Shamsaei, Fatigue behaviour of additive manufactured materials: An overview of some recent experimental studies on Ti-6Al-4V considering various processing and loading direction effects, Fatigue and Fracture of Engineering Materials and Structures 42 (5) (2019) 991–1009. doi:10.1111/ffe.13000.

[137] R. Molaei, A. Fatemi, N. Phan, Significance of hot isostatic pressing (HIP) on multiaxial deformation and fatigue behaviors of additive manufactured Ti-6Al-4V including build orientation and surface roughness effects, International Journal of Fatigue 117 (2018) 352–370. doi:10.1016/J.IJFATIGUE.2018.07.035.

[138] R. Molaei, A. Fatemi, N. Phan, Notched fatigue of additive manufactured metals under axial and multiaxial loadings, part i: Effects of surface roughness and hip and comparisons with their wrought alloys, International Journal of Fatigue 143 (2021) 106003. doi:10.1016/j.ijfatigue.2020.106003.

[139] R. Molaei, A. Fatemi, Fatigue performance of additive manufactured metals under variable amplitude service loading conditions including multiaxial stresses and notch effects: experiments and modelling, International Journal of Fatigue 145 (2021) 106002. doi:10.1016/j.ijfatigue.2020.106002.

[140] R. Molaei, A. Fatemi, N. Phan, Notched fatigue of additive manufactured metals under axial and multiaxial loadings, part ii: Data correlations and life estimations, International Journal of Fatigue 156 (2022) 106648. doi:10.1016/j.ijfatigue.2021.106648.

[141] Z. Chen, S. Cao, X. Wu, C. H. Davies, Surface roughness and fatigue properties of selective laser melted Ti–6Al–4V alloy, Additive Manufacturing for the Aerospace Industry (2019) 283–299 doi:10.1016/B978-0-12-814062-8.00015-7.

[142] E. Wycisk, A. Solbach, S. Siddique, D. Herzog, F. Walther, C. Emmelmann, Effects of Defects in Laser Additive Manufactured Ti-6Al-4V on Fatigue Properties, Physics Procedia 56 (C) (2014) 371–378. doi:10.1016/J.PHPRO.2014.08.120.

[143] H. Wan, Q. Wang, C. Jia, Z. Zhang, Multi-scale damage mechanics method for fatigue life prediction of additive manufacture structures of Ti-6Al-4V, Materials Science and Engineering A 669 (2016) 269–278. doi:10.1016/j.msea.2016.05.073.
URL http://dx.doi.org/10.1016/j.msea.2016.05.073

[144] A. Bača, R. Konečná, G. Nicoletto, L. Kunz, Influence of Build Direction on the Fatigue Behaviour of Ti6Al4V Alloy Produced by Direct Metal Laser Sintering, Materials Today: Proceedings 3 (4) (2016) 921–924. doi:10.1016/J.MATPR.2016.03.021.

[145] W. Sun, Y. E. Ma, W. Zhang, X. Qian, W. Huang, Z. Wang, Effects of the Build Direction on Mechanical Performance of Laser Powder Bed Fusion Additively Manufactured Ti6Al4V under Different Loadings, Advanced Engineering Materials 2100611 (2021) 1–12. doi:10.1002/adem.202100611.

[146] V. Tuninetti, G. Gilles, O. Milis, T. Pardoen, A.-M. Habraken, Anisotropy and tension–compression asymmetry modeling of the room temperature plastic response of ti–6al–4v, International Journal of Plasticity 67 (2015) 53–68.

[147] G. Nicoletto, Anisotropic high cycle fatigue behavior of ti–6al–4v obtained by powder bed laser fusion, International Journal of Fatigue 94 (2017) 255–262.

[148] M. Simonelli, Y. Y. Tse, C. Tuck, Effect of the build orientation on the mechanical properties and fracture modes of slm ti–6al–4v, Materials Science and Engineering: A 616 (2014) 1–11. doi:10.1016/j.msea.2014.07.086.

[149] D. Kovalchuk, V. Melnyk, I. Melnyk, D. Savvakin, O. Dekhtyar, O. Stasiuk, P. Markovsky, Microstructure and Properties of Ti-6Al-4V Articles 3D-Printed with Co-axial Electron Beam and Wire Technology, Journal of Materials Engineering and Performance 2021 30:7 30 (7) (2021) 5307–5322. doi:10.1007/S11665-021-05770-9.
URL https://link.springer.com/article/10.1007/s11665-021-05770-9

[150] A. Cutolo, C. Elangeswaran, G. K. Muralidharan, B. Van Hooreweder, On the role of building orientation and surface post-processes on the fatigue life of Ti-6Al-4V coupons manufactured by laser powder bed fusion, Materials Science and Engineering: A (2022) 142747 doi:10.1016/j.msea.2022.142747.
URL https://doi.org/10.1016/j.msea.2022.142747

[151] A. H. Chern, P. Nandwana, R. McDaniels, R. R. Dehoff, P. K. Liaw, R. Tryon, C. E. Duty, Build orientation, surface roughness, and scan path influence on the microstructure, mechanical properties, and flexural fatigue behavior of Ti–6Al–4V fabricated by electron beam melting, Materials Science and Engineering A 772 (September 2019) (2020) 138740. doi:10.1016/j.msea.2019.138740.
URL https://doi.org/10.1016/j.msea.2019.138740

[152] T. M. Mower, Degradation of titanium 6al–4v fatigue strength due to electrical discharge machining, International Journal of Fatigue 64 (2014) 84–96. doi:https://doi.org/10.1016/j.ijfatigue.2014.02.018.
URL https://www.sciencedirect.com/science/article/pii/S0142112314000772

[153] G. Nicoletto, Directional and notch effects on the fatigue behavior of as-built DMLS Ti6Al4V, International Journal of Fatigue 106 (2018) 124–131. doi:10.1016/J.IJFATIGUE.2017.10.004.

[154] L. Thijs, F. Verhaeghe, T. Craeghs, J. Van Humbeeck, J.-P. Kruth, A study of the microstructural evolution during

selective laser melting of ti–6al–4v, Acta materialia 58 (9) (2010) 3303–3312. doi:10.1016/j.actamat.2010.02.004.
[155] D. A. Renzo, E. Sgambitterra, C. Maletta, F. Furgiuele, C. A. Biffi, J. Fiocchi, A. Tuissi, Multiaxial fatigue behavior of SLM Ti6Al4V alloy under different loading conditions, Fatigue & Fracture of Engineering Materials & Structures 44 (10) (2021) 2625–2642. doi:10.1111/FFE.13518.
URL https://onlinelibrary.wiley.com/doi/full/10.1111/ffe.13518https://onlinelibrary.wiley.com/doi/abs/10.1111/ffe.13518https://onlinelibrary.wiley.com/doi/10.1111/ffe.13518
[156] A. K. Syed, X. Zhang, A. Caballero, M. Shamir, S. Williams, Influence of deposition strategies on tensile and fatigue properties in a wire + arc additive manufactured Ti-6Al-4V, International Journal of Fatigue 149 (2021) 106268. doi:10.1016/J.IJFATIGUE.2021.106268.
[157] H. Ali, H. Ghadbeigi, K. Mumtaz, Effect of scanning strategies on residual stress and mechanical properties of Selective Laser Melted Ti6Al4V, Materials Science and Engineering A 712 (October 2017) (2018) 175–187. doi:10.1016/j.msea.2017.11.103.
URL https://doi.org/10.1016/j.msea.2017.11.103
[158] P. Li, D. Warner, A. Fatemi, N. Phan, Critical assessment of the fatigue performance of additively manufactured ti–6al–4v and perspective for future research, International Journal of Fatigue 85 (2016) 130–143. doi:https://doi.org/10.1016/j.ijfatigue.2015.12.003.
URL https://www.sciencedirect.com/science/article/pii/S0142112315004399
[159] A. Fatemi, R. Molaei, S. Sharifimehr, N. Shamsaei, N. Phan, Torsional fatigue behavior of wrought and additive manufactured Ti-6Al-4V by powder bed fusion including surface finish effect, International Journal of Fatigue 99 (2017) 187–201. doi:10.1016/J.IJFATIGUE.2017.03.002.
[160] D. A. Hollander, M. Von Walter, T. Wirtz, R. Sellei, B. Schmidt-Rohlfing, O. Paar, H.-J. Erli, Structural, mechanical and in vitro characterization of individually structured ti–6al–4v produced by direct laser forming, Biomaterials 27 (7) (2006) 955–963.
[161] J. Tong, C. Bowen, J. Persson, A. Plummer, Mechanical properties of titanium-based ti–6al–4v alloys manufactured by powder bed additive manufacture, Materials Science and Technology 33 (2) (2017) 138–148.
[162] X. Shui, K. Yamanaka, M. Mori, Y. Nagata, K. Kurita, A. Chiba, Effects of post-processing on cyclic fatigue response of a titanium alloy additively manufactured by electron beam melting, Materials Science and Engineering: A 680 (2017) 239–248. doi:10.1016/J.MSEA.2016.10.059.
[163] M. Benedetti, V. Fontanari, M. Bandini, F. Zanini, S. Carmignato, Low- and high-cycle fatigue resistance of Ti-6Al-4V ELI additively manufactured via selective laser melting: Mean stress and defect sensitivity, International Journal of Fatigue 107 (2018) 96–109. doi:10.1016/j.ijfatigue.2017.10.021.
[164] Y. Choi, D. G. Lee, Correlation between surface tension and fatigue properties of Ti-6Al-4V alloy fabricated by EBM additive manufacturing, Applied Surface Science 481 (September 2018) (2019) 741–746. doi:10.1016/j.apsusc.2019.03.099.
URL https://doi.org/10.1016/j.apsusc.2019.03.099
[165] A. K. Singla, M. Banerjee, A. Sharma, J. Singh, A. Bansal, M. K. Gupta, N. Khanna, A. Shahi, D. K. Goyal, Selective laser melting of ti6al4v alloy: Process parameters, defects and post-treatments, Journal of Manufacturing Processes 64 (2021) 161–187.
[166] J. Warner, D. Celli, J. Rindler, M. H. Shen, O. Scott-Emuakpor, T. George, Fatigue assessment of porosity in electron beam melted ti-6al-4v, in: Fracture, Fatigue, Failure and Damage Evolution, Volume 3: Proceedings of the 2020 Annual Conference on Experimental and Applied Mechanics, Vol. 3, Springer Nature, 2020, p. 37.
[167] W. Zhengkai, S. Wu, J. Zhang, Z. Song, H. Y.N., G. Kang, H. Zhang, Defect induced fatigue behaviors of selective laser melted ti-6al-4v via synchrotron radiation x-ray tomography, Acta Metallurgica Sinica 55 (7) (2019) 811–820. doi:10.11900/0412.1961.2018.00408.
[168] B. Zhang, K. Ham, S. Shao, N. Shamsaei, S. M. Thompson, Effect of heat treatment and hot isostatic pressing on the morphology and size of pores in additive manufactured ti-6al-4v parts, in: Proceedings of the 28th Annual International Solid Freeform Fabrication Symposium, 2017, pp. 107–114.
[169] E. Akgun, X. Zhang, R. Biswal, Y. Zhang, M. Doré, Fatigue of wire+ arc additive manufactured ti-6al-4v in presence of process-induced porosity defects, International Journal of Fatigue 150 (2021) 106315. doi:10.1016/j.ijfatigue.2021.106315.
[170] M. Cheng, Z. Lu, J. Wu, R. Guo, J. Qiao, L. Xu, R. Yang, Effect of thermal induced porosity on high-cycle fatigue and very high-cycle fatigue behaviors of hot-isostatic-pressed Ti-6Al-4V powder components, Journal of Materials Science & Technology 98 (2022) 177–185. doi:10.1016/J.JMST.2021.04.066.
[171] S. M. J. Razavi, G. G. Bordonaro, P. Ferro, J. Torgersen, F. Berto, Fatigue behavior of porous Ti-6Al-4V made by laser-engineered net shaping, Materials 11 (2). doi:10.3390/ma11020284.
[172] T.-q. Yan, B.-q. Chen, X. Ji, S.-q. Guo, Influence of hot isostatic pressing on microstructure, properties and deformability of selective laser melting tc4 alloy, China Foundry 18 (4) (2021) 389–396.

[173] M. Benedetti, M. Cazzolli, V. Fontanari, M. Leoni, Fatigue limit of ti6al4v alloy produced by selective laser sintering, Procedia Structural Integrity 2 (2016) 3158–3167, 21st European Conference on Fracture, ECF21, 20-24 June 2016, Catania, Italy. doi:https://doi.org/10.1016/j.prostr.2016.06.394.
URL https://www.sciencedirect.com/science/article/pii/S2452321616304139

[174] R. Cunningham, A. Nicolas, J. Madsen, E. Fodran, E. Anagnostou, M. D. Sangid, A. D. Rollett, Analyzing the effects of powder and post-processing on porosity and properties of electron beam melted Ti-6Al-4V, Materials Research Letters 5 (7) (2017) 516–525. doi:10.1080/21663831.2017.1340911.

[175] P. Li, D. H. Warner, J. W. Pegues, M. D. Roach, N. Shamsaei, N. Phan, Investigation of the mechanisms by which hot isostatic pressing improves the fatigue performance of powder bed fused Ti-6Al-4V, International Journal of Fatigue 120 (October 2018) (2019) 342–352. doi:10.1016/j.ijfatigue.2018.10.015.

[176] T. Childerhouse, E. Hernández-Nava, N. Tapoglou, R. M'Saoubi, L. Franca, W. Leahy, M. Jackson, The influence of finish machining depth and hot isostatic pressing on defect distribution and fatigue behaviour of selective electron beam melted Ti-6Al-4V, International Journal of Fatigue 147 (February) (2021) 1–11. doi:10.1016/j.ijfatigue.2021.106169.

[177] K. Mertova, J. Dzugan, M. Roudnicka, Fatigue properties of SLM-produced Ti6Al4V with various post-processing processes, IOP Conference Series: Materials Science and Engineering 461 (1). doi:10.1088/1757-899X/461/1/012052.

[178] Y. Tian, W. S. Gora, A. P. Cabo, L. L. Parimi, D. P. Hand, S. Tammas-Williams, P. B. Prangnell, Material interactions in laser polishing powder bed additive manufactured ti6al4v components, Additive Manufacturing 20 (2018) 11–22.

[179] C. Ma, Y. Guan, W. Zhou, Laser polishing of additive manufactured ti alloys, Optics and Lasers in Engineering 93 (2017) 171–177. doi:10.1016/j.optlaseng.2017.02.005.

[180] S. Lee, Z. Ahmadi, J. W. Pegues, M. Mahjouri-Samani, N. Shamsaei, Laser polishing for improving fatigue performance of additive manufactured Ti-6Al-4V parts, Optics and Laser Technology 134 (October 2020). doi:10.1016/j.optlastec.2020.106639.

[181] T. Persenot, J.-Y. Buffiere, E. Maire, R. Dendievel, G. Martin, Fatigue properties of ebm as-built and chemically etched thin parts, Procedia Structural Integrity 7 (2017) 158–165. doi:10.1016/j.prostr.2017.11.073.

[182] G. Pyka, G. Kerckhofs, I. Papantoniou, M. Speirs, J. Schrooten, M. Wevers, Surface roughness and morphology customization of additive manufactured open porous ti6al4v structures, Materials 6 (10) (2013) 4737–4757. doi:10.3390/ma6104737.
URL https://www.mdpi.com/1996-1944/6/10/4737

[183] J. Lindemann, C. Buque, F. Appel, Effect of shot peening on fatigue performance of a lamellar titanium aluminide alloy, Acta Materialia 54 (4) (2006) 1155–1164. doi:10.1016/j.actamat.2005.10.043.

[184] L. Hackel, J. R. Rankin, A. Rubenchik, W. E. King, M. Matthews, Laser peening: A tool for additive manufacturing post-processing, Additive Manufacturing 24 (2018) 67–75. doi:10.1016/J.ADDMA.2018.09.013.

[185] H. Soyama, Y. Okura, H. Soyama, Y. Okura, The use of various peening methods to improve the fatigue strength of titanium alloy Ti6Al4V manufactured by electron beam melting, AIMS Materials Science 2018 5:1000 5 (5) (2018) 1000–1015. doi:10.3934/MATERSCI.2018.5.1000.
URL http://www.aimspress.com/article/doi/10.3934/matersci.2018.5.1000http://www.aimspress.com/rticle/doi/10.3934/matersci.2018.5.1000

[186] H. Soyama, F. Takeo, Effect of various peening methods on the fatigue properties of Titanium alloy Ti6Al4V manufactured by direct metal laser sintering and electron beam melting, Materials 13 (10). doi:10.3390/ma13102216.

[187] M. Benedetti, E. Torresani, M. Leoni, V. Fontanari, M. Bandini, C. Pederzolli, C. Potrich, The effect of post-sintering treatments on the fatigue and biological behavior of ti-6al-4v eli parts made by selective laser melting, Journal of the Mechanical Behavior of Biomedical Materials 71 (2017) 295–306. doi:https://doi.org/10.1016/j.jmbbm.2017.03.024.
URL https://www.sciencedirect.com/science/article/pii/S175161611730139X

[188] T. Morita, A. Miyatani, S. Kariya, M. Kumagai, S. Takesue, J. Komotori, Influences of particle collision treatments on surface characteristics and fatigue strength of Ti-6Al-4V alloy, Results in Materials 7 (2020) 100128. doi:10.1016/J.RINMA.2020.100128.

[189] S. H. Lim, Z. Zhang, D. H. L. Seng, M. Lin, S. L. Teo, F. Wei, A. K. H. Cheong, S. Wang, J. Pan, In-situ warm shot peening on Ti-6Al-4V alloy: Effects of temperature on fatigue life, residual stress, microstructure and mechanical properties, Journal of Alloys and Compounds 882 (2021) 160701. doi:10.1016/J.JALLCOM.2021.160701.

[190] C. N. Pintado, J. Vázquez, J. Domínguez, A. Periñán, M. H. García, F. Lasagni, S. Bernarding, S. Slawik, F. Mücklich, F. Boby, et al., Effect of surface treatment on the fatigue strength of additive manufactured ti6al4v alloy, Frattura ed Integrità Strutturale 14 (53) (2020) 337–344. doi:10.3221/IGF-ESIS.53.26.

[191] L. Denti, E. Bassoli, A. Gatto, E. Santecchia, P. Mengucci, Fatigue life and microstructure of additive manufactured ti6al4v after different finishing processes, Materials Science and Engineering: A 755 (2019) 1–9. doi:10.1016/j.msea.2019.03.119.

[192] S. Bagherifard, M. Guagliano, Fatigue performance of cold spray deposits: Coating, repair and additive manufacturing cases, International Journal of Fatigue 139 (2020) 105744. doi:10.1016/j.ijfatigue.2020.105744.

[193] Z. Tong, H. Liu, J. Jiao, W. Zhou, Y. Yang, X. Ren, Improving the strength and ductility of laser directed energy

deposited crmnfeconi high-entropy alloy by laser shock peening, Additive Manufacturing 35 (2020) 101417. doi:https://doi.org/10.1016/j.addma.2020.101417.
URL https://www.sciencedirect.com/science/article/pii/S2214860420307892

[194] W. Guo, R. Sun, B. Song, Y. Zhu, F. Li, Z. Che, B. Li, C. Guo, L. Liu, P. Peng, Laser shock peening of laser additive manufactured ti6al4v titanium alloy, Surface and Coatings Technology 349 (2018) 503–510.

[195] X. Jin, L. Lan, S. Gao, B. He, Y. Rong, Effects of laser shock peening on microstructure and fatigue behavior of ti–6al–4v alloy fabricated via electron beam melting, Materials Science and Engineering: A 780 (2020) 139199. doi:10.1016/j.msea.2020.139199.

[196] Y. Okura, H. Sasaki, H. Soyama, Effect of mechanical properties on fatigue life enhancement of additive manufactured titanium alloy treated by various peening methods, in: International Conference on Advanced Surface Enhancement, Springer, 2019, pp. 88–96. doi:10.1007/978-981-15-0054-1_10.

[197] I. Yeo, S. Bae, A. Amanov, S. Jeong, Effect of laser shock peening on properties of heat-treated ti–6al–4v manufactured by laser powder bed fusion, International Journal of Precision Engineering and Manufacturing-Green Technology 8 (4) (2021) 1137–1150. doi:10.1007/s40684-020-00234-2.

[198] L. Wang, L. Zhou, L. Liu, W. He, X. Pan, X. Nie, S. Luo, Fatigue strength improvement in Ti-6Al-4V subjected to foreign object damage by combined treatment of laser shock peening and shot peening, International Journal of Fatigue 155 (2022) 106581. doi:10.1016/J.IJFATIGUE.2021.106581.
URL https://linkinghub.elsevier.com/retrieve/pii/S0142112321004308

[199] X. Luo, N. Dang, X. Wang, The effect of laser shock peening, shot peening and their combination on the microstructure and fatigue properties of ti-6al-4v titanium alloy, International Journal of Fatigue 153 (2021) 106465. doi:https://doi.org/10.1016/j.ijfatigue.2021.106465.
URL https://www.sciencedirect.com/science/article/pii/S0142112321003236

[200] Q. Jiang, S. Li, C. Zhou, B. Zhang, Y. Zhang, Effects of laser shock peening on the ultra-high cycle fatigue performance of additively manufactured ti6al4v alloy, Optics & Laser Technology 144 (2021) 107391. doi:https://doi.org/10.1016/j.optlastec.2021.107391.
URL https://www.sciencedirect.com/science/article/pii/S0030399221004795

[201] S. Bagehorn, J. Wehr, H. Maier, Application of mechanical surface finishing processes for roughness reduction and fatigue improvement of additively manufactured ti-6al-4v parts, International Journal of Fatigue 102 (2017) 135–142. doi:10.1016/j.ijfatigue.2017.05.008.

[202] Y. Y. Sun, S. Gulizia, C. H. Oh, D. Fraser, M. Leary, Y. F. Yang, M. Qian, The Influence of As-Built Surface Conditions on Mechanical Properties of Ti-6Al-4V Additively Manufactured by Selective Electron Beam Melting, The Minerals, Metals & Materials Society 68. doi:10.1007/s11837-015-1768-y.

[203] K. Lu, J. Lu, Nanostructured surface layer on metallic materials induced by surface mechanical attrition treatment, Materials Science and Engineering: A 375 (2004) 38–45. doi:10.1016/j.msea.2003.10.261.

[204] C. Liu, D. Liu, X. Zhang, G. He, X. Xu, N. Ao, A. Ma, D. Liu, On the influence of ultrasonic surface rolling process on surface integrity and fatigue performance of ti-6al-4v alloy, Surface and Coatings Technology 370 (2019) 24–34. doi:10.1016/j.surfcoat.2019.04.080.

[205] K. Zhao, Y. Liu, T. Yao, B. Liu, Y. He, Surface nanocrystallization of ti–45al–7nb–0.3 w intermetallics induced by surface mechanical grinding treatment, Materials Letters 166 (2016) 59–62. doi:10.1016/j.matlet.2015.12.025.

[206] X. Yan, S. Yin, C. Chen, R. Jenkins, R. Lupoi, R. Bolot, W. Ma, M. Kuang, H. Liao, J. Lu, et al., Fatigue strength improvement of selective laser melted ti6al4v using ultrasonic surface mechanical attrition, Materials Research Letters 7 (8) (2019) 327–333. doi:10.1080/21663831.2019.1609110.

[207] Z. Cheng, X. Cao, X. Xu, Q. Shen, T. Yu, J. Jin, Effect of ultrasonic surface impact on the fatigue properties of Ti3Zr2Sn3Mo25Nb, Materials 13 (9) (2020) 2107. doi:10.3390/ma13092107.

[208] H. Zhang, R. Chiang, H. Qin, Z. Ren, X. Hou, D. Lin, G. L. Doll, V. K. Vasudevan, Y. Dong, C. Ye, The effects of ultrasonic nanocrystal surface modification on the fatigue performance of 3D-printed Ti64, International Journal of Fatigue 103 (2017) 136–146. doi:10.1016/J.IJFATIGUE.2017.05.019.

[209] M.-s. Kim, W.-j. Oh, G.-y. Baek, Y.-k. Jo, K.-y. Lee, S.-h. Park, D.-s. Shim, Ultrasonic nanocrystal surface modification of high-speed tool steel (aisi m4) layered via direct energy deposition, Journal of Materials Processing Technology 277 (2020) 116420. doi:10.1016/j.jmatprotec.2019.116420.

[210] H. Zhang, J. Zhao, J. Liu, H. Qin, Z. Ren, G. Doll, Y. Dong, C. Ye, The effects of electrically-assisted ultrasonic nanocrystal surface modification on 3D-printed Ti-6Al-4V alloy, Additive Manufacturing 22 (2018) 60–68. doi:10.1016/j.addma.2018.04.035.

[211] C. Liu, D. Liu, X. Zhang, D. Liu, A. Ma, N. Ao, X. Xu, Improving fatigue performance of ti-6al-4v alloy via ultrasonic surface rolling process, Journal of Materials Science & Technology 35 (8) (2019) 1555–1562. doi:10.1016/j.jmst.2019.03.036.

[212] A. Dekhtyar, B. Mordyuk, D. Savvakin, V. Bondarchuk, I. Moiseeva, N. Khripta, Enhanced fatigue behavior of powder

metallurgy ti-6al-4v alloy by applying ultrasonic impact treatment, Materials Science and Engineering: A 641 (2015) 348–359. doi:https://doi.org/10.1016/j.msea.2015.06.072.
URL https://www.sciencedirect.com/science/article/pii/S0921509315301283

[213] T. Mishurova, K. Artzt, J. Haubrich, G. Requena, G. Bruno, New aspects about the search for the most relevant parameters optimizing slm materials, Additive Manufacturing 25 (2019) 325–334.

[214] Q. ZHANG, J. CHEN, H. TAN, X. LIN, W. dong HUANG, Microstructure evolution and mechanical properties of laser additive manufactured Ti–5Al–2Sn–2Zr–4Mo–4Cr alloy, Transactions of Nonferrous Metals Society of China 26 (8) (2016) 2058–2066. doi:10.1016/S1003-6326(16)64300-5.

[215] J. Chi, Z. Cai, Z. Wan, H. Zhang, Z. Chen, L. Li, Y. Li, P. Peng, W. Guo, Effects of heat treatment combined with laser shock peening on wire and arc additive manufactured ti17 titanium alloy: Microstructures, residual stress and mechanical properties, Surface and Coatings Technology 396 (2020) 125908. doi:10.1016/j.surfcoat.2020.125908.

[216] W. Jinlong, P. Wenjie, Y. Jing, W. Jingsi, D. Mingchao, Z. Yuanliang, Effect of surface roughness on the fatigue failure and evaluation of tc17 titanium alloy, Materials Science and Technology 37 (3) (2021) 301–313. doi:10.1080/02670836.2021.1885777.

[217] Y. Zhuo, C. Yang, C. Fan, S. Lin, Effect of diameter and content of zirconium dioxide on the microstructure and mechanical properties of the tc17 titanium alloy repaired by wire arc additive manufacture, Journal of Alloys and Compounds (2021) 162295doi:10.1016/j.jallcom.2021.162295.

[218] H. Liu, H. Wang, Z. Zhang, Y. Liu, Z. Huang, Q. Wang, Q. Chen, Tensile and fatigue behavior of electron beam welded TC17 titanium alloy joint, International Journal of Fatigue 128 (2019) 105210. doi:10.1016/J.IJFATIGUE.2019.105210.

[219] H. Liu, J. Song, X. Cao, L. Xu, Y. Du, L. Li, Q. Wang, Q. Chen, Enhancement of fatigue resistance by direct aging treatment in electron beam welded Ti–5Al–2Sn–2Zr–4Mo–4Cr alloy joint, Materials Science and Engineering: A 829 (2022) 142168. doi:10.1016/J.MSEA.2021.142168.

[220] Q. Liu, Y. Wang, H. Zheng, K. Tang, H. Li, S. Gong, TC17 titanium alloy laser melting deposition repair process and properties, Optics & Laser Technology 82 (2016) 1–9. doi:10.1016/J.OPTLASTEC.2016.02.013.

[221] D. Xue, W. He, Y. Jiao, L. Zhou, S. Luo, Study on high cycle fatigue performance of TC17 titanium alloy improved by micro-scale laser shock processing, Laser & Infrared 46 (10) (2016) 1189–1194.

[222] Z. ZHAO, J. CHEN, Q. ZHANG, H. TAN, X. LIN, W. dong HUANG, Microstructure and mechanical properties of laser additive repaired ti17 titanium alloy, Transactions of Nonferrous Metals Society of China 27 (12) (2017) 2613–2621. doi:https://doi.org/10.1016/S1003-6326(17)60289-9.
URL https://www.sciencedirect.com/science/article/pii/S1003632617602899

[223] S. Luo, W. He, K. Chen, X. Nie, L. Zhou, Y. Li, Regain the fatigue strength of laser additive manufactured ti alloy via laser shock peening, Journal of Alloys and Compounds 750 (2018) 626–635.

[224] S. Luo, L. Zhou, X. Nie, Y. Li, W. He, The compound process of laser shock peening and vibratory finishing and its effect on fatigue strength of Ti-3.5 Mo-6.5 Al-1.5 Zr-0.25 Si titanium alloy, Journal of Alloys and Compounds 783 (2019) 828–835.

[225] C.-M. Liu, H.-M. Wang, X.-J. Tian, H.-B. Tang, D. Liu, Microstructure and tensile properties of laser melting deposited Ti–5Al–5Mo–5V–1Cr–1Fe near β titanium alloy, Materials Science and Engineering: A 586 (2013) 323–329.

[226] Y.-y. Zhu, C. Bo, H.-b. Tang, X. Cheng, H.-m. Wang, L. Jia, Influence of heat treatments on microstructure and mechanical properties of laser additive manufacturing Ti-5Al-2Sn-2Zr-4Mo-4Cr titanium alloy, Transactions of Nonferrous Metals Society of China 28 (1) (2018) 36–46.

[227] R. Sun, L. Li, W. Guo, P. Peng, T. Zhai, Z. Che, B. Li, C. Guo, Y. Zhu, Laser shock peening induced fatigue crack retardation in ti-17 titanium alloy, Materials Science and Engineering: A 737 (2018) 94–104.

[228] R. Sun, L. Li, Y. Zhu, P. Peng, Q. Li, W. Guo, Fatigue of ti-17 titanium alloy with hole drilled prior and post to laser shock peening, Optics & Laser Technology 115 (2019) 166–170.

[229] R. Sun, S. Keller, Y. Zhu, W. Guo, N. Kashaev, B. Klusemann, Experimental-numerical study of laser-shock-peening-induced retardation of fatigue crack propagation in ti-17 titanium alloy, International Journal of Fatigue 145 (2021) 106081. doi:https://doi.org/10.1016/j.ijfatigue.2020.106081.
URL https://www.sciencedirect.com/science/article/pii/S0142112320306137

[230] B. He, J. Li, X. Cheng, H.-M. Wang, Brittle fracture behavior of a laser additive manufactured near-β titanium alloy after low temperature aging, Materials Science and Engineering: A 699 (2017) 229–238. doi:https://doi.org/10.1016/j.msea.2017.05.050.
URL https://www.sciencedirect.com/science/article/pii/S0921509317306585

[231] X. Zhang, E. Liang, Metal additive manufacturing in aircraft: current application, opportunities and challenges, in: IOP Conference Series: Materials Science and Engineering, Vol. 493, IOP Publishing, 2019, p. 012032.

[232] X. Chen, P. Zhang, D. Wei, X. Huang, S. Adriana, C. Michel, F. Ding, F. Li, Structures and properties of ti-5al-5mo-5v-1cr-1fe after nb implantation, Surface and Coatings Technology 358 (2019) 676–687. doi:https://doi.org/10.1016/j.surfcoat.2018.11.093.

URL https://www.sciencedirect.com/science/article/pii/S0257897218313033
[233] Z. Li, X. J. Tian, H. B. Tang, H. M. Wang, Low cycle fatigue behavior of laser melting deposited TC18 titanium alloy, Transactions of Nonferrous Metals Society of China 23 (9) (2013) 2591–2597. doi:10.1016/S1003-6326(13)62772-7.
[234] K. Feng, J. ZHAI, K. YANG, P. ZHANG, D. GAO, Effect of Ceramic Shot Peening on Fatigue Properties of TC18 TitaniumAlloy by Laser Additive Manufacturing, Materials For Mechanical Engineering 44 (11) (2020) 92–96.
[235] K. Wang, R. Bao, B. Jiang, Y. Wu, D. Liu, C. Yan, Effect of primary $\alpha$ phase on the fatigue crack path of laser melting deposited ti–5al–5mo–5v–1cr–1fe near $\beta$ titanium alloy, International Journal of Fatigue 116 (2018) 535–542.
[236] K. Wang, R. Bao, T. Zhang, B. Liu, Z. Yang, B. Jiang, Fatigue crack branching in laser melting deposited ti–55511 alloy, International Journal of Fatigue 124 (2019) 217–226.
[237] B. Liu, K. Wang, R. Bao, F. Sui, The effects of $\alpha/\beta$ phase interfaces on fatigue crack deflections in additively manufactured titanium alloy: A peridynamic study, International Journal of Fatigue 137 (March). doi:10.1016/j.ijfatigue.2020.105622.
[238] X. HE, T. WANG, X. WANG, W. ZHAN, Y. LI, Fatigue behavior of direct laser deposited Ti-6.5Al-2Zr-1Mo-1V titanium alloy and its life distribution model, Chinese Journal of Aeronautics 31 (11) (2018) 2124–2135. doi:10.1016/j.cja.2018.09.003.
URL https://doi.org/10.1016/j.cja.2018.09.003
[239] Z. Zhan, Experiments and numerical simulations for the fatigue behavior of a novel TA2-TA15 titanium alloy fabricated by laser melting deposition, International Journal of Fatigue 121 (October 2018) (2019) 20–29. doi:10.1016/j.ijfatigue.2018.12.001.
URL https://doi.org/10.1016/j.ijfatigue.2018.12.001
[240] M. T. Hasib, H. E. Ostergaard, Q. Liu, X. Li, J. J. Kruzic, Tensile and fatigue crack growth behavior of commercially pure titanium produced by laser powder bed fusion additive manufacturing, Additive Manufacturing (2021) 102027.
[241] Y. Liu, D. Ren, S. Li, H. Wang, L. Zhang, T. Sercombe, Enhanced fatigue characteristics of a topology-optimized porous titanium structure produced by selective laser melting, Additive Manufacturing 32 (2020) 101060. doi:10.1016/j.addma.2020.101060.
[242] Q. Zhang, J. Chen, Z. Zhao, H. Tan, X. Lin, W. Huang, Microstructure and anisotropic tensile behavior of laser additive manufactured TC21 titanium alloy, Materials Science and Engineering: A 673 (2016) 204–212. doi:10.1016/J.MSEA.2016.07.040.
URL https://linkinghub.elsevier.com/retrieve/pii/S092150931630795X
[243] Q. Lanyun, J. Wu, W. Wang, C. Wang, C. Li, G. Yang, Microstructures and fatigue properties of ti-6al-2mo-2sn-2zr-2cr-2v titanium alloy fabricated using laser deposition manufacturing, Chinese Journal of Lasers 47 (10) (2020) 1002008.
[244] S. Lu, R. Bao, K. Wang, D. Liu, Y. Wu, B. Fei, Fatigue crack growth behaviour in laser melting deposited Ti-6.5Al-3.5Mo-1.5Zr-0.3Si alloy, Materials Science and Engineering A 690 (March) (2017) 378–386. doi:10.1016/j.msea.2017.03.001.
URL http://dx.doi.org/10.1016/j.msea.2017.03.001
[245] Y. Wu, R. Bao, Fatigue crack tip strain evolution and crack growth prediction under single overload in laser melting deposited ti-6.5al-3.5mo-1.5zr-0.3si titanium alloy, International Journal of Fatigue 116 (2018) 462–472. doi:https://doi.org/10.1016/j.ijfatigue.2018.07.011.
URL https://www.sciencedirect.com/science/article/pii/S0142112318302950
[246] N. Perevoshchikova, C. R. Hutchinson, X. Wu, The design of hot-isostatic pressing schemes for Ti–5Al–5Mo–5V–3Cr (Ti-5553), Materials Science and Engineering: A 657 (2016) 371–382.
[247] H. Schwab, F. Palm, U. Kühn, J. Eckert, Microstructure and mechanical properties of the near-beta titanium alloy Ti-5553 processed by selective laser melting, Materials & Design 105 (2016) 75–80. doi:https://doi.org/10.1016/j.matdes.2016.04.103.
URL https://www.sciencedirect.com/science/article/pii/S0264127516305913
[248] Y. Liu, S. Li, H. Wang, W. Hou, Y. Hao, R. Yang, T. Sercombe, L. Zhang, Microstructure, defects and mechanical behavior of beta-type titanium porous structures manufactured by electron beam melting and selective laser melting, Acta Materialia 113 (2016) 56–67. doi:10.1016/j.actamat.2016.04.029.
[249] Y. Liu, H. Wang, S. Li, S. Wang, W. Wang, W. Hou, Y. Hao, R. Yang, L. Zhang, Compressive and fatigue behavior of beta-type titanium porous structures fabricated by electron beam melting, Acta Materialia 126 (2017) 58–66. doi:10.1016/j.actamat.2016.12.052.
[250] S. Sui, Y. Chew, Z. Hao, F. Weng, C. Tan, Z. Du, G. Bi, Effect of cyclic heat treatment on the microstructure and mechanical properties of laser aided additive manufacturing ti-6al-2sn-4zr-2mo alloy, Advanced Powder Materialsdoi:10.1016/j.apmate.2021.09.002.
[251] C. Zopp, S. Blümer, F. Schubert, L. Kroll, Processing of a metastable titanium alloy (Ti-5553) by selective laser melting, Ain Shams Engineering Journal 8 (3) (2017) 475–479. doi:https://doi.org/10.1016/j.asej.2016.11.004.
URL https://www.sciencedirect.com/science/article/pii/S209044791630154X
[252] H. D. Carlton, K. D. Klein, J. W. Elmer, Evolution of microstructure and mechanical properties of selective laser

melted Ti-5Al-5V-5Mo-3Cr after heat treatments, Science and Technology of Welding and Joining 24 (5) (2019) 465–473. doi:10.1080/13621718.2019.1594589.

[253] C. Huang, Y. Zhao, S. Xin, C. Tan, W. Zhou, Q. Li, W. Zeng, High cycle fatigue behavior of ti–5al–5mo–5v–3cr–1zr titanium alloy with lamellar microstructure, Materials Science and Engineering: A 682 (2017) 107–116. doi:https://doi.org/10.1016/j.msea.2016.11.014.
URL https://www.sciencedirect.com/science/article/pii/S0921509316313648

[254] C. Huang, Y. Zhao, S. Xin, W. Zhou, Q. Li, W. Zeng, C. Tan, High cycle fatigue behavior of Ti–5Al–5Mo–5V–3Cr–1Zr titanium alloy with bimodal microstructure, Journal of Alloys and Compounds 695 (2017) 1966–1975. doi:10.1016/j.jallcom.2016.11.031.
URL http://dx.doi.org/10.1016/j.jallcom.2016.11.031

[255] Z. Liu, P. Liu, L. Wang, Y. Lu, X. Lu, Z. X. Qin, H. M. Wang, Fatigue properties of Ti-6.5Al-3.5Mo-l.5Zr-0.3Si alloy produced by direct laser deposition, Materials Science and Engineering A 716 (August 2017) (2018) 140–149. doi:10.1016/j.msea.2018.01.016.

[256] A. Mandal, M. M. Makhlouf, Chemical modification of morphology of Mg2Si phase in hypereutectic aluminium-silicon-magnesium alloys, International Journal of Cast Metals Research 23 (5) (2010) 303–309. doi:10.1179/136404610X12693537270019.

[257] E. O. Olakanmi, R. F. Cochrane, K. W. Dalgarno, A review on selective laser sintering/melting (SLS/SLM) of aluminium alloy powders: Processing, microstructure, and properties, Progress in Materials Science 74 (2015) 401–477. doi:10.1016/j.pmatsci.2015.03.002.

[258] D. Gu, H. Zhang, H. Chen, H. Zhang, L. Xi, Laser additive manufacturing of high-performance metallic aerospace components, Chinese Journal of Lasers 47 (5). doi:10.3788/CJL202047.0500002.

[259] N. E. Uzan, S. Ramati, R. Shneck, N. Frage, O. Yeheskel, On the effect of shot-peening on fatigue resistance of AlSi10Mg specimens fabricated by additive manufacturing using selective laser melting (AM-SLM), Additive Manufacturing 21 (2018) 458–464. doi:10.1016/j.addma.2018.03.030.

[260] M. H. Lee, J. J. Kim, K. H. Kim, N. J. Kim, S. Lee, E. W. Lee, Effects of HIPping on high-cycle fatigue properties of investment cast A356 aluminum alloys, Materials Science and Engineering A 340 (1-2) (2003) 123–129. doi:10.1016/S0921-5093(02)00157-0.

[261] Q. Yan, B. Song, Y. Shi, Comparative study of performance comparison of AlSi10Mg alloy prepared by selective laser melting and casting, Journal of Materials Science and Technology 41 (2020) 199–208. doi:10.1016/j.jmst.2019.08.049.

[262] E. Brandl, U. Heckenberger, V. Holzinger, D. Buchbinder, Additive manufactured AlSi10Mg samples using Selective Laser Melting (SLM): Microstructure, high cycle fatigue, and fracture behavior, Materials and Design 34 (2012) 159–169. doi:10.1016/j.matdes.2011.07.067.

[263] N. E. Uzan, R. Shneck, O. Yeheskel, N. Frage, Fatigue of AlSi10Mg specimens fabricated by additive manufacturing selective laser melting (AM-SLM), Materials Science and Engineering A 704 (2017) 229–237. doi:10.1016/j.msea.2017.08.027.

[264] N. T. Aboulkhair, I. Maskery, C. Tuck, I. Ashcroft, N. M. Everitt, Improving the fatigue behaviour of a selectively laser melted aluminium alloy: Influence of heat treatment and surface quality, Materials and Design 104 (2016) 174–182. doi:10.1016/j.matdes.2016.05.041.

[265] T. M. Mower, M. J. Long, Mechanical behavior of additive manufactured, powder-bed laser-fused materials, Materials Science and Engineering A 651 (2016) 198–213. doi:10.1016/j.msea.2015.10.068.

[266] I. Maskery, N. T. Aboulkhair, A. O. Aremu, C. J. Tuck, I. A. Ashcroft, R. D. Wildman, R. J. Hague, A mechanical property evaluation of graded density Al-Si10-Mg lattice structures manufactured by selective laser melting, Materials Science and Engineering A 670 (2016) 264–274. doi:10.1016/j.msea.2016.06.013.

[267] N. O. Larrosa, W. Wang, N. Read, M. H. Loretto, C. Evans, J. Carr, U. Tradowsky, M. M. Attallah, P. J. Withers, Linking microstructure and processing defects to mechanical properties of selectively laser melted AlSi10Mg alloy, Theoretical and Applied Fracture Mechanics 98 (2018) 123–133. doi:10.1016/j.tafmec.2018.09.011.

[268] M. H. Nasab, A. Giussani, D. Gastaldi, V. Tirelli, M. Vedani, Effect of surface and subsurface defects on fatigue behavior of AlSi10Mg alloy processed by laser powder bed fusion (L-PBF), Metals 9 (10) (2019) 7–10. doi:10.3390/met9101063.

[269] Z. M. Jian, G. A. Qian, D. S. Paolino, A. Tridello, F. Berto, Y. S. Hong, Crack initiation behavior and fatigue performance up to very-high-cycle regime of AlSi10Mg fabricated by selective laser melting with two powder sizes, International Journal of Fatigue 143 (2021) 106013. doi:10.1016/j.ijfatigue.2020.106013.

[270] J. Damon, S. Dietrich, F. Vollert, J. Gibmeier, V. Schulze, Process dependent porosity and the influence of shot peening on porosity morphology regarding selective laser melted AlSi10Mg parts, Additive Manufacturing 20 (2018) 77–89. doi:10.1016/j.addma.2018.01.001.

[271] M. Awd, S. Siddique, J. Johannsen, C. Emmelmann, F. Walther, Very high-cycle fatigue properties and microstructural damage mechanisms of selective laser melted AlSi10Mg alloy, International Journal of Fatigue 124 (2019) 55–69. doi:10.1016/j.ijfatigue.2019.02.040.

[272] M. V. Gerov, E. Y. Vladislavskaya, V. F. Terent'ev, D. V. Prosvirnin, O. S. Antonova, A. G. Kolmakov, Fatigue Strength of an AlSi10Mg Alloy Fabricated by Selective Laser Melting, Russian Metallurgy (Metally) 2019 (4) (2019) 392–397. doi:10.1134/S0036029519040098.

[273] Z. W. Xu, Q. Wang, X. S. Wang, C. H. Tan, M. H. Guo, P. B. Gao, High cycle fatigue performance of AlSi10mg alloy produced by selective laser melting, Mechanics of Materials 148. doi:10.1016/j.mechmat.2020.103499.

[274] A. Tridello, J. Fiocchi, C. A. Biffi, G. Chiandussi, M. Rossetto, A. Tuissi, D. S. Paolino, Influence of the annealing and defects on the VHCF behavior of an SLM AlSi10Mg alloy, Fatigue and Fracture of Engineering Materials and Structures 42 (12) (2019) 2794–2807. doi:10.1111/ffe.13123.

[275] W. H. Kan, Y. Nadot, M. Foley, L. Ridosz, G. Proust, J. M. Cairney, Factors that affect the properties of additively-manufactured AlSi10Mg: Porosity versus microstructure, Additive Manufacturing 29 (2019) 100805. doi:10.1016/j.addma.2019.100805.

[276] M. S. Baek, R. Kreethi, T. H. Park, Y. Sohn, K. A. Lee, Influence of heat treatment on the high-cycle fatigue properties and fatigue damage mechanism of selective laser melted AlSi10Mg alloy, Materials Science and Engineering A 819 (2021) 141486. doi:10.1016/j.msea.2021.141486.

[277] R. K. Rhein, Q. Shi, S. Arjun Tekalur, J. Wayne Jones, J. W. Carroll, Effect of direct metal laser sintering build parameters on defects and ultrasonic fatigue performance of additively manufactured AlSi10Mg, Fatigue and Fracture of Engineering Materials and Structures 44 (2) (2021) 295–305. doi:10.1111/ffe.13355.

[278] R. K. Rhein, Q. Shi, S. A. Tekalur, J. W. Jones, J. W. Carroll, Short-Crack Growth Behavior in Additively Manufactured AlSi10Mg Alloy, Journal of Materials Engineering and Performance 30 (7) (2021) 5392–5398. doi:10.1007/s11665-021-05820-2.

[279] J. Zhang, J. Li, S. Wu, W. Zhang, J. Sun, G. Qian, High-cycle and very-high-cycle fatigue lifetime prediction of additively manufactured AlSi10Mg via crystal plasticity finite element method, International Journal of Fatigue 155 (2021) 106577. doi:10.1016/j.ijfatigue.2021.106577.

[280] S. R. Ch, A. Raja, R. Jayaganthan, N. J. Vasa, M. Raghunandan, Study on the fatigue behaviour of selective laser melted AlSi10Mg alloy, Materials Science and Engineering A 781 (2020) 139180. doi:10.1016/j.msea.2020.139180.

[281] Z. Xu, A. Liu, X. Wang, Fatigue performance and crack propagation behavior of selective laser melted AlSi10Mg in 0°, 15°, 45° and 90° building directions, Materials Science and Engineering A 812. doi:10.1016/j.msea.2021.141141.

[282] S. GLODEŽ, J. KLEMENC, F. ZUPANIČ, M. VESENJAK, High-cycle fatigue and fracture behaviours of SLM AlSi10Mg alloy, Transactions of Nonferrous Metals Society of China (English Edition) 30 (10) (2020) 2577–2589. doi:10.1016/S1003-6326(20)65403-6.

[283] M. Awd, F. Stern, A. Kampmann, D. Kotzem, J. Tenkamp, F. Walther, Microstructural characterization of the anisotropy and cyclic deformation behavior of selective laser melted AlSi10Mg structures, Metals 8 (10). doi:10.3390/met8100825.

[284] K. Schnabel, J. Baumgartner, B. Möller, M. Scurria, Fatigue assessment of additively manufactured AlSi10Mg structures using effective stress concepts based on the critical distance approach, Welding in the World 65 (11) (2021) 2119–2133. doi:10.1007/s40194-021-01153-9.

[285] M. Tang, P. C. Pistorius, Oxides, porosity and fatigue performance of AlSi10Mg parts produced by selective laser melting, International Journal of Fatigue 94 (2017) 192–201. doi:10.1016/j.ijfatigue.2016.06.002.

[286] A. Pola, D. Battini, M. Tocci, A. Avanzini, L. Girelli, C. Petrogalli, M. Gelfi, Evaluation on the fatigue behavior of sand-blasted AlSi10Mg obtained by DMLS, Frattura ed Integrita Strutturale 13 (49) (2019) 775–790. doi:10.3221/IGF-ESIS.49.69.

[287] L. SUTTEY, V. SUDHAKAR, Fatigue and Fracture Behavior of Lpbf Manufactured Aluminum Alloy, European Journal of Materials Science and Engineering 4 (2) (2019) 53–60. doi:10.36868/ejmse.2019.04.02.053.

[288] M. Awd, J. Tenkamp, M. Hirtler, S. Siddique, M. Bambach, F. Walther, Comparison of microstructure and mechanical properties of Scalmalloy produced by selective laser melting and laser metal deposition, Materials 11 (1). doi:10.3390/ma11010017.

[289] S. Romano, A. Brückner-Foit, A. Brandão, J. Gumpinger, T. Ghidini, S. Beretta, Fatigue properties of AlSi10Mg obtained by additive manufacturing: Defect-based modelling and prediction of fatigue strength, Engineering Fracture Mechanics 187 (2018) 165–189. doi:10.1016/j.engfracmech.2017.11.002.

[290] P. Ferro, A. Fabrizi, F. Berto, G. Savio, R. Meneghello, S. Rosso, Defects as a root cause of fatigue weakening of additively manufactured AlSi10Mg components, Theoretical and Applied Fracture Mechanics 108 (2020) 102611. doi:10.1016/j.tafmec.2020.102611.

[291] M. Muhammad, P. D. Nezhadfar, S. Thompson, A. Saharan, N. Phan, N. Shamsaei, A comparative investigation on the microstructure and mechanical properties of additively manufactured aluminum alloys, International Journal of Fatigue 146 (2021) 106165. doi:10.1016/j.ijfatigue.2021.106165.

[292] J. G. Santos Macías, C. Elangeswaran, L. Zhao, J. Y. Buffière, B. Van Hooreweder, A. Simar, Fatigue crack nucleation and growth in laser powder bed fusion AlSi10Mg under as built and post-treated conditions, Materials and Design 210. doi:10.1016/j.matdes.2021.110084.

[293] M. H. Nasab, A. Giussani, D. Gastaldi, V. Tirelli, M. Vedani, Size-effect in Very High Cycle Fatigue: A review, International Journal of Fatigue 153 (2021) 106462. doi:10.1016/j.ijfatigue.2021.106462.

[294] F. Sajadi, J. M. Tiemann, N. Bandari, A. C. Darabi, J. Mola, S. Schmauder, Fatigue improvement of alsi10mg fabricated by laser-based powder bed fusion through heat treatment, Metals 11 (5). doi:10.3390/met11050683.

[295] G. Nicoletto, INFLUENCE OF ROUGH AS-BUILT SURFACES ON SMOOTH AND NOTCHED FATIGUE BEHAVIOR OF L-PBF AlSi10Mg, Additive Manufacturing 34 (2020) 101251. doi:10.1016/j.addma.2020.101251.

[296] W. Qian, S. Wu, Z. Wu, S. Ahmed, W. Zhang, G. Qian, P. J. Withers, In situ X-ray imaging of fatigue crack growth from multiple defects in additively manufactured AlSi10Mg alloy, International Journal of Fatigue 155 (2022) 106616. doi:10.1016/j.ijfatigue.2021.106616.

[297] J. N. Domfang Ngnekou, Y. Nadot, G. Henaff, J. Nicolai, L. Ridosz, Effect of as-built and ground surfaces on the fatigue properties of alsi10mg alloy produced by additive manufacturing, Metals 11 (9). doi:10.3390/met11091432.

[298] P. D. Nezhadfar, S. Thompson, A. Saharan, N. Phan, N. Shamsaei, Structural integrity of additively manufactured aluminum alloys: Effects of build orientation on microstructure, porosity, and fatigue behavior, Additive Manufacturing 47 (2021) 102292. doi:10.1016/j.addma.2021.102292.

[299] L. Barricelli, S. Beretta, Analysis of prospective SIF and shielding effect for cylindrical rough surfaces obtained by L-PBF, Engineering Fracture Mechanics 256 (2021) 107983. doi:10.1016/j.engfracmech.2021.107983.

[300] A. Kempf, J. Kruse, M. Madia, K. Hilgenberg, Correlation between quasistatic und fatigue properties of additively manufactured AlSi10Mg using Laser Powder Bed Fusion, Procedia Structural Integrity 38 (2021) 77–83. doi:10.1016/j.prostr.2022.03.009.

[301] Z. Wu, S. Wu, J. Bao, W. Qian, S. Karabal, W. Sun, P. J. Withers, The effect of defect population on the anisotropic fatigue resistance of AlSi10Mg alloy fabricated by laser powder bed fusion, International Journal of Fatigue 151 (2021) 106317. doi:10.1016/j.ijfatigue.2021.106317.

[302] G. Nicoletto, Fatigue Behavior of L-PBF Metals: Cost-Effective Characterization via Specimen Miniaturization, Journal of Materials Engineering and Performance 30 (7) (2021) 5227–5234. doi:10.1007/s11665-021-05717-0.

[303] N. Agenbag, C. McDuling, Fatigue Life Testing of Locally Additive Manufactured AlSilOMg Test Specimens, R&D Journal 37 (2021) 19–25. doi:10.17159/2309-8988/2019/v37a3.

[304] X. Lesperance, P. Ilie, A. Ince, Very high cycle fatigue characterization of additively manufactured AlSi10Mg and AlSi7Mg aluminium alloys based on ultrasonic fatigue testing, Fatigue and Fracture of Engineering Materials and Structures 44 (3) (2021) 876–884. doi:10.1111/ffe.13406.

[305] A. Tridello, J. Fiocchi, C. A. Biffi, G. Chiandussi, M. Rossetto, A. Tuissi, D. S. Paolino, Effect of microstructure, residual stresses and building orientation on the fatigue response up to 109 cycles of an SLM AlSi10Mg alloy, International Journal of Fatigue 137 (2020) 105659. doi:10.1016/j.ijfatigue.2020.105659.

[306] E. Maleki, S. Bagherifard, S. Razavi, M. Riccio, M. Bandini, A. du Plessis, F. Berto, M. Guagliano, Fatigue behaviour of notched laser powder bed fusion AlSi10Mg after thermal and mechanical surface post-processing, Materials Science and Engineering: A 829 (2021) 142145. doi:10.1016/j.msea.2021.142145.

[307] A. Tommasi, N. Maillol, A. Bertinetti, P. Penchev, J. Bajolet, F. Gili, D. Pullini, D. B. Mataix, Influence of surface preparation and heat treatment on mechanical behavior of hybrid aluminum parts manufactured by a combination of laser powder bed fusion and conventional manufacturing processes, Metals 11 (3) (2021) 1–10. doi:10.3390/met11030522.

[308] N. Stelzer, M. Scheerer, Z. Simon, T. Sebald, H. Gschiel, M. Hatzenbichler, B. Bonvoisin, A. C. Gmbh, A. G. Gmbh, W. Neustadt, E. S. Agency, Properties of Surface Engineered Metallic Parts Prepared by Additive Manufacturing, 15th ECSSMETdoi:10.20944/preprints202102.0128.v1.

[309] F. Del Re, V. Contaldi, A. Astarita, B. Palumbo, A. Squillace, P. Corrado, P. Di Petta, Statistical approach for assessing the effect of powder reuse on the final quality of AlSi10Mg parts produced by laser powder bed fusion additive manufacturing, International Journal of Advanced Manufacturing Technology 97 (2018) 2231–2240. doi:10.1007/s00170-018-2090-y.

[310] X. P. Li, G. Ji, Z. Chen, A. Addad, Y. Wu, H. W. Wang, J. Vleugels, J. Van Humbeeck, J. P. Kruth, Selective laser melting of nano-TiB2 decorated AlSi10Mg alloy with high fracture strength and ductility, Acta Materialia 129 (2017) 183–193. doi:10.1016/j.actamat.2017.02.062.

[311] D. D. Gu, W. Meiners, K. Wissenbach, R. Poprawe, Laser additive manufacturing of metallic components: Materials, processes and mechanisms, International Materials Reviews 57 (3) (2012) 133–164. doi:10.1179/1743280411Y.0000000014.

[312] Z. Wang, W. Wu, G. Qian, L. Sun, X. Li, J. A. Correia, In-situ SEM investigation on fatigue behaviors of additive manufactured Al-Si10-Mg alloy at elevated temperature, Engineering Fracture Mechanics 214 (2019) 149–163. doi:10.1016/j.engfracmech.2019.03.040.

[313] E. Beevers, A. D. Brandão, J. Gumpinger, M. Gschweitl, C. Seyfert, P. Hofbauer, T. Rohr, T. Ghidini, Fatigue properties and material characteristics of additively manufactured AlSi10Mg - Effect of the contour parameter on the microstructure, density, residual stress, roughness and mechanical properties, International Journal of Fatigue 117 (2018) 148–162. doi:10.1016/j.ijfatigue.2018.08.023.

[314] C. Fischer, C. Schweizer, Lifetime assessment of the process-dependent material properties of additive manufactured AlSi10Mg under low-cycle fatigue loading, MATEC Web of Conferences 326 (2020) 07003. doi:10.1051/matecconf/202032607003.

[315] N. K. Tolochko, S. E. Mozzharov, I. A. Yadroitsev, T. Laoui, L. Froyen, V. I. Titov, M. B. Ignatiev, Balling processes during selective laser treatment of powders, Rapid Prototyping Journal 10 (2) (2004) 78–87. doi:10.1108/13552540410526953.

[316] K. Kempen, L. Thijs, J. V. Humbeeck, J. P. Kruth, Processing AlSi10Mg by selective laser melting: Parameter optimisation and material characterisation, Materials Science and Technology (United Kingdom) 31 (8) (2015) 917–923. doi:10.1179/1743284714Y.0000000702.

[317] X. Zhou, X. Liu, D. Zhang, Z. Shen, W. Liu, Balling phenomena in selective laser melted tungsten, Journal of Materials Processing Technology 222 (2015) 33–42. doi:10.1016/j.jmatprotec.2015.02.032.

[318] Z. Zhan, H. Li, K. Y. Lam, Development of a novel fatigue damage model with AM effects for life prediction of commonly-used alloys in aerospace, International Journal of Mechanical Sciences 155 (2019) 110–124. doi:10.1016/j.ijmecsci.2019.02.032.

[319] E. O. Olakanmi, Selective laser sintering/melting (SLS/SLM) of pure Al, Al-Mg, and Al-Si powders: Effect of processing conditions and powder properties, Journal of Materials Processing Technology 213 (8) (2013) 1387–1405. doi:10.1016/j.jmatprotec.2013.03.009.

[320] N. Read, W. Wang, K. Essa, M. M. Attallah, Selective laser melting of AlSi10Mg alloy: Process optimisation and mechanical properties development, Materials and Design 65 (2015) 417–424. doi:10.1016/j.matdes.2014.09.044.

[321] D. Buchbinder, H. Schleifenbaum, S. Heidrich, W. Meiners, J. Bültmann, High Power Selective Laser Melting ( HP SLM ) of Aluminum Parts, Physics Procedia 12 (2011) 271–278. doi:10.1016/j.phpro.2011.03.035.

[322] M. Awd, S. Siddique, F. Walther, Microstructural damage and fracture mechanisms of selective laser melted Al-Si alloys under fatigue loading, Theoretical and Applied Fracture Mechanics 106 (2020) 102483. doi:10.1016/j.tafmec.2020.102483.

[323] M. J. Paul, Q. Liu, J. P. Best, X. Li, J. J. Kruzic, U. Ramamurty, B. Gludovatz, Fracture resistance of AlSi10Mg fabricated by laser powder bed fusion, Acta Materialia 211 (2021) 116869. doi:10.1016/j.actamat.2021.116869.

[324] K. Riener, N. Albrecht, S. Ziegelmeier, R. Ramakrishnan, L. Haferkamp, A. B. Spierings, G. J. Leichtfried, Influence of particle size distribution and morphology on the properties of the powder feedstock as well as of AlSi10Mg parts produced by laser powder bed fusion (LPBF), Additive Manufacturing 34 (2020) 101286. doi:10.1016/j.addma.2020.101286.

[325] U. Tradowsky, J. White, R. M. Ward, N. Read, W. Reimers, M. M. Attallah, Selective laser melting of AlSi10Mg: Influence of post-processing on the microstructural and tensile properties development, Materials and Design 105 (2016) 212–222. doi:10.1016/j.matdes.2016.05.066.

[326] M. Mohammadi, H. Asgari, Achieving low surface roughness AlSi10Mg 200C parts using direct metal laser sintering, Additive Manufacturing 20 (2018) 23–32. doi:10.1016/j.addma.2017.12.012.

[327] G. Qian, Z. Jian, Y. Qian, X. Pan, X. Ma, Y. Hong, Very-high-cycle fatigue behavior of AlSi10Mg manufactured by selective laser melting: Effect of build orientation and mean stress, International Journal of Fatigue 138 (2020) 105696. doi:10.1016/j.ijfatigue.2020.105696.

[328] J. N. Domfang Ngnekou, Y. Nadot, G. Henaff, J. Nicolai, L. Ridosz, Influence of defect size on the fatigue resistance of AlSi10Mg alloy elaborated by selective laser melting (SLM), Procedia Structural Integrity 7 (2017) 75–83. doi:10.1016/j.prostr.2017.11.063.

[329] J. N. Domfang Ngnekou, Y. Nadot, G. Henaff, J. Nicolai, W. H. Kan, J. M. Cairney, L. Ridosz, Fatigue properties of AlSi10Mg produced by Additive Layer Manufacturing, International Journal of Fatigue 119 (2019) 160–172. doi:10.1016/j.ijfatigue.2018.09.029.

[330] C. Zhang, H. Zhu, H. Liao, Y. Cheng, Z. Hu, X. Zeng, Effect of heat treatments on fatigue property of selective laser melting AlSi10Mg, International Journal of Fatigue 116 (2018) 513–522. doi:10.1016/j.ijfatigue.2018.07.016.

[331] N. Takata, H. Kodaira, K. Sekizawa, A. Suzuki, M. Kobashi, Change in microstructure of selectively laser melted AlSi10Mg alloy with heat treatments, Materials Science and Engineering A 704 (2017) 218–228. doi:10.1016/j.msea.2017.08.029.

[332] W. Schneller, M. Leitner, S. Springer, F. Grün, M. Taschauer, Effect of hip treatment on microstructure and fatigue strength of selectively laser melted AlSi10Mg, Journal of Manufacturing and Materials Processing 3 (1). doi:10.3390/jmmp3010016.

[333] W. Schneller, M. Leitner, S. Pomberger, S. Springer, F. Beter, F. Grün, Effect of post treatment on the microstructure, surface roughness and residual stress regarding the fatigue strength of selectively laser melted AlSi10Mg structures, Journal of Manufacturing and Materials Processing 3 (4). doi:10.3390/jmmp3040089.

[334] J. Bao, Z. Wu, S. Wu, P. J. Withers, F. Li, S. Ahmed, A. Benaarbia, W. Sun, Hot dwell-fatigue behaviour of additively manufactured AlSi10Mg alloy: Relaxation, cyclic softening and fracture mechanisms, International Journal of Fatigue 151 (2021) 106408. doi:10.1016/j.ijfatigue.2021.106408.

[335] J. G. Santos Macías, C. Elangeswaran, L. Zhao, B. Van Hooreweder, J. Adrien, E. Maire, J. Y. Buffière, W. Ludwig, P. J. Jacques, A. Simar, Ductilisation and fatigue life enhancement of selective laser melted AlSi10Mg by friction stir processing, Scripta Materialia 170 (2019) 124–128. doi:10.1016/j.scriptamat.2019.05.044.

[336] B. J. Mfusi, N. R. Mathe, L. C. Tshabalala, P. A. Popoola, The effect of stress relief on the mechanical and fatigue properties of additively manufactured AlSi10Mg parts, Metals 9 (11) (2019) 1–14. doi:10.3390/met9111216.

[337] Y. Wang, J. Wang, H. Zhang, H. Zhao, D. Ni, B. Xiao, Z. Ma, Effects of Heat Treatments on Microstructure and Mechanical Properties of AlSi10Mg Alloy Produced by Selective Laser Melting, Jinshu Xuebao/Acta Metallurgica Sinica 57 (5) (2021) 613–622. doi:10.11900/0412.1961.2020.00253.

[338] W. Zhang, H. Zhu, Z. Hu, X. Zeng, Study on the Selective Laser Melting of AlSi10Mg, Jinshu Xuebao/Acta Metallurgica Sinica 53 (8) (2017) 918–926. doi:10.11900/0412.1961.2016.00472.

[339] A. Tridello, C. A. Biffi, J. Fiocchi, P. Bassani, G. Chiandussi, M. Rossetto, A. Tuissi, D. S. Paolino, VHCF response of as-built SLM AlSi10Mg specimens with large loaded volume, Fatigue and Fracture of Engineering Materials and Structures 41 (9) (2018) 1918–1928. doi:10.1111/ffe.12830.

[340] H. Khajehmirza, A. Heydari Astaraee, S. Monti, M. Guagliano, S. Bagherifard, A hybrid framework to estimate the surface state and fatigue performance of laser powder bed fusion materials after shot peening, Applied Surface Science 567 (2021) 150758. doi:10.1016/j.apsusc.2021.150758.

[341] A. Maamoun, M. Elbestawi, S. Veldhuis, Influence of Shot Peening on AlSi10Mg Parts Fabricated by Additive Manufac- turing, Journal of Manufacturing and Materials Processing 2 (3) (2018) 40. doi:10.3390/jmmp2030040.

[342] A. Du Plessis, D. Glaser, H. Moller, N. Mathe, L. Tshabalala, B. Mfusi, R. Mostert, Pore Closure Effect of Laser Shock Peening of Additively Manufactured AlSi10Mg, 3D Printing and Additive Manufacturing 6 (5) (2019) 245–252. doi:10.1089/3dp.2019.0064.

[343] C. Cai, H. Geng, Q. Cui, S. Wang, Z. Zhang, Low cycle fatigue behavior of AlSi10Mg(Cu) alloy at high temperature, Materials Characterization 145 (2018) 594–605. doi:10.1016/j.matchar.2018.09.023.

[344] M. T. Di Giovanni, J. T. O. de Menezes, G. Bolelli, E. Cerri, E. M. Castrodeza, Fatigue crack growth behavior of a selective laser melted AlSi10Mg, Engineering Fracture Mechanics 217 (2019) 106564. doi:10.1016/j.engfracmech.2019.106564.

[345] J. Tenkamp, M. Awd, S. Siddique, P. Starke, F. Walther, Fracture-mechanical assessment of the effect of defects on the fatigue lifetime and limit in cast and additively manufactured aluminum-silicon alloys from hcf to vhcf regime, Metals 10 (7) (2020) 1–18. doi:10.3390/met10070943.

[346] P. Konda Gokuldoss, Work hardening in selective laser melted Al-12Si alloy, Material Design and Processing Communications 1 (2) (2019) 10–13. doi:10.1002/mdp2.46.

[347] S. Siddique, E. Wycisk, G. Frieling, C. Emmelmann, F. Walther, Microstructural and Mechanical Properties of Selective Laser Melted Al 4047, Applied Mechanics and Materials 752-753 (Dmd) (2015) 485–490. doi:10.4028/www.scientific.net/amm.752-753.485.

[348] R. Rashid, S. H. Masood, D. Ruan, S. Palanisamy, R. A. Rahman Rashid, J. Elambasseril, M. Brandt, Effect of energy per layer on the anisotropy of selective laser melted AlSi12 aluminium alloy, Additive Manufacturing 22 (May) (2018) 426–439. doi:10.1016/j.addma.2018.05.040.
URL https://doi.org/10.1016/j.addma.2018.05.040

[349] P. Ma, K. Prashanth, S. Scudino, Y. Jia, H. Wang, C. Zou, Z. Wei, J. Eckert, Influence of Annealing on Mechanical Properties of Al-20Si Processed by Selective Laser Melting, Metals 4 (1) (2014) 28–36. doi:10.3390/met4010028.

[350] P. Ma, Y. Jia, K. G. Prashanth, Z. Yu, C. Li, J. Zhao, S. Yang, L. Huang, Effect of Si content on the microstructure and properties of Al-Si alloys fabricated using hot extrusion, Journal of Materials Research 32 (11) (2017) 2210–2217. doi:10.1557/jmr.2017.97.

[351] S. Siddique, M. Awd, J. Tenkamp, F. Walther, High and very high cycle fatigue failure mechanisms in selective laser melted aluminum alloys, Journal of Materials Research 32 (23) (2017) 4296–4304. doi:10.1557/jmr.2017.314.

[352] S. Siddique, M. Imran, F. Walther, Very high cycle fatigue and fatigue crack propagation behavior of selective laser melted AlSi12 alloy, International Journal of Fatigue 94 (2017) 246–254. doi:10.1016/j.ijfatigue.2016.06.003.
URL http://dx.doi.org/10.1016/j.ijfatigue.2016.06.003

[353] S. Siddique, M. Imran, E. Wycisk, C. Emmelmann, F. Walther, Influence of process-induced microstructure and imperfections on mechanical properties of AlSi12 processed by selective laser melting, Journal of Materials Processing Technology 221 (2015) 205–213. doi:10.1016/j.jmatprotec.2015.02.023.
URL http://dx.doi.org/10.1016/j.jmatprotec.2015.02.023

[354] S. Siddique, M. Imran, E. Wycisk, C. Emmelmann, F. Walther, Fatigue Assessment of Laser Additive Manufactured AlSi12 Eutectic Alloy in the Very High Cycle Fatigue (VHCF) Range up to 1E9 cycles, Materials Today: Proceedings 3 (9) (2016) 2853–2860. doi:10.1016/j.matpr.2016.07.004.
URL http://dx.doi.org/10.1016/j.matpr.2016.07.004

[355] J. Richter, S. V. Sajadifar, T. Niendorf, On the influence of process interruptions during additive manufacturing on the fatigue resistance of AlSi12, Additive Manufacturing 47 (2021) 102346. doi:10.1016/j.addma.2021.102346.

URL https://doi.org/10.1016/j.addma.2021.102346
[356] S. Siddique, M. Awd, J. Tenkamp, F. Walther, Development of a stochastic approach for fatigue life prediction of AlSi12 alloy processed by selective laser melting, Engineering Failure Analysis 79 (2017) 34–50. doi:10.1016/j.engfailanal.2017.03.015.
URL http://dx.doi.org/10.1016/j.engfailanal.2017.03.015
[357] J. Suryawanshi, K. G. Prashanth, S. Scudino, J. Eckert, O. Prakash, U. Ramamurty, Simultaneous enhancements of strength and toughness in an Al-12Si alloy synthesized using selective laser melting, Acta Materialia 115 (2016) 285–294. doi:10.1016/j.actamat.2016.06.009.
URL http://dx.doi.org/10.1016/j.actamat.2016.06.009
[358] N. Shamsaei, J. Simsiriwong, Fatigue behaviour of additively-manufactured metallic parts, Procedia Structural Integrity 7 (2017) 3–10. doi:10.1016/j.prostr.2017.11.053.
URL https://doi.org/10.1016/j.prostr.2017.11.053
[359] S. Siddique, M. Imran, M. Rauer, M. Kaloudis, E. Wycisk, C. Emmelmann, F. Walther, Computed tomography for characterization of fatigue performance of selective laser melted parts, Materials and Design 83. doi:10.1016/j.matdes.2015.06.063.
URL http://dx.doi.org/10.1016/j.matdes.2015.06.063
[360] J. H. Rao, Y. Zhang, A. Huang, X. Wu, K. Zhang, Improving fatigue performances of selective laser melted Al-7Si-0.6Mg alloy via defects control, International Journal of Fatigue 129 (July) (2019) 105215. doi:10.1016/j.ijfatigue.2019.105215.
URL https://doi.org/10.1016/j.ijfatigue.2019.105215
[361] X. Lesperance, P. Ilie, A. Ince, Very high cycle fatigue characterization of additively manufactured AlSi10Mg and AlSi7Mg aluminium alloys based on ultrasonic fatigue testing, Fatigue and Fracture of Engineering Materials and Structures 44 (3) (2021) 876–884. doi:10.1111/ffe.13406.
[362] K. V. Yang, P. Rometsch, T. Jarvis, J. Rao, S. Cao, C. Davies, X. Wu, Porosity formation mechanisms and fatigue response in Al-Si-Mg alloys made by selective laser melting, Materials Science and Engineering A 712 (November 2017) (2018) 166–174. doi:10.1016/j.msea.2017.11.078.
URL https://doi.org/10.1016/j.msea.2017.11.078
[363] E. Bassoli, L. Denti, A. Comin, A. Sola, E. Tognoli, Fatigue behavior of as-built L-PBF A357.0 parts, Metals 8 (8). doi:10.3390/met8080634.
[364] L. F. L. Martins, P. R. Provencher, M. Brochu, M. Brochu, Effect of platform temperature and post-processing heat treatment on the fatigue life of additively manufactured alsi7mg alloy, Metals 11 (5). doi:10.3390/met11050679.
[365] T. Kimura, T. Nakamoto, Microstructures and mechanical properties of A356 (AlSi7Mg0.3) aluminum alloy fabricated by selective laser melting, Materials and Design 89 (2016) 1294–1301. doi:10.1016/j.matdes.2015.10.065.
URL http://dx.doi.org/10.1016/j.matdes.2015.10.065
[366] A. Aversa, M. Lorusso, F. Trevisan, E. P. Ambrosio, F. Calignano, D. Manfredi, S. Biamino, P. Fino, M. Lombardi, M. Pavese, Effect of process and post-process conditions on the mechanical properties of an A357 alloy produced via laser powder bed fusion, Metals 7 (2) (2017) 1–9. doi:10.3390/met7020068.
[367] H. Rao, S. Giet, K. Yang, X. Wu, C. H. Davies, The influence of processing parameters on aluminium alloy A357 manufactured by Selective Laser Melting, Materials and Design 109 (2016) 334–346. doi:10.1016/j.matdes.2016.07.009.
URL http://dx.doi.org/10.1016/j.matdes.2016.07.009
[368] M. Hamidi Nasab, S. Romano, D. Gastaldi, S. Beretta, M. Vedani, Combined effect of surface anomalies and volumetric defects on fatigue assessment of AlSi7Mg fabricated via laser powder bed fusion, Additive Manufacturing 34 (October 2019) (2020) 100918. doi:10.1016/j.addma.2019.100918.
URL https://doi.org/10.1016/j.addma.2019.100918
[369] L. Boniotti, S. Beretta, L. Patriarca, L. Rigoni, S. Foletti, Experimental and numerical investigation on compressive fatigue strength of lattice structures of AlSi7Mg manufactured by SLM, International Journal of Fatigue 128 (December 2018) (2019) 105181. doi:10.1016/j.ijfatigue.2019.06.041.
URL https://doi.org/10.1016/j.ijfatigue.2019.06.041
[370] M. Bonneric, C. Brugger, N. Saintier, Investigation of the sensitivity of the fatigue resistance to defect position in aluminium alloys obtained by Selective laser melting using artificial defects, International Journal of Fatigue 134 (October 2019) (2020) 105505. doi:10.1016/j.ijfatigue.2020.105505.
URL https://doi.org/10.1016/j.ijfatigue.2020.105505
[371] Z. Sajedi, R. Casati, M. Cecilia Poletti, R. Wang, F. Iranshahi, M. Vedani, Comparative thermal fatigue behavior of AlSi7Mg alloy produced by L-PBF and sand casting, International Journal of Fatigue 152 (July) (2021) 106424. doi:10.1016/j.ijfatigue.2021.106424.
[372] Tsuyoshi-Takahashi, K. Sasaki, Low cycle thermal fatigue of aluminum alloy cylinder head in consideration of changing metrology microstructure, Procedia Engineering 2 (1) (2010) 767–776. doi:10.1016/j.proeng.2010.03.083.

URL http://dx.doi.org/10.1016/j.proeng.2010.03.083

[373] Z. C, R. Casati, M. C. Poletti, M. Skalon, M. Vedani, Thermal fatigue testing of laser powder bed fusion (L-PBF) processed AlSi7Mg alloy in presence of a quasi-static tensile load, Materials Science and Engineering A 789 (May) (2020) 139617. doi:10.1016/j.msea.2020.139617.
URL https://doi.org/10.1016/j.msea.2020.139617

[374] Z. Sajedi, R. Casati, M. C. Poletti, M. Skalon, C. Sommitsch, M. Vedani, et al., Thermo-mechanical fatigue behaviour of alsi7mg alloy processed by selective laser melting, in: Euro PM2019 Conference, Proceedings of the Euro PM2019-Congress and Exhibition-Maastricht, 2019.

[375] Z. Qin, N. Kang, F. Zhang, Z. Wang, Q. Wang, J. Chen, X. Lin, W. Huang, Role of defects on the high cycle fatigue behavior of selective laser melted Al-Mg-Sc-Zr alloy, International Journal of Fracture 0. doi:10.1007/s10704-021-00593-0.
URL https://doi.org/10.1007/s10704-021-00593-0

[376] Z. Qin, N. Kang, M. El Mansori, Z. Wang, H. Wang, X. Lin, J. Chen, W. Huang, Anisotropic high cycle fatigue property of Sc and Zr-modified Al-Mg alloy fabricated by laser powder bed fusion, Additive Manufacturing 49 (August 2021) (2022) 102514. doi:10.1016/j.addma.2021.102514.
URL https://doi.org/10.1016/j.addma.2021.102514

[377] P. He, R. F. Webster, V. Yakubov, H. Kong, Q. Yang, S. Huang, M. Ferry, J. J. Kruzic, X. Li, Fatigue and dynamic aging behavior of a high strength Al-5024 alloy fabricated by laser powder bed fusion additive manufacturing, Acta Materialia 220. doi:10.1016/j.actamat.2021.117312.
URL https://doi.org/10.1016/j.actamat.2021.117312

[378] F. Lasagni, C. Galleguillos, M. Herrera, J. Santaolaya, D. Hervás, S. Gonzalez, A. Periñán, On the processability and mechanical behavior of Al-Mg-Sc alloy for PBF-LB, Progress in Additive Manufacturingdoi:10.1007/s40964-021-00216-z.
URL https://doi.org/10.1007/s40964-021-00216-z

[379] W. Schneller, M. Leitner, S. Leuders, J. M. Sprauel, F. Grün, T. Pfeifer, O. Jantschner, Fatigue strength estimation methodology of additively manufactured metallic bulk material, Additive Manufacturing 39 (November 2020) (2021) 101688. doi:10.1016/j.addma.2020.101688.
URL https://doi.org/10.1016/j.addma.2020.101688

[380] M. Awd, J. Tenkamp, M. Hirtler, S. Siddique, M. Bambach, F. Walther, Comparison of microstructure and mechanical properties of Scalmalloy produced by selective laser melting and laser metal deposition, Materials 11 (1). doi:10.3390/ma11010017.

[381] P. Chernyshova, T. Guraya, S. Singamneni, T. Zhu, Z. W. Chen, Fatigue Crack Growth Behavior of Al-4.5Mg-0.6Sc-0.3Zr Alloy Processed by Laser Powder Bed Fusion, Journal of Materials Engineering and Performance 30 (9) (2021) 6743–6751. doi:10.1007/s11665-021-05989-6.

[382] A. B. Spierings, K. Dawson, M. Voegtlin, F. Palm, P. J. Uggowitzer, Microstructure and mechanical properties of as-processed scandium-modified aluminium using selective laser melting, CIRP Annals - Manufacturing Technology 65 (1) (2016) 213–216. doi:10.1016/j.cirp.2016.04.057.
URL http://dx.doi.org/10.1016/j.cirp.2016.04.057

[383] A. B. Spierings, K. Dawson, T. Heeling, P. J. Uggowitzer, R. Schäublin, F. Palm, K. Wegener, Microstructural features of Sc- and Zr-modified Al-Mg alloys processed by selective laser melting, Materials and Design 115 (2017) 52–63. doi:10.1016/j.matdes.2016.11.040.
URL http://dx.doi.org/10.1016/j.matdes.2016.11.040

[384] H. Zhang, D. Gu, J. Yang, D. Dai, T. Zhao, C. Hong, A. Gasser, R. Poprawe, Selective laser melting of rare earth element Sc modified aluminum alloy: Thermodynamics of precipitation behavior and its influence on mechanical properties, Additive Manufacturing 23 (July) (2018) 1–12. doi:10.1016/j.addma.2018.07.002.

[385] A. B. Spierings, K. Dawson, P. J. Uggowitzer, K. Wegener, Influence of SLM scan-speed on microstructure, precipitation of Al3Sc particles and mechanical properties in Sc- and Zr-modified Al-Mg alloys, Materials and Design 140 (2018) 134–143. doi:10.1016/j.matdes.2017.11.053.

[386] A. B. Spierings, K. Dawson, K. Kern, F. Palm, K. Wegener, SLM-processed Sc- and Zr- modified Al-Mg alloy: Mechanical properties and microstructural effects of heat treatment, Materials Science and Engineering A 701 (May) (2017) 264–273. doi:10.1016/j.msea.2017.06.089.

[387] R. Li, H. Chen, H. Zhu, M. Wang, C. Chen, T. Yuan, Effect of aging treatment on the microstructure and mechanical properties of Al-3.02Mg-0.2Sc-0.1Zr alloy printed by selective laser melting, Materials and Design 168 (2019) 107668. doi:10.1016/j.matdes.2019.107668.
URL https://doi.org/10.1016/j.matdes.2019.107668

[388] S. Griffiths, M. D. Rossell, J. Croteau, N. Q. Vo, D. C. Dunand, C. Leinenbach, Effect of laser rescanning on the grain microstructure of a selective laser melted Al-Mg-Zr alloy, Materials Characterization 143 (2018) 34–42. doi:10.1016/j.matchar.2018.03.033.

[389] R. Li, M. Wang, Z. Li, P. Cao, T. Yuan, H. Zhu, Developing a high-strength Al-Mg-Si-Sc-Zr alloy for selective laser

melting: Crack-inhibiting and multiple strengthening mechanisms, Acta Materialia 193. doi:10.1016/j.actamat.2020.03.060.

[390] L. Cordova, T. Bor, M. de Smit, S. Carmignato, M. Campos, T. Tinga, Effects of powder reuse on the microstructure and mechanical behaviour of Al-Mg-Sc-Zr alloy processed by laser powder bed fusion (LPBF), Additive Manufacturing 36. doi:10.1016/j.addma.2020.101625.

[391] C. Xie, S. Wu, Y. Yu, H. Zhang, Y. Hu, M. Zhang, G. Wang, Defect-correlated fatigue resistance of additively manufactured Al-Mg4.5Mn alloy with in situ micro-rolling, Journal of Materials Processing Technology 291 (2021) 117039. doi:10.1016/j.jmatprotec.2020.117039.
URL https://doi.org/10.1016/j.jmatprotec.2020.117039

[392] Z. Liao, B. Yang, S. Xiao, G. Yang, T. Zhu, Fatigue crack growth behaviour of an Al-Mg4.5Mn alloy fabricated by hybrid in situ rolled wire + arc additive manufacturing, International Journal of Fatigue 151 (February) (2021) 106382. doi:10.1016/j.ijfatigue.2021.106382.
URL https://doi.org/10.1016/j.ijfatigue.2021.106382

[393] J. Gu, X. Wang, J. Bai, J. Ding, S. Williams, Y. Zhai, K. Liu, Deformation microstructures and strengthening mechanisms for the wire+arc additively manufactured Al-Mg4.5Mn alloy with inter-layer rolling, Materials Science and Engineering A 712 (November 2017) (2018) 292–301. doi:10.1016/j.msea.2017.11.113.
URL https://doi.org/10.1016/j.msea.2017.11.113

[394] W. Reschetnik, J. P. Brüggemann, M. E. Aydinöz, O. Grydin, K. P. Hoyer, G. Kullmer, H. A. Richard, Fatigue crack growth behavior and mechanical properties of additively processed en AW-7075 aluminium alloy, Procedia Structural Integrity 2 (2016) 3040–3048. doi:10.1016/j.prostr.2016.06.380.
URL http://dx.doi.org/10.1016/j.prostr.2016.06.380

[395] J. Shin, T. Kim, D. E. Kim, D. Kim, K. Kim, Castability and mechanical properties of new 7xxx aluminum alloys for automotive chassis/body applications, Journal of Alloys and Compounds 698 (2017) 577–590. doi:10.1016/j.jallcom.2016.12.269.
URL http://dx.doi.org/10.1016/j.jallcom.2016.12.269

[396] W. Stopyra, K. Gruber, I. Smolina, T. Kurzynowski, B. Kuźnicka, Laser powder bed fusion of AA7075 alloy: Influence of process parameters on porosity and hot cracking, Additive Manufacturing 35 (April). doi:10.1016/j.addma.2020.101270.

[397] L. Li, R. Li, T. Yuan, C. Chen, Z. Zhang, X. Li, Microstructures and tensile properties of a selective laser melted Al-Zn-Mg-Cu (Al7075) alloy by Si and Zr microalloying, Materials Science and Engineering A 787 (April) (2020) 139492. doi:10.1016/j.msea.2020.139492.
URL https://doi.org/10.1016/j.msea.2020.139492

[398] B. Dong, X. Cai, S. Lin, X. Li, C. Fan, C. Yang, H. Sun, Wire arc additive manufacturing of Al-Zn-Mg-Cu alloy: Microstructures and mechanical properties, Additive Manufacturing 36 (January) (2020) 101447. doi:10.1016/j.addma.2020.101447.
URL https://doi.org/10.1016/j.addma.2020.101447

[399] B. Dong, X. Cai, Y. Xia, S. Lin, C. Fan, F. Chen, Effects of interlayer temperature on the microstructures of wire arc additive manufactured Al-Zn-Mg-Cu alloy: Insights into texture responses and dynamic precipitation behaviors, Additive Manufacturing 48 (PB) (2021) 102453. doi:10.1016/j.addma.2021.102453.
URL https://doi.org/10.1016/j.addma.2021.102453

[400] Z. Zhu, F. L. Ng, H. L. Seet, W. Lu, C. H. Liebscher, Z. Rao, D. Raabe, S. Mui Ling Nai, Superior mechanical properties of a selective-laser-melted AlZnMgCuScZr alloy enabled by a tunable hierarchical microstructure and dual-nanoprecipitation, Materials Today xxx (xx). doi:10.1016/j.mattod.2021.11.019.
URL https://doi.org/10.1016/j.mattod.2021.11.019

[401] P. J. Morais, B. Gomes, P. Santos, M. Gomes, R. Gradinger, M. Schnall, S. Bozorgi, T. Klein, D. Fleischhacker, P. Warczok, A. Falahati, E. Kozeschnik, Characterisation of a High-Performance Additive Manufacturing, Materials 13 (1610) (2020) 1–17.

[402] B. C. White, W. A. Story, L. N. Brewer, J. B. Jordon, Fatigue behavior of freestanding AA2024 and AAA7075 cold spray deposits, International Journal of Fatigue 112 (December 2017) (2018) 355–360. doi:10.1016/j.ijfatigue.2018.03.007.
URL https://doi.org/10.1016/j.ijfatigue.2018.03.007

[403] D. Z. Avery, B. J. Phillips, C. J. Mason, M. Palermo, M. B. Williams, C. Cleek, O. L. Rodriguez, P. G. Allison, J. B. Jordon, Influence of Grain Refinement and Microstructure on Fatigue Behavior for Solid-State Additively Manufactured Al-Zn-Mg-Cu Alloy, Metallurgical and Materials Transactions A: Physical Metallurgy and Materials Science 51 (6) (2020) 2778–2795. doi:10.1007/s11661-020-05746-9.
URL https://doi.org/10.1007/s11661-020-05746-9

[404] M. Yuqing, K. Liming, H. Chunping, L. Fencheng, L. Qiang, Formation characteristic, microstructure, and mechanical performances of aluminum-based components by friction stir additive manufacturing, International Journal of Advanced Manufacturing Technology 83 (9-12) (2016) 1637–1647. doi:10.1007/s00170-015-7695-9.

[405] T. Klein, M. Schnall, B. Gomes, P. Warczok, D. Fleischhacker, P. J. Morais, Wire-arc additive manufacturing of a novel high-performance Al-Zn-Mg-Cu alloy: Processing, characterization and feasibility demonstration, Additive Manufacturing 37 (October 2020) (2021) 101663. doi:10.1016/j.addma.2020.101663.
URL https://doi.org/10.1016/j.addma.2020.101663

[406] R. Joey Griffiths, D. T. Petersen, D. Garcia, H. Z. Yu, Additive friction stir-enabled solid-state additive manufacturing for the repair of 7075 aluminum alloy, Applied Sciences (Switzerland) 9 (17). doi:10.3390/app9173486.

[407] H. Zhang, H. Zhu, X. Nie, T. Qi, Z. Hu, X. Zeng, Fabrication and heat treatment of high strength Al-Cu-Mg alloy processed using selective laser melting, Laser 3D Manufacturing III 9738 (April 2016) (2016) 97380X. doi:10.1117/12.2211362.

[408] H. Zhang, H. Zhu, X. Nie, J. Yin, Z. Hu, X. Zeng, Effect of Zirconium addition on crack, microstructure and mechanical behavior of selective laser melted Al-Cu-Mg alloy, Scripta Materialia 134 (2017) 6–10. doi:10.1016/j.scriptamat.2017.02.036.
URL http://dx.doi.org/10.1016/j.scriptamat.2017.02.036

[409] X. Nie, H. Zhang, H. Zhu, Z. Hu, L. Ke, X. Zeng, Effect of Zr content on formability, microstructure and mechanical properties of selective laser melted Zr modified Al-4.24Cu-1.97Mg-0.56Mn alloys, Journal of Alloys and Compounds 764 (2018) 977–986. doi:10.1016/j.jallcom.2018.06.032.
URL https://doi.org/10.1016/j.jallcom.2018.06.032

[410] B. A. Rutherford, D. Z. Avery, B. J. Phillips, H. M. Rao, K. J. Doherty, P. G. Allison, L. N. Brewer, J. Brian Jordon, Effect of thermomechanical processing on fatigue behavior in solid-state additive manufacturing of Al-Mg-Si alloy, Metals 10 (7) (2020) 1–17. doi:10.3390/met10070947.

[411] C. E. Roberts, D. Bourell, T. Watt, J. Cohen, A novel processing approach for additive manufacturing of commercial aluminum alloys, Physics Procedia 83 (2016) 909–917. doi:10.1016/j.phpro.2016.08.095.

[412] S. Z. Uddin, L. E. Murr, C. A. Terrazas, P. Morton, D. A. Roberson, R. B. Wicker, Processing and characterization of crack-free aluminum 6061 using high-temperature heating in laser powder bed fusion additive manufacturing, Additive Manufacturing 22 (June) (2018) 405–415. doi:10.1016/j.addma.2018.05.047.
URL https://doi.org/10.1016/j.addma.2018.05.047

[413] Y. Yin, Q. Tan, M. Bermingham, N. Mo, J. Zhang, M. X. Zhang, Laser additive manufacturing of steels, International Materials Reviews 0 (0) (2021) 1–87. doi:10.1080/09506608.2021.1983351.
URL https://doi.org/10.1080/09506608.2021.1983351

[414] P. Bajaj, A. Hariharan, A. Kini, P. Kürnsteiner, D. Raabe, E. A. Jägle, Steels in additive manufacturing: A review of their microstructure and properties, Materials Science and Engineering A 772 (October 2019). doi:10.1016/j.msea.2019.138633.

[415] M. Zhang, H. Li, X. Zhang, D. Hardacre, Review of the fatigue performance of stainless steel 316L parts manufactured by selective laser melting, Proceedings of the International Conference on Progress in Additive Manufacturing Part F1290 (2016) 563–568.

[416] S. Mirzababaei, S. Pasebani, A review on binder jet additive manufacturing of 316l stainless steel, Journal of Manufacturing and Materials Processing 3 (3) (2019) 82.

[417] H. Fayazfar, M. Salarian, A. Rogalsky, D. Sarker, P. Russo, V. Paserin, E. Toyserkani, A critical review of powder-based additive manufacturing of ferrous alloys: Process parameters, microstructure and mechanical properties, Materials and Design 144 (2018) 98–128. doi:10.1016/j.matdes.2018.02.018.
URL https://doi.org/10.1016/j.matdes.2018.02.018

[418] J. C. Lippold, D. J. Kotecki, Welding metallurgy and weldability of stainless steels, Wiley, 2005.

[419] K. M. Bertsch, G. Meric de Bellefon, B. Kuehl, D. J. Thoma, Origin of dislocation structures in an additively manufactured austenitic stainless steel 316L, Acta Materialia 199 (2020) 19–33. doi:10.1016/j.actamat.2020.07.063.
URL https://doi.org/10.1016/j.actamat.2020.07.063

[420] J. Suryawanshi, K. G. Prashanth, U. Ramamurty, Mechanical behavior of selective laser melted 316L stainless steel, Materials Science and Engineering A 696 (January) (2017) 113–121. doi:10.1016/j.msea.2017.04.058.
URL http://dx.doi.org/10.1016/j.msea.2017.04.058

[421] Y. Zhong, L. E. Rännar, L. Liu, A. Koptyug, S. Wikman, J. Olsen, D. Cui, Z. Shen, Additive manufacturing of 316L stainless steel by electron beam melting for nuclear fusion applications, Journal of Nuclear Materials 486 (2017) 234–245. doi:10.1016/j.jnucmat.2016.12.042.

[422] M. J. Holzweissig, A. Taube, F. Brenne, M. Schaper, T. Niendorf, Microstructural Characterization and Mechanical Performance of Hot Work Tool Steel Processed by Selective Laser Melting, Metallurgical and Materials Transactions B: Process Metallurgy and Materials Processing Science 46 (2) (2015) 545–549. doi:10.1007/s11663-014-0267-9.

[423] R. Mertens, B. Vrancken, N. Holmstock, Y. Kinds, J. P. Kruth, J. Van Humbeeck, Influence of powder bed preheating on microstructure and mechanical properties of H13 tool steel SLM parts, Physics Procedia 83 (2016) 882–890. doi:10.1016/j.phpro.2016.08.092.

[424] M. Mazur, P. Brincat, M. Leary, M. Brandt, Numerical and experimental evaluation of a conformally cooled H13 steel injection mould manufactured with selective laser melting, International Journal of Advanced Manufacturing Technology 93 (1-4) (2017) 881–900. doi:10.1007/s00170-017-0426-7.

[425] J. Mazumder, J. Choi, K. Nagarathnam, J. Koch, D. Hetzner, The direct metal deposition of H13 tool steel for 3-D components, Jom 49 (5) (1997) 55–60. doi:10.1007/BF02914687.

[426] L. Xue, J. Chen, S. H. Wang, Freeform Laser Consolidated H13 and CPM 9V Tool Steels, Metallography, Microstructure, and Analysis 2 (2) (2013) 67–78. doi:10.1007/s13632-013-0061-0.

[427] Y. M. Wang, T. Voisin, J. T. McKeown, J. Ye, N. P. Calta, Z. Li, Z. Zeng, Y. Zhang, W. Chen, T. T. Roehling, et al., Additively manufactured hierarchical stainless steels with high strength and ductility, Nature materials 17 (1) (2018) 63–71.

[428] S. Gorsse, C. Hutchinson, M. Gouné, R. Banerjee, Additive manufacturing of metals: a brief review of the characteristic microstructures and properties of steels, ti-6al-4v and high-entropy alloys, Science and Technology of advanced MaTerialS 18 (1) (2017) 584–610.

[429] F. Hengsbach, P. Koppa, K. Duschik, M. J. Holzweissig, M. Burns, J. Nellesen, W. Tillmann, T. Tröster, K.-P. Hoyer, M. Schaper, Duplex stainless steel fabricated by selective laser melting-microstructural and mechanical properties, Materials & Design 133 (2017) 136–142.

[430] Z. Sun, X. Tan, S. B. Tor, C. K. Chua, Simultaneously enhanced strength and ductility for 3d-printed stainless steel 316l by selective laser melting, NPG Asia Materials 10 (4) (2018) 127–136.

[431] D. Karlsson, C.-Y. Chou, N. H. Pettersson, T. Helander, P. Harlin, M. Sahlberg, G. Lindwall, J. Odqvist, U. Jansson, Additive manufacturing of the ferritic stainless steel ss441, Additive Manufacturing 36 (2020) 101580.

[432] F. Deirmina, N. Peghini, B. AlMangour, D. Grzesiak, M. Pellizzari, Heat treatment and properties of a hot work tool steel fabricated by additive manufacturing, Materials Science and Engineering: A 753 (2019) 109–121.

[433] J. Mazumder, A. Schifferer, J. Choi, Direct materials deposition: designed macro and microstructure, Material Research Innovations 3 (3) (1999) 118–131.

[434] J. Sander, J. Hufenbach, L. Giebeler, H. Wendrock, U. Kühn, J. Eckert, Microstructure and properties of fecrmovc tool steel produced by selective laser melting, Materials & Design 89 (2016) 335–341.

[435] I. Yadroitsev, A. Gusarov, I. Yadroitsava, I. Smurov, Single track formation in selective laser melting of metal powders, Journal of Materials Processing Technology 210 (12) (2010) 1624–1631.

[436] L. Zumofen, C. Beck, A. Kirchheim, H.-J. Dennig, Quality related effects of the preheating temperature on laser melted high carbon content steels, in: International Conference on Additive Manufacturing in Products and Applications, Springer, 2017, pp. 210–219.

[437] R. Mahshid, H. N. Hansen, K. L. Højbjerre, Strength analysis and modeling of cellular lattice structures manufactured using selective laser melting for tooling applications, Materials & Design 104 (2016) 276–283.

[438] Z. Xiao, Y. Yang, R. Xiao, Y. Bai, C. Song, D. Wang, Evaluation of topology-optimized lattice structures manufactured via selective laser melting, Materials & Design 143 (2018) 27–37.

[439] M. Ahmed Obeidi, S. M. Uí Mhurchadha, R. Raghavendra, A. Conway, C. Souto, D. Tormey, I. U. Ahad, D. Brabazon, Comparison of the porosity and mechanical performance of 316L stainless steel manufactured on different laser powder bed fusion metal additive manufacturing machines, Journal of Materials Research and Technology 13 (2021) 2361–2374. doi:10.1016/j.jmrt.2021.06.027.

[440] H. Zhang, M. Xu, Z. Liu, C. Li, P. Kumar, Z. Liu, Y. Zhang, Microstructure, surface quality, residual stress, fatigue behavior and damage mechanisms of selective laser melted 304L stainless steel considering building direction, Additive Manufacturing 46 (June) (2021) 102147. doi:10.1016/j.addma.2021.102147.
URL https://doi.org/10.1016/j.addma.2021.102147

[441] F. Stern, J. Kleinhorst, J. Tenkamp, F. Walther, Investigation of the anisotropic cyclic damage behavior of selective laser melted AISI 316L stainless steel, Fatigue and Fracture of Engineering Materials and Structures 42 (11) (2019) 2422–2430. doi:10.1111/ffe.13029.

[442] F. Stern, J. Tenkamp, F. Walther, Non-destructive characterization of process-induced defects and their effect on the fatigue behavior of austenitic steel 316L made by laser-powder bed fusion, Progress in Additive Manufacturing 5 (3) (2020) 287–294. doi:10.1007/s40964-019-00105-6.
URL https://doi.org/10.1007/s40964-019-00105-6

[443] P. Kumar, R. Jayaraj, J. Suryawanshi, U. R. Satwik, J. McKinnell, U. Ramamurty, Fatigue strength of additively manufactured 316L austenitic stainless steel, Acta Materialia 199 (2020) 225–239. doi:10.1016/j.actamat.2020.08.033.
URL https://doi.org/10.1016/j.actamat.2020.08.033

[444] R. Shrestha, J. Simsiriwong, N. Shamsaei, Fatigue behavior of additive manufactured 316L stainless steel under axial versus rotating-bending loading: Synergistic effects of stress gradient, surface roughness, and volumetric defects, International Journal of Fatigue 144 (November 2020) (2021) 106063. doi:10.1016/j.ijfatigue.2020.106063.
URL https://doi.org/10.1016/j.ijfatigue.2020.106063

[445] R. Shrestha, J. Simsiriwong, N. Shamsaei, Fatigue behavior of additive manufactured 316L stainless steel parts: Effects of layer orientation and surface roughness, Additive Manufacturing 28 (2019) 23–38. doi:10.1016/j.addma.2019.04.011.
URL https://doi.org/10.1016/j.addma.2019.04.011

[446] C. Yu, Y. Zhong, P. Zhang, Z. Zhang, C. Zhao, Z. Zhang, Z. Shen, W. Liu, Effect of Build Direction on Fatigue Performance of L-PBF 316L Stainless Steel, Acta Metallurgica Sinica (English Letters) 33 (4) (2020) 539–550. doi:10.1007/s40195-019-00983-3.
URL https://doi.org/10.1007/s40195-019-00983-3

[447] B. o. t. i. o. a. s.-r. h. t. by using efficient short-time procedures Blinn, additively manufactured surface on the fatigue behavior of selectively laser melted AISI 316L, F. Krebs, M. Ley, R. Teutsch, T. Beck, Determination of the influence of a stress-relief heat treatment and additively manufactured surface on the fatigue behavior of selectively laser melted AISI 316L by using efficient short-time procedures, International Journal of Fatigue 131 (October 2019) (2020) 105301. doi:10.1016/j.ijfatigue.2019.105301.
URL https://doi.org/10.1016/j.ijfatigue.2019.105301

[448] B. Blinn, P. Lion, O. Jordan, S. Meiniger, S. Mischliwski, C. Tepper, C. Gläßner, J. C. Aurich, M. Weigold, T. Beck, Process-influenced fatigue behavior of AISI 316L manufactured by powder- and wire-based Laser Direct Energy Deposition, Materials Science and Engineering A 818. doi:10.1016/j.msea.2021.141383.

[449] M. Zhang, C. N. Sun, X. Zhang, P. C. Goh, J. Wei, H. Li, D. Hardacre, Elucidating the Relations Between Monotonic and Fatigue Properties of Laser Powder Bed Fusion Stainless Steel 316L, Jom 70 (3) (2018) 390–395. doi:10.1007/s11837-017-2640-z.

[450] D. Kotzem, S. Kleszczynski, F. Stern, A. Elspaß, J. Tenkamp, G. Witt, F. Walther, Impact of single structural voids on fatigue properties of aisi 316l manufactured by laser powder bed fusion, International Journal of Fatigue 148 (2021) 106207.

[451] A. Riemer, S. Leuders, M. Thöne, H. A. Richard, T. Tröster, T. Niendorf, On the fatigue crack growth behavior in 316L stainless steel manufactured by selective laser melting, Engineering Fracture Mechanics 120 (2014) 15–25. doi:10.1016/j.engfracmech.2014.03.008.
URL http://dx.doi.org/10.1016/j.engfracmech.2014.03.008

[452] O. Fergani, A. Bratli Wold, F. Berto, V. Brotan, M. Bambach, Study of the effect of heat treatment on fatigue crack growth behaviour of 316L stainless steel produced by selective laser melting, Fatigue and Fracture of Engineering Materials and Structures 41 (5) (2018) 1102–1119. doi:10.1111/ffe.12755.

[453] J. Kluczyński, L. Śniezek, K. Grzelak, J. Torzewski, I. Szachogłuchowicz, M. Wachowski, J. Łuszczek, Crack growth behavior of additively manufactured 316L steel-influence of build orientation and heat treatment, Materials 13 (15). doi:10.3390/MA13153259.

[454] G. Strano, L. Hao, R. M. Everson, K. E. Evans, Surface roughness analysis, modelling and prediction in selective laser melting, Journal of Materials Processing Technology 213 (4) (2013) 589–597. doi:10.1016/J.JMATPROTEC.2012.11.011.

[455] X. Wang, J. A. Muñiz-Lerma, M. Attarian Shandiz, O. Sanchez-Mata, M. Brochu, Crystallographic-orientation-dependent tensile behaviours of stainless steel 316L fabricated by laser powder bed fusion, Materials Science and Engineering A 766 (September) (2019) 138395. doi:10.1016/j.msea.2019.138395.
URL https://doi.org/10.1016/j.msea.2019.138395

[456] Effects of building orientation and heat treatment on fatigue behavior of selective laser melted 17-4 PH stainless steel, International Journal of Fatigue 94 (2017) 218–235. doi:10.1016/j.ijfatigue.2016.03.014.
URL http://dx.doi.org/10.1016/j.ijfatigue.2016.03.014

[457] R. I. Stephens, A. Fatemi, R. R. Stephens, H. O. Fuchs, Metal fatigue in engineering, John Wiley & Sons, 2000.

[458] P. Wood, T. Libura, Z. L. Kowalewski, G. Williams, A. Serjouei, Influences of horizontal and vertical build orientations and post-fabrication processes on the fatigue behavior of stainless steel 316l produced by selective laser melting, Materials 12 (24). doi:10.3390/ma1224203.

[459] M. Zhang, C. N. Sun, X. Zhang, P. C. Goh, J. Wei, D. Hardacre, H. Li, Fatigue and fracture behaviour of laser powder bed fusion stainless steel 316L: Influence of processing parameters, Materials Science and Engineering A 703 (2017) 251–261. doi:10.1016/j.msea.2017.07.071.
URL https://doi.org/10.1016/j.msea.2017.07.071

[460] D. Wang, C. Song, Y. Yang, Y. Bai, Investigation of crystal growth mechanism during selective laser melting and mechanical property characterization of 316L stainless steel parts, Materials and Design 100 (2016) 291–299. doi:10.1016/j.matdes.2016.03.111.
URL http://dx.doi.org/10.1016/j.matdes.2016.03.111

[461] A. R. A. Dezfoli, W.-S. Hwang, W.-C. Huang, T.-W. Tsai, Determination and controlling of grain structure of metals after laser incidence: Theoretical approach, Scientific reports 7 (1) (2017) 1–11.

[462] P. Nezhadfar, N. Shamsaei, N. Phan, Enhancing ductility and fatigue strength of additively manufactured metallic materials by preheating the build platform, Fatigue & Fracture of Engineering Materials & Structures 44 (1) (2021)

257–270.

[463] K. Solberg, F. Berto, What is going on with fatigue of additively manufactured metals?, Material Design & Processing Communications 1 (5) (2019) e84.

[464] E. Liverani, S. Toschi, L. Ceschini, A. Fortunato, Effect of selective laser melting (SLM) process parameters on microstructure and mechanical properties of 316L austenitic stainless steel, Journal of Materials Processing Technology 249 (May) (2017) 255–263. doi:10.1016/j.jmatprotec.2017.05.042.
URL http://dx.doi.org/10.1016/j.jmatprotec.2017.05.042

[465] H. D. Carlton, A. Haboub, G. F. Gallegos, D. Y. Parkinson, A. A. MacDowell, Damage evolution and failure mechanisms in additively manufactured stainless steel, Materials Science and Engineering A 651 (2016) 406–414. doi:10.1016/j.msea.2015.10.073.
URL http://dx.doi.org/10.1016/j.msea.2015.10.073

[466] N. P. Lavery, J. Cherry, S. Mehmood, H. Davies, B. Girling, E. Sackett, S. G. Brown, J. Sienz, Effects of hot isostatic pressing on the elastic modulus and tensile properties of 316L parts made by powder bed laser fusion, Materials Science and Engineering A 693 (March) (2017) 186–213. doi:10.1016/j.msea.2017.03.100.
URL http://dx.doi.org/10.1016/j.msea.2017.03.100

[467] A. Aversa, A. Saboori, E. Librera, M. de Chirico, S. Biamino, M. Lombardi, P. Fino, The role of Directed Energy Deposition atmosphere mode on the microstructure and mechanical properties of 316L samples, Additive Manufacturing 34 (February) (2020) 101274. doi:10.1016/j.addma.2020.101274.
URL https://doi.org/10.1016/j.addma.2020.101274

[468] D. K. Kim, W. Woo, E. Y. Kim, S. H. Choi, Microstructure and mechanical characteristics of multi-layered materials composed of 316L stainless steel and ferritic steel produced by direct energy deposition, Journal of Alloys and Compounds 774 (2019) 896–907. doi:10.1016/j.jallcom.2018.09.390.
URL https://doi.org/10.1016/j.jallcom.2018.09.390

[469] A. Yadollahi, N. Shamsaei, S. M. Thompson, D. W. Seely, Effects of process time interval and heat treatment on the mechanical and microstructural properties of direct laser deposited 316L stainless steel, Materials Science and Engineering A 644 (2015) 171–183. doi:10.1016/j.msea.2015.07.056.
URL http://dx.doi.org/10.1016/j.msea.2015.07.056

[470] M. Ghayoor, K. Lee, Y. He, C. hung Chang, B. K. Paul, S. Pasebani, Selective laser melting of 304L stainless steel: Role of volumetric energy density on the microstructure, texture and mechanical properties, Additive Manufacturing 32 (October 2019) (2020) 101011. doi:10.1016/j.addma.2019.101011.
URL https://doi.org/10.1016/j.addma.2019.101011

[471] K. Guan, Z. Wang, M. Gao, X. Li, X. Zeng, Effects of processing parameters on tensile properties of selective laser melted 304 stainless steel, Materials and Design 50 (2013) 581–586. doi:10.1016/j.matdes.2013.03.056.
URL http://dx.doi.org/10.1016/j.matdes.2013.03.056

[472] L. E. Murr, E. Martinez, J. Hernandez, S. Collins, K. N. Amato, S. M. Gaytan, P. W. Shindo, Microstructures and properties of 17-4 PH stainless steel fabricated by selective laser melting, Journal of Materials Research and Technology 1 (3) (2012) 167–177. doi:10.1016/S2238-7854(12)70029-7.
URL http://dx.doi.org/10.1016/S2238-7854(12)70029-7

[473] T. LeBrun, T. Nakamoto, K. Horikawa, H. Kobayashi, Effect of retained austenite on subsequent thermal processing and resultant mechanical properties of selective laser melted 17-4 PH stainless steel, Materials and Design 81 (2015) 44–53. doi:10.1016/j.matdes.2015.05.026.
URL http://dx.doi.org/10.1016/j.matdes.2015.05.026

[474] L. Facchini, N. Vicente, I. Lonardelli, E. Magalini, P. Robotti, M. Alberto, Metastable austenite in 17-4 precipitation-hardening stainless steel produced by selective laser melting, Advanced Engineering Materials 12 (3) (2010) 184–188. doi:10.1002/adem.200900259.

[475] J. Vishwakarma, K. Chattopadhyay, N. C. Santhi Srinivas, Effect of build orientation on microstructure and tensile behaviour of selectively laser melted M300 maraging steel, Materials Science and Engineering A 798 (August) (2020) 140130. doi:10.1016/j.msea.2020.140130.
URL https://doi.org/10.1016/j.msea.2020.140130

[476] C. Tan, K. Zhou, W. Ma, P. Zhang, M. Liu, T. Kuang, Microstructural evolution, nanoprecipitation behavior and mechanical properties of selective laser melted high-performance grade 300 maraging steel, Materials and Design 134 (2017) 23–34. doi:10.1016/j.matdes.2017.08.026.
URL https://doi.org/10.1016/j.matdes.2017.08.026

[477] R. Casati, J. N. Lemke, A. Tuissi, M. Vedani, Aging behaviour and mechanical performance of 18-Ni 300 steel processed by selective laser melting, Metals 6 (9). doi:10.3390/met6090218.

[478] T. H. Becker, D. DImitrov, The achievable mechanical properties of SLM produced Maraging Steel 300 components, Rapid Prototyping Journal 22 (3) (2016) 487–494. doi:10.1108/RPJ-08-2014-0096.

[479] K. Kempen, E. Yasa, L. Thijs, J. P. Kruth, J. Van Humbeeck, Microstructure and mechanical properties of selective laser melted 18Ni-300 steel, Physics Procedia 12 (PART 1) (2011) 255–263. doi:10.1016/j.phpro.2011.03.033.
URL http://dx.doi.org/10.1016/j.phpro.2011.03.033

[480] C. Elangeswaran, A. Cutolo, G. K. Muralidharan, K. Vanmeensel, B. Van Hooreweder, Microstructural analysis and fatigue crack initiation modelling of additively manufactured 316L after different heat treatments, Materials and Design 194 (2020) 108962. doi:10.1016/j.matdes.2020.108962.
URL https://doi.org/10.1016/j.matdes.2020.108962

[481] C. Elangeswaran, A. Cutolo, G. K. Muralidharan, C. de Formanoir, F. Berto, K. Vanmeensel, B. Van Hooreweder, Effect of post-treatments on the fatigue behaviour of 316L stainless steel manufactured by laser powder bed fusion, International Journal of Fatigue 123 (February) (2019) 31–39. doi:10.1016/j.ijfatigue.2019.01.013.
URL https://doi.org/10.1016/j.ijfatigue.2019.01.013

[482] A. Polishetty, G. Littlefair, Heat Treatment Effect on the Fatigue Characteristics of Additive Manufactured Stainless Steel 316L, International Journal of Materials, Mechanics and Manufacturing 7 (2) (2019) 114–118. doi:10.18178/ijmmm.2019.7.2.442.

[483] N. Kalentics, M. O. V. de Seijas, S. Griffiths, C. Leinenbach, R. E. Logé, 3D laser shock peening - A new method for improving fatigue properties of selective laser melted parts, Additive Manufacturing 33 (2020) 101112. doi:10.1016/J.ADDMA.2020.101112.

[484] P. Merot, F. Morel, L. Gallegos Mayorga, E. Pessard, P. Buttin, T. Baffie, Observations on the influence of process and corrosion related defects on the fatigue strength of 316L stainless steel manufactured by Laser Powder Bed Fusion (L-PBF), International Journal of Fatigue 155 (July 2021) (2022) 106552. doi:10.1016/j.ijfatigue.2021.106552.
URL https://doi.org/10.1016/j.ijfatigue.2021.106552

[485] M. Braun, E. Mayer, I. Kryukov, C. Wolf, S. Böhm, A. Taghipour, R. E. Wu, S. Ehlers, S. Sheikhi, Fatigue strength of PBF-LB/M and wrought 316L stainless steel: effect of post-treatment and cyclic mean stress, Fatigue and Fracture of Engineering Materials and Structures 44 (11) (2021) 3077–3093. doi:10.1111/ffe.13552.

[486] S. Leuders, T. Lieneke, S. Lammers, T. Tröster, T. Niendorf, On the fatigue properties of metals manufactured by selective laser melting–the role of ductility, Journal of Materials Research 29 (17) (2014) 1911–1919.

[487] A. B. Spierings, T. L. Starr, K. Wegener, Fatigue performance of additive manufactured metallic parts, Rapid prototyping journal.

[488] J. Kluczyński, L. Śniezek, K. Grzelak, J. Torzewski, I. Szachogłuchowicz, M. Wachowski, J. Łuszczek, Crack Growth Behavior of Additively Manufactured 316L Steel—Influence of Build Orientation and Heat Treatment, Materials 13 (15). doi:10.3390/MA13153259.

[489] B. Voloskov, S. Evlashin, S. Dagesyan, S. Abaimov, I. Akhatov, I. Sergeichev, Very high cycle fatigue behavior of additively manufactured 316L stainless steel, Materials 13 (15) (2020) 1–11. doi:10.3390/ma13153293.

[490] B. Blinn, M. Klein, C. Gläßner, M. Smaga, J. C. Aurich, T. Beck, An investigation of the microstructure and fatigue behavior of additively manufactured aisi 316l stainless steel with regard to the influence of heat treatment, Metals 8 (4) (2018) 220.

[491] M. M. Parvez, Y. Chen, J. W. Newkirk, F. F. Liou, Comparison of fatigue performance between additively manufactured and wrought 304U stainless steel using a novel fatigue test setup, Solid Freeform Fabrication 2019: Proceedings of the 30th Annual International Solid Freeform Fabrication Symposium - An Additive Manufacturing Conference, SFF 2019 (2019) 353–363.

[492] J. Gordon, J. Hochhalter, C. Haden, D. G. Harlow, Enhancement in fatigue performance of metastable austenitic stainless steel through directed energy deposition additive manufacturing, Materials and Design 168 (2019) 107630. doi:10.1016/j.matdes.2019.107630.
URL https://doi.org/10.1016/j.matdes.2019.107630

[493] J. V. Gordon, C. V. Haden, H. F. Nied, R. P. Vinci, D. G. Harlow, Fatigue crack growth anisotropy, texture and residual stress in austenitic steel made by wire and arc additive manufacturing, Materials Science and Engineering A 724 (March) (2018) 431–438. doi:10.1016/j.msea.2018.03.075.
URL https://doi.org/10.1016/j.msea.2018.03.075

[494] S. Lee, J. W. Pegues, N. Shamsaei, Fatigue behavior and modeling for additive manufactured 304L stainless steel: The effect of surface roughness, International Journal of Fatigue 141 (July). doi:10.1016/j.ijfatigue.2020.105856.

[495] A. Yadollahi, M. Mahmoudi, A. Elwany, H. Doude, L. Bian, J. C. Newman, Fatigue-life prediction of additively manufactured material: Effects of heat treatment and build orientation, Fatigue and Fracture of Engineering Materials and Structures 43 (4) (2020) 831–844. doi:10.1111/ffe.13200.

[496] A. Soltani-Tehrani, J. Pegues, N. Shamsaei, Fatigue behavior of additively manufactured 17-4 PH stainless steel: The effects of part location and powder re-use, Additive Manufacturing 36 (2020) 101398. doi:10.1016/J.ADDMA.2020.101398.

[497] P. D. Nezhadfar, E. Burford, K. Anderson-Wedge, B. Zhang, S. Shao, S. R. Daniewicz, N. Shamsaei, Fatigue crack growth behavior of additively manufactured 17-4 PH stainless steel: Effects of build orientation and microstructure,

International Journal of Fatigue 123 (February) (2019) 168–179. doi:10.1016/j.ijfatigue.2019.02.015.
URL https://doi.org/10.1016/j.ijfatigue.2019.02.015

[498] L. Carneiro, B. Jalalahmadi, A. Ashtekar, Y. Jiang, Cyclic deformation and fatigue behavior of additively manufactured 17-4 PH stainless steel, International Journal of Fatigue 123 (January) (2019) 22–30. doi:10.1016/j.ijfatigue.2019.02.006.
URL https://doi.org/10.1016/j.ijfatigue.2019.02.006

[499] P. D. Nezhadfar, R. Shrestha, N. Phan, N. Shamsaei, Fatigue behavior of additively manufactured 17-4 PH stainless steel: Synergistic effects of surface roughness and heat treatment, International Journal of Fatigue 124 (2019) 188–204. doi:10.1016/j.ijfatigue.2019.02.039.
URL https://doi.org/10.1016/j.ijfatigue.2019.02.039

[500] M. Akita, Y. Uematsu, T. Kakiuchi, M. Nakajima, R. Kawaguchi, Defect-dominated fatigue behavior in type 630 stainless steel fabricated by selective laser melting, Materials Science and Engineering: A 666 (2016) 19–26.

[501] S. Romano, P. D. Nezhadfar, N. Shamsaei, M. Seifi, S. Beretta, High cycle fatigue behavior and life prediction for additively manufactured 17-4 PH stainless steel: Effect of sub-surface porosity and surface roughness, Theoretical and Applied Fracture Mechanics 106 (November 2019). doi:10.1016/j.tafmec.2020.102477.

[502] A. Yadollahi, M. Mahmoudi, A. Elwany, H. Doude, L. Bian, J. C. Newman, Effects of crack orientation and heat treatment on fatigue-crack-growth behavior of AM 17-4 PH stainless steel, Engineering Fracture Mechanics 226 (January) (2020) 106874. doi:10.1016/j.engfracmech.2020.106874.
URL https://doi.org/10.1016/j.engfracmech.2020.106874

[503] S. Sarkar, C. S. Kumar, A. K. Nath, Effects of heat treatment and build orientations on the fatigue life of selective laser melted 15-5 PH stainless steel, Materials Science and Engineering A 755 (April) (2019) 235–245. doi:10.1016/j.msea.2019.04.003.
URL https://doi.org/10.1016/j.msea.2019.04.003

[504] D. Croccolo, M. De Agostinis, S. Fini, G. Olmi, N. Bogojevic, S. Ciric-Kostic, Effects of build orientation and thickness of allowance on the fatigue behaviour of 15-5 PH stainless steel manufactured by DMLS, Fatigue and Fracture of Engineering Materials and Structures 41 (4) (2018) 900–916. doi:10.1111/ffe.12737.

[505] S. Sarkar, C. S. Kumar, A. K. Nath, Effects of different surface modifications on the fatigue life of selective laser melted 15-5 PH stainless steel, Materials Science and Engineering A 762 (2019) 138109. doi:10.1016/j.msea.2019.138109.
URL https://doi.org/10.1016/j.msea.2019.138109

[506] J. Damon, T. Hanemann, S. Dietrich, G. Graf, K. H. Lang, V. Schulze, Orientation dependent fatigue performance and mechanisms of selective laser melted maraging steel X3NiCoMoTi18-9-5, International Journal of Fatigue 127 (November 2018) (2019) 395–402. doi:10.1016/j.ijfatigue.2019.06.025.
URL https://doi.org/10.1016/j.ijfatigue.2019.06.025

[507] G. Meneghetti, D. Rigon, D. Cozzi, W. Waldhauser, M. Dabalà, Influence of build orientation on static and axial fatigue properties of maraging steel specimens produced by additive manufacturing, Procedia Structural Integrity 7 (2017) 149–157. doi:10.1016/j.prostr.2017.11.072.
URL https://doi.org/10.1016/j.prostr.2017.11.072

[508] G. Meneghetti, D. Rigon, C. Gennari, An analysis of defects influence on axial fatigue strength of maraging steel specimens produced by additive manufacturing, International Journal of Fatigue 118 (August 2018) (2019) 54–64. doi:10.1016/j.ijfatigue.2018.08.034.
URL https://doi.org/10.1016/j.ijfatigue.2018.08.034

[509] D. Croccolo, M. De Agostinis, S. Fini, G. Olmi, A. Vranic, S. Ciric-Kostic, Influence of the build orientation on the fatigue strength of EOS maraging steel produced by additive metal machine, Fatigue and Fracture of Engineering Materials and Structures 39 (5) (2016) 637–647. doi:10.1111/ffe.12395.

[510] K. Solberg, E. W. Hovig, K. Sørby, F. Berto, Directional fatigue behaviour of maraging steel grade 300 produced by laser powder bed fusion, International Journal of Fatigue 149. doi:10.1016/j.ijfatigue.2021.106229.

[511] T. Tezel, V. Kovan, Heat treatment effect on fatigue behavior of 3D-printed maraging steels, Rapid Prototyping Journal 28 (1) (2022) 175–184. doi:10.1108/RPJ-03-2021-0069.

[512] L. F. Van Swam, R. M. Pelloux, N. J. Grant, Fatigue behavior of maraging steel 300, Metallurgical Transactions A 6 (1) (1975) 45–54. doi:10.1007/BF02673669.

[513] S. Ćirić-Kostić, D. Croccolo, M. De Agostinis, S. Fini, G. Olmi, L. Paiardini, F. Robusto, Z. Šoškić, N. Bogojević, Fatigue response of additively manufactured maraging stainless steel cx and effects of heat treatment and surface finishing, Fatigue & Fracture of Engineering Materials & Structures 45 (2) (2022) 482–499.

[514] L. M. Santos, L. P. Borrego, J. A. Ferreira, J. de Jesus, J. D. Costa, C. Capela, Effect of heat treatment on the fatigue crack growth behaviour in additive manufactured AISI 18Ni300 steel, Theoretical and Applied Fracture Mechanics 102 (April) (2019) 10–15. doi:10.1016/j.tafmec.2019.04.005.
URL https://doi.org/10.1016/j.tafmec.2019.04.005

[515] R. Dörfert, J. Zhang, B. Clausen, H. Freiße, J. Schumacher, F. Vollertsen, Comparison of the fatigue strength between additively and conventionally fabricated tool steel 1.2344, Additive Manufacturing 27 (February) (2019) 217–223. doi:10.1016/j.addma.2019.01.010.
URL https://doi.org/10.1016/j.addma.2019.01.010

[516] M. Pellizzari, B. AlMangour, M. Benedetti, S. Furlani, D. Grzesiak, F. Deirmina, Effects of building direction and defect sensitivity on the fatigue behavior of additively manufactured H13 tool steel, Theoretical and Applied Fracture Mechanics 108 (May) (2020) 102634. doi:10.1016/j.tafmec.2020.102634.
URL https://doi.org/10.1016/j.tafmec.2020.102634

[517] L. Tshabalala, O. Sono, W. Makoana, J. Masindi, O. Maluleke, C. Johnston, S. Masete, Axial fatigue behaviour of additively manufactured tool steels, Materials Today: Proceedings 38 (2021) 789–792. doi:10.1016/j.matpr.2020.04.548.
URL https://doi.org/10.1016/j.matpr.2020.04.548

[518] H. Qi, M. Azer, A. Ritter, Studies of standard heat treatment effects on microstructure and mechanical properties of laser net shape manufactured inconel 718, Metallurgical and Materials Transactions A 40 (10) (2009) 2410–2422.

[519] H. Attia, S. Tavakoli, R. Vargas, V. Thomson, Laser-assisted high-speed finish turning of superalloy Inconel 718 under dry conditions, CIRP Annals - Manufacturing Technology 59 (1) (2010) 83–88. doi:10.1016/j.cirp.2010.03.093.
URL http://dx.doi.org/10.1016/j.cirp.2010.03.093

[520] J. P. Costes, Y. Guillet, G. Poulachon, M. Dessoly, Tool-life and wear mechanisms of CBN tools in machining of Inconel 718, International Journal of Machine Tools and Manufacture 47 (7-8) (2007) 1081–1087. doi:10.1016/j.ijmachtools.2006.09.031.

[521] T. Kellner, New plant will assemble world's first passenger jet engine with 3d printed fuel nozzles, next-gen materials (2014).

[522] C. E. Leshock, J. N. Kim, Y. C. Shin, Plasma enhanced machining of Inconel 718: Modeling of workpiece temperature with plasma heating and experimental results, International Journal of Machine Tools and Manufacture 41 (6) (2001) 877–897. doi:10.1016/S0890-6955(00)00106-1.

[523] R. M. Nunes, D. Pereira, T. Clarke, T. K. Hirsch, Delta phase characterization in inconel 718 alloys through x-ray diffraction, ISIJ International 55 (11) (2015) 2450–2454. doi:10.2355/isijinternational.ISIJINT-2015-111.

[524] J. Stroßner, M. Terock, U. Glatzel, Mechanical and Microstructural Investigation of Nickel-Based Superalloy IN718 Manufactured by Selective Laser Melting (SLM), Advanced Engineering Materials 17 (8) (2015) 1099–1105. doi:10.1002/adem.201500158.

[525] T. Brynk, Z. Pakiela, K. Ludwichowska, B. Romelczyk, R. M. Molak, M. Plocinska, J. Kurzac, T. Kurzynowski, E. Chlebus, Fatigue crack growth rate and tensile strength of Re modified Inconel 718 produced by means of selective laser melting, Materials Science and Engineering A 698 (October 2016) (2017) 289–301. doi:10.1016/j.msea.2017.05.052.
URL http://dx.doi.org/10.1016/j.msea.2017.05.052

[526] H. Xie, K. Yang, F. Li, C. Sun, Z. Yu, Investigation on the Laves phase formation during laser cladding of IN718 alloy by CA-FE, Journal of Manufacturing Processes 52 (2020) 132–144.

[527] S. Sui, J. Chen, E. Fan, H. Yang, X. Lin, W. Huang, The influence of Laves phases on the high-cycle fatigue behavior of laser additive manufactured Inconel 718, Materials Science and Engineering A 695 (February) (2017) 6–13. doi:10.1016/j.msea.2017.03.098.
URL http://dx.doi.org/10.1016/j.msea.2017.03.098

[528] S. Findlay, N. Harrison, Why aircraft fail, Materials today 5 (11) (2002) 18–25.

[529] S. Gribbin, J. Bicknell, L. Jorgensen, I. Tsukrov, M. Knezevic, Low cycle fatigue behavior of direct metal laser sintered Inconel alloy 718, International Journal of Fatigue 93 (2016) 156–167. doi:10.1016/j.ijfatigue.2016.08.019.
URL http://dx.doi.org/10.1016/j.ijfatigue.2016.08.019

[530] A. S. Johnson, S. Shao, N. Shamsaei, S. M. Thompson, L. Bian, Microstructure, Fatigue Behavior, and Failure Mechanisms of Direct Laser-Deposited Inconel 718, Jom 69 (3) (2017) 597–603. doi:10.1007/s11837-016-2225-2.

[531] S. Shao, M. M. Khonsari, S. Guo, W. J. Meng, N. Li, Overview: Additive Manufacturing Enabled Accelerated Design of Ni-based Alloys for Improved Fatigue Life, Additive Manufacturing 29 (July). doi:10.1016/j.addma.2019.100779.

[532] M. Muhammad, P. Frye, J. Simsiriwong, S. Shao, N. Shamsaei, An investigation into the effects of cyclic strain rate on the high cycle and very high cycle fatigue behaviors of wrought and additively manufactured Inconel 718, International Journal of Fatigue 144 (September 2020) (2021) 106038. doi:10.1016/j.ijfatigue.2020.106038.
URL https://doi.org/10.1016/j.ijfatigue.2020.106038

[533] C. Pei, D. Shi, H. Yuan, H. Li, Assessment of mechanical properties and fatigue performance of a selective laser melted nickel-base superalloy Inconel 718, Materials Science and Engineering A 759 (April) (2019) 278–287. doi:10.1016/j.msea.2019.05.007.
URL https://doi.org/10.1016/j.msea.2019.05.007

[534] H. Mughrabi, R. Wang, K. Differt, U. Essmann, Fatigue crack initiation by cyclic slip irreversibilities in high-cycle fatigue,

in: Fatigue mechanisms: advances in quantitative measurement of physical damage, Vol. 811, ASTM International, 1983, p. 1.

[535] A. S. Johnson, S. Shuai, N. Shamsaei, S. M. Thompson, L. Bian, Fatigue behavior and failure mechanisms of direct laser deposited Inconel 718, Solid Freeform Fabrication 2016: Proceedings of the 27th Annual International Solid Freeform Fabrication Symposium - An Additive Manufacturing Conference, SFF 2016 (2016) 499–511.

[536] P. D. Nezhadfar, A. S. Johnson, N. Shamsaei, Fatigue behavior and microstructural evolution of additively manufactured Inconel 718 under cyclic loading at elevated temperature, International Journal of Fatigue 136 (January) (2020) 105598. doi:10.1016/j.ijfatigue.2020.105598.
URL https://doi.org/10.1016/j.ijfatigue.2020.105598

[537] S. Kim, H. Choi, J. Lee, S. Kim, Room and elevated temperature fatigue crack propagation behavior of Inconel 718 alloy fabricated by laser powder bed fusion, International Journal of Fatigue 140 (June) (2020) 105802. doi:10.1016/j.ijfatigue.2020.105802.
URL https://doi.org/10.1016/j.ijfatigue.2020.105802

[538] R. Konečna, L. Kunz, G. Nicoletto, A. Bača, Long fatigue crack growth in Inconel 718 produced by selective laser melting, International Journal of Fatigue 92 (2016) 499–506. doi:10.1016/j.ijfatigue.2016.03.012.

[539] R. Konečná, G. Nicoletto, L. Kunz, A. Bača, Microstructure and directional fatigue behavior of Inconel 718 produced by selective laser melting, Procedia Structural Integrity 2 (2016) 2381–2388. doi:10.1016/j.prostr.2016.06.298.
URL http://dx.doi.org/10.1016/j.prostr.2016.06.298

[540] G. Osinkolu, G. Onofrio, M. Marchionni, Fatigue crack growth in polycrystalline in 718 superalloy, Materials Science and Engineering: A 356 (1) (2003) 425–433. doi:https://doi.org/10.1016/S0921-5093(03)00156-4.
URL https://www.sciencedirect.com/science/article/pii/S0921509303001564

[541] M. Clavel, A. Pineau, Frequency and wave-form effects on the fatigue crack growth behavior of alloy 718 at 298 k and 823 k, Metallurgical transactions A 9 (4) (1978) 471–480.

[542] W. Ye, J. Akram, L. T. Mushongera, Fatigue Behavior of Additively Manufactured IN718 with Columnar Grains, Advanced Engineering Materials 23 (3). doi:10.1002/adem.202001031.

[543] J. Denny, A. N. Jinoop, C. P. Paul, R. Singh, K. S. Bindra, Fatigue crack propagation behaviour of inconel 718 structures built using directed energy deposition based laser additive manufacturing, Materials Letters 276 (2020) 128241. doi:10.1016/j.matlet.2020.128241.
URL https://doi.org/10.1016/j.matlet.2020.128241

[544] X. Yu, X. Lin, H. Tan, Y. Hu, S. Zhang, F. Liu, H. Yang, W. Huang, Microstructure and fatigue crack growth behavior of Inconel 718 superalloy manufactured by laser directed energy deposition, International Journal of Fatigue 143 (July 2020) (2021) 106005. doi:10.1016/j.ijfatigue.2020.106005.
URL https://doi.org/10.1016/j.ijfatigue.2020.106005

[545] Y. W. Luo, B. Zhang, X. Feng, Z. M. Song, X. B. Qi, C. P. Li, G. F. Chen, G. P. Zhang, Pore-affected fatigue life scattering and prediction of additively manufactured Inconel 718: An investigation based on miniature specimen testing and machine learning approach, Materials Science and Engineering A 802 (December 2020). doi:10.1016/j.msea.2020.140693.

[546] M. M. Kirka, F. Medina, R. Dehoff, A. Okello, Mechanical behavior of post-processed Inconel 718 manufactured through the electron beam melting process, Materials Science and Engineering A 680 (October 2016) (2017) 338–346. doi:10.1016/j.msea.2016.10.069.

[547] S. C. Lee, S. H. Chang, T. P. Tang, H. H. Ho, J. K. Chen, Improvement in the microstructure and tensile properties of inconel 718 superalloy by HIP treatment, Materials Transactions 47 (11) (2006) 2877–2881. doi:10.2320/matertrans.47.2877.

[548] K. Moussaoui, W. Rubio, M. Mousseigne, T. Sultan, F. Rezai, Effects of Selective Laser Melting additive manufacturing parameters of Inconel 718 on porosity, microstructure and mechanical properties, Materials Science and Engineering A 735 (August) (2018) 182–190. doi:10.1016/j.msea.2018.08.037.
URL https://doi.org/10.1016/j.msea.2018.08.037

[549] A. Kaletsch, S. Qin, S. Herzog, C. Broeckmann, Influence of high initial porosity introduced by laser powder bed fusion on the fatigue strength of Inconel 718 after post-processing with hot isostatic pressing, Additive Manufacturing 47 (September) (2021) 102331. doi:10.1016/j.addma.2021.102331.
URL https://doi.org/10.1016/j.addma.2021.102331

[550] W. Tillmann, C. Schaak, J. Nellesen, M. Schaper, M. E. Aydinöz, K. P. Hoyer, Hot isostatic pressing of IN718 components manufactured by selective laser melting, Additive Manufacturing 13 (2017) 93–102. doi:10.1016/j.addma.2016.11.006.
URL http://dx.doi.org/10.1016/j.addma.2016.11.006

[551] S. Tammas-Williams, P. J. Withers, I. Todd, P. Prangnell, Porosity regrowth during heat treatment of hot isostatically pressed additively manufactured titanium components, Scripta Materialia 122 (2016) 72–76.

[552] C. Yu, Z. Huang, Z. Zhang, J. Shen, J. Wang, Z. Xu, Influence of post-processing on very high cycle fatigue resistance of Inconel 718 obtained with laser powder bed fusion, International Journal of Fatigue 153 (August) (2021) 106510.

doi:10.1016/j.ijfatigue.2021.106510.
URL https://doi.org/10.1016/j.ijfatigue.2021.106510

[553] S. Holland, X. Wang, J. Chen, W. Cai, F. Yan, L. Li, Multiscale characterization of microstructures and mechanical properties of Inconel 718 fabricated by selective laser melting, Journal of Alloys and Compounds 784 (2019) 182–194. doi:10.1016/j.jallcom.2018.12.380.
URL https://doi.org/10.1016/j.jallcom.2018.12.380

[554] J. Huang, Z. Huang, H. Du, J. Zhang, Effect of Aging Temperature on Microstructure and Tensile Properties of Inconel 718 Fabricated by Selective Laser Melting, Transactions of the Indian Institute of Metalsdoi:10.1007/s12666-021-02487-0.
URL https://doi.org/10.1007/s12666-021-02487-0

[555] E. Chlebus, K. Gruber, B. Kuźnicka, J. Kurzac, T. Kurzynowski, Effect of heat treatment on the microstructure and mechanical properties of Inconel 718 processed by selective laser melting, Materials Science and Engineering A 639 (2015) 647–655. doi:10.1016/j.msea.2015.05.035.

[556] S. Periane, A. Duchosal, S. Vaudreuil, H. Chibane, A. Morandeau, M. A. Xavior, R. Leroy, Influence of heat treatment on the fatigue resistance of inconel 718 fabricated by selective laser melting (slm), Materials Today: Proceedings 46 (2021) 7860–7865.

[557] A. A. Popovich, V. S. Sufiiarov, I. A. Polozov, E. V. Borisov, Microstructure and mechanical properties of Inconel 718 produced by SLM and subsequent heat treatment, Key Engineering Materials 651-653 (2015) 665–670. doi:10.4028/www.scientific.net/KEM.651-653.665.

[558] Y. Karabulut, Y. Kaynak, S. Sharif, M. A. Suhaimi, Effect of machining and drag finishing on the surface integrity and mechanical properties of Inconel 718 alloys fabricated by laser powder bed fusion additive manufacturing, Materialwissenschaft und Werkstofftechnik 53 (1) (2022) 109–118. doi:10.1002/mawe.202100066.

[559] A. B. Spierings, T. L. Starr, K. Wegener, Fatigue performance of additive manufactured metallic parts, Rapid Prototyping Journal 19 (2) (2013) 88–94. doi:10.1108/13552541311302932.

[560] P. F. Kelley, A. Saigal, J. K. Vlahakis, A. Carter, Tensile and fatigue behavior of direct metal laser sintered (DMLS) inconel 718, in: ASME International Mechanical Engineering Congress and Exposition, Proceedings (IMECE), Vol. 2A-2015, 2015, pp. 1–9. doi:10.1115/IMECE2015-50937.

[561] C. A. Kantzos, R. W. Cunningham, V. Tari, A. D. Rollett, Characterization of metal additive manufacturing surfaces using synchrotron X-ray CT and micromechanical modeling, Computational Mechanics 61 (5) (2018) 575–580. doi:10.1007/s00466-017-1531-z.
URL https://doi.org/10.1007/s00466-017-1531-z

[562] D. B. Witkin, D. Patel, T. V. Albright, G. E. Bean, T. McLouth, Influence of surface conditions and specimen orientation on high cycle fatigue properties of Inconel 718 prepared by laser powder bed fusion, International Journal of Fatigue 132 (August 2019) (2020) 105392. doi:10.1016/j.ijfatigue.2019.105392.
URL https://doi.org/10.1016/j.ijfatigue.2019.105392

[563] H. Y. Wan, Y. W. Luo, B. Zhang, Z. M. Song, L. Y. Wang, Z. J. Zhou, C. P. Li, G. F. Chen, G. P. Zhang, Effects of surface roughness and build thickness on fatigue properties of selective laser melted Inconel 718 at 650 C, International Journal of Fatigue 137 (April). doi:10.1016/j.ijfatigue.2020.105654.

[564] S. Lee, S. Shao, D. N. Wells, M. Zetek, M. Kepka, N. Shamsaei, Fatigue behavior and modeling of additively manufactured IN718: The effect of surface treatments and surface measurement techniques, Journal of Materials Processing Technology 302 (December 2021) (2022) 117475. doi:10.1016/j.jmatprotec.2021.117475.
URL https://doi.org/10.1016/j.jmatprotec.2021.117475

[565] E. Maleki, S. Bagherifard, M. Bandini, M. Guagliano, Surface post-treatments for metal additive manufacturing: Progress, challenges, and opportunities, Additive Manufacturing 37 (May 2020) (2021) 101619. doi:10.1016/j.addma.2020.101619.
URL https://doi.org/10.1016/j.addma.2020.101619

[566] K. Kobayashi, K. Yamaguchi, M. Hayakawa, M. Kimura, Grain size effect on high-temperature fatigue properties of alloy 718, Materials Letters 59 (2-3) (2005) 383–386. doi:10.1016/j.matlet.2004.09.029.

[567] Q. Chen, N. Kawagoishi, H. Nisitani, Evaluation of fatigue crack growth rate and life prediction of Inconel 718 at room and elevated temperatures, Materials Science and Engineering A 277 (1-2) (2000) 250–257. doi:10.1016/S0921-5093(99)00555-9.

[568] D. Gustafsson, J. Moverare, S. Johansson, M. Hörnqvist, K. Simonsson, S. Sjöström, B. Sharifimajda, Fatigue crack growth behaviour of Inconel 718 with high temperature hold times, Procedia Engineering 2 (1) (2010) 1095–1104. doi:10.1016/j.proeng.2010.03.118.
URL http://dx.doi.org/10.1016/j.proeng.2010.03.118

[569] C. Brinkman, G. Korth, Strain fatigue and tensile behavior of inconel 718 from room temperature to 650$^{0}$ c, Tech. rep., Aerojet Nuclear Co., Idaho Falls, Idaho (USA) (1973).

[570] A. S. Johnson, R. Shrestha, P. D. Nezhadfar, N. Shamsaei, Fatigue behavior of laser beam directed energy deposited

inconel 718 at elevated temperature, Solid Freeform Fabrication 2019: Proceedings of the 30th Annual International Solid Freeform Fabrication Symposium - An Additive Manufacturing Conference, SFF 2019 (2019) 584–590.

[571] N. Kawagoishi, Q. Chen, H. Nisitani, Fatigue strength of Inconel 718 at elevated temperatures, Fatigue and Fracture of Engineering Materials and Structures 23 (3) (2000) 209–216.

[572] X. Ma, H. Zhai, L. Zuo, W. Zhang, S. Rui, Q. Han, J. Jiang, C. Li, G. Chen, G. Qian, et al., Fatigue short crack propagation behavior of selective laser melted Inconel 718 alloy by in-situ SEM study: Influence of orientation and temperature, International Journal of Fatigue 139 (May). doi:10.1016/j.ijfatigue.2020.105739.

[573] X. Zhu, C. Gong, Y.-F. Jia, R. Wang, C. Zhang, Y. Fu, S.-T. Tu, X.-C. Zhang, Influence of grain size on the small fatigue crack initiation and propagation behaviors of a nickel-based superalloy at 650 c, Journal of Materials Science & Technology 35 (8) (2019) 1607–1617.

[574] X. Ma, H.-J. Shi, On the fatigue small crack behaviors of directionally solidified superalloy dz4 by in situ sem observations, International journal of fatigue 35 (1) (2012) 91–98.

[575] A. Shyam, J. E. Allison, C. J. Szczepanski, T. M. Pollock, J. W. Jones, Small fatigue crack growth in metallic materials: A model and its application to engineering alloys, Acta Materialia 55 (19) (2007) 6606–6616.

[576] T. Trosch, J. Strößner, R. Völkl, U. Glatzel, Microstructure and mechanical properties of selective laser melted Inconel 718 compared to forging and casting, Materials Letters 164 (2016) 428–431. doi:10.1016/j.matlet.2015.10.136.
URL http://dx.doi.org/10.1016/j.matlet.2015.10.136

[577] Y. L. Kuo, S. Horikawa, K. Kakehi, Effects of build direction and heat treatment on creep properties of Ni-base superalloy built up by additive manufacturing, Scripta Materialia 129 (2017) 74–78. doi:10.1016/j.scriptamat.2016.10.035.
URL http://dx.doi.org/10.1016/j.scriptamat.2016.10.035

[578] Y. Yamashita, T. Murakami, R. Mihara, M. Okada, Y. Murakami, Defect analysis and fatigue design basis for Ni-based superalloy 718 manufactured by selective laser melting, International Journal of Fatigue 117 (April) (2018) 485–495. doi:10.1016/j.ijfatigue.2018.08.002.
URL https://doi.org/10.1016/j.ijfatigue.2018.08.002

[579] Z. Zhou, X. Hua, C. Li, G. Chen, The effect of texture on the low cycle fatigue property of Inconel 718 by selective laser melting, MATEC Web of Conferences 165 (2018) 0–3. doi:10.1051/matecconf/201816502007.

[580] F. Uriati, G. Nicoletto, A comparison of Inconel 718 obtained with three L-PBF production systems in terms of process parameters, as-built surface quality, and fatigue performance, International Journal of Fatigue 162 (April) (2022) 107004. doi:10.1016/j.ijfatigue.2022.107004.
URL https://doi.org/10.1016/j.ijfatigue.2022.107004

[581] X. Hu, Z. Xue, T. Ren, Y. Jiang, C. Dong, F. Liu, On the fatigue crack growth behaviour of selective laser melting fabricated inconel 625: Effects of build orientation and stress ratio, Fatigue & Fracture of Engineering Materials & Structures 43 (4) (2020) 771–787.

[582] Q. Chen, N. Kawagoishi, H. Nisitani, Evaluation of notched fatigue strength at elevated temperature by linear notch mechanics, International Journal of Fatigue 21 (9) (1999) 925–931. doi:10.1016/S0142-1123(99)00081-X.

[583] K. Solberg, J. Torgersen, F. Berto, Fatigue Behaviour of Additively Manufactured Inconel 718 Produced by Selective Laser Melting., Procedia Structural Integrity 13 (2018) 1762–1767. doi:10.1016/j.prostr.2018.12.371.
URL https://doi.org/10.1016/j.prostr.2018.12.371

[584] K. Solberg, F. Berto, The effect of defects and notches in quasi-static and fatigue loading of Inconel 718 specimens produced by selective laser melting, International Journal of Fatigue 137 (March) (2020) 105637. doi:10.1016/j.ijfatigue.2020.105637.
URL https://doi.org/10.1016/j.ijfatigue.2020.105637

[585] D. B. Witkin, D. N. Patel, G. E. Bean, Notched fatigue testing of Inconel 718 prepared by selective laser melting, Fatigue and Fracture of Engineering Materials and Structures 42 (1) (2019) 166–177. doi:10.1111/ffe.12880.

[586] K. Solberg, F. Berto, Notch-defect interaction in additively manufactured Inconel 718, International Journal of Fatigue 122 (2019) 35–45. doi:10.1016/j.ijfatigue.2018.12.021.
URL https://linkinghub.elsevier.com/retrieve/pii/S0142112318307382

[587] R. Konečná, G. Nicoletto, E. Riva, Notch fatigue behavior of Inconel 718 produced by selective laser melting, Procedia Structural Integrity 17 (2019) 138–145. doi:10.1016/j.prostr.2019.08.019.
URL https://doi.org/10.1016/j.prostr.2019.08.019

[588] G. Nicoletto, Smooth and notch fatigue behavior of selectively laser melted Inconel 718 with as-built surfaces, International Journal of Fatigue 128 (July) (2019) 105211. doi:10.1016/j.ijfatigue.2019.105211.
URL https://doi.org/10.1016/j.ijfatigue.2019.105211

[589] J. P. Oliveira, A. D. LaLonde, J. Ma, Processing parameters in laser powder bed fusion metal additive manufacturing, Materials and Design 193 (2020) 1–12. doi:10.1016/j.matdes.2020.108762.

[590] G. P. Dinda, A. K. Dasgupta, J. Mazumder, Laser aided direct metal deposition of Inconel 625 superalloy: Microstructural evolution and thermal stability, Materials Science and Engineering A 509 (1-2) (2009) 98–104. doi:10.1016/j.msea.

[591] G. P. Dinda, A. K. Dasgupta, J. Mazumder, Texture control during laser deposition of nickel-based superalloy, Scripta Materialia 67 (5) (2012) 503–506. doi:10.1016/j.scriptamat.2012.06.014.
URL http://dx.doi.org/10.1016/j.scriptamat.2012.06.014

[592] H. Y. Wan, Z. J. Zhou, C. P. Li, G. F. Chen, G. P. Zhang, Effect of scanning strategy on grain structure and crystallographic texture of Inconel 718 processed by selective laser melting, Journal of Materials Science and Technology 34 (10) (2018) 1799–1804. doi:10.1016/j.jmst.2018.02.002.

[593] B. B. Ravichander, K. Mamidi, V. Rajendran, B. Farhang, A. Ganesh-Ram, M. Hanumantha, N. S. Moghaddam, A. Amerinatanzi, Experimental investigation of laser scan strategy on the microstructure and properties of Inconel 718 parts fabricated by laser powder bed fusion, Materials Characterization 186 (June 2021) (2022) 111765. doi:10.1016/j.matchar.2022.111765.
URL https://doi.org/10.1016/j.matchar.2022.111765

[594] H. Y. Wan, Z. J. Zhou, C. P. Li, G. F. Chen, G. P. Zhang, Effect of scanning strategy on mechanical properties of selective laser melted Inconel 718, Materials Science and Engineering A 753 (December 2018) (2019) 42–48. doi:10.1016/j.msea.2019.03.007.

[595] S. A. Farzadfar, M. J. Murtagh, N. Venugopal, Impact of IN718 bimodal powder size distribution on the performance and productivity of laser powder bed fusion additive manufacturing process, Powder Technology 375 (2020) 60–80. doi:10.1016/j.powtec.2020.07.092.
URL https://doi.org/10.1016/j.powtec.2020.07.092

[596] M. Guzman-Tapia, G. M. Dominguez Almaraz, B. Bermudez Reyes, I. F. Zuñiga Tello, J. A. Ruiz Vilchez, Failure analysis on pre-corroded specimens of Inconel alloy 718, under ultrasonic fatigue tests at room temperature, Engineering Failure Analysis 120 (2021) 105064. doi:10.1016/j.engfailanal.2020.105064.
URL https://linkinghub.elsevier.com/retrieve/pii/S1350630720315880

[597] F. Xu, Y. Lv, Y. Liu, F. Shu, P. He, B. Xu, Microstructural Evolution and Mechanical Properties of Inconel 625 Alloy during Pulsed Plasma Arc Deposition Process, Journal of Materials Science and Technology 29 (5) (2013) 480–488. doi:10.1016/j.jmst.2013.02.010.
URL http://dx.doi.org/10.1016/j.jmst.2013.02.010

[598] Ö. Özgün, H. Özkan Gülsoy, R. Yilmaz, F. Findik, Injection molding of nickel based 625 superalloy: Sintering, heat treatment, microstructure and mechanical properties, Journal of Alloys and Compounds 546 (2013) 192–207. doi:10.1016/j.jallcom.2012.08.069.
URL http://dx.doi.org/10.1016/j.jallcom.2012.08.069

[599] N.-b. Alloy, Z. Tian, C. Zhang, D. Wang, W. Liu, X. Fang, A Review on Laser Powder Bed Fusion of Inconel 625, Applied Scien.

[600] V. Shankar, K. Bhanu Sankara Rao, S. L. Mannan, Microstructure and mechanical properties of Inconel 625 superalloy, Journal of Nuclear Materials 288 (2-3) (2001) 222–232. doi:10.1016/S0022-3115(00)00723-6.

[601] K. D. Ramkumar, W. S. Abraham, V. Viyash, N. Arivazhagan, A. M. Rabel, Investigations on the microstructure, tensile strength and high temperature corrosion behaviour of Inconel 625 and Inconel 718 dissimilar joints, Journal of Manufacturing Processes 25 (2017) 306–322. doi:10.1016/j.jmapro.2016.12.018.
URL http://dx.doi.org/10.1016/j.jmapro.2016.12.018

[602] D. Verdi, M. A. Garrido, C. J. Múnez, P. Poza, Microscale evaluation of laser cladded Inconel 625 exposed at high temperature in air, Materials and Design 114 (2017) 326–338. doi:10.1016/j.matdes.2016.11.014.
URL http://dx.doi.org/10.1016/j.matdes.2016.11.014

[603] M. Karmuhilan, S. Kumanan, A Review on Additive Manufacturing Processes of Inconel 625, Journal of Materials Engineering and Performancedoi:10.1007/s11665-021-06427-3.
URL https://doi.org/10.1007/s11665-021-06427-3

[604] J. Nguejio, F. Szmytka, S. Hallais, A. Tanguy, S. Nardone, M. Godino Martinez, Comparison of microstructure features and mechanical properties for additive manufactured and wrought nickel alloys 625, Materials Science and Engineering A 764 (July) (2019) 138214. doi:10.1016/j.msea.2019.138214.
URL https://doi.org/10.1016/j.msea.2019.138214

[605] A. Kreitcberg, V. Brailovski, S. Turenne, Effect of heat treatment and hot isostatic pressing on the microstructure and mechanical properties of Inconel 625 alloy processed by laser powder bed fusion, Materials Science & Engineering A 689 (January) (2017) 1–10. doi:10.1016/j.msea.2017.02.038.
URL http://dx.doi.org/10.1016/j.msea.2017.02.038

[606] F. A. List, R. R. Dehoff, L. E. Lowe, W. J. Sames, Properties of Inconel 625 mesh structures grown by electron beam additive manufacturing, Materials Science and Engineering A 615 (2014) 191–197. doi:10.1016/j.msea.2014.07.051.
URL http://dx.doi.org/10.1016/j.msea.2014.07.051

[607] W. Yangfan, C. Xizhang, S. Chuanchu, Microstructure and mechanical properties of Inconel 625 fabricated by wire- arc

additive manufacturing, Surface & Coatings Technology 374 (May) (2019) 116–123. doi:10.1016/j.surfcoat.2019.05.079.
URL https://doi.org/10.1016/j.surfcoat.2019.05.079

[608] D. Z. Avery, O. G. Rivera, C. J. Mason, B. J. Phillips, J. B. Jordon, J. Su, N. Hardwick, P. G. Allison, Fatigue Behavior of Solid-State Additive Manufactured Inconel 625, Jom 70 (11) (2018) 2475–2484. doi:10.1007/s11837-018-3114-7.

[609] P. Ganesh, R. Kaul, C. P. Paul, P. Tiwari, S. K. Rai, R. C. Prasad, L. M. Kukreja, Fatigue and fracture toughness characteristics of laser rapid manufactured Inconel 625 structures, Materials Science and Engineering A 527 (29-30) (2010) 7490–7497. doi:10.1016/j.msea.2010.08.034.
URL http://dx.doi.org/10.1016/j.msea.2010.08.034

[610] A. Anam, Microstructure and mechanical properties of selective laser melted superalloy inconel 625 ., Ph.D. thesis, University of Louisville (2018).

[611] A. Yadollahi, N. Shamsaei, S. M. Thompson, A. Elwany, L. Bian, Effects of building orientation and heat treatment on fatigue behavior of selective laser melted 17-4 PH stainless steel, International Journal of Fatigue 94 (2017) 218–235. doi:10.1016/j.ijfatigue.2016.03.014.
URL http://dx.doi.org/10.1016/j.ijfatigue.2016.03.014

[612] A. E. Abela, High cycle fatigue of additively manufactured inconel 625, Ph.D. thesis, Georgia Institute of Technology (2020).

[613] J. R. Poulin, V. Brailovski, P. Terriault, Long fatigue crack propagation behavior of Inconel 625 processed by laser powder bed fusion: Influence of build orientation and post-processing conditions, International Journal of Fatigue 116 (April) (2018) 634–647. doi:10.1016/j.ijfatigue.2018.07.008.
URL https://doi.org/10.1016/j.ijfatigue.2018.07.008

[614] J. R. Poulin, A. Kreitcberg, P. Terriault, V. Brailovski, Long fatigue crack propagation behavior of laser powder bed-fused inconel 625 with intentionally-seeded porosity, International Journal of Fatigue 127 (June) (2019) 144–156. doi:10.1016/j.ijfatigue.2019.06.008.
URL https://doi.org/10.1016/j.ijfatigue.2019.06.008

[615] J. R. Poulin, A. Kreitcberg, V. Brailovski, Effect of hot isostatic pressing of laser powder bed fused Inconel 625 with purposely induced defects on the residual porosity and fatigue crack propagation behavior, Additive Manufacturing 47 (June) (2021) 102324. doi:10.1016/j.addma.2021.102324.
URL https://doi.org/10.1016/j.addma.2021.102324

[616] X. Hu, Z. Xue, T. T. Ren, Y. Jiang, C. L. Dong, F. Liu, On the fatigue crack growth behaviour of selective laser melting fabricated Inconel 625: Effects of build orientation and stress ratio, Fatigue and Fracture of Engineering Materials and Structures 43 (4) (2020) 771–787. doi:10.1111/ffe.13181.

[617] D. Ren, Z. Xue, Y. Jiang, X. Hu, Y. Zhang, Influence of single tensile overload on fatigue crack propagation behavior of the selective laser melting inconel 625 superalloy, Engineering Fracture Mechanics 239 (September 2019) (2020) 107305. doi:10.1016/j.engfracmech.2020.107305.
URL https://doi.org/10.1016/j.engfracmech.2020.107305

[618] J. R. Poulin, A. Kreitcberg, P. Terriault, V. Brailovski, Fatigue strength prediction of laser powder bed fusion processed Inconel 625 specimens with intentionally-seeded porosity: Feasibility study, International Journal of Fatigue 132 (November 2019) (2020) 105394. doi:10.1016/j.ijfatigue.2019.105394.
URL https://doi.org/10.1016/j.ijfatigue.2019.105394

[619] D. B. Witkin, P. Adams, T. Albright, Microstructural evolution and mechanical behavior of nickel-based superalloy 625 made by selective laser melting, Laser 3D Manufacturing II 9353 (March 2015) (2015) 93530B. doi:10.1117/12.2083699.

[620] I. Koutiri, E. Pessard, P. Peyre, O. Amlou, T. De Terris, Influence of SLM process parameters on the surface finish, porosity rate and fatigue behavior of as-built Inconel 625 parts, Journal of Materials Processing Technology 255 (June 2017) (2018) 536–546. doi:10.1016/j.jmatprotec.2017.12.043.
URL https://doi.org/10.1016/j.jmatprotec.2017.12.043

[621] A. Mostafaei, S. H. V. R. Neelapu, C. Kisailus, L. M. Nath, T. D. Jacobs, M. Chmielus, Characterizing surface finish and fatigue behavior in binder-jet 3D-printed nickel-based superalloy 625, Additive Manufacturing 24 (September) (2018) 200–209. doi:10.1016/j.addma.2018.09.012.
URL https://doi.org/10.1016/j.addma.2018.09.012

[622] D. B. Witkin, D. N. Patel, H. Helvajian, L. Steffeney, A. Diaz, Surface Treatment of Powder-Bed Fusion Additive Manufactured Metals for Improved Fatigue Life, Journal of Materials Engineering and Performance 28 (2) (2019) 681–692. doi:10.1007/s11665-018-3732-9.
URL https://doi.org/10.1007/s11665-018-3732-9

[623] K. S. Kim, T. H. Kang, M. E. Kassner, K. T. Son, K. A. Lee, High-temperature tensile and high cycle fatigue properties of inconel 625 alloy manufactured by laser powder bed fusion, Additive Manufacturing 35 (May) (2020) 101377. doi:10.1016/j.addma.2020.101377.

URL https://doi.org/10.1016/j.addma.2020.101377
[624] A. Theriault, L. Xue, J. R. Dryden, Fatigue behavior of laser consolidated IN-625 at room and elevated temperatures, Materials Science and Engineering A 516 (1-2) (2009) 217–225. doi:10.1016/j.msea.2009.03.056.
[625] Y. Zhang, X. A. Hu, Y. Jiang, Study on the Microstructure and Fatigue Behavior of a Laser-Welded Ni-Based Alloy Manufactured by Selective Laser Melting Method, Journal of Materials Engineering and Performance 29 (5) (2020) 2957–2968. doi:10.1007/s11665-020-04844-4.
URL https://doi.org/10.1007/s11665-020-04844-4
[626] W. Philpott, M. A. Jepson, R. C. Thomson, Comparison of the effects of a conventional heat treatment between cast and selective laser melted IN939 alloy, Advances in Materials Technology for Fossil Power Plants - Proceedings from the 8th International Conference (2016) 735–746.
[627] G. Marchese, S. Parizia, A. Saboori, D. Manfredi, M. Lombardi, P. Fino, D. Ugues, S. Biamino, The influence of the process parameters on the densification and microstructure development of laser powder bed fused inconel 939, Metals 10 (7) (2020) 1–19. doi:10.3390/met10070882.
[628] P. Kanagarajah, F. Brenne, T. Niendorf, H. J. Maier, Inconel 939 processed by selective laser melting: Effect of microstructure and temperature on the mechanical properties under static and cyclic loading, Materials Science and Engineering A 588 (2013) 188–195. doi:10.1016/j.msea.2013.09.025.
URL http://dx.doi.org/10.1016/j.msea.2013.09.025
[629] D. Tomus, P. A. Rometsch, M. Heilmaier, X. Wu, Effect of minor alloying elements on crack-formation characteristics of Hastelloy-X manufactured by selective laser melting, Additive Manufacturing 16 (2017) 65–72. doi:10.1016/j.addma.2017.05.006.
URL http://dx.doi.org/10.1016/j.addma.2017.05.006
[630] F. Wang, Mechanical property study on rapid additive layer manufacture Hastelloy X alloy by selective laser melting technology, International Journal of Advanced Manufacturing Technology 58 (5-8) (2012) 545–551. doi:10.1007/s00170-011-3423-2.
[631] Q. Han, R. Mertens, M. L. Montero-Sistiaga, S. Yang, R. Setchi, K. Vanmeensel, B. Van Hooreweder, S. L. Evans, H. Fan, Laser powder bed fusion of Hastelloy X: Effects of hot isostatic pressing and the hot cracking mechanism, Materials Science and Engineering A 732 (July) (2018) 228–239. doi:10.1016/j.msea.2018.07.008.
URL https://doi.org/10.1016/j.msea.2018.07.008
[632] Z. H. Jiao, L. M. Lei, H. C. Yu, F. Xu, R. D. Xu, X. R. Wu, Experimental evaluation on elevated temperature fatigue and tensile properties of one selective laser melted nickel based superalloy, International Journal of Fatigue 121 (May 2018) (2019) 172–180. doi:10.1016/j.ijfatigue.2018.12.024.
[633] P. D. Enrique, A. Keshavarzkermani, R. Esmaeilizadeh, S. Peterkin, H. Jahed, E. Toyserkani, N. Y. Zhou, Enhancing fatigue life of additive manufactured parts with electrospark deposition post-processing, Additive Manufacturing 36 (August) (2020) 101526. doi:10.1016/j.addma.2020.101526.
URL https://doi.org/10.1016/j.addma.2020.101526
[634] R. Esmaeilizadeh, A. Keshavarzkermani, U. Ali, B. Behravesh, A. Bonakdar, H. Jahed, E. Toyserkani, On the effect of laser powder-bed fusion process parameters on quasi-static and fatigue behaviour of Hastelloy X: A microstructure/defect interaction study, Additive Manufacturing 38 (November 2020) (2021) 101805. doi:10.1016/j.addma.2020.101805.
URL https://doi.org/10.1016/j.addma.2020.101805
[635] J. Saarimäki, M. Lundberg, H. Brodin, J. J. Moverare, Dwell-fatigue crack propagation in additive manufactured Hastelloy X, Materials Science and Engineering A 722 (March) (2018) 30–36. doi:10.1016/j.msea.2018.02.091.
URL https://doi.org/10.1016/j.msea.2018.02.091
[636] I. Lopez-Galilea, B. Ruttert, J. He, T. Hammerschmidt, R. Drautz, B. Gault, W. Theisen, Additive manufacturing of CMSX-4 Ni-base superalloy by selective laser melting: Influence of processing parameters and heat treatment, Additive Manufacturing 30 (August) (2019) 100874. doi:10.1016/j.addma.2019.100874.
URL https://doi.org/10.1016/j.addma.2019.100874
[637] C. Körner, M. Ramsperger, C. Meid, D. Bürger, P. Wollgramm, M. Bartsch, G. Eggeler, Microstructure and Mechanical Properties of CMSX-4 Single Crystals Prepared by Additive Manufacturing, Metallurgical and Materials Transactions A: Physical Metallurgy and Materials Science 49 (9) (2018) 3781–3792. doi:10.1007/s11661-018-4762-5.
[638] G. Bi, C. N. Sun, H. chi Chen, F. L. Ng, C. C. K. Ma, Microstructure and tensile properties of superalloy IN100 fabricated by micro-laser aided additive manufacturing, Materials and Design 60 (2014) 401–408. doi:10.1016/j.matdes.2014.04.020.
URL http://dx.doi.org/10.1016/j.matdes.2014.04.020
[639] R. Acharya, S. Das, Additive Manufacturing of IN100 Superalloy Through Scanning Laser Epitaxy for Turbine Engine Hot-Section Component Repair: Process Development, Modeling, Microstructural Characterization, and Process Control, Metallurgical and Materials Transactions A: Physical Metallurgy and Materials Science 46 (9) (2015) 3864–3875. doi:10.1007/s11661-015-2912-6.

[640] A. Basak, S. Das, Additive Manufacturing of Nickel-Base Superalloy IN100 Through Scanning Laser Epitaxy, Jom 70 (1) (2018) 53–59. doi:10.1007/s11837-017-2638-6.

[641] F. Weng, Y. Liu, Y. Chew, X. Yao, S. Sui, C. Tan, F. L. Ng, G. Bi, IN100 Ni-based superalloy fabricated by micro-laser aided additive manufacturing: Correlation of the microstructure and fracture mechanism, Materials Science and Engineering A 788 (March) (2020) 139467. doi:10.1016/j.msea.2020.139467.
URL https://doi.org/10.1016/j.msea.2020.139467

[642] J. Hirsch, T. Al-Samman, Superior light metals by texture engineering: Optimized aluminum and magnesium alloys for automotive applications, Acta Materialia 61 (3) (2013) 818–843.

[643] R. Allavikutty, P. Gupta, T. S. Santra, J. Rengaswamy, Additive manufacturing of mg alloys for biomedical applications: Current status and challenges, Current Opinion in Biomedical Engineering (2021) 100276.

[644] Y. Li, H. Jahr, X. Zhang, M. Leeflang, W. Li, B. Pouran, F. Tichelaar, H. Weinans, J. Zhou, A. Zadpoor, Biodegradation-affected fatigue behavior of additively manufactured porous magnesium, Additive Manufacturing 28 (2019) 299–311. doi:10.1016/j.addma.2019.05.013.

[645] Q. Deng, Y. Wu, Q. Wu, Y. Xue, Y. Zhang, L. Peng, W. Ding, Microstructure evolution and mechanical properties of a high-strength mg-10gd-3y-1zn-0.4 zr alloy fabricated by laser powder bed fusion, Additive Manufacturing (2021) 102517 doi:10.1016/j.addma.2021.102517.

[646] B. Köhler, H. Bomas, W. Leis, L. Kallien, Endurance limit of die-cast magnesium alloys am50hp and az91hp depending on type and size of internal cavities, International journal of fatigue 44 (2012) 51–60. doi:10.1016/j.ijfatigue.2012.05.011.

[647] Y. Wang, H. Huang, G. Jia, H. Zeng, G. Yuan, Fatigue and dynamic biodegradation behavior of additively manufactured Mg scaffolds, Acta Biomaterialia doi:10.1016/j.actbio.2021.08.040.

[648] B. Cantor, I. Chang, P. Knight, A. Vincent, Microstructural development in equiatomic multicomponent alloys, Materials Science and Engineering: A 375 (2004) 213–218.

[649] J.-W. Yeh, S.-K. Chen, S.-J. Lin, J.-Y. Gan, T.-S. Chin, T.-T. Shun, C.-H. Tsau, S.-Y. Chang, Nanostructured high-entropy alloys with multiple principal elements: novel alloy design concepts and outcomes, Advanced Engineering Materials 6 (5) (2004) 299–303. doi:10.1002/adem.200300567.

[650] Y. Ye, Q. Wang, J. Lu, C. Liu, Y. Yang, High-entropy alloy: challenges and prospects, Materials Today 19 (6) (2016) 349–362. doi:10.1016/j.mattod.2015.11.026.

[651] Y.-K. Kim, M.-S. Baek, S. Yang, K.-A. Lee, In-situ formed oxide enables extraordinary high-cycle fatigue resistance in additively manufactured cocrfemnni high-entropy alloy, Additive Manufacturing 38 (2021) 101832.

[652] K. Liu, S. Nene, M. Frank, S. Sinha, R. Mishra, Metastability-assisted fatigue behavior in a friction stir processed dual-phase high entropy alloy, Materials Research Letters 6 (11) (2018) 613–619. doi:10.1080/21663831.2018.1523240.

[653] P. Agrawal, S. Thapliyal, S. S. Nene, R. S. Mishra, B. A. McWilliams, K. C. Cho, Excellent strength-ductility synergy in metastable high entropy alloy by laser powder bed additive manufacturing, Additive Manufacturing 32 (2020) 101098. doi:10.1016/J.ADDMA.2020.101098.

[654] P. Agrawal, R. S. Haridas, S. Thapliyal, S. Yadav, R. S. Mishra, B. A. McWilliams, K. C. Cho, Metastable high entropy alloys: An excellent defect tolerant material for additive manufacturing, Materials Science and Engineering: A 826 (2021) 142005. doi:10.1016/J.MSEA.2021.142005.

[655] Y. O. Kuzminova, D. G. Firsov, S. A. Dagesyan, S. D. Konev, S. N. Sergeev, A. P. Zhilyaev, M. Kawasaki, I. S. Akhatov, S. A. Evlashin, Fatigue behavior of additive manufactured CrFeCoNi medium-entropy alloy, Journal of Alloys and Compounds 863 (2021) 158609. doi:10.1016/J.JALLCOM.2021.158609.

[656] M. A. Hemphill, T. Yuan, G. Y. Wang, J. W. Yeh, C. W. Tsai, A. Chuang, P. K. Liaw, Fatigue behavior of Al0.5CoCrCuFeNi high entropy alloys, Acta Materialia 60 (16) (2012) 5723–5734. doi:10.1016/J.ACTAMAT.2012.06.046.

[657] T. Fujieda, H. Shiratori, K. Kuwabara, T. Kato, K. Yamanaka, Y. Koizumi, A. Chiba, First demonstration of promising selective electron beam melting method for utilizing high-entropy alloys as engineering materials, Materials Letters 159 (2015) 12–15. doi:10.1016/j.matlet.2015.06.046.

[658] K. Kuwabara, H. Shiratori, T. Fujieda, K. Yamanaka, Y. Koizumi, A. Chiba, Mechanical and corrosion properties of alcocrfeni high-entropy alloy fabricated with selective electron beam melting, Additive Manufacturing 23 (2018) 264–271. doi:10.1016/j.addma.2018.06.006.

[659] S. Chen, Y. Tong, P. K. Liaw, Additive manufacturing of high-entropy alloys: a review, Entropy 20 (12) (2018) 937. doi:10.3390/e20120937.

[660] F. Bahadur, K. Biswas, N. Gurao, Micro-mechanisms of microstructural damage due to low cycle fatigue in cocufemnni high entropy alloy, International Journal of Fatigue 130 (2020) 105258. doi:https://doi.org/10.1016/j.ijfatigue.2019.105258.
URL https://www.sciencedirect.com/science/article/pii/S0142112319303627

[661] Z. Tang, T. Yuan, C.-W. Tsai, J.-W. Yeh, C. D. Lundin, P. K. Liaw, Fatigue behavior of a wrought al0.5cocrcufeni two-phase high-entropy alloy, Acta Materialia 99 (2015) 247–258. doi:https://doi.org/10.1016/j.actamat.2015.07.004.

URL https://www.sciencedirect.com/science/article/pii/S1359645415004668

[662] P. Chen, C. Lee, S.-Y. Wang, M. Seifi, J. J. Lewandowski, K. A. Dahmen, H. Jia, X. Xie, B. Chen, J.-w. Yeh, et al., Fatigue behavior of high-entropy alloys: A review, Science China Technological Sciences 61 (2) (2018) 168–178. doi:10.1007/s11431-017-9137-4.

[663] K. Liu, S. Nene, M. Frank, S. Sinha, R. Mishra, Extremely high fatigue resistance in an ultrafine grained high entropy alloy, Applied Materials Today 15 (2019) 525–530. doi:10.1016/j.apmt.2019.04.001.

[664] S. Thapliyal, S. S. Nene, P. Agrawal, T. Wang, C. Morphew, R. S. Mishra, B. A. McWilliams, K. C. Cho, Damage-tolerant, corrosion-resistant high entropy alloy with high strength and ductility by laser powder bed fusion additive manufacturing, Additive Manufacturing 36 (2020) 101455. doi:10.1016/j.addma.2020.101455.

[665] Y. Tian, S. Sun, H. Lin, Z. Zhang, Fatigue behavior of cocrfemnni high-entropy alloy under fully reversed cyclic deformation, Journal of materials science & technology 35 (3) (2019) 334–340. doi:10.1016/j.jmst.2018.09.068.

[666] Z. Tong, X. Ren, J. Jiao, W. Zhou, Y. Ren, Y. Ye, E. A. Larson, J. Gu, Laser additive manufacturing of fecrcomnni high-entropy alloy: Effect of heat treatment on microstructure, residual stress and mechanical property, Journal of Alloys and Compounds 785 (2019) 1144–1159.

[667] Y. O. Kuzminova, E. A. Kudryavtsev, J.-K. Han, M. Kawasaki, S. A. Evlashin, Phase and structural changes during heat treatment of additive manufactured crfeconi high-entropy alloy, Journal of Alloys and Compounds 889 (2022) 161495. doi:10.1016/j.jallcom.2021.161495.

[668] J. Joseph, P. Hodgson, T. Jarvis, X. Wu, N. Stanford, D. M. Fabijanic, Effect of hot isostatic pressing on the microstructure and mechanical properties of additive manufactured alxcocrfeni high entropy alloys, Materials Science and Engineering: A 733 (2018) 59–70. doi:10.1016/j.msea.2018.07.036.

[669] J.-T. Liang, K.-C. Cheng, S.-H. Chen, Effect of heat treatment on the phase evolution and mechanical properties of atomized alcocrfeni high-entropy alloy powders, Journal of Alloys and Compounds 803 (2019) 484–490. doi:10.1016/j.jallcom.2019.06.301.

[670] N. Kalentics, E. Boillat, P. Peyre, S. Ćirić-Kostić, N. Bogojević, R. E. Logé, Tailoring residual stress profile of selective laser melted parts by laser shock peening, Additive Manufacturing 16 (2017) 90–97.

[671] N. Kalentics, M. O. V. de Seijas, S. Griffiths, C. Leinenbach, R. E. Loge, 3d laser shock peening–a new method for improving fatigue properties of selective laser melted parts, Additive Manufacturing 33 (2020) 101112.

[672] A. Yadollahi, M. Mahtabi, A. Khalili, H. Doude, J. Newman Jr, Fatigue life prediction of additively manufactured material: Effects of surface roughness, defect size, and shape, Fatigue & Fracture of Engineering Materials & Structures 41 (7) (2018) 1602–1614. doi:10.1111/ffe.12799.

[673] C. Zhang, K. Feng, H. Kokawa, B. Han, Z. Li, Cracking mechanism and mechanical properties of selective laser melted CoCrFeMnNi high entropy alloy using different scanning strategies, Materials Science and Engineering A 789 (February) (2020) 139672. doi:10.1016/j.msea.2020.139672.
URL https://doi.org/10.1016/j.msea.2020.139672

[674] M. Jin, A. Piglione, B. Dovgyy, E. Hosseini, P. A. Hooper, S. R. Holdsworth, M. S. Pham, Cyclic plasticity and fatigue damage of CrMnFeCoNi high entropy alloy fabricated by laser powder-bed fusion, Additive Manufacturing 36 (2020) 101584. doi:10.1016/J.ADDMA.2020.101584.

[675] Z. Sun, X. Tan, M. Descoins, D. Mangelinck, S. Tor, C. Lim, Revealing hot tearing mechanism for an additively manufactured high-entropy alloy via selective laser melting, Scripta Materialia 168 (2019) 129–133. doi:https://doi.org/10.1016/j.scriptamat.2019.04.036.
URL https://www.sciencedirect.com/science/article/pii/S1359646219302477

[676] S. Luo, P. Gao, H. Yu, J. Yang, Z. Wang, X. Zeng, Selective laser melting of an equiatomic AlCrCuFeNi high-entropy alloy: Processability, non-equilibrium microstructure and mechanical behavior, Journal of Alloys and Compounds 771 (2019) 387–397. doi:10.1016/j.jallcom.2018.08.290.
URL https://doi.org/10.1016/j.jallcom.2018.08.290

[677] Y. Wang, R. Li, P. Niu, Z. Zhang, T. Yuan, J. Yuan, K. Li, Microstructures and properties of equimolar AlCoCrCuFeNi high-entropy alloy additively manufactured by selective laser melting, Intermetallics 120 (February) (2020) 106746. doi:10.1016/j.intermet.2020.106746.
URL https://doi.org/10.1016/j.intermet.2020.106746

[678] Z. Sun, X. Tan, C. Wang, M. Descoins, D. Mangelinck, S. B. Tor, E. A. Jägle, S. Zaefferer, D. Raabe, Reducing hot tearing by grain boundary segregation engineering in additive manufacturing: example of an AlxCoCrFeNi high-entropy alloy, Acta Materialia 204 (2021) 116505. doi:10.1016/j.actamat.2020.116505.
URL https://doi.org/10.1016/j.actamat.2020.116505

[679] H. Peng, S. Xie, P. Niu, Z. Zhang, T. Yuan, Z. Ren, X. Wang, Y. Zhao, R. Li, Additive manufacturing of Al0.3CoCrFeNi high-entropy alloy by powder feeding laser melting deposition, Journal of Alloys and Compounds 862 (2021) 158286. doi:10.1016/j.jallcom.2020.158286.
URL https://doi.org/10.1016/j.jallcom.2020.158286

[680] Y. Cai, L. Zhu, Y. Cui, J. Han, Manufacturing of FeCoCrNi + FeCoCrNiAl laminated high-entropy alloy by laser melting deposition (LMD), Materials Letters 289 (2021) 129445. doi:10.1016/j.matlet.2021.129445.
URL https://doi.org/10.1016/j.matlet.2021.129445

[681] P. Niu, R. Li, S. Zhu, M. Wang, C. Chen, T. Yuan, Hot cracking, crystal orientation and compressive strength of an equimolar CoCrFeMnNi high-entropy alloy printed by selective laser melting, Optics and Laser Technology 127 (July 2019) (2020) 106147. doi:10.1016/j.optlastec.2020.106147.
URL https://doi.org/10.1016/j.optlastec.2020.106147

[682] Y. Chew, G. J. Bi, Z. G. Zhu, F. L. Ng, F. Weng, S. B. Liu, S. M. Nai, B. Y. Lee, Microstructure and enhanced strength of laser aided additive manufactured CoCrFeNiMn high entropy alloy, Materials Science and Engineering A 744 (December 2018) (2019) 137–144. doi:10.1016/j.msea.2018.12.005.

[683] A. Amar, J. Li, S. Xiang, X. Liu, Y. Zhou, G. Le, X. Wang, F. Qu, S. Ma, W. Dong, Q. Li, Additive manufacturing of high-strength CrMnFeCoNi-based High Entropy Alloys with TiC addition, Intermetallics 109 (December 2018) (2019) 162–166. doi:10.1016/j.intermet.2019.04.005.
URL https://doi.org/10.1016/j.intermet.2019.04.005

[684] M. Song, R. Zhou, J. Gu, Z. Wang, S. Ni, Y. Liu, Nitrogen induced heterogeneous structures overcome strength-ductility trade-off in an additively manufactured high-entropy alloy, Applied Materials Today 18 (2020) 1–6. doi:10.1016/j.apmt.2019.100498.

[685] S. Gorsse, C. Hutchinson, M. Gouné, R. Banerjee, Additive manufacturing of metals: a brief review of the characteristic microstructures and properties of steels, ti-6al-4v and high-entropy alloys, Science and Technology of advanced MaTerialS 18 (1) (2017) 584–610. doi:10.1080/14686996.2017.1361305.

[686] Y. Min, C. Ke, L. Chenguang, Z. Liucheng, Y. Yangyiwei, Y. Xin, X. Baixiang, Computational study of evolution and fatigue dispersity of microstructures by additive manufacturing, Chinese Journal of Theoretical and Applied Mechanic 53 (12) (2021) 3265–3275. doi:10.6052/0459-1879-21-389.

[687] P. Liu, Z. Wang, Y. Xiao, R. A. Lebensohn, Y. Liu, M. F. Horstemeyer, X. Cui, L. Chen, Integration of phase-field model and crystal plasticity for the prediction of process-structure-property relation of additively manufactured metallic materials, International Journal of Plasticity 128 (2020) 102670. doi:10.1016/j.ijplas.2020.102670.

[688] M. Yi, K. Chang, C. Liang, Z. Liucheng, Y. Yangyiwei, X. Yi, B. Xu, Computational study of evolution and fatigue dispersity of microstructures by additive manufacturing, Chinese Journal of Theoretical and Applied Mechanics 53 (12) (2021) 3265–3275. doi:10.6052/0459-1879-21-389.

[689] S. Afazov, A. Serjouei, G. J. Hickman, R. Mahal, D. Goy, I. Mitchell, Defect-based fatigue model for additive manufacturing, Progress in Additive Manufacturing (2022) 1–8.

[690] Z. Zhan, W. Hu, Q. Meng, Data-driven fatigue life prediction in additive manufactured titanium alloy: A damage mechanics based machine learning framework, Engineering Fracture Mechanics 252 (April) (2021) 107850. doi:10.1016/j.engfracmech.2021.107850.
URL https://doi.org/10.1016/j.engfracmech.2021.107850

[691] H. Bao, S. Wu, Z. Wu, G. Kang, X. Peng, P. J. Withers, A machine-learning fatigue life prediction approach of additively manufactured metals, Engineering Fracture Mechanics 242 (2021) 107508.